# Analiza regenerării pădurii

## Perspective statistice și informatice

Ciprian Palaghianu

# Cuvânt înainte

*„În cercetarea științifică modelele sunt uneori atât de întinse la limită încât cedează. Când se întâmplă acest lucru, trebuie să fim satisfăcuți întrucât se generează **cunoaștere**. Modelul este apoi modificat și forțat din nou până când cedează în alt mod.”*

(Garcia, 2008)

Lucrarea de față se bazează pe rezultatele cercetărilor efectuate în perioada de pregătire a tezei de doctorat și are ca temă aspecte privitoare la analiza regenerării arboretelor prin mijloace statistice și informatice.

Desigur, teza de doctorat a fost finalizată acum mai bine de cinci ani, dar efervescența acelor observații și investigații nu s-a stins, astfel că această carte conține atât fragmente din teza de doctorat cât și informații care s-au sedimentat și consolidat în următorii ani.

Tema se înscrie în tendințele actuale manifestate în cercetarea silvică, fiind propuse metode de analiză a regenerării arboretelor care integrează tehnici informatice și statistice moderne ce oferă posibilități superioare de investigare, prelucrare, interpretare și prezentare a datelor și rezultatelor.

Doresc să mulțumesc în primul rând domnului profesor universitar dr. ing. Ioan MILESCU, conducătorul științific al lucrării de doctorat pentru faptul că a acceptat să-mi ghideze pașii spre cercetarea silvică. Sprijinul manifestat prin sfaturi, îndemnuri și încurajări atunci când nu adoptam soluția cea mai convenabilă de urmat m-au ajutat să găsesc cu ajutorul domniei sale soluții la problemele ivite pe parcursul conceperii tezei. Răbdarea și înțelegerea de care a dat dovadă au constituit de asemenea un sprijin important, mobilizându-mă să continui munca în momentele dificile.

În aceeași măsură adresez mulțumiri domnului prof. univ. dr. ing. Radu CENUȘĂ, nu doar pentru sugestiile utile în finalizarea lucrării ci și pentru faptul ca de-a lungul anilor a fost pentru mine, din punct de vedere profesional, un adevărat părinte. De asemenea, pe parcursul carierei didactice am fost întotdeauna sprijinit de doamna prof. univ. dr. ing. Filofteia NEGRUȚIU, față de care îmi exprim toată considerația și recunoștința.

Mulțumesc prietenilor și colegilor mei care m-au ajutat la culegerea datelor de teren și ulterior mi-au oferit indicații utile în conceperea lucrării de doctorat: șef lucr. dr. ing. Ionuț BARNOAIEA și conf. dr. ing. Florin CLINOVSCHI.

Apreciez, de asemenea, observațiile și sprijinul din partea distinșilor mei colegi: conf. dr. ing. Sergiu Andrei HORODNIC, decanul Facultății de Silvicultură Suceava, conf. dr. ing. Marian DRĂGOI, conf. dr. ing. Ovidiu IACOBESCU, conf. dr. ing. Laura BOURIAUD, șef de lucrări dr. ing. Daniel AVĂCĂRIȚEI, șef de lucrări ing. dr. Alexei SAVIN, șef de lucrări dr. ing. Cătălin ROIBU și dr. ing. Marius TEODOSIU.

Activitatea de prelucrare a datelor a fost mult ușurată de utilizarea unor aplicații informatice și sunt îndatorat autorilor acestora pentru faptul că mi-au furnizat aceste soluții software și mi-au permis folosirea lor: domnului cercetător principal gr. I dr. ing. Ionel POPA (*SilvaSTAT*), domnului George PERRY (*SpPack*) și domnului Peter HAASE (*SPPA*).

În final, dar nu în ultimul rând, mulțumesc soției mele, Delia, pentru înțelegerea manifestată, sprijinul și încurajările permanente și prietenului și fiului meu, David, cu siguranța un viitor cercetător.

Ciprian Palaghianu,

Suceava, 1 octombrie 2015

# CUPRINS

# PARTEA I
## ASPECTE TEORETICE ÎN RELAŢIA PĂDURE - CIBERNETICĂ

# Capitolul 1
# Considerații referitoare la regenerarea pădurii

## 1.1. Regenerarea pădurii - legătură între generații succesive de arbori și arborete

Pădurea reprezintă o resursă naturală remarcabilă pentru om și societate. Importanța și valoarea acesteia este sporită de faptul că față de alte resurse naturale ca petrolul, gazele naturale, cărbunii, pădurea se caracterizează prin capacitatea de reproducere, ca urmare a unui proces natural de reînnoire a generațiilor de arbori.

Permanența pădurilor, în scopul receptării continue a beneficiilor izvorâte din funcțiile de producție și protecție atribuite, presupune o perpetuă înnoire a biocenozei forestiere la nivelul indivizilor. Viața limitată, deși îndelungată, a arborilor determină o înlocuire treptată a acestora pentru ca biocenoza în ansamblul ei să-și păstreze, între anumite limite, structura în scopul realizării funcțiilor sale. Așadar, viața pădurii ca întreg nu este determinată de durata de viață a elementelor ce o compun.

Reînnoirea în cazul ecosistemelor forestiere este un produs natural, dictat de legile firii, ce se produce continuu sau periodic. Similar acestui proces, în pădurea cultivată vorbim despre regenerarea arboretelor, care presupune înlocuirea arborilor ajunși la o anumită vârstă, ce se extrag prin tăiere, cu exemplare tinere obținute prin procese generative sau vegetative.

Procesul de înlocuire a vechii generații de arbori printr-una nouă, tânără, este cunoscut sub numele de regenerare. Regenerarea pădurii derivă din proprietatea fundamentală, generală a lumii vegetale și animale de autoreproducere.

Dată fiind importanța regenerării arboretelor, silvicultorul trebuie să acționeze cu maximă responsabilitate în vederea conducerii acestui proces deoarece acesta condiționează existența arboretului și implicit realizarea eficientă a funcțiilor sale. Momentul acesta, de maximă importanță în viața pădurii, trebuie pregătit cu multă grijă și presupune analize atente atât în ceea ce privește caracteristicile și exigențele speciilor cu care se lucrează cât și în ceea ce privește însușirile stațiunii care asigură suportul fizic al regenerării.

Acțiunea de regenerare nu presupune în mod neapărat intervenția umană. În pădurile virgine acest proces s-a derulat și se derulează în mod natural, sub influența exclusivă a factorilor naturali. Regenerarea se produce în acest caz în mod neregulat în timp și spațiu, în funcție de apariția unor condiții favorabile. Procesul se desfășoară lent, extinderea vegetației forestiere este întâmplătoare, la fel ca și structura din punct de vedere al compoziției și consistenței.

Evoluția unui arboret este dictată la un moment dat de două procese ce se derulează concomitent și oarecum antagonic: un proces negativ de slăbire fiziologică a elementelor mature și un altul pozitiv, de instalare și ulterior de dezvoltare a unor noi semințișuri. Prin urmare, premisele declanșării procesului de regenerare naturală sunt asigurate de apariția semnelor de slăbire fiziologică a pădurii, combinată cu procesul de fructificație al arborilor.

Studiul regenerării în pădurea virgină permite decelarea legităților după care se realizează regenerarea în condiții naturale. Cunoașterea acestor legități naturale ghidează silvicultorii în acțiunea de regenerare a pădurii cultivate în direcția unei aplicări corecte a diverselor metode tehnice de lucru. În pădurea cultivată acest lucru nu presupune imitarea modului de regenerare al pădurii virgine, ci desprinderea unor concluzii utile, cu valabilitate generală. Modul de acțiune al naturii contravine de foarte multe ori intereselor silvicultorului care dorește ca acest proces să se realizeze cât mai repede, cât mai uniform și cu o reușită cât mai bună. Se desprinde astfel ideea că, în pădurea cultivată, acțiunea aleatoare a naturii este înlocuită cu intervenția conștientă, planificată și sistematică a silvicultorului.

Intervenția specialistului presupune determinarea momentului și modalității de întrerupere a producției arboretului bătrân, concomitent cu adoptarea unor măsuri care să favorizeze instalarea și dezvoltarea unei noi generații. Dirijarea procesului de regenerare impune reglarea desfășurării în timp a acestuia prin lichidarea arboretului bătrân într-un ritm impus de cerințele dezvoltării semințișurilor instalate cărora trebuie să li se asigure condiții optime de lumină, căldură și umiditate. În opinia cunoscutului silvicultor N. Constantinescu (1973), ″regenerarea în pădurea cultivată încetează a mai fi un proces pur natural″. Se remarcă o nouă idee importantă, că în pădurea cultivată regenerarea nu se mai produce în urma eliminării ca proces natural, ci este o consecință firească a unui proces artificial necesar: exploatarea arborilor.

În pădurea cultivată se deosebesc două tipuri de intervenții în ceea ce privește regenerarea: regenerarea pe cale naturală și regenerarea artificială.

Regenerarea naturală presupune obținerea în urma efectuării de tăieri de regenerare a unei noi generații din sămânța diseminată de arbori sau pe cale vegetativă. Se poate evidenția că ceea ce numim regenerare naturală în pădurea cultivată ″este un proces influențat puternic de om, dirijat de acesta″ (Constantinescu, 1973). Denumirea de regenerare ″naturală″ este în acest caz convențională, subliniind asemănarea cu procesul natural de regenerare. Este însă evidentă antropizarea acestui proces prin dirijarea convenabilă a regenerării din punct de vedere temporal (privitor la momentul în care se realizează), spațial (privitor la locația și modul în care avansează), compozițional etc. Silvicultorul este parte activă a acestui proces prin efectuarea unor lucrări de stimulare a fructificației, de favorizare a instalării și dezvoltării semințișului, de promovare a anumitor specii sau de eliminare de la regenerare a altora.

Regenerarea în pădurea cultivată se diferențiază de regenerarea în pădurea virgină atât prin ritm cât și prin calitatea arboretului obținut. În multe cazuri productivitatea arboretelor din pădurea virgină este inferioară celei din arboretele pădurii cultivate, iar sortimentele de lemn obținute nu corespund de cele mai multe ori nevoilor economice ale societății.

Apar situații în care, din diverse motive, regenerarea naturală nu este posibilă sau nu este avantajoasă din punct de vedere economic, caz în care se recurge la cealaltă modalitate de regenerare a pădurii cultivate: regenerarea artificială. Aceasta presupune instalarea unui nou arboret cu material de reproducere produs după o tehnologie specifică. Instalarea se poate face artificial prin semănături directe, prin plantații sau prin butășiri directe.

Capacitatea naturală a arborilor de a se reproduce conduce la ideea firească potrivit căreia, în mod normal, perpetuarea pădurilor se asigură prin regenerarea naturală. În acest context, regenerarea artificială apare ca o acțiune de refacere a unui dezechilibru provocat de diverse calamități naturale sau de intervenții antropice neinspirate și păgubitoare. În multe cazuri impactul dezgolirii unor mari suprafețe (din dorința de a concentra tăierile) a condus la înlocuirea regenerării naturale cu cea artificială. Rucăreanu (1962) susține că ″*parchetația reprezintă o metodă de amenajare cu totul necorespunzătoare,* […] *pentru că posibilitatea pe suprafață obligă la tăieri rase. În aceste condiții regenerarea pe cale naturală este practic exclusă…*″.

Desigur, regenerarea artificială nu este doar consecința exploatării în general, ci este de multe ori impusă de efectele negative generate de exploatarea nerațională. Extragerile concentrate, reducerea puternică a consistenței, înțelenirea solului prin pășunat abuziv, înmlăștinarea, vătămarea semințișului natural sau chiar imposibilitatea dezvoltării acestuia au condus treptat la folosirea regenerării artificiale drept unic mijloc de regenerare a anumitor arborete.

Examinarea argumentelor prezentate mai sus, ne permite a evidenția faptul că procesul de regenerare a pădurilor, în general, constituie un țel în sine, de a cărui reușită se leagă așteptările și nevoile societății umane și, desigur, însăși existența și continuitatea în timp a pădurii.

În acest cadru, regenerarea se impune ca o condiție a perenității pădurii și continuității producției de lemn, ca o legătură între generații succesive de arbori.

## 1.2. Evoluția concepțiilor privitoare la regenerarea arboretelor

Evoluția societății omenești a fost în mod evident influențată de existența pădurilor. La început se putea vorbi despre o relație între om, ca individ, și pădure, urmând ca apoi această relație să se maturizeze și să evolueze în relația pragmatică dintre societatea umană și pădure.

Această trecere a pădurii de la obiect al interesului primar individual la statutul de resursă naturală regenerabilă, de interes public a cunoscut numeroase etape care au condus în final la o situație îngrijorătoare.

În prezent se estimează că aproape jumătate din pădurile care existau cu circa 10000 de ani în urmă au dispărut. Potrivit unui studiu elaborat de World Resources Institute 46% din pădurile existente acum 8000 de ani au dispărut (Bryant, 1997). Ba mai mult, aproximativ 75% din acest deficit s-a înregistrat în ultimele două secole (Mygatt, 2006). Practic se estimează că din cele aproximativ 8 miliarde hectare de pădure existente acum 8000 de ani au mai rămas în prezent 3,952 miliarde hectare (FAO, 2007), doar 22% din pădurile inițiale ale globului fiind situate în zone naturale nefragmentate (Bryant, 1997). Până la mijlocul secolului trecut, pădurile tropicale au rămas în mare parte intacte: între anii 1960 și 1990 au fost defrișate aproximativ 450 de milioane de hectare, respectiv 20% din totalitatea pădurilor tropicale (Bryant, 1997).

Declinul suprafețelor ocupate de pădure nu reprezintă o problematică recentă, a ultimilor ani. Istoria ne arată că societatea umană a ales o modalitate greșită de management al resurselor forestiere încă de timpuriu. Astfel, până în anul 1000 î.H. majoritatea pădurilor Chinei de Est au fost defrișate în vederea extinderii plantațiilor agricole. În Europa, acum circa 2000 de ani grecii și romanii au distrus majoritatea suprafețelor ocupate de pădure din zona mediteraneană, iar în Evul Mediu această situație s-a agravat în toată Europa Occidentală.

La nivelul continentelor se apreciază că procentul suprafeței ocupate de pădure în prezent față de suprafața inițială este destul de scăzut (Bryant, 1997):

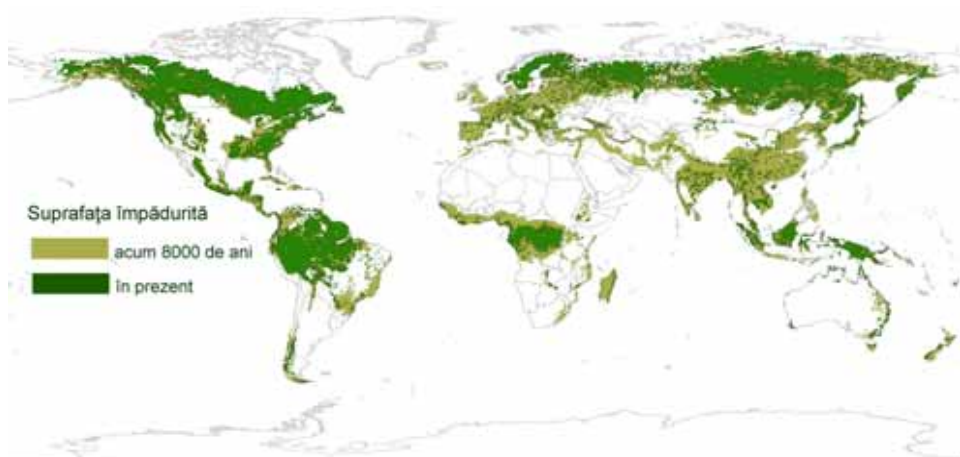

**Figura 1.1. Dinamica reducerii pădurii la nivel mondial**
(adaptare după hărțile oferite on-line de *WRI*)

Asia păstrează doar 25% din pădurile inițiale, Africa 34%, Europa cu Federația Rusă 58%, America de Nord si Centrală 75%, America de Sud 70% iar Oceania 65%.

Revenind la istoricul relației dintre om și pădure, primele interacțiuni nu vizau în mod prioritar obținerea de material lemnos ci erau axate pe un consum al produselor accesorii ale pădurii. Omul primitiv folosea pădurea ca pe un mijloc de a-și procura hrana. Dezvoltarea societății umane a făcut ca pădurea să fie supusă primelor presiuni. Expansiunea terenurilor agricole, nevoia de terenuri de pășunat în cazul creșterii animalelor, apariția orașelor și a rețelelor de drumuri, dezvoltarea navigației reprezintă fenomene ce au determinat primele defrișări masive de păduri (Milescu, 1969, 1990).

În funcție de nevoile locale, oamenii extrăgeau arborii care corespundeau dimensional sau calitativ intereselor curente, fără a manifesta nici un interes pentru operația de regenerare a pădurii. În cazul în care extragerile de material lemnos erau relativ dispersate și reduse din punct de vedere cantitativ, pădurea se regenera în mod natural, fără nici un ajutor uman. Procesul de regenerare se caracteriza printr-o dinamică aleatoare în timp și în spațiu. În acest stadiu se poate vorbi despre un așa-zis ″*grădinărit primitiv*″, ce nu constituie încă un mod de gospodărire dar reprezintă primul indiciu de folosire al pădurii ca resursă de material lemnos. De altfel, acest prim mod de folosire a pădurii, a fost îndelung utilizat până în secolele XV-XVI sau în unele regiuni de pe glob, chiar până în timpurile moderne (Milescu, 1990, 1997).

Continua evoluție socială și economică a umanității a determinat o creștere a consumului de lemn. Lemnul era apreciat deja ca un produs, ca o marfă, ceea ce a determinat extrageri intensive, concentrate ce au condus la efecte negative neașteptate: înmlăștinarea sau înțelenirea solului, rărirea excesivă a unor zone, afectarea regenerării naturale prin distrugerea semințișurilor sau crearea unor condiții improprii dezvoltării acestora. În multe din țările europene starea pădurilor a fost agravată și de un pășunat intens sau de o extindere agresivă a zonelor destinate agriculturii.

Degradarea rapidă și accentuată a pădurilor accesibile a condus chiar la dispariția unor zone întinse de pădure. Efectele degradării pădurilor, ca de altfel și dezvoltarea economică rapidă, au determinat confruntarea țărilor europene cu o lipsă a lemnului. Pentru prima oară societatea umană a realizat că păstrarea suprafețelor împădurite este o condiție esențială în asigurarea continuității pădurii ca resursă economică. În această perioadă apar primele reglementări cu caracter de gospodărire a pădurilor care nu urmăreau să asigure regenerarea ci erau mai degrabă axate pe restricționarea exploatărilor. Aceste prime reglementări au limitat evoluția procesului de degradare dar nu au rezolvat problema asigurării continuității pădurii.

Primele idei privind regenerarea pădurilor au apărut în Europa, unde dezvoltarea economică din secolele XVII-XVIII a determinat confruntarea, de altfel inevitabilă, cu o criză a lemnului, ce impunea soluții rapide. Atunci s-a sesizat diferența între lemn ca resursă regenerabilă și celelalte resurse naturale neregenerabile. În urma unei oarecare experiențe forestiere, s-a realizat că pădurea dispune de o caracteristică excepțională și anume capacitatea de regenerare. Astfel s-au cristalizat câteva principii care vizau direct regenerarea pădurilor în scopul asigurării continuității acestei resurse naturale   de valori ecologice și economice deosebite (Milescu, 1982, 2002).

Aceste principii s-au conturat treptat, în baza observațiilor empirice ale fenomenelor ce au loc în urma exploatării unei păduri. A fost sesizată capacitatea de regenerare naturală vegetativă a unor specii ceea ce a dus la folosirea unui prim regim de gospodărire - regimul crângului.

Efectele negative ale tăierilor în crâng au fost observate și s-a trecut la înlocuirea unor cioate epuizate cu exemplare ce proveneau din sămânță care să permită obținerea de lemn gros și să contribuie la realizarea unei regenerări mixte din lăstari și din sămânță.

Dorința de a asigura continuitatea producției de lemn a marcat apariția modului de exploatare în  parchete anuale prin tăieri rase.

Acest pas a determinat folosirea regimului de codru care presupune rezolvarea problemei regenerării prin alte mijloace decât pe cale vegetativă. Regenerarea suprafețelor exploatate se realiza fie natural, prin păstrarea unor arbori seminceri, fie artificial  prin instalarea arboretului prin semănături sau plantații.

Experiența acumulată în aceste operații a determinat chiar aplicarea unor tipuri de regenerare sub masiv.

Începutul secolului al XVIII-lea a găsit deja bine conturate principiile exploatării și regenerării pădurilor, principii concretizate în unele cazuri în diverse reglementări ale vremii - *Ordonanța lui Colbert* din 1669 din Franța sau diverse alte reglementări regionale în Germania (Constantinescu, 1976).

Lucrările și cercetările lui *Tristan de Rostaing – „grand maître des Eaux & Forests de France"* din 1520 (Troup, 1928)*, Varenne de Fenille* în Franța între 1790-1791 (Constantinescu, 1976) sau ale lui *Hartig* (1796) în Prusia sunt deschizătoare de noi orizonturi și au marcat începuturi timide ale silviculturii ca știință. Până atunci se evidențiase strânsa legătură dintre regenerare și exploatare, cele două procese considerându-se două componente ale aceluiași sistem. Se acceptase faptul că tehnicile de regenerare trebuie să țină cont pe cât posibil de eficiența exploatării iar aceasta din urmă să se deruleze în funcție de mersul regenerării. Munca specialiștilor forestieri de la sfârșitul secolului al XVIII-lea și începutul secolului al XIX-lea a introdus un nou element în relația dintre regenerare și exploatare, și anume lucrările de conducere și îngrijire. Astfel s-a încercat realizarea de sisteme integrate de gospodărire a arboretelor care să ofere soluții complete practicii silvice. Această evoluție a fost determinată atât de interesul societății în maximizarea producției de lemn cât și creșterea interesului privind calitatea lemnului.

Eforturile privind elaborarea și implementarea unor concepte cu referire directă la regenerarea arboretelor a unor cercetători merită îndeosebi amintite.

*Hartig* (1796) propunea în lucrarea sa „*Anweisung zur Holtzzucht fur Forster*" ca modul de regenerare a arboretelor să se realizeze prin aplicarea tăierilor succesive. La începutul secolului al XIX-lea, *Cotta* (1821) recomandă în „*Anweisung zum Waldbau*" un sistem de avansare a regenerării care prin distribuția tăierilor în timp și spațiu să asigure o structură normală. Influențați de *Hartig* și *Cotta*, silvicultorii francezi *Lorentz* și *Parade* (1837) dogmatizează ulterior în lucrarea „*Cours elementare de culture des bois*" tratamentul codrului regulat cu tăieri succesive pe care îl numesc ″*la methode de reensemcement natural et des eclaircises periodiques*″. Ei recomandă păstrarea unei consistențe strânse care să evite înțelenirea solului și instalarea semințișurilor înainte de momentul tăierilor de regenerare.

Perioada secolului al XIX-lea a fost una deosebit de prolifică în apariția de noi idei și concepte cu privire la regenerarea pădurilor și în general cu privire la gospodărirea  arboretelor, experiența silvicultorilor fiind lărgită de nenumăratele metode și tehnici de lucru aplicate în care s-au obținut rezultate considerate satisfăcătoare. La baza dezvoltării unor metode de regenerare au stat și noile concepții economice ale respectivei perioade. Astfel, efecte deosebit de nefavorabile asupra gospodăririi pădurilor a avut concepția economică privind renta maximă a solului. Potrivit acestei concepții terenul reprezintă un mijloc de producție ce trebuie să genereze o rentă maximă. Această idee încuraja cultivarea de specii repede crescătoare care să necesite un minim de cheltuieli de îngrijire și care să permită recoltarea cât mai rapidă. În Germania acest mod de gospodărire a pădurilor a fost utilizat un timp, beneficiind și de justificarea sa prin sistemul economic bazat pe exploatabilitatea financiară descris de *Judeich* în 1871 (Dragoi, 2000).

Acest sistem a condus la o gospodărire a pădurilor ce utiliza ca mod de exploatare tăierile rase urmate de o regenerare pe cale artificială a arboretelor. Pentru a putea folosi acest sistem se utilizau ca specii de cultură molidul și pinul, extinzându-se cultura lor în arborete pure. Deceniile ce au urmat au evidențiat consecințele negative ale unei astfel de gospodăriri defectuoase asupra productivității și stabilității pădurilor. O reacție violentă împotriva acestor concepții de gospodărire a avut-o *Gayer* (1868), ce a promovat ideea respectului față de legitățile naturale în practica forestieră. *Gayer* a concretizat aceste idei în conceperea unui nou sistem de regenerare: ″*regenerarea prin tăieri cvasigrădinărite*″.

Anii ce au urmat au marcat și apariția altor metode sau a unor variante, procedee sau pur și simplu îmbunătățiri ale metodelor existente deja. Experiența silvicultorilor s-a îmbogățit în mod continuu, pădurea fiind privită nu doar ca o resursă naturală exploatabilă ci ca un obiectiv de maximă importanță a cărei perpetuare în timp și spațiu este vitală pentru societatea umană.

Regenerarea pădurilor nu mai este considerată drept o consecință a acțiunii de exploatare ci reprezintă o preocupare majoră a silvicultorilor ce dispune de o fundamentare teoretică și metodologică și este rezultatul eforturilor îndelungate ale practicii și cercetării științifice în acest domeniu.

Metodele de regenerare ale pădurii cultivate sunt în mod normal derivate din observațiile asupra modului de regenerare în pădurea virgină dar au evoluat și s-au diversificat în funcție de natura pădurii cultivate, de progresul științific și tehnic, de experiența și profesionalismul silvicultorilor.

Concepțiile privitoare la regenerarea arboretelor au suferit modificări în decursul timpului și chiar în aceeași perioadă au variat de la o regiune la alta în funcție de stadiul dezvoltării sociale și economice și de gradul de dezvoltare al științelor silvice, complexitatea acestor concepții fiind și rezultatul nevoilor societății.

În prezent, suprafața ocupată de pădure la nivel global este de 3,952 miliarde de hectare – adică 30,3% din suprafața uscatului. Repartizarea resurselor forestiere este neuniformă, atât la nivelul continentelor cât și la nivelul statelor. Astfel, sunt 5 state care dețin împreună mai mult de jumătate din resursele forestiere mondiale (53%) – Rusia (808,79 milioane hectare), Brazilia (477,698 milioane hectare), Canada (310,134 milioane hectare), SUA (303,089 milioane hectare) și China (197,29 milioane hectare) (FAO, 2007).

Referitor la dinamica suprafeței ocupate de pădure, rata anuală a despăduririlor la nivel global, în urma exploatării de lemn, a involuat de la 16 milioane hectare în anul 1995 la 13 milioane hectare în anul 2005 (FAO, 2007). În urma eforturilor de regenerare și împădurire se obține un deficit anual de 7,3 milioane hectare. Practic suprafața pădurii la nivel mondial scade anual cu aceasta valoare sau se poate spune că suprafața pădurii pe Terra scade zilnic cu circa 20 de mii de hectare pe zi. În urma ritmului alert al despăduririlor în ultimii 15 ani, suprafața ocupată de păduri s-a diminuat cu 3%. Această tendință este posibil chiar să se accentueze datorită faptului că valoarea consumului de lemn și produse lemnoase crește an de an și este preconizată o dublare a acestuia în următorii 10 ani (Mygatt, 2006).

În ultimii ani - perioada 2000-2005 - se constată un interes sporit acordat gestiunii resurselor lemnoase, fiind remarcate progrese în acest sens la nivelul majorității continentelor. Daca Europa reprezintă în continuare exemplul cel mai elocvent al posibilității de creștere a suprafeței împădurite, America de Sud rămâne singurul continent care nu înregistrează progrese, având un deficit anual de 4,483 milioane hectare, chiar și Africa înregistrând o scădere a deficitului anual la circa 4,040 milioane hectare față de perioada 1995-2000 (FAO, 2003, 2005, 2007).

Europa, are o creștere anuală a suprafețelor ocupate de pădure constantă și considerabilă, suprafața acoperită de pădure crescând în medie cu 0,845 milioane de hectare în perioada 1990-2000 (o rată de creștere de 0,46% pe an), respectiv cu 0,756 milioane de hectare în perioada 2000-2005 (o rată de creștere de 0,40% pe an) (Palaghianu, 2007).

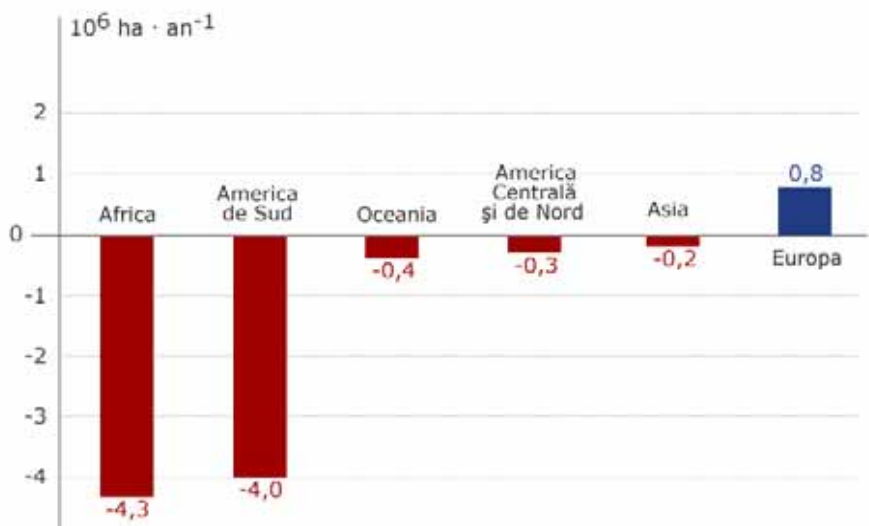

**Figura 1.2. Media anuală a modificării suprafeței acoperite de pădure**
Perioada **1995-2000** (conform datelor FAO, 2003, 2005, 2007)

Europa este continentul care a luat măsurile cele mai drastice și mai timpurii. Programe de refacere a suprafeței împădurite au fost demarate încă din secolul al XVIII-lea. Astfel Danemarca avea în 1810 suprafața împădurită de 4% iar in 2005 de 10%, Elveția 18% în 1810 și 29% în 2005, Franța 14% în 1830 și 27% în 2005 iar Scoția 5% în 1920 și 15% în 2005. Și în prezent există țări extrem de preocupate de creșterea suprafeței împădurite - mai ales în condițiile reducerii suprafețelor agricole. Se înregistrează ritmuri de creștere foarte bune pentru Spania (296 mii ha/an), Italia (peste 100 mii ha/an), precum și Franța, Portugalia, Turcia, Grecia (figura 1.3).

Uniunea Europeană s-a implicat în activitatea de management al resurselor forestiere încă din 1990 prin Politica Comună pentru Agricultură. Din anul 2000, conform directivei 1257/1999 s-a oferit suport pentru dezvoltare rurală, în domeniul forestier acest suport referindu-se la împăduriri și alte măsuri silviculturale. Au fost acordate prin Fondul European de Orientare și Garantare în Agricultură (EAGGF) contribuții financiare importante. Spania a beneficiat de circa 1,5 miliarde de euro (din care 650 milioane de euro doar pentru împăduriri),

Italia de 900 de milioane de euro (550 milioane doar pentru împăduriri), Portugalia de 700 de milioane de euro (Zanchi, 2007). În prezent toate aceste măsuri sunt finanţate prin Fondul European pentru Agricultură şi Dezvoltare Rurală (EARFD), ce a acţionat iniţial în perioada 2007-2013.

O altă tendinţă observată în ultimii ani în ţările dezvoltate, constă în faptul că ritmul alert de pierdere a folosinţei agricole a terenurilor determinat de "declinul rural" şi abandonarea unor terenuri de către agricultură îşi găseşte soluţii prin împădurirea terenurilor neproductive (Milescu, 2002).

Acest transfer al folosinţei terenurilor este chiar  încurajat  de către stat prin diverse facilităţi şi avantaje fiscale. În condiţiile actuale, tot mai adesea se apelează la valenţele pe care le deţine pădurea în rezolvarea crizelor legate de mediu – diminuarea biodiversităţii, încălzirea globală, creşterea concentraţiei de $CO_2$ din atmosferă.

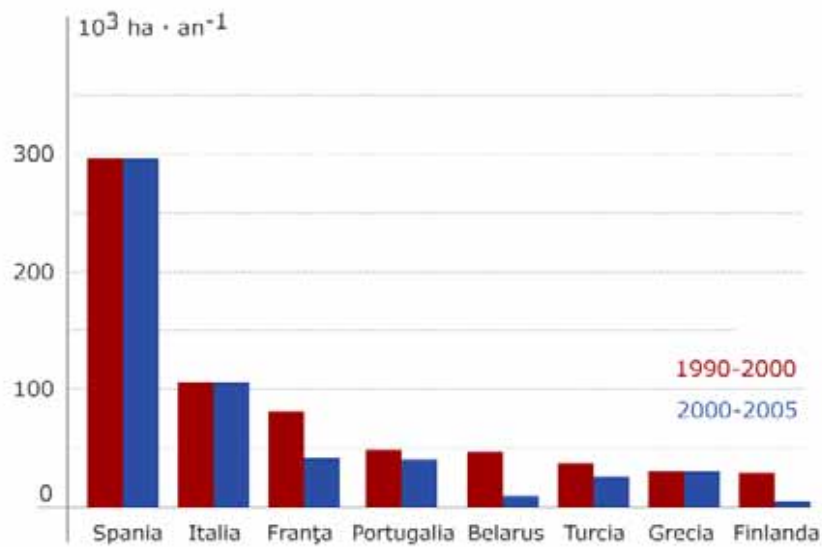

**Figura 1.3. Ritmul anual de creştere al suprafeţei acoperite de pădure în Europa**
(conform datelor FAO, 2003, 2005, 2007)

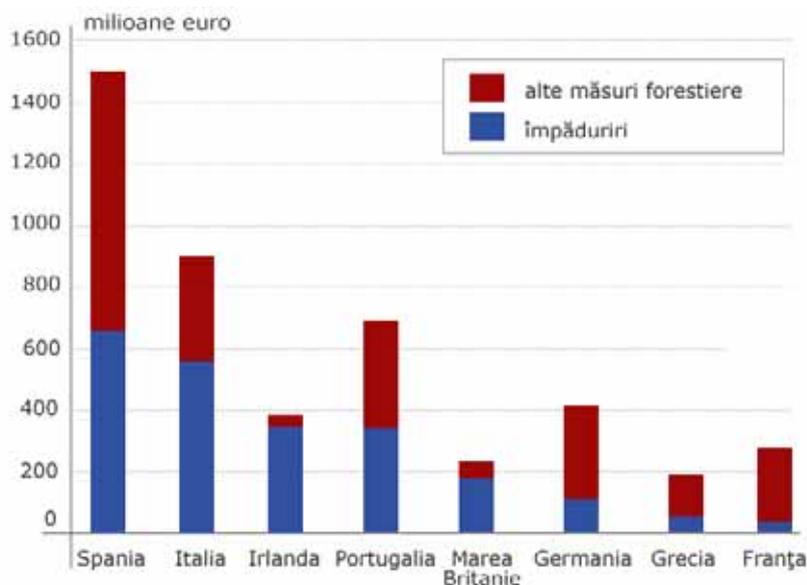

**Figura 1.4. Contribuții acordate de UE prin fondul EAGGF între anii 2000-2006**
(adaptare după Zanchi, 2007)

Regenerarea pădurilor constituie un deziderat ce implică în aceeași măsură acțiunea rațională, conștientă, planificată și sistematică a silvicultorului precum și acțiunea aleatoare dar generatoare de viață a naturii. Regenerarea în arboret este practic o acțiune concertată, o legătură între doi parteneri, o convenție prin care silvicultorul încearcă să dirijeze acest acord cu hazardul în limite acceptate de cerințele economice și culturale ale societății.

Trebuie găsită însă o cale prin care acest parteneriat să funcționeze, în prezent silvicultorul fiind de multe ori nevoit, datorită greșelilor din trecutul mai mult sau mai puțin îndepărtat, să suplinească prin intervențiile sale acțiunea naturii. Pentru că, după cum spunea și G. Pinchot (1905), "*adevărata cale de a salva pădurile nu este reprezentată de plantarea de noi arbori ci de protejarea și folosirea judicioasă a pădurilor existente*".

# Capitolul 2
# Scopul şi obiectivele cercetărilor

## 2.1. Scopul cercetărilor

Pe plan internaţional, în domeniul cercetării, se manifestă tendinţe care arată o aplecare intensă asupra studiului mecanismelor interioare ale proceselor naturale. Pădurea este generoasă în a oferi nenumărate astfel de procese, constituind un veritabil laborator în aer liber în care natura experimentează. Capacitatea deosebită a arborilor de a se reproduce şi implicit capacitatea pădurii de a se perpetua a suscitat în mod constant interesul cercetătorilor, numeroase studii şi lucrări de cercetare încadrându-se în această arie de interes. Lucrarea de faţă se înscrie şi ea în acest arie de interes, vizând problematica regenerării arboretelor.

Scopul cercetărilor a fost reprezentat de identificarea unor particularităţi ale procesului de regenerare, în general, şi a caracteristicilor de structură relaţională din cadrul seminţişurilor rezultate în urma regenerării, în special. S-a acordat un interes sporit procesului competiţional declanşat în fazele iniţiale de constituire a unui nou arboret, competiţia fiind unul din elementele ecologice dominante, active şi definitorii în populaţiile tinere de arbori, modelând structura viitorului arboret.

În aceeaşi măsură, analizele au urmărit să determine anumite tipare ale modului de distribuţie în spaţiu a indivizilor, relaţiile intra şi interspecifice conducând în timp la realizarea anumitor modele de organizare. S-a dorit identificarea acelor atribute ale seminţişului instalat pe cale naturală care să poată fi folosite în acţiunea de modelare a regenerării. În final, s-a intenţionat realizarea unei caracterizări sintetice a regenerării în baza căreia să se poată identifica parametrii necesari unui model şi relaţiile funcţionale din interiorul acestuia.

Conceperea unui astfel de model ar permite testarea anumitor ipoteze și elaborarea unor previziuni privind dinamica suprafețelor aflate în faza de regenerare.

Toate analizele din cadrul lucrării implică utilizarea de instrumente informatice moderne, ca urmare a multiplelor avantaje ce derivă din folosirea acestora: putere de calcul sporită, prelucrare rapidă a unui volum mare de date, creșterea complexității analizelor efectuate, caracterul modern, adaptat cerințelor actuale ale cercetării științifice. Mijloacele informatice oferă posibilități superioare de prelucrare, analiză, interpretare și prezentare a datelor și rezultatelor, fiind folosite în cadrul lucrării pentru a demonstra capacitatea și potențialul acestora în activitatea de cercetare silvică.

## 2.2. Obiective urmărite

Luând în considerare tendințele actuale manifestate pe plan internațional în cercetarea silvică, prin lucrarea de față s-a propus ca direcție principală de studiu evaluarea regenerării arboretelor.

Din perspectiva științifică a lucrării, evaluarea regenerării nu se referă la operația din practica silvică de estimare cantitativă și calitativă a rezultatului procesului de regenerare, operație realizată în teren prin mijloace expeditive. Evaluarea regenerării, ca și temă abordată în lucrare, are un caracter dictat de tendințele internaționale înregistrate în studiile de acest tip. Cercetările din ultimii ani care abordează această problemă fac frecvent referire la relațiile competiționale dintre indivizi, la tiparele de distribuție în spațiu a puieților, la omogenitatea structurală a semințișurilor sau la determinarea distribuțiilor specifice acestei faze de dezvoltare. Foarte multe lucrări urmăresc integrarea acestor informații în modele individuale de creștere care să permită simularea dinamicii regenerării arboretelor.

Prin prisma acestor considerații, evaluarea este percepută în lucrare drept operația de identificare a anumitor particularități și caracteristici ale regenerării care să poată fi cuantificate prin analize complexe ale datelor de teren. Deși lucrarea nu a avut ca obiectiv fundamentarea unui model de regenerare, acesta putând fi realizat doar în urma studierii dinamicii regenerării pe o perioadă îndelungată, se dorește ca informațiile obținute să fie utile și să poată fi integrate pe viitor într-un astfel de model.

Din punctul de vedere al obiectivelor propuse se definesc patru niveluri de evaluare care vor fi urmărite în lucrare și pot fi suprapuse peste **obiectivele principale**:

▪ **evaluarea structurii semințișului** – primul nivel de evaluare a unui sistem este nivelul structural. La acest punct se urmărește determinarea particularităților de structură pe orizontală și pe verticală a semințișului, a caracteristicilor distribuțiilor principalelor elemente structurale și a relațiilor de dependență dintre acestea;

▪ **evaluarea diversității structurale dimensionale** – heterogenitatea dimensională este una din caracteristicile definitorii ale semințișului. Se propune stabilirea gradului de diversitate dimensională și a implicațiilor pe care le are acesta asupra structurii și relațiilor individuale din cadrul suprafețelor analizate;

▪ **evaluarea particularităților de organizare în spațiu a puieților** – informația spațială are un rol determinant în fundamentarea modelelor individuale dependente de distanță. Se urmărește să se stabilească cât de „întâmplătoare" (aleatorie) este distribuția în spațiu a puieților, dacă există posibilitatea identificării unor tipare de agregare a puieților sau dacă se observă tendința de asociere sau repulsie a speciilor sau categoriilor dimensionale de puieți;

▪ **evaluarea relațiilor de competiție în semințișuri** – se propune estimarea intensității acestor relații în suprafețele studiate, precum și decelarea influenței competiției asupra creșterii puieților.

Un **obiectiv secundar** ataşat fiecărui obiectiv principal enunţat anterior constă în identificarea acelor metode de evaluare care să surprindă cât mai fidel particularităţile nivelului studiat.

În scopul realizării obiectivelor de cercetare propuse s-a dorit folosirea unor mijloace informatice specifice. A fost preferat acest mod de abordare datorită faptului că în ultimii ani au fost dezvoltate ramuri ale informaticii, statisticii şi matematicii care oferă noi posibilităţi de explorare şi prelucrare a datelor. De asemenea, cantitatea foarte mare de date şi complexitatea prelucrărilor impune utilizarea unor tehnici adecvate care să ofere rezultate rapide şi de încredere. Nu în ultimul rând, apariţia unor tehnologii moderne, materializate în diverse instrumente de prelevare a datelor – telemetre laser, clupe informatizate, hipsometre ultrasonice sau laser, soluţii integrate de inventariere de tip FieldMap, impune folosirea unor mijloace informatice adecvate în activitatea ulterioară de exploatare a datelor obţinute.

În conformitate cu aspectele menţionate anterior au fost propuse **obiective adiţionale** care se referă la conceperea unor aplicaţii informatice dedicate, capabile să prelucreze datele colectate şi să ofere rezultate şi soluţii adecvate obiectivelor anterior formulate.

Prin abordarea şi soluţionarea tuturor obiectivelor menţionate s-a dorit descifrarea unor aspecte legate de relaţiile inter şi intraspecifice ce se stabilesc între indivizii ce vor alcătui un viitor arboret. Cu cât înţelegem mai bine pădurea şi mecanismele sale interne, cu atât mai uşor putem să găsim soluţiile potrivite pentru păstrarea şi perpetuarea acesteia.

## 2.3. Locul cercetărilor. Materialul şi metodologia de cercetare

Pentru alegerea suprafeţelor de studiu au fost căutate arborete aflate în faza de regenerare, cu seminţişul instalat cât mai compact şi uniform, eliberat în proporţie cât mai mare de protecţia arboretului matur. Criteriile au urmărit omogenitatea din punctul de vedere al condiţiilor microstaţionale pentru a se reduce din factorii care determină variaţia caracteristicilor puieţilor.

Cercetările au fost localizate pe raza Ocolului silvic Flămânzi, Direcţia silvică Botoşani, în UA 50A, UP I Flămânzi, în apropierea localităţii Cotu. Arboretul analizat are o suprafaţă de 21,5 ha, având compoziţia specifică conform ultimului amenajament (2005): 3Go 2St 3Ca 1Te 1Fr. Tipul de pădure este 5514 - *Şleau de deal cu gorun şi stejar pedunculat de productivitate mijlocie* iar tipul de staţiune 7420 - *Deluros de cvercete cu stejar pe platouri şi versanţi slab moderat înclinaţi, cu soluri cenuşii, cenuşii – brune (faeoziom greic-eutricambosol, preluvosol), +/- brune slab luvice, edafic mare-mijlociu, Bm.*

Tăierile de regenerare aplicate în acest arboret au fost specifice tratamentului tăierilor progresive, seminţişul instalându-se în perioada 2001-2002 odată cu primele tăieri de deschidere a ochiurilor. Amenajamentul întocmit în anul 2005 recomandă tăieri cvasigrădinărite, iar în anul 2007 a fost realizată o tăiere de lărgire a ochiurilor şi racordare. Compoziţia seminţişului instalat este conform controlului anual al regenerărilor efectuat în anul 2007: 7Go 1Ca 1Fr 1Dt.

În cadrul acestui arboret a fost identificată o suprafaţă regenerată având un caracter compact şi omogen, cu o proporţie de participare mare a cvercineelor, în care a fost amplasată o reţea de 10 suprafeţe de probă cu caracter permanent (figura 2.2).

Proiectarea modului de amplasare al reţelei a fost realizată utilizând o hartă în format GIS şi extensia DNR Sampling Tool (v 2.8) pentru aplicaţia ArcView Gis. Suprafeţele rectangulare de 7 x 7 m au fost poziţionate în teren cu ajutorul unui dispozitiv GPS, materializate cu ajutorul unor ţăruşi şi inventariate integral.

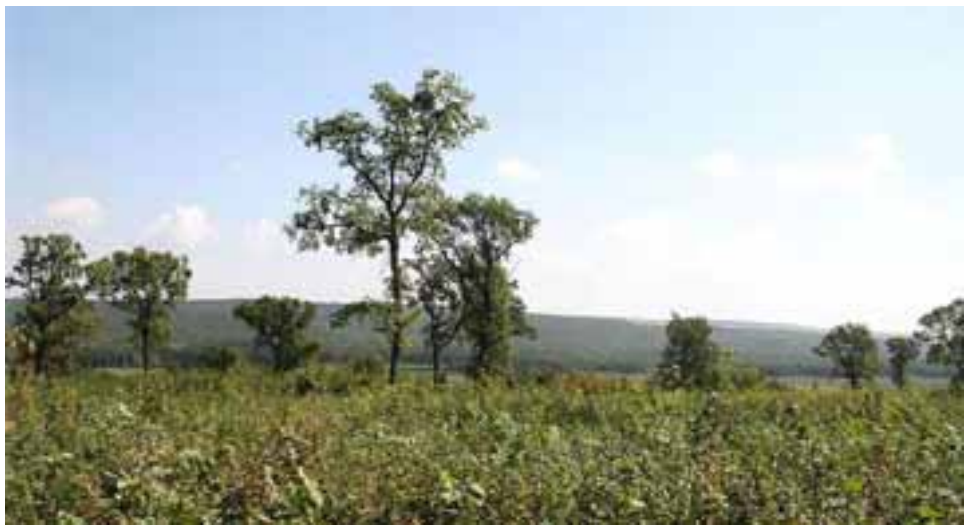

**Figura 2.1.  Aspect din suprafaţa regenerată a arboretului din UA 50 A, UP I Flămânzi**

În vederea constituirii unei baze de date consistente, care să permită prelucrările statistico-informatice ulterioare, au fost prelevate date primare cu caracter general pentru fiecare suprafaţă (altitudinea, panta, expoziţia, date referitoare la pătura erbacee).

S-a inventariat un număr de 7253 de puieţi, pentru fiecare individ fiind prelevate următoarele date:

- specia;
- poziţia – coordonatele x,y într-un sistem cartezian, cu o precizie de 1 cm;
- diametrul la colet, cu o precizie de 1 mm;
- înălţimea, cu precizia de 1 cm;
- înălţimea până la prima ramură, cu precizia de 1 cm;
- două diametre ale coroanei, măsurate pe direcţia N-S, respectiv E-V, cu o precizie de 5 cm;
- observaţii cu privire la eventuale particularităţi ale exemplarului.

Suplimentar, pentru minim 3 puieţi pe metrul pătrat, a fost măsurată ultima creştere în înălţime, cu o precizie de 1 cm.

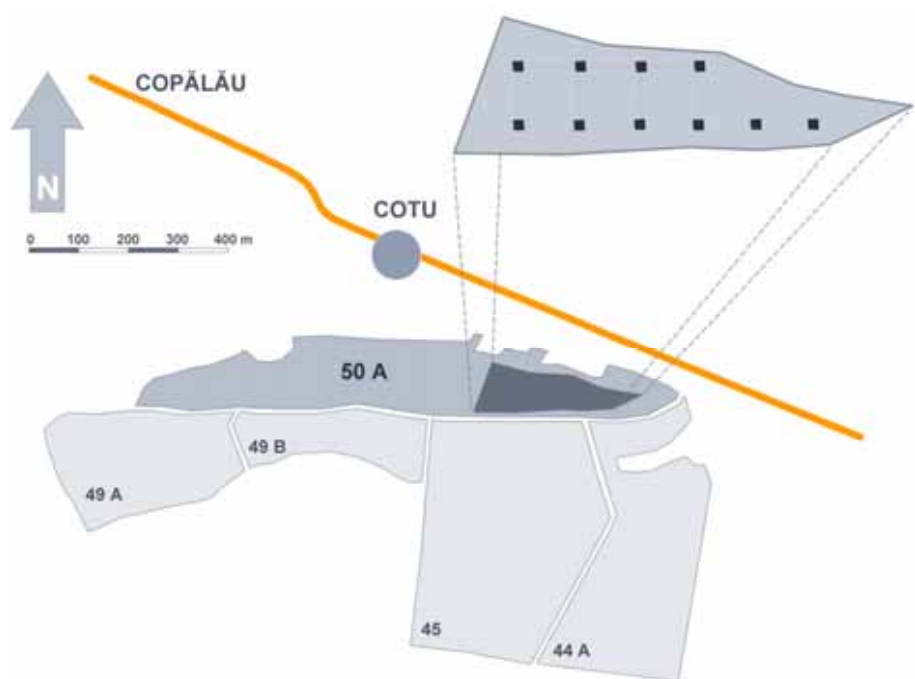

**Figura 2.2.  Localizarea cercetărilor în UA 50 A, UP I Flămânzi şi modul de amplasare al suprafeţelor de probă**

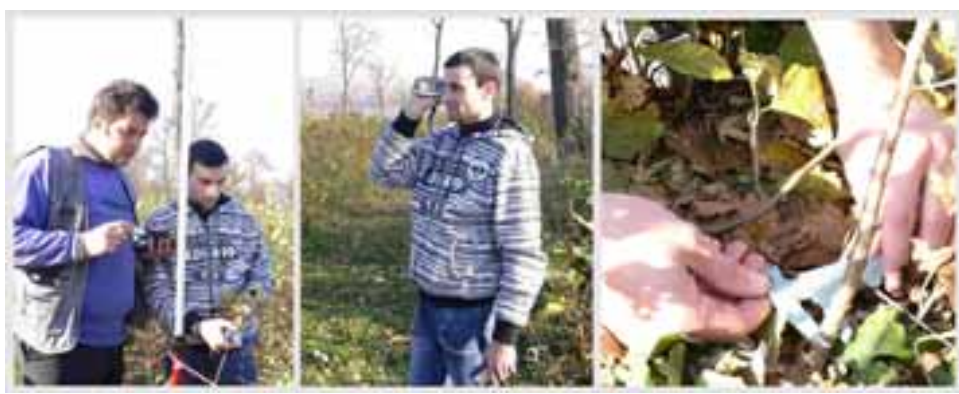

**Figura 2.3.  Instrumentele utilizate în prelevarea datelor**
(telemetrul laser tip Leica DISTO şi hipsometrul ultrasonic tip Haglof Vertex IV)

A fost completată o fişă separată ce cuprinde date privitoare la cioatele şi arborii maturi de pe o rază de 40 de metri faţă de centrul fiecărei pieţe. În cazul cioatelor s-au determinat specia, poziţia (distanţa faţă de centrul suprafeţei şi orientarea) şi două diametre perpendiculare. Pentru arborii maturi s-a determinat specia, poziţia, două diametre perpendiculare ale fusului, înălţimea totală şi înălţimea până la prima ramură a coroanei.

În ceea ce priveşte instrumentele utilizate în teren pentru prelevarea datelor –pentru măsurarea distanţelor au fost folosite rulete şi un telemetru laser tip Leica DISTO A6, pentru diametrele la colet - şublere, pentru înălţimile puieţilor - stadii gradate, pentru diametrele coroanei şi creşteri – rulete, pentru diametrele arborilor maturi – clupe, iar pentru înălţimile arborilor maturi – hipsometre ultrasonice tip Haglof Vertex IV (figura 2.3).

Datorită volumului mare de date, precum şi complexităţii prelucrărilor statistice, matematice şi informatice au fost utilizate programe specializate de calcul statistic şi analiză a datelor – Microsoft Excel, StatSoft Statistica, Stand Visualization System, ESRI ArcView, SpPack (Perry, 2004), SPPA (Haase, 1995), SILVASTAT (Popa, 1999). Pentru prelucrări suplimentare ale datelor, necesare atingerii obiectivelor urmărite, au fost realizate soluţii software proprii (CARTOGRAMA, BIODIV, DIFDOM, IDIV, SPATIAL, VORONOI, SVS Export, Hegyi, Schutz), utilizând mediul de dezvoltare şi programare *Microsoft Visual Basic*.

Lucrarea de faţă se bazează şi pe culegerea şi prelucrarea datelor secundare, a materialului bibliografic. Au fost consultate numeroase lucrări de specialitate originale apărute pe plan naţional şi internaţional, fiind acordată o importanţă crescută celor apărute recent în publicaţii cu un factor de impact ridicat conform *Thomson ISI - Journal Citation Report*.

# Capitolul 3
# Mijloace informatice folosite în modelarea forestieră

### 3.1. Elemente generale ale procesului de modelare cibernetică

**Modelarea** rezidă într-un proces complex prin care se reprezintă, prin metode variate, funcționarea unui sistem sau a unui fenomen, urmând a oferi previziuni sau prognoze asupra comportamentului posibil al entității studiate.

În acest mod modelarea oferă informații viabile asupra procesului studiat, informații care altfel s-ar putea obține doar prin realizarea unor teste reale. Testele experimentale conduc la rezultate concludente, având avantajul că se fac direct asupra sistemului real.

De multe ori însă, realizarea acestor teste este tardivă, erorile detectate prea târziu fiind extrem de costisitoare și capabile să compromită eforturi îndelungate de cercetare.

Drept răspuns la această problemă stringentă, a fost elaborat procesul de simulare, ce prezintă anumite avantaje certe:

- modelarea se poate efectua în faza de proiectare, înainte de demararea unui proiect care s-ar putea să se dovedească neviabil;
- în general modelarea nu implică costuri deosebite și chiar elimină costurile unor erori detectabile ulterior;
- modelarea se poate efectua în unele cazuri, cum poate fi și cel al silviculturii, într-un timp mult mai scurt decât cel necesar testării în condiții reale.

**Simularea** reproduce procesele şi interacţiunile unui sistem real prin intermediul unui model. Acest model va „imita" comportamentul corespondentului din realitate. Practic, se studiază comportamentul unui sistem real folosind un sistem înlocuitor, abstractizat.

**Modelele** sunt *„aproximaţii structurate, selective şi repetabile ale lumii reale, create şi folosite pentru a sintetiza şi analiza datele şi informaţiile"* (Drăgoi, 2000). Ele pot fi definite drept reprezentări abstractizate, matematice sau fizice, ale realităţii.

Aceste reprezentări trebuie să îndeplinească anumite cerinţe:

- să fie clare;
- să fie consistente (ne-ambigue);
- să fie cât mai complete.

Operatorul uman ce realizează modelarea trebuie să fie capabil să elimine contradicţiile, ambiguităţile şi omisiunile în vederea obţinerii unui model viabil. De asemenea, pentru realizarea unui model cât mai performant, va fi nevoie de o descompunere a sistemului real în componente funcţionale elementare şi de o analiză atentă şi pertinentă a relaţiilor şi interacţiunilor ce se realizează la nivelul acestor componente.

Indiferent de metoda folosită, procesul de modelare presupune parcurgerea anumitor etape, în vederea obţinerii rezultatelor:

- fixarea obiectivului;
- colectarea datelor;
- analiza preliminară a datelor;
- alegerea sau generarea unui model;
- estimarea inputurilor;
- relaţiilor procedurale între datele de intrare şi ieşire;
- validarea şi/sau calibrarea modelului (simularea unor situaţii anterioare şi a unor situaţii cu consecinţe previzibile);
- planificarea experimentelor şi simularea propriu-zisă;
- analiza şi interpretarea rezultatelor.

În funcție de caracterul procesului de modelare, se poate vorbi, despre metode:

- *deterministe;*
- *probabilistice (stohastice).*

*Metodele deterministe* – se bazează exclusiv pe identificarea unor relații funcționale între componentele unui sistem analizat, excluzând elementele cu un anumit grad de incertitudine. În acest sens se poate utiliza modelarea euristică, programarea liniară sau programarea dinamică.

*Metodele stohastice* – în realitate, sistemele au o complexitate crescută, ce poate fi mai greu surprinsă prin metode de modelare deterministă. Drept urmare, în foarte multe cazuri, se apelează la metodele stohastice care pot opera cu elemente ce au un caracter probabil. Aceste metode (e.g. metoda simulării Monte Carlo, a probabilităților condiționate Bayes, lanțurile Markov, modelele matriceale etc.) permit elaborarea unor modele statistico-matematice capabile să exprime mai fidel atributele și relațiile unui sistem.

Cel mai frecvent însă, modelele sunt o mixtiune avantajoasă între metodele deterministe și cele stohastice, formându-se practic o nouă categorie de metode: metodele mixte.

Trebuie surprinsă esența funcțională a procesului real studiat, pentru ca la rândul ei, simularea, prin exploatarea modelului creat, să ofere o procedură logică, consistentă de raționament. Erorile ce se strecoară în modelare sunt provocate, în marea majoritate a cazurilor, nu de descrierea incompletă a componentelor sistemului ci de interacțiunile neprevăzute, ce nu au fost surprinse de model.

Se disting trei categorii de modele:

- *descriptive* – folosite în scopul familiarizării cu un sistem greu, incomplet sau deloc observabil în realitate;
- *predictive* – folosite în elaborarea prognozelor pentru diferite sisteme;
- *de adoptare a deciziilor* – ce oferă informații cu privire la posibilele stări optime ale unui sistem.

Modelele, în sine, nu reprezintă ipoteze, dar ele pot confirma sau infirma o serie de ipoteze prin teste riguroase: „*un model nu este nici ipoteză, nici teorie*" (Giurgiu, 1974).

Scopul modelului este acela de a evidenția relațiile și interacțiunile dintre numeroasele variabile ale unui sistem. Reproducerea abstractă a proceselor prin intermediul modelelor și exploatarea acestora prin intermediul simulărilor a fost mult impulsionată de apariția calculatoarelor. Puterea mare de prelucrare a unui volum impresionant de date, precum și rapiditatea de care dau dovadă în efectuarea unor operații complexe au făcut din calculatoare instrumente perfect adecvate pentru realizarea proceselor de simulare. În ultimele decenii calculatoarele s-au impus drept o „prelungire" a creierului uman, iar informatica, prin tehnici și metode performante de prelucrare a datelor, a devenit tot mai prezentă în activitatea cercetătorilor.

Beneficiind de acest impuls tehnologic major, modelarea, ca proces științific, a devenit extrem de utilă în domenii foarte variate. Ea poate fi aplicată, cu tot atât de mult succes, în proiectare, științe biologice, sociologie, medicină sau ecologie.

Importanța modelării se datorează nu doar versatilității și flexibilității ei, ci și rezultatelor deosebite obținute în urma folosirii unor tehnici și metode speciale de simulare, prezentate ulterior în cuprinsul acestei lucrări.

„*Simularea matematică reprezintă un mijloc experimental eficace la îndemâna silvicultorului; prin imitarea proceselor și fenomenelor silvice, folosind modele matematice de simulare, se pot efectua cele mai diverse experimentări în toate domeniile de activitate silvică*" (Giurgiu, 1974).

## 3.2. Particularități ale procesului de modelare a activităților din silvicultură

Un model reprezintă totalitatea conexiunilor logice, a dependențelor formale și a formulelor matematice care fac posibilă studierea unui obiect real (Gertsev & Gertseva, 2004).

Obiectele studiului în cazul silviculturii sunt reprezentate de indivizi, populații și comunități de arbori și ecosistemele forestiere ca întreg. Silvicultorii au identificat arborii, respectiv arboretele drept unități ecologice ce pot fi ușor integrate în modele. Efectuarea unor experimente asupra acestor entități este dificilă sau chiar imposibilă datorită faptului că aceste activități pot conduce la schimbări nedorite ale structurii sau chiar distrugeri permanente ale elementelor analizate. În această situație este evident avantajul folosirii modelelor matematice și a simulărilor, a mijloacelor nedistructive de cercetare.

Un alt avantaj apreciat al modelării în silvicultură se referă la posibilitatea efectuării modelării într-un timp mult mai scurt decât cel necesar testării ipotezelor formulate în condiții reale, folosind arboretul drept obiect al experimentelor. Pădurea se constituie într-un sistem complex care se dezvoltă într-un ritm propriu, mult prea lent pentru a fi convenabil din punctul de vedere al cercetătorilor silvici. Acest ritm încet de dezvoltare reprezintă neajunsul primordial al tuturor cercetărilor ce se derulează în mediul forestier.

Pentru a obține informații viabile despre acest ecosistem complex ar trebui să se facă analize, observații și experimente care s-ar întinde pe un număr de ani ce ar depăși probabil speranța medie de viață a uni om. În mod evident, se folosește experiența a numeroase generații de silvicultori, dar chiar și așa, complexitatea evenimentelor și cazurilor ce pot să intervină în gestionarea unei păduri poate să depășească acest cumul de experiență. De multe ori este nevoie de decizii rapide în conducerea unui arboret, decizii care pot influența în mod benefic dezvoltarea arboretului sau, din contra, pot compromite munca susținută a două-trei generații de silvicultori. Pentru rezolvarea unor astfel de probleme, abordarea unor procese de simulare și modelare se dovedește a fi o soluție potrivită.

Un aspect secundar al folosirii modelării în domeniul forestier, dar extrem de util în activitatea silviculturală, se referă la estimarea unor caracteristici ce nu pot fi ușor măsurate sau determinate (e.g. cazul înălțimilor sau a volumului).

Managementul forestier trebuie să includă criterii multiple și adesea neconvergente în stabilirea obiectivelor sale, drept urmare scenariile bazate pe modelări și simulări sunt instrumente utile de planificare în cazul unor elemente schimbătoare sau caracterizate de incertitudine. Cu ajutorul modelelor se pot efectua estimări predictive și proiecții sau pot fi analizate efectele unor scenarii alternative. Modelarea permite reproducerea relațiilor și interacțiunilor din mediul complex al pădurii prin intermediul unei abstractizări ce va imita evoluția naturală a ecosistemului forestier.

Calitatea simulării este influențată direct de calitatea modelului, care trebuie să surprindă relațiile și interacțiunile dintre numeroasele variabile ale unui sistem. Modelarea presupune „*determinarea valorilor unui set de variabile dependente, considerând că acestea sunt influențate concomitent de mai multe caracteristici factoriale*" (Drăgoi, 1996). Aceste caracteristici factoriale vor fi reprezentate de coeficienții modelului folosit în activitatea de simulare. Deducerea modelului utilizat va presupune calculul coeficienților modelului prin metode ale regresiei matematice (e.g. metoda celor mai mici pătrate, a verosimilității maxime) sau metode neparametrice (metoda celor mai apropiați sau mai similari vecini).

Vanclay și Skovsgaard (1997) consideră că evaluarea modelelor forestiere este o operație care trebuie să țină cont de anumite caracteristici intrinseci ale acestora:

- complexitatea metodelor de estimare a parametrilor modelului;
- simplitatea modelului și a modului de selectare a variabilelor;
- realismul relațiilor și proceselor biologice;
- compatibilitatea submodelelor utilizate;
- încrederea acordată estimărilor furnizate.

Aspectele care trebuie soluționate cel mai adesea în ceea ce privește modelele adaptate mediului forestier se referă la procesele principale desfășurate în cadrul arboretelor – dinamica extinderii în spațiu, creșterea arborilor, mortalitatea și rata de supraviețuire, efectele unor factori perturbatori, respectiv structura și regenerarea arboretelor (Dincă, 2004; Ciubotaru & Păun, 2014). Procesele amintite anterior pot fi modelate într-o manieră foarte diferită – de la simplu la complex, pornind de la simple tabele de producție și ajungând la utilizarea inteligenței artificiale, modelării geometrice sau a tehnicilor GIS.

Modelele au fost întotdeauna privite cu interes în silvicultură datorită dificultăților de a învăța din experiență sau din experimente, evoluând în timp de la simple tabele de producție la simulatoarele cibernetice complexe din prezent. Originea modelelor se situează în secolul al XVII-lea, odată cu dezvoltarea de către chinezi a unor tabele de producție rudimentare (Vuokila, 1965, citat de Ek, Dudek, 1980).

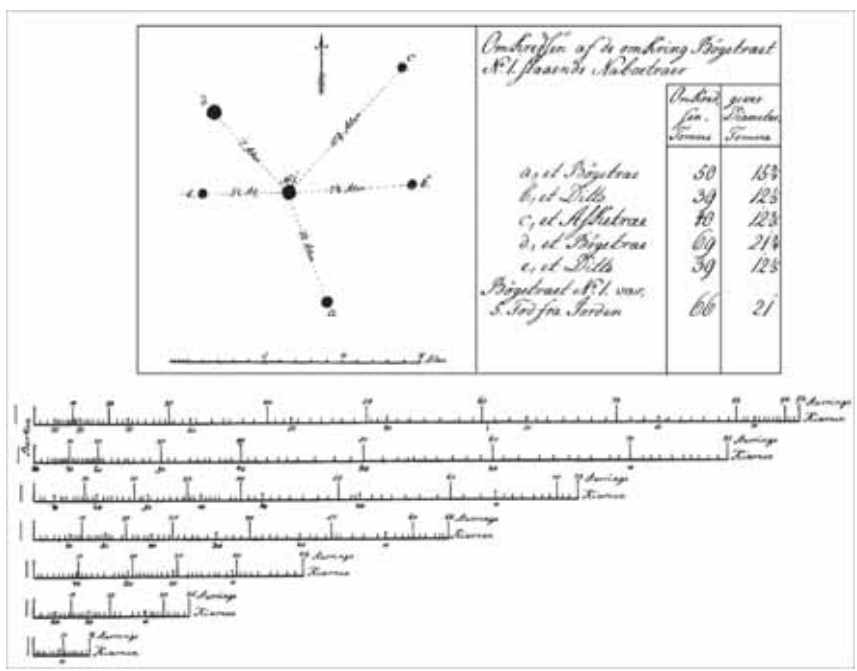

**Figura 3.1. Fisa de descriere a unui arbore în cazul cercetărilor efectuate de Roventlow**
(după Skovsgaard, 2004)

Un alt studiu de pionierat este cel al contelui Roventlow, care a efectuat în Danemarca anului 1793 primele măsurători dendrometrice moderne. Cercetările inițiate în arborete de stejar și fag, au urmărit construirea unui model al creșterii și dezvoltării arborilor. S-a înregistrat poziția fiecărui arbore și dimensiunile arborilor competitori, s-au efectuat măsurători ale înălțimilor și ale creșterilor radiale după patru direcții (figura 3.1), iar în final s-a reconstituit istoricul creșterilor în înălțime prin analiza secțiunilor fusului (Vlad et al., 1997; Skovsgaard, 2004;).

Se consideră că la originea modelelor actuale se regăsesc cercetări ale silvicultorilor germani de la sfârșitul secolului al XVIII–lea, materializate sub forma tabelelor de producție și creștere a arboretelor (Assman, 1970; Lexer et al., 2000; Garcia 2008). Interesul pentru această formă primară de modelare a fost deosebit, în Germania fiind publicate până în anul 1880 peste 1000 de lucrări privitoare la tabele forestiere de producție (Kangas, 2004). Tabelele de producție sunt folosite și în prezent oferind estimări viabile, în special pentru arboretele echiene monospecifice, în cazul în care condițiile climatice și staționale rămân neschimbate. Din păcate, în cazul unor modificări majore rezultatele furnizate sunt imprecise (Kimmins et al., 2008). Drept urmare au fost dezvoltate numeroase alte tipuri de modele, în ordine cronologică  menționând modelele de creștere, modelele ecologice tip „gap models", modelele de procese sau modelele hibride.

Modelele moderne ale dinamicii ecosistemelor forestiere au apărut în America de Nord, majoritatea fiind concepute pentru arborete de conifere, monospecifice, echiene. Perioada cea mai prolifică de apariție a modelelor a fost 1960-1990 (Newnham, 1964 – citat de Gratzer et al., 2004; Botkin et al., 1972; Stage, 1973; Ek, Monserud, 1974; Shugart, West, 1977; Wykoff et al., 1982; Hilt, 1985; Urban, 1990, Pacala et al., 1993), perioadă ce coincide cu dezvoltarea științelor informatice și folosirea pe scară largă a calculatoarelor. Ulterior, între anii 1990-2000, modelele au cunoscut o dezvoltare rapidă și în nordul și centrul Europei, cu precădere în Finlanda, Germania, Austria (Pukkala, 1989; Pukkala, Kolstrom, 1992; Pretzch, 1992; Hasenauer, 1994; Nagel 1996; Schweiger, Sterba, 1997; Pukkala, Miina, 1997; Monserud, Sterba, 1996).

În prezent modelarea este un domeniu al cercetării silvice care preocupă specialiști din toată lumea, cu o arie foarte largă de specializări. Modelarea ecosistemelor forestiere a devenit un instrument util în analiza dinamicii pădurii și practic în anticiparea modificărilor apărute la nivelul întregului ecosistem forestier.

Tendințele moderne indică trecerea la o nouă paradigmă de modelare prin deplasarea interesului cercetătorilor de la modelele adresate arboretului spre cele de creștere individuală, ale arborilor – așa numitele IBM (*Individual based models*) (Teuffel, 2006). Structura generală a unui model individual de creștere cuprinde, după Hasenauer (2006), următoarele elemente:

- funcții de incrementare a diametrului și înălțimii - în funcție de specie, relații de competiție, tipul stațiunii, geometria coroanei;
- indici de competiție dependenți sau independenți de distanțe;
- submodele ale dezvoltării coroanei - bazate pe indici de formă și înălțimea până la prima ramură verde a coroanei;
- submodele ale estimării mortalității – se calculează pentru fiecare arbore o variabilă binară (0 sau 1) ce exprimă probabilitatea mortalității în funcție de diverși parametri;
- submodele de regenerare – ce oferă detalii privitoare la instalarea și dezvoltarea semințișului.

În căutarea unor soluții cât mai performante și mai adaptate mediului forestier, Monserud (2001) sugerează o paradigmă a modelului ideal, menționând condițiile care ar trebui să le îndeplinească acesta:

- să fie bazat pe cauze fundamentale ale productivității și creșterii;
- să fie întocmit în acord cu cerințele managementului forestier;
- incertitudinea să fie reprezentată rezonabil;
- să fie verificabil în limite satisfăcătoare;
- să fie capabil să prezinte caracteristici noi, originale;
- să fie ușor de înțeles;
- sa fie flexibil, adaptabil unor situații diferite.

## 3.3. Tipuri și categorii de modele

Modelele folosite în activitatea de cercetare din silvicultură au fost clasificate în timp de un număr foarte mare de cercetători (Botkin, 1993, 2001; Bugmann, 2001; Dudek, Ek, 1980; Edminster, 1978; Gadow, Hui, 1999; Gertsev et al., 2004; Gratzer et al., 2004; Kangas, 2004; Kimmins et al., 1999, 2008; Lexer et al., 2000; Matsushita et al., 2004; Mladenoff, Baker, 1999; Monserud et al., 1996, 2001; Munro, 1974; Pacala et al., 1996; Peng 2000 a,b, 2006; Pretzch et al., 2006; Shugart, 1980, 2002), criteriile folosite fiind în general tehnica de modelare și rezoluția spațială a modelului. Clasificările sunt de regulă formale, întocmite mai mult pentru a reliefa direcția de acțiune a unui model. În realitate modelele sunt greu de clasificat datorită complexității lor și multitudinii de aspecte implicate.

Gertsev (2004) consideră că modelele folosite în ecologie sunt din punct de vedere matematic homomorfice, gnoseologice și în general dependente de timp și face o clasificare a acestora în funcție de:

- momentul final surprins: *modele de prognoză sau de reconstrucție;*
- reprezentarea temporală a dinamicii obiectului modelat: *modele continue sau discrete;*
- modul de determinare a parametrilor: *modele deterministe sau stohastice;*
- identificarea tipului de dependență a parametrilor: *modele analitice sau numerice;*
- momentul întocmirii modelului față de perioada investigării sistemului real: *modele dominante sau subdominante;*

Monserud (2001) acceptă patru clase de modele folosite în silvicultură:

- *Modele de estimare a creșterii și producției* (*growth and yield models*);
- *Modelele ecologice* - de tipul modelului ochiurilor (*gap models*) sau a compartimentelor ecologice (*ecological compartment models*);
- *Modele de procese / mecaniciste* (*process / mechanistic models*);
- *Modele hibride* (*hybrid models*);

*a) Modelele de estimare a creșterii și producției* - sunt cele mai vechi modele, datând din secolele XVIII (Lexer et al., 2000) și urmăresc prognoza producției în anumite condiții de gospodărire date. În general s-a modelat relația dintre desime și creșterea arborilor în diametru și înălțime, obținându-se modele dependente de specie. Ulterior, cercetările privitoare la mortalitate au îmbunătățit prognozele privind dinamica arboretelor. Monserud și Sterba (1999) au modelat mortalitatea printr-o funcție dependentă de specie, diametru, înălțime, dimensiunile coroanei și desimea arboretului. În prezent, folosirea calculatoarelor a făcut posibilă modelarea individuală a arborilor. Ca măsură a productivității, se folosește un indice al stațiunii (*site index*) ce se exprimă prin valoarea înălțimii superioare la o vârstă dată.

Avantajul acestui tip de model constă în prognozele precise cu privire la valoarea producției arboretului, fiind folosite de managementul forestier - unele modele de creștere au fost incorporate în sistemele naționale de inventariere a resurselor forestiere – cazul FVS (Wykoff et al., 1982) sau PROGNAUS (Monserud, Sterba, 1996).

Dezavantajul fundamental se referă la faptul că nu integrează cauzele productivității (nutrienții, umiditatea, clima), fiind adesea numite modele empirice. Exemple de astfel de modele: PROGNOSIS (Stage, 1973), CACTOS (Wensel, Biging, 1987), SILVA (Pretzsch, 1992, 2002), MOSES (Hasenauer, 1994), BWIN (Nagel, 1996), STAND (Pukkala, Miina, 1997), ORGANON (Hester et al., 1998; Hann, 2006), CORKFITS (Ribeiro et al., 2001, 2006), DRYMOS (Chatziphilippidis, Spyroglou, 2006).

*b1) Modelele ecologice ale golurilor (ochiurilor)* - au fost folosite intensiv în ultimii ani în modelarea ecologică a succesiunilor populațiilor; se bazează pe studiul efectului produs într-o populație de apariția unui ochi (gol) datorat dispariției unui arbore dominant (Botkin et al., 1972; Shugart, West, 1977; Pacala 1993, 1996; Didion, 2009). Modelele de tip ”gap models” sunt modele individuale, deterministe care simulează dezvoltarea ecosistemelor forestiere pe baza teoriei dinamicii ochiurilor - "*shifting-mosaic dynamics*" (Watt, 1947).

Toate modelele de acest tip derivă dintr-un model "părinte" : JABOWA (Botkin et al., 1972), ulterior fiind luat în calcul şi FORET (Shugart, West, 1977). Acestea au determinat o proliferare a acestui tip, modele similare fiind adoptate în foarte multe zone ale lumii: Australia, Asia, Europa Centrală (Bugmann, 1996).

Creşterea în diametru se apreciază în funcţie de o creştere potenţială a diametrului, calculată printr-o funcţie ce depinde de diametrul curent şi diametrul maxim observat în populaţia studiată. Regenerarea în ochiuri va fi apreciată prin intermediul unei funcţii dependente de cantitatea de lumină ce ajunge la nivelul solului şi nu va depinde de existenţa unor surse de seminţe disponibile. În schimb ea va fi influenţată de temperatura şi umiditatea din sol. Mortalitatea este apreciată probabilistic, considerând că un individ are 2% şanse să ajungă la vârsta maximă observată în populaţie. Mortalitatea va creşte pentru arborii cu o creştere redusă.

Succesul modelului ochiurilor (*gap model*) s-a datorat mai ales protocoalelor simple de estimare a parametrilor specifici, ce permit aplicarea în condiţii extrem de variate. Din păcate, predicţiile bazate pe acest tip de modele se referă la următorii 100 -1000 de ani, adică mult peste înregistrările efectuate până acum. Din acest motiv, validarea acestor modele rămâne o problemă dificilă şi discutabilă, nefiind dezvoltate metode adecvate de testare (Bugmann, 2001).

Desigur au apărut şi completări şi îmbunătăţiri ale acestui tip de modele – un exemplu ar fi modelul SORTIE (Pacala et al., 1993), considerat o „*alternativă promiţătoare*" (Monserud, 2001), ForClim (Bugmann, 1996, 2001; Wehrli, 2005), ZELIG (Urban, 1990, 1991) sau FORSKA (Prentice et al., 1993).

*b2) Modelul compartimentelor ecologice -* acest model ecologic mai este numit şi modelul fluxului de resurse deoarece urmăreşte prognoza fluxurilor ce se realizează între compartimentele unui sistem. Aceste modele au fost mai rar folosite în managementul forestier, dar totuşi au existat cazuri de folosire a lor.

*c) Modele mecaniciste (de procese) -* au fost utilizate mai ales în fiziologie şi biochimie, ele sacrificând capacitatea predictivă în favoarea explicării raţionale, ştiinţifice a proceselor cauzale care determină productivitatea ecosistemelor: fotosinteza, respiraţia, ciclul nutrienţilor, regimul hidric etc. Prin aceste modele se încearcă reprezentarea dependenţei dintre variabilele unui proces, motiv pentru care se consideră destul de generale. Din cauza faptului că aceste modele au o bază fiziologică solidă, se apreciază că oferă rezultate satisfăcătoare. Merită menţionate modelele ECOPHYS (Rauscher et al., 1990), TREGRO (Weinstein, Beloin, 1990), PnET (Aber, Federer, 1992), TREE- BGC (Korol et al., 1995), TRAGIC (Hauhs et al., 1995), PipeSTEM (Valentine et al., 1997).

*d) Modele hibride* - în ultimii ani, foarte mulţi specialişti au subliniat nevoia de adaptare a modelele nevoilor şi problemelor forestiere actuale, una din soluţii fiind combinarea celor mai bune calităţi ale modelor fiziologice cu cele ale modelelor empirice (Valentine, Makela, 2005). Rezultatul este un aşa numit model hibrid, mult mai adaptat specificului mediului forestier.

Sunt numeroase astfel de modele, unul dintre primele modelele hibride menţionate în literatura de specialitate fiind 3-PG (*Physiological Processes Predicting Growth*), dezvoltat de Landsberg şi Waring (1977). Acesta este capabil să calculeze carbonul fixat prin activitatea de fotosinteză a coronamentului unui arboret, rezultatele corectându-se în funcţie de secetă, deficitul de umiditate atmosferică sau vârsta arboretului. Modelul este încă folosit şi a fost testat în Europa (Suedia), Australia, Noua Zeelandă, coasta de nord-vest a Americii de Nord, sudul Africii.

Modelele hibride sunt cele mai promiţătoare deoarece înglobează caracteristici folositoare din alte tipuri de modele şi elimină anumite caracteristici considerate inadecvate mediului forestier. Alte exemple de modele hibride sunt: PipeQual şi CROBAS (Makela, 1997), LINKAGES (Post, Pastor, 1986, 1996), FORECAST (Kimmins et al., 1999, 2008), FOREST 5 (Robinson, Ek, 2003), EFIMOD (Chertov et al., 2006).

Munro (1974) distinge trei "filozofii de modelare" ce caracterizează modelele forestiere:

- cea în care unitatea de modelare este reprezentată de individ (arbore), bazată pe relaţiile spaţiale dintre aceştia – *modele individuale dependente de distanţă;*

- cea în care individul este în continuare unitatea de modelare dar nu este necesară includerea datelor privitoare la distanţe - *modele individuale independente de distanţă*;

- cea în care consideră arboretul ca fiind unitatea primară de modelare – *modele ale arboretelor*.

*Modelele individuale dependente de distanţă* pot oferi rezultate foarte detaliate cu privire la structura şi dinamica arboretelor, luând în considerare relaţiile de competiţie ce se stabilesc între arbori (relaţiile dintre arbori vecini, întrepătrunderea coroanelor sau diferenţele de statut).

Dezavantajele acestora sunt reprezentate de nevoia înregistrării poziţiei fiecărui arbore şi modalitatea greoaie şi complexă de calcul, ce reclamă utilizarea unor aplicaţii informatice în activitatea de prelucrare a datelor (Munro, 1974; Edminster, 1978).

*Modelele individuale independente de distanţă* sunt mai eficiente în studierea efectelor tratamentelor silviculturale, fiind preferate datorită flexibilităţii şi simplităţii în utilizare, precum şi eliminării calculelor complexe necesare evaluării competiţiei (Botkin et al., 1972; Stage 1973; Shugart, West 1977; Wykoff et al., 1982).

În ceea ce priveşte această clasificare, Dudek şi Ek (1980) remarcă o împărţire echilibrată a acestora, din cele 47 de modele analizate 21 fiind dependente şi 22 independente de distanţă, restul de 4 fiind dificil de încadrat în unul din cele două tipuri.

*Modelele adresate arboretelor* sunt independente de distanţă, fiind familiare silvicultorilor prin modul simplu de înregistrare a parametrilor specifici. Sunt folosite valori medii ale diametrului, înălţimii, volumului, vârstei, suprafeţei de baza, numărului de arbori la suprafaţă iar rezultatele oferite se referă la producţia arboretelor în condiţiile unor strategii diferite de conducere (Edminster, 1978).

Shugart (1980) recunoaşte trei tipuri de modele forestiere de simulare, oarecum similare celor identificate de Munro (1974):

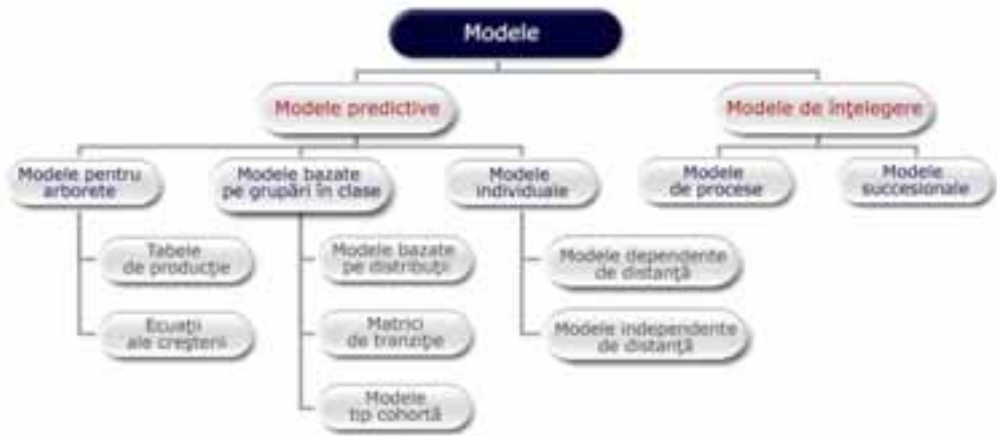

**Figura 3.2.  Clasificarea modelelor forestiere - adaptare după Kangas (2004)**

- modele individuale (*tree models*) axate pe modelarea din perspectiva individului;

- modele ale ochiurilor (*gap models*)- ce simulează dinamica forestieră pe baza atributelor structurale ale unui spațiu restrâns (gol, ochi);

- modele ale arboretelor (*forest models*)- ce își focalizează interesul în modelare asupra arboretului ca întreg.

Gratzer (2004, a) identifică trei clase de modele ale dinamicii forestiere:

- Modelul pâlcurilor (*patch models*) cu două subcategorii: automatul cu număr finit de stări (Shugart, 2002) și modelele ce urmăresc evoluția și creșterea indivizilor de la instalare și până la dispariție, inspirate de conceptul JABOWA (Botkin et al., 1972) și FORET (Shugart, West, 1977). Acest tip a fost des utilizat, în timp, modelele fiind rafinate (Bugmann, 1996, 2001; Price et al., 2001; Urban , 2005).

- Modelele individuale spațiale (*spatially explicit individual-based models*). Interesul se îndreaptă de la dinamica pâlcurilor spre interacțiunile și relațiile dintre entități. Modelele de acest tip: ZELIG (Urban, 1990, 1991), SILVA (Pretzsch, 1992) sau SORTIE (Pacala et al., 1993) consideră arboretul ca fiind o mulțime de indivizi între care intervine așa-numita "*dinamică de vecinătate*" (*neighbourhood dynamics*) și în care acționează procese cu caracter spațial. Succesul acestor modele a fost determinat și de dezvoltarea unor noi metode și instrumente de prelevare și prelucrare a datelor, precum și de folosirea extensivă a tehnicii de calcul. Dezavantajele folosirii acestor metode rezidă în complexitatea lor și în caracterul stohastic al determinărilor.

- Modelele bazate pe momente (*moment-based models*) sunt modele ce apelează la noțiuni specifice recunoașterii formelor, transformărilor matriceale și analizei spațiale (Ripley, 1997). Se face trecerea de la descrierea statică a structurilor spațiale la o caracterizare a dinamicii acestora în timp.

Indiferent de tipul modelelor folosite, cercetarea silvică are nevoie de perfecționarea unor elemente ale modelelor, care se referă la:

- îmbunătățirea reprezentării comportamentului ecosistemelor;
- creșterea acurateței prognozelor și a nivelului de detaliu al rezultatelor;
- înglobarea unor elemente naturale de risc - doborâturi de vânt, incendii, boli sau atacuri de insecte;
- adăugarea unor elemente în modelare care să țină cont de calitatea materialului de regenerare (materiale ameliorate genetic sau naturale);
- estimarea și prognoza unor atribute non-cantitative (calitatea lemnului, produsele accesorii, potențialul recreativ).

Interesul acordat activității de modelare în silvicultură s-a materializat și în România atât sub forma unor lucrări de sinteză privind folosirea modelelor forestiere (Giurgiu, 1974; Tamaș, 1983; Dincă, 2004, Ciubotaru & Păun, 2014) cât și a unor lucrări de cercetare importante (Giurgiu, 1968; Dissescu, 1973; Seceleanu, 1975, a,b; Cenușă, 1992, 1996 b, 2002 b; Leahu, 1988; Barbu, Cenușă, 2001; Barbu et al., 2002; Dragoi, 1996, 2002; Giurgiu, Drăghiciu, 2004; Popa, 1998, 2005; Vlad, 2003, a, 2006). Acest fapt demonstrează încă odată versatilitatea, flexibilitatea și potențialul deosebit al modelării în domeniul forestier.

## 3.4. Tendințe în activitatea de modelare

Pacala (1996) consideră că motivul pentru care Watt (1947) a conceput comunitatea forestieră ca fiind un mozaic de pâlcuri ("*mosaic of patches*") a fost unul pragmatic – dificultatea descrierii comunității la nivel individual și a relațiilor spațiale din cadrul acesteia. În prezent, avansul tehnologic din domeniul ecologiei, biologiei, matematicii, informaticii, teledetecției, permite soluționarea celor două aspecte.

Au apărut însă alte provocări cu care se confruntă modelarea în silvicultură. După Gratzer (2004, a) acestea vizează trei aspecte principale:

- caracterizarea dinamicii și structurii spațiale a vegetației;
- identificarea proceselor relevante care determină o anumită structură spațială (dispersia semințelor, interacțiunile inter-specifice, perturbările cu caracter aleatoriu ale mediului);
- înțelegerea consecințelor unei anumite structuri spațiale asupra dinamicii comunității.

O altă tendință manifestată recent în activitatea de modelare a ecosistemelor terestre - a apărut ideea necesității unui sistem de modele "în cascadă", specifice ecologiei peisajului, care să acopere diferitele rezoluții spațiale - să opereze fiecare la o scară diferită (Chertov et al., 2006; Urban et al., 2005).

În cazul ecosistemelor forestiere nu se poate vorbi despre o modelare exhaustivă, completă, și prin urmare nu se pot oferi garanții absolute. Pentru a compensa acest neajuns au fost dezvoltate metode și tehnici noi care să ofere un suport solid în activitatea științifică de modelare și simulare și care să prezinte anumite avantaje specifice, în funcție de modul în care vor fi folosite.

Kangas (2004) apreciază că odată cu creșterea complexității unui model scade acuratețea estimărilor furnizate de acesta, recomandând concentrarea eforturilor cercetătorilor asupra calității datelor de intrare. În acest mod, noile modele vor fi mai flexibile și mai adaptabile diferitelor situații întâlnite în practica forestieră.

În ultimele decenii au început să fie folosite metode şi tehnici speciale, impuse cu precădere de apariţia unor sisteme hardware şi software tot mai performante. Astfel, metodele de analiză a modului de organizare spaţială, tehnicile moderne de GIS şi teledetecţie, modelare geometrică algoritmică sau tehnicile inteligenţei artificiale nu mai reprezintă noutăţi, dar cu toate acestea au adus un suflu nou în activitatea forestieră de modelare şi simulare. Aceste noi tehnici pot fi în general greu de încadrat; dacă analiza spaţială şi modelarea geometrică s-ar apropia mai mult de metodele deterministe, inteligenţa artificială cu greu s-ar putea spune în ce categorie ar trebui încadrată. În mod cert însă, indiferent de metodele şi tehnicile folosite, interesează în primul rând obţinerea unor rezultate utile.

### 3.4.1. Inteligenţa artificială

Inteligenţa artificială (IA) este un termen care în sens larg exprimă abilitatea unui sistem sau a unei entităţi lipsite de viaţă de a reproduce raţionamentul uman. Termenul de IA a fost introdus în anul 1956 cu prilejul Conferinţei de la Colegiul Dartmouth, iar părinţii acestui nou domeniu sunt consideraţi John McCarthy, Alen Newell, Marvin Minsky si Herbert Simon.

Conceptul de inteligenţă artificială a evoluat şi a condus în timp la cristalizarea anumitor opinii conform cărora căreia IA urmăreşte studierea şi crearea unor software-uri şi sisteme de calcul ce posedă o formă de inteligenţă: entităţi care sunt capabile să raţioneze, să asimileze şi să deducă concepte diferite, să înţeleagă limbajul natural sau să perceapă reprezentări grafice, pe scurt entităţi capabile să reproducă elemente de inteligenţă specific umană (Holban et al, 1994).

Inteligenţa artificială este un domeniu extrem de vast ce cuprinde subdomenii ca: reprezentarea cunoaşterii, recunoaşterea formelor, demonstrarea teoremelor, calculul inteligent, prelucrarea limbajului natural, sisteme expert, învăţarea automată etc. În continuare vor fi prezentate succint câteva subdomenii care prezintă un interes din punctul de vedere al simulării şi modelării.

Sistemele expert reprezintă „*o clasă particulară de sisteme informatice bazate pe inteligența artificială, având drept scop reproducerea cu ajutorul calculatorului a cunoștințelor și raționamentelor experților umani într-un domeniu bine definit*" (Turcu, 2003). Un sistem expert este capabil nu doar să imite raționamentul unui expert uman folosind raționamente artificiale ci să multiplice experiența acestuia și să justifice raționamentul efectuat. Utilitatea sistemelor expert rezidă mai ales în faptul că acestea sunt capabile să multiplice și să transfere experiența unui specialist uman.

Elementele constituente ale unui sistem expert sunt: *baza de cunoștințe* (baza de reguli), *baza de fapte*, *motorul de inferențe* (mecanismul rezolutiv), *modulul explicativ* (pentru verificarea coerenței bazei de cunoștințe), *modulul de achiziție a cunoștințelor* (destinat preluării cunoștințelor), *interfața cu utilizatorul* (asigură dialogul dintre utilizator și sistemul expert), fiind prezentate în figura 3.3.

Sistemele expert se utilizează într-o sferă extrem de variată de domenii încă din deceniul al șaselea al secolului XX. Primele sisteme expert au apărut în domenii foarte variate: DENDRAL (1967) în domeniul chimiei, MYCIN (1976) în medicină, PROSPECTOR (1979) pentru prospecții geologice (Turcu, 2003). De atunci au apărut numeroase sisteme expert, în majoritatea domeniilor care necesitau pregătirea unor experți umani.

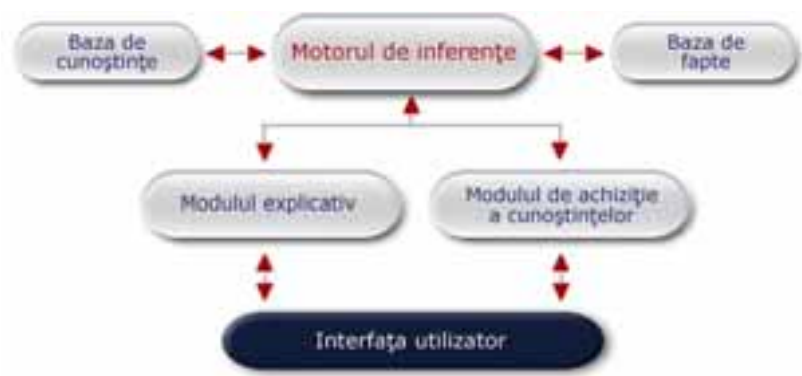

**Figura 3.3. Arhitectura unui sistem expert**
(adaptare după Turcu, 2003)

**Recunoaşterea formelor** „*constituie o ramură a inteligenţei artificiale şi are ca scop identificarea unor relaţii de asemănare între reprezentările abstracte ale unor obiecte sau fenomene*" (Pentiuc, 1996). În fapt, recunoaşterea formelor reprezintă un conglomerat de metode şi tehnici ce aparţin  unor domenii fundamentale variate (statistica, teoria deciziei, teoria limbajelor formale, teoria mulţimilor vagi, reţele neuronale) ce oferă instrumente de analiză performante ale unor procese abstractizate. Putem să vorbim despre o modelare a realităţii prin intermediul conceptului de formă şi a relaţiilor ce se pot stabili între forme.

Forma reprezintă o „*entitate abstractă, descrisă printr-un număr finit de atribute numite caracteristici care constituie reprezentarea matematică a unui obiect sau fenomen din lumea reală*" (Pentiuc, 1996). Această abstractizare a formei permite aplicarea formalismului matematic într-o gamă variată de activităţi.

Forma este descrisă matematic printr-un set de variabile sau caracteristici. Procesul de analiză a formei include prelucrări matematice cum ar fi: normalizări, transformări liniare şi neliniare, ordonarea, reducerea şi selectarea caracteristicilor. Pentru reprezentarea şi analiza relaţiilor dintre forme se utilizează două măsuri reale: *distanţa* şi *coeficientul de similaritate*.

Elementele unui sistem de recunoaştere a formelor sunt: *translatorul*, *preprocesarea*, *analiza*, *învăţarea, modulul de decizie* (rezultatul procesului de învăţare). Modulul de învăţare trebuie conceput să lucreze în două contexte: de învăţarea supravegheată şi nesupravegheată.

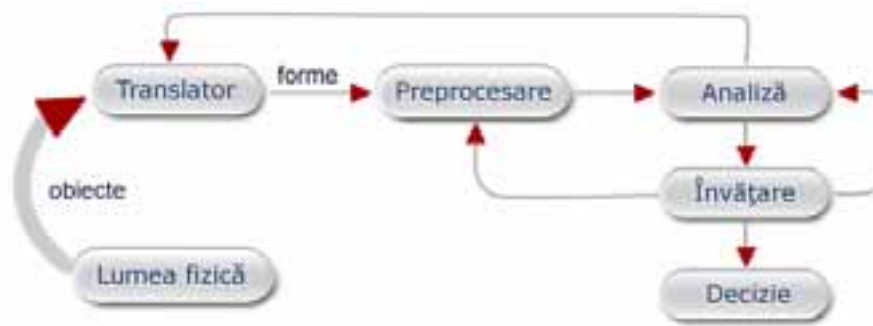

**Figura 3.4.  Schema sistemului de recunoaştere a formelor**
(adaptare după Pentiuc, 1996)

**Calculul inteligent** aparține sferei IA, grupând tehnici speciale de soluționare a unor probleme greu structurabile, care nu pot fi caracterizate de un model matematic clar sau a căror soluționare prin metode clasice presupune folosirea unor algoritmi foarte costisitori. În general, este acceptată din această perspectivă gruparea problemelor în două categorii: *probleme „corect puse"* – care pot fi modelate matematic și rezolvate printr-o metodă cu caracter algoritmic, respectiv *probleme „incorect puse"* – categorie care conține probleme pentru care definirea completă printr-un model formal nu este posibilă și nu se poate evidenția o relație de asociere între datele de intrare și cele de ieșire. Rezolvarea se efectuează în acest caz prin calcul inteligent, pe baza unor exemple, prin adaptare sau învățare.

Există mai multe direcții tradiționale de folosire a calculului inteligent: calculul neuronal, calculul evolutiv și calculul fuzzy. Există și alte direcții moderne, aflate în primele faze de dezvoltare: calculul molecular, cu membrane sau cuantic.

*Calculul neuronal* se utilizează în special în soluționarea unor probleme de asociere și se bazează pe identificarea, prin învățare, a unui tipar plecând de la diferite exemple. Din punct de vedere structural, rețelele neuronale sunt caracterizate de flexibilitate și adaptabilitate, imită funcționarea creierului și sunt formate dintr-un ansamblu de unități interconectate, care sunt capabile de calcule foarte imple. Calculul neuronal reprezintă o alternativă viabilă la paradigma programării algoritmice secvențiale.

*Calculul evolutiv* reprezintă o formă de calcul inspirată de principiile evoluționismului darwinist. Acest sistem oferă instrumente de căutare a soluției (algoritmi genetici, programare genetică sau evolutivă) utilizând o populație de căutători, ce va fi supusă unui proces virtual de evoluție prin mecanisme specifice similare celor din evoluția naturală: selecție, încrucișare și mutație.

*Calculul fuzzy* este folosit în rezolvarea problemelor a căror descriere este caracterizată de un anumit grad de incertitudine (*fuzziness*). Prelucrările datelor vor urmări înlocuirea valorilor exacte cu valori fuzzy descrise prin funcții de apartenență. În acest mod conceptele vagi pot fi transformate, abordate și prelucrate prin tehnici matematice convenționale.

În silvicultură metodele inteligenței artificiale au fost aplicate odată cu dezvoltarea unor sisteme expert de management a resurselor naturale (Coulson et al., 1987). Ulterior tehnicile de procesare paralelă distribuită, specifice rețelelor neuronale, au devenit o alternativă la clasicele tehnici statistice aplicate în modelarea comportamentului sistemelor neliniare (Peng, Wen, 1999).

În prezent se apreciază că modelele de estimare ce utilizează rețele neuronale sunt mai flexibile și oferă rezultate superioare ca precizie modelelor liniare (Guan, Gertner, 1991; Hasenauer et al., 2001; Hasenauer, Kindermann, 2002).

Tehnicile de inteligență artificială sunt în prezent utilizate într-o gamă variată de activități ale cercetării silvice:

- Clasificarea și evaluarea caracteristicilor ecosistemelor forestiere prin folosirea imaginilor satelitare (Peng, 2000; Franklin, 2001; Foody, Cutler, 2006; Kuplich, 2006). Aplicația software Definiens eCognition reprezintă un exemplu de implementare a tehnicilor specifice inteligenței artificiale – un sistem expert de clasificare a structurilor pe baza imaginilor satelitare, ce înglobează sisteme de învățare supravegheată, metode ale recunoașterii formelor, algoritmi genetici. Sistemul a dat rezultate foarte bune în clasificarea ecosistemelor forestiere (Shiba, Itaya, 2006);

- Modelarea dinamicii arboretelor – studii privind mortalitatea (Guan, Gertner, 1991; Hasenauer et al., 2001), prognoza compoziției specifice (Hasenauer, Merkl, 2001), competiția (Richards et al., 2008), respectiv regenerarea (Hasenauer, Kindermann, 2002; Kuplich, 2006) și producția arboretelor (Porras, 2007);

- Evaluarea și modelarea structurii spațiale (Aitkenhead et al., 2004);

- Identificarea arboretelor susceptibile la acțiuni de vătămare produse de boli și insecte (Park, Chung, 2006) sau provocate de vânt (Hanewinkal et al., 2004);

- Managementul forestier – aplicații ale mulțimilor fuzzy în modelele de management forestier (Boyland et al., 2006), sisteme expert (Dorn, Mitterbock, 1998), modelarea datelor necesare proceselor decizionale (Scrinzi et al., 2007).

### 3.4.2. Analiza spațială

Informația spațială în silvicultură este folosită în multe activități specifice – de la planificarea inventarierilor sau dezvoltarea unor modele ale dinamicii ecosistemelor forestiere, până la rezolvarea problemelor legate de regenerare sau de aplicare a lucrărilor silvotehnice. Gradul de heterogenitate al suprafețelor împădurite este puternic influențat de scara la care se desfășoară analiza unui fenomen sau proces. Dinamica spațio-temporală a unui atribut analizat este adesea influențată de scara spațială sau temporală a cercetării (Hurlbert, 1990), respectiv de mărimea suprafeței elementare de prelevare a datelor (Dale, 2004; Haining, 2004; Tokola, 2004). Pentru a elimina o parte din aceste influențe se recomandă stratificarea ierarhică a informațiilor pe niveluri diferite: al arborelui, arboretului, unități de administrare sau regiunii.

La *nivelul arborilor* – cel mai simplu model de proces spațial aplicabil în cazul studierii modului de distribuție în suprafață este modelul punctiform Poisson, cu cele trei variante de tipare: neomogene, grupate și uniforme.

Tiparele grupate sau aglomerate sunt frecvent folosite în cadrul modelelor de simulare în scopul determinării poziției puieților într-o regenerare naturală. Procesele uniforme sunt potrivite pentru analiza spațială a distribuției puieților în plantații. Procesele punctiforme de tip Cox, Neyman-Scott, Gibbs sau Markov sunt utilizate în caracterizarea dinamicii arboretelor deoarece cu ajutorul acestora pot fi incluse în modele şi caracteristici probabilistice.

La *nivelul landşaftului* mozaicul habitatelor influenţează diversitatea şi dinamica ecosistemelor forestiere. Eforturile de cuantificare a heterogenităţii landşaftului au început abia în ultimele trei decenii (Tokola, 2004), dar în ultimii ani s-au înregistrat progrese considerabile ce au condus la apariţia a numeroşi indicatori de evaluare a diferitelor aspecte ale heterogenităţii peisajului. Imaginile satelitare, tehnologia GPS sau tehnicile GIS au constituit instrumente utile în realizarea acestor progrese. Cercetările actuale în acest domeniu urmăresc determinarea unor indicatori caracteristici: compoziţia sau structura, configuraţia spaţială, gradul de conectivitate, contagiozitatea sau dimensiunea fractală.

Analiza proceselor punctiforme este o tehnică flexibilă şi modernă de studiere a modului de organizare spaţială a proceselor naturale. Punctul este unul din cele mai populare unelte aflate la îndemâna specialiştilor în diverse domenii. Versatilitatea sa se bazează pe simplitatea şi claritatea descrierii unor distribuţii spaţiale. El caracterizează în general entităţi a căror mărime, deşi nu este foarte redusă, în comparaţie cu distanţele dintre ele poate fi ignorată. Odată obţinută o hartă a punctelor pentru o populaţie de interes, se stabileşte un model pentru distribuţia spaţială respectivă. În operaţia de căutare a modelelor trebuie să se ţină cont de faptul că fiecare distribuţie este rezultatul interacţiunii a numeroase procese la un anumit moment din timp şi într-un anumit spaţiu, cercetătorul surprinzând în aşa numita fereastră de observaţie (Illian et al., 2008) un fragment al realităţii.

Uneori punctele identificate pentru un anumit proces sunt atât de numeroase încât se impune folosirea unor metode de sondaj. În analiza spaţială există două metode de selecţie a probelor, fiecare necesitând ulterior tehnici specifice de analiză: metoda bazată pe prelevarea datelor din suprafeţe de probă elementare (*quadrat sampling*) şi metoda bazată pe prelevarea datelor în funcţie de distanţe (distance sampling).

Prima metoda (*quadrat sampling*) (Gleason, 1920) se bazează pe stabilirea unor suprafețe de probă reprezentative pentru o populație, în care se vor număra câți indivizi (entități) apar în fiecare suprafață. Cea de a doua metodă (*distance sampling*) este caracterizată de măsurarea distanțelor până la indivizi sau a distanțelor dintre aceștia. Prin folosirea acestor metode se urmărește detectarea tipului de model (randomizat, agregat sau uniform) și/sau a modului de asociere dintre două sau mai multe grupuri.

În cadrul metodei de analiză a quadratelor (*quadrat sampling*) se înregistrează frecvența evenimentelor în fiecare quadrat și apoi se compară raportul dintre varianța frecvențelor experimentale și medie cu raportul unei distribuții randomizate. Metoda prezintă anumite puncte slabe – mărimea quadratului influențează rezultatele obținute; se măsoară dispersia bazată pe densitate și nu se identifică relațiile dintre elemente; încadrarea într-un anumit tipar se face pentru toată suprafața, nefiind identificate variațiile din interiorul acesteia.

Metoda de prelevare a datelor în funcție de distanțe (*distance sampling)* se fundamentează pe comparația distanței medii dintre puncte și vecini cu distanța medie caracteristică unei distribuții aleatoare. În acest fel se elimină o parte din problemele metodei precedente (influența mărimii quadratului), dar pot să apară erori cauzate de stabilirea limitelor. Această tehnică a fost dezvoltată inițial pentru studierea relațiilor din cadrul comunităților vegetale, la începutul anilor '50 apărând numeroase modele bazate pe distanța dintre indivizi (Pielou 1959; Holgate, 1965; Clark și Evans 1954). Majoritatea modelelor au fost concepute pe baza distanței până la cel mai apropiat vecin, fiind acceptată ipoteza conform căreia, într-o populație distribuită uniform într-o suprafață, distribuția distanțelor va fi o distribuție Poisson. În cadrul acestei paradigme sunt acceptate diferite teste de detectare a modelelor spațiale nealeatorii: indicele Fisher (et al., 1922), indicele de agregare (Clark, Evans, 1954), indicele de ne-randomizare (Pielou, 1959), coeficientul de agregare (Skellam, 1952) și raportul Holgate (1965).

Rezultate bune în stabilirea semnificației abaterilor de la modelul aleatoriu al unei distribuții s-au obținut și folosind tehnici de simulare Monte Carlo (Besag și Diggle, 1977). O formă nouă de analiză, numită procedura permutațiilor multi-răspuns (MRPP) (Mielke, 1986) a fost aplicată, de asemenea, cu succes.

Metodele bazate pe tehnica distanțelor sau a quadratelor, numite și metode de analiză spațială de ordinul I au fost aplicate în cercetarea forestieră de mulți ani, studiile fiind numeroase în acest domeniu (Daniels, 1978; Gratzer, Rai 2004; Houle, Duchesne, 1999; Pommerening, 2002; Stoyan, Penttienen, 2000).

În ultimii ani au devenit tot mai frecvent folosite *metodele de analiză spațială de ordinul al II-lea,* în special cele bazate pe funcția *K-Ripley* sau a variantelor acesteia (Ripley, 1977, 1981; Besag, 1977; Diggle, 1983; Getis, Franklin, 1987).

Aceste tehnici au fost utilizate în numeroase ercetări silvice care au urmărit stabilirea modelelor de organizare spațială a arboretelor (Awada, et al., 2004; Camarero et al., 2005; Dixon, 2002; Fajardo et al., 2006;  Goureaud, 2000; Dimov, 2004; Hofmeister et al., 2008; Maltez-Mouro et al., 2007; McDonald et al., 2003; Montes et al., 2007; Popa, 2006; Roibu, Popa, 2007; Wolf, 2005; Woodal, Graham, 2004).

Avantajul major al folosirii funcțiilor din gama *K Ripley* constă în faptul că acestea oferă o paletă extinsă de informații privitoare la organizarea spațială, comparativ cu statistica bazată pe analiza distanțelor dintre cei mai apropiați vecini. Practic sunt descrise caracteristicile proceselor punctiforme pentru o gamă largă de distanțe, o proprietate deosebit de importantă considerând faptul că rezultatele analizei proceselor ecologice sunt sensibil influențate de scara la care se desfășoară studiul.

Tot în scopul identificării caracteristicilor de organizare în spațiu se pot folosi indici care sunt utilizați în general pentru analiza autocorelației: indicele Moran (1948, 1950); indicele Geary (1954) sau testul Mantel (1967). Paluch (2005, 2007) a determinat caracterul spațial agregat al puieților folosind aceste tehnici de analiză.

Majoritatea tehnicilor de analiză spațială de ordinul I și II presupun adoptarea unor metode de corectare a erorilor provocate de efectul de margine, fiind propuse mai multe soluții pentru această problemă (Ripley, 1981; Besag, 1977; Getis, Franklin, 1987; Stoyan, Stoyan, 1994; Haase, 1995; Goreaud, Pellisier, 1999; Pommerening, Stoyan, 2006).

Metodele de corectare a efectului de margine sunt selectate în funcție de tipul și metodologia de prelevare a datelor, respectiv în funcție de metodele de analiză spațială folosite (Dale, 2004; Fortin, Dale, 2005).

În unele lucrări (Illian et al., 2008) se admite că în cazul unor dimensiuni mari ale *ferestrei de observație* și a unui număr mare de puncte analizate, efectul de margine poate fi ignorat, afectând în foarte mică măsură rezultatele obținute.

**Tipuri de modele de organizare spațială**

CSR (*Complete spatial randomness*) reprezintă ipoteza centrală a teoriei analizei proceselor punctiforme bazate pe distribuția omogenă Poisson. Devațiile de la ipoteza distribuției complet randomizate conduc la modele uniforme (distanța dintre evenimente este mai mare decât cea corespunzătoare distribuției CSR) sau la modele agregate (distanța dintre evenimente este mai mică).

Cercetările și studiile efectuate până în prezent au arătat că majoritatea populațiilor biologice se abat de la ipoteza distribuției CSR (Reich, Davis, 1998; Dale, 2004; Fortin, Dale, 2005). Distribuția spațială a acestora este determinată de principalele elemente ale mediului de viață, cum ar fi condițiile staționale și microclimatice, relațiile stabilite cu ceilalți indivizi ai populației, etc. Ecologiștii recunosc în prezent trei modele distincte de organizare spațială a populațiilor: modelul randomizat, modelul agregat și modelul uniform.

**Modelul aleatoriu (randomizat)** (*Random Spatial Pattern*)

Acest model se acceptă în caracterizarea unei distribuții spațiale atunci când pozițiile individuale ale entităților studiate sunt independente de pozițiile altor obiecte sau entități de pe o arie infinit de largă (Reich, Davis, 1998). Practic nu poate fi identificat nici un gradient între regiunile dense ale populației și regiunile mai puțin populate. Pentru un model spațial aleatoriu, distanța medie dintre punctele individuale este dată de relația:

$$\overline{d} = \frac{1}{2\sqrt{\lambda}} \qquad (3.1)$$

unde λ reprezintă densitatea punctelor pe unitatea de suprafață.

**Modelul agregat** (*Agregated Spatial Pattern*)

În cazul unei populații a cărei distribuție respectă modelul agregat, indivizii sunt grupați în aglomerări de densitate și mărime variabilă. Acest mod de distribuție este foarte întâlnit în natură (Pielou, 1961). S-au formulat mai multe teorii ce au încercat să explice acest mod de repartizare în spațiu a entităților. Cole (1946) a lansat ipoteza agregării ca formă a unui „*proces contagios*".

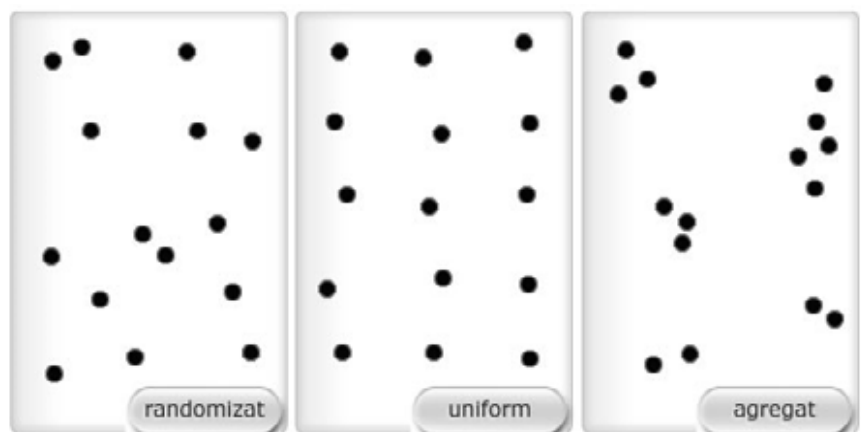

**Figura 3.5  Modele de organizare spațială (model randomizat, uniform și agregat)**

Această ipoteză susținea că prezența unuia sau a mai multe organisme într-o anumită locație influențează probabilitatea apariției altor organisme în respectivul spațiu. Pielou (1961) a explicat agregarea prin prisma particularităților procesului de reproducere al speciilor și a mecanismelor de dispersie caracteristice fiecărei specii. Thompson (1958) susține această idee, subliniind  faptul ca agregarea depinde de puterea de diseminare a speciilor vegetale.

Pe de altă parte, o mare importanță o au desigur și condițiile staționale ce pot influența abundența relativă a unei specii într-un anumit spațiu dat (Pielou, 1961). Metode ale statisticii spațiale aplicate în pedologie au arătat că soluri forestiere situate în zone aparent omogene pot să posede caracteristici foarte diferite (Jarvinen et al., 1993), fapt ce ar putea explica unele tipuri de agregare (Fox et al., 2007, a).

Daniels (1978) a analizat modelul spațial în cazul regenerării naturale a unor arborete de *Pinus taeda* și a arătat că gradul de agregare nu depinde de metoda de regenerare, de vârstă sau de desime, dar poate fi influențat de modalitatea de pregătire a terenului, de condițiile pedologice, de vegetația competitoare și de variația condițiilor micro-staționale. Wolf (2005) a identificat caracterul spațial agregat al regenerării arboretelor în urma unui studiu desfășurat pe o perioadă de 50 de ani în pădurile din Danemarca, multe alte cercetări indicând același fenomen, indiferent de metoda de analiză folosită (Awada, et al., 2004; Camarero et al., 2005; Fajardo et al., 2006;  Gratzer, Rai, 2004; Hofmeister et al., 2008; Maltez-Mouro et al., 2007; McDonald et al., 2003; Montes et al., 2007; Nigh, 1997; Paluch, 2005, 2007).

**Modelul uniform** (*Regular Spatial Pattern*)

Acest model este rar întâlnit în cazul proceselor naturale, fiind caracterizat printr-o distribuție uniformă a entităților într-o suprafață dată. Această repartizare în spațiu  poate fi reprezentată mai simplu printr-o rețea de triunghiuri, pătrate sau hexagoane.

S-au făcut numeroase speculații în ceea ce privește apariția în natură a acestui tip de model. Pielou (1961) sugera că acest mod de distribuție caracterizează o suprafață în care se desfășoară o competiție acerbă între indivizi, adeseori acest model fiind numit inhibitiv sau repulsiv. Cole (1946) descrie acest mod de distribuție ca fiind rezultatul unui *„proces contagios negativ"*. Anumite procese naturale *„resping"* alți indivizi cu o asemenea intensitate încât aceștia nu pot exista pe raza de acțiune a respectivului proces, generând distribuții regulate, uniforme.

În mod evident, această împărțire a distribuției în trei categorii (aleatoare, agregată și uniformă) nu are un caracter stabil și permanent. Este dificil de diferențiat un model agregat de unul aleatoriu sau unul uniform de unul aleatoriu, cu atât mai mult cu cât se cunoaște că această diferențiere este influențată de cele mai multe ori de scara la care se face analiza (Hurlbert, 1990; Dale, 2004; Haining, 2004; Tokola, 2004).

În plus, se apreciază că există o anumită dinamică a distribuției spațiale în funcție de stadiul de dezvoltare a arboretului, determinată se pare de intensitatea competiției și variația în timp a condițiilor microstaționale. Numeroase cercetări au arătat că modelele uniforme, în cazul arboretelor, converg spre modele randomizate dacă extragerile sunt la rândul lor aleatoare, iar modelele agregate, specifice în general regenerărilor naturale și arboretelor tinere, tind să se stabilizeze în timp (Gratzer, Rai, 2004; Houle, Duchesne, 1999; McDonald et al., 2003; Wolf, 2005).

Modul de organizare spațială influențează o gamă largă de procese - intensitatea competiției, dinamica creșterilor, modul de dezvoltare al regenerării, răspunsul la  influența unor factori perturbatori (Goreaud, 2000; Little, 2002; Fajardo et al., 2006; Xi et al., 2008). În acest context tiparele de organizare spațială permit explicarea și motivarea ecologică a unor procese intra sau interspecifice, mai ales că odată cu dezvoltarea modelelor  individuale de caracterizare a dinamicii ecosistemelor forestiere s-a acordat o importanță sporită datelor  privitoare la distribuția în spațiu.

Utilitatea informației spațiale în conceperea modelelor individuale de creștere este controversată. Unii cercetători susțin că informația privitoare la organizarea spațială este neconcludentă în activitatea de modelare (Biging, Dobbertin, 1995), afirmații determinate de performanța scăzută a indicilor de competiție dependenți de distanță. Sunt însă studii care au demonstrat că modelele individuale de creștere pot să beneficieze de pe urma informațiilor privitoare la organizarea spațială (Pukkala, Kolstrom, 1991; Pretzsch, 1997; Nigh, 1997, Aitkenhead et al., 2004; Little, 2002; Fox et al., 2007, a, b), arătând că doar folosirea incorectă sau nepotrivită a indicilor poate să conducă la rezultate nesatisfăcătoare.

Și în România structura și organizarea spațială a arboretelor a suscitat interesul cercetătorilor, fiind efectuate numeroase studii în acest domeniu, cu rezultate notabile (Cenușă, 1986, 1992, 1996 a, 2000; Iacob A.L., 1995; Iacob C., 1998; Barbu, Cenușă, 2001; Cenușă et al., 2002; Popa, 2001, 2006; Vlad, 2003; Roibu, Popa, 2007, Popa C., 2007).

Analiza modelului spațial de organizare a unei populații poate să ofere informații foarte utile cercetătorului silvic, dar trebuie să se țină cont de evenimentele ce au avut loc și de relațiile cauză-efect ce intervin la nivelul populației studiate. În condițiile în care mecanismele de interacțiune mediu-populație sunt extrem de complexe și prea puțin cunoscute, se impune o segregare a principalilor factori de mediu și o analiză a efectelor independente ale acestora în același timp, deoarece modelul spațial observat este, în fapt, un răspuns la întregul sistem de factori (Reich & Davis, 1998).

### 3.4.3. Modelarea geometrică și vizualizarea tridimensională

Modelarea geometrică computerizată a apărut în ultimele decenii, grație dezvoltării deosebite a sistemelor de calcul și reprezintă ansamblul de tehnici și procese de construire a unei descrieri matematice pentru un obiect fizic sau virtual. Practic, modelarea geometrică codifică geometria unui obiect prin intermediul unor structuri simbolice ce caracterizează aspectul spațial al corpului studiat. Această codificare implică construirea unei baze de date care să memoreze toate caracteristicile modelului.

Chiar dacă la prima vedere se poate bănui că ar exista o legătură între geometria analitică clasică și modelarea geometrică, între cele două concepte există o distincție clară. Dacă prima folosește reprezentări prin ecuații ale curbelor sau suprafețelor, cea de-a doua utilizează reprezentări procedurale pentru definirea acestor elemente. Reprezentarea procedurală este o schemă recursivă, care pornind de la un număr finit de puncte, numite puncte de control, poate să genereze puncte adiționale pe o curbă sau suprafață. În acest mod se pot genera forme cu o geometrie complexă, în baza unei scheme simple.

Faptul că se lucrează cu o cantitate enormă de date face practic imposibilă utilizarea modelării geometrice fără ajutorul unui sistem de calcul performant. Astfel au apărut numeroase aplicații software care oferă specialiștilor instrumente utile de generare, modificare și manipulare a modelelor geometrice, reducând în acest mod efortul depus pentru crearea unor scene complexe.

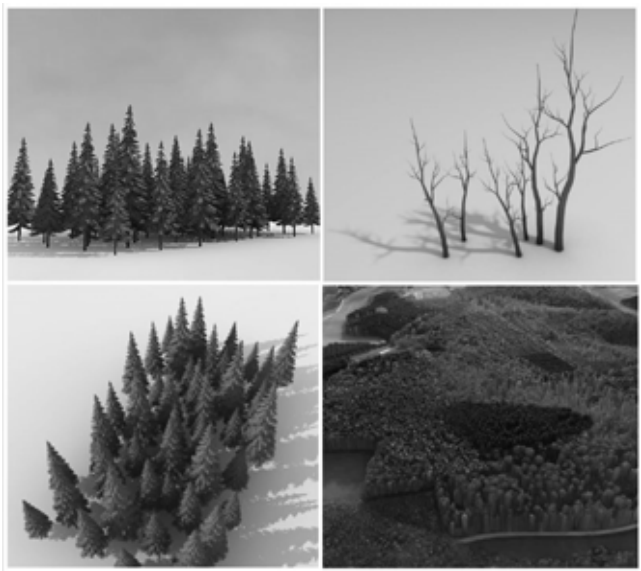

**Figura 3.6. Exemple de modelare computerizată a arborilor sau a peisajului**
(colaj de imagini - www.rssgmbh.de; www.fallingpixel.com; www.3dexport.com)

În principal, acest ansamblu de tehnici este folosit cu precădere în proiectarea inginerească – CAD (*Computer-Aided Design*), dar flexibilitatea și ușurința simulării unor procese prin modelare geometrică a condus la lărgirea semnificativă a sferei de aplicabilitate. În prezent, acest tip de modelare poate fi folosit cu succes în orice simulare a evenimentelor și fenomenelor ce implică o componentă a vizualizării efectelor.

În cadrul modelării geometrice există două paradigme fundamentale:

- cea bazată pe modelarea prin folosirea unor suprafețe NURBS (*Non-Uniform Rational B-Splines*);
- cea bazată pe modelarea prin tehnici de subdivizare.

Rezultatul acțiunii de modelare poate fi încadrat, în funcție de modul de prezentare în unul din tipurile următoare:

- modele de tip cadru de sârmă (*wire-frame*);
- modele de suprafață;
- modele solide (de volum).

Indiferent de tehnica sau sistemul de modelare adoptat, gama foarte variată de domenii în care se pot folosi modelele precum și flexibilitatea, complexitatea și claritatea simulărilor fac din acest ansamblu de tehnici și metode ale modelării geometrice un instrument indispensabil ingineriei moderne.

În ultimele două decenii, tehnicile moderne de prezentare vizuală computerizată a datelor, cum ar fi modelarea 3D, animația sau realitatea virtuală, și-au găsit locul în galeria de instrumente a cercetătorului silvic.

Domeniul forestier a fost interesat de valorificarea avantajului folosirii informației vizuale, mai ales datorită ușurinței perceperii și interpretării datelor prezentate. Primele încercări de modelare vizuală computerizată în acest domeniu au avut loc cu circa trei decenii în urmă și foloseau reprezentări abstracte de tip *wire-frame* (figura 3.7). Ulterior acestea au fost îmbunătățite cu ajutorul unor tehnici de pre-randare și a simulărilor bazate pe capturi de imagine (Weber, Penn, 1995; Max, 1996; Chiba et al., 1997).

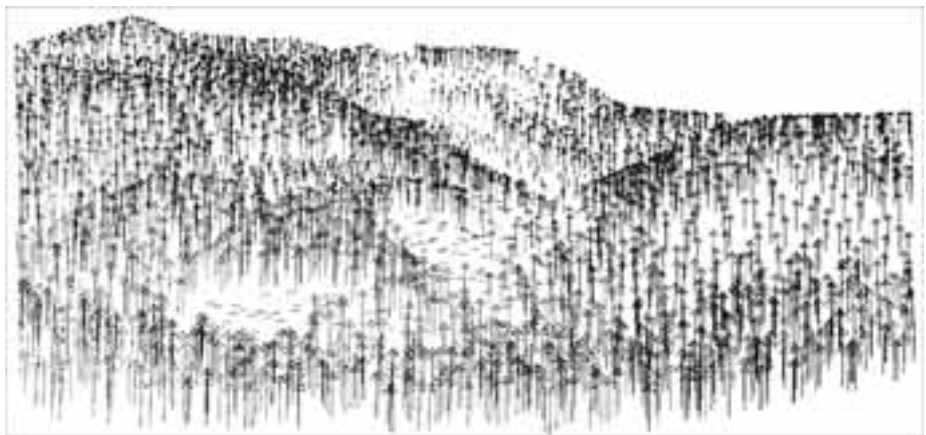

**Figura 3.7 Exemplu de modelare 3D tip wire-frame folosit în anii '80 în programul PREVIEW**

(Sheppard, Salter, 2004)

Song (2008) identifică trei tipuri de abordare în cazul prezentărilor vizuale computerizate ale ecosistemelor forestiere:

- folosirea figurilor geometrice elementare pentru reprezentarea coroanelor arborilor şi a unui număr mare de inventarieri forestiere pentru determinarea parametrilor de reprezentare (caracteristici dimensionale, poziția în spațiu). Această tehnică este puțin practică din perspectiva modului greoi de prelevare a atributelor de reprezentare. Este realistă, dar calitatea vizuală este în general redusă – cazul SVS (*Stand Visualization System*), EnVision, SmartForest.

- utilizarea unor tehnici de editare a fotografiilor pentru a obține texturi foarte detaliate (Bishop et al., 2003; Orland, 1994). În acest mod se obține o scenă vizuală la o calitate fotorealistă. Metoda se aplică cu succes în cazul simulării peisajelor, a suprafețelor întinse, dar folosește puține informații ecologice cantitative (compoziția specifică, desimea, caracteristicile dimensionale).

▪ combinarea unor elemente specifice tehnicilor anterioare - se folosesc informații referitoare la modelul digital al terenului și se inserează în pozițiile cunoscute ale arborilor reprezentări fotorealiste ce țin cont de informațiile calitative și cantitative prelevate din teren.

În anii '90 ai secolului trecut, utilizarea tot mai frecventă a calculatoarelor și a modelelor spațiale pentru arborete a determinat includerea în acestea din urmă a unor module de prezentare vizuală a structurilor modelate. Prezentarea vizuală computerizată a permis rularea unor scenarii alternative și observarea în timp real a efectului deciziilor luate, pe când în mod normal, consecințele lucrărilor de conducere a unui arboret sunt observabile după câteva decenii și au un caracter ireversibil.

Unul dintre primele programe informatice destinate modelării și vizualizării 3D a arborilor a fost **Sylvan Display Program** elaborat în cadrul Universității din Missouri – SUA (http://oak.snr.missouri.edu/qsl/display/ sylvan.display.html).

Programul a fost conceput de către Mark Davison în 1994 pentru a fi integrat în modelul de management silvicultural Sylvan (Larsen, 1994) și a introdus concepte inovative care aveau să fie ulterior preluate de majoritatea aplicațiilor de acest tip - vizualizarea grafică a structurii arboretului în plan orizontal, profil vertical sau tridimensional, analiza distribuțiilor și posibilitatea de marcare a arborilor.

Probabil cel mai cunoscut și utilizat sistem de reprezentare vizuală a datelor este **SVS** - ***Stand Visualization System*** (McGaughey, 1997), conceput la Universitatea din Washington pentru USDA Forest Service (http://forsys.cfr.washington.edu/svs.html).

Acest program se bazează pe modelarea individuală a arborelui și este capabil să reprezinte într-o formă grafică computerizată componentele unui arboret: arborii, arbuștii și lemnul mort.

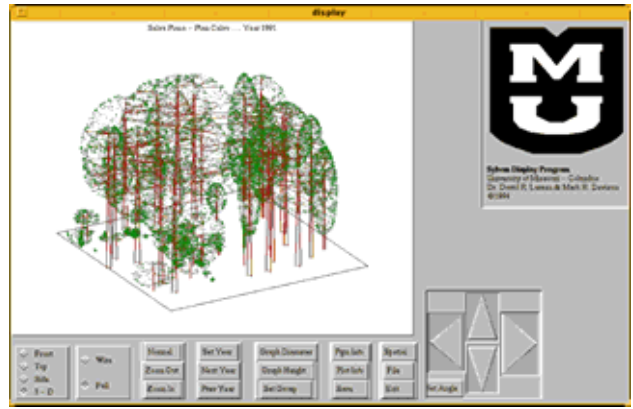

**Figura 3.8 Interfața aplicației Sylvan Display Program**

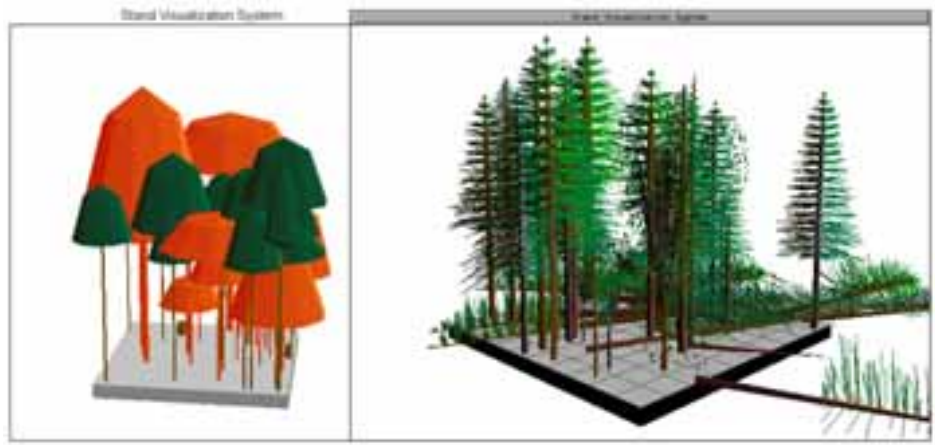

**Figura 3.9  Modelare 3D cu ajutorul aplicației SVS (*Stand Visualization System*)**

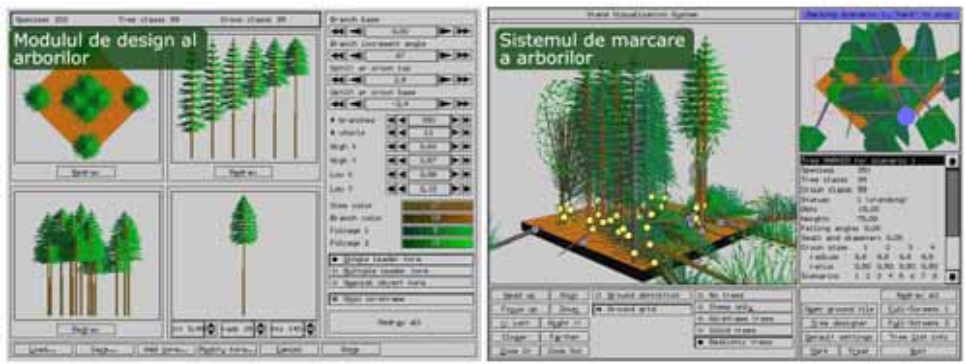

**Figura 3.10 Modulul de design și sistemul de marcare a arborilor în  aplicația SVS**

Aplicația SVS - Stand Visualization System este prevăzută cu un modul de design al arborilor care permite definirea formei de reprezentare a indivizilor dintr-o anumită specie sau clasă. Ulterior reprezentarea grafică tridimensională se realizează utilizând caracteristicile individuale introduse: specia, poziția în spațiu (coordonatele x,y,z), diametrul de bază, înălțimea, tipul de coroană, dimensiunile și raportul de formă al coroanei (razele coroanei după patru direcții).

Buna integrare cu modelele de simulare FVS - *Forest Vegetation Simulator* (Stage, 1973; Wykoff et al., 1982), LMS - *Landscape Management System* (McCarter, 1997) sau ORGANON (Hester, Hann, 1989), un sistem interactiv de marcare a arborilor în scopul simulării unor lucrări tehnice silvice sau a unor tratamente, precum și determinarea distribuțiilor principalelor caracteristici analizate au făcut din SVS unul din cele mai frecvent utilizate aplicații software de vizualizare computerizată a pădurii. Datorită modului flexibil de preluare a datelor, sunt numeroase modele forestiere nevizuale, care exportă rezultatele în fișiere compatibile cu aplicația SVS.

McGaughey (1997) este de asemenea, creatorul unui alt program de vizualizare computerizată numit **EnVision** (*Environmental Visualization System*), un software foarte versatil, cu suport GIS, proiectat pentru a oferi prezentări 3D la scări diferite – imagini ale arboretului, respectiv ale peisajului (http://forsys.cfr.washington.edu/envision.html).

Proiectul a demarat în 1998 și a fost finalizat un an mai târziu, preluând concepte originale ale sistemului Vantage Point Visualization dar integrând și facilitățile de structură spațială ale SVS.

Aplicația importă modelul digital al terenului, pe care aplică texturile caracteristice fiecărui tip de suprafață, reprezintă individual elementele de vegetație, ține cont de poziția sursei de lumină și aplică și efecte atmosferice variate. Imaginile produse sunt de o calitate superioară, fiind folosite pentru randare librăriile OpenGL.

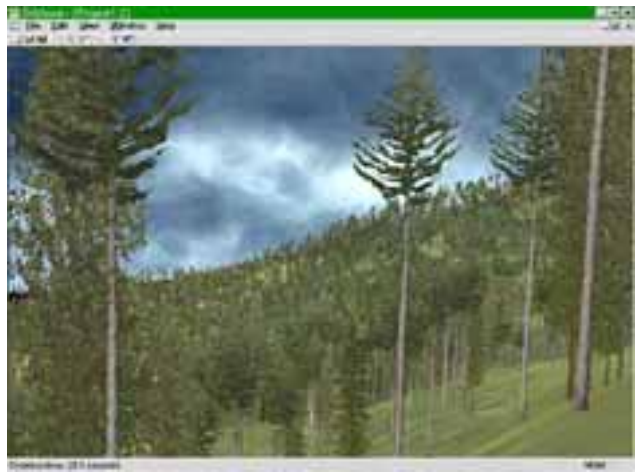

**Figura 3.11 Simulare a unui arboret cu ajutorul programului EnVision**

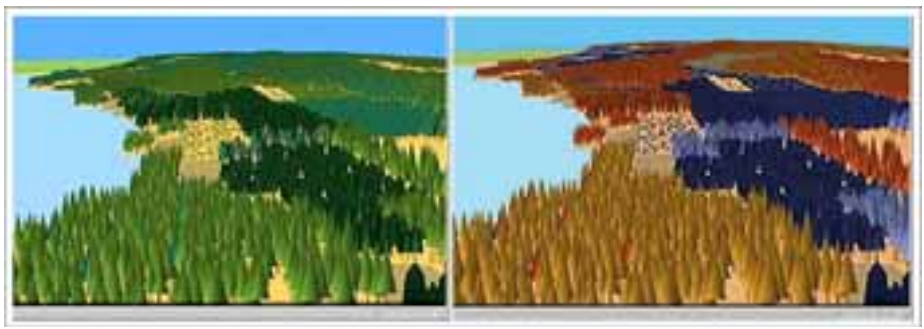

**Figura 3.12 Imagini generate de aplicația SmartForest**

**SMARTFOREST** (Orland, 1997; Uusitalo, Orland, 1997) este un program informatic de prezentare vizuală computerizată, dezvoltat la Universitatea Penn State (SUA) în cadrul IMLAB - *The Imaging Systems Laboratory* (www.imlab.psu.edu/ smartforest). Acesta este un mediu tridimensional interactiv de vizualizare a structurii  arboretelor și a peisajului, ce integrează  date spațiale georeferențiate, fiind compatibil cu ESRI ArcView. Sistemul reprezintă arborii  în baza atributelor acestora (tip, dimensiuni, poziție) și este capabil să  ofere imagini de ansamblu ale arboretelor cu o întindere de maxim 30 x 30 km.  Datorită modului intuitiv de prezentare a informațiilor într-o formă grafică de calitate, aplicația este folosită adeseori în tandem cu modele ecologice ale peisajului.

Virtual Forest  (Buckely et al., 1998) este un produs software creat de Innovative GIS Solutions, capabil să îmbine funcțiile de vizualizare cu cele ale bazelor de date forestiere în format GIS. Aplicația integrează un motor grafic 3D pentru generarea în timp real a imaginilor ecosistemelor forestiere analizate. Utilizatorul poate folosi anumite tipuri de teme predefinite care includ modelul digital al terenului, efectele atmosferice, iluminarea în funcție de poziția soarelui, variante alternative de modelare și texturare. Folosirea programului presupune doua etape – modelarea  tridimensională a arborilor cu ajutorul modulului *Tree Designer* și apoi  randarea peisajelor prin includerea obiectelor 3D create anterior, utilizând modulul *Landscape Viewer*.

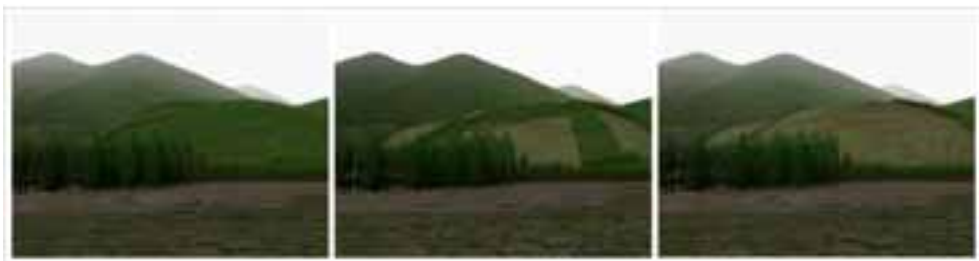

**Figura 3.13 Simulare vizuală a lucrărilor de exploatare pe un versant cu ajutorul Virtual Forest  (Buckely et al., 1998)**

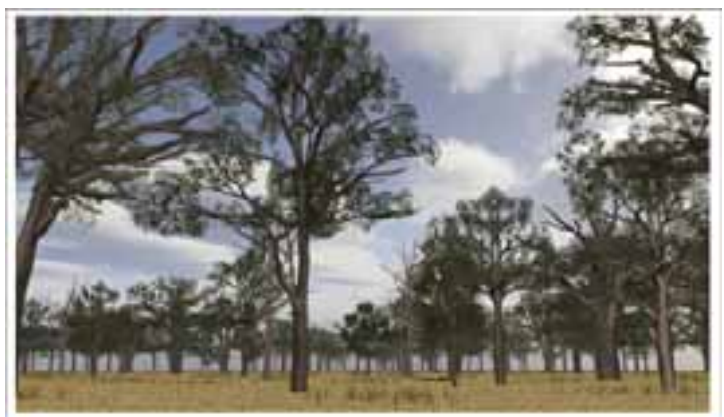

**Figura 3.14 Simulare a peisajului forestier cu ajutorul Red Gum Forest 3D**

Red Gum Forest 3D (Chandler, Saw, 2008) este un proiect prototip, dezvoltat în cadrul laboratorului de realitate virtuală al Universității Monash, Australia (http://vlab.infotech.monash.edu.au/simulations/virtual-reality/red-gum-3d), care urmărește să identifice modul în care prezentarea computerizată 3D poate să ajute cercetătorii și publicul în general să interacționeze cu ecosistemele forestiere.

În România a fost dezvoltată pînă în prezent o singură aplicație software de prezentare vizuală computerizată, PROARB (Popa, 1999). Aplicația a fost intens folosită de cercetătorii de la noi din țară pentru reprezentarea structurii orizontale, verticale și tridimensionale a arboretelor.

Programul a fost elaborat în cadrul Stațiunii Experimentale de Cultura Molidului Câmpulung Moldovenesc și a suferit numeroase îmbunătățiri ale conținutului în versiunile ulterioare. Pe baza datelor de intrare - coordonatele spațiale și parametrii biometrici ai fiecărui arbore, elemente ce pot fi importate direct din foile de calcul Microsoft Excel sau dintr-o bază de date de tip Access, se generează o reprezentare grafică a structurii arboretului.

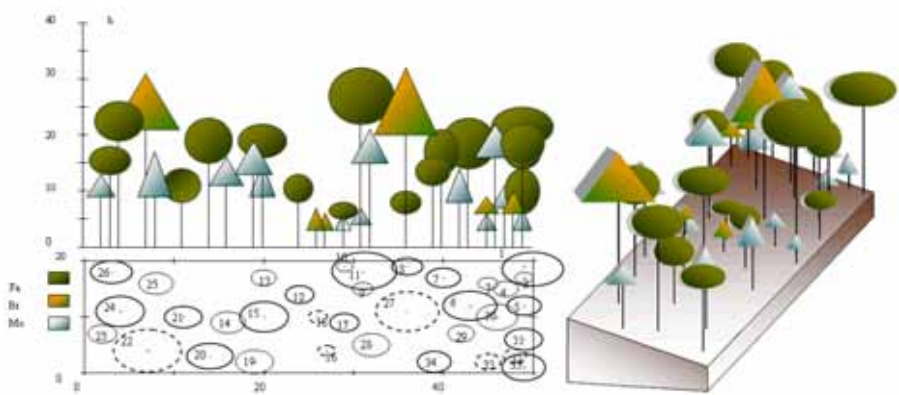

**Figura 3.15  Prezentarea structurii pădurii cu ajutorul programului PROARB**
(Popa, 1999)

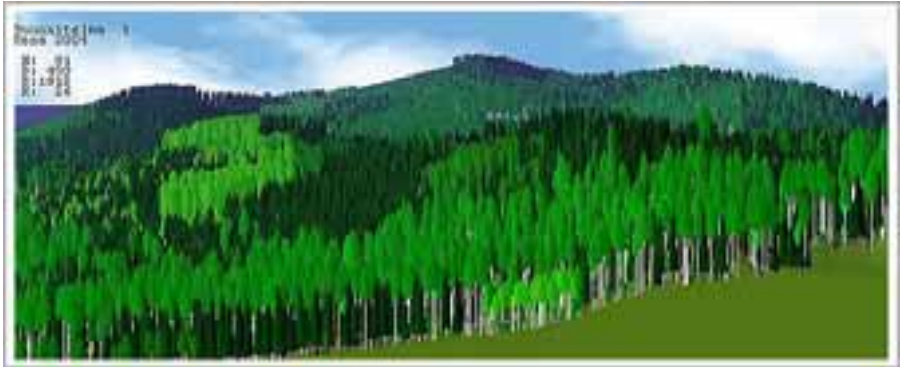

**Figura 3.16 Imagine generată de simulatorul MONSU**

(Pukkala, 2004)

Desigur există și alte aplicații software care se încadrează în categoria celor amintite în acest subcapitol, dar care rezolvă doar parțial problema vizualizării, consideră acest aspect de importanță secundară sau răspund unor nevoi particulare, fiind folosite de comunități științifice restrânse.

În această secțiune au fost menționate doar acele programe informatice care realizează exclusiv prezentarea grafică computerizată. Sunt însă numeroase aplicațiile de modelare a structurii, producției sau dinamicii forestiere care integrează module proprii de vizualizare: JABOWA (Botkin et al., 1972), SILVA (Pretzsch, 1992), MOSES (Hasenauer, 1994), BWIN-Pro (Nagel, 1996), LANDIS (Mladenoff et al., 1996), MONSU (Pukkala, 2004), SIBYLA (Fabrika, Iursky, 2006). În general, aceste module integrate oferă o calitate grafică inferioară aplicațiilor specializate menționate anterior, dar asigură un plus de informație utilizatorului, într-o formă explicită și intuitivă, ce poate fi accesibilă chiar și unui observator nespecialist.

Tehnicile de prezentare computerizată nu sunt instrumente perfecte. Performanța acestora trebuie apreciată prin prisma utilității și calității lor. Prezentările realizate în condiții improprii, cu dispozitive tehnice inadecvate, cele lipsite de realism, cele foarte complexe sau animațiile prea rapide pot altera transmiterea informației, creând confuzii sau chiar conduc la interpretări incorecte din partea observatorilor.

Sheppard (2004) avertizează asupra unor riscuri în folosirea acestor tehnici:

- crearea unor aşteptări nerealiste în rândul observatorilor care nu sunt familiarizaţi cu aceste tehnici;
- caracterul modern al prezentării poate să distragă atenţia observatorului şi să ascundă conţinutul real al mesajului;
- tehnologia sofisticată poate să îndepărteze anumiţi observatori, să le inducă o percepţie negativă care să conducă la refuzul acestora de a recepta informaţia, indiferent de acurateţea ei.

Karjalainen (2002) recomandă impunerea unor criterii obiective de apreciere a modelelor vizuale, care se referă la realismul, gradul de libertate oferit privitorului, capacitatea de a surprinde dinamica ecosistemelor forestiere şi nu în ultimul rând costul şi efortul de producţie al acestora.

Noile clase de instrumente de explorare vizuală vor corecta neajunsurile existente, vor beneficia de o putere de calcul superioară şi vor fi mai interactive, intuitive, dinamice, cu o interfaţă grafică ergonomică şi un realism superior, ajutând specialistul silvic să ia decizii corecte şi obiective în acţiunea de conducere a arboretelor.

Odată cu avansul tehnologic important înregistrat în ultimii ani în cazul graficii computerizate, atât la nivel hardware cât şi software (implementarea tehnicilor *OpenGL, Direct3D, VRML, X3D, WebGL, Vulkan),* preocupările în domeniul simulării vizuale a structurii sau dinamicii arboretelor au devenit tot mai numeroase (Bishop et al., 2003; Lewis et al., 2004;  Meitner, 2005; Song et al., 2006, 2008; Stoltman, 2007; Zamuda, 2007), fapt ce confirmă potenţialul ridicat al acestor tehnici în activitatea managerială şi de cercetare ştiinţifică a resurselor forestiere.

### 3.3.4. Folosirea imaginilor satelitare şi a tehnicilor GIS

Dacă primele încercări de modelare au vizat arborele (Botkin et al., 1972; Ek, Monserud 1974), apoi grupurile de arbori şi arboretele, odată cu trecerea timpului s-a încercat predicţia proceselor la o scară mai mare - comportamentul unui ecosistem sau chiar a unui grup dintr-o anumită regiune (Botkin et al., 1972; Urban et al., 2005). Nici un sistem nu poate fi corect apreciat în afara contextului în care se află. Relevanţa compoziţiei, structurii şi funcţiilor sistemului ca întreg nu reiese doar din analiza unui ecosistem izolat ci şi din utilizarea unui factor de scară mai mare. Aşa cum o pădure este mai mult decât o simplă aglomerare de arbori, landşaftul reprezintă mai mult decât o sumă de arborete (Bragg, 2004).

Trei tehnologii de caracterizare spaţială a datelor – GIS (*Geographical Information System*), GPS (*Global Positioning System*) şi teledetecţia au schimbat modul în care societatea percepe, evaluează şi administrează resursele naturale (Bolstad, 2003).

Fotografiile aeriene sunt folosite în evaluarea resurselor forestiere încă din 1930 în SUA, în prezent constituind o tehnologie matură, cu multiple aplicaţii în silvicultură (Bolstad, 2003). Imaginile satelitare au fost utilizate în domeniul forestier începând cu anul 1972, odată cu lansarea primului satelit tip LANDSAT (Bolstad, 2003; Barrett, Fried, 2004). De atunci au fost lansaţi şi alţi sateliţi comerciali ce furnizează imagini: SPOT, GeoCover, Globe 15, RADARSAT. În ultimul deceniu au apărut sateliţi care furnizează imagini de înaltă rezoluţie, cu o dimensiune redusă a pixelilor: IKONOS (lansat în anul 1999), cu o rezoluţie în pancromatic de 0,8 m, QuickBird (2001) - 0,6 m, ORBVIEW-3 (2003) - 1 m, WorldView (2007/2014) - 0,5 m/0,31 m, GeoEye-1 (2008) - 0,4 m.

Datele georeferenţiate oferite de teledetecţie sunt uşor de integrat în bazele de date GIS, creând posibilitatea aplicării modelelor pe suprafeţe întinse, lucru deosebit de important pentru generalizarea anumitor observaţii sau pentru identificarea unor tendinţe în dinamica pădurilor. Imaginile satelitare şi fotografiile aeriene sunt capabile să ofere observaţii rapide privind zone foarte largi, ce ar putea fi cu greu înregistrate în altă manieră. Apariția dronelor a simplificat și mai mult procesul, permițând obținerea de material imagistic într-un mod flexibil, rapid și la un cost mult mai redus.

Un alt avantaj al folosirii materialului imagistic se referă la sensibilitatea înregistrării anumitor caracteristici. Filmul fotografic este limitat la un interval îngust al sensibilității electromagnetice (400-900 nm), în timp ce senzorii optici digitali moderni pot opera într-un interval mai larg (400-1400 nm) (Wulder, 2004). Imaginile preluate folosind filtre speciale, nu doar în spectrul vizibil, pot să conțină informații în mod normal ascunse ochiului uman, care pot fi folosite în domenii extrem de variate – de la simpla inventariere sau delimitare a arboretelor cu structuri, compoziții sau vârste diferite, până la aprecierea riscului apariției unor evenimente nedorite în viața pădurii (poluare, incendii, doborâturi, atacuri de insecte etc.) sau la aprecierea productivității arboretelor. De exemplu, pentru diferențierea tipurilor de vegetație sunt folosite în general fotografiile în infraroșu (Bolstad, 2003).

Datele digitale preluate de senzori nu oferă direct evaluări ale atributelor forestiere, fiind înregistrate intensitatea radianței electromagnetice în diferite benzi spectrale sau lungimi de undă. Ulterior se efectuează prelucrări ale acestor date pentru extragerea informațiilor. Couturier (2009) subliniază importanța folosirii unor imagini adecvate obiectivelor propuse, precum și a unor sisteme sau programe performante de clasificare a modelelor de răspuns spectral.

Scopul acestei clasificări urmărește stabilirea atributelor specifice pădurii. Interesează de asemenea identificarea acelor atribute care sunt considerate cele mai reprezentative pentru o suprafață forestieră. Atributele de interes forestier (compoziția specifică, consistența, vârsta, productivitatea, elemente biometrice etc.), odată determinate vor ajuta la procesul de clasificare a modelelor de răspuns spectral. Pentru clasificarea cvasi-automată a ecosistemelor forestiere pot fi folosite aplicații software ca: ERDAS Imagine, ITT Envi sau Definiens eCognition (Shiba, Itaya, 2006).

Franklin (2001) identifica drept posibile domenii de utilizare a analizei imaginilor satelitare sau a fotografiilor aeriene:

- clasificarea tipurilor de ecosisteme forestiere;
- inventarierea forestieră şi managementul resurselor;
- detectarea modificărilor din ecosistemele forestiere provocate de factori perturbatori naturali sau antropici;
- modelarea forestieră – mai cu seamă integrarea în modelele de procese şi în modelele specifice ecologiei landşaftului.

**Figura 3.17 Exemplu de hartă de clasificare a vegetaţiei forestiere pe baza imaginilor satelitare** (Giriraj, 2008)

**Figura 3.18 Răspunsul spectral în funcţie de vârsta pădurii**

(adaptare după Franklin, 2001)

Au fost dezvoltate modele flexibile, capabile să acopere rezoluții spațiale diferite, fiind aplicabile la scara arboretului, la scară locală sau chiar regională – e.g. modelul EFIMOD (Chertov et al., 2006), care este axat pe 3 niveluri:

- nivelul arboretului - implementat pentru analiza efectelor mediului asupra ecosistemului forestier și pentru compararea diferitelor tipuri de rărituri;

- nivel local - aplicat pentru simulări pe perioade îndelungate ale efectului aplicării diferitelor regimuri silviculturale în arborete din Rusia Centrală;

- nivel regional - determinarea pe o arie întinsă a rolului solurilor forestiere în circuitul carbonului.

În ceea ce privește regenerarea, tehnica GIS creează posibilitatea unor analize pertinente cu privire la amplasarea în spațiu, respectiv cu privire la dinamica acestui proces. Devine astfel mult mai facilă studierea atentă a dinamicii spațiale și a evoluției calitative a suprafețelor regenerate.

Imaginile satelitare nu au fost extrem de folosite în studiul regenerării, dar există cercetări în acest domeniu care arată că se pot identifica zonele de regenerare și chiar diversele faze de dezvoltare ale regenerării, prin analiza reflectanței spectrale (Peterson, Nilson, 1993; Kuplich, 2006). Peterson și Nilson (1993) au introdus conceptul de traiectorie a reflectanței unui arboret, pe baza căruia se consideră că se poate prognoza evoluția dinamicii unui arboret, în ceea ce privește stadiile de dezvoltare.

Reprezentând un complex de tehnici foarte moderne, tehnicile GIS și teledetecția sunt capabile să realizeze, de multe ori, interacțiuni cu metode și tehnologii cel puțin la fel de moderne, care pot să îmbunătățească rezultatele cercetărilor efectuate. Peng (2000), recomandă utilizarea inteligenței artificiale, mai precis a rețelelor neuronale, în activitatea de prelucrare a datelor oferite de teledetecție, și alți cercetători surprinzând creșteri semnificative ale capacităților decizionale în cazul folosirii tandemului rețele neuronale – imagini satelitare (Franklin, 2001, Kuplich, 2006).

O altă tehnologie emergentă se referă la scanările de tip LIDAR (*Light Detection and Ranging*) terestre sau aeriene, cu aplicații în studiile de analiză a coronamentului, suprafeței foliare, biomasei (Lee et al., 2007; Hudak et al., 2008). Perfecționarea acestei tehnologii va permite, odată cu scăderea costurilor de utilizare, folosirea ei mai frecventă în activitatea de inventariere forestieră. Dezvoltarea continuă a tehnologiilor computaționale a condus de asemenea la prelucrări mai rapide și mai facile ale complexului nor de puncte generat de scanările LIDAR. Posibilitățile care se deschid prin folosirea acestei tehnici în inventarul forestier și în analiza detaliată a parametrilor structurali ai pădurii au fost deja intuite și folosite în ultimii ani în domeniul silvic (Parker & Evans, 2009; Van Leeuwen, & Nieuwenhuis, 2010).

În ultimii ani și în România au fost realizate cercetări cu privire la potențialul folosirii imaginilor satelitare (Gancz, 2005; Barnoaiea, 2011) sau a tehnologiei LIDAR (Apostol, 2015) în silvicultură, studii de monitorizare a pădurii și a stării ei fitosanitare (Zoran et al., 2003; Barnoaiea & Iacobescu, 2008), de clasificare a arboretelor și analiză a structurii (Vorovencii, 2006; Barnoaiea, 2010) și de determinare a volumului acestora pe baza imaginilor satelitare (Barnoaiea, 2007).

Studiile de până acum demonstrează că tehnicile LIDAR, GIS și teledetecția sunt instrumente foarte utile în acumularea unor cantități importante de observații, ce pot fi utilizate cu succes în scopul analizei regenerării arboretelor (Bollandsas et al., 2008; Brooks, Merenlander, 1998; Kuplich, 2006; Qiu et al., 2008). Informațiile rezultate în urma folosirii acestor metode pot fi de mare importanță pentru managerii forestieri în procesul de instalare a pădurii și ulterior de conducere a acesteia.

# Capitolul 4
# Considerații cu privire la modelarea procesului de regenerare a arboretelor

## 4.1. Caracteristici ale modelării regenerării

Permanența pădurilor, în scopul receptării continue a beneficiilor izvorâte din funcțiile de producție și protecție atribuite, presupune o perpetuă înnoire a biocenozei forestiere la nivelul indivizilor. Reînnoirea în cazul ecosistemelor forestiere este un proces natural, dictat de legile firii, ce se produce continuu sau periodic. Similar acestui proces, în pădurea cultivată vorbim despre regenerarea arboretelor, care presupune înlocuirea arborilor ajunși la o anumită vârstă, ce se extrag prin tăiere, cu exemplare tinere obținute prin procese generative sau vegetative. Termenul de regenerare se referă atât la procesul în sine de perpetuare a pădurii cât și la noua generație de arbori instalați.

Pentru a nu lua decizii inadecvate, care să compromită procesul de regenerare al pădurii, s-a simțit nevoia simulării acestui fenomen complex, în baza unor condiții date, a unor scenarii particulare, care să ofere anumite răspunsuri asupra posibilităților de evoluție a arboretului. În acest fel există și avantajul legat de posibilitatea efectuării simulării într-un timp mult mai scurt decât cel necesar realizării unor teste în condiții reale.

În ceea ce privește procesul de regenerare al arboretelor, obiectivele urmărite sunt dominate de trei aspecte:

- *dinamica spațială a regenerării* – analizată în general prin modele deterministe, dependente de distanță, care doresc să determine posibilitățile de dezvoltare spațială a regenerării;

- *dinamica compozițională* – ce urmărește realizarea unei prognoze asupra evoluției ponderii de participare a speciilor în alcătuirea arboretului (se folosesc de regulă metode stohastice – în special modelarea prin lanțurile Markov);

- *aspecte privitoare la eventuale avantaje economice* – în acest caz se folosesc cu precădere modele de creștere și evaluare a producției sau sisteme expert care să ofere informații și prognoze utile.

În timp ce procesele ce se desfășoară în etajul arborilor maturi sunt relativ bine cunoscute, multe procese caracteristice regenerării arboretelor sunt incomplet înțelese deoarece este dificil și uneori chiar imposibil de evaluat toți factorii ce condiționează regenerarea. În plus apar numeroase procese și fenomene aleatoare în timp și spațiu care influențează apariția, creșterea, integritatea și mortalitatea plantulelor.

Modelarea regenerării privită în detaliu, implică aspecte extrem de variate, care țin nu doar de biologia și ecologia speciilor ci și de mediul în care se desfășoară procesul de regenerare.

În general modelarea regenerării include procesele de producție, diseminare și germinare a semințelor, precum și a estimării numărului de plantule/puieți (ținând cont de supraviețuirea plantulelor în primul sezon de vegetație și de mortalitatea puieților). De exemplu producția de semințe și dispersia este legată de specie, vârstă, mărimea și vigoarea arborelui semincer, condiții staționale, în timp ce procesul de germinație a semințelor are un caracter mai degrabă stohastic. Instalarea semințișului este, de asemenea, un proces complex, fiind implicate procese ecofiziologice, condițiile de mediu, precum și diverse evenimente aleatoare ce pot perturba instalarea.

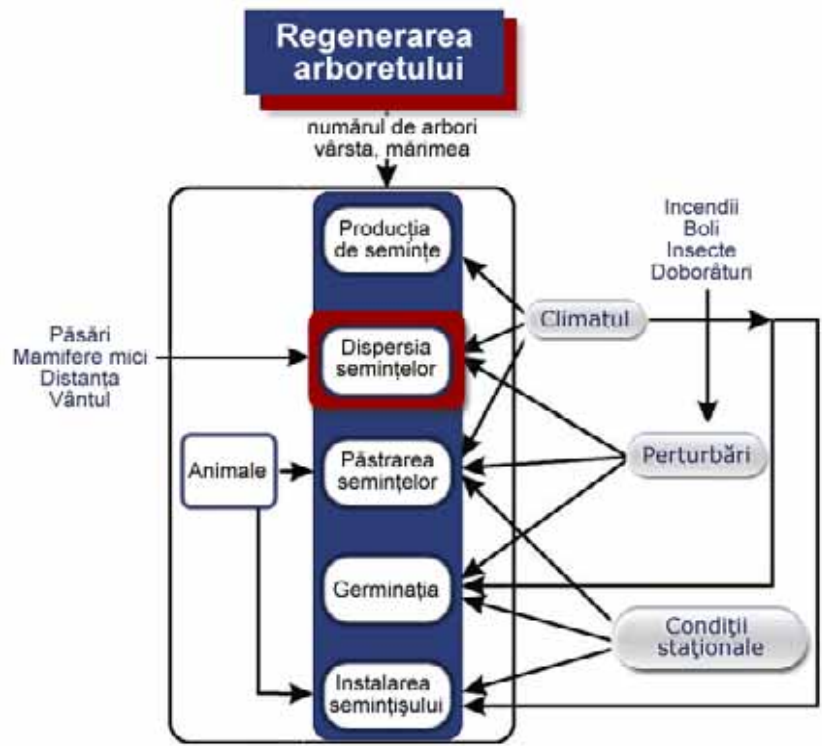

**Figura 4.1 Componentele procesului de regenerare a arboretelor**

(adaptare după Price et al., 2001)

Majoritatea modelelor, cu mici diferențe, fac referire la următoarele procese:

- producția de semințe;
- dispersia semințelor;
- eventualele condiții de păstrare a semințelor (până la germinare);
- germinarea semințelor;
- instalarea semințișului,
- mortalitatea plantulelor și puieților,
- competiția intra și interspecifică.

Fiecare dintre procesele enumerate sunt influențate de nenumărați factori, care la rândul lor sunt evaluați stohastic sau determinist, în funcție de posibilitatea de a determina relații funcționale între elementele sistemului.

   **Producția de semințe** – în general este modelată în funcție de anumite caracteristici, cum ar fi: specia, vârsta și dimensiunile arborelui semincer. De exemplu, Dey (1995) a folosit pentru evaluarea producției de ghindă a cinci specii de stejar o funcție dependentă de diametrul de bază și aria proiecției coroanei arborelui. Pennanen a modelat producția de semințe în baza unei relații determinate empiric (Pennanen et al., 2004):

$$Q_o = 3067 \cdot B \cdot m^{-0.58} \qquad (4.1)$$

   unde Qo – cantitatea de semințe, B – suprafața proiecției coroanei, m – masa medie a unei semințe.

   Unii cercetători consideră că modelarea empirică a producției de semințe doar în baza caracteristicilor biometrice nu oferă întotdeauna cele mai bune rezultate (Burgos et al., 2008). Calama et al., (2008) susține aceasta opinie, obținând cele mai bune rezultate în evaluarea producției prin folosirea unui model empiric ecologic ce include în procesul de determinare și altitudinea, tipul solului, date climatice și geologice.

   Rezultatele determinărilor producției de semințe sunt frecvent afectate și de periodicitatea de fructificație (Du et al., 2007), fiind dificil de integrat în modele acest fenomen complex. Abordarea stochastică a periodicității și sincronizării fructificațiilor abundente pare să ofere rezultate interesante (Palaghianu și Drăgoi, 2015), ce pot fi integrate în viitoare modele.

   **Dispersia semințelor** – diseminarea reprezintă principala formă de migrare a plantelor. Acest proces determină dinamica spațială a pădurilor, dinamică influențată de numeroase caracteristici: vârsta la care fructifică specia respectivă, cantitatea de semințe produsă și calitatea acestora, distanța pe care semințele se pot împrăștia, anumiți factori locali ce se pot erija în vectori ai semințelor (vântul, mamifere mici, păsări, cursuri de apă etc.) precum și condițiile de germinație (patul de germinație, condițiile microclimatice) și particularitățile biologice ale germinației.

   Dispersia deține un rol important în procesul de regenerare, fiind frecvent modelată în funcție de distanța la care pot ajunge semințele, care la rândul ei este influențată de viteza de cădere a semințelor, înălțimea de cădere, viteza și direcția vântului, perioada și durata diseminării.

Cercetări în acest sens au fost efectuate de către Pukkala și Kolstrom (1987, 1992) respectiv Kuuluvainen și Pukkala (1989). Pennanen et al., (2004) a modelat, de asemenea, dispersia în funcție de producția de semințe și distanța pe care sămânța o poate parcurge, corectată cu viteza vântului și înălțimea coroanei.

He și Mladenoff (1999) modelează dispersia ca o funcție dependentă de distanța maximă MD (distanța la care sămânța are probabilitatea de a ajunge de $p < 0,001$) și distanța efectivă de dispersie ED (distanța la care sămânța are cea mai mare probabilitate de a ajunge, $p > 0,95$). În cazul unei distanțe oarecare $x$, cu MD $> x >$ ED, probabilitatea $P$ ca sămânța să ajungă la respectiva distanță se calculează conform formulei:

$$P = e^{-b(\frac{x}{MD})} \tag{4.2}$$

unde $b$ este un coeficient de formă ce depinde de tipul dispersiei.

Se recomandă ca rezultatele evaluărilor să includă și dispersia secundară produsă de păsări, mici rozătoare sau alte mamifere (Xiao et al., 2004; Iida, 2006; Gomez și Hodar, 2008).

În ultimii ani modelarea dispersiei a cunoscut maniere diferite și moderne de abordare – prin tehnici de analiză spațială (Sagnard et al., 2007), tehnici de teledetecție și GIS (Qiu et al., 2008) sau folosirea de markeri genetici (Chen et al., 2008).

**Germinația** – în acest caz modelarea pune accentul pe particularitățile biologice ale speciilor (perioada de germinație), pe substratul asigurat germinării (caracteristicile patului de germinație, pregătirea terenului, caracteristicile solului, eventuala lucrare a acestuia) și pe condițiile de mediu (condițiile staționale, de microclimat).

Sunt puține cercetări privitoare la modelarea acestui proces, literatura de specialitate recomandând în acest sens tehnici de abordare mecaniciste (fiziologice) (Price et al., 2001).

**Instalarea semințișului** – modelarea în acest caz se referă la supraviețuirea plantulelor în primul an de vegetație, respectiv la mortalitatea puieților. Interesează în final numărul de puieți instalați la unitatea de suprafață.

Greene și Johnson (1998) (citați de Pennanen et al., 2004) modelează stohastic supraviețuirea plantulelor, în funcție de caracteristicile patului de germinație, folosind ecuații cu coeficienți determinați empiric. Bravo-Oviedo (2006) folosește un model logistic condiționat de competiție, dimensiunea arborilor și bonitatea stațiunii. Datorită dificultăților decizionale, unele cercetări folosesc pentru îmbunătățirea rezultatelor obținute în procesul de modelare a mortalității tehnici specifice rețelelor neuronale (Guan, Gertner, 1991; Hasenauer et al., 2001; Hasenauer, Kindermann, 2002).

Indiferent de metodele folosite în modelele ce caracterizează populațiile de arbori, mortalitatea este un eveniment discret, ce poate lua doar două valori - 0 (*mort*) sau 1 (*viu*), extrem de variabil și greu de prognozat, fiind caracterizat din punct de vedere spațial de un aspect *uniform* (ca proces aleator desprins din competiția dintre cei mai apropiați vecini), respectiv *neuniform* (rezultat al unor factori perturbatori sau limitativi).

Modelarea mortalității comportă în general o discuție privind două aspecte ale caracterului acesteia (Botkin, 1993):

- ▪ *mortalitatea intrinsecă* – ce are loc în condiții de mediu favorabile și nu este influențată de prezența sau absența competiției. Ea este modelată stohastic, printr-o funcție ce depinde de vârsta maximă observată pentru o specie, într-o populație dată. Majoritatea modelelor ecologice (e.g. cele de tip *gap models*) se bazează în cazul mortalității pe o probabilitate constantă de 1% - 2% (din numărul total de indivizi) ca un arbore să ajungă la limita vârstei biologice. Din cauza faptului că sunt și alte surse ale mortalității, mai puțin de 1- 2% dintre arbori ajung efectiv la maximul de vârstă al speciei lor. Unii cercetători (Bigler, Bugmann, 2004; Wunder et al., 2006) critică folosirea acestor funcții teoretice de evaluare a mortalității, recomandând în vederea obținerii unor performanțe superioare de estimare folosirea unor funcții empirice obținute pe baza creșterilor anuale în diametru.

- *mortalitatea dependentă de creștere* – este mortalitatea provocată de competiție, datorită creșterilor slabe induse de concurența acerbă pentru resurse. Acest tip de mortalitate are un caracter uniform din punct de vedere spațial și se modelează stohastic printr-o funcție ce depinde de creșterea în diametru. Se bazează pe ipoteza potrivit căreia arborii cu creșterile cele mai slabe vor fi cel mai probabil primii eliminați. Botkin (1993) afirmă că o creștere radială anuală mai mică de 0,1 mm conduce la o creștere a mortalității arborilor în următorii 10 ani (aprecia că 30% din arborii cu creșteri reduse, de sub 0,1 mm, sunt eliminați în următorul deceniu).

- Keane et al., (2001) indică și un al treilea aspect al mortalității: *mortalitatea exogenă*, neuniformă din punct de vedere spațial, ce rezultă din acțiunea unui factor extern disturbant (incendii, doborâturi provocate de vânt, atacuri de insecte, agenți fitopatogeni).

**Competiția** - este un proces complex în care numeroase elemente biotice și abiotice interacționează, ce caracterizează lupta pentru resurse a arborilor. Datorită faptului că nu poate fi măsurată direct se utilizează indici sintetici capabili să ofere o imagine relevantă asupra relațiilor dintre arbori. Indicii de competiție furnizează informații utile în activitatea de modelare, fiind de mult timp integrați în modelele de creștere a arboretelor.

În general în literatura de specialitate se acceptă clasificarea după Munro (1974) a indicilor de competiție:

- indici independenți de distanță;
- indici dependenți de distanță.

*Indicii independenți de distanță* – folosesc în analiza competiției comparații între caracteristicile biometrice (suprafața de bază, volumul coroanei, suprafața proiecției coroanei, înălțimea etc.) ale arborilor, care exprimă de fapt diferența între pozițiile sociale ale indivizilor. Acești indici sunt mai ușor de folosit deoarece nu conțin informații cu privire la poziționarea arborilor dar se consideră că oferă rezultate inferioare ca precizie comparativ cu celălalt tip de indici. Biging și Dobbertin (1995) arată însă că noile clase de indici de acest tip furnizează rezultate îndeajuns de precise, efortul de prelevare a datelor spațiale nefiind justificat întotdeauna de sporul de precizie obținut.

*Indicii dependenți de distanță* – includ nu doar de caracteristicile biometrice, ci și pozițiile arborilor raportate la vecinii lor. Acești indici sunt capabili să surprindă mai fidel relațiile de competiție dintre indivizi și prin urmare sunt preferați indicilor independenți de distanță. Practic, se evidențiază influența arborilor vecini și se permite selectarea competitorilor direcți ai unui arbore studiat. Pentru selecția competitorilor au fost definite metode variate – pe baza unei distanțe limită sau a unui număr maxim de vecini, respectiv pe baza unor expresii matematice în care sunt implicate distanțele dintre arbori și caracteristicile biometrice ale acestora – diametrul de bază, diametrul coroanei, înălțime. Marea varietate a acestor indici a impus clasificarea acestora în mai multe categorii în funcție de modul de apreciere al procesului competițional. Astfel, după Holmes și Reed (1991) se deosebesc trei clase:

- indici bazați pe suprapunerea ariilor de influență (*influence zone overlap*) – aria de influență fiind suprafața necesară unui arbore pentru a se dezvolta. Acești indici se bazează pe ideea potrivit căreia competiția dintre doi arbori vecini este caracterizată de suprapunerea *ariilor de influență*, concept introdus de Staebler (1951, citat de Schaer, 1981). După determinarea ariilor de influență și identificarea zonelor de suprapunere se trece la calcularea indicelui de competiție conform unei formule care diferă de la indice la indice. Cei mai cunoscuți indici de acest tip sunt: indicele Staebler (1951), Bella (1971) (citați de Schaer, 1981), Arney (1973), Opie (1968) (citat de Smith, 1987).

- indici ce utilizează relațiile dintre dimensiunile arborilor (*size ratio*) – aceştia sunt cei mai utilizați indici, datorită uşurinței cu care pot fi calculați şi rezultatelor bune obținute. Ei se bazează pe ipoteza conform căreia concurența este invers proporțională cu distanța dintre arbori şi direct proporțională cu raportul dintre dimensiunile arborilor vecini şi dimensiunile arborelui studiat. Indicii cei mai cunoscuți: indicele Hegyi (1974, citat Schaer, 1981), Lorimer (1983), Pukkala şi Kolstrom (1987), Schutz (1989, citat de Ung et al., 1997). Aceşti indici, deşi prezintă un mod de calcul complex, au avantajul uşurinței în algoritmizare, fiind frecvent folosiți în determinarea relațiilor de competiție dintre arbori.

- indici dependenți de spațiul de dezvoltare (*growing space*) – sunt fundamentați teoretic pe conceptul de arie potențial disponibilă (*area potentially aveilable*) (Brown, 1965, citat Kenkel et al., 1989) ce poate fi definită drept suprafața pe care un arbore o are la dispoziție pentru creştere şi dezvoltare. Cu cât această suprafață caracteristică unui arbore este mai mare, cu atât va fi mai mică valoarea indicelui de competiție pentru acel arbore. Cercetările lui Mark şi Esler (1970), Moore et al., (1973), Smith (1987) sau Mercier şi Baujard (1997) indică faptul că această abordare, deşi puțin folosită oferă rezultate interesante.

O modalitate modernă de evaluare a intensității relațiilor competiționale o reprezintă analiza fotografiilor hemisferice şi determinarea indicelui LAI (*leaf area index*), rezultatele obținute în cazul cercetărilor efectuate situând această metodă foarte aproape de indicii dependenți de distanță, din punctul de vedere al preciziei de estimare (Maily et al., 2003; Coomes & Allen, 2007).

În România studiul proceselor competiționale prin indicatori sintetici a fost abordat în ultimii ani în numeroase lucrări de cercetare – Barbu și Cenușă (2001), Hapca (2003), Popa I. (2004), Avăcăriței (2005), Popa C. (2007), Palaghianu (2009, 2012), Duduman et al. (2010).

Studiul relațiilor competiționale dintre indivizi prin folosirea indicilor de competiție permite extragerea unor concluzii ce pot determina identificarea unor tendințe și legități în evoluția arboretului, de mare interes pentru conducerea procesului de regenerare.

**Prognoza compoziției specifice** - în general se folosesc modele matematice bazate pe proiecții și transformări Markov. Modelele Markov sunt folosite pentru a caracteriza schimbările ce apar în timp în compoziția specifică a unui arboret (Mladenoff, Baker, 1999). Aceste modele utilizează o matrice a probabilităților de transformare, determinate empiric, având capacitatea de a preconiza modul de înlocuire a speciilor unui arboret (Ek, Monserud, 1974; Arévalo et al., 1999).

O altă modalitate o reprezintă folosirea tehnicilor specifice inteligenței artificiale – de pildă a rețelelor neuronale ce permit identificarea structurii unui proces dintr-un set de date cu un grad redus de structurare și relevanță. Rețelele neuronale au capacitatea de a izola un semnal în cazul unor procese caracterizate de zgomot puternic, factori aleatori și structură puțin cunoscută. Ele acționează ca niște funcții estimatoare neparametrice și neliniare, comportamentul lor fiind rezultatul unui proces de învățare pe baza prezentării unor situații specifice.

Hasenauer și Merkl (2001) au folosit rețele neuronale pentru a estima dinamica numărului de puieți și a proporției speciilor într-un arboret plurien de amestec folosind drept variabile de intrare diametrul maxim, numărul de arbori pe specii, suprafața de baza și tipul de humus. Procesul de învățare a fost realizat folosind tehnica *percepteron multi-strat* (MLP) și un algoritm de învățare supravegheată de tip *Rprop (Resilient back propagation)*. Rezultatele obținute au indicat că procedura de estimare a fost mai precisă decât tehnicile statistice bazate pe analiza regresiei.

## 4.2. Aplicații software folosite în simularea regenerării arborilor și arboretelor

Activitatea de modelare, precum și cea de exploatare a modelelor prin intermediul simulărilor a fost mult impulsionată de apariția calculatoarelor. Puterea mare de prelucrare a datelor și rapiditatea efectuării unor operații complexe recomandă calculatoarele drept instrumente adaptate proceselor de simulare. Calculatoarele facilitează efectuarea rapidă a simulărilor și garantează corectitudinea prelucrărilor. Desigur, ele nu garantează și corectitudinea modelelor folosite în această operație.

În ultimele decenii au apărut numeroase aplicații software, gama acestora fiind extrem de variată atât ca și concepție cât și ca domeniu în care pot fi utilizate. În ceea ce privește aplicațiile folosite în activitatea de simulare a regenerării, acestea au apărut de foarte mult timp și se împart în două mari categorii:

- simulatoare;
- sisteme expert sau alte sisteme de asistare a deciziei.

a) *Simulatoarele* – sunt aplicații software complexe, concepute în baza unui model, capabile să prognozeze stările finale posibile ale unui sistem, în condițiile unui scenariu dat. Ele integrează mai multe subrutine care concură la furnizarea unor informații complexe privind procesul analizat. În general sunt aplicații dezvoltate în limbaje de nivel înalt, capabile să folosească tehnici de programare moderne (e.g. programarea orientată pe obiecte) în vederea descrierii proceselor. De cele mai multe ori integrează baze de date vaste pentru a păstra informații cât mai complete asupra sistemelor implicate în simulare. Foarte multe dintre ele dispun și de un motor grafic capabil să redea vizual transformările prin care trece sistemul analizat.

Primul simulator forestier a fost JABOWA (Botkin et al., 1972), urmat apoi de Prognosis-FVS (Stage, 1973), FOREST (Ek, Monserud, 1974) și FORET (Shugart, West, 1977). Aceste simulatoare ale anilor '70 sunt considerate „părinții" simulatoarelor moderne apărute după anul 1990 – ZELIG (Urban, 1990), SILVA (Pretzsch, 1992), SORTIE (Pacala et al., 1993), FORSKA (Prentice et al., 1993), MOSES (Hasenauer, 1994), BWIN (Nagel, 1996), ForClim (Bugmann, 1996), LANDIS (Mladenoff et al., 1996), PROGNAUS (Monserud, Sterba, 1996), CORKFITS (Ribeiro, 2001), BALANCE (Grote, Pretzsch, 2002), DRYMOS (Chatziphilippidis și Spyroglou, 2006) sau SIBYLA (Fabrika, Iursky, 2006). Acestea nu sunt simulatoare destinate exclusiv regenerării arboretelor ci sunt bazate în general pe modele de creștere și evaluare a producției și înglobează elemente privitoare la modelarea regenerării.

Trebuie remarcat că majoritatea simulatoarelor se ocupă în principal de creșterea indivizilor din stratul arborilor și nu de faza de regenerare. Unele chiar ignoră regenerarea sau folosesc modelări simpliste (utilizează ecuații de regresie pentru a prognoza regenerarea în funcție de condițiile staționale și de caracteristicile arboretului).

Brunner et al., (2004) consideră că în cazul regenerării arboretelor există trei tipuri de abordare a problemei: modele empirice ce descriu regenerarea prin variabile simple, submodele de regenerare incluse în modele de creștere sau în modele ale dinamicii ecosistemelor forestiere și modele ce descriu elemente componente ale procesului de regenerare (e.g. producția sau dispersia semințelor).

În continuare vor fi prezentate pentru unele din cele mai cunoscute simulatoare o parte din caracteristicile esențiale în ceea ce privește procesul de modelare a regenerării.

Simulatorul **JABOWA** (Botkin et al., 1972) este un model independent de distanță, ce evaluează dinamica pădurii prin patru factori de mediu: lumina, temperatura, cantitatea de azot din sol și umiditatea solului. Instalarea semințișului în ochiurile de regenerare se apreciază prin intermediul unei funcții dependente de cantitatea de lumină ce ajunge la nivelul solului și nu depinde de existența unor surse de semințe disponibile. Speciile se includ în baza unei liste predefinite de specii ale căror exigențe corespund stațiunii respective. Germinația este dependentă de temperatura și umiditatea solului iar creșterea ulterioară a plantulelor și puieților este modelată diferit în funcție de temperamentul speciei față de lumină.

Mortalitatea este indusă de competiție, fiind apreciată printr-o funcție stohastică dependentă de creșterea în diametru.

**Prognosis** (actual **FVS**) (Stage, 1973; Wykoff et al., 1982; Crookston, Dixon, 2005) este un model individual de creștere, independent de distanță folosit în Statele Unite în asistarea deciziilor de management. În anii '80, modelul Prognosis a fost ales drept model național platformă de către Serviciul Forestier al Departamentului de Stat pentru Agricultură, fiind redenumit **FVS** (*Forest Vegetation Simulator*). Modelul se bazează pe codul sursă scris în FORTRAN, oferă utilizatorilor o interfață grafică numită *Suppose* și este capabil să modeleze creșterea arboretului prin intermediul unui set de ecuații cu coeficienți deduși empiric, în funcție de creșterea în diametru și înălțime, dimensiunile relative ale coroanei, regenerare și mortalitate.

În ceea ce privește instalarea semințișului, modelul determină numărul, dimensiunea și specia puieților în funcție de tehnologia de pregătire a terenului și a compoziției specifice existente. Utilizatorul are la dispoziție și un editor de evenimente (*Event Monitor*) prin care poate defini reguli de simulare a procesului de regenerare specifice unei anumite regiuni. FVS este foarte versatil, Monserud (2003) considerându-l drept cel mai adaptabil model dintr-o lungă serie de modele analizate. FVS înglobează un model propriu de creștere în diametru și înălțime; trei modele ale mortalității (*Prognosis, SDI, Twigs*); două modele de regenerare; cinci modele ale dinamicii coroanei (*Prognosis, Weibull, GENGYM, Twigs, BGC*). Acest model a stat la baza a numeroase variante folosite în toată lumea, cea mai cunoscută dintre ele fiind **PROGNAUS** (*PROGNosis for AUStria*) (Monserud, Sterba 1996, 1999).

**FOREST** (Ek, Monserud, 1974) – simulator dezvoltat în SUA, bazat pe un model dependent de distanță, ce oferă analize foarte detaliate privind regenerarea, creșterea și mortalitatea arborilor. Modelul este foarte complex, ținând cont de producția de semințe, dispersia și germinația acestora, precum și de creșterea, competiția și mortalitatea juvenilă. FOREST a fost unul dintre cele mai complexe modele ale anilor '70 dar necesita introducerea unui număr foarte mare de parametri și avea cerințe exagerate față de resursele tehnice de calcul ale acelei perioade (Mladenoff, Baker, 1999).

**FORET** (Shugart, West, 1977) este un model derivat din JABOWA, care aduce unele îmbunătățiri legate de modul de regenerare a speciilor. Perceperea pădurii ca un mozaic de porțiuni închise și goluri în coronament a condus la ideea că influența luminii este determinată în succesul anumitor specii. Spre deosebire de JABOWA, FORET poate să urmărească relațiile dintre piesele acestui mozaic și să le includă în calculul competiției. O altă diferență constă în faptul că instalarea semințișului este influențată de arborii maturi din zona dar are în continuare un caracter uniform.

Versiunile moderne inspirate de acest model au suferit îmbunătățiri substanțiale. De exemplu simulatorul **ZELIG** (Urban, 1990) este un model individual în care regenerarea, instalarea semințișului și mortalitatea sunt evaluate stohastic iar competiția este influențată de înălțimea, suprafața foliară și biomasa fiecărui individ.

**SORTIE** (Pacala et al 1993, 1996) este un model de simulare dependent de distanță, adresat arboretelor, care oferă date privitoare la structura și dinamica pădurilor naturale. Prin volumul și complexitatea datelor, SORTIE amintește de modelul FOREST dar structura și mecanismul de simulare sunt inspirate din modelele din categoria JABOWA-FORET. Spre deosebire de acestea folosește procese modelate mecanicist și în consecință relațiile sunt mai veridice și mai bine definite. Parametrii folosiți în modelare sunt: diametrul, înălțimea, dimensiunea coroanei arborilor, funcții de modelare a creșterii, mortalității, dispersiei și o funcție de apreciere a cantității de lumină interceptate de coroană.

**SILVA** (Pretzsch, 1992) este un simulator conceput în Germania ce funcționează pe baza unui model individual de creștere, dependent de distanță. Parametri individuali folosiți sunt: specia, diametrul, înălțimea, dimensiunile coroanei și coordonatele arborelui. Se calculează un indice de competiție (*indicele KKL*) pentru fiecare arbore, competitorii fiind selectați prin metoda conului inversat. În evaluarea relațiilor competiționale se ține cont și de tipul de organizare spațială a arborilor. Dată fiind flexibilitatea acestui simulator au fost dezvoltate modele regionale ale acestuia – modelul **CORKFITS** (Ribeiro, 2001) în Portugalia sau **SIBYLA** (Fabrika, Iursky, 2006) în Slovacia.

Simulatorul austriac pentru regenerarea molidului dezvoltat de Schweiger și Sterba (1997) a fost încorporat în două modele europene importante – **MOSES** (Hasenauer, 1994) și **PROGNAUS** (varianta austriacă a modelului *FVS*) (Monserud, Sterba, 1996). Simulatorul este coordonat de un model spațial individual pentru a cărui dezvoltare au fost folosite două tipuri de ecuații: ecuații ce prognozează probabilitatea prezenței regenerării naturale a molidului și ecuații ce prognozează probabilitatea regenerării naturale a molidului care să prezinte o anumită înălțime și un anumit număr de tulpini.

**MOSES** (Hasenauer, 1994) este un simulator dezvoltat la Universitatea Boku, ce are la bază un model individual de creștere dependent de distanță, fundamentat teoretic prin conceptul de creștere potențială. Simulatorul include un modul independent de regenerare care prevede probabilitatea de regenerare a unei suprafețe pentru intervale elementare de 5 ani, proporția speciilor și desimea puieților. Calibrarea modelului s-a realizat pentru majoritatea speciilor din Austria și Elveția, fiind realizat și un simulator pentru pădurile din Grecia  - DRYMOS (Chatziphilippidis, Spyroglou, 2006).

**LANDIS** (Mladenoff et al., 1996) este un simulator forestier ce utilizează un model stohastic dependent de distanță. Modelul se remarcă prin tehnica de concepere utilizată (*designul orientat pe obiect*), ceea ce îi permite să simuleze dinamica pădurii pe suprafețe foarte întinse și pentru perioade îndelungate de timp. Instalarea semințișului ține cont de capacitatea de înmulțire vegetativă a speciilor și de dispersia semințelor. Evaluarea modului de dispersie a semințelor se realizează în funcție de doi parametri: distanța maximă și distanța efectivă de dispersie. Ulterior instalarea generației noi de arbori este condiționată de accesul la lumină, temperamentul speciilor față de acest factor și condițiile de sol.

Mai poate fi amintit simulatorul **ACORn** (Dey, 1995) – conceput în scopul simulării regenerării speciilor de stejar și a modelării producției de ghindă în funcție de diametrul de bază și aria coroanei, simulatorul finlandez *Young Stand Simulation System* (Valkonen et al., 2002) bazat pe un model de creștere juvenilă pentru pin, molid și mesteacăn  și simulatorul german **BWIN** (Nagel, 1996) ce a dezvoltat recent un modul pentru analiza regenerării arboretelor capabil să prevadă instalarea, dezvoltarea și structura semințișului (Schmidt et al., 2006).

Aplicațiile software specializate exclusiv în analiza regenerării sunt foarte puține, aici putând să amintim simulatorul **FOREGEN** (Solomon și Leak, 2002) sau programul informatic dezvoltat în cadrul proiectului *Nat-Man*, **REGENERATOR** (Brunner et al., 2004). Pentru o mai bună înțelegere a funcționării acestor aplicații, se prezintă unul din simulatoarele capabile să modeleze regenerarea arboretelor: simulatorul FOREGEN.

**FOREGEN** este un simulator care modelează stohastic regenerarea suprafețelor exploatate – este capabil să lucreze cu ochiuri (goluri) create de extrageri individuale (de doar câțiva metri pătrați) dar și cu suprafețe tăiate ras cu o arie de până la 36 de hectare. Modelul pe care se bazează a fost conceput pentru speciile de esență tare ale continentului nord-american. Acest model încorporează efectele probabilistice legate de răspândirea semințelor, germinație, instalarea semințișului, condițiile climatice, caracteristicile regenerării, etc. Utilizatorul introduce datele de intrare ce caracterizează arboretul studiat, iar rezultatele vor viza prognoza compoziției specifice după trei ani de la exploatare. Deși există mari variații datorită elementelor stohastice încorporate în model, FOREGEN produce rezultate realiste, care corespund așteptărilor în ceea ce privește regenerarea speciilor pentru care a fost conceput.

Aplicația a fost scrisă în Fortran 90 și include foarte multe date referitoare la specificul regenerării speciilor, în locația deja amintită. Ea poate fi folosită și ca simulator de sine stătător, dar și în combinație cu alte simulatoare de creștere și dezvoltare. Variabilele primare folosite se introduc într-o matrice și urmează un proces de calcul similar lanțurilor Markov.

Aceste variabile primare sunt constituite de:

- producția de semințe – va fi apreciată prin prisma unei scări a fructificației cu 5 trepte (foarte bună, bună, medie, slabă și foarte slabă), separate pe trei clase de vârstă ale arboretului (30-50, 50-90, >90 de ani). Datele se corelează cu pierderi provocate de insecte, boli sau factori climatici (în funcție de datele precizate în literatura de specialitate pentru speciile studiate);

- dispersia semințelor – se ține cont de perioada în care are loc exploatarea, de viteza de cădere a semințelor, înălțimea de la care cad, viteza și direcția vântului, perioada și durata diseminării;

- patul de germinație – se ține cont de substrat, de o eventuală pregătire a terenului sau lucrare a solului;

- germinația și supraviețuirea plantulelor – odată semințele diseminate, trebuie să se determine șansele ca acestea să germineze, respectiv ca plantulele să supraviețuiască în primul sezon de vegetație. Aceste aspecte vor fi apreciate în funcție de patul de germinație, expoziție, precipitațiile din primăvară, regimul de temperatură și lumină.

Simulatorul are nevoie de o serie de date de intrare: compoziția inițială a arboretului, vârsta, perioada de exploatare, numărul de arbori la unitatea de suprafață, dimensiunile suprafeței exploatate, numărul de seminceri care se lasă pe picior și diametrul mediu al acestora, procentul din suprafață pe care se lucrează solul, caracteristicile solului, date climatice.

FOREGEN, conform autorilor săi, este un model „umil", care nu are pretenția de a fi surprins toți factorii ce influențează procesul complex al regenerării. Totuși acesta poate simula situații reale, oferind informații interesante în ceea ce privește evoluția regenerării. Autorii recomandă acest simulator mai ales pentru activitățile didactice.

Pot fi menționate în cadrul simulatoarelor și *aplicațiile destinate simulării exclusiv vizuale*, care nu înglobează modele de creștere sau analiză a datelor, ci pur și simplu generează seturi de imagini sau secvențe 3D, pe baza unor date reprezentative pentru o situație dată.

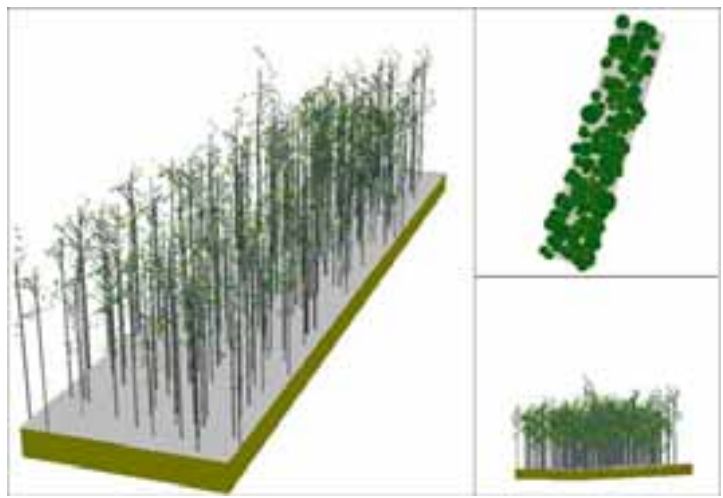

**Figura 4.2 Simulare grafică a unei suprafețe regenerate folosind aplicația SVS**

Acestea pot să ajute în formarea unei idei asupra modului de regenerare a arboretului, în cazul folosirii unor date adecvate, furnizate eventual de alte simulatoare de creștere și dezvoltare. Merită amintite aici programele: SVS - EnVision (McGaughey, 1997), SMARTFOREST (Orland, 1997), Virtual Forest (Buckely et al., 1998) sau PROARB (Popa, 1999).

Legat de utilizarea unor modele în țara noastră, merită menționată aplicarea cu succes a modelului ForClim (Bugmann, 1996) în zona de nord a României, în cadrul proiectului internațional FP7 MOTIVE: "Models for Adaptive Forest Management", proiect în care a fost implicată și Facultatea de Silvicultură din cadrul Universității Ștefan cel Mare Suceava.

b) *Sistemele expert* – sunt sisteme informatice ce utilizează metode specifice inteligenței artificiale și urmăresc reproducerea computerizată a cunoștințelor și raționamentelor experților umani într-un anumit domeniu. Datorită neliniarității sunt asociate ca funcționare principiului „cutiei negre", dar compensează lipsa transparenței prin posibilitatea justificării raționamentelor efectuate. Există câteva sisteme forestiere de acest tip care pot fi folosite şi în domeniul regenerării arboretelor.

Sistemul expert **ForEx** - sistem expert de management forestier dezvoltat în Austria, destinat reabilitării ecosistemelor forestiere cu molid. Constituirea bazei de cunoştinţe a fost dificilă, mai ales datorită teoriilor de multe ori contradictorii ale experţilor umani. Un prototip al bazei de cunoştinţe a fost dezvoltat în PROLOG, iar mecanismul de achiziţie a cunoştinţelor a utilizat formulare de tip HTML (Dorn, Mitterbock, 1998)

Saarenmaa (1996) a conceput în Finlanda un sistem expert de asistare a deciziilor privitoare la regenerare. Bazele de date au inclus 47 de variabile, sortarea fiind făcută pe baza variabilelor discrete (tipul de staţiune, specie, calitatea solului, tratament, metoda de pregătire a solului, numărul de puieţi existenţi înainte de plantare) prin intermediul unui algoritm genetic.

Tot în Finlanda preocupările lui Pukkala au fost concretizate în două sisteme de asistare a deciziei - STAND (Pukkala, Miina, 1997) şi MONSU (Pukkala, 2004).

În ultimii ani au apărut o nouă tendinţă de oferire a unor pachete complete de management forestier care înglobează modele, simulatoare şi sisteme expert. Un bun exemplu de soluţie integrată de management este sistemul REMSOFT SPATIAL PLANNING (Remsoft, 1996) ce utilizează drept model un nucleu *Spatial Woodstock.* Sistemul REMSOFT reprezintă în prezent o soluţie comercială completă de management al resurselor forestiere, oferind posibilitatea de a integra informaţiile inclusiv în format GIS. Pachetul software conţine modele analitice şi predictive capabile să identifice eventuale riscuri şi oportunităţi, oferind suport pentru planificarea: operaţională pe termen scurt, tactică pe termen mediu şi strategică pe termen lung. Acest gen de soluţii pot reprezenta o bună modalitate de a integra într-o modalitate facilă şi extrem de orientată spre partea practică a modelării, noilor tehnologii şi a tehnicii de calcul moderne în domeniul forestier.

# PARTEA a II-a
# CERCETĂRI EFECTUATE ȘI REZULTATE OBȚINUTE

# Capitolul 5
# Caracterizarea şi modelarea elementelor structurale ale regenerării

## 5.1. Structura în raport cu numărul de puieţi pe specii

Arboretul reprezintă un sistem complex al ecosistemului forestier, iar structura şi dinamica sa sunt reglate prin mecanisme biocibernetice specifice. Identificarea atributelor structurale specifice este una din acţiunile ce caracterizează orice studiu sau cercetare a pădurii, în vederea stabilirii particularităţilor populaţiilor de arbori din zona analizată. Unul din obiectivele acestei lucrări se referă chiar la identificarea acestor particularităţi, foarte utile în interpretarea analizelor ulterioare. Caracteristicile structurii orizontale şi verticale sunt prezentate prin metode adecvate acestui tip de analiză, fiind expuse informaţii privitoare la desime, compoziţia pe specii, principalele atribute biometrice şi variaţia spaţială a acestora.

Un element esenţial al structurii este reprezentat de desimea stratului de puieţi, acesta fiind un parametru de bază în caracterizarea şi evaluarea regenerării arboretelor. În cele 10 suprafeţe de probă au fost inventariaţi 7253 de puieţi, datele privitoare la numărul de puieţi pe metrul pătrat fiind prezentate sintetic în  tabelul 5.1.

Tabelul 5.1

**Situația pe specii a numărului de puieți la metrul pătrat**

| Număr puieți/m² | Ca | St | Te | Fr | Ju | Ci | Pam | Total |
|---|---|---|---|---|---|---|---|---|
| Suprafața nr. 1 | 6,37 | 7,59 | 1,47 | 0,47 | 0,55 | 0,35 | 0,10 | 16,94 |
| Suprafața nr. 2 | 6,61 | 3,33 | 0,47 | 0,08 | 0,41 | 0,08 | 0,53 | 11,57 |
| Suprafața nr. 3 | 7,39 | 0,39 | 1,61 | 0,49 | 0,47 | 0,10 | 0,08 | 10,82 |
| Suprafața nr. 4 | 10,53 | 0,37 | 1,63 | 0,29 | 0,80 | 0,41 | 0,12 | 14,29 |
| Suprafața nr. 5 | 5,41 | 2,67 | 0,57 | 0,33 | 0,98 | 0,16 | 0,00 | 10,37 |
| Suprafața nr. 6 | 13,73 | 0,08 | 0,78 | 0,37 | 0,04 | 0,04 | 0,00 | 15,14 |
| Suprafața nr. 7 | 6,39 | 0,84 | 4,00 | 6,59 | 0,63 | 0,24 | 0,06 | 18,94 |
| Suprafața nr. 8 | 9,47 | 3,67 | 0,90 | 0,24 | 0,69 | 0,12 | 0,33 | 15,65 |
| Suprafața nr. 9 | 8,20 | 2,47 | 1,55 | 0,51 | 3,00 | 0,43 | 0,00 | 16,16 |
| Suprafața nr. 10 | 9,41 | 1,76 | 1,80 | 3,08 | 1,10 | 0,59 | 0,02 | 18,14 |
| Media nr. puieți/m2 | **8,35** | **2,32** | **1,48** | **1,24** | **0,87** | **0,25** | **0,12** | **14,80** |
| Coef. de variație | 30% | 97% | 68% | 166% | 93% | 72% | 139% | 20% |
| % participare | 56% | 16% | 10% | 8% | 6% | 2% | 1% | - |

În cuprinsul celor 10 suprafețe analizate au fost identificate 13 specii lemnoase, în tabelul 5.1 fiind prezentate doar datele pentru speciile cu o proporție de participare de peste 1% - carpen (*Carpinus betulus* – 56%), stejar pedunculat (*Quercus robur* – 16%), tei pucios (*Tilia cordata* – 10%), frasin comun (*Fraxinus excelsior* – 8%), jugastru (*Acer campestre* – 6%), cireș (*Prunus avium* – 2%) și paltin de munte (*Acer pseudoplatanus* – 1%). Celelalte specii identificate, în ordinea procentului de participare, sunt: măceș (*Rosa canina* - 0,70%), sânger (*Cornus sanguinea* - 0,29%), păducel (*Crataegus monogyna* - 0,04%), salbă râioasă (*Euonymus verrucosa* - 0,04%), soc (*Sambucus nigra* - 0,03%) și ulm de câmp (*Ulmus minor* - 0,03%).

Se observă că valorile procentului de participare a speciilor identificate în suprafața analizată sunt diferite de cele precizate în situațiile întocmite în urma controlului anual al regenerărilor efectuat în anul 2007: 7Go 1Ca 1Fr 1Dt. Acest lucru se explică atât prin faptul că a fost analizată doar o porțiune a unității amenajistice 50, UP I Flămânzi, cât și prin precizia slabă a determinărilor efectuate cu prilejul controalelor anuale ale regenerărilor.

O situație interesantă se prezintă și în cazul relației dintre compoziția semințișului și compoziția arboretului matur, care este conform ultimului amenajament (2005): 3Go 2St 3Ca 1Te 1Fr. În compoziția semințișului nu se regăsește gorunul, nefiind identificați cu ocazia inventarierii puieți de gorun. Acest fapt poate fi explicat prin prezența unui număr mic de arbori maturi de gorun, problemele de fructificație ale speciei în regiunea respectivă (ultimul an de fructificație abundentă la gorun în UP I Flămânzi fiind înregistrat în 1986), dificultăți de instalare a semințișului, precum și posibilității unor eventuale înregistrări eronate ale speciei.

Carpenul este specia care deține proporția de participare cea mai mare, fiind cunoscută capacitatea de instalare deosebită a acestei specii, considerată chiar invadantă. Carpenul este și cea mai uniformă specie din punct de vedere al răspândirii în cele 10 piețe de probă, cu cel mai mic coeficient de variație - 30%. Teiul și cireșul manifestă o uniformitate în spațiu mai redusă decât carpenul, având coeficienți de variație apropiați de 70%, dar superioară stejarului și cireșului (97%, respectiv 93%). Frasinul este specia cu cea mai mare inconstanță în ceea ce privește răspândirea în spațiu (coeficient de variație de 166%), fiind întâlnit preponderent în două suprafețe de probă în care s-a înregistrat o umiditate a solului mai ridicată (apreciată prin identificarea unor specii erbacee higrofite).

Valoarea numărului mediu de puieți la metrul pătrat (14,80) este o valoare adecvată stadiului de dezvoltare al semințișului și tipului de regenerare, fiind înregistrat un coeficient de variație de 20% al acestui parametru pentru suprafețele studiate. Dat fiind faptul că a fost aleasă pentru studiu o suprafața de regenerare compactă și omogenă din punct de vedere al desimii, se aștepta obținerea unei valori reduse a coeficientului de variație.

Nu au fost efectuate analize ale desimii puieților pe specii în funcție de expoziție, altitudine sau pantă deoarece nu s-au înregistrat diferențe majore între valorile acestora în suprafețele analizate. În toate suprafețele analizate expoziția a fost nord-estică, panta a avut valori de 2-5°, iar ecartul altitudinal a fost cuprins între 135 și 159 m.

Pentru a realiza o analiză mai fină a desimii puieţilor pe specii s-a conceput şi implementat un algoritm de încadrare a puieţilor unei suprafeţe de probă în suprafeţe elementare cu dimensiuni variabile, specificate de utilizator. Necesitatea acestei subrutine de calcul este reliefată de modalitatea dificilă de calcul şi analiză, precum şi de volumul mare al datelor prelucrate. Scopul constă în determinarea unor caracteristici ale răspândirii speciilor în suprafeţele considerate. Este cunoscut faptul că analizele care implică răspândirea în spaţiu a proceselor sau fenomenelor sunt afectate de scara la care se face analiza (Hurlbert, 1990; Dale, 2004; Haining, 2004). De aceea aplicaţia oferă posibilitatea utilizatorului să testeze mai multe variante de încadrare, în funcţie de dimensiunile suprafeţei elementare de analiză.

Calculele au fost realizate pentru 7 variante de încadrare: 0,5x0,5 m; 1x1 m; 1,4x1,4 m; 1,75x1,75 m; 2,33x2,33 m; 3,5x3,5 m şi 7x7 m. Astfel fiecare suprafaţă de probă a fost împărţită în 196, 49, 25, 16, respectiv 4 suprafeţe elementare. Ulterior s-a efectuat o pivotare a datelor pe specii şi suprafeţe elementare, obţinându-se desimea tuturor speciilor în ariile de analiză. Pasul următor a constat în calcularea coeficienţilor de corelaţie între desimile speciilor la aceeaşi scară de studiu, pentru a aprecia care este modalitatea de asociere a acestora. Au fost analizate în total 147 de legături corelative, între principalele şapte specii (cele cu procentul de participare de peste 1%), pentru cele şapte variante dimensionale menţionate anterior. Rezultatele sunt prezentate sintetic în tabelul 5.2, fiind identificate o serie de legături corelative pozitive sau negative, în general slabe ca intensitate, cu un grad de semnificaţie acoperit statistic. Unele relaţii dintre specii par să confirme menţiunile privitoare la modul de asociere din literatura de specialitate – cazul relaţiei negative stejar–frasin (Clinovschi, 2005).

Corelaţiile pozitive pot fi explicate prin gradul similar de accesibilitate la resurse şi exigenţe ecologice apropiate faţă de condiţiile microstaţionale, dar întrucât nu există corelaţii directe pozitive între toate speciile, ba mai mult există unele corelaţii negative, se poate formula chiar o ipoteză referitoare la tendinţele de asociere a speciilor în suprafeţele studiate, ipoteză ce va fi analizată prin mijloace ale analizei spaţiale în capitolul următor.

Tabelul 5.2

## Matricea corelațiilor între desimile (nr. puieți/m2) pe specii ale puieților

|  | Ca | St | Te | Fr | Ju | Ci | Pam |
|---|---|---|---|---|---|---|---|
| **Ca** | *0,5 x 0,5* | -0,121 *** | -0,105 | -0,018 | -0,066 * | -0,029 | -0,028 |
|  | *1 x 1* | -0,050 | -0,028 | 0,021 | -0,030 | -0,034 | -0,032 |
|  | *1,4 x 1,4* | -0,068 | -0,087 | 0,016 | -0,037 | -0,065 | -0,040 |
|  | *1,75 x* | -0,086 | 0,004 | 0,014 | -0,040 | -0,060 | -0,041 |
|  | *2,33 x* | -0,140 | -0,045 | 0,009 | -0,015 | 0,003 | -0,106 |
|  | *3,5 x 3,5* | -0,200 | -0,007 | -0,004 | -0,108 | 0,075 | 0,013 |
|  | *7 x 7* | -0,469 | -0,194 | -0,201 | -0,161 | -0,020 | -0,172 |
| **St** | -0,121 | *0,5 x 0,5* | -0,096 | -0,033 | -0,001 | -0,003 | 0,003 |
|  | -0,050 | *1 x 1* | -0,109 * | -0,058 | 0,057 | 0,026 | 0,113* |
|  | -0,068 | *1,4 x 1,4* | -0,133 * | -0,084 | 0,066 | 0,075 | 0,110 |
|  | -0,086 | *1,75 x 1,75* | -0,154 * | -0,109 | 0,118 | 0,116 | 0,139 |
|  | -0,140 | *2,33 x 2,33* | -0,138 | -0,131 | 0,120 | 0,151 | 0,167 |
|  | -0,200 | *3,5 x 3,5* | -0,189 | -0,171 | 0,068 | 0,098 | 0,220 |
|  | -0,469 | *7 x 7* | -0,247 | -0,255 | 0,059 | 0,114 | 0,290 |
| **Te** | -0,105 | -0,096 *** | *0,5 x 0,5* | 0,153 *** | -0,053 | 0,022 | -0,030 |
|  | -0,028 | -0,109 * | *1 x 1* | 0,341 *** | -0,006 | 0,055 | -0,029 |
|  | -0,087 | -0,133 * | *1,4 x 1,4* | 0,427 *** | 0,029 | 0,032 | -0,022 |
|  | 0,004 | -0,154 * | *1,75 x* | 0,506 *** | 0,053 | 0,163 * | -0,025 |
|  | -0,045 | -0,138 | *2,33 x* | 0,659 *** | 0,114 | 0,273 ** | -0,097 |
|  | -0,007 | -0,189 | *3,5 x 3,5* | 0,716 *** | 0,088 | 0,284 | -0,206 |
|  | -0,194 | -0,247 | *7 x 7* | 0,902 *** | 0,084 | 0,350 | -0,340 |
| **Fr** | -0,018 | -0,033 | 0,153 *** | *0,5 x 0,5* | 0,005 | -0,008 | -0,036 |
|  | 0,021 | -0,058 | 0,341 *** | *1 x 1* | -0,007 | 0,069 | -0,053 |
|  | 0,016 | -0,084 | 0,427 *** | *1,4 x 1,4* | 0,010 | 0,083 | -0,096 |
|  | 0,014 | -0,109 | 0,506 *** | *1,75 x* | -0,005 | 0,193 * | -0,115 |
|  | 0,009 | -0,131 | 0,659 *** | *2,33 x* | -0,008 | 0,243 * | -0,159 |
|  | -0,004 | -0,171 | 0,716 *** | *3,5 x 3,5* | -0,010 | 0,254 | -0,179 |
|  | -0,201 | -0,255 | 0,902 *** | *7 x 7* | -0,030 | 0,276 | -0,260 |
| **Ju** | -0,066 * | -0,001 | -0,053 | 0,005 | *0,5 x 0,5* | 0,000 | -0,038 |
|  | -0,030 | 0,057 | -0,006 | -0,007 | *1 x 1* | 0,124 ** | -0,041 |
|  | -0,037 | 0,066 | 0,029 | 0,010 | *1,4 x 1,4* | 0,176 ** | -0,082 |
|  | -0,040 | 0,118 | 0,053 | -0,005 | *1,75 x* | 0,167 * | -0,124 |
|  | -0,015 | 0,120 | 0,114 | -0,008 | *2,33 x* | 0,283 ** | -0,132 |
|  | -0,108 | 0,068 | 0,088 | -0,010 | *3,5 x 3,5* | 0,254 | -0,168 |
|  | -0,161 | 0,059 | 0,084 | -0,030 | *7 x 7* | 0,559 | -0,305 |
| **Ci** | -0,029 | -0,003 | 0,022 | -0,008 | 0,000 | *0,5 x 0,5* | -0,037 |
|  | -0,034 | 0,026 | 0,055 | 0,069 | 0,124 ** | *1 x 1* | -0,024 |
|  | -0,065 | 0,075 | 0,032 | 0,083 | 0,176 ** | *1,4 x 1,4* | -0,064 |
|  | -0,060 | 0,116 | 0,163 * | 0,193 * | 0,167 * | *1,75 x* | -0,076 |
|  | 0,003 | 0,151 | 0,273 ** | 0,243 ** | 0,283 ** | *2,33 x* | -0,114 |
|  | 0,075 | 0,098 | 0,284 | 0,254 | 0,254 | *3,5 x 3,5* | -0,262 |
|  | -0,020 | 0,114 | 0,350 | 0,276 | 0,559 | *7 x 7* | -0,390 |
| **Pam** | -0,028 | 0,003 | -0,030 | -0,036 | -0,038 | -0,037 | *0,5 x 0,5* |
|  | -0,032 | 0,113* | -0,029 | -0,053 | -0,041 | -0,024 | *1 x 1* |
|  | -0,040 | 0,110 | -0,022 | -0,096 | -0,082 | -0,064 | *1,4 x 1,4* |
|  | -0,041 | 0,139 | -0,025 | -0,115 | -0,124 | -0,076 | *1,75 x 1,75* |
|  | -0,106 | 0,167 | -0,097 | -0,159 | -0,132 | -0,114 | *2,33 x 2,33* |
|  | 0,013 | 0,220 | -0,206 | -0,179 | -0,168 | -0,262 | *3,5 x 3,5* |
|  | -0,172 | 0,290 | -0,340 | -0,260 | -0,305 | -0,390 | *7 x 7* |

**Notă:** coeficienții de corelație sunt prezentați pentru cele șapte variante de alegere a dimensiunii suprafeței elementare de analiză (începând de la varianta de 0,5x0,5 m și apoi 1x1 m, 1,4x1,4 m, 1,75x1,75 m, 2,33x2,33 m, 3,5x3,5 m, 7x7 m). Valorile coeficienților de corelație care depășesc pragul de semnificație de 10% sunt tipărite folosind stilul aldin. Se folosesc de asemenea notațiile consacrate de marcare a semnificației (*, **, ***).

Corelațiile negative pot fi determinate de cerințe ecologice diferite, de variația pe suprafețe mici a caracteristicilor solului (Jarvinen et al., 1993), de competiție sau chiar de scara dimensională folosită. În plus relațiile dintre desimile speciilor sunt influențate de o serie de factori aleatori, greu predictibili – variabilitatea fructificației, dispersia semințelor, condiții microclimatice, diverși dăunători.

Pentru a simplifica reprezentarea relațiilor dintre desimile speciilor a fost întocmită și o reprezentare grafică a acestora, doar pe baza legăturilor corelative semnificative, foarte semnificative și distinct semnificative.

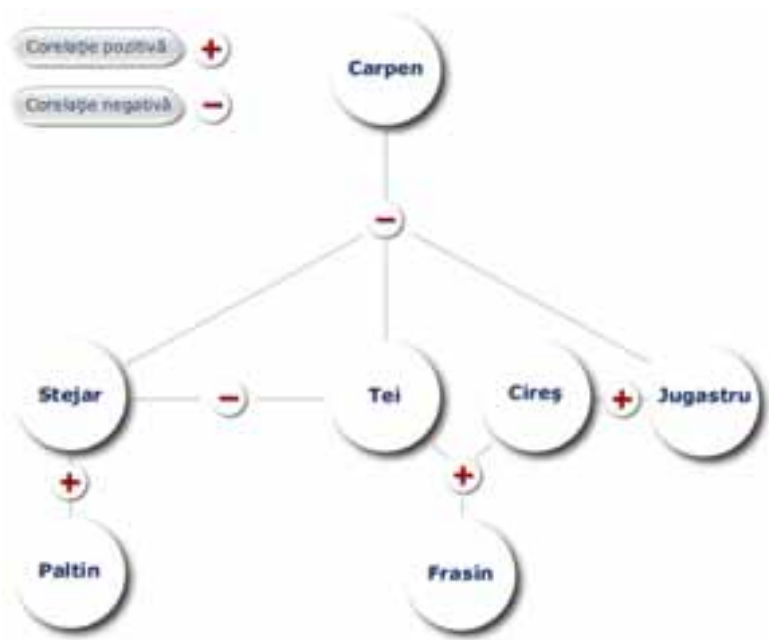

**Figura 5.1 Sinteza grafică a relațiilor corelative dintre desimile puieților pe specii**

Relațiile dintre unele specii își păstrează acoperirea statistică în aproape toate variantele analizate, indiferent de valoarea dimensiunilor suprafeței elementare (relațiile Te-Fr, Ju-Ci), altele devin semnificative doar în anumite variante dimensionale (relațiile Ca-St, Ca-Te, Ca-Ju, St-Te pentru variantele dimensionale reduse, respectiv Fr-Ci, Te-Ci pentru variantele dimensionale medii și mari). Analiza mai multor variante dimensionale a permis studierea relațiilor dintre specii în condițiile raportării la diferite mărimi ale suprafeței de analiză, relațiile dintre indivizi fiind în mod cert influențate de variația distanței dintre aceștia. Deși au fost efectuate calcule și pentru suprafețe elementare mai mici de 0,5x0,5 m, rezultatele (corelații negative între aproape toate speciile) au arătat că sub acest prag competiția pe spații reduse între puieți influențează negativ relevanța informațiilor.

Studiul posibilei asocieri a speciilor în suprafețele studiate a fost realizat și folosind tehnica analizei în componente principale (Horodnic, 2004), pentru prelucrarea datelor fiind folosită aplicația software *StatSoft STATISTICA*. Această metodă este o formă de analiză factorială ce poate reduce complexitatea datelor, evidenția asocieri ascunse dintre variabile, respectiv determina variabile latente. Rezultatele analizei se constituie într-o mulțime de axe principale generate de vectorii asociați la valorile proprii ale variabilelor, aranjate în ordinea descrescătoare din matricea de corelații. Pentru a vizualiza rezultatele se folosește un cerc al corelației, o proiecție a sferei unitate pe două plane factoriale. Distanțele dintre punctele reprezentate în cercul corelațiilor sunt invers proporționale cu valorile corelațiilor dintre variabilele corespunzătoare. Stabilirea variabilelor pozitiv corelate se face prin identificarea grupărilor de puncte pe grafic. Trebuie acordată o atenție sporită eventualelor erori provocate de efectul de perspectivă, pentru formularea unor concluzii fiind necesară vizualizarea proiecțiilor grafice după mai multe planuri factoriale. În cazul de față, pentru toate cele șapte variante dimensionale analizate, s-a optat pentru prezentarea a două alternative de proiecție – în planul dintre primul și al doilea, respectiv al doilea și al treilea factor, ordinea factorilor fiind dată de tabelul valorilor proprii (*eigenvalues*), primii trei factori explicând în cea mai mare proporție varianța datelor. Rezultatele, prezentate în figurile 5.2 - 5.8, susțin formularea unor ipoteze de asociere a speciilor menționate anterior – relațiile pozitive (în ordinea intensității) Fr-Te, Ci-Ju, chiar și St-Pam (intensitate slabă), respectiv cele negative: Ca-St, de cea mai mare intensitate și relațiile de intensitate mai slabă Ca-Ju, Ca-Pam. Formularea unor concluzii este însă inoportună datorită proporției scăzute din varianța totală pe care principalele componente o explică.

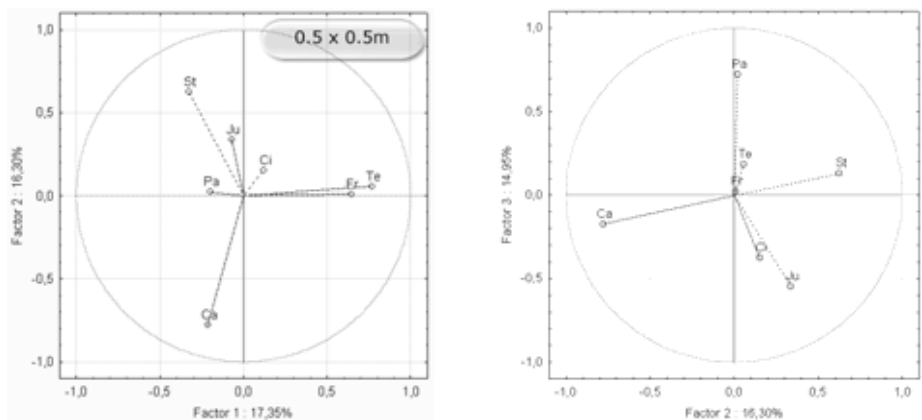

**Figura 5.2 Analiza în componente principale a desimii pe specii pentru varianta suprafețelor elementare de 0,5 x 0,5 m**

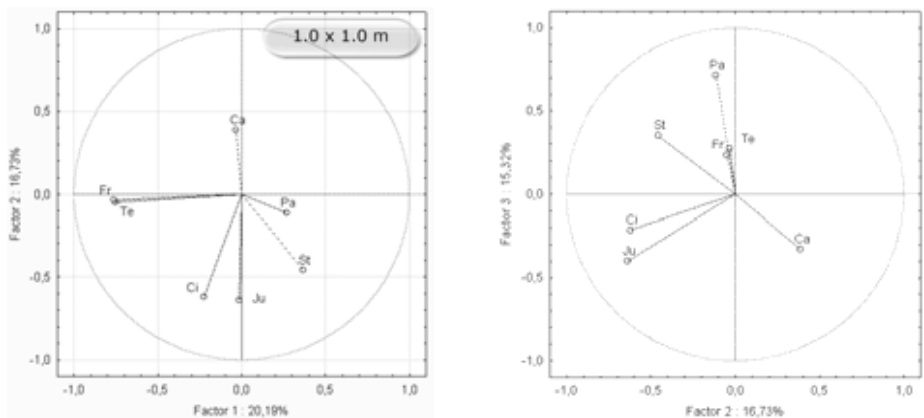

**Figura 5.3 Analiza în componente principale a desimii pe specii pentru varianta suprafețelor elementare de 1,0 x 1,0 m**

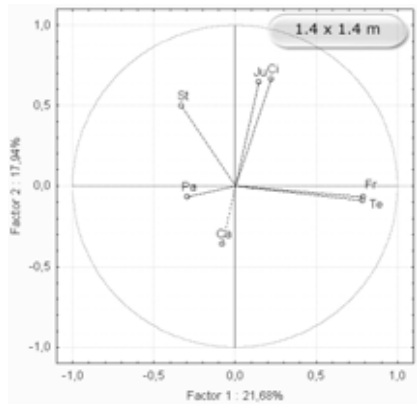 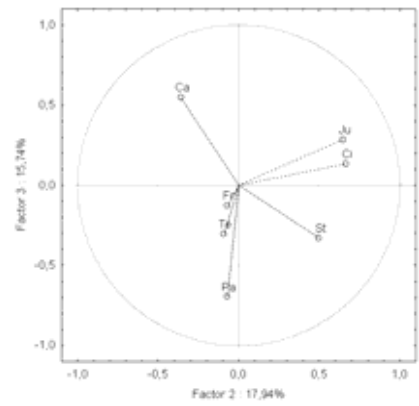

**Figura 5.4 Analiza în componente principale a desimii pe specii (1,4 x 1,4 m)**

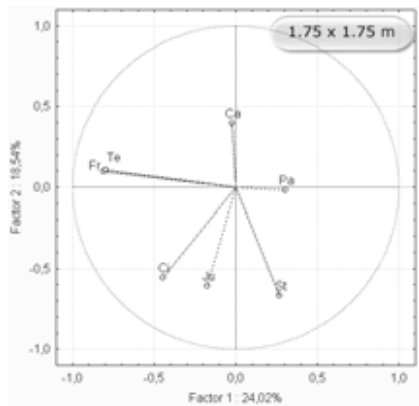 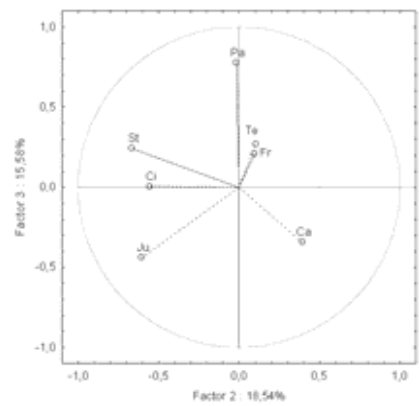

**Figura 5.5 Analiza în componente principale a desimii pe specii ( 1,75 x 1,75 m)**

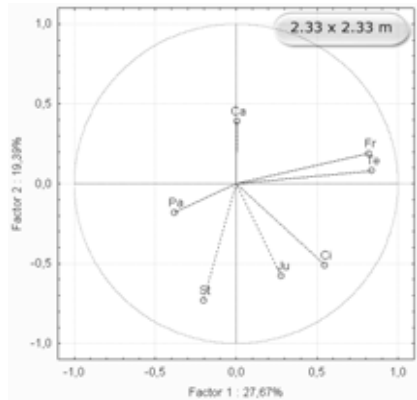 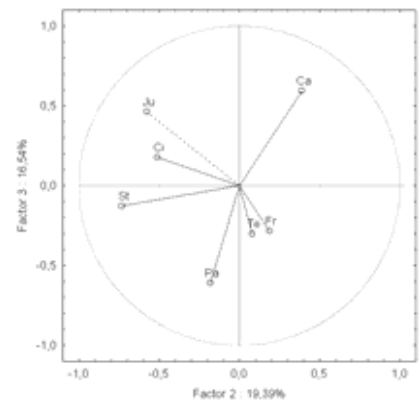

**Figura 5.6 Analiza în componente principale a desimii pe specii (2,33 x 2,33 m)**

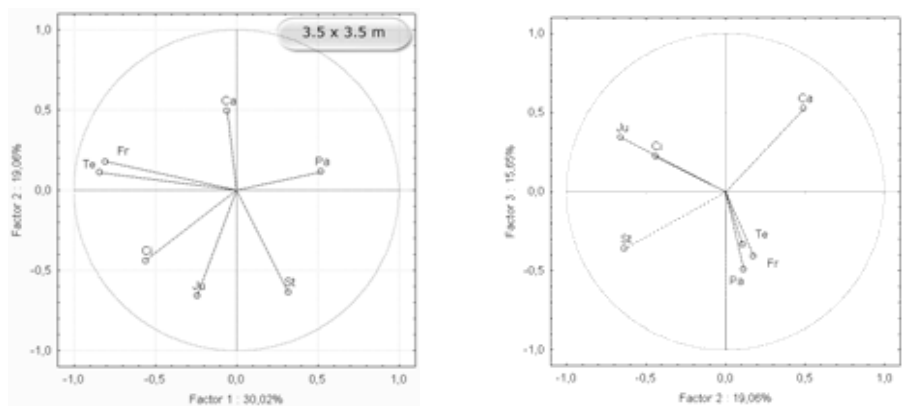

**Figura 5.7 Analiza în componente principale a desimii pe specii (3,5 x 3,5 m)**

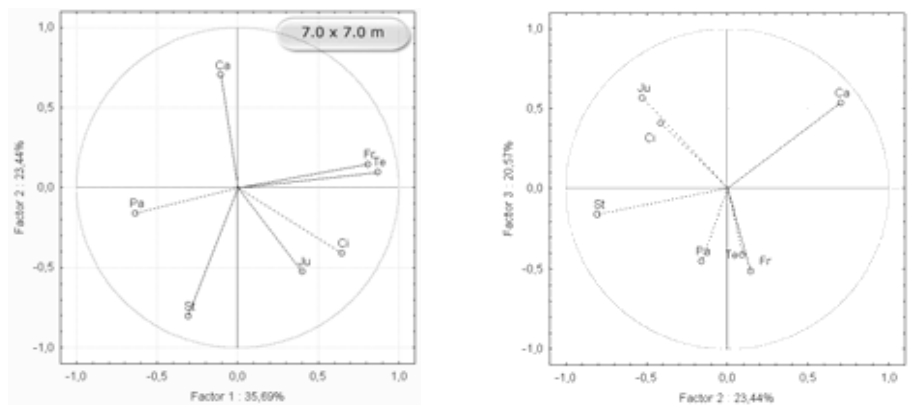

**Figura 5.8 Analiza în componente principale a desimii pe specii (7,0 x 7,0 m)**

Rumsey (2007) afirma „*Nu mergeți niciodată atât de departe încât să spuneți că prelucrările statistice dovedesc ceva*". Chiar dacă identificarea unor tendințe de asociere și distribuție spațială a speciilor are o importanță deosebită pentru practica silvică, deocamdată nu se poate trage o concluzie fermă privitoare la aceste aspecte doar din analiza corelației dintre desimi, respectiv din analiza în componente principale. În spatele acestor procese se află factori încă necunoscuți și fenomene bio-ecologice care trebuie deslușite. Elucidarea raporturilor concurențiale, precum și analiza organizării spațiale și a caracteristicilor spațiului de dezvoltare pot însă conduce la completarea informațiilor și formularea unor opinii mai clare în capitolele următoare.

Un alt aspect important privitor la desimea puieților care este analizat în acest subcapitol se referă la variația spațială a acestui parametru în suprafețele analizate, în figurile 5.9 – 5.18 fiind prezentate grafic aceste informații.

Reprezentările grafice care exprimă variația spațială a caracteristicilor populațiilor de arbori (cartogramele) sunt frecvent folosite în lucrările de cercetare forestieră deoarece oferă o modalitate ușoară și intuitivă de percepere a informației. În acest scop sunt folosite o serie de aplicații informatice performante și foarte complexe (e.g. *Golden Software Surfer, ESRI ArcView*). În unele cazuri aceste aplicații devin dificil de folosit tocmai datorită complexității lor sau oferă prelucrări ce sunt adesea interpretate eronat. De exemplu, în cazul aplicației *Surfer*, prelucrările oferite implică folosirea unor metode de interpolare între valorile rețelei de date, foarte utile în cazul acoperirii unor zone lipsite de informații, dar care pot să conducă la reprezentări eronate datorită alterării rezultatelor.

Inconvenientele menționate anterior au determinat conceperea unei aplicații software proprii, care să ofere o modalitate rapidă și simplă de obținere a unor rezultate grafice nealterate prin metode de interpolare. Programul CARTOGRAMA a fost creat folosind mediul de dezvoltare *Microsoft Visual Basic,* rezultatul fiind un fișier executabil independent. Aplicația realizează citirea datelor (coordonatele evenimentelor) din foi de calcul de tip *Microsoft Excel*, efectuează o serie de prelucrări statistice și generează obiecte grafice în care este reprezentată frecvența de apariție a unui eveniment în cadrul unei suprafețe elementare ale cărei dimensiuni sunt particularizabile de către utilizator.

Cu ajutorul aplicației CARTOGRAMA s-au generat graficele următoare și s-au calculat parametrii statistici corespunzători fiecărei suprafețe. Pentru a avea un grad crescut de consistență vizuală și pentru a asigura posibilitatea comparării rezultatelor obținute a fost folosită aceeași scară de intensitate coloristică pentru toate suprafețele analizate, cu minimul corespunzător valorii 0 și maximul corespunzător valorii 56 (valorile extreme înregistrate în cele 490 de suprafețe elementare analizate).

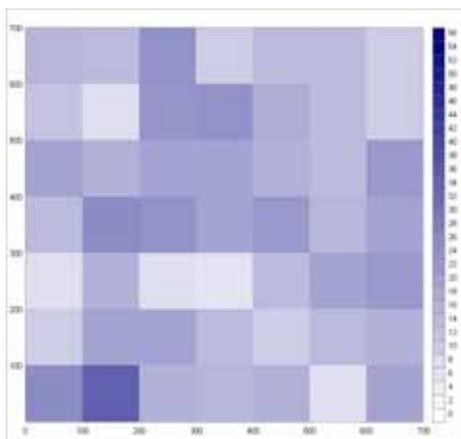

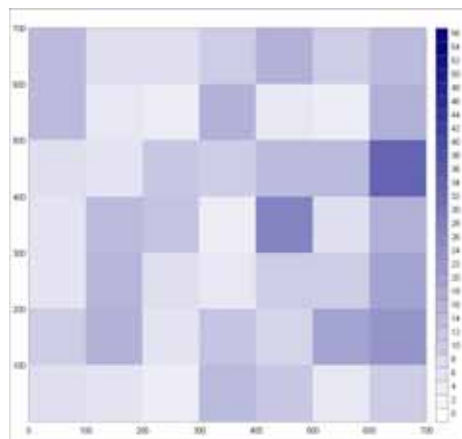

Media frecvențelor:     Frecv. minimă: 6
16,94                 Frecv. maximă: 35
Coef. de variație:
33,40%

**Figura 5.9 Repartiția spațială a frecvenței puieților pe metrul pătrat în suprafața nr. 1**

Media frecvențelor:     Frecv. minimă: 3
11,57                 Frecv. maximă: 34
Coef. de variație:
55,22%

**Figura 5.10 Repartiția spațială a frecvenței puieților pe metrul pătrat în suprafața nr. 2**

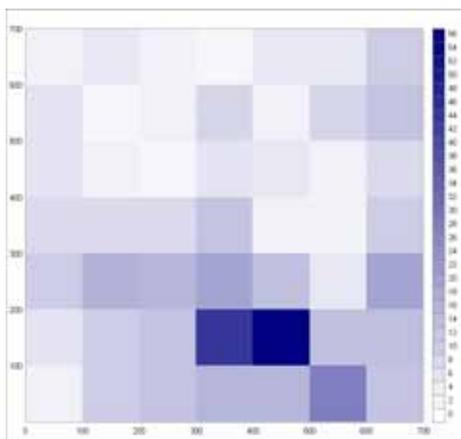

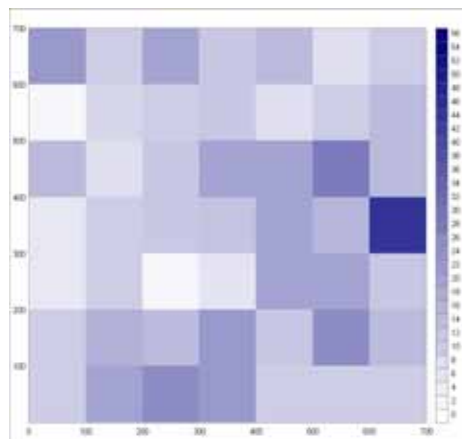

Media frecvențelor: 10,82     Frecv. minimă: 2
Coef. de variație: 2,35%      Frecv. maximă: 56

**Figura 5.11 Repartiția spațială a frecvenței puieților pe metrul pătrat în suprafața nr. 3**

Media frecvențelor: 14,92     Frecv. minimă: 2
Coef. de variație: 2,75%      Frecv. maximă: 45

**Figura 5.12 Repartiția spațială a frecvenței puieților pe metrul pătrat în suprafața nr. 4**

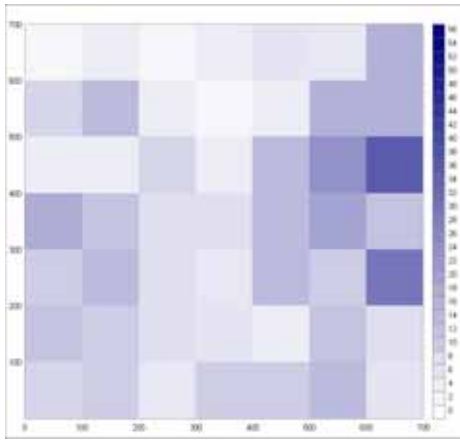

Media frecvențelor: 10,37    Frecv. minimă: 1
Coef. de variație: 69,71%    Frecv. maximă: 36

**Figura 5.13 Repartiția spațială a frecvenței puieților pe metrul pătrat în suprafața nr. 5**

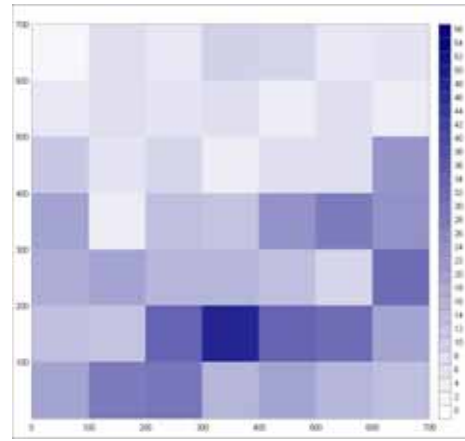

Media frecvențelor: 15,14    Frecv. minimă: 2
Coef. de variație: 68,24%    Frecv. maximă: 48

**Figura 5.14 Repartiția spațială a frecvenței puieților pe metrul pătrat în suprafața nr. 6**

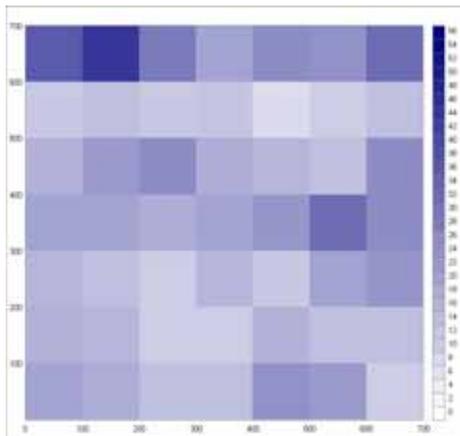

Media frecvențelor: 18,94    Frecv. minimă: 8
Coef. de variație: 38,10%    Frecv. maximă: 44

**Figura 5.15 Repartiția spațială a frecvenței puieților pe metrul pătrat în suprafața nr. 7**

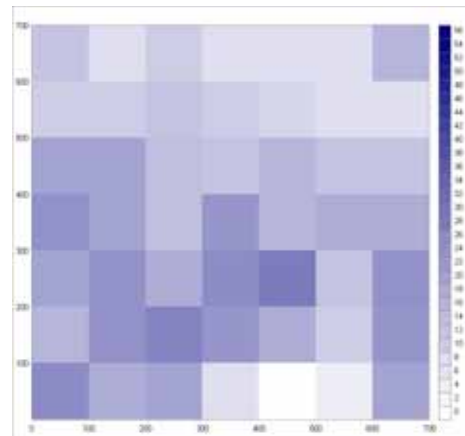

Media frecvențelor: 15,65    Frecv. minimă: 0
Coef. de variație: 43,18%    Frecv. maximă: 29

**Figura 5.16 Repartiția spațială a frecvenței puieților pe metrul pătrat în suprafața nr. 8**

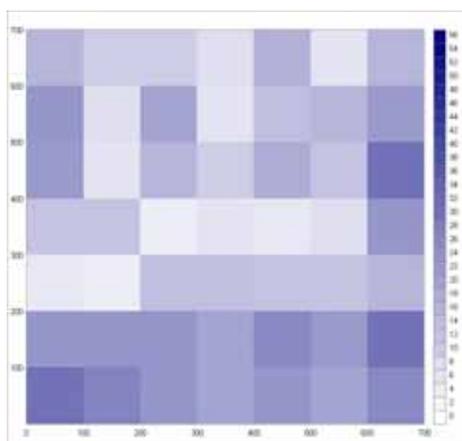

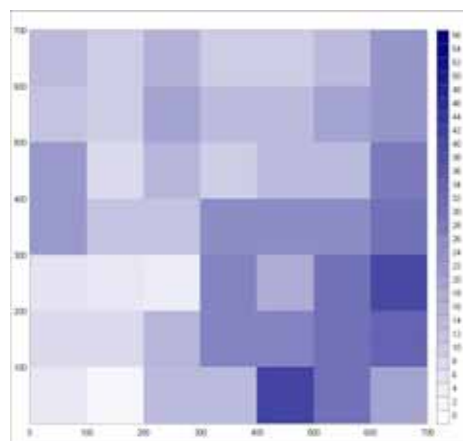

Media frecvențelor: 16,16   Frecv. minimă: 3

Coef. de variație: 47,39%   Frecv. maximă: 31

Media frecvențelor: 18,14   Frecv. minimă: 2

Coef. de variație: 51,51%   Frecv. maximă: 41

**Figura 5.17 Repartiția spațială a frecvenței puieților pe metrul pătrat în suprafața nr. 9**

**Figura 5.18 Repartiția spațială a frecvenței puieților pe metrul pătrat în suprafața nr. 10**

**Tabelul 5.3**

**Date statistice privitoare la frecvența puieților în suprafețele elementare de 1x1 m**

| Media | 14,80 | Coeficientul de variație | 57% |
|---|---|---|---|
| Abaterea standard a mediei | 0,38 | Valoarea minimă | 0 |
| Abaterea standard | 8,38 | Valoarea maximă | 56 |
| Varianța | 70,25 | Volumul probei | 490 |

Analiza cartogramelor evidențiază variația gradului de omogenitate în ceea ce privește repartiția spațială a desimii puieților. Se confirmă astfel importanța rezoluției spațiale de analiză în evaluarea informațiilor. Dacă la analiza desimii pe cele 10 suprafețe de probă a fost calculat un coeficient de variație 20% (la o scară de analiză de 7x7 m), în cazul reducerii rezoluției la suprafețe elementare de 1x1 m, coeficientul de variație obținut pentru cele 490 valori ale desimii a crescut la 57%, gradul de heterogenitate fiind mult mai mare pe suprafețe mici. Dimensiunea de 1x1 m a suprafeței elementare a fost aleasă datorită faptului că un număr mare de studii privitoare la regenerarea arboretelor folosesc drept etalon această suprafață în raportările privitoare la desime. Avantajul folosirii acestei metode de analiză constă în faptul că se pot afla informații detaliate asupra unui proces. În acest caz, al analizei desimii, a fost posibilă determinarea coeficienților de variație din interiorul fiecărei suprafețe de probă, evidențiindu-se valori cuprinse între un minim de 33,40% (corespunzător suprafeței nr. 1) și 92,35% (în cazul suprafeței nr. 3), ceea ce arată un grad ridicat de neomogenitate.

S-a determinat o legătură puternică între coeficienții de variație ai desimii pe suprafața de probă și desimea puieților din respectiva suprafață, caracterizată de un coeficient de corelație negativ, distinct semnificativ (r=-0,753 **). Această relație care arată că omogenitatea crește odată cu desimea puieților, aparent este contradictorie. În mod normal apare tentația de a afirma că la o desime redusă puieții vor reuși să valorifice mai bine aria potențială de dezvoltare, repartizându-se cât mai uniform în suprafață. În realitate apare un fenomen de aglomerare pe anumite spații restrânse, vizibil în cartogramele prezentate anterior, care determină o omogenitate redusă. După cum se afirmă în multe lucrări de ecologie a plantelor (Cole, 1946; Pielou, 1961), concurența determină o valorificare superioară a spațiului printr-o răspândire a indivizilor ce tinde să urmeze un model uniform de organizare spațială. Acest fapt conduce la o creștere a omogenității din punct de vedere al desimii și implicit la reducerea coeficientului de variație.

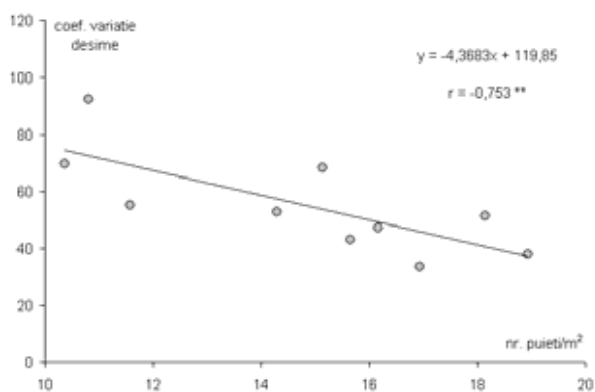

**Figura 5.19 Relația dintre coeficientul de variație al desimii în suprafața (calculat pentru o împărțire în suprafețe de 1x1 m) și numărul de puieți la metrul pătrat**

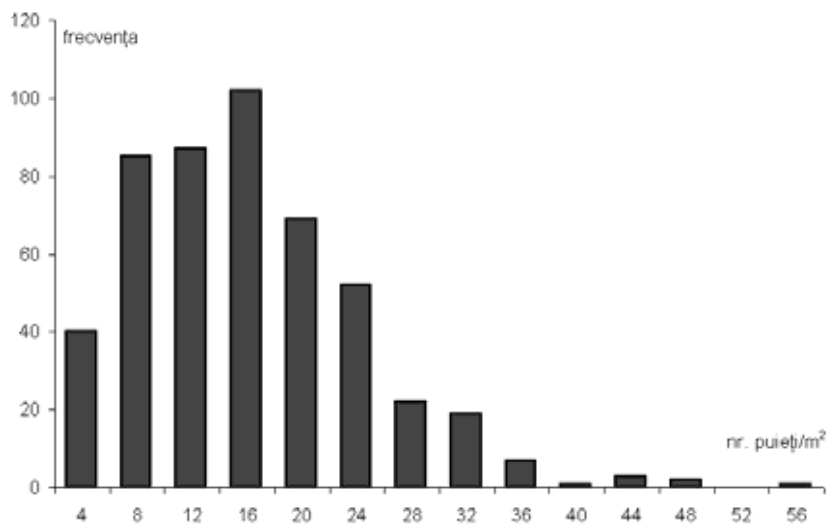

Parametrii statistici ai distribuției experimentale calculați în urma grupării în clase:
media: 14,21;  abaterea standard a mediei: 0,38;  abaterea standard: 8,31
modulul: 13,26;  mediana: 13,30;  coeficientul de variație: 58,48;
indicele de asimetrie:0,97;  indicele excesului: 1,55

**Figura 5.20 Distribuția numărului suprafețelor elementare (1x1 m) pe categorii
ale numărului de puieți**

În finalul analizei structurii după numărul de arbori, se prezintă distribuția numărului suprafețelor elementare (de 1x1 m) pe categorii ale numărului de puieți. Au fost formate 14 clase ale numărului de puieți, cu mărimea clasei egală cu 4. Se observă o distribuție unimodală, cu modulul situat în clasa cu limita maximă 16 (Mo=13,26), în care are loc o descreștere a numărului de suprafețe odată cu creșterea desimii (nr. puieți/m$^2$). Distribuția experimentală este caracterizată de o asimetrie pozitivă, de stânga (A=0,97) ce arată proporția ridicată a suprafețelor elementare cu un număr de 5-16 puieți/m$^2$ și de un exces pozitiv (E=1,55; curbă leptokurtică), valori obținute în urma calculelor efectuate pentru valorile individuale grupate în clase (fig. 5.20).

În urma analizei caracteristicilor distribuției experimentale au fost alese pentru ajustarea acesteia distribuțiile teoretice Beta, Gamma și Weibull, frecvent folosite în modelarea caracteristicilor arboretelor, ceea din urmă fiind recomandată în caracterizarea structurii semințișului (Miina et al., 2006; Siipilehto, 2006).

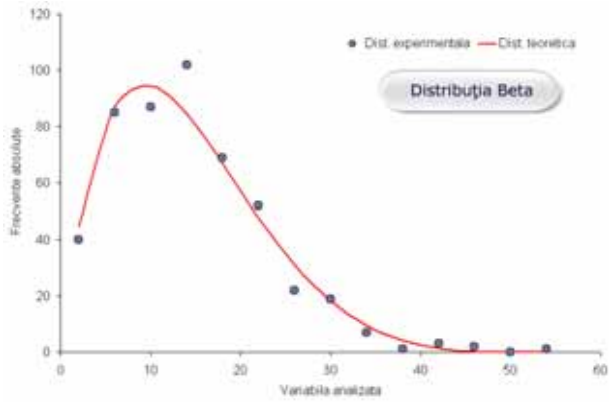

a
Parametrii distribuției teoretice Beta:
$\alpha = 1,93$  $\gamma = 5,67$  $k = 1,889 \times 10^{-7}$

$\chi^2_{exp} = 8,98$

$\chi^2_{teor} = 11,07$  q=0,05

$\chi^2_{exp} < \chi^2_{teor}$

$D_{exp} = 0,031$

$D_{teor} = 0,258$

$D_{exp} < D_{teor}$  q=0,05

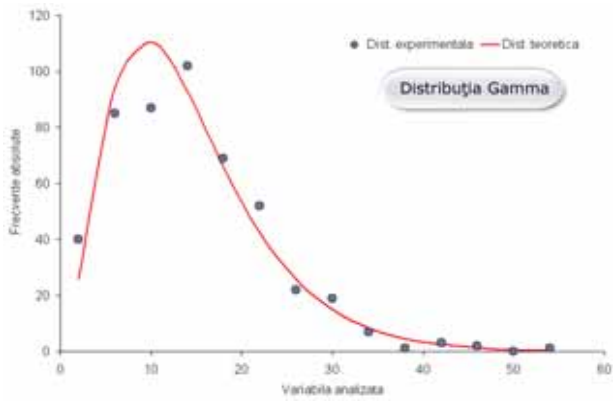

b
Parametrii distribuției teoretice Gamma:
$\theta = 14,21$  $k = 2,93$

$\chi^2_{exp} = 18,35$

$\chi^2_{teor} = 16,92$  q=0,05

$\chi^2_{exp} > \chi^2_{teor}$

$D_{exp} = 0,037$

$D_{teor} = 0,258$

$D_{exp} < D_{teor}$  q=0,05

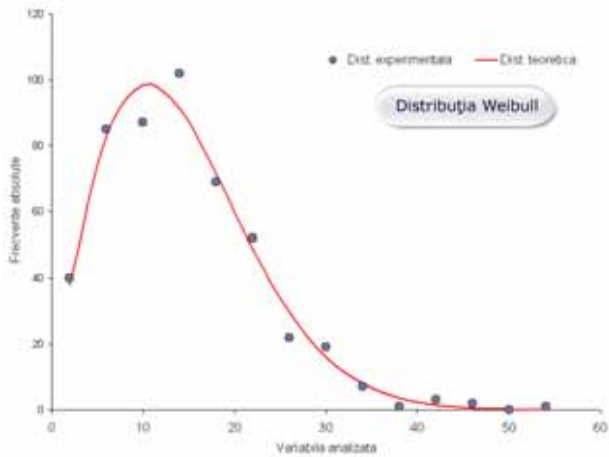

c
Parametrii distribuției teoretice Weibull:
$\alpha = 1,839$  $\beta = 16,081$  $\gamma = 1944$

$\chi^2_{exp} = 5,95$

$\chi^2_{teor} = 16,92$  q=0,05

$\chi^2_{exp} < \chi^2_{teor}$

$D_{exp} = 0,014$

$D_{teor} = 0,258$

$D_{exp} < D_{teor}$  q=0,05

**Figura 5.21 Ajustarea distribuției experimentale a numărului suprafețelor elementare pe categorii ale numărului de puieți cu ajutorul distribuțiilor teoretice Beta (a), Gamma (b) și Weibull (c)**

Calculul valorilor funcțiilor teoretice Beta și Gamma s-a realizat cu ajutorul macro-ului de tip Excel SILVASTAT (Popa, 1999). Parametrii distribuției teoretice Weibull au fost determinați separat, folosind un algoritm bazat pe metoda celor mai mici pătrate. Ulterior pentru verificarea ipotezei nule prin testul $\chi^2$ au fost contopite clasele cu frecvențe absolute mai mici decât 5, s-a calculat valoarea $\chi^2_{exp}$ pentru un prag de semnificație de 5% și s-a comparat cu valoarea $\chi^2_{teoretic}$. Suplimentar, în urma constatării unor deficiențe ale testului de ajustare $\chi^2$, s-a calculat și testul Kolmogorov–Smirnov (D) pentru pragul de semnificație de 5%.

În cazul ajustării distribuției experimentale a frecvențelor suprafețelor elementare pe categorii ale numărului de puieți cu ajutorul distribuției teoretice Beta s-a obținut $\chi^2_{exp} < \chi^2_{teoretic}$ (pentru q=5%) ceea ce conduce la validarea ipotezei nule, având o bună ajustare (fig. 5.21 a).

Distribuția teoretică Gamma se dovedește a fi mai puțin flexibilă decât Beta, în cazul acestui tip de ajustare a distribuției experimentale obținându-se $\chi^2_{exp} > \chi^2_{teoretic}$ (pentru q=5%) ceea ce conduce la invalidarea ipotezei nule (fig. 5.21 b). Totuși și în acest caz valoarea apropiată a lui $\chi^2_{exp}$ de $\chi^2_{teoretic}$, precum și analiza vizuală arată un anumit grad de similaritate a celor două distribuții.

În cazul folosirii distribuției Weibull (fig. 5.21 c) a fost acceptată ipoteza nulă, obținându-se cea mai mică valoare a lui $\chi^2_{exp}$ (5,95) și implicit cea mai bună ajustare a distribuției experimentale. Acest fapt atestă încă odată capacitatea deosebită a acestei distribuții teoretice de a surprinde caracteristicile structurale ale ecosistemelor forestiere.

Modelul matematic al distribuției numărului de suprafețe elementare pe categorii ale numărului de puieți oferă informații cu privire la răspândirea în spațiu a indivizilor și poate fi integrat într-un model de regenerare. Analiza structurii după numărul de indivizi este utilă în evaluarea organizării spațiale a populațiilor, în cazul regenerărilor arboretelor importanța acesteia fiind augmentată de numărul mare al puieților pe spații restrânse, precum și de variabilitatea mare a acestui parametru structural.

## 5.2. Structura în raport cu parametrii biometrici înregistrați

Garcia (2008) consideră că distribuțiile parametrilor structurali ai unui arboret reprezintă primele forme de modelare a unui sistem forestier. Modelul Optimal (Cenușă, 1996 b) constituie un exemplu în acest sens, fiind modelată dinamica arboretelor de molid prin intermediul distribuțiilor parametrilor biometrici și a relațiilor dintre aceștia și celelalte atribute structurale.

În cazul de față, în suprafețele analizate au fost prelevate datele biometrice ale tuturor puieților în vederea folosirii acestora la modelarea matematică a distribuțiilor, a evidențierii legăturilor dintre elementele biometrice ale puieților, precum și a analizelor relațiilor dintre aceștia. Pentru fiecare atribut structural analiza a presupus determinarea indicatorilor statistici, modelarea matematică a distribuțiilor experimentale și analiza variabilității în spațiu. Anterior determinării indicatorilor statistici și a distribuțiilor experimentale s-a folosit un criteriu de eliminarea a observațiilor extreme. Acest fapt este motivat de prezența în suprafețele studiate a unui număr redus de puieți instalați anterior celorlalți. Acești puieți au fost înregistrați deoarece anumite analize și prelucrări impun folosirea datelor respective, dar numărul mic al acestora (o proporție mai mică de 3‰ din numărul total de puieți) nu a permis constituirea unei populații statistice diferite. Pentru a nu avea prelucrările statistice alterate de caracteristicile biometrice ale acestor puieți a fost efectuată eliminarea valorilor extreme conform criteriului *Grubbs* (Horodnic, 2004). În vederea automatizării operației de identificare a observațiilor aberante a fost realizat un model algoritmic de calcul, implementat într-o foaie de lucru tip *Microsoft Excel.*

Distribuțiile teoretice utilizate în vederea modelării matematice a distribuțiilor experimentale a parametrilor biometrici sunt Beta, Gamma și Weibull. Au fost testate și alte funcții de densitate dar rezultatele obținute au fost inferioare calitativ celor amintite. Pentru a testa ajustarea distribuțiilor a fost folosit criteriul $\chi^2$ cu un prag de semnificație de 1%.

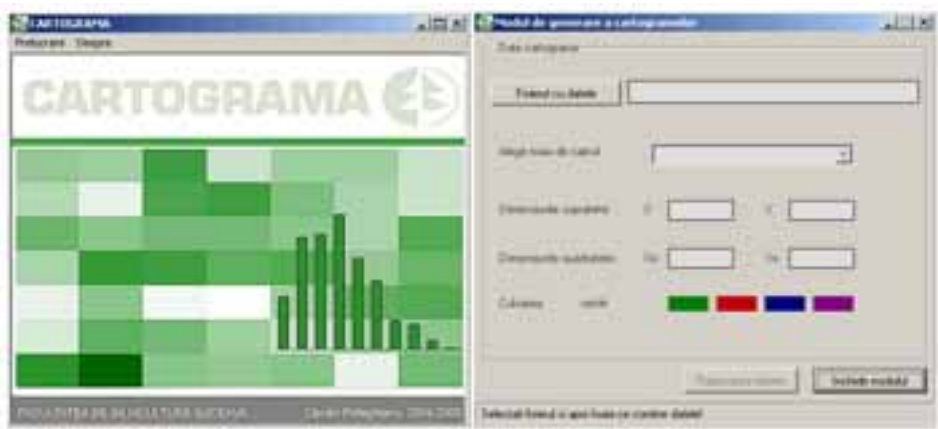

**Figura 5.22 Interfața utilizator a aplicației CARTOGRAMA**

Variabilitatea parametrilor biometrici în interiorul suprafețelor studiate a fost analizată cu ajutorul aplicației software CARTOGRAMA. Aceasta este o aplicație proprie, creată pentru a facilita studiul distribuirii în spațiu a evenimentelor prin împărțirea unei suprafețe în mai multe suprafețe elementare. Programul preia datele din foi de calcul de tip *Microsoft Excel*, calculează o serie de indicatori statistici (media, frecvența, abaterea standard, coeficientul de variație – pentru toată suprafața și pentru fiecare suprafață elementară în parte) și generează grafice ale frecvenței de apariție, respectiv ale mediei valorilor unui eveniment pentru fiecare suprafață elementară. Pentru a asigura posibilitatea comparării vizuale a rezultatelor obținute la analiza unui parametru s-a folosit aceeași scară de intensitate coloristică pentru toate suprafețele studiate.

## 5.2.1. Structura în raport cu diametrul puieților

Particularitățile suprafețelor regenerate natural impun adoptarea unor tipuri de distribuție diferite de cele folosite în modelarea structurii arboretelor mature. Gradul mare de neomogenitate al valorilor diametrelor redat de coeficienții de variație, precum și distribuția neomogenă în spațiu, face ca analiza structurii în raport cu diametrul puieților să se desfășoare într-o manieră aparte.

O altă particularitate a studiului regenerărilor arboretelor se referă la folosirea diametrului la colet, stabilitatea acestui parametru fiind mult mai redusă decât a diametrului de bază datorită erorilor produse de inconsecvența în poziționare la efectuarea măsurătorilor. Chiar dacă pentru măsurarea diametrului la colet s-a folosit un șubler care oferă o bună precizie de măsurare, poziționarea diferită de la un puiet la altul poate induce erori.

Tabelul 5.4

Indicatorii statistici ai diametrului puieților în suprafețele studiate

| | Medie (mm) | Abatere standard | Coef. de variație | Minim (mm) | Maxim (mm) | Volum probă |
|---|---|---|---|---|---|---|
| Suprafața nr. 1 | 8,27 | 4,50 | 54% | 2 | 22 | 830 |
| Suprafața nr. 2 | 7,73 | 4,23 | 55% | 2 | 23 | 567 |
| Suprafața nr. 3 | 10,83 | 7,44 | 69% | 2 | 41 | 525 |
| Suprafața nr. 4 | 9,94 | 4,77 | 48% | 1 | 30 | 698 |
| Suprafața nr. 5 | 9,11 | 5,50 | 60% | 1 | 32 | 506 |
| Suprafața nr. 6 | 8,10 | 6,13 | 76% | 2 | 33 | 736 |
| Suprafața nr. 7 | 7,73 | 4,71 | 61% | 1 | 28 | 923 |
| Suprafața nr. 8 | 9,98 | 4,30 | 43% | 2 | 27 | 767 |
| Suprafața nr. 9 | 9,36 | 6,14 | 66% | 2 | 36 | 792 |
| Suprafața nr. 10 | 9,04 | 4,43 | 49% | 1 | 27 | 888 |
| **Total** | 8,94 | 5,30 | 59% | 2 | 41 | 7232 |

Tabelul 5.5

Indicatorii statistici ai diametrului puieților pe specii

| | Ca | Ci | Fr | Ju | Pam | St | Te |
|---|---|---|---|---|---|---|---|
| Medie (mm) | 9,26 | 10,23 | 7,01 | 8,36 | 11,56 | 7,01 | 11,99 |
| Abatere standard | 5,32 | 5,84 | 4,37 | 4,29 | 5,47 | 4,29 | 5,91 |
| Coef. de variație | 57% | 57% | 62% | 51% | 47% | 61% | 49% |
| Valoare minimă | 1 | 2 | 2 | 1 | 3 | 1 | 1 |
| Valoare maximă | 39 | 41 | 34 | 28 | 23 | 34 | 35 |
| Volumul probei | 4075 | 123 | 610 | 425 | 61 | 1132 | 724 |

Din analiza datelor prezentate în tabelele anterioare se remarcă un grad redus de omogenitate în privinţa valorilor diametrului, atât în ceea ce priveşte situaţia referitoare la suprafeţele analizate cât şi pe specii. Diametrul la colet mediu al puieţilor se situează la valoarea de 8,94 mm dacă luăm în calcul indivizii din toate pieţele de probă. În cadrul suprafeţelor aceste valori variază de la valoarea minimă de 7,73 mm în suprafaţa 2 la 10,38 mm în suprafaţa 3. Omogenitatea redusă a acestui parametru este subliniată şi de valorile ridicate ale coeficienţilor de variaţie pe suprafaţă - între 43% şi 76%. În ceea ce priveşte situaţia pe specii, puieţii de paltin şi tei înregistrează cele mai mari medii ale diametrului, cu peste 30% mai mari faţă de media totală. Aceste specii manifestă şi cea mai mare omogenitate, fiind singurele specii ce înregistrează coeficienţi de variaţie de sub 50%. De cealaltă parte, stejarul şi frasinul înregistrează cele mai mici medii, cu peste 20% sub media totală, dar şi cei mai mari coeficienţi de variaţie (peste 60%).

În cazul analizei distribuţiei numărului de puieţi pe categorii de diametre (fig. 5.24), la nivelul suprafeţelor se înregistrează în general distribuţii unimodale, cu o tendinţă de deplasare spre stânga şi o oarecare variabilitate a formei.

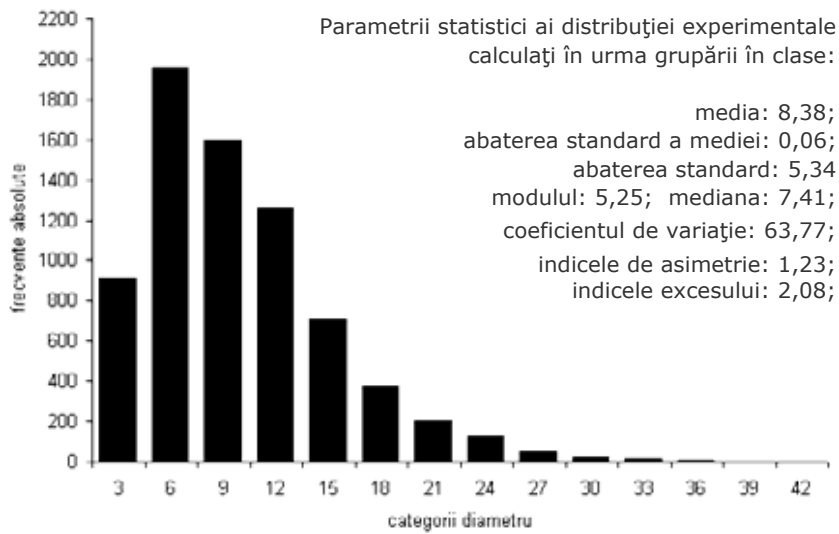

Figura 5.23 Histograma frecvenţelor tuturor puieţilor pe categorii ale diametrului la colet

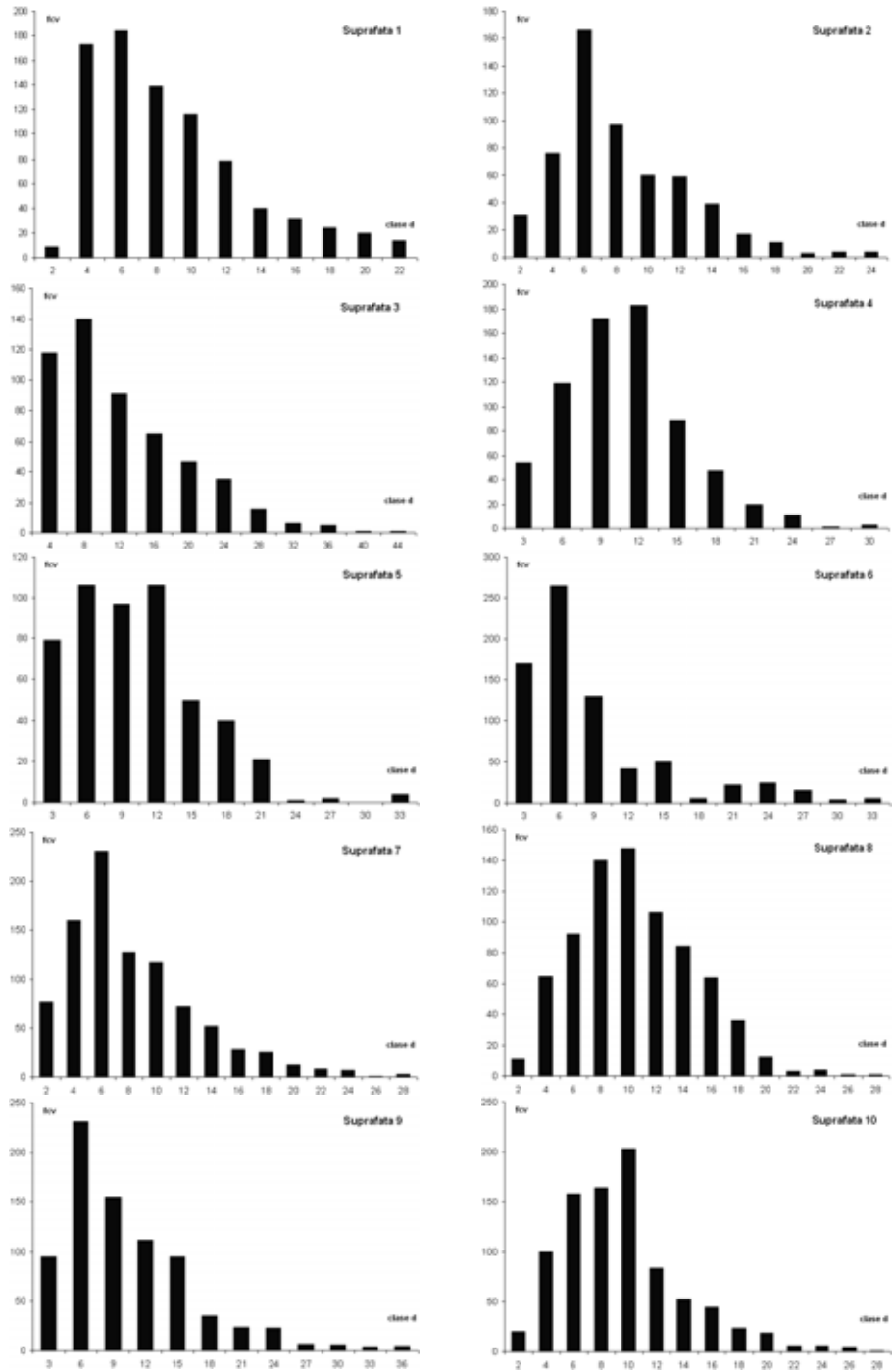

**Figura 5.24 Histograma frecvenţelor puieţilor pe categorii ale diametrului la colet**

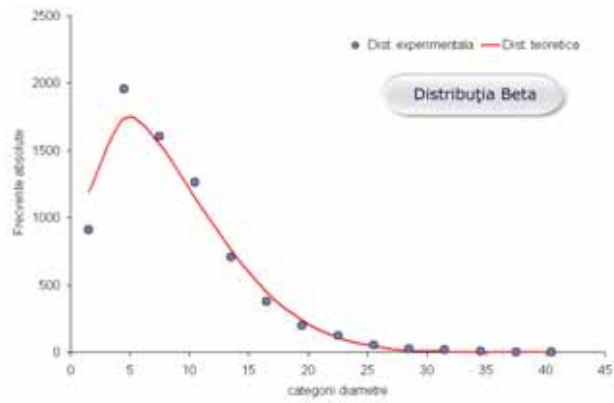

a

Parametrii distribuției teoretice Beta:

$\alpha = 1{,}77$  $\gamma = 7{,}10$  $k = 1{,}36 \times 10^{-7}$

$\chi^2_{exp} = 191{,}79$

$\chi^2_{teor} = 15{,}09$  q=0,01

$\boldsymbol{\chi^2_{exp} > \chi^2_{teor}}$

$D_{exp} = 0{,}039$

$D_{teor} = 0{,}258$  $D_{exp} < D_{teor}$  q=0,05

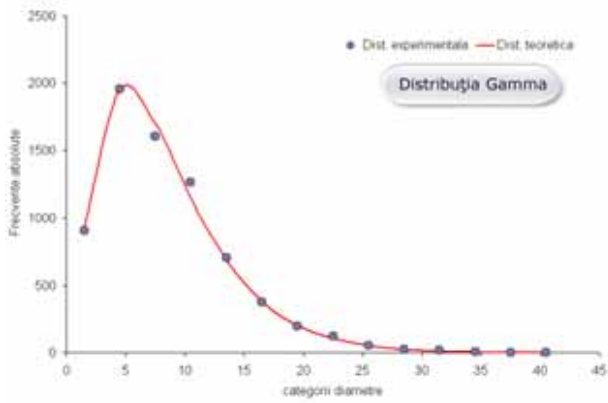

b

Parametrii distribuției teoretice Gamma:

$\theta = 8{,}38$  $k = 2{,}46$

$\chi^2_{exp} = 24{,}42$

$\chi^2_{teor} = 24{,}73$  q=0,01

$\boldsymbol{\chi^2_{exp} < \chi^2_{teor}}$

$D_{exp} = 0{,}018$

$D_{teor} = 0{,}242$  $D_{exp} < D_{teor}$  q=0,05

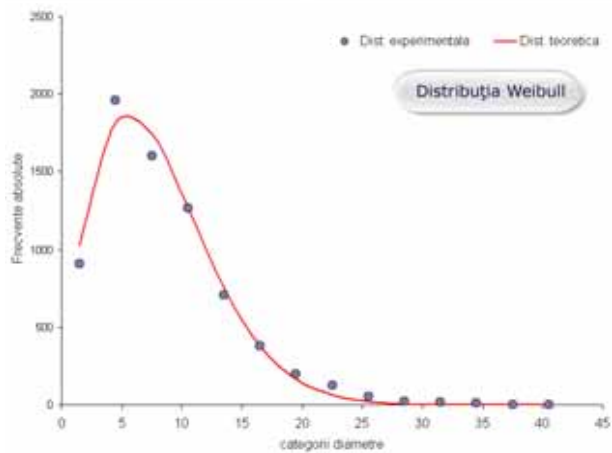

c

Parametrii distribuției teoretice Weibull:

$\alpha = 1{,}745$  $\beta = 9{,}092$  $\gamma = 21402$

$\chi^2_{exp} = 356{,}08$

$\chi^2_{teor} = 21{,}67$  q=0,01

$\boldsymbol{\chi^2_{exp} > \chi^2_{teor}}$

$D_{exp} = 0{,}023$

$D_{teor} = 0{,}258$  $D_{exp} < D_{teor}$  q=0,05

**Figura 5.25 Ajustarea distribuției experimentale a numărului de puieți pe categorii ale diametrului la colet cu ajutorul distribuțiilor Beta (a), Gamma (b) și Weibull (c)**

Distribuția tuturor puieților pe categorii de diametre devine mult mai stabilă (fig. 5.23), fiind unimodală, caracterizată de o puternică asimetrie de stânga (A=1,23) și un exces pozitiv (E=2,08).

Modelarea matematică a distribuției numărului de puieți pe clase de diametre este unul din obiectivele majore ale conceperii unui model al regenerării, ulterior estimându-se cu ajutorul unei matrice de tranziție creșterea și mortalitatea pentru fiecare clasă considerată (Miina et al., 2006). Parametrii funcțiilor de ajustare se pot determina, în baza unui număr suficient de mare de măsurători, prin metode cunoscute - metoda verosimilității maxime, metoda momentelor sau metoda celor mai mici pătrate.

Din analiza vizuală a figurii 5.25, se observă că toate distribuțiile teoretice folosite reușesc să surprindă caracteristicile distribuției experimentale.

Cu toate acestea doar pentru distribuția Gamma s-a obținut statistic o bună ajustare, cu $\chi^2_{exp} < \chi^2_{teoretic}$ (pentru q=1%) ceea ce a condus la validarea ipotezei nule. Distribuțiile teoretice Beta și Weibull se dovedesc în această situație mai puțin flexibile decât Gamma, obținându-se $\chi^2_{exp} > \chi^2_{teoretic}$ (pentru q=1%) ceea ce conduce la respingerea ipotezei nule. Cu toate că aparent (în urma analizei vizuale) aceste distribuții sunt foarte apropiate de cea experimentală, s-au obținut valori ridicate ale lui $\chi^2_{exp}$ (191,79 și 356,08), datorate în special frecvențelor foarte mari înregistrate în unele clase (peste 1000-1500), unde și cele mai mici diferențe generează valori însemnate ce se adaugă lui $\chi^2_{exp}$.

În diagramele din figura 5.26, generate de aplicația CARTOGRAMA, se prezintă variabilitatea spațială a diametrului mediu la colet, pe suprafețe elementare de 1x1m. Se poate observa atât variabilitatea înregistrată între suprafețele analizate cât și în interiorul fiecărei suprafețe.

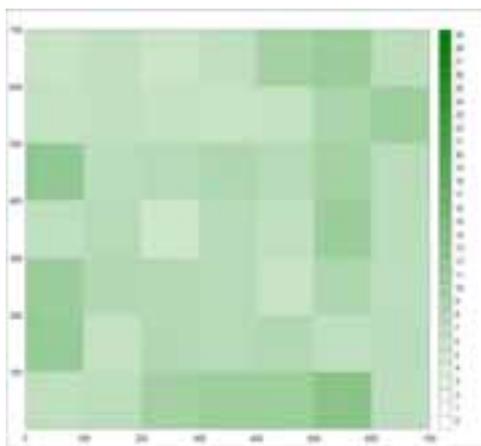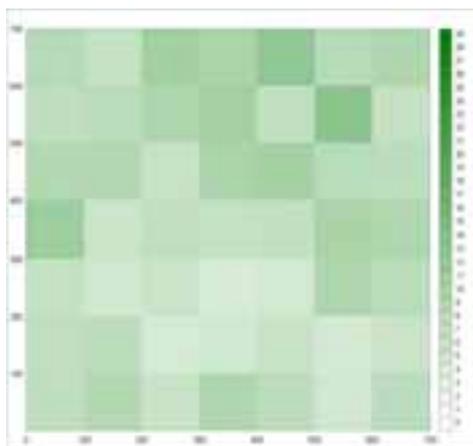

Suprafața 1                    Suprafața 2

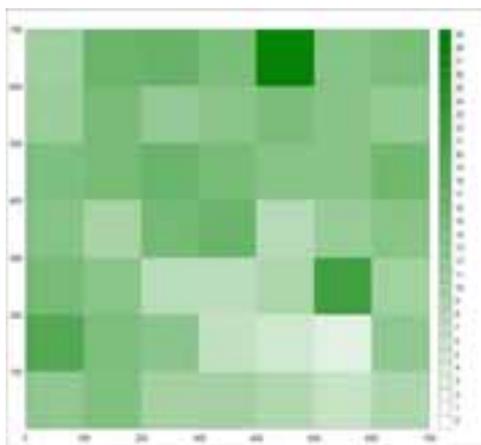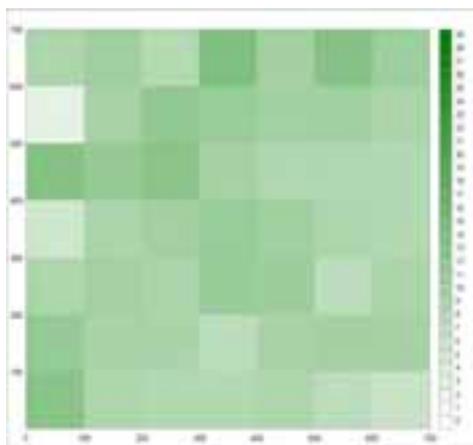

Suprafața 3                    Suprafața 4

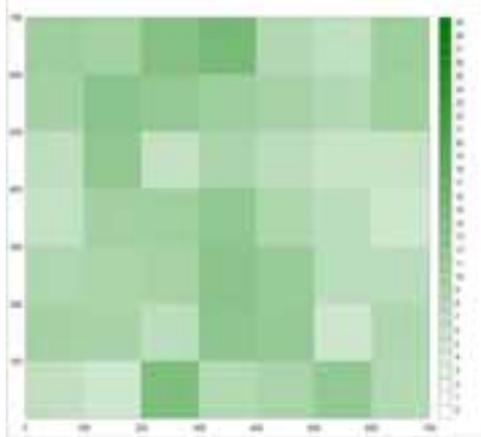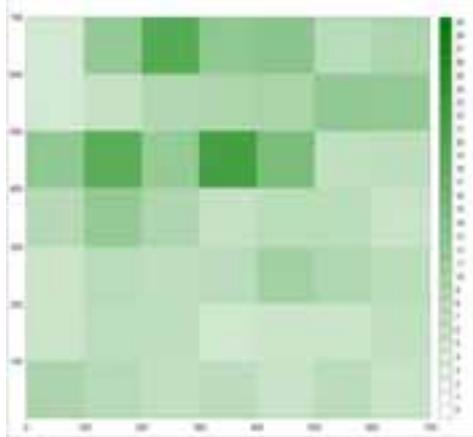

Suprafața 5                    Suprafața 6

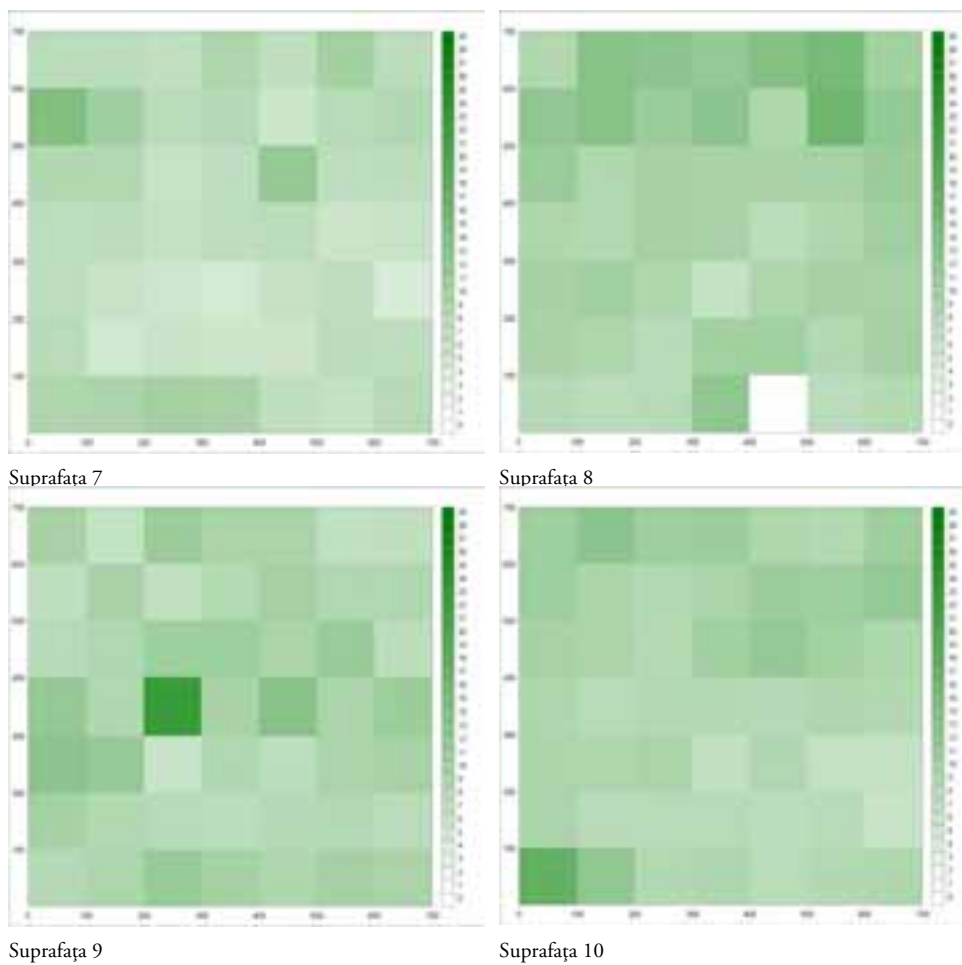

Suprafaţa 7                    Suprafaţa 8

Suprafaţa 9                    Suprafaţa 10

**Figura 5.26 Variabilitatea spaţială a diametrului la colet în suprafeţele studiate**

### 5.2.2. Structura în raport cu înălţimea puieţilor

Datorită rolului determinant al luminii şi competiţiei intense în stratul puieţilor, înălţimea poate fi considerată parametrul central în analiza structurală a seminţişului. Din analiza indicatorilor statistici se remarcă un grad de omogenitate în interiorul suprafeţelor mai ridicat în cazul înălţimii, comparativ cu situaţia diametrelor. Înălţimea medie a puieţilor se situează la 98,01 cm dacă luăm în calcul toţi indivizii din cele zece pieţe de probă. În cadrul suprafeţelor aceste valori variază de la valoarea minimă de 53,07 cm în suprafaţa 2 la 119,73 cm în suprafaţa 9. Coeficienţii de variaţie pe suprafaţă variază între 35% şi 68%. În ceea ce priveşte situaţia pe specii, puieţii de tei, cireş şi jugastru înregistrează cele mai mari medii ale înălţimii, cu 15-20% mai mari faţă de media tuturor puieţilor. Aceste specii manifestă şi cea mai mare omogenitate, fiind singurele specii ce înregistrează coeficienţi de variaţie de sub 50%. De cealaltă parte, frasinul înregistrează cea mai mică medie a înălţimii, cu peste 30% sub media totală, dar şi cel mai ridicat coeficient de variaţie (69%), acest lucru reducând din relevanţa mediei.

Tabelul 5.6

**Indicatorii statistici ai înălţimii puieţilor în suprafeţele studiate**

|  | Medie (cm) | Abatere standard | Coef. de variaţie | Minim (cm) | Maxim (cm) | Volum probă |
|---|---|---|---|---|---|---|
| Suprafaţa nr. 1 | 88,46 | 43,22 | 49% | 10 | 235 | 830 |
| Suprafaţa nr. 2 | 53,07 | 31,61 | 60% | 7 | 163 | 563 |
| Suprafaţa nr. 3 | 119,13 | 64,18 | 54% | 15 | 368 | 528 |
| Suprafaţa nr. 4 | 94,57 | 43,61 | 46% | 6 | 260 | 699 |
| Suprafaţa nr. 5 | 95,30 | 52,12 | 55% | 7 | 288 | 506 |
| Suprafaţa nr. 6 | 79,90 | 54,39 | 68% | 15 | 290 | 718 |
| Suprafaţa nr. 7 | 102,03 | 60,88 | 60% | 7 | 295 | 928 |
| Suprafaţa nr. 8 | 119,65 | 42,28 | 35% | 21 | 244 | 767 |
| Suprafaţa nr. 9 | 119,73 | 53,83 | 45% | 20 | 280 | 792 |
| Suprafaţa nr. 10 | 99,51 | 42,27 | 42% | 11 | 229 | 889 |
| **Total** | 98,01 | 52,94 | 54% | 6 | 368 | 7220 |

Tabelul 5.7

**Indicatorii statistici ai înălţimii puieţilor pe specii**

|  | Ca | Ci | Fr | Ju | Pam | St | Te |
|---|---|---|---|---|---|---|---|
| Medie (cm) | 102,69 | 111,94 | 65,32 | 108,42 | 105,76 | 77,62 | 122,16 |
| Abatere standard | 51,25 | 55,31 | 45,04 | 50,22 | 61,94 | 48,50 | 54,28 |
| Coef. de variaţie | 50% | 49% | 69% | 46% | 59% | 62% | 44% |
| Valoare minimă | 6 | 23 | 10 | 15 | 20 | 7 | 7 |
| Valoare maximă | 368 | 335 | 250 | 262 | 250 | 295 | 280 |
| Volumul probei | 4064 | 124 | 610 | 425 | 59 | 1132 | 724 |

În cazul analizei distribuției numărului de puieți pe categorii de înălțimi (fig. 5.28), la nivelul celor zece suprafețe analizate se înregistrează o variabilitate mai mare a gamelor de distribuții întâlnite. Competiția pentru factorul lumină determină în cazul înălțimilor o separare mai clară în interiorul fiecărei suprafețe a categoriilor de înălțimi, fapt ce poate conduce și la apariția mai multor maxime locale ale frecvenței. Competiția pentru lumină, ce acționează mult mai intens în cazul acestei faze de creștere, explică și reducerea asimetriei de stânga și a excesului distribuției tuturor puieților pe categorii de înălțimi comparativ cu distribuția pe categorii de diametre. În acest caz indicele de asimetrie are valoarea 0,59 (față de 1,23 în cazul diametrelor), iar indicele excesului este egal cu 0,05 (față de 2,08 în cazul diametrelor). Din comparația distribuțiilor pentru cei doi parametri biometrici se deduce că puieții reușesc să utilizeze mult mai bine spațiul aerian, acest fapt determinând răspândirea mai echilibrată a frecvențelor pe clase de înălțimi. Distribuția pe categorii de înălțimi se reglează în cazul luării în calcul a tuturor puieților (fig. 5.27), devenind mai uniformă. Este caracterizată de o curbă unimodală mezokurtică (E≈0), cu o moderată asimetrie de stânga (A=0,59).

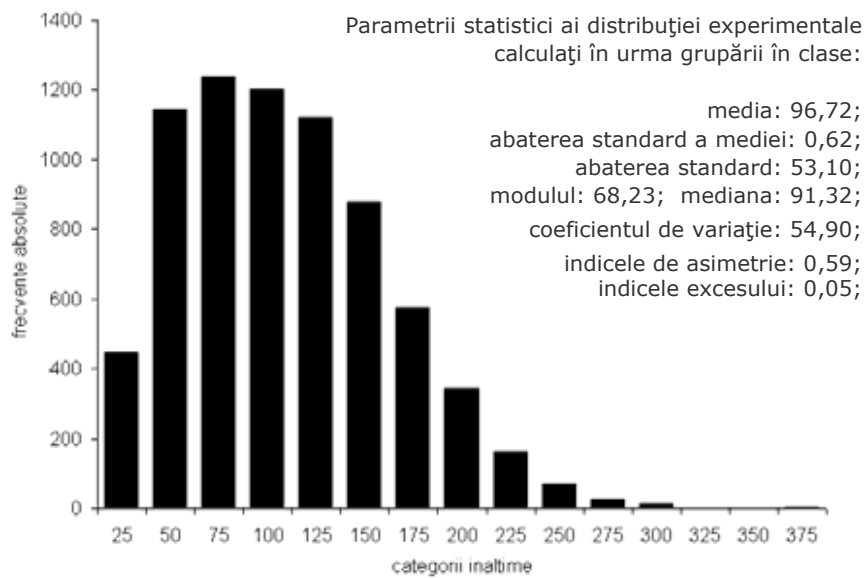

**Figura 5.27 Histograma frecvențelor tuturor puieților pe categorii ale înălțimii**

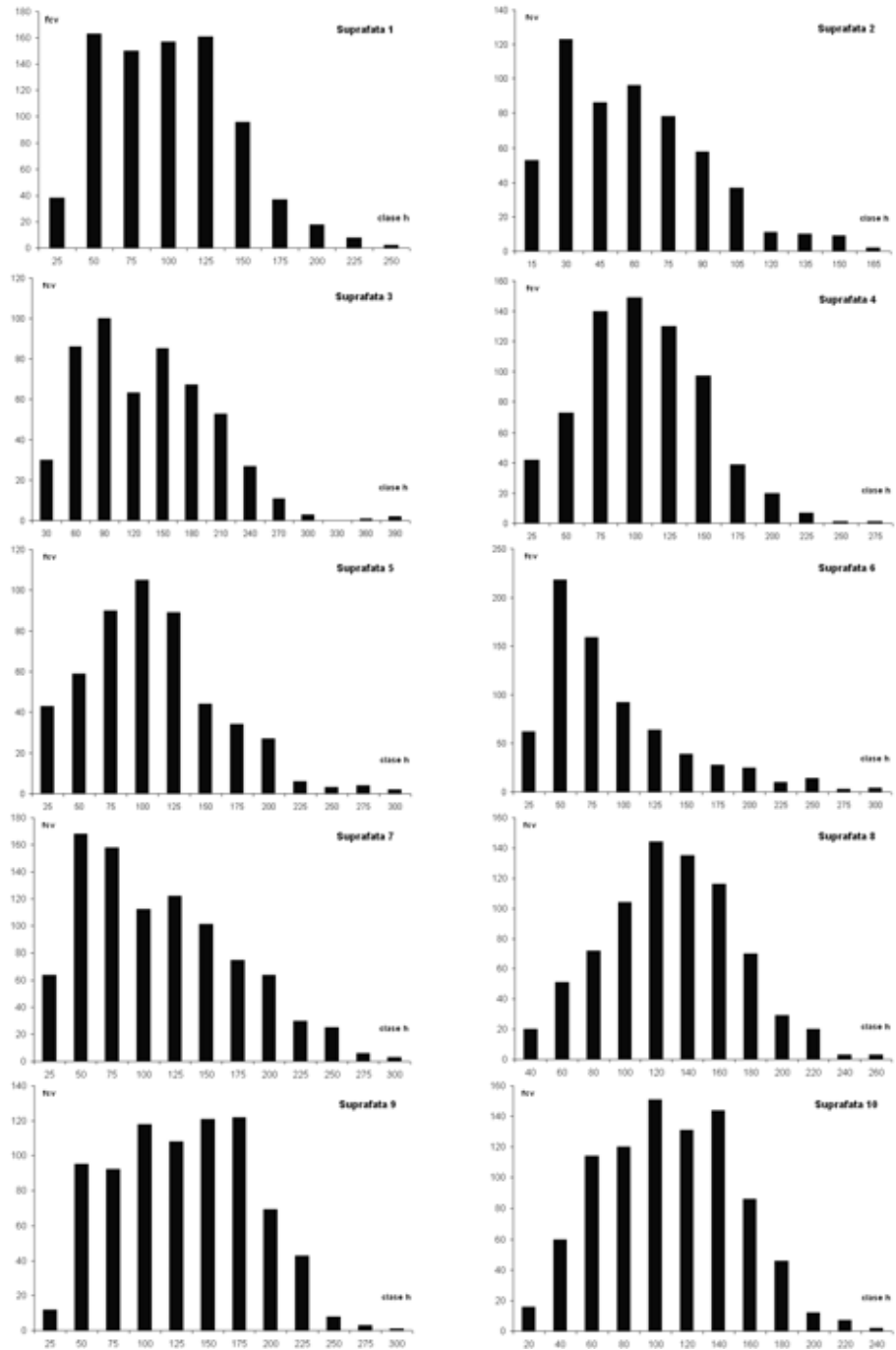

**Figura 5.28 Histograma frecvenţelor puieţilor pe categorii ale înălţimii**

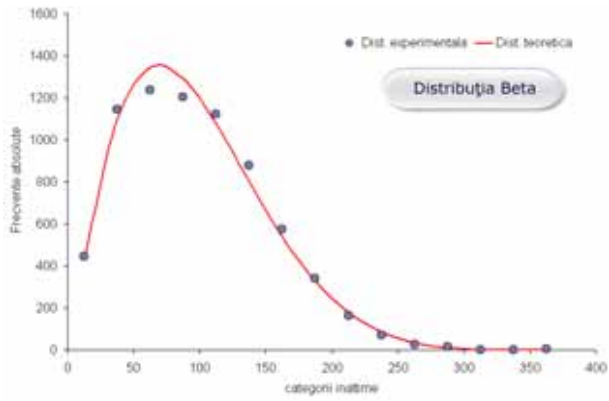

a
Parametrii distribuţiei teoretice Beta:

$\alpha$ = 2,20   $\gamma$ = 6,34   $k$ = 0,4 x 10⁻⁷

$\chi^2_{exp}$ = 36,44

$\chi^2_{teor}$ = 18,48   q=0,01

$\boldsymbol{\chi^2_{exp} > \chi^2_{teor}}$

$D_{exp}$ = 0,019

$D_{teor}$ = 0,242   $\boldsymbol{D_{exp} < D_{teor}}$

q=0,05

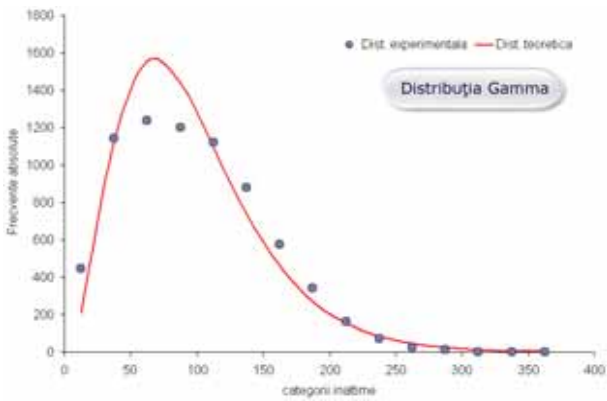

b
Parametrii distribuţiei teoretice Gamma:

$\theta$ = 96,72   $k$ = 3,32

$\chi^2_{exp}$ = 483,83

$\chi^2_{teor}$ = 24,73   q=0,01

$\boldsymbol{\chi^2_{exp} > \chi^2_{teor}}$

$D_{exp}$ = 0,041

$D_{teor}$ = 0,242   $\boldsymbol{D_{exp} < D_{teor}}$

q=0,05

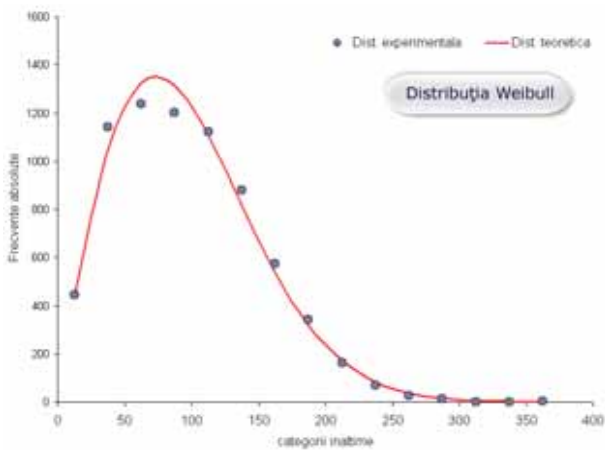

c
Parametrii distribuţiei teoretice Weibull:

$\alpha$ = 1,883   $\beta$ = 109,788

$\gamma$ =179504

$\chi^2_{exp}$ = 43,94

$\chi^2_{teor}$ = 24,73   q=0,01

$\boldsymbol{\chi^2_{exp} > \chi^2_{teor}}$

$D_{exp}$ = 0,014

$D_{teor}$ = 0,242   $\boldsymbol{D_{exp} < D_{teor}}$

q=0,05

**Figura 5.29 Ajustarea distribuţiei experimentale a numărului de puieţi pe categorii ale înălţimii cu ajutorul distribuţiilor Beta (a), Gamma (b) şi Weibull (c)**

Modelarea distribuţiei experimentale s-a realizat cu aceleaşi distribuţii teoretice folosite anterior (Beta, Gamma, Weibull). Parametrii funcţiilor şi valorile necesare testului de semnificaţie a gradului de ajustare sunt prezentate în figura 5.29.

Analiza vizuală a variantelor prezentate în figura 5.29 arată că toate distribuţiile teoretice reuşesc să surprindă o mare parte din caracteristicile distribuţiei experimentale. Cu toate acestea pentru nici una din ele nu s-a obţinut din punct de vedere al asigurării statistice o bună ajustare, în toate cele trei cazuri având $\chi^2_{exp} > \chi^2_{teoretic}$ (pentru q=1%) ceea ce a condus la respingerea ipotezei nule. Distribuţiile teoretice Beta şi Weibull se dovedesc în această situaţie mai flexibile decât Gamma, obţinându-se valori mult mai mici ale lui $\chi^2_{exp}$ (36,44; respectiv 43,94 faţă de 483,83).

Se poate considera că modelarea matematică a repartizării numărului de puieţi pe categorii de înălţimi poate fi realizată cu una din cele două funcţii de densitate. Şi în acest caz frecvenţele foarte mari înregistrate în unele clase (peste 1000), determină creşterea valorii $\chi^2_{exp}$ chiar şi pentru diferenţe mici înregistrate între distribuţii.

În diagramele incluse în figura 5.30, generate de aplicaţia software CARTOGRAMA, se prezintă variabilitatea spaţială a înălţimii, prin reprezentarea grafică a mediei înălţimilor pe suprafeţe elementare de 1x1m. Pentru a se oferi posibilitatea comparării variabilităţii parametrului analizat nu doar în interiorul suprafeţelor de 7x7 m ci şi între aceste suprafeţe s-a folosit aceeaşi scară de reprezentare (de la 0 la 310 cm) pentru toate cele zece pieţe.

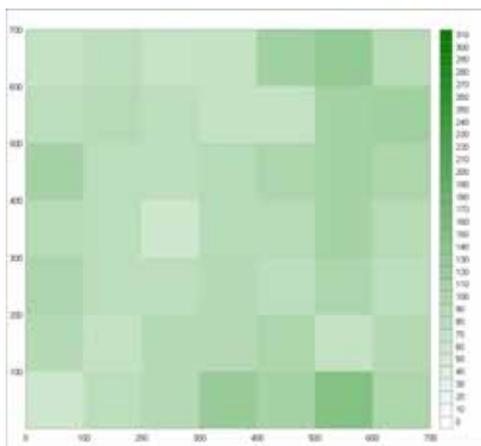

Suprafaţa 1

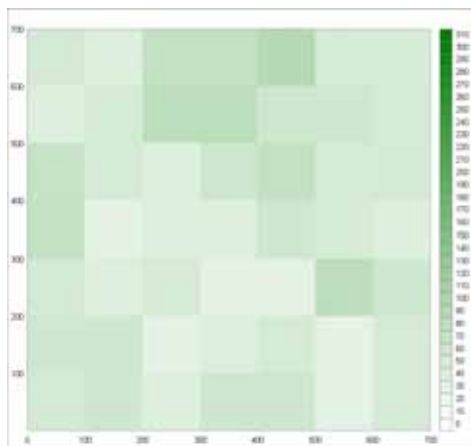

Suprafaţa 2

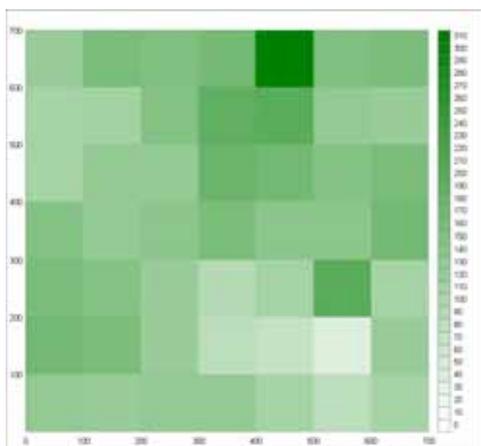

Suprafaţa 3

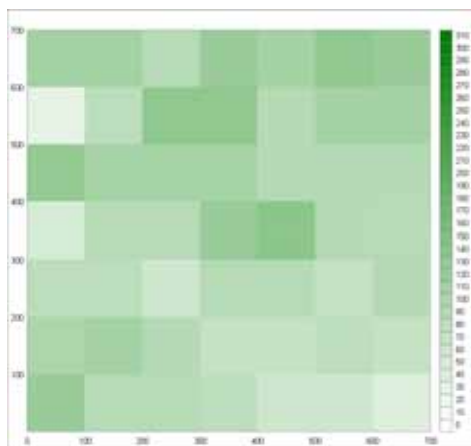

Suprafaţa 4

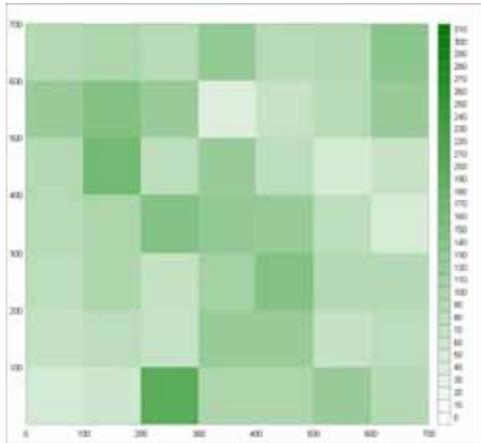

Suprafaţa 5

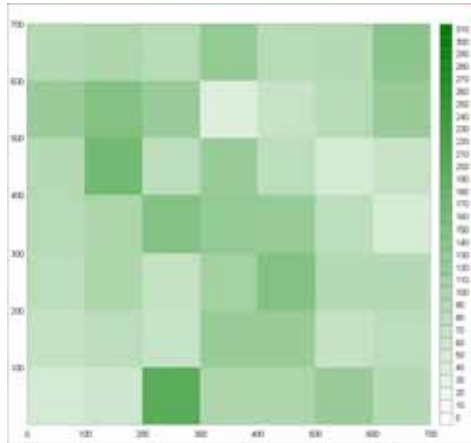

Suprafaţa 6

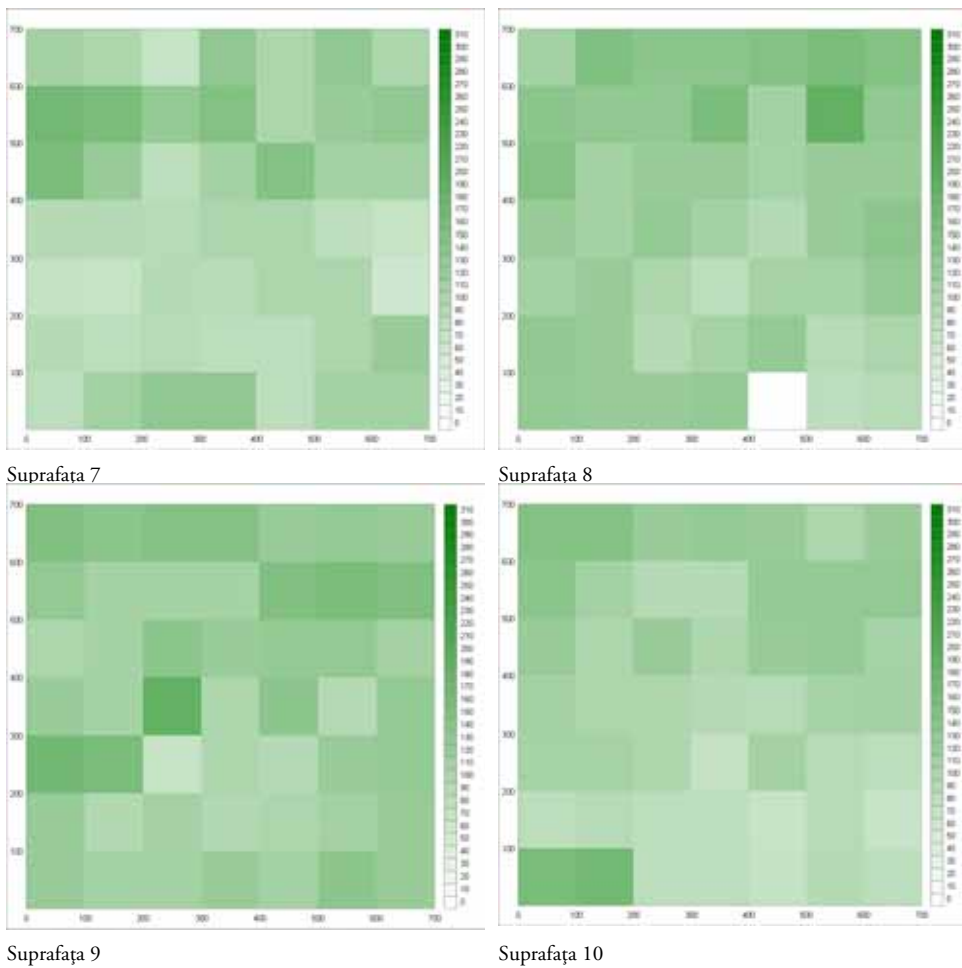

Suprafaţa 7    Suprafaţa 8

Suprafaţa 9    Suprafaţa 10

**Figura 5.30 Variabilitatea spaţială a înălţimii puieţilor în suprafeţele studiate**

Luând în considerare importanţa deosebită a înălţimii în stabilirea relaţiilor dintre puieţi, pentru acest parametru biometric au fost efectuate o serie de analize suplimentare. Pentru a simplifica efectuarea acestor analize s-au definit patru straturi de încadrare a puieţilor după înălţimea totală. Aceste straturi sunt:

- stratul 1 – puieţii cu înălţimea cuprinsă între 0 şi 25 cm;
- stratul 2 – puieţii cu înălţimea cuprinsă între 26 şi 75 cm;
- stratul 3 – puieţii cu înălţimea cuprinsă între 76 şi 150 cm;
- stratul 4 – puieţii cu înălţimea mai mare de 150 cm.

Datele privitoare la încadrarea puieților în aceste straturi, pe specii, sunt prezentate procentual, sub formă grafică, făcându-se referire doar la speciile care au un procent de participare în compoziția semințișului mai mare de 1%.

S-a constatat că structura pe straturile de înălțimi a semințișului instalat natural este următoarea: 6 % din totalul puieților sunt încadrați în primul strat, 33 % în stratul 2, 44 % în stratul 3 și 17 % în stratul 4.

Numărul mic de indivizi din primul strat (al plantulelor), arată că instalarea unor noi generații de puieți se realizează tot mai greu în condițiile în care generațiile anterioare au atins dimensiuni considerabile, care le permit să elimine cea mai mare parte din vegetația lemnoasă proaspăt instalată. Se observă că puieții sunt echilibrat repartizați în celelalte categorii, valorificând la maxim spațiul pe verticală, după cum s-a remarcat și din distribuția pe clase de înălțimi.

În cazul analizei pe specii se disting două aspecte – cel al procentului de participare al speciilor la formarea fiecărui strat și a modului de repartizare al puieților unei specii în aceste categorii.

Carpenul, fiind specia cu cel mai mare procent de participare în toată suprafața, este specia majoritară în toate straturile, cu un procent situat în jurul valorii de 50% (figura 5.31). În stratul 1 o reprezentare consistentă o au și stejarul (26%) și frasinul (21%), păstrată și în cel de al doilea strat (22%, respectiv 14%). În straturile superioare, după carpen, teiul are cea mai bună reprezentare, cu un procent de 11% în stratul al treilea și 19% în stratul al patrulea.

Modul de repartizare al puieților unei specii în fiecare strat (figura 5.32), poate să surprindă strategia de dezvoltare pe verticală a unei specii, precum și rezultatul concurenței exercitate de celelalte specii. Frasinul și stejarul sunt speciile care au cel mai bine reprezentat strat 1 în structura lor pe înălțimi, faptul că au o pondere mare a numărului de indivizi în stratul plantulelor dovedind o bună „vigoare" de regenerare. 15% din totalul puieților de frasin și 10% din cei de stejar sunt încadrați în primul strat, în condițiile în care celelalte specii au ponderea acestui strat de circa 2-5%. Aceste două specii sunt și singurele care au maximul de puieți încadrat în stratul 2.

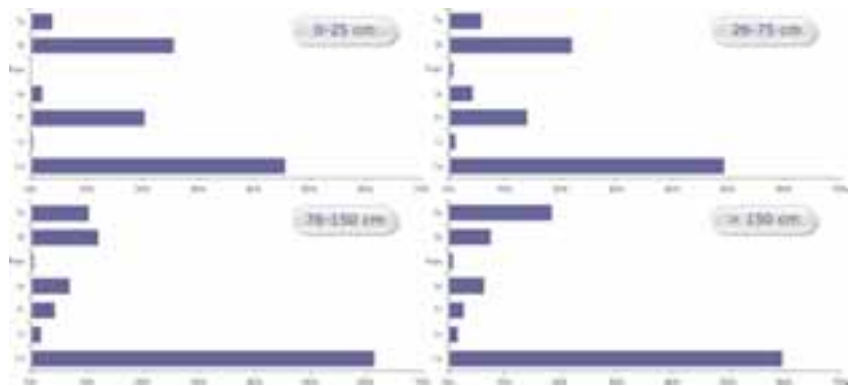

**Figura 5.31 Procentul de participare a speciilor în alcătuirea straturilor după înălțimea puieților**

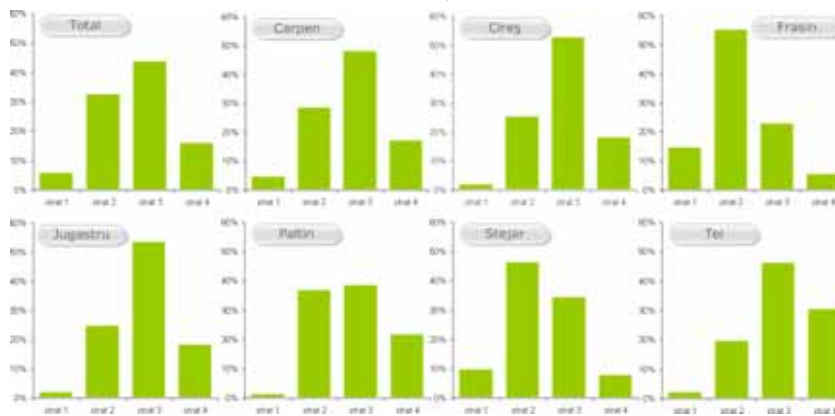

**Figura 5.32 Încadrarea puieților fiecărei specii în cele patru straturi după înălțime**

Ponderea mare a indivizilor acestor specii în straturile inferioare (frasinul are peste 70% din puieți în primele două straturi) indică un ritm al creșterilor mai puțin activ comparativ cu celelalte specii de amestec. La polul opus, dovedind creșteri active, se află teiul cu peste 30% din indivizi situați în stratul 4 și 78% situați în straturile superioare (3 și 4). Jugastrul și cireșul sunt singurele specii care mai au peste 70% dintre puieți situați în straturile superioare.

Analiza structurii înălțimii pe specii localizează poziția pe verticală a fiecărei specii din compoziția semințișului, oferind informații utile în interpretarea rezultatelor unor analize ce vor fi efectuate în capitolele următoare.

### 5.2.3. Structura în raport cu înălțimea până la prima ramură verde

Înălțimea până la prima ramură verde este un indicator ce oferă informații utile privitoare la structura arboretelor. Datorită concurenței pentru accesul la lumină, acest parametru biometric este puternic influențat de relațiile competiționale ale puietului cu vecinii săi și de desime.

În aceeași măsură înălțimea până la prima ramură verde este un factor determinant indirect al volumului coroanei, ce influențează creșterea și dezvoltarea fiecărui individ, fiind frecvent folosită drept variabilă în modelele de creștere (Stage, 1973; Zumrawi, Hann, 1989; Bachmann, 1996).

Ținând cont de aceste considerente s-a considerat oportună includerea acestui parametru în analiza structurală a semințișului. Trebuie făcută observația că s-a forțat folosirea termenului de ”coroană”, practic neputând vorbi încă de această structură în cazul puieților.

Analiza coeficienților de variație indică un grad de omogenitate mai ridicat în cazul înălțimii până la prima ramură verde a coroanei (notată de acum hv), comparativ cu situația înălțimilor și a diametrelor.

Valoarea medie a hv se situează la 26,79 cm dacă luăm în calcul toți indivizii din cele zece piețe de probă. În cadrul suprafețelor valori le hv variază de la valoarea minimă de 20,25 cm în suprafața 2 la 34,60 cm în suprafața 8. Coeficienții de variație pe suprafață variază între 39% și 63%.

În ceea ce privește situația pe specii, puieții de tei, cireș și paltin înregistrează cele mai mari valori medii ale hv, iar cei de frasin și jugastru cele mai mici. Situația puieților de jugastru este interesantă deoarece, deși înregistrează valori peste medie ale înălțimii totale, valorile hv sunt reduse, având și cel mai mic procent al hv din înălțimea totală (26%), în condițiile în care media acestui procent calculată pentru toți puieții este de 33% (tabelul 5.9).

Tabelul 5.8

**Indicatorii statistici ai înălţimii până la prima ramură verde în suprafeţele studiate**

|  | Medie (cm) | Abatere standard | Coef. de variaţie | Minim (cm) | Maxim (cm) | Volum probă |
|---|---|---|---|---|---|---|
| Suprafaţa nr. 1 | 23,64 | 11,19 | 47% | 3 | 68 | 823 |
| Suprafaţa nr. 2 | 20,25 | 9,51 | 47% | 5 | 55 | 555 |
| Suprafaţa nr. 3 | 25,06 | 12,54 | 50% | 5 | 77 | 526 |
| Suprafaţa nr. 4 | 21,93 | 13,87 | 63% | 2 | 75 | 697 |
| Suprafaţa nr. 5 | 21,49 | 10,38 | 48% | 5 | 65 | 505 |
| Suprafaţa nr. 6 | 33,89 | 18,99 | 56% | 5 | 115 | 734 |
| Suprafaţa nr. 7 | 29,68 | 15,34 | 52% | 3 | 96 | 926 |
| Suprafaţa nr. 8 | 34,60 | 9,99 | 39% | 7 | 75 | 762 |
| Suprafaţa nr. 9 | 21,99 | 10,44 | 47% | 9 | 61 | 791 |
| Suprafaţa nr. 10 | 30,36 | 15,85 | 52% | 3 | 95 | 887 |
| **Total** | 26,79 | 14,31 | 53% | 2 | 115 | 7206 |

Tabelul 5.9

**Indicatorii statistici ai înălţimii până la prima ramură verde a coroanei puieţilor pe specii**

|  | Ca | Ci | Fr | Ju | Pam | St | Te |
|---|---|---|---|---|---|---|---|
| Medie (cm) | 26,69 | 30,70 | 22,81 | 24,38 | 33,41 | 24,35 | 34,23 |
| Abatere standard | 14,16 | 17,11 | 12,63 | 11,72 | 19,00 | 12,20 | 16,53 |
| Coef. de variaţie | 53% | 56% | 55% | 48% | 57% | 50% | 48% |
| Valoare minimă | 3 | 8 | 3 | 5 | 5 | 3 | 5 |
| Valoare maximă | 115 | 95 | 90 | 76 | 80 | 90 | 100 |
| Volumul probei | 4075 | 122 | 608 | 424 | 58 | 1132 | 708 |
| **Date privitoare la procentul hv din înălţimea totală** (media acestui procent calculată pentru toţi puieţii - 33%) | | | | | | | |
| Procentul hv din h | 31% | 30% | 42% | 26% | 35% | 39% | 32% |
| Coef. de variaţie | 57% | 45% | 43% | 54% | 43% | 51% | 52% |

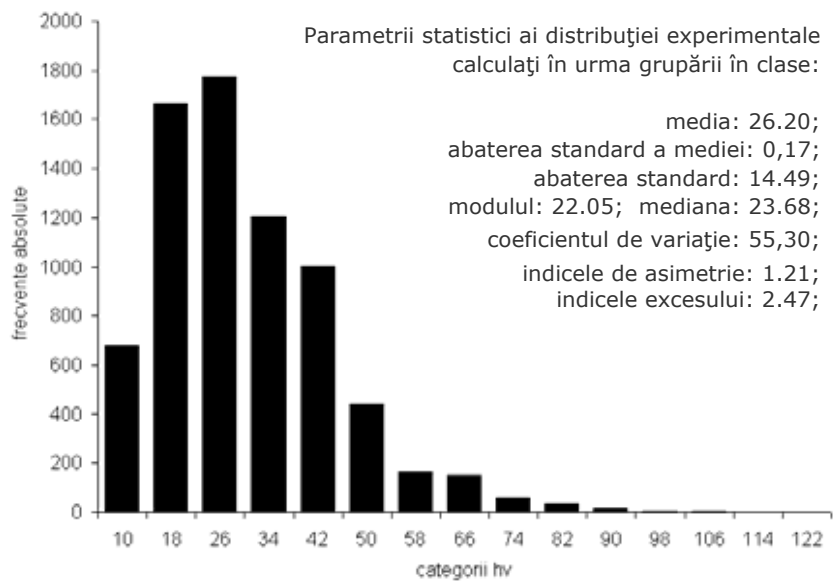

Figura 5.33 Histograma frecvențelor tuturor puieților pe categorii ale înălțimii până la prima ramură verde a coroanei

Analiza distribuției numărului de puieți pe categorii de înălțimi până la prima ramură verde a coroanei (fig. 5.34) arată că la nivelul celor zece suprafețe de probă se înregistrează o variabilitate scăzută a gamelor de distribuții întâlnite, acestea fiind similare cu distribuția generală a tuturor puieților.

Competiția determină în cazul hv o asimetrie puternică de stânga, aglomerarea în clasele mici ale hv (în condițiile în care înălțimea totală este mult mai uniform distribuită) reprezentând tendința puieților de a valorifica cât mai bine lumina la nivelul coroanei prin extinderea acesteia la maxim în plan vertical.

Parametrii statistici ai distribuției tuturor puieților pe categorii hv sunt prezentați în figura 5.33. Distribuția este caracterizată de o curbă unimodală leptokurtică (E=2,47), cu o puternică asimetrie pozitivă (A=1,21).

Modelarea distribuției experimentale s-a realizat cu distribuțiile teoretice folosite anterior: Beta, Gamma, Weibull. Parametrii funcțiilor și valorile necesare testului de semnificație a gradului de ajustare sunt prezentate în figura 5.35

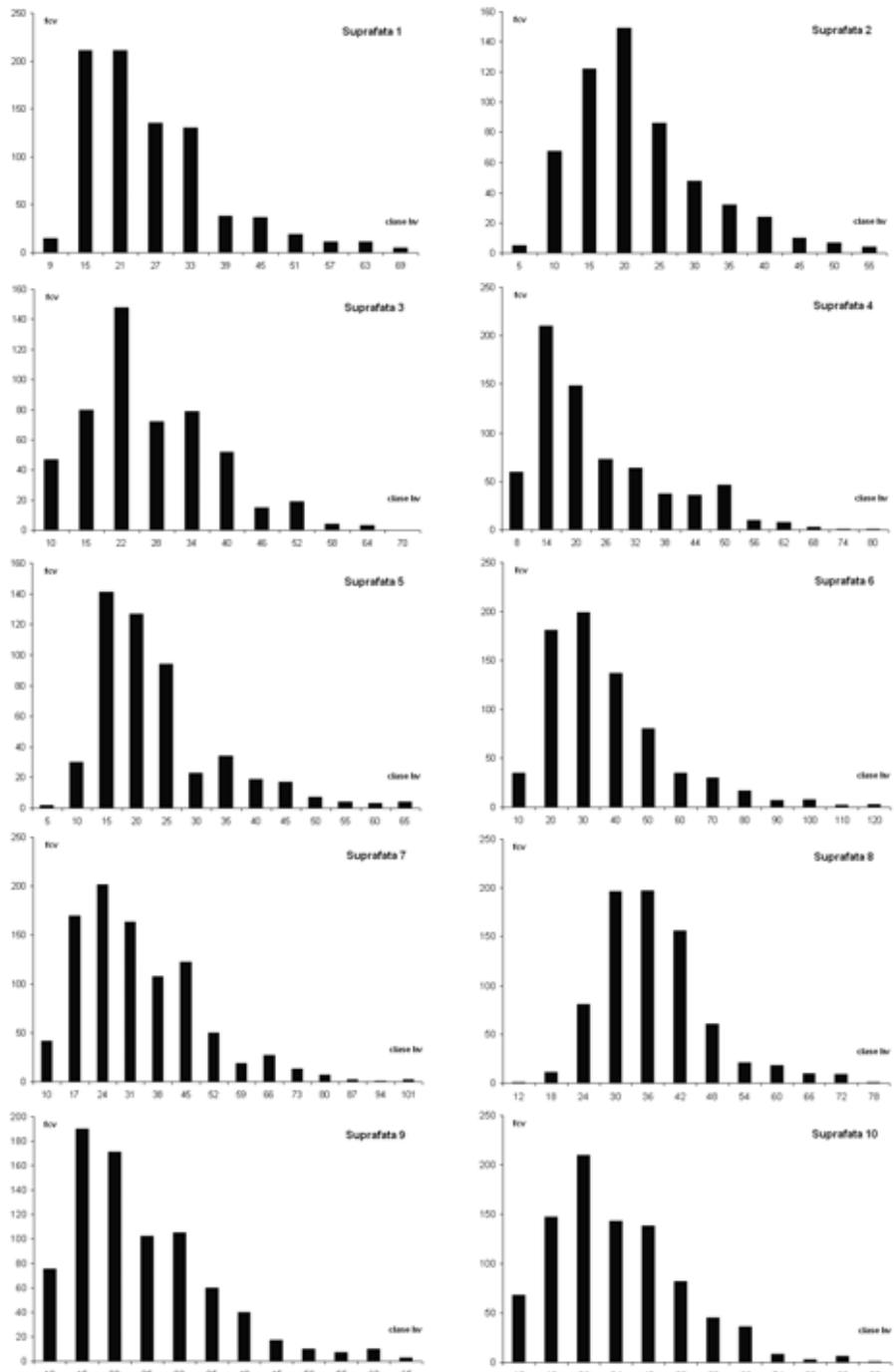

**Figura 5.34 Histograma frecvenţelor puieţilor pe categorii ale înălţimii până la prima ramură verde a coroanei în suprafeţele studiate**

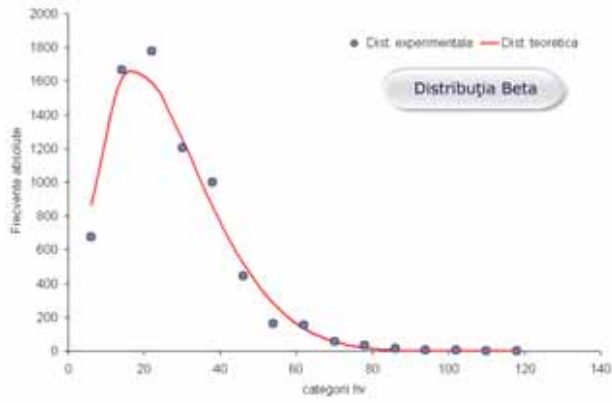

a
Parametrii distribuției teoretice Beta:
$\alpha$ = 2,03 $\gamma$ = 8,02 $k$= 0,7 x $10^{-7}$
$\chi^2_{exp}$ = 163,71
$\chi^2_{teor}$ = 16,81 q=0,01
**$\chi^2_{exp} > \chi^2_{teor}$**

$D_{exp}$ = 0,026
$D_{teor}$ = 0,249 $D_{exp} < D_{teor}$
q=0,05

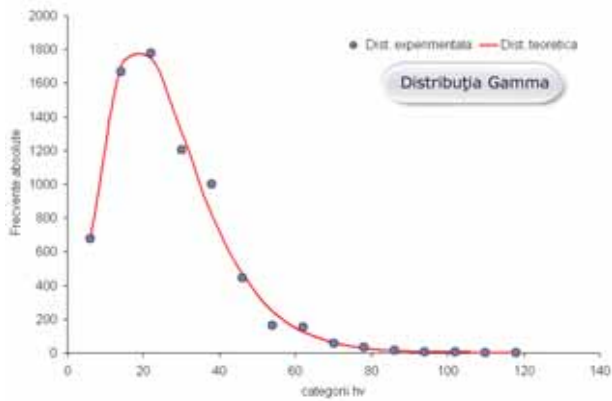

b
Parametrii distribuției teoretice Gamma:
$\theta$ = 26,20 $k$= 3,27
$\chi^2_{exp}$ = 84,27
$\chi^2_{teor}$ = 24,73 q=0,01
**$\chi^2_{exp} > \chi^2_{teor}$**

$D_{exp}$ = 0,015
$D_{teor}$ = 0,242 $D_{exp} < D_{teor}$
q=0,05

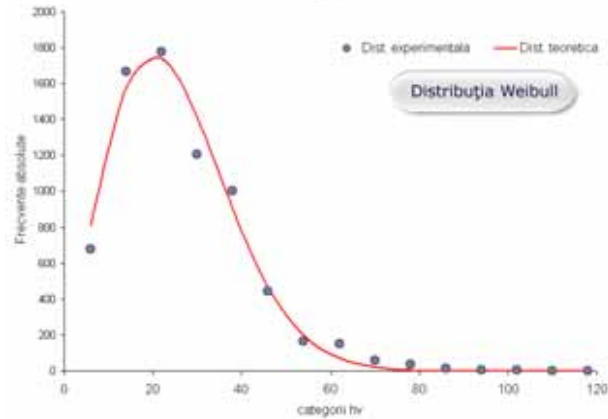

c
Parametrii distribuției teoretice Weibull:
$\alpha$ = 2,017 $\beta$ = 28,268 $\gamma$ =57544.5
$\chi^2_{exp}$ = 799,46
$\chi^2_{teor}$ = 21,67 q=0,01
**$\chi^2_{exp} > \chi^2_{teor}$**

$D_{exp}$ = 0,030
$D_{teor}$ = 0,258 $D_{exp} < D_{teor}$
q=0,05

**Figura 5.35 Ajustarea distribuției experimentale a numărului de puieți pe categorii ale înălțimii până la prima ramură verde**

Din analiza vizuală a variantelor prezentate în figura 5.35 reiese că distribuțiile teoretice utilizate reușesc să surprindă caracteristicile distribuției experimentale. Cu toate acestea nu s-a obținut din punct de vedere al asigurării statistice o bună ajustare, în toate cele trei cazuri având relația $\chi^2_{exp} > \chi^2_{teoretic}$ (pentru q=1%) ceea ce a condus la respingerea ipotezei nule. Pentru distribuția teoretică Gamma s-a obținut cea mai mică valoare a $\chi^2_{exp}$ (84,27), în vederea modelării matematice a repartizării numărului de puieți pe categorii ale hv funcția de densitate Gamma reprezentând cea mai bună opțiune disponibilă.

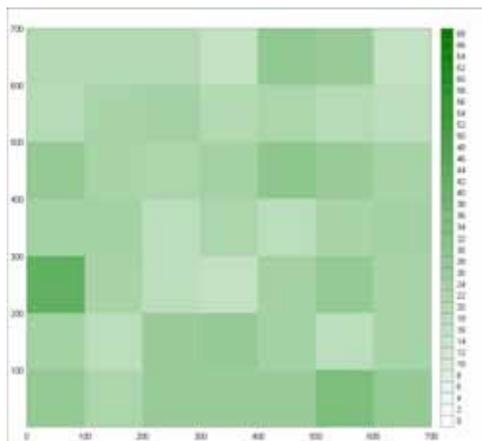

Suprafața 1

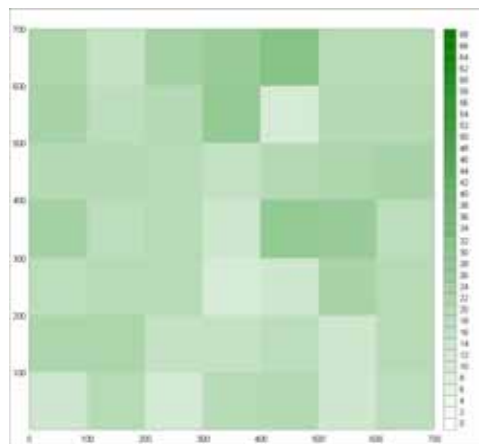

Suprafața 2

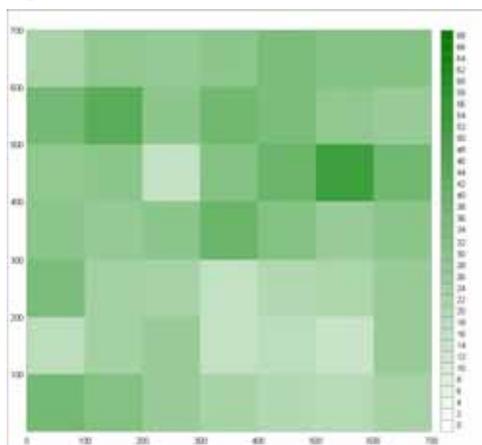

Suprafața 3

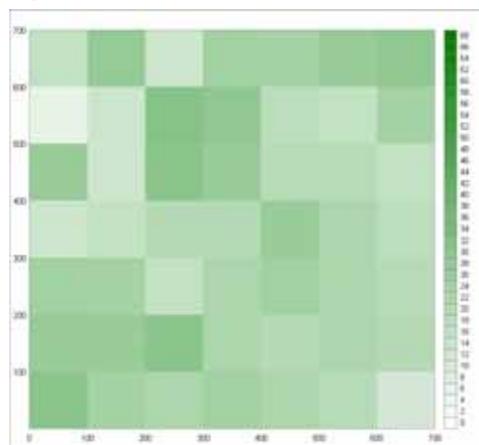

Suprafața 4

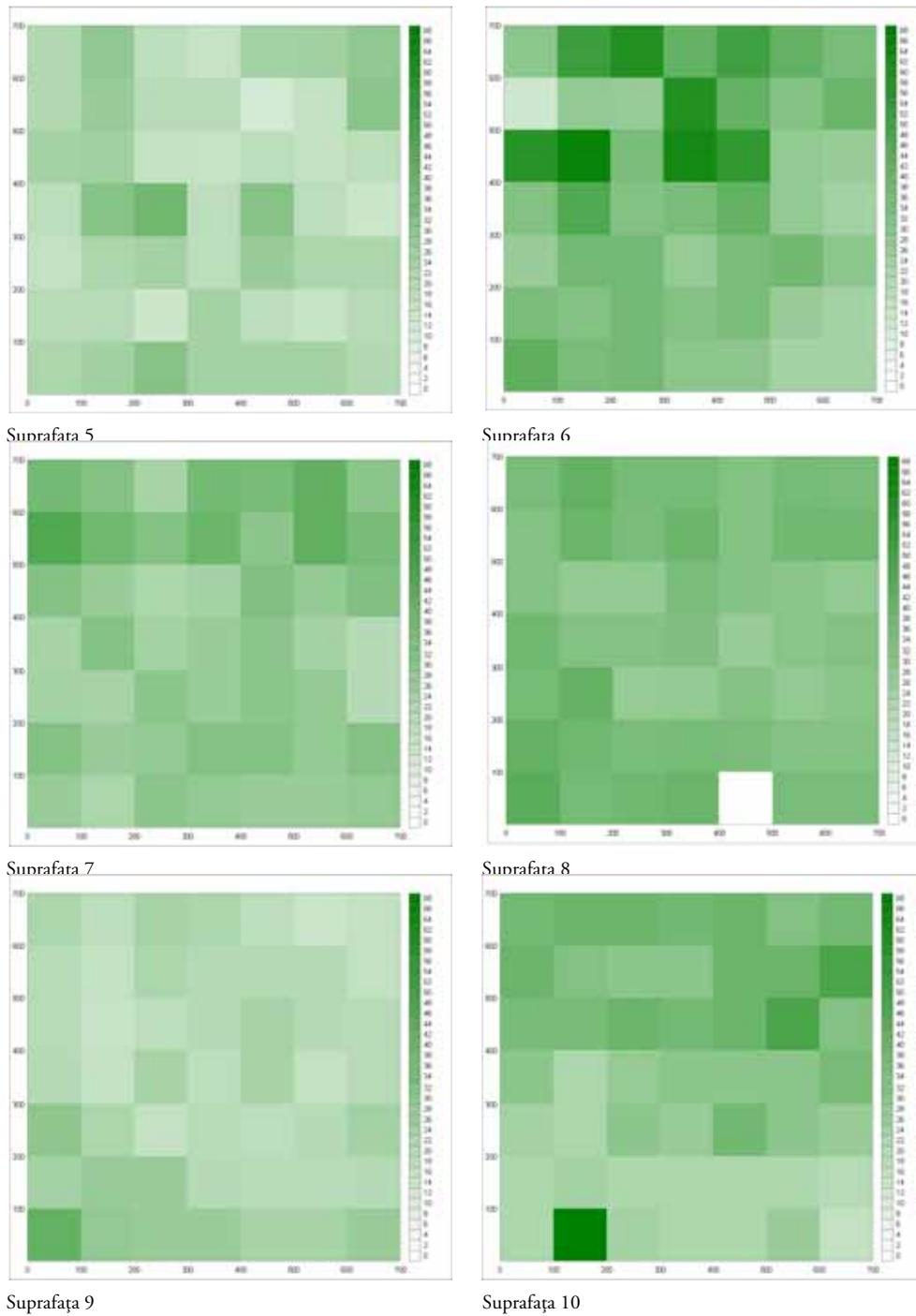

Suprafaţa 5    Suprafaţa 6

Suprafaţa 7    Suprafaţa 8

Suprafaţa 9    Suprafaţa 10

**Figura 5.36 Variabilitatea spaţială a înălţimii până la prima ramură verde**

În diagramele din figura 5.36, generate de aplicația software CARTOGRAMA, se prezintă variabilitatea spațială a înălțimii, prin reprezentarea grafică a mediei valorilor hv pe suprafețe elementare de 1x1m. În vederea comparării variabilității parametrului analizat atât în interiorul suprafețelor de 7x7 m cât și între aceste suprafețe s-a folosit aceeași scară de reprezentare (de la 0 la 68 cm) pentru toate cele zece piețe.

## 5.2.4. Structura în raport cu dimensiunile coroanei

Coroana unui arbore îndeplinește trei funcții fiziologice importante – asimilația, respirația și transpirația. Drept urmare, dimensiunea coroanei este strâns corelată cu creșterea arborilor, fiind un bun indicator al vigorii și al stării de sănătate. Dificultatea măsurătorilor biometrice directe ale coroanei arborilor a condus la proceduri de estimare ale dimensiunilor acesteia în baza unor caracteristici ale arborilor mai ușor măsurabile sau apreciabile: diametrul de bază, suprafața de bază, înălțimea sau înălțimea până la prima ramură verde a coroanei. Dimensiunile coroanei sunt variabile frecvent incluse în modelele de creștere, de apreciere a competiției sau de dinamică a arboretului. Parametrii ce caracterizează coroana au fost incluși în modele ale competiției (Schutz, 1989, citat de Ung et al., 1997; Biging, Dobbertin, 1995; Bachmann, 1996), modele de predicție a mortalității (Hasenauer, Monserud, 1996; Hasenauer et al., 2001) sau modele de creștere și dezvoltare (Monserud, Sterba, 1996).

Coroanele arborilor sunt caracterizate în general de un mare grad de variabilitate al formelor și dimensiunilor, lucru ce este evidențiat și în suprafețele studiate. Din analiza coeficienților de variație se remarcă un grad de neomogenitate ridicat în interiorul suprafețelor - caracterizat de o medie de 60% și o valoare maximă de 77%. Valoarea medie a diametrului mediu al coroanei se situează la valoarea de 24,90 cm pentru toate cele zece piețe de probă, cu o amplitudine descrisă de valorile extreme: 5 cm – 107,5 cm. În ceea ce privește situația pe specii, puieții de cireș, tei și stejar înregistrează cele mai mari valori medii ale diametrului coroanei (cireşul cu 30% peste media generală), iar cei de frasin cele mai mici (cu 34% sub media generală). Variabilitatea acestui parametru pe specii este similară, exceptând jugastrul (cu un coeficient de variaţie 67%).

<div align="right">Tabelul 5.10</div>

**Indicatorii statistici ai diametrului mediu al coroanei puieţilor în suprafeţele studiate**

|  | Medie (cm) | Abatere standard | Coef. de variaţie | Minim (cm) | Maxim (cm) | Volum probă |
|---|---|---|---|---|---|---|
| Suprafaţa nr. 1 | 27,37 | 12,18 | 45% | 5 | 70 | 830 |
| Suprafaţa nr. 2 | 26,33 | 17,87 | 68% | 5 | 97,5 | 567 |
| Suprafaţa nr. 3 | 23,38 | 15,87 | 68% | 5 | 90 | 518 |
| Suprafaţa nr. 4 | 22,60 | 12,89 | 57% | 5 | 75 | 688 |
| Suprafaţa nr. 5 | 29,39 | 18,48 | 63% | 5 | 107,5 | 504 |
| Suprafaţa nr. 6 | 20,47 | 15,80 | 77% | 5 | 87,5 | 734 |
| Suprafaţa nr. 7 | 24,92 | 11,94 | 48% | 5 | 75 | 927 |
| Suprafaţa nr. 8 | 26,86 | 11,81 | 44% | 5 | 70 | 767 |
| Suprafaţa nr. 9 | 27,08 | 19,05 | 70% | 5 | 92,5 | 792 |
| Suprafaţa nr. 10 | 21,78 | 12,52 | 57% | 5 | 75 | 888 |
| **Total** | 24,90 | 14,98 | 60% | 5 | 107,5 | 7215 |

<div align="right">Tabelul 5.11</div>

**Indicatorii statistici ai diametrului mediu al coroanei puieţilor pe specii**

|  | Ca | Ci | Fr | Ju | Pam | St | Te |
|---|---|---|---|---|---|---|---|
| Medie (cm) | 25,54 | 30,31 | 16,33 | 22,72 | 25,57 | 26,68 | 26,51 |
| Abatere standard | 15,41 | 16,20 | 8,52 | 15,17 | 12,22 | 15,23 | 13,63 |
| Coef. de variaţie | 60% | 53% | 52% | 67% | 48% | 57% | 51% |
| Valoare minimă | 5 | 5 | 5 | 5 | 7,5 | 5 | 5 |
| Valoare maximă | 105 | 75 | 60 | 75 | 70 | 107,5 | 85 |
| Volumul probei | 4068 | 123 | 609 | 424 | 61 | 1128 | 720 |

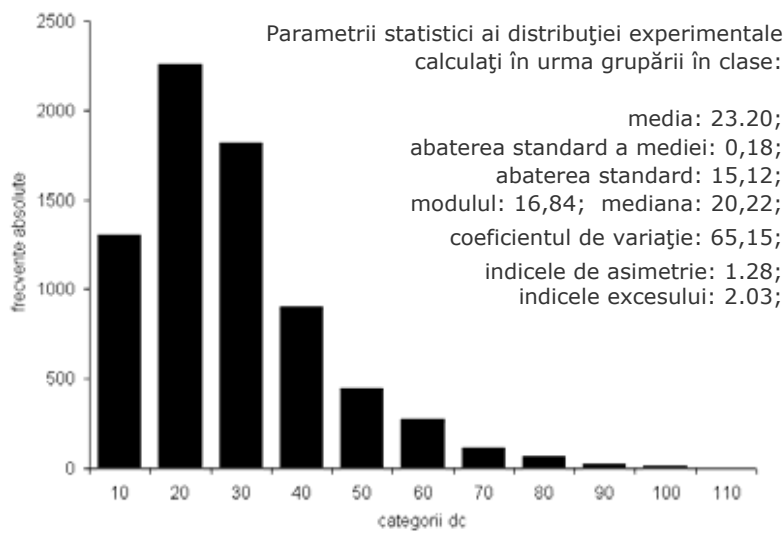

**Figura 5.37 Histograma frecvenţelor tuturor puieţilor pe categorii ale diametrului mediu al coroanei**

Analiza distribuţiei numărului de puieţi pe categorii ale diametrului mediu al coroanei (fig. 5.38) arată că la nivelul celor zece suprafeţe de probă se întâlnesc distribuţii unimodale, cu asimetrie pozitivă (de stânga) şi o variabilitate relativ redusă a formei.

Distribuţia tuturor puieţilor pe categorii ale diametrului mediu al coroanei devine mai stabilă, după cum se observă în figura 5.37. Aceasta prezintă un modul situat la valoarea 16,84 cm, o puternică asimetrie de stânga (A=1,28) şi un exces pozitiv (E=2,08) caracteristic unei distribuţii leptokurtice.

Modelarea distribuţiei experimentale s-a realizat şi în această situaţie cu ajutorul distribuţiilor teoretice folosite anterior: Beta, Gamma şi Weibull. Parametrii funcţiilor şi valorile necesare efectuării testului de semnificaţie a gradului de ajustare sunt prezentate în figura 5.39.

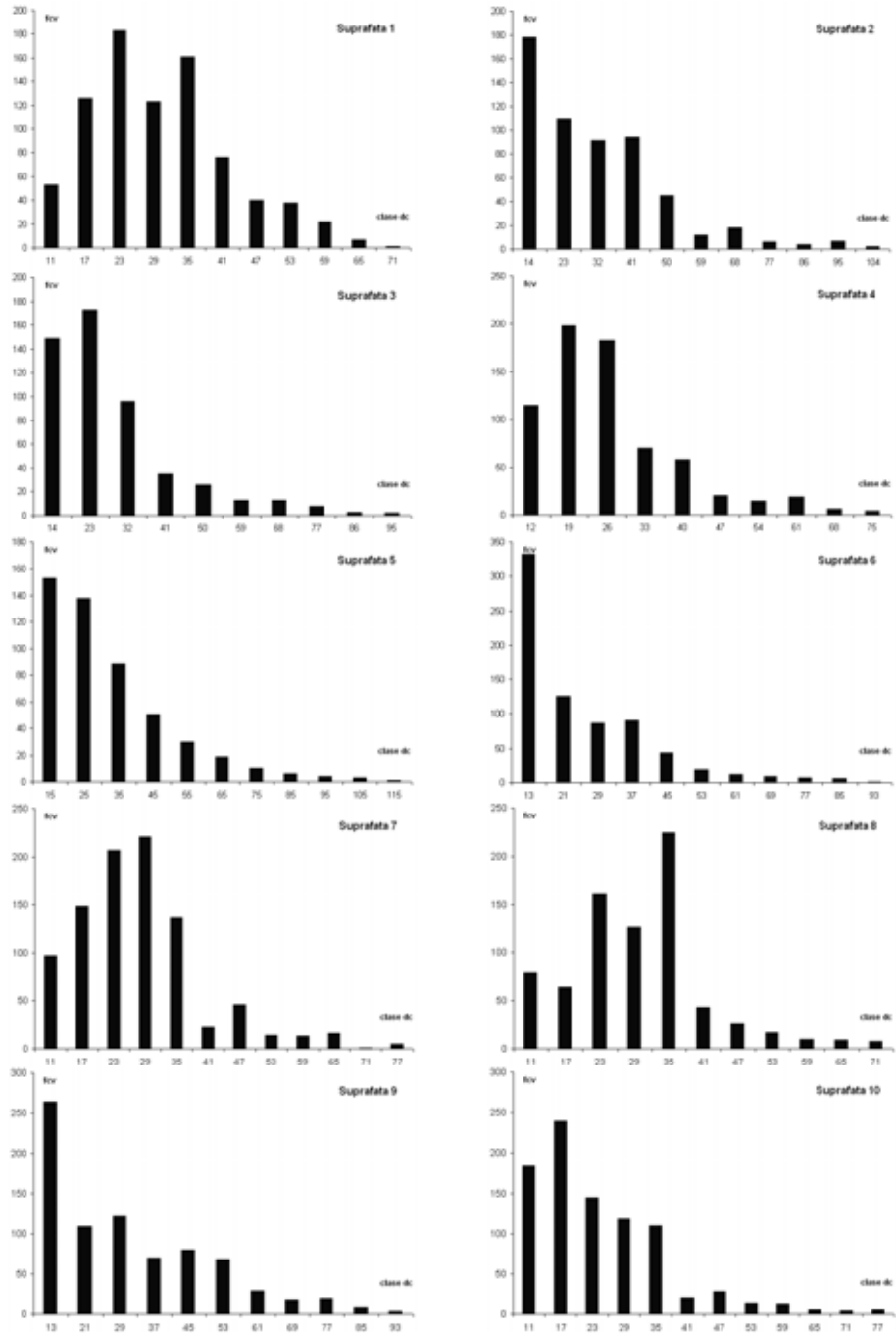

**Figura 5.38 Histograma frecvențelor puieților pe categorii ale diametrului mediu al coroanei în suprafețele studiate**

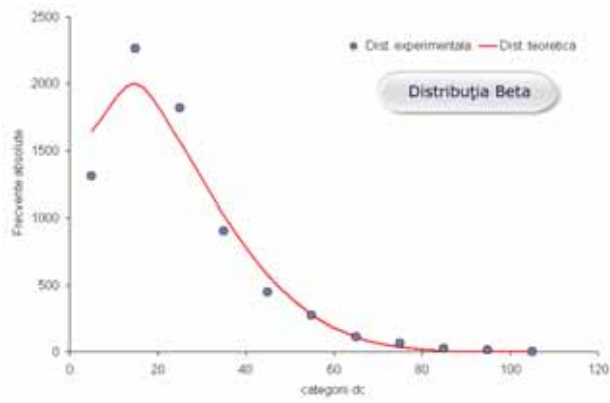

a

Parametrii distribuției teoretice Beta:

$\alpha = 1,65$  $\gamma = 6,16$  $k = 0,21 \times 10^{-7}$

$\chi^2_{exp} = 352,22$

$\chi^2_{teor} = 13,28$  q=0,01

$\chi^2_{exp} > \chi^2_{teor}$

$D_{exp} = 0,046$

$D_{teor} = 0,271$

$D D_{exp} < D_{teor}$  q=0,05

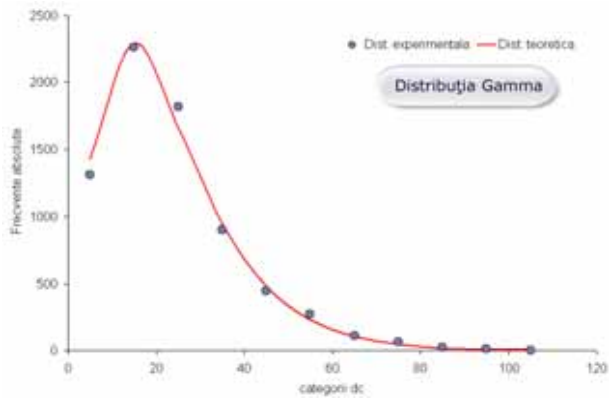

b

Parametrii distribuției teoretice Gamma:

$\theta = 23,20$  $k = 2,36$

$\chi^2_{exp} = 45,12$

$\chi^2_{teor} = 21,67$  q=0,01

$\chi^2_{exp} > \chi^2_{teor}$

$D_{exp} = 0,020$

$D_{teor} = 0,258$

$D_{exp} < D_{teor}$  q=0,05

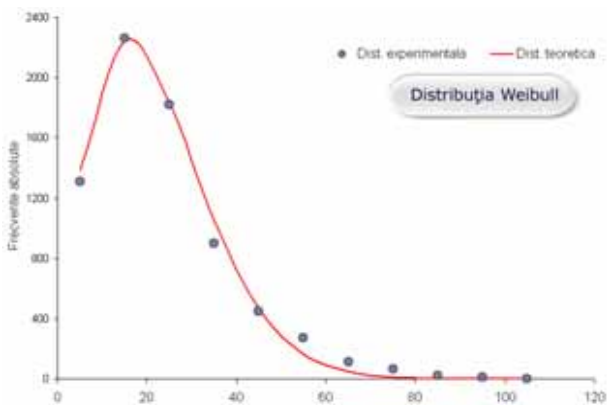

c

Parametrii distribuției teoretice Weibull:

$\alpha = 1,759$  $\beta = 25,017$

$\gamma = 70780$

$\chi^2_{exp} = 648,16$

$\chi^2_{teor} = 18,46$  q=0,01

$\chi^2_{exp} > \chi^2_{teor}$

$D_{exp} = 0,035$

$D_{teor} = 0,285$

$D_{exp} < D_{teor}$  q=0,05

**Figura 5.39 Ajustarea distribuției experimentale a numărului de puieți pe categorii ale diametrului mediu al coroanei**

Analiza vizuală a variantelor prezentate în figura 5.39 arată că distribuțiile teoretice surprind caracteristicile distribuției experimentale. Se întâlnește aceeași situație remarcată anterior cu privire la testarea semnificației cu ajutorul testului $\chi^2$ - analiza statistică nu confirmă analiza vizuală, numărul mare de observații alterând rezultatele testului. În toate cele trei cazuri s-a obținut relația $\chi^2_{exp} > \chi^2_{teoretic}$ (pentru q=1%) ceea ce a condus la respingerea ipotezei nule. Pentru distribuția teoretică Gamma s-a obținut cea mai mică valoare a $\chi^2_{exp}$ (45,12), în vederea modelării matematice a repartizării numărului de puieți pe categorii ale diametrului mediu al coroanei funcția de densitate Gamma reprezentând cea mai bună opțiune disponibilă.

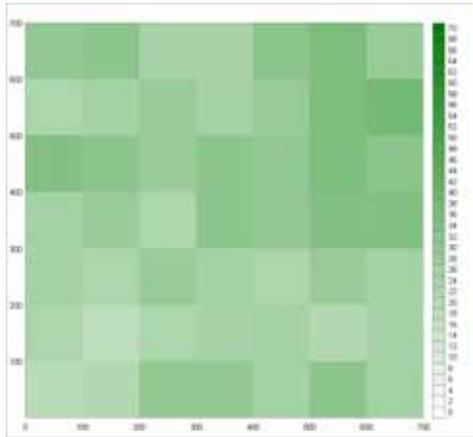

Suprafața 1

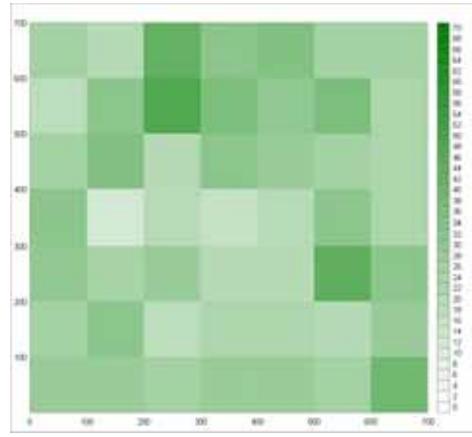

Suprafața 2

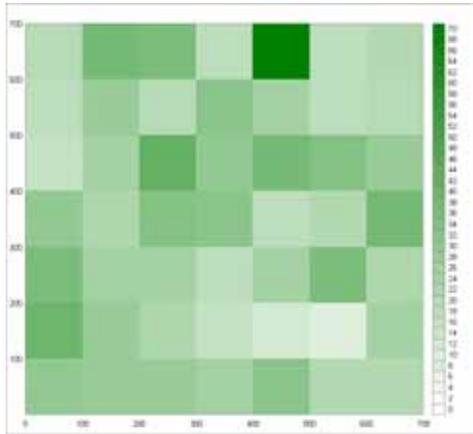

Suprafața 3

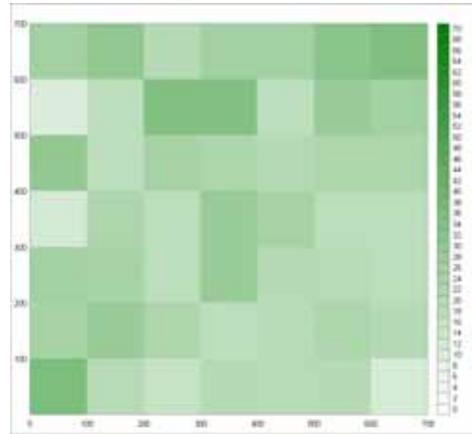

Suprafața 4

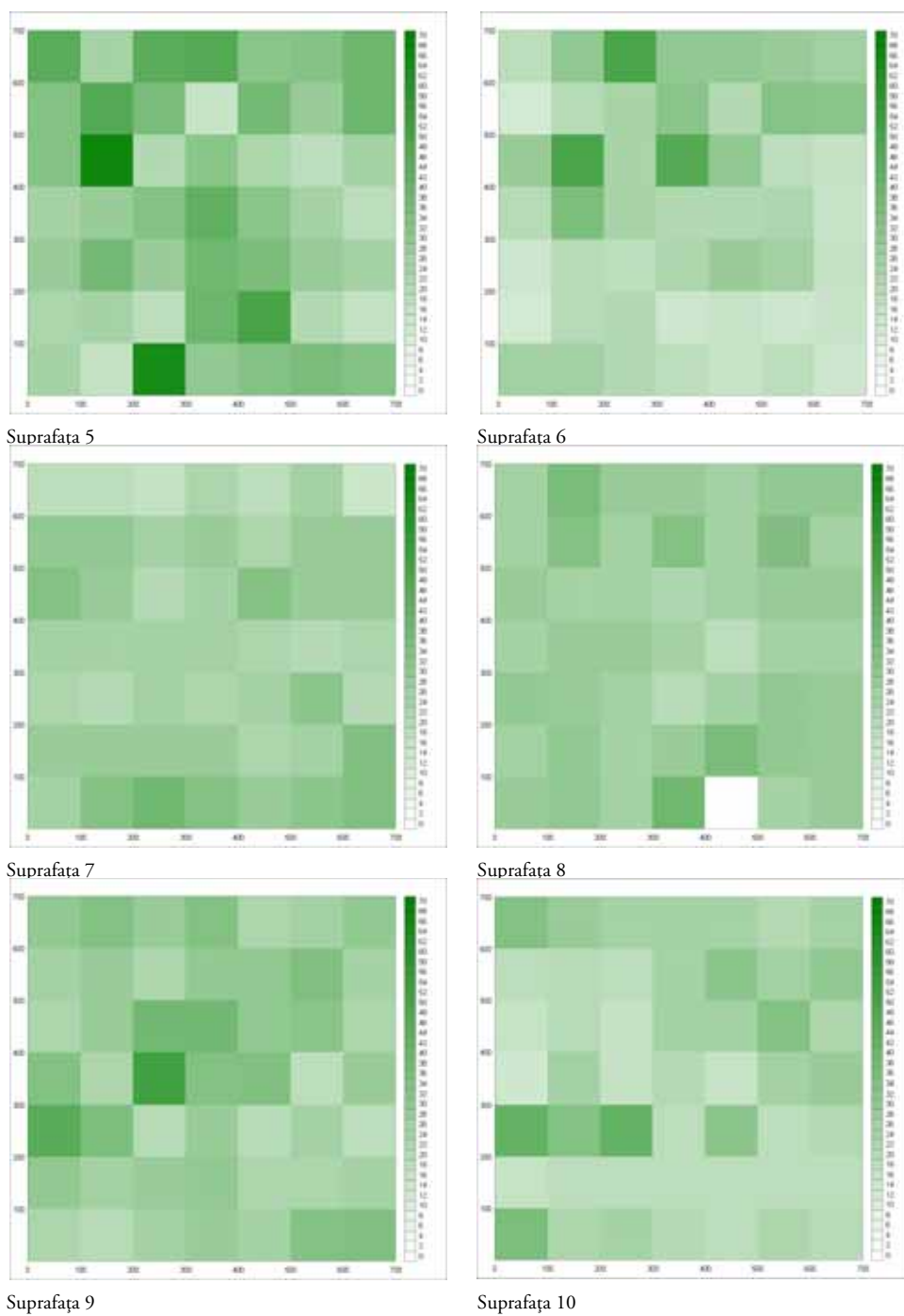

Suprafața 5      Suprafața 6

Suprafața 7      Suprafața 8

Suprafața 9      Suprafața 10

**Figura 5.40 Variabilitatea spațială a diametrului mediu al coroanei puieților**

În diagramele din figura 5.40, generate de aplicația CARTOGRAMA, se prezintă variabilitatea spațială a diametrului mediu al coroanei, pe suprafețe elementare de 1x1m. În vederea comparării variabilității parametrului analizat atât în interiorul suprafețelor de 7x7 m cât și între aceste suprafețe s-a folosit aceeași scară de reprezentare (de la 0 la 70 cm) pentru toate cele zece piețe.

## 5.2.5. Profilul tridimensional al suprafețelor studiate

Structura arboretelor este descrisă frecvent prin intermediul unor variabile care nu înglobează direct caracterul orizontal și vertical al dispunerii arborilor în spațiu. Desimea, suprafața de bază sau distribuția indivizilor pe categorii de diametre reprezintă parametri care ignoră caracterul spațial al structurii ecosistemelor forestiere. Pentru a completa lipsa informațiilor spațiale este recomandată folosirea unor metode de analiză moderne, adecvate cercetărilor forestiere, care să utilizeze parametri relevanți ai dispunerii în spațiu (Zenner, Hibbs, 2000).

Determinarea coordonatelor fiecărui puiet, a dimensiunilor coroanei, a înălțimii totale precum și a înălțimii de inserție a coroanei au permis reconstituirea modelului tridimensional al suprafețelor studiate. Modelarea geometrică tridimensională reprezintă o formă modernă și intuitivă de apreciere unitară a complexității structurale a ecosistemelor forestiere. Analiza modelului se poate efectua chiar și fără ajutorul unui calculator, prin schițe explicite - metoda profilelor spațiale (Cenușă, 1992), dar în prezent este recomandată utilizarea reprezentărilor computerizate. Calculatorul a devenit o unealtă esențială în cercetarea inginerească, iar tehnicile moderne de prezentare vizuală computerizată a datelor, cum ar fi, de exemplu, modelarea 3D, facilitează perceperea și interpretarea corectă a informațiilor prezentate.

În cercetarea silvică sunt folosite numeroase aplicații informatice destinate modelării tridimensionale a structurii ecosistemelor forestiere (prezentate în subcapitolul 3.4.3), chiar și la noi în țară existând soluții software specializate – programul PROARB (Popa, 1999).

În cazul de față s-a preferat modelarea tridimensională cu ajutorul aplicației SVS *Stand Visualization System* (McGaughey, 1997). Alegerea a fost determinată de faptul că acest program este cel mai frecvent utilizat la nivel internațional, fiind foarte complex și versatil.

Modulul de design al arborilor a permis definirea formei de reprezentare a puieților fiecărei specii, în figura 5.41 fiind prezentate simbolurile grafice utilizate în profilurile tridimensionale și orizontale. Deși aplicația permite și vizualizarea fotorealistă a suprafețelor s-a optat pentru reprezentarea arborilor prin modele solide considerându-se a fi mai explicită în cazul unei desimi mari a puieților.

Datorită prelucrării unui volum mare de date a fost conceput un program propriu - SVS Export, de transfer automat a datelor din foile de calcul de tip Microsoft Excel în fișierele de date specifice aplicației SVS. Formatul relativ criptic al fișierelor datelor de intrare, precum și numărul mare de variabile care trebuie să respecte un tipar specific reprezintă un impediment în folosirea SVS, dar care poate fi soluționat prin utilizarea programului SVS Export.

În figurile 5.42 – 5.51 sunt prezentate structurile caracteristice ale celor zece suprafețe studiate. Pentru fiecare suprafață de probă s-a determinat atât profilul tridimensional cât și cel orizontal. În acest fel informațiile se completează reciproc, pentru ca în final structura să fie percepută și apreciată corect.

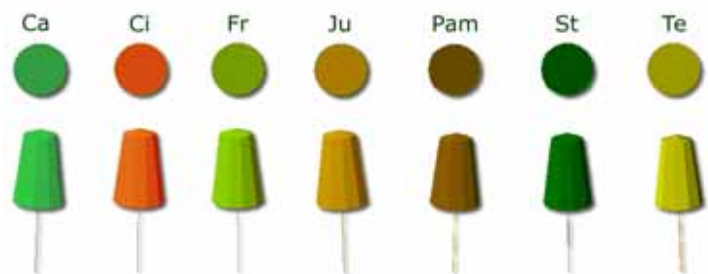

**Figura 5.41 Simbolurile grafice folosite în reprezentarea profilelor spațiale**

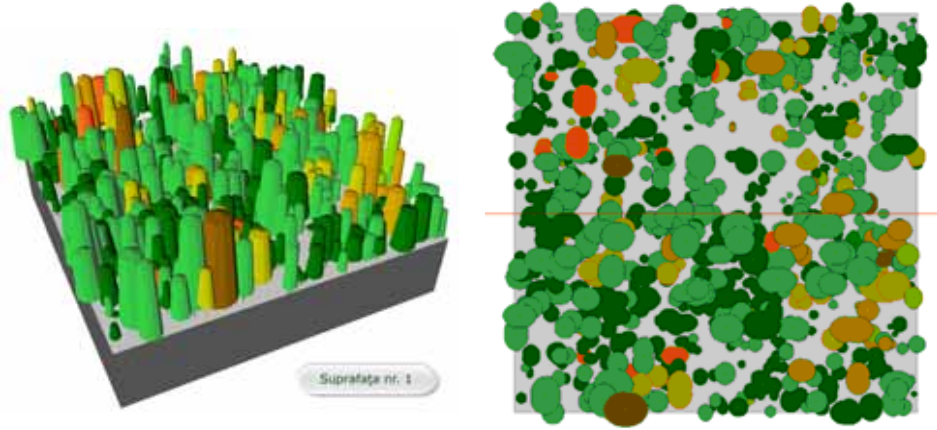

**Figura 5.42 Profilul tridimensional şi orizontal al suprafeţei nr. 1**

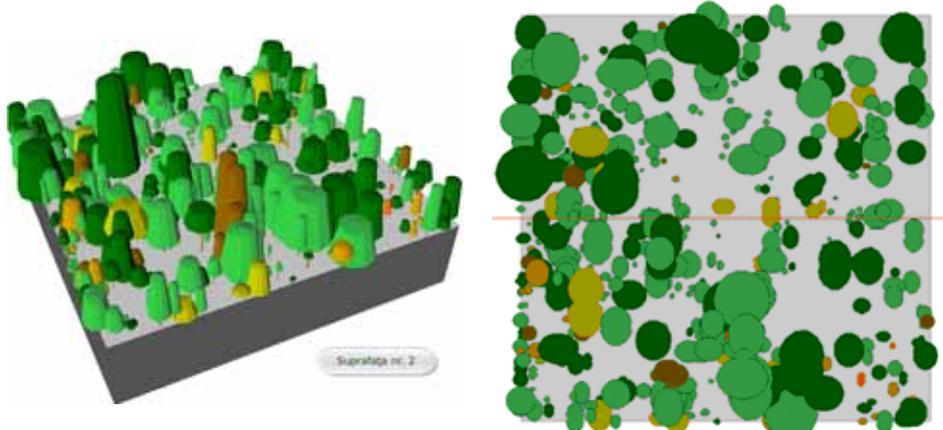

**Figura 5.43 Profilul tridimensional şi orizontal al suprafeţei nr. 2**

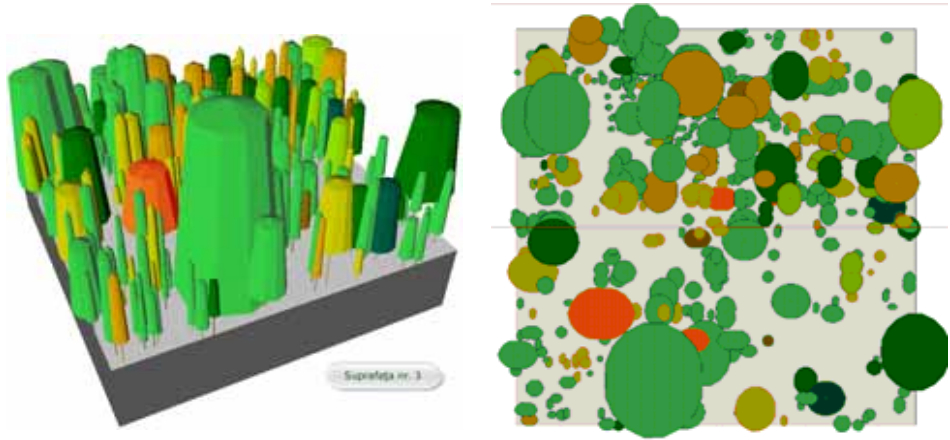

**Figura 5.44 Profilul tridimensional şi orizontal al suprafeţei nr. 3**

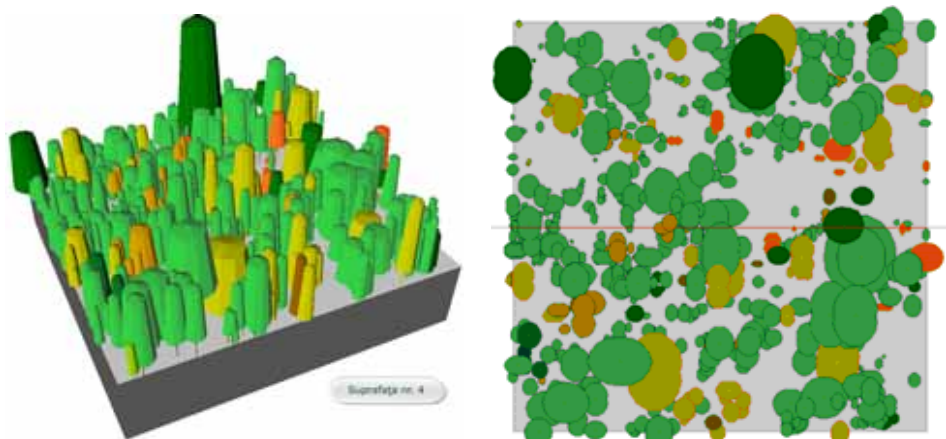

**Figura 5.45 Profilul tridimensional şi orizontal al suprafeţei nr. 4**

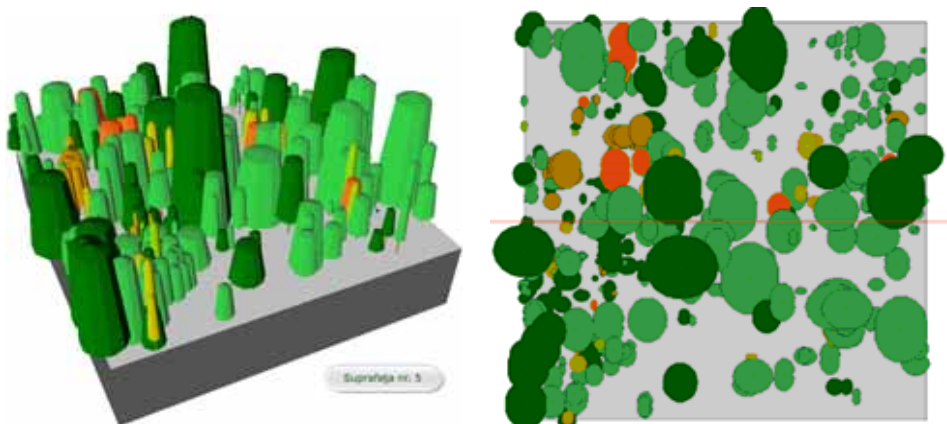

**Figura 5.46 Profilul tridimensional şi orizontal al suprafeţei nr. 5**

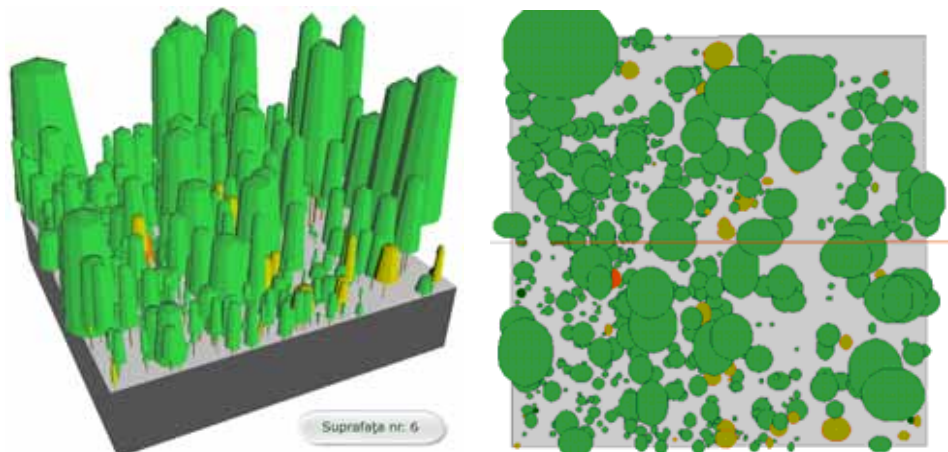

**Figura 5.47 Profilul tridimensional şi orizontal al suprafeţei nr. 6**

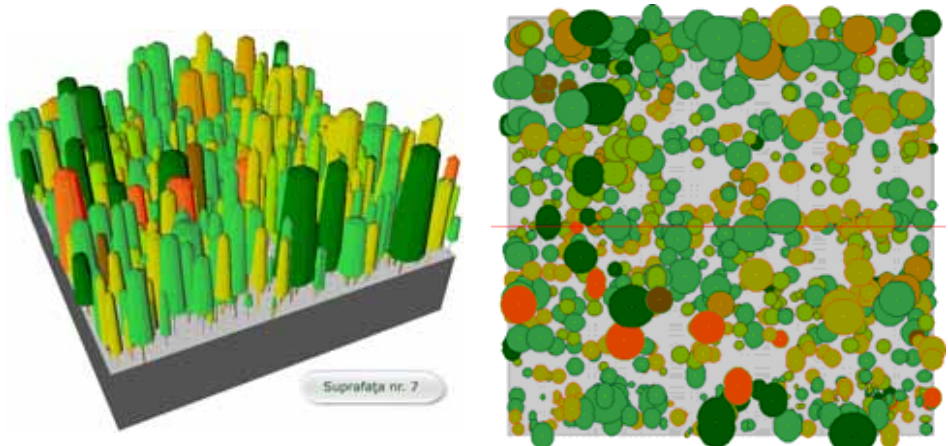

**Figura 5.48 Profilul tridimensional şi orizontal al suprafeţei nr. 7**

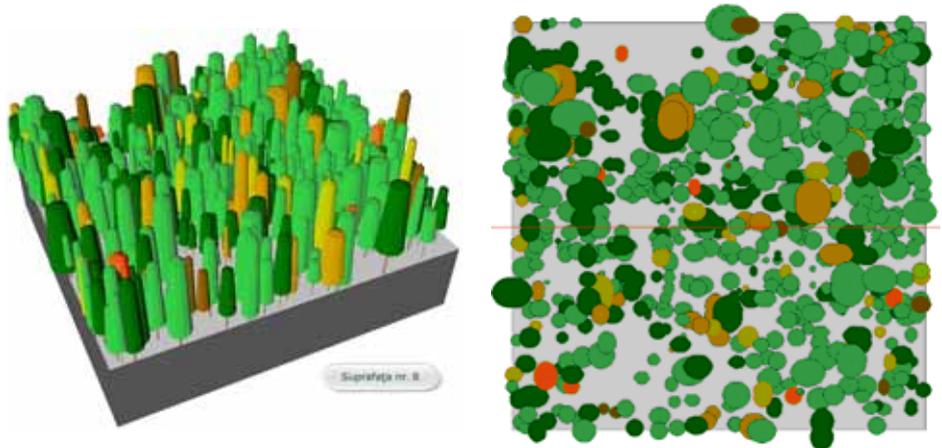

**Figura 5.49 Profilul tridimensional şi orizontal al suprafeţei nr. 8**

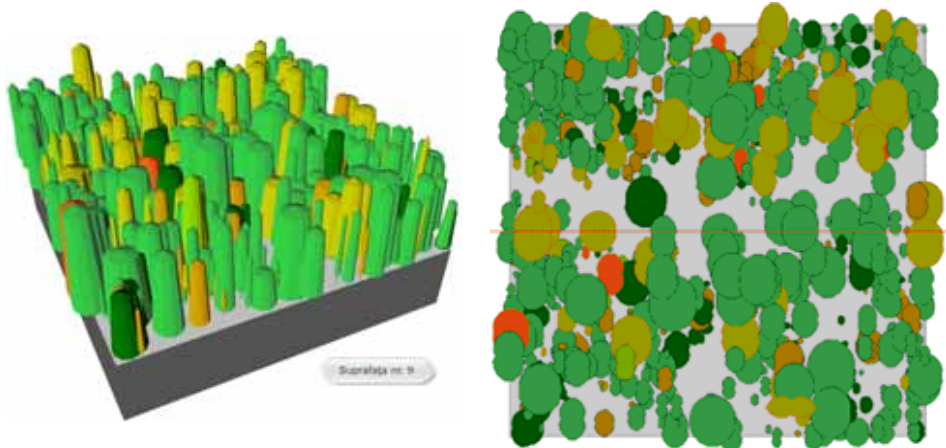

**Figura 5.50 Profilul tridimensional şi orizontal al suprafeţei nr. 9**

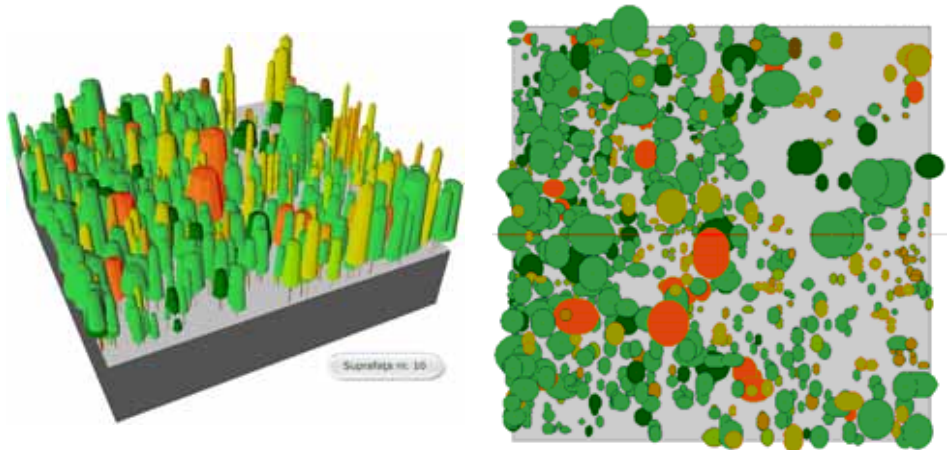

**Figura 5.51 Profilul tridimensional și orizontal al suprafeței nr. 10**

Analiza informațiilor vizuale ale structurii tridimensionale și orizontale confirmă și completează informațiile descriptive din subcapitolele anterioare privitoare la numărul de arbori pe specii, distribuțiile caracteristicilor biometrice și gradul de neomogenitate al parametrilor studiați. Datele profilelor spațiale permit de asemenea interpretarea unor aspecte ce vor fi prezentate ulterior, referitoare la modul de organizare spațială a puieților, spațiul potențial de dezvoltare și relațiile de competiție ce se stabilesc între indivizi.

## 5.3. Modelarea relațiilor dintre caracteristicile structurale ale semințișului

Parametrii structurali ai unui arboret reprezintă caracteristici importante care pot fi utilizate în modele de creștere și dezvoltare. Analiza intensității și tipului relațiilor stabilite între diversele variabile poate să conducă la identificarea atributelor cheie, determinante în fundamentarea matematică a modelului propus.

În literatura de specialitate sunt menționate relațiile stabilite între caracteristicile biometrice ale arborilor – diametru, înălțime, înălțimea până la punctul de inserție al coroanei, dimensiunile coroanei. În cazul regenerării arboretelor aceste relații prezintă anumite particularități care vor fi evidențiate în subcapitolele următoare.

### 5.3.1. Relațiile dintre parametrii biometrici ai puieților

În cazul arborilor maturi diametrul de bază este considerat parametrul central al structurii, fiind în același timp caracteristica cea mai ușor măsurabilă. În cazul semințișurilor de cele multe ori nu se poate preleva diametrul de bază datorită înălțimii reduse a puieților, astfel că se utilizează în caracterizarea structurii diametrul la colet. Acest parametru este mult mai puțin stabil decât diametrul de bază, fiind dificil de măsurat și oferind o precizie adesea insuficientă. În cadrul acestui studiu diametrul la colet a fost prelevat cu ajutorul șublerului, utilizând o precizie de 1 mm. Cu toate că această precizie este aparent satisfăcătoare, o analiză mai atentă dezvăluie că procentual marja de eroare a măsurătorilor efectuate la cei 7253 de puieți este cuprinsă între 0,95% și 50%. Având în vedere că o creștere a preciziei de măsurare la valori mai mici de 1 mm nu este practic posibilă datorită dificultăților de măsurare și erorilor de poziționare a șublerului la nivelul coletului, reprezentativitatea acestui parametru scade în special la valori reduse ale sale.

Înălțimea reprezintă parametrul cel mai ușor măsurabil în cazul semințișului. Precizia determinărilor este de asemenea procentual superioară evaluării oricărui alt parametru biometric, marja de eroare datorată rotunjirii la precizia de 1 cm fiind în cazul acestui studiu cuprinsă între 0,12% și 8,33%. Înălțimea poate fi considerată un parametru mult mai stabil statistic și cu o sensibilitate de înregistrare a relațiilor dintre puieți deosebită datorită competiției intense ce caracterizează această faza de dezvoltare. Unele cercetări efectuate în semințișuri naturale (Siipilehto, 2006) confirmă importanța înălțimii în caracterizarea regenerării arboretelor, recomandând folosirea acesteia drept parametru central de evaluare structurală.

În continuare se va face o analiză a relațiilor dintre caracteristicile biometrice prelevate. Scopul analizei îl reprezintă conceperea unor modele matematice simple, bazate pe analiza regresiei, care în funcție de un număr minim de variabile să conducă la determinarea suficient de precisă a unor parametri biometrici dificil de estimat, în vederea simplificării metodologiei de prelevare a datelor în semințișuri.

Prima etapă în vederea selectării variabilelor modelului este reprezentată de analiza corelațiilor dintre variabilele candidate. Pentru a simplifica prezentarea datelor, se va folosi următoarea convenție de notare a variabilelor: d - diametrul la colet, h – înălțimea totală, hv – înălțimea punctului de inserție a coroanei, dc – diametrul mediu al coroanei.

Valorile matricei corelațiilor (tabelul 5.12) arată legături foarte semnificative între toți parametrii studiați. Privitor la intensitatea acestora, se remarcă legătura puternică dintre *d* și *h* și de intensitate medie dintre *h* și *hv*, *dc* și *d*, respectiv *dc* și *h*. Înălțimea până la punctul de inserție a coroanei *(hv)* stabilește legăturile de intensitatea cea mai scăzută. În figurile 5.52-5.57 sunt prezentate grafic câmpurile de corelație a elementelor, din analiza acestora observându-se natura și forma legăturii.

Pentru aprecierea relațiilor dintre parametri, precum și pentru a stabili importanța acestora în explicarea variabilității datelor s-a folosit tehnica analizei în componente principale, pentru prelucrarea datelor fiind folosită aplicația *StatSoft STATISTICA*. În figura 5.58 sunt prezentate două alternative de proiecție – în planul dintre primul și al doilea, respectiv al doilea și al treilea factor, ordinea factorilor fiind dată de tabelul valorilor proprii (*eigenvalues*), primii trei factori explicând variația datelor în proporție de 94,4%. Din cele două proiecții se identifică legătura puternică dintre *h* și *d*, precum și cea de intensitate medie dintre primele două amintite și *dc*. Și în acest caz se observă legătura slabă a *hv* cu celelalte elemente.

Relația dintre diametru și înălțime reprezintă unul dintre elementele definitorii ale structurii unui arboret, fiind esențială în modelarea creșterii și dezvoltării arborilor (Botkin et al., 1972). De la funcția exponențială prezentată de Meyer (1940)

$$h = 1.3 + b_1 \cdot (1 - e^{b_2 d}) \qquad (5.1)$$

unde *h*-înălțimea, *d*-diametrul de bază iar $b_1$, $b_2$–coeficienți, au fost concepute numeroase modele care descriu această relație (Prodan, 1965; Curtis, 1967; Wykoff et al., 1982; Larsen, Hann, 1987; Huang et al., 1992).

Tabelul 5.12

Matricea corelațiilor dintre parametrii biometrici analizați

|  | d | h | hv | dc |
|---|---|---|---|---|
| **d** | 1 | | | |
| **h** | 0,761 *** | 1 | | |
| **hv** | 0,388 *** | 0,512 *** | 1 | |
| **dc** | 0,654 *** | 0,644 *** | 0,303 *** | 1 |

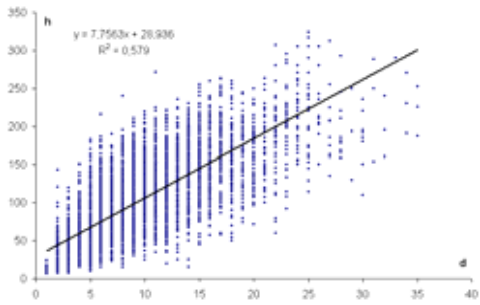

Figura 5.52 Legătura corelativă d-h

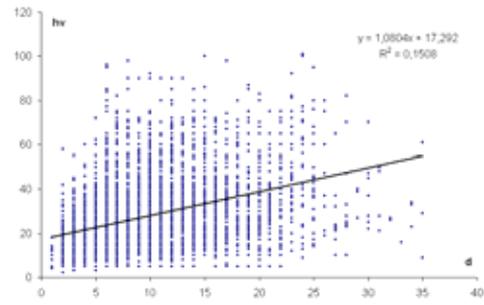

Figura 5.53 Legătura corelativă d-hv

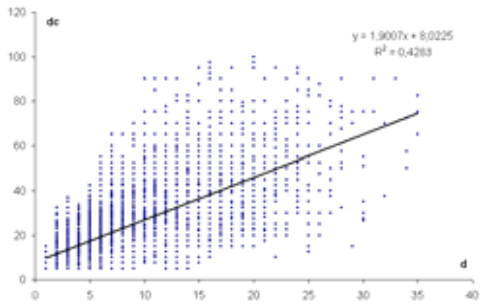

Figura 5.54 Legătura corelativă d-dc

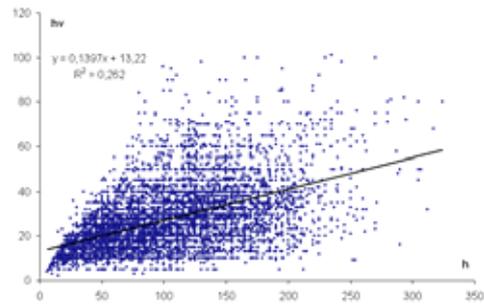

Figura 5.55 Legătura corelativă h-hv

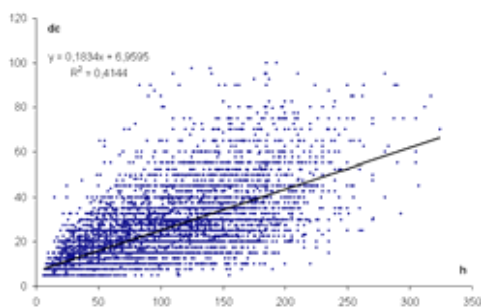 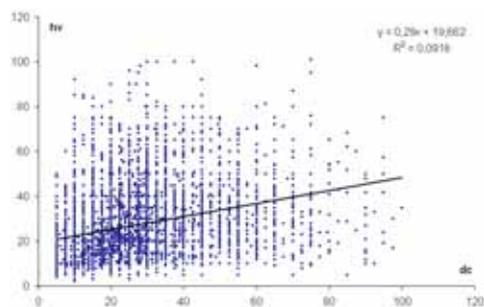

**Figura 5.56 Legătura corelativă h-dc**   **Figura 5.57 Legătura corelativă dc-hv**

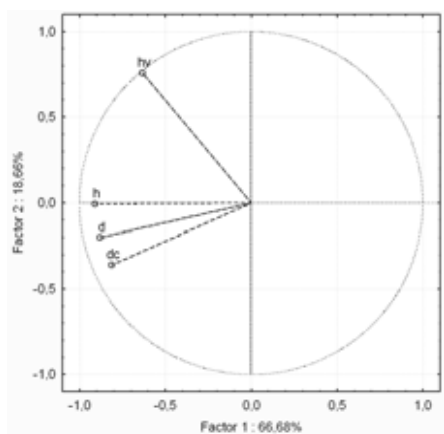 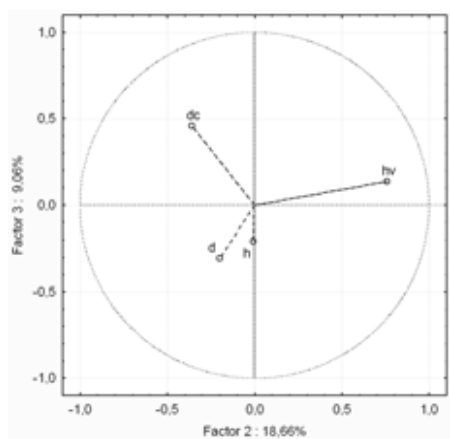

**Figura 5.58 Analiza în componente principale a legăturii dintre parametri biometrici**

Relaţia diametru-înălţime prezintă anumite particularităţi în cazul seminţişurilor naturale, comparativ cu situaţia din arboretele mature. Din analiza graficului ce descrie legătura corelativă dintre cele două elemente (fig. 5.52) se observă că tendinţa asimptotică specifică arboretelor mature, ce constrânge înălţimea, limitându-i valorile spre dimensiunile maxime ale diametrului este mult mai puţin evidentă. În cazul puieţilor nu apare încă fenomenul de limitare a valorilor maxime ale înălţimii. O altă particularitate se referă la raportul dintre cele două elemente – dacă în cazul arborilor maturi se preferă măsurarea diametrului de bază şi determinarea în funcţie de acesta a înălţimii (dificil şi costisitor de apreciat în cazul unui număr mare de indivizi), în cazul seminţişurilor se inversează procedura – este de preferat măsurarea înălţimii şi determinarea diametrului la colet în funcţie de aceasta. Recent, această modalitate de determinare a diametrului în funcție de celelalte caracteristici biometrice (e.g. înălțime și diametrul mediu al coroanei) a devenit folosită și în cazul arboretelor mature datorită determinării ușoare a parametrilor necesari folosind imagini satelitare de înaltă rezoluție (Kalliovirta, Tokola, 2005).

Dimensiunile coroanei sunt de asemenea importante în modelarea creșterii arborilor, numeroase cercetări stabilind funcții de determinare a acestora în raport de diametrul de bază sau de înălțime. În general componenta orizontală a coroanei – diametrul mediu al coroanei (*dc*) se determină în funcție de diametrul de bază (Krajicek et al., 1961; Arney, 1973; Paine, Hann, 1982; Meng et al., 2007), iar componenta verticală – înălțimea coroanei, se determină prin intermediul înălțimii punctului de inserție a coroanei (*hv*), în funcție de înălțimea totală (Ritchie, Hann, 1987; Zumrawi, Hann, 1989).

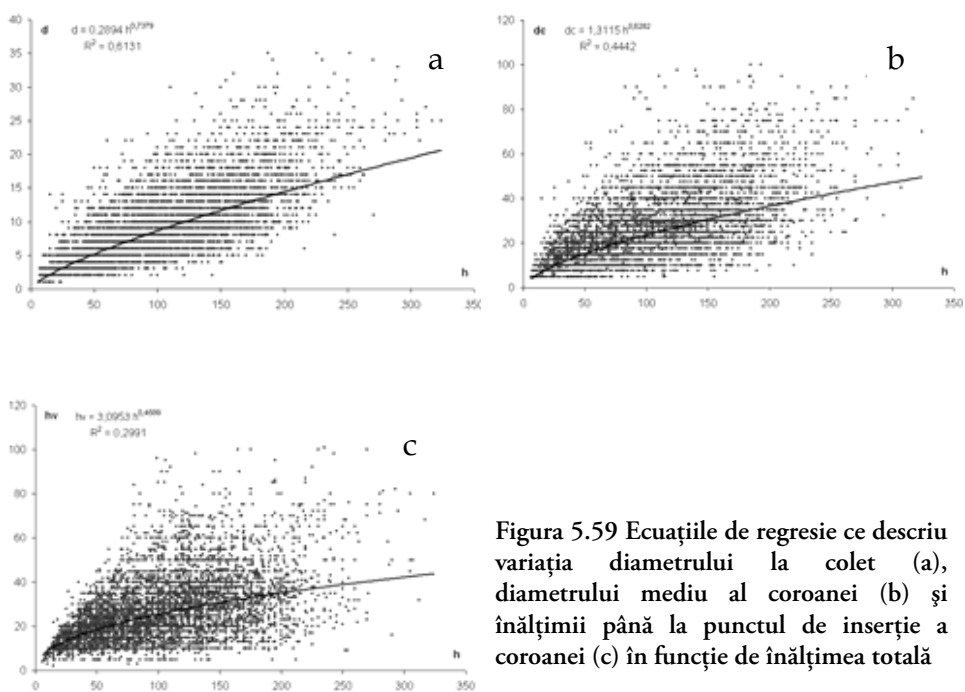

**Figura 5.59 Ecuațiile de regresie ce descriu variația diametrului la colet (a), diametrului mediu al coroanei (b) și înălțimii până la punctul de inserție a coroanei (c) în funcție de înălțimea totală**

În urma analizei relaţiilor dintre elemente s-a optat pentru modelarea caracteristicilor biometrice în funcţie de înălţimea totală, fiind considerată un parametru esenţial al structurii seminţişului. A fost testată o gamă largă de ecuaţii de regresie, selectându-se ecuaţiile pentru care au fost obţinute cele mai mari valori ale coeficientului de determinare ($R^2$) – figura 5.59. În cazul ecuaţiei de regresie de determinare a diametrului la colet, variabila independentă (*h*) explică 61,31% din variaţia diametrelor (*d*). În cazul determinării diametrului mediu al coroanei (*dc*), este explicată 44,42% din variaţie, iar pentru înălţimea punctului de inserţie a coroanei (*hv*) 29,91%. Rezultatele se înscriu în limitele aşteptate, gradul de incertitudine fiind indus de către comportamentul şi reacţia diferită a multitudinii de specii din compoziţia seminţişului, precum şi de competiţia intensă. Krajicek (et al., 1961) consideră că relaţiile dintre parametrii biometrici ai aceleiaşi specii nu sunt afectate de vârstă sau bonitatea staţiunii, competiţia dintre arbori fiind singurul proces care alterează rezultatele determinărilor.

Pentru a îmbunătăţi capacităţile predictive ale modelelor matematice s-a efectuat o stratificare a datelor pe specii, şi au fost determinaţi coeficienţii ecuaţiilor de regresie separat pe fiecare specie, graficele fiind prezentate în figurile 5.60-5.64.

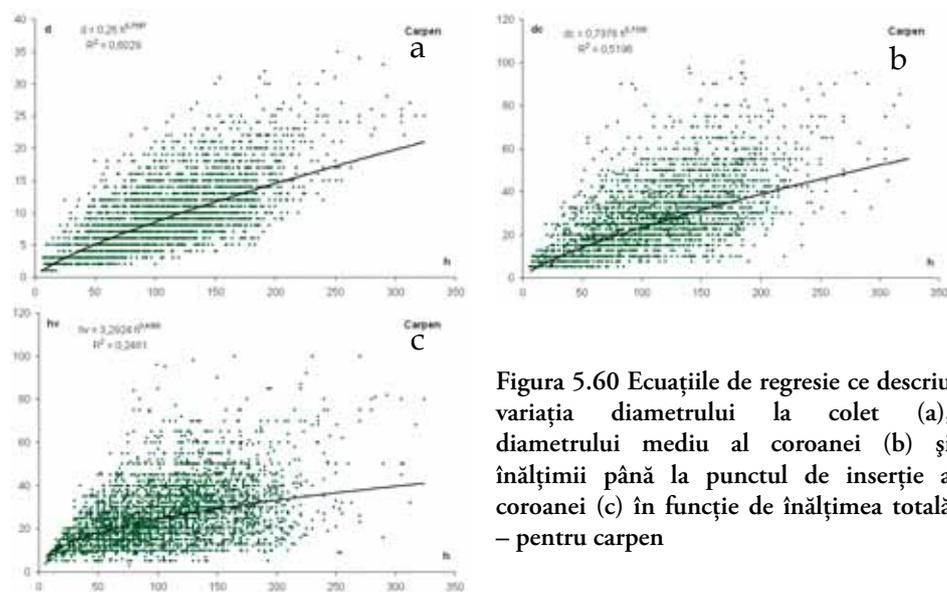

**Figura 5.60 Ecuaţiile de regresie ce descriu variaţia diametrului la colet (a), diametrului mediu al coroanei (b) şi înălţimii până la punctul de inserţie a coroanei (c) în funcţie de înălţimea totală – pentru carpen**

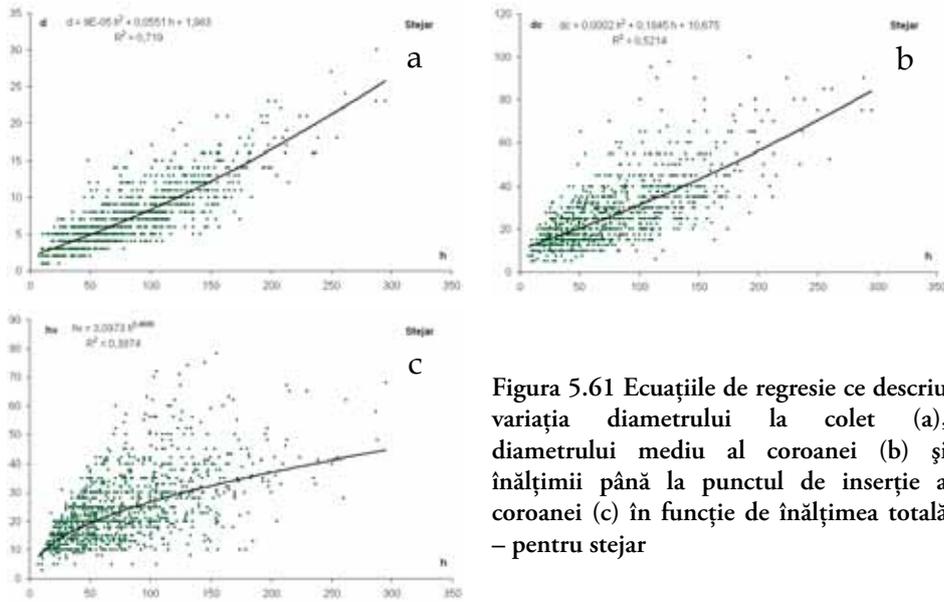

Figura 5.61 Ecuaţiile de regresie ce descriu variaţia diametrului la colet (a), diametrului mediu al coroanei (b) şi înălţimii până la punctul de inserţie a coroanei (c) în funcţie de înălţimea totală – pentru stejar

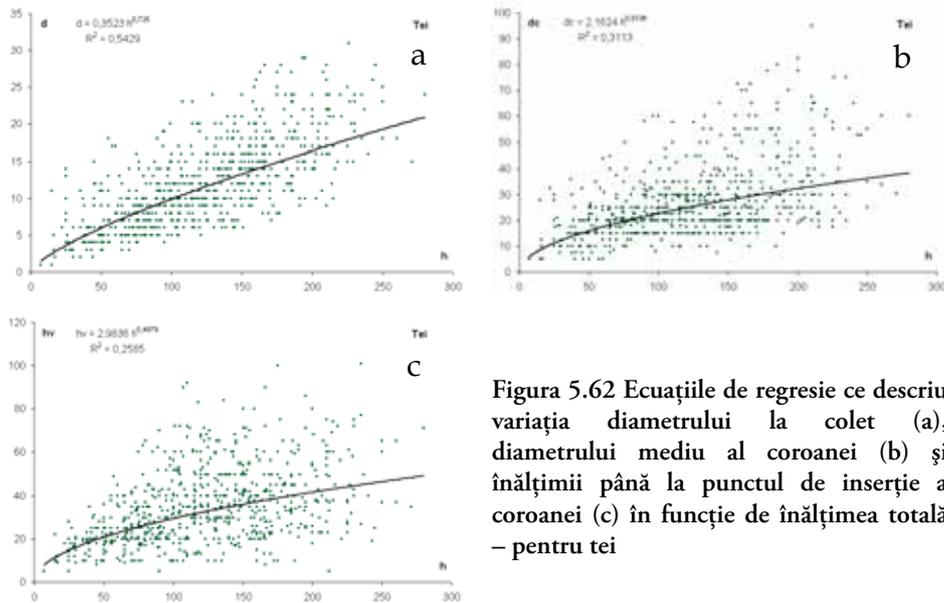

Figura 5.62 Ecuaţiile de regresie ce descriu variaţia diametrului la colet (a), diametrului mediu al coroanei (b) şi înălţimii până la punctul de inserţie a coroanei (c) în funcţie de înălţimea totală – pentru tei

Calculele au fost efectuate doar pentru speciile cu un procent de participare de peste 5% (carpen, stejar, tei, frasin şi jugastru), pentru care există un volum substanţial de date.

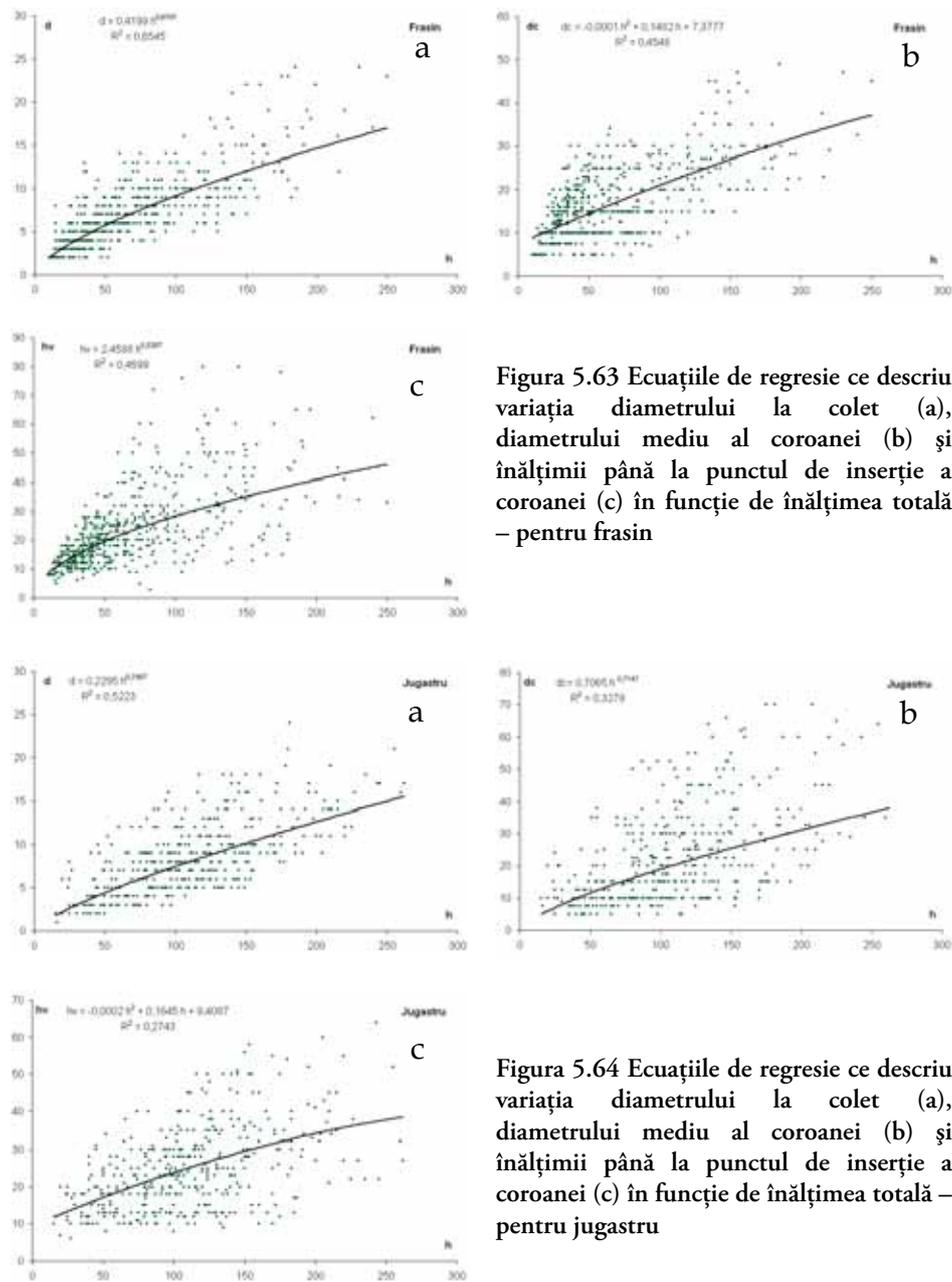

**Figura 5.63 Ecuaţiile de regresie ce descriu variaţia diametrului la colet (a), diametrului mediu al coroanei (b) şi înălţimii până la punctul de inserţie a coroanei (c) în funcţie de înălţimea totală – pentru frasin**

**Figura 5.64 Ecuaţiile de regresie ce descriu variaţia diametrului la colet (a), diametrului mediu al coroanei (b) şi înălţimii până la punctul de inserţie a coroanei (c) în funcţie de înălţimea totală – pentru jugastru**

În vederea evaluării acurateţei de estimare a modelelor folosite s-au analizat valorile reziduale obţinute prin diferenţele dintre valorile determinate de ecuaţiile de regresie și valorile individuale experimentale observate. Pentru fiecare ecuație de regresie a fost calculată valoarea abaterii pătratice medii (RMSD), media procentuală a abaterilor absolute reziduale (MPEA%), și media procentuală a abaterilor reziduale (MPE%),conform metodologiei de apreciere a acurateței modelelor prezentate de Uzoh și Oliver (2008):

$$RMSD = \sqrt{\frac{\sum_{i=1}^{n}(Y_i - \hat{Y}_i)^2}{n}} \quad MPEA_{\%} = \frac{1}{n}\sum_{i=1}^{n}\frac{\left|Y_i - \hat{Y}_i\right|}{Y_i} \quad MPE_{\%} = \frac{1}{n}\sum_{i=1}^{n}\frac{Y_i - \hat{Y}_i}{Y_i} \quad (5.2)$$

unde n reprezintă numărul de observații, $Y_i$ valorile individuale experimentale iar $\hat{Y}_i$ valorile individuale estimate de ecuațiile de regresie. Situația este prezentată sintetic în tabelul următor.

Tabelul 5.12

**Acuratețea ecuațiilor de regresie folosite per total și pe specii - variabila independentă h**

|  | Total | Ca | St | Te | Fr | Ju |
|---|---|---|---|---|---|---|
| **determinare d** | | | | | | |
| R² | 0,613 | 0,603 | 0,719 | 0,543 | 0,655 | 0,522 |
| RMSD | 3,9464 | 3,5615 | 2,1850 | 4,0949 | 2,3783 | 3,0081 |
| MPEA% | 29% | 31% | 28% | 30% | 26% | 32% |
| MPE% | 8% | 7% | 6% | 7% | 5% | 7% |
| **determinare dc** | | | | | | |
| R² | 0,444 | 0,520 | 0,521 | 0,311 | 0,455 | 0,341 |
| RMSD | 11,9221 | 11,7133 | 10,5097 | 12,1535 | 6,1691 | 12,8195 |
| MPEA% | 39% | 36% | 34% | 34% | 38% | 48% |
| MPE% | 11% | 10% | 9% | 9% | 18% | 15% |
| **determinare hv** | | | | | | |
| R² | 0,299 | 0,248 | 0,387 | 0,259 | 0,470 | 0,274 |
| RMSD | 12,5850 | 12,5104 | 9,9293 | 15,6810 | 9,8285 | 9,4271 |
| MPEA% | 40% | 41% | 33% | 41% | 31% | 38% |
| MPE% | 12% | 12% | 8% | 13% | 8% | 17% |

Rezultatele obţinute arată că la majoritatea speciilor se îmbunătăţesc rezultatele predictive faţă de modelul general în care au fost utilizate observaţiile de la puieţii aparţinând tuturor speciilor. Speciile la care se determină cele mai apropiate valori de cele experimentale sunt stejarul şi frasinul, la polul opus aflându-se carpenul şi jugastrul, pentru care determinările specifice scad chiar sub precizia oferită de modelul general.

În ceea ce priveşte comparaţia între parametrii biometrici, abaterile cele mai reduse de la valorile experimentale se înregistrează în cazul diametrului, explicabil datorită faptului că modelele ce utilizează înălţimea drept variabilă predictivă reuşesc să explice o mare parte din variaţia datelor experimentale – în cazul stejarului este explicată în proporţie de 72%. O altă observaţie se poate face referitor la diametrul mediu al coroanei – este parametrul la care s-au constatat îmbunătăţiri ale determinărilor în cazul  tuturor speciilor (mai puţin în cazul jugastrului), ceea ce conduce la ideea că dimensiunea coroanei în plan orizontal este un parametru cu un grad mare de specificitate.

Înălţimea până la punctul de inserţie al coroanei rămâne parametrul cel mai dificil de modelat matematic, surprinzându-se foarte puţin din variaţia datelor experimentale.

Îmbunătăţirea rezultatelor predictive ale unui model se poate realiza prin introducerea în modelul matematic a unor variabile independente suplimentare. Precizia modelului de determinare a valorilor variabilei dependente poate să crească, dar implicit creşte şi complexitatea modelului, adăugarea de variabilele adiţionale presupunând un volum de muncă crescut în vederea prelevării acestora. Acest efort suplimentar poate fi justificat doar de obţinerea unei precizii mult sporite. Pentru a studia posibilitatea şi oportunitatea integrării unor variabile suplimentare în determinarea parametrilor biometrici s-a efectuat o analiză a regresiei multiple. În acest scop s-a utilizat aplicaţia software *StatSoft Statistica*. Indicatorii statistici obţinuţi în urma analizei regresiei multiple sunt prezentaţi în tabelul 5.13. Indicele β reprezintă valoarea coeficientului de regresie multiplă standardizat, măsurat în unităţi ale abaterii standard şi arată cât de mult influenţează una din variabilele independente pe cea dependentă. Valoarea coeficientului de corelaţie multiplă (R) indică intensitatea legăturii dintre variabile.

Coeficientul de determinare $R^2$ arată proporţia din variaţia variabilei dependente ce este explicată de modelul matematic propus. Datorită faptului că această valoare tinde să supraestimeze capacitatea predictivă a modelului, ea se ajustează în funcţie de numărul de variabile independente şi observaţii. Pentru fiecare variantă s-a precizat valoarea testului F, respectiv p – nivelul de semnificaţie, care trebuie să aibă o valoare situată sub 5% pentru a considera viabil modelul propus.

Rezultatele analizei arată că se înregistrează îmbunătăţiri ale predicţiei valorilor diametrului la colet, respectiv a diametrului mediu al coroanei dacă se utilizează pe lângă înălţimea totală o altă variabilă predictivă.

În cazul modelării diametrului la colet folosirea înălţimii totale şi a diametrului mediu al coroanei explică 62,5% din variaţia variabilei dependente, în condiţiile în care nivelul de semnificaţie este sub pragul de 5%. Valorile coeficientului de regresie multiplă β arată că înălţimea influenţează într-o măsură mult mai mare valoarea diametrului la colet în comparaţie cu diametrul mediu al coroanei. În cel de al doilea caz, folosirea înălţimii totale şi a diametrului la colet explică doar 47,9% din variaţia diametrului mediu al coroanei, în condiţiile în care nivelul de semnificaţie este sub pragul de 5%. Valorile coeficientului de regresie multiplă β arată că cele două variabile independente influenţează valoarea diametrului mediu al coroanei în proporţii sensibil egale, cu o uşoară diferenţă în favoarea diametrului la colet.

Indicatorii statistici ai analizei susţin ideea că în acest caz folosirea unor modele matematice predictive bazate pe două variabile independente conduce la îmbunătăţiri de evaluare a variabilei dependente, dar într-o măsură redusă. Rezultatele pot fi însă afectate de fenomenul de multicoliniaritate datorită folosirii ca variabile independente a unui set de variabile între care există corelaţie (atenţie în special la folosirea concomitentă a înălţimii şi diametrului între care există o corelaţie puternică).

**Tabelul 5.13**

**Indicatori statistici ai analizei regresiei multiple**

| Variabila dependentă | Variabilele independente | β | R | $R^2$ | $R^2$ ajustat | F | p |
|---|---|---|---|---|---|---|---|
| **d** | h | 0,580 | 0,791 | 0,6252 | 0,6251 | 6006,323 | <0,05 |
| | dc | 0,281 | | | | | |
| **dc** | h | 0,346 | 0,692 | 0,4788 | 0,4786 | 3306,251 | <0,05 |
| | d | 0,391 | | | | | |

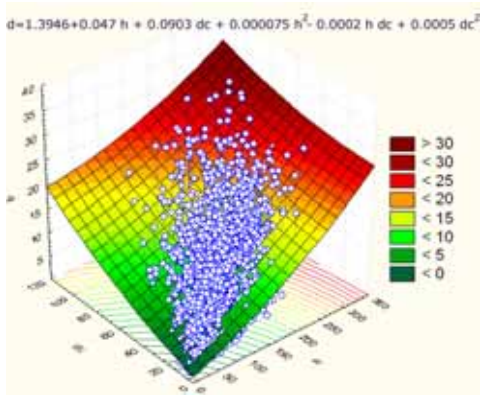

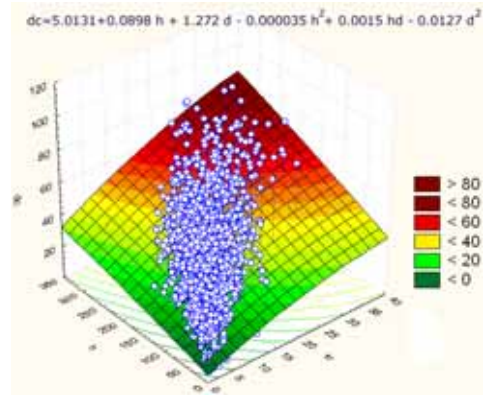

**Figura 5.65 Modelul matematic al dependenței variabilei *d* de *h* și *dc***

**Figura 5.66 Modelul matematic al dependenței variabilei *dc* de *h* și *d***

În concluzie se poate aprecia că îmbunătățirile predictive minore nu justifică pe deplin folosirea unui parametru biometric suplimentar înălțimii. Folosirea unei a doua variabile conduce în practică la prelevarea unui volum sporit de date, dificil de măsurat – atât în cazul diametrului la colet cât și a diametrului mediu al coroanei. Drept urmare se poate opta pentru modelarea tuturor celorlalte caracteristici biometrice doar în funcție de înălțime, fără pierderi majore de precizie.

Există și alți parametri biometrici importanți care pot fi evaluați în baza datelor prelevate. Volumul coroanei este o caracteristică folosită adesea în caracterizarea structurii forestiere (Meng et al., 2007; Rautiainen et al., 2008),

Assman (1970) acordându-i o mare importanța în lucrarea „*The Principles of Forest Yield Study*". Assman folosește volumul și aria suprafeței exterioare a coroanei în fundamentarea eficienței creșterii arborilor, considerând că suprafața exterioară a coroanei este un parametru cu o sensibilitate superioară volumului în explicarea proceselor de creștere și dezvoltare, beneficiind și de o justificare fiziologică mai bună. El recomandă folosirea unor corpuri simple pentru aprecierea volumului și a suprafeței exterioare în procesul de modelare a coroanei – con, cilindru, paraboloid, elipsoid, dar admite că există diferențe semnificative între acestea. Studii comparative recente efectuate de Rautiainen (et al., 2008) au evidențiat că folosirea conului drept corp de modelare a coroanei, chiar și în cazul rășinoaselor, generează abateri negative majore, fiind recomandată folosirea elipsoidului.

În baza acestor idei s-a calculat pentru fiecare puiet volumul coroanei și aria suprafeței exterioare, folosind elipsoidul drept model geometric de aproximare. În cazul volumului s-a folosit formula:

$$Volum\_elipsoid = \frac{4}{3}\pi \cdot a \cdot b \cdot c \qquad (5.3)$$

unde *a*, *b*, și *c* sunt razele elipsoidului. Datorită faptului că aria suprafeței elipsoidului nu poate fi exprimată exact prin funcții elementare (folosește integralele eliptice incomplete de ordinul I și II, dificil de algoritmizat) s-a optat pentru folosirea formulei de aproximare a suprafeței exterioare a elipsoidului propusă de Cantrell (2004) ce furnizează rezultate afectate de o eroare mai mică de 0,13%.

$$Suprafata = 4\pi \cdot \frac{15 \cdot Q^3 - 7 \cdot P \cdot Q \cdot R - 27 \cdot R^2}{15 \cdot (2 \cdot Q^2 + P \cdot R)} \qquad (5.4)$$

unde *P, Q, R* sunt polinoamele simetrice elementare, *P=a+b+c; Q=ab+bc+ca; R=abc,* iar *a*, *b*, și *c* sunt razele elipsoidului.

Modelarea acestor doi noi parametri a presupus o analiză prealabilă în vederea alegerii celei mai bune variabile predictive. În literatura de specialitate se menționează determinarea volumului prin ecuații de regresie ce folosesc drept variabilă predictivă diametrul de bază (Meng et al., 2007). Diametrul de bază este în general puternic corelat cu diametrul coroanei (Paine, Hann, 1982), iar acesta din urmă intră direct în calculul volumului prin cele două dimensiuni în plan orizontal. În cazul semințișului am propus ipoteza relevanței scăzute a diametrului la colet în predicția celorlalte elemente biometrice, comparativ cu înălțimea totală. Astfel că în cazul volumului coroanei și a suprafeței exterioare a coroanei se va face o comparație între acuratețea de determinare a ecuațiilor de regresie bazate pe diametrul la colet și a celor bazate pe înălțimea totală. Analiza preliminară constă în compararea coeficienților de corelație dintre parametri. Coeficienții de corelație sunt foarte semnificativi, indicând legături de intensitate medie și chiar puternică.

Tabelul 5.14

Legătura corelativă a volumului coroanei și suprafeței exterioare a coroanei
cu înălțimea totală și diametrul la colet

|  | înălțimea totală | diametrul la colet |
|---|---|---|
| volumul coroanei | 0,63 *** | 0,61 *** |
| suprafața exterioară a coroanei | 0,80 *** | 0,72 *** |

Valorile coeficienților de corelație sunt inferiori în cazul diametrului, comparativ cu cei corespunzători înălțimii. Se remarcă de asemenea faptul că suprafața exterioară a coroanei pare un element mai expresiv față de volumul coroanei, atât în relația cu diametrul cât și cu înălțimea.

Nu se poate deocamdată formula o concluzie privind superioritatea valorii predictive a înălțimii deoarece dependența dintre elementele analizate nu este neapărat liniară, caz în care coeficienții de corelație nu sunt foarte sugestivi.

Pasul următor al analizei a constat în alegerea dintr-un set de ecuații de regresie a variantei pentru care s-a obținut cel mai bun coeficient de determinare ($R^2$). Acuratețea modelelor folosite va fi studiată prin analiza valorilor reziduale. Ecuațiile de regresie de determinare a volumului și suprafeței exterioare a coroanei prin intermediul înălțimii totale și a diametrului la colet sunt prezentate în figurile 5.67-5.70. Pentru a simplifica prezentarea datelor, se va folosi următoarea convenție de notare a variabilelor: d - diametrul la colet, h – înălțimea totală, vol_c – volumul coroanei, sup_c – suprafața exterioară a coroanei.

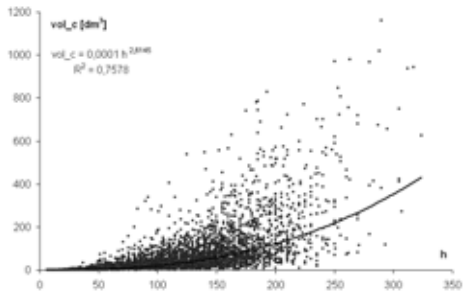

**Figura 5.67** Ecuația de regresie de determinare a vol_c pe baza h

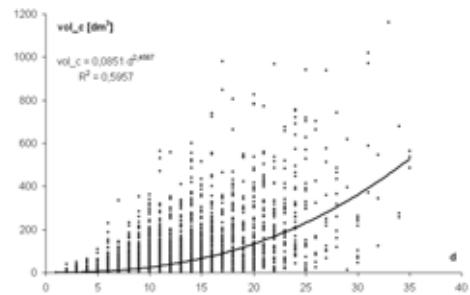

**Figura 5.68** Ecuația de regresie de determinare a vol_c pe baza d

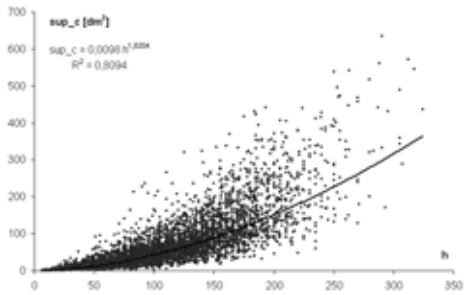

**Figura 5.69** Ecuația de regresie de determinare a sup_c pe baza h

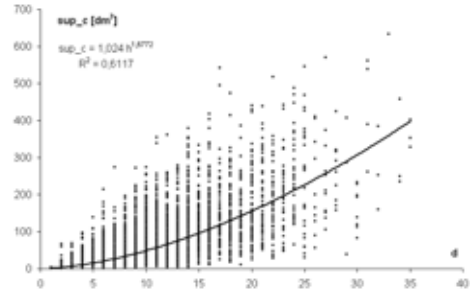

**Figura 5.70** Ecuația de regresie de determinare a sup_c pe baza d

Tabelul 5.15

Comparație între acuratețea ecuațiilor de regresie de determinare a volumului coroanei și suprafeței exterioare a coroanei

| | Variabila predictivă: h | Variabila predictivă: d |
|---|---|---|
| **Determinarea volumului coroanei** | | |
| $R^2$ | 0,758 | 0,596 |
| RMSD | 68,70 | 72,76 |
| $MPEA_\%$ | 113% | 202% |
| $MPE_\%$ | 67% | 145% |
| **Determinarea suprafeței exterioare a coroanei** | | |
| $R^2$ | 0,809 | 0,612 |
| RMSD | 39,48 | 48,07 |
| $MPEA_\%$ | 51% | 90% |
| $MPE_\%$ | 19% | 47% |

În tabelul 5.15 precizia modelelor de determinare a volumului şi suprafeţei exterioare a coroanei este analizată prin comparaţia valorilor următorilor indicatori statistici: coeficientul de determinare ($R^2$), valoarea abaterilor pătratice medii (RMSD), media procentuală a abaterilor absolute reziduale (MPEA%) şi media procentuală a abaterilor reziduale (MPE%). Diferenţele dintre valorile obţinute pentru cele două variante de determinare – prin folosirea înălţimii totale, respectiv a diametrului la colet ca variabile predictive, ne conduc spre concluzia că înălţimea reprezintă variabila care asigură cea mai mare precizie de determinare a elementelor de structură specifice coroanei. Modelele bazate pe înălţimea totală explică în cel mai mare grad variaţia volumului (76%) şi suprafeţei exterioare a coroanei (81%), fiind în acelaşi timp caracterizate şi de cele mai mici abateri reziduale.

Pentru a stabili dacă se obţin îmbunătăţiri în predicţia volumului şi suprafeţei exterioare a coroanei faţă de modelul general în care calculele au fost efectuate pentru puieţii aparţinând tuturor speciilor, s-au determinat şi ecuaţiile de regresie pentru fiecare specie în parte, fiind prezentate în figurile 5.71-5.75.

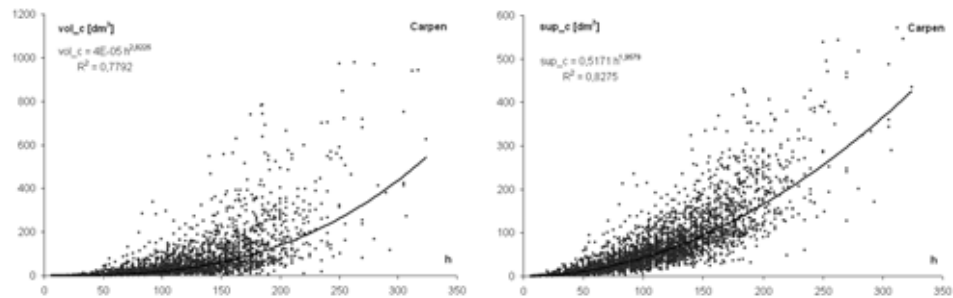

**Figura 5.71 Regresie de determinare a** *vol_c* **şi** *sup_c* **pe baza** *h* **la carpen**

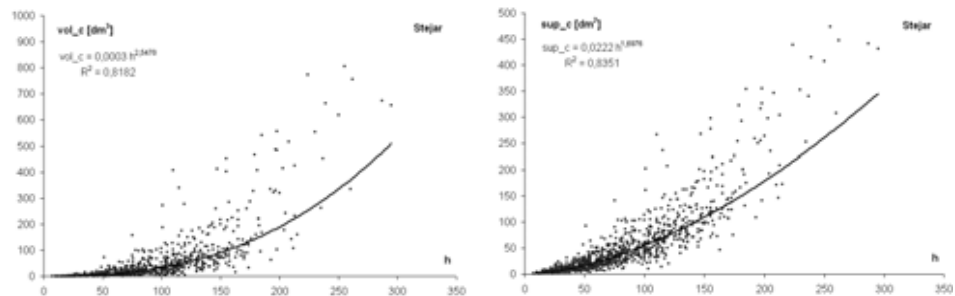

**Figura 5.72 Regresie de determinare a** *vol_c* **şi** *sup_c* **pe baza** *h* **la stejar**

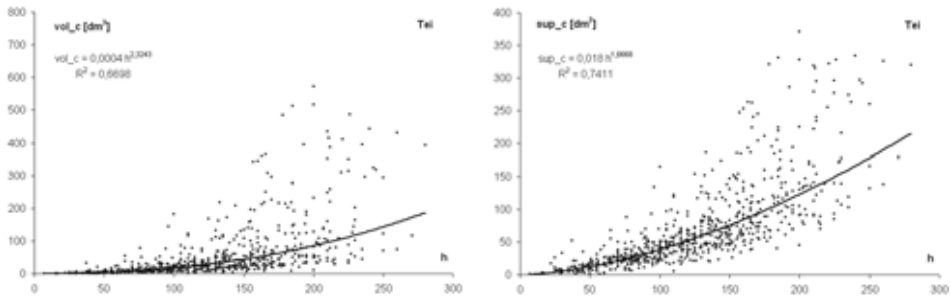

**Figura 5.73 Regresie de determinare a *vol_c* şi *sup_c* pe baza *h* la tei**

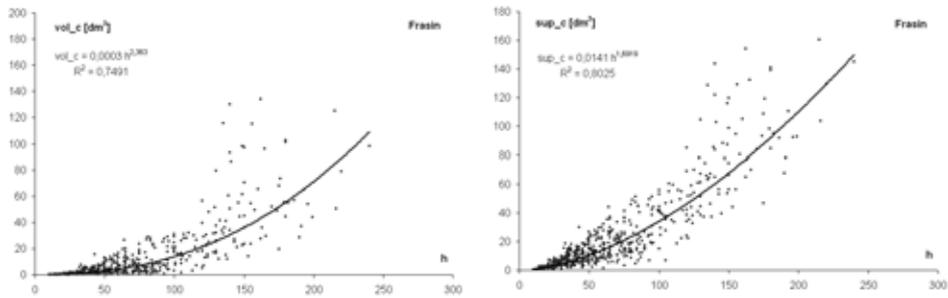

**Figura 5.74 Regresie de determinare a *vol_c* şi *sup_c* pe baza *h* la frasin**

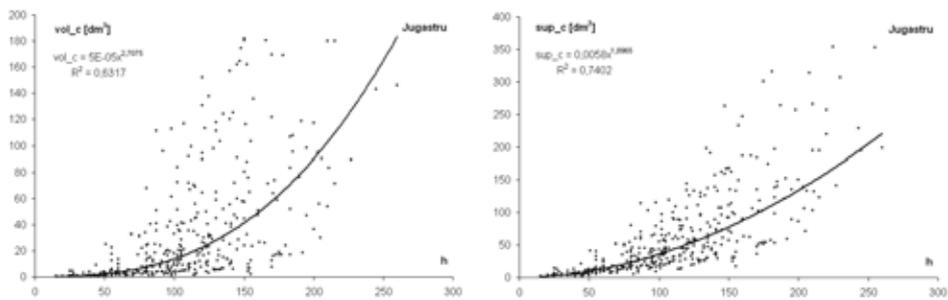

**Figura 5.75 Regresie de determinare a *vol_c* şi *sup_c* pe baza *h* la jugastru**

Ecuaţiile de regresie folosesc exclusiv înălţimea totală ca variabilă, în urma analizei anterioare stabilindu-se superioritatea predictivă a acestui parametru. Din analiza graficelor 5.71-5.75 se observă că speciile îşi modifică volumul coroanei în proporţii diferite la aceeaşi variaţie a înălţimii, în acest fel confirmându-se unele observaţii efectuate în teren cu prilejul prelevării datelor – de exemplu la puieţii de tei a fost observată variaţia mică a volumului coroanei în funcţie de înălţimea puieţilor.

În vederea stabilirii eventualelor îmbunătățiri ale predicției volumului și suprafeței exterioare a coroanei în cazul folosirii datelor caracteristice fiecărei specii, se prezintă sintetic în tabelul 5.16 indicatorii statistici ai analizei valorilor reziduale, obținuți pentru fiecare model matematic. Valorile indică faptul că ecuațiile de regresie specifice aduc un plus de a curatețe determinării volumului și suprafeței exterioare a coroanei doar în cazul puieților de stejar. Stejarul înregistrează și în cazul acestor parametri abaterile cele mai reduse în determinarea parametrilor biometrici în funcție de înălțime.

Comparația dintre modelele folosite pentru determinarea volumului și cele de determinare a suprafeței exterioare a coroanei subliniază diferențe legate de expresivitatea parametrilor determinați. Atât modelul general cât și modelele specifice explică un procent mult mai mare din varianța suprafeței exterioare a coroanei (74-84%) comparativ cu cea explicată pentru volumul coroanei (63-82%), fapt ce conduce la ideea că primul parametru menționat are o dependență mai mare de înălțime.

Tabelul 5.16

**Comparație între acuratețea ecuațiilor de regresie de determinare a volumului coroanei și suprafeței exterioare a coroanei pe specii (cu variabilă predictivă înălțimea totală)**

|  | Total | Ca | St | Te | Fr | Ju |
|---|---|---|---|---|---|---|
| **Determinarea volumului coroanei** | | | | | | |
| $R^2$ | 0,76 | 0,78 | 0,82 | 0,67 | 0,75 | 0,63 |
| RMSD | 68,69 | 71,32 | 67,27 | 66,04 | 14,93 | 59,75 |
| MPEA$_\%$ | 113% | 111% | 60% | 124% | 135% | 196% |
| MPE$_\%$ | 67% | 69% | 14% | 91% | 99% | 161% |
| **Determinarea suprafeței exterioare a coroanei** | | | | | | |
| $R^2$ | 0,81 | 0,83 | 0,84 | 0,74 | 0,80 | 0,74 |
| RMSD | 39,48 | 40,81 | 37,45 | 41,75 | 14,19 | 39,82 |
| MPEA$_\%$ | 51% | 50% | 38% | 54% | 61% | 66% |
| MPE$_\%$ | 19% | 21% | 16% | 31% | 37% | 41% |

## 5.3.2. Relațiile dintre parametrii biometrici și desimea puieților

Numărul de indivizi la unitatea de suprafață reprezintă un factor structural de mare importanță în caracterizarea ecosistemelor forestiere. În cazul regenerării arboretelor numărul de puieți la metrul pătrat oferă informații utile privitoare la spațiere, cu implicații asupra competiției, creșterii și dezvoltării tinerelor exemplare.

Desimea este considerată un factor important al structurii și datorită influenței puternice exercitate asupra caracteristicilor dimensionale ale indivizilor. În cazul semințișurilor naturale numărul de puieți la metrul pătrat poate fi determinat relativ ușor, ceea ce poate constitui un avantaj în folosirea acestui factor drept variabilă predictivă în determinarea altor parametri structurali. Oportunitatea folosirii desimii în modele matematice de evaluare a unor caracteristici structurale va fi analizată în continuare. În literatura de specialitate sunt menționate în special cazuri în care desimea, folosită ca variabilă predictivă adițională, a condus la îmbunătățirea acurateței modelelor – Mcalpine și Hobbs (1994) au modelat înălțimea până la punctul de inserție al coroanei în funcție de înălțimea totală și desime; Kanazawa (et al., 1985) a determinat lungimea coroanei în funcție de înălțime și desime; Sharma și Zhang (2004), respectiv Newton și Amponsah (2007) au inclus desimea în modelarea relației dintre diametre și înălțime.

Prima etapă a analizei constă în determinarea coeficienților de corelație dintre elementele biometrice și desimea puieților și prezentarea grafică a câmpurilor de corelație. În vederea creșterii numărului de observații pentru aprecierea corelațiilor s-a făcut o împărțire a suprafețelor de probă în suprafețe elementare de formă pătrată, fiind propuse trei variante dimensionale – suprafețe de 1 x 1 m; 1,75 x 1,75 m și 3,5 x 3,5 m (tabelul 5.17).

Cu ajutorul aplicației CARTOGRAMA s-a efectuat repartizarea puieților pe suprafețe elementare în funcție de coordonatele pozițiilor acestora și a fost calculată pentru fiecare quadrat desimea și valoarea medie a fiecărui parametru biometric. Coeficienții de corelație au fost calculați pentru aceste valori medii și desimi, obținute pentru fiecare suprafață elementară în parte.

Au fost alese trei variante dimensionale diferite de analiză pentru a nu avea eventuale alterări ale rezultatelor produse de o alegere nepotrivită.

Tabelul 5.17

**Valorile coeficienților de corelație dintre elementele biometrice și desimea puieților**

|  | Varianta 1x1m | Varianta 1,75x1,75m | Varianta 3,5x3,5m |
|---|---|---|---|
| Diametrul la colet | -0,418 *** | -0,577 *** | -0,638 *** |
| Înălțimea totală | -0,230 *** | -0,266 *** | -0,343 ** |
| Înălțimea pct. de inserție a coroanei | -0,049 | -0,082 | -0,018 |
| Diametrul mediu al coroanei | -0,404 *** | -0,521 *** | -0,617 *** |
| Volumul coroanei | -0,304 *** | -0,434 *** | -0,599 *** |
| Suprafața exterioară a coroanei | -0,337 *** | -0,475 *** | -0,556 *** |
| Număr de observații | 490 | 160 | 40 |

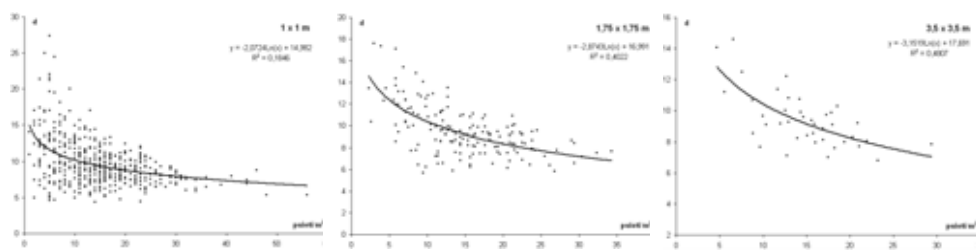

**Figura 5.76 Legătura corelativă dintre diametrul la colet și desimea puieților**

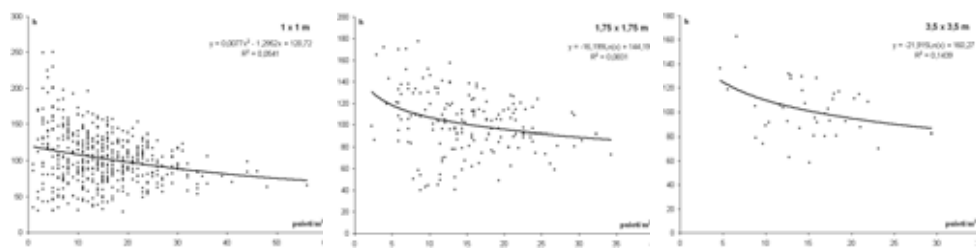

**Figura 5.77 Legătura corelativă dintre înălțimea totală și desimea puieților**

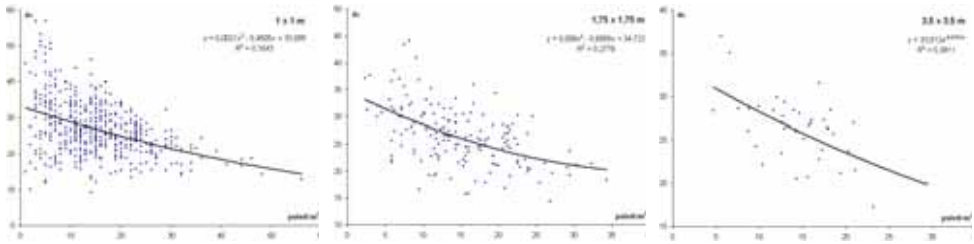

**Figura 5.78 Legătura corelativă dintre diametrul mediu al coroanei și desime**

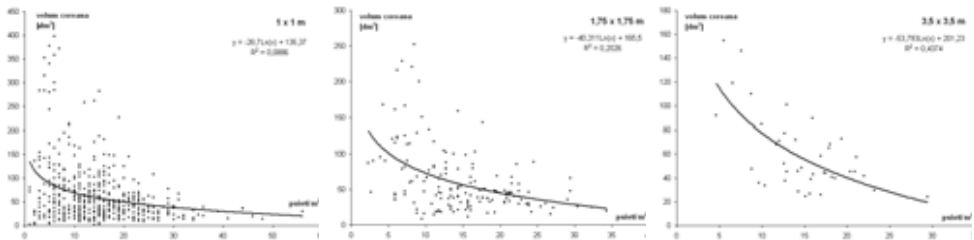

**Figura 5.79 Legătura corelativă dintre volumul coroanei și desimea puieților**

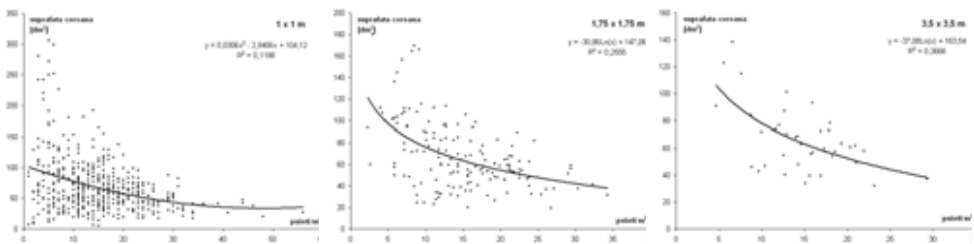

**Figura 5.80 Legătura corelativă dintre suprafața coroanei și desimea puieților**

Se observă că dimensiunile suprafețelor elementare de analiză nu modifică natura relațiilor dintre elementele dimensionale și desime dar apare o modificare a intensității în sensul creșterii acesteia odată cu lărgirea ariei suprafețelor elementare. Acest lucru este explicabil datorită faptului că raportarea la suprafețe mai largi determină o creștere a omogenității observațiilor (Zenner, 2005), caz în care o mare parte din „zgomotul" informațional se reduce.

În toate variantele de analiză se constată că desimea este un factor al structurii ce influențează caracteristicile dimensionale ale puieților. Pentru toți parametri biometrici (mai puțin pentru înălțimea până la punctul de inserție al coroanei) au fost obținuți coeficienți de corelație negativi, foarte semnificativi.

Desimea influențează în cea mai mare măsură, în ordine: diametrul la colet și diametrul mediu al coroanei (intensitate medie), suprafața exterioară și volumul coroanei (intensitate slabă spre medie), respectiv înălțimea (intensitate slabă). Este surprinzător că nu s-a evidențiat o legătură între desime și înălțimea până la punctul de inserție al coroanei, acest parametru fiind în general puțin expresiv, în analizele anterioare înregistrând cele mai slabe corelații cu celelalte caracteristici biometrice.

Faptul că înălțimea totală stabilește cea mai slabă legătură cu desimea prezintă un avantaj în regresia multiplă, la folosirea concomitentă a celor două variabile predictive în modelare reducându-se riscul ca determinările să fie afectate de fenomenul de multicoliniaritate. În vederea studierii oportunității de integrare a desimii ca variabilă suplimentară în determinarea parametrilor biometrici s-a efectuat o analiză a regresiei multiple cu ajutorul aplicației software StatSoft Statistica. Indicatorii statistici obținuți în urma analizei regresiei multiple sunt prezentați în tabelul 5.18. Notațiile parametrilor în vederea simplificării prezentării datelor sunt cele folosite anterior în conținutul lucrării. Datele au fost preluate pentru varianta dimensională a suprafețelor de 3,5 x 3,5 m.

Tabelul 5.18

**Indicatori statistici ai analizei regresiei multiple**

| Variabila dependentă | Variabilele independente | β | R | $R^2$ | $R^2$ ajustat | F | p |
|---|---|---|---|---|---|---|---|
| d | h | 0,663 | 0,892 | 0,795 | 0,784 | 71,889 | <0,05 |
| | desime | -0,410 | | | | | |
| dc | h | 0,265 | 0,665 | 0,442 | 0,412 | 14,658 | <0,05 |
| | desime | -0,530 | | | | | |
| vol_c | h | 0,455 | 0,736 | 0,541 | 0,517 | 21,843 | <0,05 |
| | desime | -0,440 | | | | | |
| sup_c | h | 0,720 | 0,876 | 0,767 | 0,754 | 61,038 | <0,05 |
| | desime | -0,310 | | | | | |

**Tabelul 5.19**

**Comparație între coeficienții corelației simple și cei ai corelației multiple**

|  | Coef. de corelație cu înălțimea | Coef. de corelație cu desimea | Coef. de corelație multiplă |
|---|---|---|---|
| d | 0,804 | -0,638 | 0,892 |
| dc | 0,445 | -0,617 | 0,665 |
| vol_c | 0,607 | -0,599 | 0,736 |
| sup_c | 0,826 | -0,556 | 0,876 |

Rezultatele analizei arată că se înregistrează îmbunătățiri ale predicției tuturor parametrilor dacă se utilizează desimea ca variabilă predictivă suplimentară față de înălțimea totală.

În cazul modelării diametrului la colet și a suprafeței exterioare a coroanei este explicată peste 75% din variația variabilei dependente, în condițiile în care nivelul de semnificație este sub pragul de 5%. Valorile coeficientului de regresie multiplă β arată că înălțimea influențează într-o măsură mult mai mare valoarea diametrului la colet și suprafața coroanei în comparație cu desimea.

Desimea puieților influențează într-o proporție superioară înălțimii valorile diametrului coroanei, iar în cazul volumului coroanei influența variabilelor predictive este sensibil egală.

În tabelul 5.19 sunt prezentate valorile coeficienților de corelație dintre elementele supuse modelării și înălțime, respectiv desime. În ultima coloană sunt afișate valorile coeficientului de corelație multiplă, considerând drept variabile independente înălțimea și desimea. Toți indicatorii statistici sunt calculați pentru varianta împărțirii suprafețelor în quadrate 3,5 x 3,5 m, având drept variabile media caracteristicilor biometrice înregistrate în fiecare quadrat.

În figurile 5.81-5.84 sunt expuse grafic modelele matematice ale relațiilor dintre parametrii biometrici și variabilele independente reprezentate de înălțimea totală și desimea puieților.

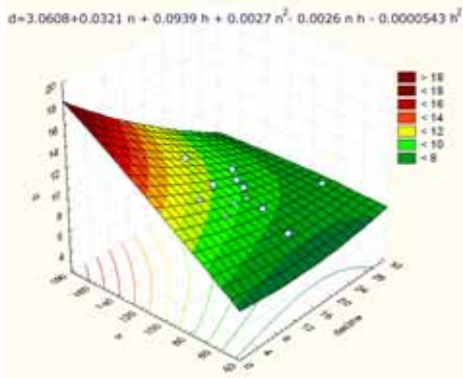

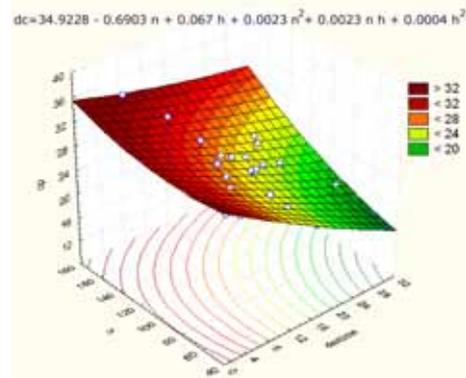

**Figura 5.81 Modelul matematic al dependenței variabilei *d* de *h* și *desime***

**Figura 5.82 Modelul matematic al dependenței variabilei *dc* de *h* și *desime***

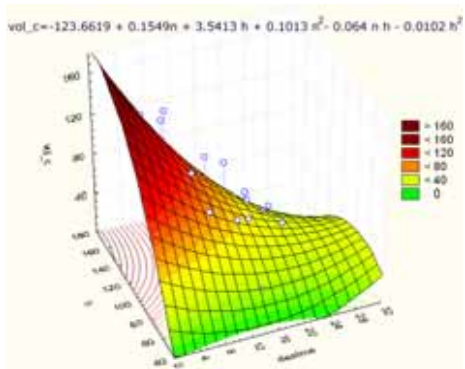

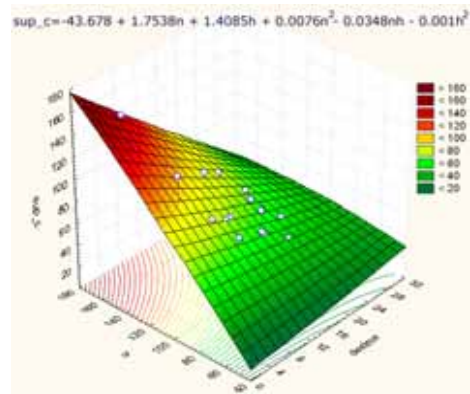

**Figura 5.83 Modelul matematic al dependenței variabilei *vol_c* de *h* și *desime***

**Figura 5.84 Modelul matematic al dependenței variabilei *sup_c* de *h* și *desime***

Indicatorii statistici obținuți în cadrul analizei regresiei multiple, susțin oportunitatea folosirii unor modele matematice predictive bazate pe înălțimea totală și desime, utilizarea celor două variabile independente conducând la îmbunătățiri ale determinărilor.

# Capitolul 6
# Diversitatea structurală a seminţişului

## 6.1. Caracterizarea diversităţii structurale prin indicatori sintetici

Diversitatea, ca măsură a variabilităţii unor atribute, poate fi măsurată la niveluri şi scări diferite în cadrul ecosistemelor forestiere. Datorită diversităţii specifice relativ scăzute a arboretelor din zona temperată, diversitatea dimensională (numită adesea diversitate structurală) a suscitat un interes superior cercetătorilor silvici. În vederea stabilirii unor criterii comune de apreciere şi comparare a evaluărilor, a fost concepută o gamă largă de indicatori sintetici de apreciere a diversităţii structurale, care permit surprinderea fidelă a caracteristicilor unui arboret la un anumit moment din timp. Folosirea acestora alături de indicii convenţionali (diametrul, înălţimea, suprafaţa de bază, vârsta, desimea) determină obţinerea unui nivel de detaliu superior în aprecierea complexităţii structurale a arboretelor.

Aplicaţiile cuantificării diversităţii dimensionale se regăsesc în comparaţii ale acestui parametru între arborete în vederea stabilirii stabilităţii structurale, aspecte privitoare la dinamica structurii în timp şi spaţiu sau implicaţii ale influenţei tratamentelor silviculturale asupra structurii. O parte a indicilor sintetici folosiţi în aprecierea diversităţii structurale au fost integraţi în modele de creştere şi dezvoltare a arboretelor.

Deşi cercetări în acest domeniu au fost efectuate începând cu mijlocul secolului trecut, în ultimul deceniu interesul cercetătorilor în domeniul biodiversităţii şi diversităţii forestiere a crescut.

În cazul diversității structurale a arboretelor, preocupările s-au materializat în numeroase lucrări științifice (Zenner, Hibbs, 2000; Staudhammer, LeMay, 2001; Pommerening, 2002; Aguirre et al., 2003; Lexerod, Eid, 2005; Vorcak et al., 2006; Davies, Pommerening, 2008).

La noi în țară primele lucrări privitoare la tematica diversității structurale forestiere au apărut încă de acum trei decenii, Ștefania Leahu (1978) introducând o serie de noțiuni împrumutate din teoria informației. Botnariuc și Vădineanu prezintă în tratatul de ecologie din 1982 o scurtă analiză a unor indicatori sintetici de apreciere a diversității, modalitatea de calcul a acestora și modul de interpretare a rezultatelor obținute. Ulterior, interesul pentru acest aspect particular al analizei structurii pădurii a fost concretizat în numeroase cercetări (e.g. Leahu, I., 1988; Cenușă 1992, 1996 a, 2002; Avăcăriței, 2005; Duduman, 2009; Dănilă, 2009).

În continuare se prezintă o analiză a diversității structurale a semințișului din suprafețele studiate. Sunt calculați principalii indicatori ai diversității recomandați de literatura de specialitate, în raport cu diverși parametri biometrici, pe specii și la nivel general. Scopul analizei îl reprezintă determinarea expresivității indicilor, a particularităților distribuțiilor dimensionale, respectiv a diferențelor pe care acestea le înregistrează la nivel de specie. Clarificarea acestor aspecte poate să conducă la includerea informațiilor diversității structurale în modele forestiere de creștere și dezvoltare. Datorită faptului că până acum nu a fost stabilită superioritatea absolută a unui anumit indicator sintetic, cercetările ce urmăresc această temă folosesc o gamă largă de indici de evaluare a diversității, în vederea surprinderii cât mai multor particularități.

Prelucrările numerice ce trebuie efectuate în vederea stabilirii valorii acestor indici sunt relativ complexe, operațiile fiind dificile în cazul unui volum mare de date. Calculatorul este în acest caz un instrument util, deoarece permite algoritmizarea și automatizarea prelucrărilor. Dezvoltarea unor aplicații de calcul tabelar a ușurat într-o oarecare măsură calculul acestor indici, dar chiar și cu ajutorul unei aplicații de calcul tabelar (e.g. *Microsoft Excel*), multe din aceste prelucrări ale datelor ar fi greu de efectuat fără cunoștințe de programare.

Pentru a veni în întâmpinarea cercetătorilor care doresc să efectueze studii ecologice dar nu posedă toate cunoştinţele necesare prelucrării datelor, ca de altfel şi tuturor celor care doresc o formă rapidă de a afla informaţii privitoare la biodiversitatea sau diversitatea structurală a unei zone analizate s-a realizat o aplicaţie de calcul a celor mai utilizaţi indicatori sintetici, numită BIODIV (Palaghianu & Avăcăriţei, 2006; Palaghianu, 2014).

Aplicaţia, dezvoltată în *Microsoft Visual Basic*, preia datele de intrare dintr-o foaie de calcul de tip *Microsoft Excel* şi calculează valorile pentru: indicele Simpson, indicele Shannon şi echitatea pentru acesta (indicele Pielou), indicele Brillouin, indicele Berger-Parker, indicele McIntosh, indicele Margalef, indicele Menhinick şi coeficientul Gleason.

Folosirea aplicaţiei în activitatea de cercetare prezintă anumite avantaje - introducerea datelor primare este facilă, se obţine o creştere a productivităţii muncii, se obţin rapid rezultate şi este extrem de uşor de folosit datorită interfeţei grafice. Se poate considera că aplicaţia BIODIV poate fi de un real folos cercetătorilor din domeniul silviculturii, ecologiei, biologiei (figura 6.1).

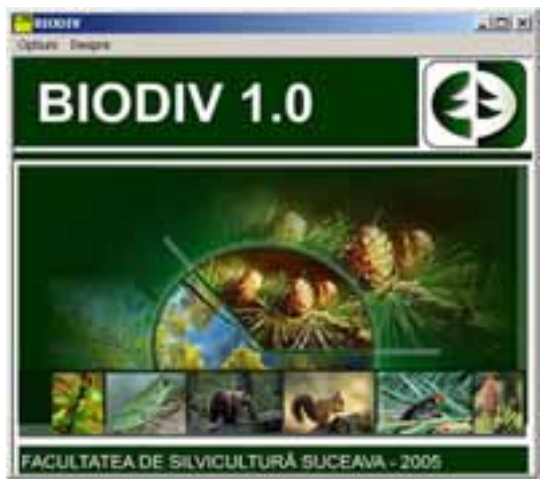

**Figura 6.1  Fereastra aplicaţiei BIODIV**

Indicatorii calculați de aplicația software sunt:

**Indicele Shannon (H)**

Acesta este unul dintre cei mai utilizați indici de apreciere a diversității, având originea în conceptul de entropie a informației dezvoltat de Shannon (1948) în lucrarea „*A Mathematical Theory of Communication*". Adesea este incorect menționat sub denumirea de indicele Shannon-Weaver sau Shannon-Wiener (deși nici unul din autorii ulterior alăturați numelui indicelui nu au contribuit la fundamentarea sa). Valoarea indicelui reprezintă gradul de organizare/ dezorganizare al unui sistem dat. Acesta se calculează cu următoarea relație:

$$H = -\sum_{i=1}^{S} p_i \ln p_i , \quad p_i = n_i / N \tag{6.1}$$

$p_i$ - proporția de reprezentare a unei categorii de caracteristici observate

$n_i$ - numărul de observații pentru o categorie

$S$ – numărul de categorii analizate

N - numărul total de observații în perimetrul analizat

Valoarea minimă a indicelui este 0 și se obține atunci când toate observațiile se găsesc într-o singură categorie. Valoarea maximă este egală cu *ln (1/S)* și se obține în cazul în care toate observațiile sunt repartizate în mod egal între categoriile analizate.

**Echitatea (E)** arată relațiile dintre abundențele categoriilor – în cazul unor abundențe relative similare echitatea va avea o valoare unitară iar în cazul în care majoritatea observațiilor aparțin unei singure categorii ea tinde spre valoarea zero. Echitatea reprezintă o formă de standardizare a indicelui Shannon, ca raport între diversitatea observată și diversitatea maximă (Pielou, 1969; citat de Lexerod, Eid, 2006).

$$E = \frac{H}{\ln(S)} \tag{6.2}$$

H    - valoarea indicelui Shannon

$S$ – numărul de categorii analizate

**Indicele Brillouin (HB)**

În anumite condiții acest indice surprinde mai bine diversitatea, dar în general valoarea acestui indice este foarte apropiată de cea a indicelui Shannon și întotdeauna inferioară (Magurran, 2004). Indicele este considerat din punct de vedere matematic superior indicelui Shannon ca măsură a diversității, folosirea sa fiind recomandată de către numeroși cercetători. Cu toate acestea, modul de calcul greoi descurajează folosirea sa, fiind mai puțin utilizat comparativ cu indicele Shannon. Dependența de volumul eșantionului poate conduce de asemenea la rezultate neașteptate în anumite situații.

$$HB = \frac{\ln(N!) - \sum_{i=1}^{S} \ln(n_i!)}{N} \tag{6.3}$$

$n_i$ - numărul de observații pentru o categorie

$S$ – numărul de categorii analizate

N - numărul total de observații în perimetrul analizat

**Indicele Simpson** – este un indice care ține cont nu doar de numărul categoriilor ci și de proporția fiecăreia. A fost prezentat de Simpson (1949) în lucrarea „*Measurement of Diversity*", fiind un instrument deosebit de util aflat la îndemâna specialiștilor în ecologie. Este puternic influențat de abundența categoriilor și puțin sensibil la numărul de categorii. În publicațiile de specialitate se prezintă în general trei variante ale acestui indice de evaluare a biodiversității:

**Indicele Simpson (D)**

$$D = \sum p_i^2 \, , \; p_i = n_i / N \tag{6.4}$$

$p_i$ - proporția de reprezentare a unei categorii de observații

$n_i$ - numărul de observații pentru o categorie

N - numărul total de observații în perimetrul analizat

În această variantă indicele Simpson (D) exprimă probabilitatea ca două observații extrase întâmplător din perimetrul analizat să aparțină aceleiași categorii.

Valoarea minimă este egală cu 1/S (unde $S$ – numărul de categorii analizate), iar maximul este 1.

**Indicele complementar de diversitate Simpson (1 – D)** exprimă probabilitatea ca două observații efectuate aleatoriu, să aparțină unor categorii diferite. În acest caz valoarea minimă este 0 iar maximul 1-(1/S).

**Indicele reciproc Simpson (1/D)** exprimă numărul de categorii cu pondere ridicată ce conduc la o anumită valoare a indicelui Simpson D. Valoarea minimă este 1, iar valoarea maximă numărul de categorii.

**Indicele Berger-Parker (d)**

Acest indice (Berger, Parker, 1970) este o formă simplă de calcul a dominanței, reprezentând proporția maximă din cadrul categoriilor analizate. Este un indice independent de numărul categoriilor, fiind însă cel mai influențat de către echitate. Valoarea maximă este 1 în cazul în care toate observațiile sunt grupate în aceeași categorie și minimă 1/S (unde $S$ – numărul de categorii analizate) în cazul unei distribuiri uniforme a observațiilor între categorii.

$$d = p_{max}(\forall i : p_{max} \geq p_i) \qquad (6.5)$$

$p_i$ - proporția de reprezentare a unei categorii de observații

**Indicele McIntosh ($D_{MI}$)**

Este un indice puțin folosit datorită modului complex de calcul și interpretării dificile. Este o altă formă de calcul a dominanței propusă de McIntosh (1967), indicele fiind independent de numărul de categorii. Pielou (1969, citat de Magurran, 2004) a conceput o formă de standardizare a indicelui, ca măsură a echității.

$$D_{MI} = \frac{N - \sqrt{\sum_{i=1}^{S} n_i^{\,2}}}{N - \sqrt{N}} \qquad (6.6)$$

$n_i$ - numărul de observații pentru o categorie

N - numărul total de observații în perimetrul analizat

**Indicele Margalef ($D_{Mg}$)**

Este un indice frecvent folosit datorită modului simplu de calcul, conceput în forma inițială de Margalef (1958) (citat de Margalef, 1994), inspirat de indicele Gleason. Valoarea sa nu este influențată de abundența relativă a categoriilor ci de numărul acestora. Prezintă și o variantă alternativă de calcul, ca raport între numărul de categorii a diametrelor și logaritmul natural al sumei suprafețelor de bază pentru toate clasele de diametre (Lexerod, Eid, 2006). Valoarea minimă este 0 – când toți arborii se încadrează în aceeași clasă dimensională analizată.

$$D_{Mg} = \frac{S-1}{\ln(N)}$$ (6.7)

*S* – numărul de categorii analizate

N - numărul total de observații în perimetrul analizat

**Indicele Menhinick ($D_{Mn}$)**

Este un indice conceput de Menhinick (1964) în urma unui studiu comparativ al indicilor de diversitate apăruți până la acea dată (indicele Gleason, Margalef, ș.a.). Autorul a încercat realizarea unui indice superior ca expresivitate care să fie influențat cât mai puțin de scara la care se face analiza.

$$D_{Mn} = \frac{S}{\sqrt{N}}$$ (6.8)

*S* – numărul de categorii analizate

N - numărul total de observații în perimetrul analizat

**Indicele Gleason ($K_{gl}$)**

Este unul din primii indici ai diversității, inspirând o mare parte din indicii concepuți ulterior (e.g. indicele Margalef). A fost conceput de către Gleason (1922), ca urmare a preocupărilor legate de studiul relației dintre numărul de specii dintr-o arie dată. Se consideră că valoarea sa este în mică măsură influențată de către scara de analiză, datorită formei sale logaritmice.

$K_{gl}$ = (S− 1) / log(N) (6.9)

N - numărul total de observații în perimetrul analizat

*S* – numărul de categorii analizate

**Tabelul 6.1**

**Valorile indicilor sintetici de apreciere a diversității pe categorii de diametre**

| | total | Ca | St | Te | Fr | Ju |
|---|---|---|---|---|---|---|
| **indicele Simpson (D)** | 0,240 | 0,238 | 0,300 | 0,201 | 0,315 | 0,258 |
| indicele Simpson (1-D) | 0,760 | 0,762 | 0,700 | 0,799 | 0,685 | 0,742 |
| indicele Simpson (1/D) | 4,164 | 4,207 | 3,339 | 4,976 | 3,171 | 3,876 |
| **indicele Shannon** | 1,610 | 1,628 | 1,381 | 1,753 | 1,350 | 1,470 |
| echitatea (Shannon) | 0,610 | 0,617 | 0,629 | 0,798 | 0,614 | 0,756 |
| indicele Brillouin | 1,606 | 1,620 | 1,365 | 1,725 | 1,322 | 1,439 |
| indicele Berger-Parker | 0,358 | 0,354 | 0,405 | 0,291 | 0,439 | 0,348 |
| indicele McIntosh | 0,516 | 0,521 | 0,467 | 0,573 | 0,457 | 0,517 |
| indicele Margalef | 1,462 | 1,563 | 1,137 | 1,215 | 1,247 | 0,991 |
| indicele Menhinick | 0,164 | 0,219 | 0,267 | 0,334 | 0,364 | 0,340 |
| indicele Gleason | 1,014 | 1,083 | 0,788 | 0,842 | 0,865 | 0,687 |

**Tabelul 6.2**

**Valorile indicilor sintetici de apreciere a diversității pe categorii de înălțime**

| | total | Ca | St | Te | Fr | Ju |
|---|---|---|---|---|---|---|
| **indicele Simpson (D)** | 0,157 | 0,159 | 0,195 | 0,148 | 0,239 | 0,170 |
| indicele Simpson (1-D) | 0,843 | 0,841 | 0,805 | 0,852 | 0,761 | 0,830 |
| indicele Simpson (1/D) | 6,351 | 6,278 | 5,136 | 6,775 | 4,182 | 5,880 |
| **indicele Shannon** | 1,973 | 1,971 | 1,806 | 2,011 | 1,645 | 1,907 |
| echitatea (Shannon) | 0,729 | 0,728 | 0,727 | 0,873 | 0,749 | 0,868 |
| indicele Brillouin | 1,967 | 1,962 | 1,783 | 1,978 | 1,614 | 1,861 |
| indicele Berger-Parker | 0,199 | 0,206 | 0,297 | 0,198 | 0,379 | 0,252 |
| indicele McIntosh | 0,610 | 0,610 | 0,576 | 0,640 | 0,533 | 0,618 |
| indicele Margalef | 1,575 | 1,683 | 1,564 | 1,367 | 1,247 | 1,322 |
| indicele Menhinick | 0,176 | 0,234 | 0,356 | 0,372 | 0,364 | 0,437 |
| indicele Gleason | 1,092 | 1,167 | 1,084 | 0,947 | 0,865 | 0,916 |

**Tabelul 6.3**

**Valorile indicilor sintetici de apreciere a diversităţii pe categorii de volum a coroanei**

|  | total | Ca | St | Te | Fr | Ju |
|---|---|---|---|---|---|---|
| **indicele Simpson (D)** | 0,974 | 0,968 | 0,965 | 0,986 | 0,997 | 0,991 |
| indicele Simpson (1-D) | 0,026 | 0,032 | 0,035 | 0,014 | 0,003 | 0,009 |
| indicele Simpson (1/D) | 1,027 | 1,033 | 1,036 | 1,014 | 1,003 | 1,009 |
| **indicele Shannon** | 0,084 | 0,100 | 0,104 | 0,041 | 0,012 | 0,033 |
| echitatea (Shannon) | 0,036 | 0,043 | 0,065 | 0,060 | 0,018 | 0,030 |
| indicele Brillouin | 0,082 | 0,097 | 0,099 | 0,039 | 0,011 | 0,028 |
| indicele Berger-Parker | 0,987 | 0,984 | 0,982 | 0,993 | 0,998 | 0,995 |
| indicele McIntosh | 0,013 | 0,016 | 0,018 | 0,007 | 0,002 | 0,005 |
| indicele Margalef | 1,012 | 1,082 | 0,569 | 0,152 | 0,156 | 0,330 |
| indicele Menhinick | 0,117 | 0,156 | 0,148 | 0,074 | 0,081 | 0,146 |
| indicele Gleason | 0,702 | 0,750 | 0,394 | 0,105 | 0,108 | 0,229 |

**Tabelul 6.4**

**Valorile indicilor sintetici de apreciere a diversităţii pe categorii ale suprafeţei exterioare a coroanei**

|  | total | Ca | St | Te | Fr | Ju |
|---|---|---|---|---|---|---|
| **indicele Simpson (D)** | 0,785 | 0,759 | 0,838 | 0,746 | 0,961 | 0,766 |
| indicele Simpson (1-D) | 0,215 | 0,241 | 0,162 | 0,254 | 0,039 | 0,234 |
| indicele Simpson (1/D) | 1,273 | 1,317 | 1,193 | 1,340 | 1,040 | 1,306 |
| **indicele Shannon** | 0,461 | 0,507 | 0,387 | 0,499 | 0,102 | 0,463 |
| echitatea (Shannon) | 0,186 | 0,204 | 0,186 | 0,360 | 0,093 | 0,334 |
| indicele Brillouin | 0,458 | 0,503 | 0,377 | 0,489 | 0,097 | 0,449 |
| indicele Berger-Parker | 0,881 | 0,865 | 0,914 | 0,856 | 0,980 | 0,868 |
| indicele McIntosh | 0,115 | 0,131 | 0,087 | 0,141 | 0,020 | 0,131 |
| indicele Margalef | 1,237 | 1,323 | 0,995 | 0,456 | 0,312 | 0,496 |
| indicele Menhinick | 0,141 | 0,188 | 0,237 | 0,149 | 0,121 | 0,194 |
| indicele Gleason | 0,858 | 0,917 | 0,690 | 0,316 | 0,216 | 0,344 |

<div align="right">Tabelul 6.5</div>

**Valorile coeficienților de corelație dintre indicii de apreciere a diversității structurale**

|  | 1-D | H | HB | d | DMi | DMg | DMn | Kgl |
|---|---|---|---|---|---|---|---|---|
| indicele Simpson 1-D | 1 | | | | | | | |
| indicele Shannon H | 0,995 | 1 | | | | | | |
| indicele Brillouin HB | 0,994 | 1,000 | 1 | | | | | |
| indicele Berger-Parker d | -0,994 | -0,996 | -0,996 | 1 | | | | |
| indicele McIntosh Dmi | 0,997 | 0,998 | 0,997 | -0,998 | 1 | | | |
| indicele Margalef DMg | 0,785 | 0,798 | 0,801 | -0,774 | 0,776 | 1 | | |
| indicele Menhinick DMn | 0,778 | 0,768 | 0,761 | -0,760 | 0,782 | 0,573 | 1 | |
| indicele Gleason Kgl | 0,785 | 0,798 | 0,801 | -0,774 | 0,776 | 1,000 | 0,573 | 1 |

Din analiza modului de calcul al indicilor sintetici, a valorilor diversității obținute pentru elementele biometrice precum și a corelațiilor dintre indici se poate face o încadrare a indicilor în două mari grupe, fiecare grupă având o expresivitate similară – prima grupă formată din indicele Shannon, Simpson, Brillouin, Berger-Parker, McIntosh, iar cea de a doua este reprezentată de indicii Gleason, Margalef și Menhinick.

Se remarcă de asemenea similaritatea valorilor unor indici – indicele Brillouin are nu doar o relație aproape funcțională cu Shannon (r=0,994***), dar și valori individuale absolute atât de apropiate încât grafic nu pot fi deosebite. De altfel, între toți indicii din prima grupă s-au obținut coeficienți de corelație cu valori mai mari de 0,99***. În cadrul celei de-a doua grupe se observă relația funcțională dintre indicele Gleason și Margalef, formula celor doi indici deosebindu-se doar prin baza logaritmică folosită. Toți indicii din cea de a doua grupă au fost influențați de observațiile lui Gleason (1922), formula lor de calcul neținând cont de proporția categoriilor ci doar de raportul dintre numărul de categorii și numărul total de observații.

Pentru facilitarea observării diferențelor dintre indici, au fost reprezentate grafic valorile acestora, separat pentru fiecare grupă, folosindu-se pentru reprezentare diagrame multi-axă, separate pe elementele dimensionale în funcție de care au fost calculate valorile indicilor. S-a ales acest tip de comparaţie în defavoarea folosirii unor valori aşa-zis „relative" (utilizate uneori în comparaţii) care elimină informaţii importante (fig. 6.2-6.3).

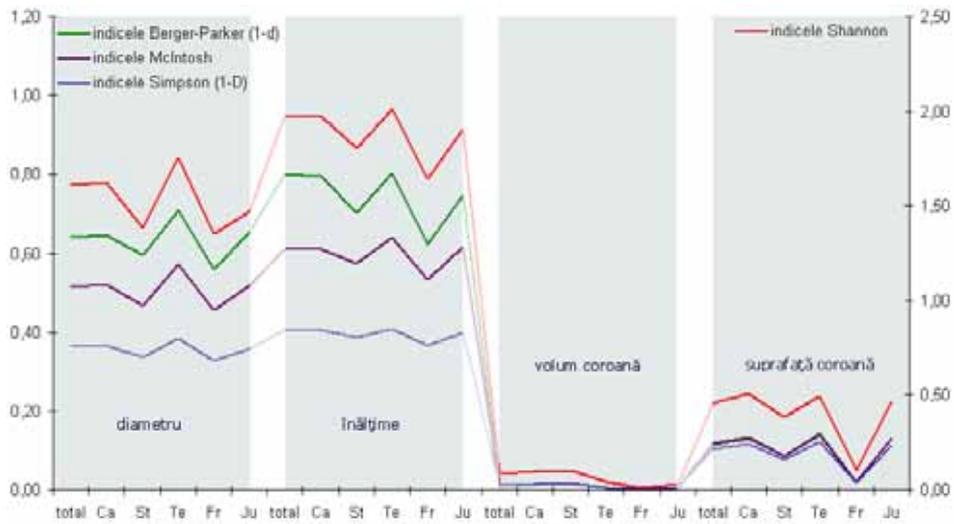

**Figura 6.2 Variaţia grafică a indicatorilor sintetici de evaluare a diversităţii pe specii şi elemente biometrice – prima grupă de indici**

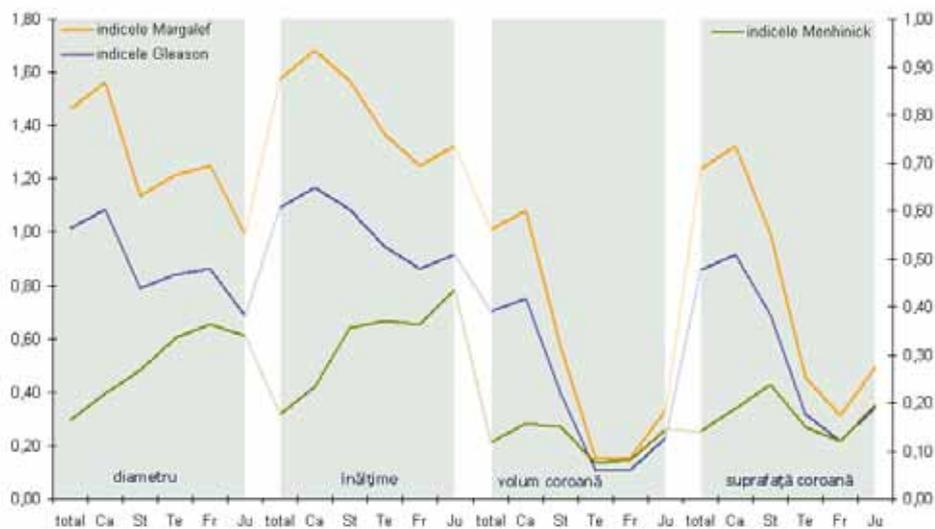

**Figura 6.3 Variaţia grafică a indicatorilor sintetici de evaluare a diversităţii pe specii şi elemente biometrice – grupa a doua de indici**

În graficul din figura 6.2 s-a reprezentat un singur indice din gama Simpson (1-D), iar datorită faptului că indicele Berger-Parker (d) este singurul corelat negativ cu ceilalţi din grupa sa, s-a utilizat pentru reprezentare valoarea 1-d.

Diferenţele înregistrate între valorile indicilor din grupa I sunt nesemnificative din punctul de vedere al sensibilităţii, toţi înregistrând aceleaşi fluctuaţii ale valorilor. Singurele diferenţe se înregistrează la nivelul valorilor absolute, indicele Shannon înregistrând constant valori superioare celorlalţi, fapt ce-i poate aduce fals plus de expresivitate în cazul reprezentării la aceeaşi scară cu ceilalţi indici, chiar şi în cazul folosirii unor valori relative. În cazul celui de al doilea grup (fig. 6.3), diferenţele între indici se remarcă doar la nivelul indicelui Menhinick, între indicele Gleason şi Margalef existând o relaţie funcţională.

Chiar dacă relaţiile de calcul ale tuturor celor trei indici sunt în mare măsură similare, indicele Menhinick înregistrează un comportament mult diferit de al celorlalţi doi. Comparaţia între valorile primei şi celei de a doua grupe de indici arată o expresivitate superioară a indicilor primei grupe, justificată de includerea în formula de calcul nu doar a numărului de categorii şi observaţii ci şi a proporţiei categoriilor.

Cu toate că au fost efectuate numeroase comparaţii între diverşi indici ai diversităţii structurale (Lexerod, Eid, 2005; Davies, Pommerening, 2008), nu s-a oferit un verdict clar privitor la superioritatea unui anumit indice. Consider însă că se poate aprecia că în vederea evaluării diversităţii dimensionale este suficientă folosirea unui singur tip de indice din prima grupă şi a unuia din cea de a doua sau chiar a unui singur indice din prima grupă, datorită expresivităţii superioare a acestora. Este recomandabilă folosirea din prima grupă a indicilor Shannon sau Simpson datorită numeroaselor cercetări care utilizează aceşti indici, în acest fel existând posibilitatea efectuării unor comparaţii între rezultatele obţinute. Dacă se intenţionează compararea rezultatelor cu ale altor cercetări este bine să se precizeze valorile standardizate (relative) ale indicilor (e.g. echitatea determinată pentru unii indici ca raport între diversitatea observată şi diversitatea maximă).

Diversitatea structurală a parametrilor biometrici arată un maxim în cazul înălțimii, iar apoi valorile descresc în ordine pentru diametru, suprafața coroanei și volumul coroanei.  Având în vedere că diversitatea maximă se realizează la o repartizare cât mai echilibrată în clase, se poate afirma că înălțimea este, în cazul semințișului din suprafețele studiate, un parametru mai stabil decât diametrul și de asemenea suprafața coroanei în relație directă cu volumul coroanei.

În ceea ce privește analiza gradului de organizare pe specii se constată frecvent valori minime pentru frasin și stejar și maxime pentru tei și carpen. Structurarea superioară pe orizontală și verticală a carpenului și teiului exprimă caracterul invadant al acestor specii, determinat de creșterile viguroase din primele etape de dezvoltare.

## 6.2. Aprecierea dominanței și diferențierii dimensionale

Efortul silvicultorilor de a regla funcționalitatea și stabilitatea ecosistemelor forestiere este dublat de numeroase cercetări ce urmăresc stabilirea relațiilor dintre structura arboretelor și funcțiile atribuite acestora. Heterogenitatea structurii orizontale și verticale a unei păduri este frecvent asociată cu o stabilitate ecologică superioară a acesteia. Informațiile privitoare la diversitatea structurală sunt utile în modelarea și managementul forestier, putând fi folosite în optimizarea anumitor decizii cu implicații economice sau ecologice (Pommerening, 2006).

În literatura de specialitate (Gadow, Hui, 1999; Pommerening, 2002) se acceptă că din punct de vedere al modalității de calcul majoritatea indicatorilor de evaluare a diversității structurale se împart în două mari grupuri: independenți de distanță și dependenți de distanță (sau spațiali). Primul grup nu ține cont de caracteristicile distribuției în spațiu a evenimentelor și are o utilitate restrânsă în caracterizarea ecosistemelor forestiere datorită imposibilității de a efectua analize la un nivel de detaliu satisfăcător. Cel de al doilea grup folosește informațiile privitoare la repartiția în spațiu a evenimentelor, în funcție de modalitatea de calcul sau de scara la care sunt prezentate rezultatele deosebind (Pommerening, 2002):

- indici individuali – percep diferenţe structurale la scară redusă; rezultatele se referă la indivizii populaţiei analizate, fiind stabilite în funcţie de relaţiile dintre arborii vecini;
- indicatori ai diversităţii la nivel de suprafaţă – valorile indicilor se referă la o scară mai largă (suprafeţe de probă, arborete);
- funcţii de evaluare a diversităţii – funcţii continue, specifice analizei proceselor punctiforme sau geostatisticii; folosirea acestora în cercetarea forestieră a luat amploare datorită faptului că pădurea oferă numeroase clase de tipare ale modului de organizare spaţială.

În general sunt preferaţi indicii individuali deoarece aceştia au avantajul de a putea fi calculaţi relativ uşor folosind date care se înregistrează în mod curent în inventarierile forestiere. În unele situaţii prelucrarea rezultatelor oferite de indicatorii individuali oferă informaţii privitoare la diversitatea unei întregi suprafeţe studiate.

Evaluarea diversităţii structurale spaţiale presupune conform „şcolii Göttingen" - von Gadow şi Hui (1999) trei aspecte majore: poziţionarea indivizilor (se pot efectua filtrări după specie sau caracteristici dimensionale), asocierea (specifică, dimensională sau calitativă) şi diferenţierea (între caracteristicile cantitative – parametrii biometrici).

În continuare se prezintă modul în care poate fi evaluată diferenţierea parametrilor biometrici prin intermediul unor indici frecvent menţionaţi în literatura de specialitate. Evaluarea diferenţelor dintre atributele dimensionale ale arborilor se realizează în general prin două metode distincte: aprecierea dominanţei (indicele sintetic este apreciat prin intermediul unui contor care se autoincrementează la înregistrarea unui raport de dominanţă) sau determinarea diferenţierii (indicele indică în acest caz raportul subunitar dintre atributele dimensionale comparate).

Diametrul şi înălţimea sunt parametrii biometrici cei mai frecvent folosiţi în calculul acestui tip de indici, în situaţia de faţă optându-se doar pentru calculul diversităţii structurale a înălţimilor, ca urmare a importanţei acesteia în seminţişuri.

Gadow și Fuldner (citați de Gadow 1997) au introdus în 1992 indicele de diferențiere dimensională T (*Dimensionsdifferenzierung*), pe care l-au definit și calculat pentru circumferință și înălțime. Ulterior variante ale aceluiași indice au fost prezentate de Gadow (1993), Fuldner (1995 b, citat de Kint et al., 2000) și Pommerening (2002).

Formula indicelui de diferențiere dimensională *T* pentru arborele *i* este:

$$T_i = 1 - \frac{1}{n} \sum_{j=1}^{n} \frac{\min(p_i, p_j)}{\max(p_i, p_j)} \tag{6.10}$$

*n* – numărul de vecini față de care se calculează raportul

*i* – indicele arborelui de referință

*j* - indicele unui arbore considerat a fi în primii *n* vecini

$p_i$, $p_j$ – valoarea atributului dimensional analizat al arborelui *i* și *j*

Variantele acestei formule vizează numărul de vecini pentru care se calculează raportul de diferențiere. Pommerening (2002) utilizează doar cel mai apropiat vecin în calculul acestui indice (notat $T_1$), în vreme ce în alte cercetări au fost folosiți 2, 3 ,4 sau 5 vecini (Kint et al., 2000; Pommerening, 2006).

Valoarea minimă a indicelui T este 0 în cazul în care toți arborii vecini au aceleași dimensiuni cu cele ale arborelui de referință și crește odată cu majorarea mediei diferenței dimensionale dintre arborii vecini și arborele referinței. Valoarea maximă menționată în literatură este 1, dar este o valoare spre care se tinde, deoarece nu poate fi atinsă nici practic și nici teoretic, implicând dimensiuni nule.

Pentru a ușura interpretarea indicilor, Pommerening (et al., 2000) propune o scară de apreciere cu 4 trepte, frecvent folosită:

- diferențiere slabă (*schwache differenzierung*) (0,0 ≤ T < 0,3) – cea mai mică dimensiune a parametrului analizat reprezintă minim 70% din dimensiunea arborilor vecini;
- diferențiere medie (*mittlere differenzierung*) (0,3 ≤ T < 0,5) – dimensiunea minimă este de 50 - 70% din dimensiunea vecinilor;
- diferențiere puternică (*starke differenzierung*) (0,5 ≤ T < 0,7) – dimensiunea minimă este de 30 - 50% din dimensiunea vecinilor;

- diferențiere foarte puternică (*sehr starke differenzierung*)  (0,7 ≤ T < 1) – dimensiunea minimă este mai mică de 30% din dimensiunea arborilor vecini.

Aguirre (et al., 1998) sugerează folosirea în comparații a unei scări de apreciere cu 5 trepte, cu ecartul fiecărui interval de 0,2.

Hui (et al., 1998) consideră că diferențierea dimensională nu este suficientă în aprecierea diversității structurale la scară redusă, deoarece nu se înregistrează sensul superiorității dimensionale dintre vecini. Astfel, menționează un nou indice care contorizează raporturile de dominanță și surprinde media numărului de vecini cu dimensiuni superioare unui arbore de referință (Hui et al., 1998). Ulterior Gadow și Hui  (1999), respectiv  Aguirre (et al., 2003) au considerat că dominanța se poate aprecia și prin înregistrarea raporturilor favorabile arborelui de referință, calculându-se media numărului de indivizi cu dimensiuni inferioare arborelui de referință.

Relația de calcul a indicelui de dominanță este:

$$U_i = \frac{1}{n} \sum_{j=1}^{n} v_j \tag{6.11}$$

unde: $vj = \begin{cases} 1, p_j > p_i \\ 0, altfel \end{cases}$  pentru varianta propusă de Hui et al. (1998)

și $\qquad vj = \begin{cases} 1, p_i > pj \\ 0, altfel \end{cases}$  conform Gadow și Hui  (1999),  Aguirre et al.

(2003)

$n$ – numărul de vecini pentru care se calculează valoarea  $vj$

$i$ – indicele arborelui de referință

$j$ - indicele unui arbore considerat a fi în primii $n$ vecini

$p_i$, $p_j$ – valoarea atributului dimensional analizat al arborelui $i$ și $j$

Valoarea indicelui de dominanță dimensională U este cuprinsă între 0 și 1, în funcție de proporția arborilor vecini cu dimensiuni superioare/inferioare arborelui de referință.

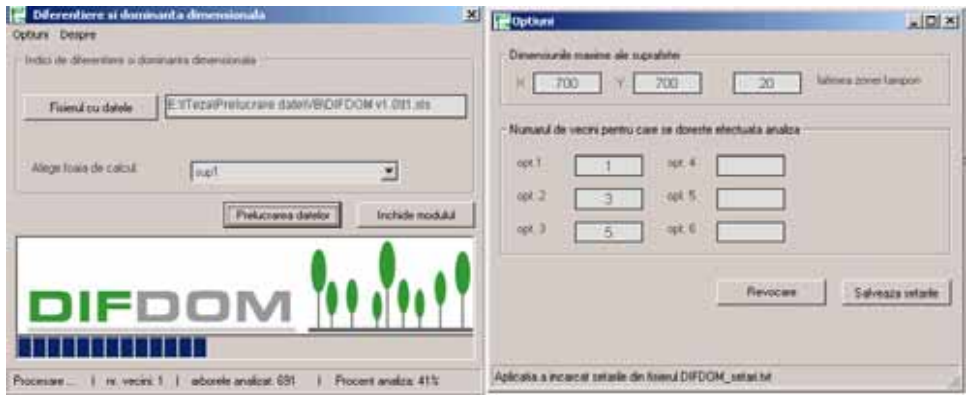

**Figura 6.4 Interfața utilizator a aplicației DIFDOM**

În cadrul acestei lucrări, diversitatea structurală a înălțimilor a fost analizată folosind indicii de diferențiere și dominanță dimensională prezentați anterior. Calculele au fost efectuate cu ajutorul unei aplicații informatice proprii, DIFDOM. Aplicația a fost creată în mediul de dezvoltare *Microsoft Visual Basic*, fiind un program independent ce importă datele din foi de calcul tabelar tip *Microsoft Excel*. Scopul dezvoltării acesteia a fost automatizarea calculelor ce devin complexe pentru un volum mare de indivizi, singura soluție alternativă de calcul a acestor indici fiind CRANCOD, un program realizat de către o echipă de cercetători ai Universității Bangor (Țara Galilor). Acest program este dezvoltat în *Java*, fiind dificil de instalat și utilizat de operatori nefamiliarizați.

Aplicația DIFDOM calculează indicii în baza unui număr al celor mai apropiați vecini introdus de către utilizator și include și un modul de corecție a erorilor generate de efectul de margine. Efectul de margine poate afecta rezultatele obținute datorită neincluderii în calcule a unor evenimente situate în afara ferestrei de observație, în literatura de specialitate fiind menționate mai multe metode de soluționare a acestui aspect (Diggle, 1983; Stoyan, Stoyan, 1994). În cazul de față s-a folosit metoda zonei tampon, fiind una din soluțiile recomandate de către Pommerening și Stoyan (2006) pentru eliminarea erorilor de calcul a acestui tip de indici. S-a folosit o zonă tampon cu o lățime de 30 cm (dimensiunea poate fi definită de către utilizator în aplicația DIFDOM). Această valoare a fost aleasă în urma unei analize a distanțelor medii dintre arbori și cei mai apropiați vecini. S-au determinat distanțele medii față de cei mai apropiați 5 vecini, media acestui parametru fiind de 21,98 cm. A fost luată în calcul nu doar media, ci si variația individuală a distanței față de cei mai apropiați 5 vecini, ajungând în final la valoarea de 30 cm. Valoarea indicatorilor s-a calculat în funcție de un număr variabil al celor mai apropiați $k$ vecini ($k$ = 1 - 5), fiind notat explicit pentru fiecare indice numărul de vecini considerați (e.g. T3 – indicele T calculat în funcție de cei mai apropiați 3 vecini).

Indicii de diferențiere și dominanță dimensională vor fi analizați atât prin prisma valorilor individuale (a distribuției acestora), cât și prin valorile medii înregistrate în fiecare suprafață de probă.

Distribuția indicelui de diferențiere a înălțimilor (T) se prezintă grafic în figura 6.5, pentru cele 5 variante ale numărului de vecini luați în calcul.

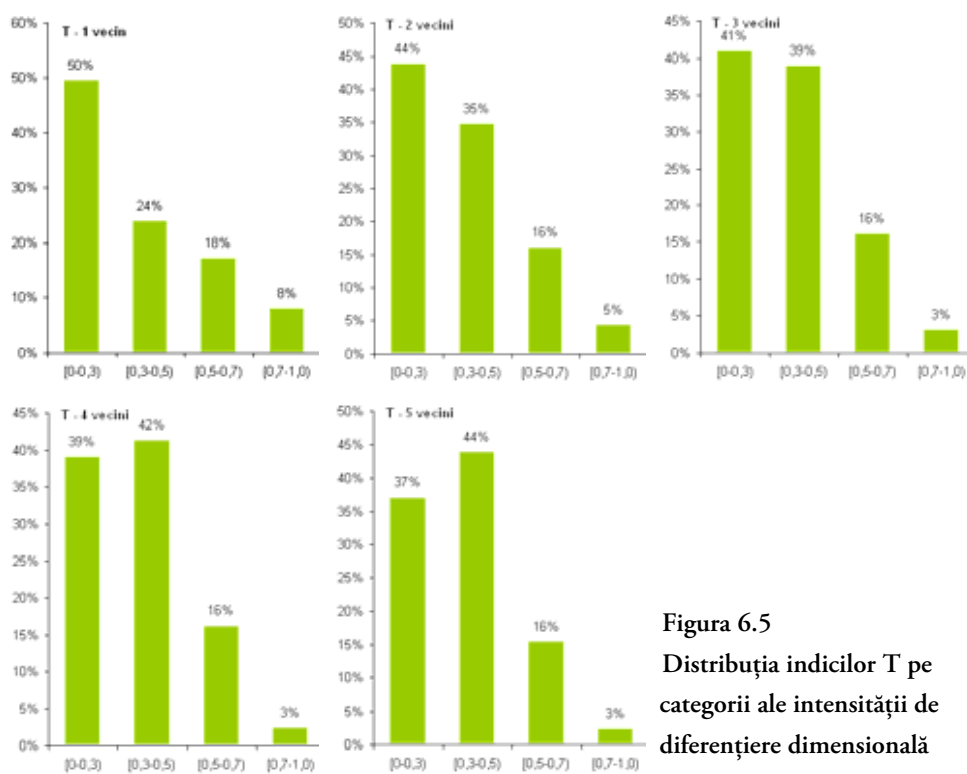

**Figura 6.5**
**Distribuția indicilor T pe categorii ale intensității de diferențiere dimensională**

Distribuțiile au fost realizate folosind scara de interpretare menționată de Pommerening (2000), prezentată anterior în cuprinsul acestui subcapitol.

Se remarcă faptul că ultimele două categorii (diferențiere puternică și foarte puternică) sunt foarte puțin influențate de numărul de vecini luat în calcul, proporția puieților încadrați în această categorie rămânând constantă. În grupurile celor mai apropiați 3-6 indivizi, proporția puieților cu dimensiunea minimă de sub 50% din dimensiunea vecinilor rămâne constant, la valori de 19-21%, nefiind influențată de dimensiunea grupului.

Putem vorbi astfel de o dinamică a proporției de reprezentare în cazul primelor două categorii – cu diferențierea slabă $(0 \leq T < 0,3)$ și diferențiere medie $(0,3 \leq T < 0,5)$. Proporția puieților încadrați în cea dintâi categorie scade odată cu creșterea numărului de vecini.

Odată cu creșterea numărului de puieți din grup proporția celor cu dimensiunea minimă de peste 70% din dimensiunea vecinilor se reduce (de la 50% în cazul unui grup de 2 puieți la 37% în cadrul unui grup de 6 puieți), crescând practic diversitatea structurală. Această reducere a proporției primei categorii de diferențiere se regăsește  în creșterea reprezentării celei de a doua categorii (diferențiere medie). Procentul puieților cu dimensiuni de 50-70% din dimensiunea puieților vecini crește de la 24% în cazul comparației cu cel mai apropiat vecin la 44% în cazul luării în calcul a celor mai apropiați 5 vecini. Se poate concluziona că per total diferențierea dimensională majoritară este slabă spre medie în grupurile de 2-6 indivizi, crescând odată cu mărirea grupului, dar nedepășind intensitatea medie.

În ceea ce privește valorile individuale ale indicilor de dominanță U var 1 (Hui et al., 1998) și U var 2 (Gadow și Hui, 1999;  Aguirre et al., 2003), acestea indică poziționarea unui arbore în ierarhia dimensională a grupului format dintr-un număr dat de puieți.

Cele două variante ale indicelui de dominanță sunt aproape complementare, diferența până la întreg fiind dată de puieții cu dimensiuni egale cu ale vecinilor.

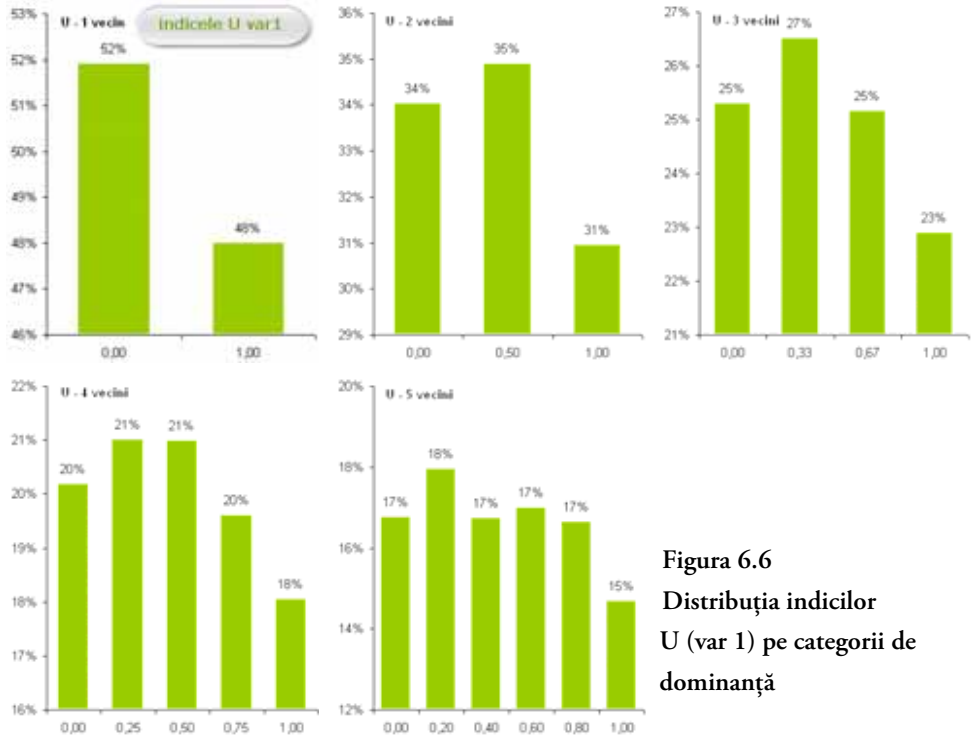

**Figura 6.6**
**Distribuția indicilor**
**U (var 1) pe categorii de**
**dominanță**

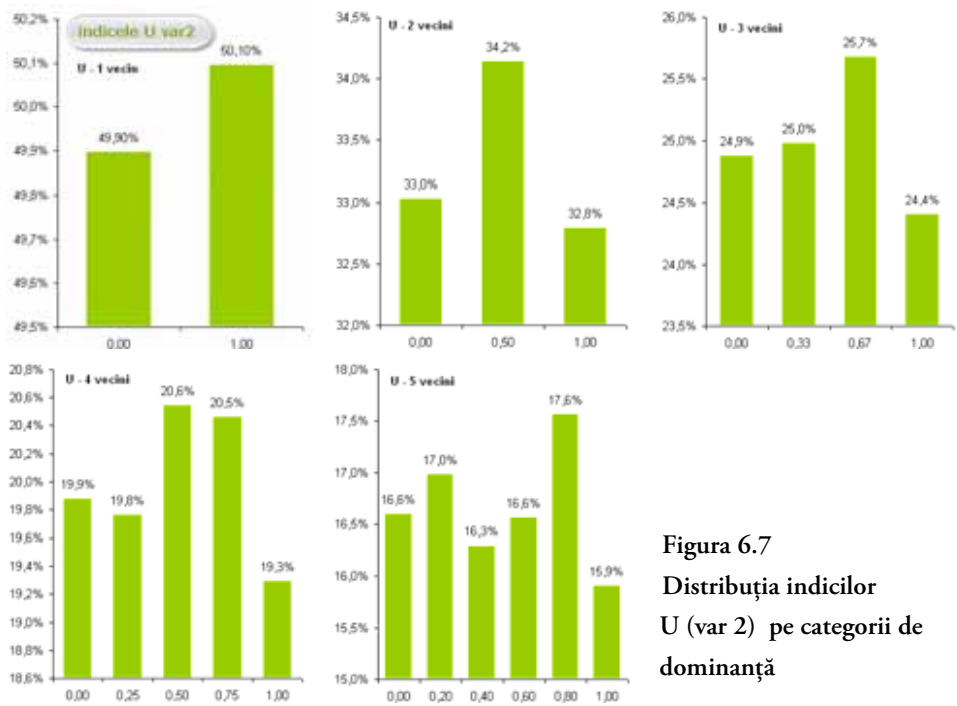

**Figura 6.7**
**Distribuția indicilor**
**U (var 2) pe categorii de**
**dominanță**

Indicele de dominanţă U var 1 indică procentual proporţia indivizilor vecini cu dimensiuni superioare puietului de referinţă iar U var 2 proporţia indivizilor vecini cu dimensiuni inferioare puietului de referinţă.

Valoarea informaţională a indicelui U de dominanţă este foarte ridicată mai ales la nivel individual, însă în cazul distribuţiilor sau a valorilor medii pe suprafaţă situaţia foarte echilibrată este greu interpretabilă (figura 6.6 şi 6.7).

În cazul primei variante a indicelui, analiza grafică dezvăluie o proporţie inferioară a situaţiilor în care puietul de referinţă este mai mic decât majoritatea vecinilor săi, în grupurile de 2-6 indivizi. Situaţia nu se oglindeşte într-o complementaritate perfectă faţă de distribuţia variantei a doua a indicelui, distribuţiile valorilor U var 2 fiind mult mai echilibrate, avantajul procentual în favoarea situaţiilor în care puietul de referinţă este mai mare decât majoritatea vecinilor fiind infinitezimal. Rezultatele indică o diversitate structurală după înălţime redusă, fapt ce confirmă observaţiile analizei diferenţierii dimensionale.

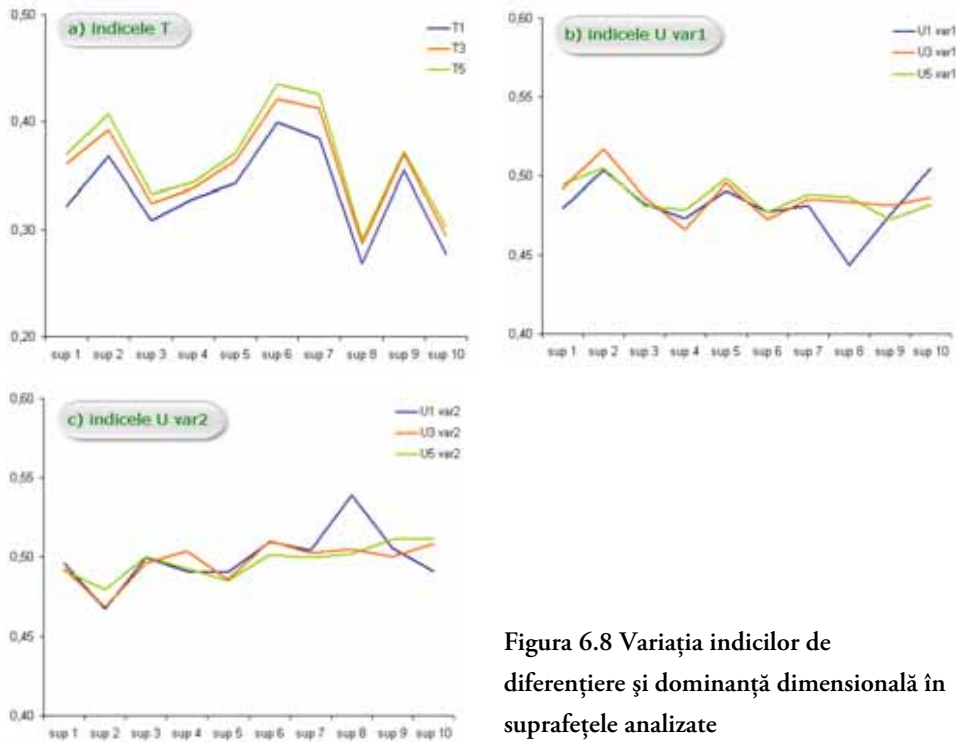

**Figura 6.8 Variaţia indicilor de diferenţiere şi dominanţă dimensională în suprafeţele analizate**

Au fost calculate și valorile medii pe fiecare suprafață, pentru a vedea în ce măsură indicii T și U pot să le caracterizeze, datele calculate pentru un număr de 1-5 vecini fiind prezentate detaliat în anexa 1.

Din analiza grafică a variației indicelui de diferențiere dimensională (fig. 6.8 a) în suprafețele studiate se observă că acesta este puțin influențat de numărul de vecini luat în calcul, dinamica valorilor T fiind similară. Sporirea numărului de vecini determină o creștere a valorilor indicelui, nedeterminând însă o încadrare într-o categorie diferită a intensității de diferențiere – se observă în cadrul majorității suprafețelor o încadrare în categoria de diferențiere medie, în suprafețele 8 și 10 chiar slabă.

Indicele de dominanță U este mult mai sensibil la schimbarea numărului de vecini considerați, numărul de puieți vecini având o pondere mult mai mare în relația de calcul a acestuia în ambele din cele două variante prezentate. Se remarcă în cazul indicelui de dominanță valorile mult diferite obținute pentru modul de calcul bazat doar pe cel mai apropiat vecin față de cazul utilizării unui număr superior de vecini. În acest sens se recomandă pentru creșterea valorii informaționale a acestui indice includerea unui număr mai mare de vecini, dar care să nu depășească numărul mediu de puieți dintr-o aglomerare. Cu ajutorul metodelor specifice analizei spațiale se poate determina pentru fiecare suprafață distanța pe care acționează fenomenele de agregare, și în baza distanței medii dintre puieți se poate evalua numărul mediu al puieților dintr-un pâlc. Această analiză va fi efectuată în următorul capitol al lucrării.

Pentru a explica  variația valorilor indicilor în cele zece suprafețe, s-au calculat coeficienții de corelație ai acestora cu înălțimea medie și desimea puieților pe suprafață (anexa 2). Rezultatele arată că indicele de diferențiere dimensională T nu este influențat de desimea puieților dar stabilește o relație negativă de intensitate medie (semnificativă doar la valorile calculate pentru 5 vecini: -0,564*) cu înălțimea medie.

În cazul indicelui de dominanță U rezultatele sunt destul de diferite pentru cele două variante folosite. Prima variantă a indicelui (U var 1) are o relație slabă cu desimea, fără a avea o semnificație statistică sub pragul de semnificație de 10% și o relație instabilă (foarte variabilă la valori diferite ale numărului de vecini) cu înălțimea medie a puieților – relație de intensitate medie, negativă, distinct semnificativă la utilizarea a 2 (-0,702**) și 5 vecini (-0,605**).

Cea de a doua variantă a indicelui (U var 2) se dovedește mult mai stabilă în relațiile cu cei doi parametri. Se înregistrează relații de intensitate medie, pozitive atât cu desimea cât și cu înălțimea medie a puieților (valori semnificative și distinct semnificative).

Dominanța este influențată mai intens atât de desime cât și de înălțimea medie a puieților, dar valorile dominanței pe suprafață trebuie interpretate cu precauție, valorile individuale fiind mai reprezentative. Diferențierea surprinde mai bine diversitatea structurală, fiind înregistrate corelații de intensitate slabă și medie cu principalii indici de evaluare a diversității structurale din prima grupă (Shannon, Simpson, Berger-Parker - anexa 3), dar include în plus o parte din informația spațială, ce poate fi completată cu datele referitoare la dominanță.

Majoritatea cercetărilor referitoare la diferențierea și dominanța dimensională se referă la arborete mature, una din puținele referințe bibliografice cu referire la analiza regenerării arboretelor fiind lucrarea lui Vorcak (et al., 2006). Ținând cont de această lipsă a informațiilor, cu atât mai mult se simte nevoia efectuării unui număr mai mare de cercetări în cazul regenerărilor, pentru a se putea realiza comparații ale indicilor obținuți și pentru a se identifica caracteristici specifice stadiilor tinere de dezvoltare în ceea ce privește diferențierea și dominanța dimensională.

## 6.3. Fundamentarea unui indice de apreciere a diversității structurale care include poziția în spațiu a puieților - IDIV

Majoritatea indicilor sintetici de apreciere a diversității structurale se bazează pe datele oferite de distribuțiile dimensionale ale arborilor. Deși au fost efectuate numeroase cercetări în domeniul variabilității spațiale a omogenității dimensionale (Kint et al., 2000; Vorcak et al., 2006; Mason et al., 2007; Davies, Pommerening, 2008), puțini autori au sugerat modalități de evaluare care să înglobeze direct nu doar caracteristicile biometrice ci și distribuția în spațiu a acestora. Dificultatea conceperii unor indicatori sintetici care să țină cont de caracterul spațial al datelor biometrice este determinată de modalitatea anevoioasă de prelevare a datelor spațiale (coordonatele arborilor într-un sistem de poziționare) precum și de prelucrările complexe ale datelor.

Pentru evaluarea diversității se folosesc în prezent analize bazate pe indicatori sintetici (e.g. Shannon), urmată de includerea rezultatelor obținute în modele de prelucrare care utilizează tehnici geostatistice sau specifice statisticii spațiale de ordinul doi (funcții tip K-Ripley). Unii cercetători (Gadow, 1993, 1997; Hui et al., 1998; Gadow, Hui, 1999; Pommerening, 2002; Aguirre et al., 2003) au încercat să soluționeze aspectele variabilității spațiale prin folosirea comparațiilor dintre vecini, pe baza raportului dintre dimensiunile acestora determinându-se un indice de diferențiere sau dominanță. Există variante diferite ale acestei metode, în funcție de modul de prelevare și prelucrare a datelor - determinat de folosirea coordonatelor sau a distanțelor dintre arbori. Dezavantajul acestei tehnici constă în faptul că nu se include direct în calculul indicilor componenta spațială.

O abordare superioară din punctul de vedere al fundamentării și justificării variabilității spațiale o au Zenner și Hibbs (2000). Modelul conceput de aceștia se bazează pe construirea unei rețele tridimensionale de triunghiuri obținute prin unirea vârfurilor arborilor, transpunerea în plan orizontal a acestei rețele fiind o triangulație Delauney. A fost definit un indice de complexitate structurală (SCI), ca raport între suma ariilor triunghiurilor care aparţin reţelei tridimensionale şi suma ariilor triunghiurilor din plan orizontal. SCI integrează foarte bine atât componenta orizontală cât şi cea verticală. Cu cât diferenţele în înălţime dintre arborii care formează triunghiurile sunt mai mari, cu atât raportul care furnizează valoarea indicelui sintetic va fi mai mare. SCI va înregistra valoarea minimă (egală cu 1) doar în cazul în care toţi arborii au aceeaşi înălţime. Utilitatea acestui indice este deosebită deoarece permite realizarea unor comparaţii ale structurii spaţiale între arborete (sau suprafeţe analizate).

Obiecţiile cu privire la folosirea acestui model de evaluare a diversităţii fac referire în primul rând la dificultatea calculelor. Algoritmii sunt dificil de implementat, soluţia nu oferă rezultate rapide, autorii nu au furnizat o aplicaţie software care să efectueze aceste prelucrări complexe, ceea ce face metoda inaccesibilă majorităţii cercetătorilor.

În vederea obţinerii rapide a informaţiilor privitoare la diversitatea spaţială a structurii, s-a fundamentat un nou indicator sintetic, numit *IDIV*. Modelul teoretic al acestuia se bazează pe descompunerea variabilităţii structurale în două componente IDIV_Ox, IDIV_Oy corespunzătoare celor două axe din planul orizontal după care se reprezintă poziţia arborilor.

Modelul teoretic prezentat în continuare consideră înălţimea variabila pentru care se doreşte calcularea diversităţii structurale, dar acestei variabile i se poate asocia orice parametru biometric.

Analiza se efectuează prin intermediul profilului vertical al arborilor, având drept referinţe două din cele patru laturi perpendiculare ale suprafeţei de probă (axa Ox şi Oy). Se unesc vârfurile arborilor învecinaţi prin segmente de dreaptă, se adună distanţele elementare obţinute şi se raportează la proiecţia distanţei dintre primul şi ultimul arbore din profil. Determinarea gradului de îndepărtare a variabilei înălţime de la omogenitatea în spaţiu se realizează prin intermediul raportului calculat anterior. În condiţiile omogenităţii perfecte, toţi arborii au aceeaşi înălţime, caz în care, indiferent de valoarea acesteia, se obţine valoarea 1 pentru indicele IDIV.

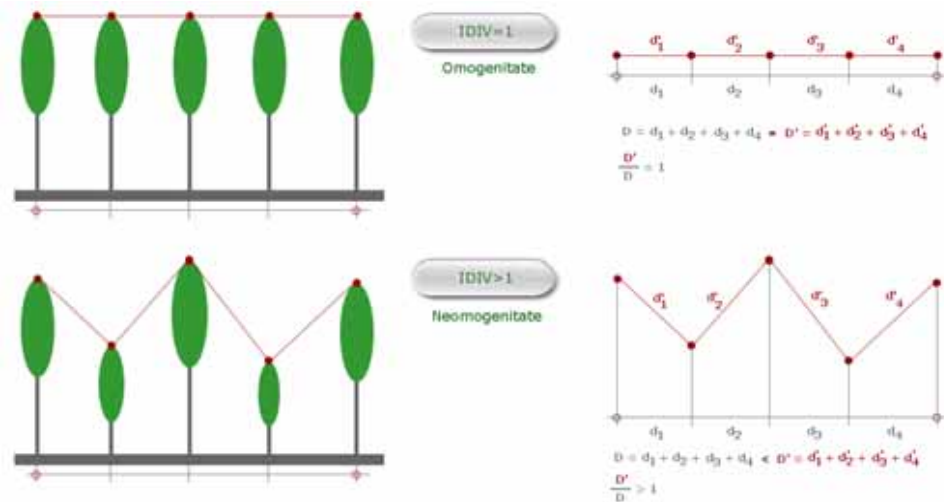

**Figura 6.9 Explicarea modalităţii de calcul a indicelui *IDIV***

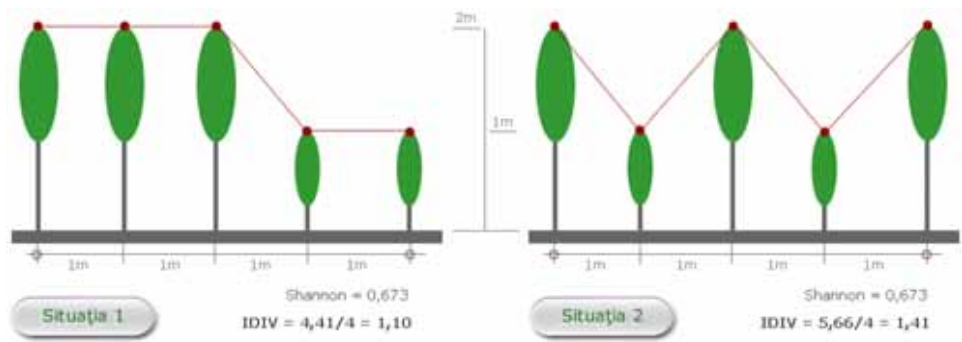

**Figura 6.10 Exemplu al influenţei poziţiei în spaţiu asupra diversităţii structurale**

În acest caz suma distanţelor obţinute prin unirea vârfurilor arborilor este egală cu suma distanţelor dintre arbori (se iau în calcul distanţele dintre arbori proiectate pe una din laturile suprafeţei). În cazul înregistrării unor diferenţe între înălţimile arborilor învecinaţi distanţele dintre vârfurile arborilor vor fi superioare distanţelor dintre arbori şi implicit raportul dintre cele două sume va fi supraunitar. Cu cât diferenţele dintre înălţimile arborilor învecinaţi vor fi mai mari, cu atât mai mare va fi şi raportul IDIV, indicând o diversitate structurală superioară ca valoare (figura 6.9).

Indicele IDIV nu este sensibil doar la diferențele dintre valorile biometrice ale arborilor, în formula sa intervenind și diferența dintre poziția arborilor. În figura 6.10 se prezintă două situații ipotetice, în care 5 puieți sunt poziționați diferit în spațiu. Dimensiunile lor sunt identice în ambele situații, având trei puieți cu înălțimea de 2m și doi puieți cu înălțimea de 1m, iar distanța dintre puieți este de 1m. Pentru ambele situații prezentate, indicii sintetici de estimare a diversității structurale (în exemplu s-a folosit indicele Shannon) indică aceeași valoare, neidentificând diferențe între cele două cazuri. Indicele IDIV face o distingere între cele două situații prezentate, indicând valori diferite ale diversității structurale – 1,10, respectiv 1,41. Situația a doua este identificată ca având o diversitate structurală spațială superioară primei variante, datorită modului de poziționare în spațiu a puieților.

Influența distanței dintre puieți se observă și din formula de calcul a indicelui IDIV:

$$IDIV = \frac{\sum_{i=1}^{n-1}\sqrt{(\Delta h_{i,i+1}^2 + \Delta s_{i,i+1}^2)}}{\left| s_n - s_1 \right|} = \frac{\sum_{i=1}^{n-1}\sqrt{((h_i - h_{i+1})^2 + (s_i - s_{i+1})^2)}}{\left| s_n - s_1 \right|} \qquad (6.12)$$

unde:

n – numărul de puieți analizat

$\Delta h$ – diferența de înălțime dintre doi puieți vecini

$h_i$ – înălțimea puietului i

$\Delta s$ – distanța dintre puieți proiectată pe una din axele sistemului de coordonate

$s_i$ – fiind proiecția distanței față de originea sistemului de coordonate a puietului i

Valoarea teoretică minimă a IDIV este 1 și se obține în cazul în care toți puieții au aceeași înălțime. În cazul în care se admit variații ale parametrului analizat (diferențe între înălțimile puieților) în baza relației de calcul a indicelui IDIV putem defini minimul teoretic spre care se poate tinde, în funcție de înălțimea maximă $h_{max}$ și cea minimă $h_{min}$ ca fiind :

$$IDIV\_MIN = \frac{\sqrt{(h_{max} - h_{min})^2 + (s_n - s_1)^2}}{|s_n - s_1|} \qquad (6.13)$$

Valoarea minimă teoretică ar putea fi obținută doar în cazul în care toți puieții având înălțimea egală cu $h_{max}$ sunt situați la o extremitate și toți puieții având înălțimea egală cu $h_{min}$ sunt situați la cealaltă extremitate a suprafeței considerate.

În mod similar se poate defini maximul teoretic spre care se poate tinde, în funcție de înălțimea maximă $h_{max}$ și cea minimă $h_{min}$ ca fiind :

$$IDIV\_MAX = (n-2) \cdot (h_{max} - h_{min}) + \frac{\sqrt{(h_{max} - h_{min})^2 + (s_n - s_1)^2}}{|s_n - s_1|} \qquad (6.14)$$

În ambele formule se folosesc aceleași notații folosite la prezentarea modului de calcul al indicelui IDIV.

Valoarea maximă teoretică este tot o valoare ideală ce ar putea fi obținută doar în cazul în care jumătate din puieți ar avea înălțimea egală cu $h_{max}$ iar cealaltă jumătate înălțimea egală cu $h_{min}$. În plus ar fi nevoie ca în una din extremități să se găsească un singur puiet iar toți ceilalți să se găsească în extremitatea opusă. Valorile extreme teoretice ale indicelui IDIV sunt în realitate greu sau chiar imposibil de atins, aceste valori fiind obținute prin ignorarea poziției reale a puieților și plasarea acestora în poziții convenabile pentru maximizarea sau minimizarea unei valori. Ele oferă însă indicii asupra limitelor de variație a indicelui IDIV, ținând cont de distanța dintre puieții cei mai îndepărtați și de înălțimea minimă și maximă a puieților din suprafața studiată.

Pentru a reduce intervalul de variație al indicelui se poate calcula un minim, respectiv maxim, care să includă pozițiile individuale reale ale puieților și valorile individuale corespunzătoare înălțimii și în acest fel să exprime mai bine extremele posibile ale diversității structurale pentru parametrul studiat. În vederea calculului limitelor diversității în acest mod se face o rearanjare a puieților, schimbând poziția acestora între ei pentru a obține diferențe ale variabilei analizate minime, respectiv maxime între puieții vecini. Valoarea minimă (IDIV_min) se obține în acest caz prin ordonarea vectorului înălțimilor astfel încât să se minimizeze diferența dintre oricare doi puieți vecini, iar valoarea maximă (IDIV_max) se obține prin plasarea pe poziții vecine în vectorul înălțimilor a valorilor extreme pentru a obține cele mai mari diferențe, atașând modelului de calcul cele mai mari diferențe de înălțime celor mai mici diferențe de poziție. Valorile extreme indică diversitatea posibilă de obținut în condițiile păstrării acelorași valori ale înălțimii și al aceluiași model de organizare spațială a indivizilor.

Indicele IDIV este sensibil la spațialitate, valoarea sa înglobând influența desimii puieților. În condițiile unor distanțe reduse între puieți efectul diferențierii dimensionale se resimte mai puternic, fapt ce corespunde realităților ecologice din interiorul unei populații de arbori. Datorită acestei proprietăți se poate aprecia că IDIV este capabil să surprindă intensitatea proceselor concurențiale la nivelul unei întregi suprafețe (fig. 6.6).

Analiza presupune aplicarea aceleiași metode și pentru cealaltă axă, în final fiind obținute două valori ale diversității structurale spațiale (IDIV_Ox și IDIV_Oy) ce caracterizează aceeași suprafață, pe cele două componente din planul orizontal.

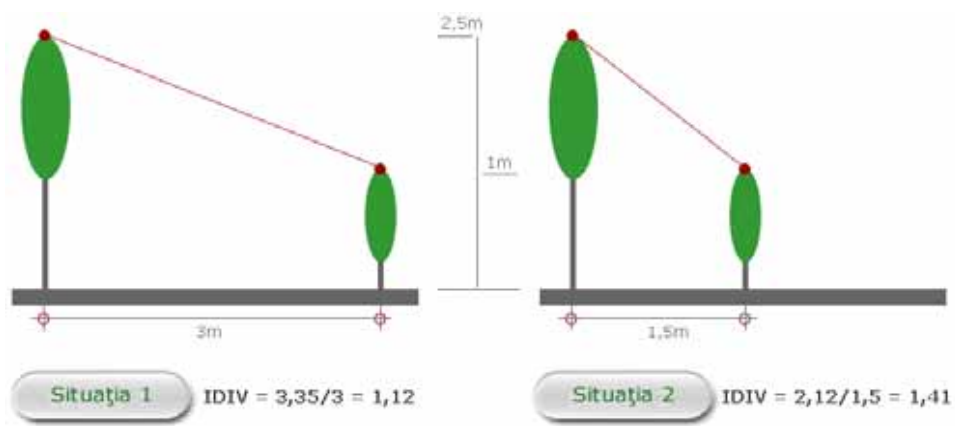

**Figura 6.11 Exemplu al influenței distanței dintre indivizi asupra diversității structurale**

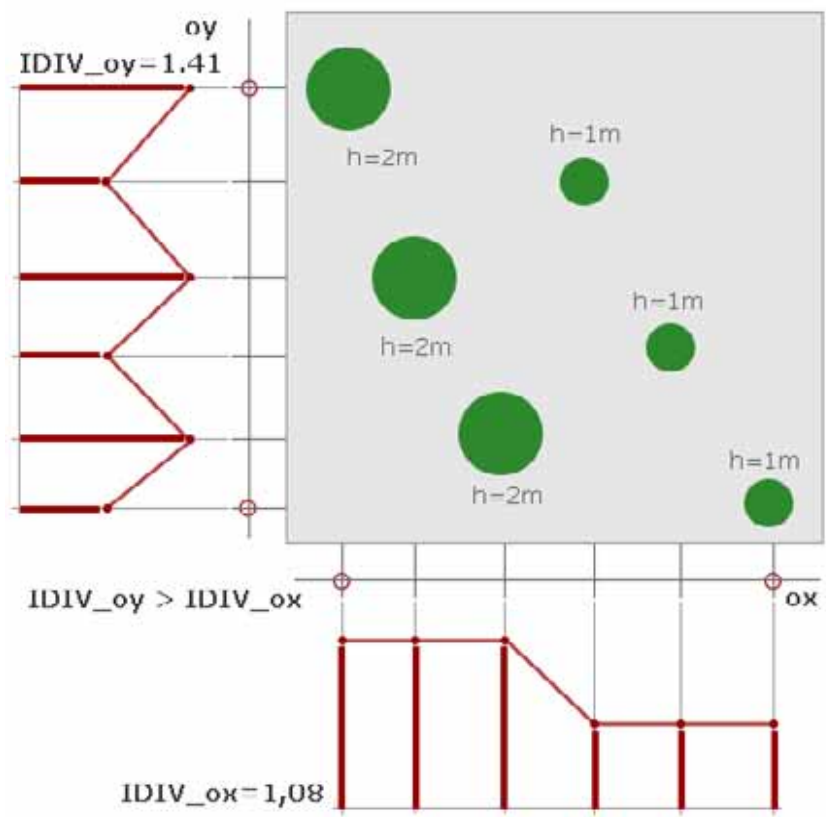

**Figura 6.12 Analiza fenomenelor direcționale prin compararea valorilor IDIV obținute pe fiecare din cele două axe**

În cazul în care se constată diferențe majore între cele două componente se poate considera că în suprafața respectivă acționează fenomene direcționale care provoacă dezechilibre între valorile obținute pe fiecare dintre axe (exemplu în figura 6.12).

În vederea aprecierii indicelui IDIV pentru o întreagă suprafață, pentru fiecare componentă Ox, respectiv Oy se va proceda la o împărțire în benzi de aceeași lățime. Scopul acestei împărțiri constă în evitarea comparațiilor între arbori foarte îndepărtați dar care ar avea proiecțiile pe una din axe foarte apropiate și ar putea fi considerați vecini deși ei au apropiate valorile pozițiilor doar pe una din axele Ox sau Oy.

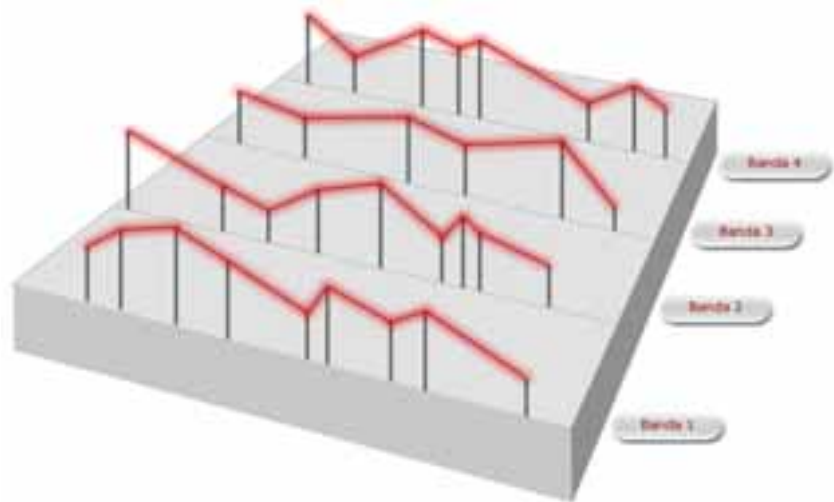

**Figura 6.13 Împărţirea unei suprafeţe în benzi în vederea calculării IDIV**

Metoda presupune adunarea sumelor distanţelor dintre vârfurile arborilor pe benzi şi raportarea la valoarea însumată a sumelor distanţelor dintre puieţi proiectate pe axa pe care se face analiza. În acest mod se obţine un indice capabil să caracterizeze o suprafaţă întreagă.

Un alt avantaj al acestei împărţiri constă în posibilitatea efectuării unor comparaţii între diferitele benzi, astfel că se pot determina diferenţe ale diversităţii structurale chiar în interiorul aceleiaşi suprafeţe.

Determinarea pasului optim folosit pentru împărţirea suprafeţei în benzi de lăţime egală nu se poate face prin introducerea unei relaţii de calcul. O lăţime prea mică a benzilor (un pas mic) poate determina neînregistrarea unor particularităţi observabile doar în cazul utilizării unor lăţimi mai mari ale benzilor. Pe de altă parte o lăţime prea mare poate determina o uniformizare a diversităţii la nivelul suprafeţei analizate sau face imposibilă analiza diferenţelor structurale din interiorul suprafeţei. De aceea pasul de observaţie trebuie ales în funcţie de scopul cercetărilor sau se poate opta pentru o analiză diferenţiată pe valori diferite ale pasului.

Un punct de plecare în alegerea unei mărimi adecvate a pasului îl constituie distanța medie dintre puieți. Aceasta poate fi calculată ca medie a distanței medii față de un număr limitat de vecini (în general în analizele bazate pe distanța medie dintre indivizi sunt folosiți cei mai apropiați 3-5 vecini). În cazul unor suprafețe în care s-a remarcat un grad ridicat de omogenitate pasul optim poate fi ales acel pas pentru care s-a obținut o diferență minimă între componentele axiale ale IDIV.

În final, se poate aprecia că această metodă de apreciere a diversității structurale prezintă avantaje reale:

- eliminarea aprecierii bazate pe încadrarea în clase, un proces de multe ori inconsistent ce poate altera rezultatele în cazul folosirii unor indicatori clasici (e.g. *Shannon, Simpson*);

- modelul clar, ușor algoritmizabil;

- posibilitatea efectuării analizelor atât în interiorul suprafețelor studiate cât și a comparației între suprafețe diferite;

- posibilitatea identificării unor fenomene direcționale;

- oferirea de informații cu privire la competiția din interiorul suprafeței studiate

- adaptabilitatea modelului poate conduce la folosirea unor parametri biometrici diferiți drept variabilă pe ordonată (diametrul, diametrul coroanei etc.);

- aplicabilitatea extinsă a modelului (aplicabil nu doar puieților ci și arborilor).

## 6.4. Analiza diversităţii structurale a seminţişului din suprafeţele studiate cu ajutorul indicelui IDIV

În conformitate cu obiectivele cercetărilor, de integrare a mijloacelor cibernetice în studiul regenerării, s-a conceput un program informatic care realizează prelucrarea automată a datelor în vederea obţinerii valorii indicelui IDIV, fundamentat teoretic în subcapitolul anterior.

Aplicaţia software IDIV (figura 6.14) este o soluţie proprie, concepută pentru analiza diversităţii structurale în spaţiu conform unei metode proprii. Programul foloseşte datele introduse în foi de calcul de tip *Microsoft Excel* şi calculează în baza acestora o serie de indicatori - valoarea *IDIV* pe fiecare componentă axială (Ox, Oy) şi pe fiecare bandă, valoarea minimă (IDIV_min) şi maximă (IDIV_max), valorile teoretice extreme, coeficienţii de variaţie ai *IDIV* pe suprafaţă. Se calculează şi media distanţelor medii ale puieţilor faţă de un număr de 3, 5 şi 7 vecini, în vederea alegerii unui pas optim al analizei.

Datele de intrare sunt coordonatele puieţilor şi înălţimea totală a fiecărui puiet. În locul înălţimii puieţilor se poate folosi orice alt parametru biometric – diametrul la colet, diametrul, volumul sau suprafaţa exterioară a coroanei. Utilizatorul trebuie să specifice mărimea suprafeţei analizate, precum şi paşii la care se doreşte efectuarea analizei, date care pot fi salvate într-un fişier de setări iniţiale.

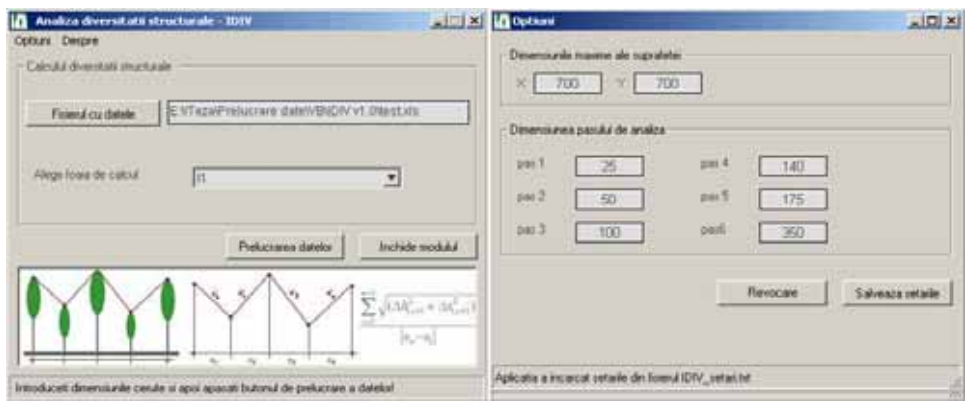

**Figura 6.14 Interfaţa utilizator a aplicaţiei IDIV**

Programul IDIV a fost creat folosind mediul de dezvoltare *Microsoft Visual Basic*, rezultatul fiind un fișier executabil independent. Integrarea aplicației cu o soluție performantă de calcul tabelar, interfața grafică intuitivă, ușurința în utilizare, precum și rapiditatea prelucrării unui volum mare de date sunt câteva din avantajele folosirii programului IDIV în procesul de analiză a datelor.

Analiza diversității structurale în spațiu s-a realizat pentru toate cele 10 suprafețe, fiind folosită aplicația software IDIV pentru calcularea valorilor indicatorilor sintetici. S-a efectuat atât o analiză a diversității structurii în interiorul suprafețelor (prin examinarea valorilor benzilor) cât și o comparație între suprafețele de probă.

În ceea ce privește parametrul considerat a fi studiat prin această metodă, s-a apreciat că înălțimea este variabila care reprezintă cea mai potrivită alegere. Deși metoda fundamentată în subcapitolul precedent se pretează la analiza unor variabile diferite, s-a optat pentru înălțime datorită analizelor desfășurate anterior în cadrul lucrării care au evidențiat importanța acesteia în definirea relațiilor dintre puieți.

Au fost calculați indicatorii specifici pentru fiecare componentă axială, utilizându-se trei variante ale pasului de observație, la o lățime a benzii de 25, 50 și 100 de cm. Efectuarea unei analize în pași multipli permite decelarea unor aspecte legate de influența scării de studiu asupra aprecierii diversității structurale a unui parametru. Se poate stabili, de asemenea, un pas optim, adecvat obiectivelor cercetărilor. Valoarea pașilor a fost adoptată în urma calculului pentru fiecare suprafață a mediei distanței medii a puieților față de primii 3, 5 și 7 puieți vecini. Numărul de vecini adoptat de regulă în cercetările ce implică calculul distanței dintre indivizi este situat între valorile menționate anterior (Pommerening, 2002; Davies, Pommerening, 2008). Distanțele medii până la primii 3 și 5 vecini sunt în general situate în intervalul 20-25 cm, în timp ce pentru 7 vecini distanța medie depășește 25 de cm. Datorită acestui fapt sunt prezentate în acest subcapitol doar graficele și analizele pentru un pas de 25 și 50 cm, în anexa 4 fiind prezentate graficele individuale, iar în anexa 5 valorile indicatorilor pentru fiecare suprafață, în cele trei variante ale lățimii benzii (figura 6.15).

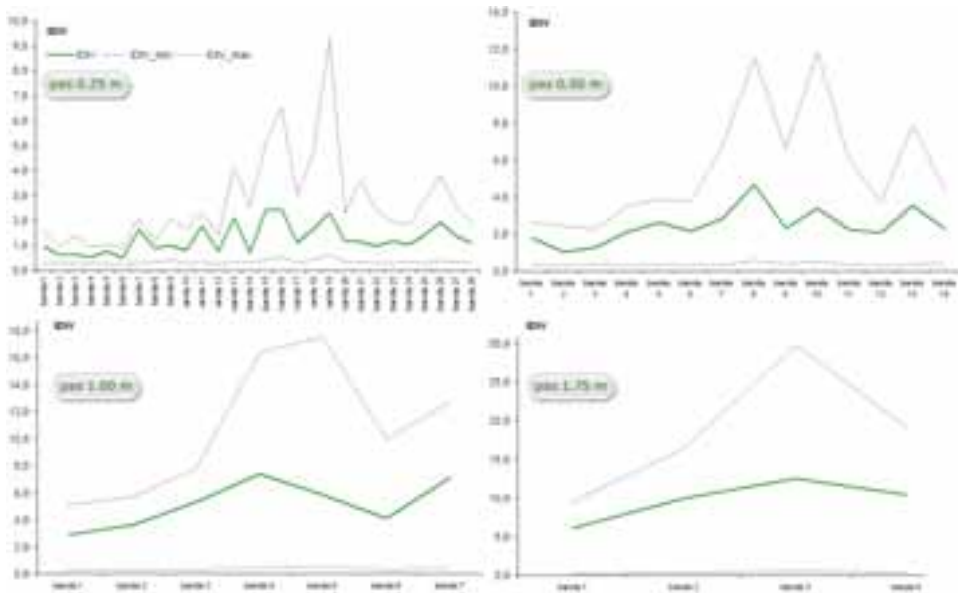

**Figura 6.15 Exemplu privind efectul creşterii pasului de analiză
asupra valorilor IDIV**

La creşterea pasului de analiză se observă o stabilizare (reducere a diferenţelor) a valorilor IDIV pe fiecare bandă, în cadrul unei suprafeţe (în primul rând datorită reducerii numărului de benzi), dar şi o creştere a valorilor IDIV. Valorile limitate ale diversităţii înregistrate pentru lăţimile mici ale benzilor conduc la ipoteza că pe distanţe reduse apare o agregare a puieţilor de dimensiuni apropiate. În cazul creşterii distanţelor sporeşte şi posibilitatea de a înregistra diferenţe dimensionale superioare. Ipoteza agregării puieţilor este testată şi prezentată în cuprinsul capitolului 7 prin tehnici specifice analizei spaţiale.

Nu s-a realizat calculul valorilor individuale ale IDIV pentru fiecare specie deoarece s-a considerat că analiza ar forţa modelul propus. S-a apreciat că valorile ce ar putea fi astfel obţinute nu ar fi relevante, neavând acoperire în realitate datorită faptului că puieţii de aceeaşi specie, de cele mai multe ori nu sunt vecini, deşi în cadrul analizei ar fi consideraţi astfel. În această situaţie indicele IDIV ar putea oferi informaţii asupra diversităţii structurale specifice, dar componenta spaţială a acestuia ar fi inconsistentă.

a) **Analiza diversităţii structurale în interiorul suprafeţelor studiate**

În continuare sunt prezentate graficele variaţiei valorii indicelui IDIV pentru cele 10 suprafeţe analizate, folosind paşii de analiză de 25, respectiv 50 de cm. Este prezentată situaţia separat pe cele două componente axiale (IDIV_Ox şi IDIV_Oy) (figura 6.16, 6.17).

În grafice sunt prezentate valorile indicelui IDIV pe fiecare componentă axială, precum şi valorile minime şi maxime (IDIV_min, IDIV_max) care se pot obţine în condiţiile în care se face o rearanjare a puieţilor folosind aceleaşi coordonate spaţiale (se schimbă poziţia puieţilor între ei). Valorile minime şi maxime oferă posibilitatea stabilirii gradului de apropiere a indicelui IDIV faţă de valorile extreme ale diversităţii structurale, oglindind influenţa distribuţiei spaţiale asupra diversităţii.

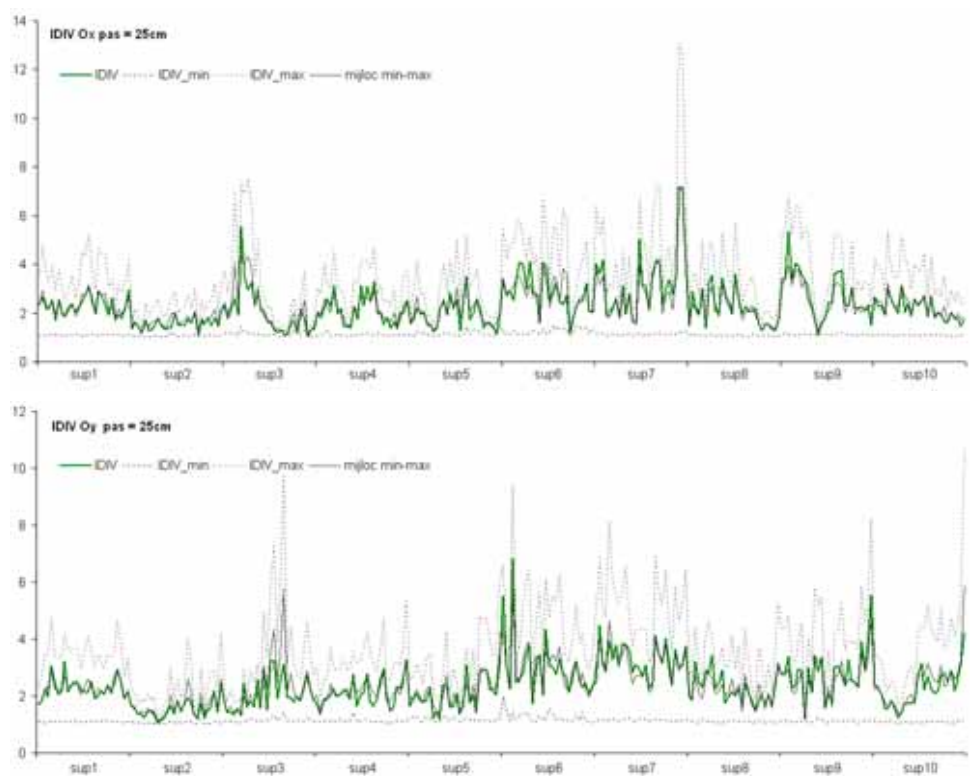

**Figura 6.16 Variaţia valorilor IDIV în cadrul benzilor din suprafeţele analizate**
(lăţimea benzilor = 25 cm)

Atât în cazul calculării IDIV_min cât şi a IDIV_max sunt folosiţi aceeaşi puieţi, cu aceleaşi caracteristici dimensionale, diferenţa dintre valorile calculate fiind atribuită exclusiv distribuţiei în spaţiu.

Apropierea IDIV de valoarea minimă sau maximă s-a studiat atât grafic, cât şi prin intermediul unor indicatori. Se apreciază că dacă valoarea IDIV depăşeşte valoarea medie a intervalului dintre minim şi maxim se apropie mai mult de valoarea IDIV_max, iar dacă este sub acest prag se apropie mai mult de IDIV_min. Abaterea IDIV de la referinţa considerată s-a calculat pentru toate valorile individuale. S-a definit un raport (notat β) între valoarea IDIV şi valoarea referinţă (jumătatea intervalului minim-maxim) ce exprimă nu doar apropierea de una din cele două extremităţi, ci şi gradul de apropiere.

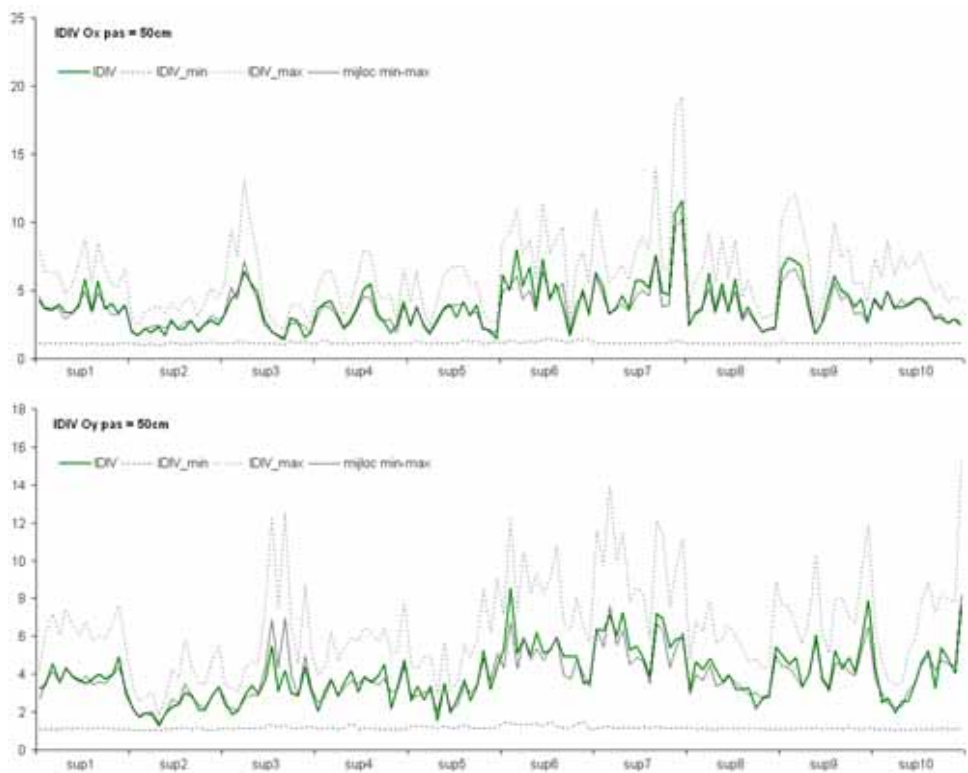

**Figura 6.17 Variaţia valorilor IDIV în cadrul benzilor din suprafeţele analizate**
**(lăţimea benzilor = 50 cm)**

**Tabelul 6.6**

**Valori utile în determinarea gradului de apropiere a indicelui IDIV de extreme**

| | Valori mai apropiate de IDIV_max | Valori mai apropiate de IDIV_min | Media abaterilor individuale de la referință | Raportul β |
|---|---|---|---|---|
| pas 25 cm comp. Ox | 136 (49%) | 144 (51%) | - 0.0118 | 0.990 |
| pas 25 cm comp. Oy | 125 (45%) | 155 (55%) | - 0.0369 | 0.986 |
| pas 50 cm comp. Ox | 90 (64%) | 50 (36%) | 0.1860 | 1.037 |
| pas 50 cm comp. Oy | 85 (61%) | 55 (39%) | 0.1002 | 1.022 |

Valorile obținute în cazul variantei lățimii benzii de 25 cm arată în general o situație echilibrată, cu o ușoară tendință de deplasare spre minim - 55%, respectiv 51% din indicii IDIV corespunzători fiecărei componente axiale fiind mai apropiați de minim. La utilizarea unui pas de 50 cm se constată o tendință mai pronunțată de apropiere de valoarea maximă, peste 60% din totalul indicilor IDIV fiind mai apropiați de maxim în cazul ambelor componente axiale. Această tendință se păstrează și în cazul utilizării unei lățimi a benzii de 100 cm (anexa 4 c). Rezultatele dovedesc încă odată influența scării de analiză în cercetările ce implică distribuția spațială a evenimentelor (Hurlbert, 1990; Dale, 2004; Tokola, 2004; Zenner, 2005). Tendința de maximizare a diversității structurale odată cu creșterea suprafeței de analiză poate fi explicată prin aceeași ipoteză de agregare a puieților pe spații reduse și creștere a heterogenității datorită diferențelor dintre aceste grupuri odată cu lărgirea perspectivei de analiză.

În ceea ce privește omogenitatea valorilor IDIV calculate pentru benzile aceleiași componente axiale se remarcă o variabilitate relativ redusă. În cazul pasului de analiză de 25 cm, majoritatea coeficienților de variație sunt situați între 20-25% (anexa 5), doar patru suprafețe înregistrând valori medii ce depășesc cu puțin pragul de 30%, suprafața numărul 3 înregistrând cea mai crescută heterogenitate (37%). Situaţia este similară şi în cazul utilizării unui pas de 50 cm, respectiv de 100 cm. Acest fapt rezolvă problema reprezentativităţii mediei valorilor individuale ale IDIV, indicele IDIV pentru o componentă axială (Ox, Oy) fiind foarte apropiat ca valoare şi modalitate de calcul de media valorilor obţinute pentru fiecare bandă a suprafeţei studiate.

Un alt aspect se referă la diferenţele dintre valorile IDIV înregistrate pentru aceeaşi suprafaţă, dar pentru cele două componente axiale. Sunt diferenţe majore între componenta Ox şi cea Oy? Se poate utiliza media celor două valori pentru a caracteriza o suprafaţă? În figura 6.18 sunt prezentate diferenţele procentuale dintre indicii IDIV pe componente axiale, în varianta paşilor de analiză de 25 şi 50 cm.

Diferenţele procentuale înregistrate între cele două componente axiale sunt reduse, în unele suprafeţe (nr. 2, 4, 7, 9) fiind sub 2%, pentru ambele variante ale pasului de analiză. În suprafeţele nr. 5 şi 10 se realizează maximul diferenţelor dintre componentele axiale, fiind singurele suprafeţe cu un decalaj de peste 5% între IDIV_Ox şi IDIV_Oy. Rezultatele obţinute permit considerarea drept reprezentativă a mediei dintre valorile IDIV pentru cele două componente axiale, cel puţin în cazul suprafeţelor studiate.

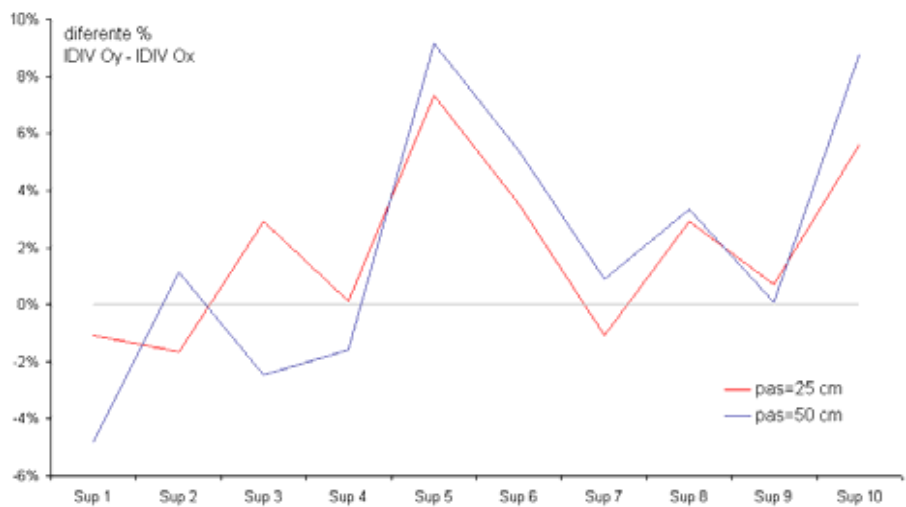

**Figura 6.18 Diferenţele procentuale între valorile IDIV ale componentelor axiale**

Înregistrarea unor diferențe mari între indicii componentelor axiale ale aceleiași suprafețe poate să conducă la ipoteza surprinderii unor fenomene direcționale. Referitor la această posibilitate nu au fost observate cu ocazia prelevării datelor astfel de fenomene și nici diferențele dintre valorile IDIV_Ox și IDIV_Oy nu permit formularea unei astfel de ipoteze. Trebuie menționat că în teren coordonatele corespunzătoare axei Ox au fost plasate pe direcția N-S iar cele corespunzătoare Oy au fost plasate pe direcția E-V.

Un ultim aspect abordat în cadrul analizei diversității structurale în interiorul suprafețelor de probă se referă la legătura dintre indicele IDIV și desimea puieților. Factorul spațial poate acționa asupra structurii semințișului și prin desime, iar valoarea indicelui IDIV ar trebui să surprindă acest fapt. Drept urmare se prezintă valorile coeficienților de corelație dintre indicii individuali IDIV, caracteristici fiecărei benzi și desimea puieților din banda corespunzătoare (tabelul 6.7).

Au fost calculați coeficienții de corelație dintre desime și valoarea IDIV, IDIV_min și IDIV_max pentru a vedea în ce măsură desimea puieților afectează nu doar diversitatea structurală reală ci și cea potențială, exprimată prin limitele minime și maxime. Este interesant de observat că deși se înregistrează corelații pozitive, de intensitate medie și chiar puternică între desime și IDIV, respectiv IDIV_max (coeficienți de corelație foarte semnificativi), nu există o corelație nici măcar slabă cu IDIV_min.

Tabelul 6.7

**Coeficienții de corelație dintre valorile individuale IDIV și desimea puieților**

|  | Ox, pas 25 cm | Oy, pas 25 cm | Ox, pas 50 cm | Oy, pas 50 cm |
|---|---|---|---|---|
| IDIV - desime | 0,770 *** | 0,650 *** | 0,819 *** | 0,718 *** |
| IDIV_min - desime | -0,005 | -0,007 | -0,008 | -0,058 |
| IDIV_max - desime | 0,786 *** | 0,739 *** | 0,802 *** | 0,750 *** |

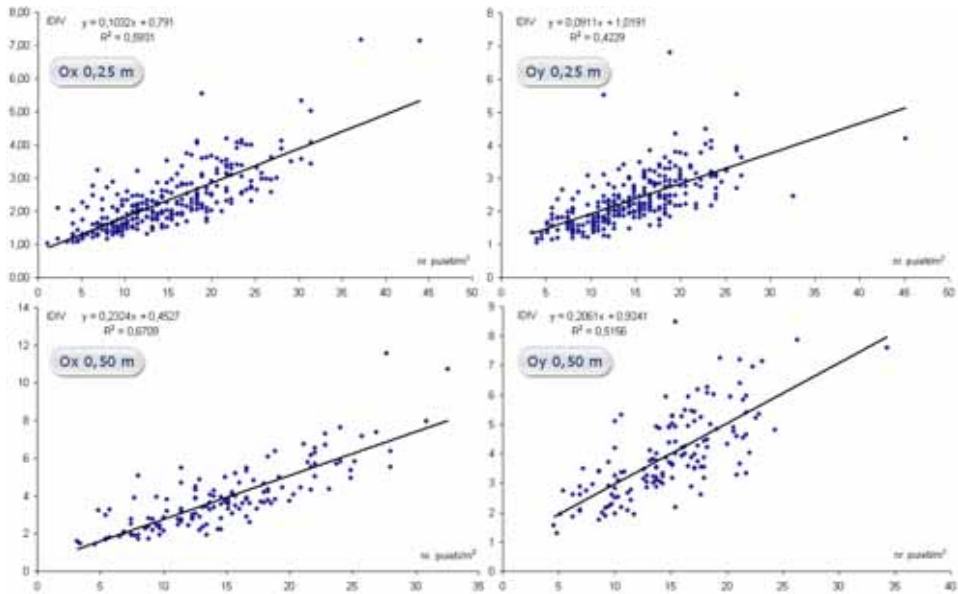

**Figura 6.19 Legătura corelativă dintre valorile pe fiecare bandă ale IDIV şi desimea puieţilor**

Este evident că IDIV are o componentă neafectată de desime – la aceeaşi diversitate structurală (aceleaşi valori ale înălţimilor) şi aceeaşi dispunere spaţială (implicit aceeaşi desime), se pot obţine valori ale IDIV diferite cuprinse între IDIV_min şi IDIV_max. Totuşi, valoarea limită minimă a IDIV, deşi ca model de calcul ţine cont de poziţia în spaţiu şi înălţimea aceloraşi puieţi, este insensibilă la variaţia desimii. Tendinţa de reducere a diferenţelor dintre înălţimile indivizilor conduce la acordarea unei ponderi mai mici distanţelor dintre puieţi (desimii).

Din graficele 6.20 şi 6.21 se observă că valorile IDIV_min prezintă o anumită variabilitate determinată de banda şi suprafaţa pentru care sunt calculate, graficele fiind mult mai expresive atunci când sunt analizate folosind o scară proprie şi nu o scară comună şi valorilor IDIV_max. Această observaţie ne permite practic să apreciem că se poate folosi pentru comparaţiile între benzi sau suprafeţe capacitatea discriminatorie a indicelui IDIV_min, în toate cazurile în care se doreşte eliminarea influenţei desimii puieţilor.

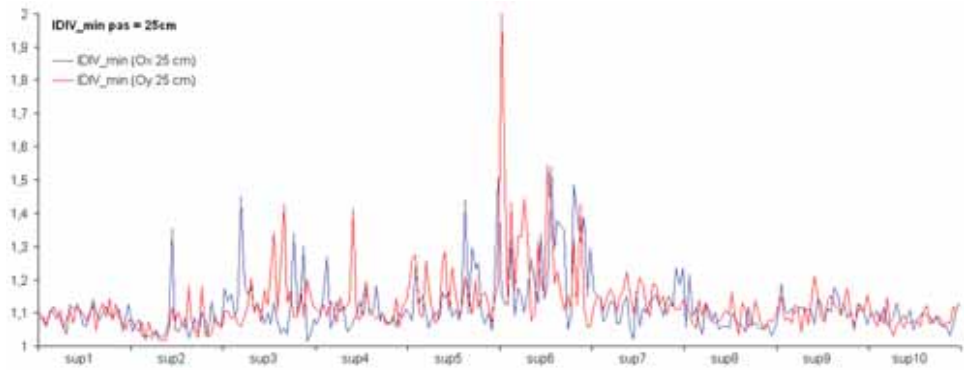

**Figura 6.20 Variația IDIV_min în cadrul benzilor din suprafețele analizate (pas 25 cm)**

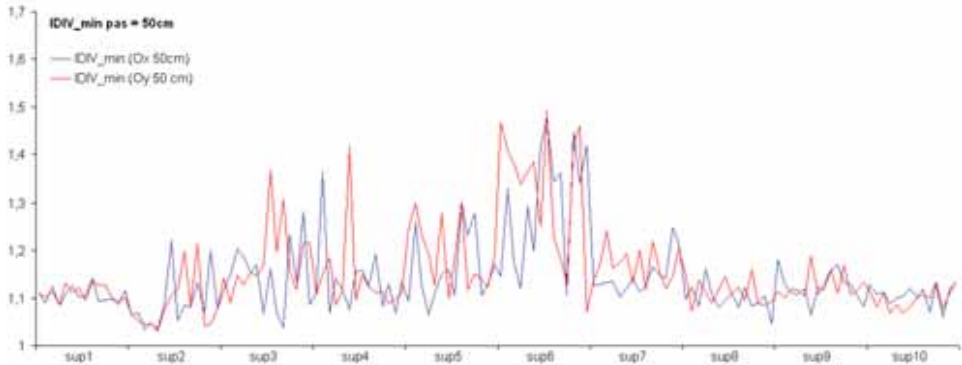

**Figura 6.21 Variația IDIV_min în cadrul benzilor din suprafețele analizate (pas 50 cm)**

Valoarea minimă potențială a IDIV oferă o evaluare a diversității structurale a unui parametru dar omite o mare parte din informațiile legate de distribuția în spațiu, reprezentând un compromis care poate fi acceptat de către cercetător în funcție de obiectivele analizelor efectuate.

**b) Comparația între valorile diversității structurale ale suprafețelor**

Diferențele mici înregistrate între valorile IDIV pentru cele două componente axiale ale unei suprafețe au permis caracterizarea fiecărei piețe de probă printr-un indice calculat drept medie între IDIV_Ox și IDIV_Oy. Acest fapt a permis efectuarea unei analize comparative între valorile diversității structurale ale suprafețelor. S-au analizat situațiile în care pasul de analiză este de 25, respectiv 50 cm pentru ca analiza comparativă să nu fie bazată exclusiv pe folosirea unui singur pas, considerând că lățimea benzilor poate conduce la diferențe semnificative.

În figura 6.22 se prezintă grafic variația valorilor IDIV în cele zece suprafețe și se poate remarca faptul că pasul analizei nu influențează studiul diferențelor dintre suprafețe. Chiar dacă sunt înregistrate decalaje între valorile obținute pentru aceeași suprafață la  pași de analiză diferiți, reprezentarea la o altă scară oferă grafic aceleași informații referitoare la diferențele dintre suprafețe. Corelațiile dintre valorile IDIV, IDIV_min și IDIV_max obținute la diferite lățimi ale benzilor sunt foarte puternice (coeficienți de corelație de peste 0,99 ***), astfel că folosirea unei anumite valori a pasului de analiză nu conduce la concluzii diferite în ceea ce privește comparațiile dintre suprafețe, fiind necesară doar consecvența în folosirea aceleiași valori a pasului pentru toate suprafețele.

Din analiza diagramei 6.22 se observă că diversitatea structurală a înălțimii, exprimată prin IDIV, variază în intervalul 1,550 - 3,287 (coeficient de variație: 21,9%) în cazul pasului de 25 cm, respectiv 2,273 - 5,967 (coeficient de variație: 27,6%) pentru pasul de 50 cm, în ambele variante observându-se același tip de variație, cu minime în suprafețele 2 și 5, respectiv maxime în suprafețele 6 și 7. O explicație a modului în care valorile IDIV variază în suprafețele analizate se referă la influența desimii puieților. Și la nivelul suprafețelor s-a observat o corelație directă pozitivă între valorile indicilor IDIV, IDIV_max și desime, de intensitate medie (r=0,688** și r=0,674** pentru pasul de 25 cm, respectiv r=0,725** și r=0,713** pentru pasul de 50 cm).

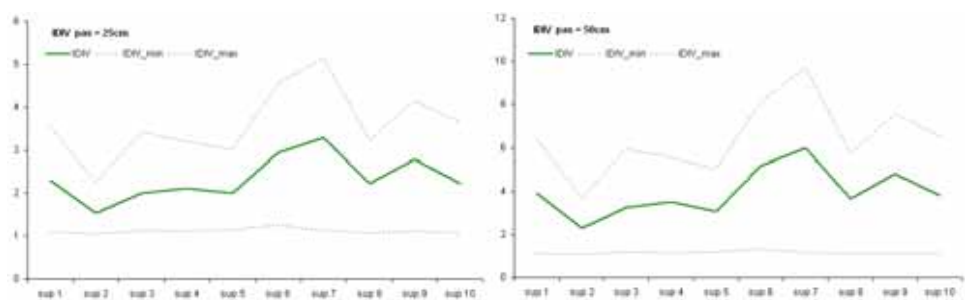

**Figura 6.22 Variația valorilor medii ale IDIV în suprafețele analizate (pas 25 și 50 cm)**

În cazul IDIV_min coeficientul de corelație calculat (r=-0,116 pentru ambele variante ale pașilor de analiză) nu indică o influență a desimii asupra acestuia, fapt constatat și în cazul studierii acestei legături corelative la nivelul fiecărei benzi.

Au fost calculate și valorile principalilor indicatori ai diversității structurale pentru distribuțiile înălțimilor în suprafețele studiate, datele fiind prezentate tabelar în anexa 6. S-a urmărit determinarea unei relații între respectivii indicatori și IDIV, pentru a vedea în ce măsură valorile IDIV surprind aceeași diversitatee structurală.

În tabelul 6.8 sunt prezentate valorile coeficienților de corelație dintre IDIV, IDIV_min, IDIV_max și restul indicilor. Se remarcă legătura slabă pozitivă dintre IDIV și indicele Shannon și Brillouin și negativă cu indicele Menhinick, pragul de semnificație al coeficienților de corelație depășind însă valoarea de 10%. Aceleași tip și intensitate a relațiilor s-a consemnat și pentru IDIV_max.

**Tabelul 6.8**

**Coeficienții de corelație ai legăturii dintre IDIV și indicatorii clasici
ai diversității structurale**

| | pasul: 25 cm | | | pasul: 50 cm | | |
|---|---|---|---|---|---|---|
| | IDIV | IDIV_min | IDIV_max | IDIV | IDIV_min | IDIV_max |
| *indicele Simpson (D)* | -0,036 | **0,533 \*** | -0,061 | -0,034 | **0,576 \*** | -0,099 |
| *indicele Simpson (1-D)* | 0,036 | **-0,533 \*** | 0,061 | 0,034 | **-0,576 \*** | 0,099 |
| *indicele Simpson (1/D)* | 0,131 | -0,473 | 0,157 | 0,130 | -0,516 | 0,194 |
| *indicele Shannon* | 0,225 | -0,254 | 0,248 | 0,214 | -0,286 | 0,272 |
| *echitatea Shannon* | 0,044 | **-0,667 \*\*** | 0,052 | 0,056 | **-0,695 \*\*** | 0,109 |
| *indicele Brillouin* | 0,267 | -0,273 | 0,288 | 0,260 | -0,304 | 0,315 |
| *indicele Berger-Parker* | 0,065 | **0,732 \*\*** | 0,047 | 0,053 | **0,788 \*\*\*** | -0,004 |
| *indicele McIntosh* | -0,017 | -0,507 | 0,011 | -0,024 | **-0,551 \*** | 0,045 |
| *indicele Margalef* | 0,004 | 0,405 | 0,043 | -0,042 | 0,384 | -0,001 |
| *indicele Menhinick* | -0,408 | 0,309 | -0,370 | -0,456 | 0,293 | -0,418 |
| *indicele Gleason* | 0,004 | 0,405 | 0,043 | -0,042 | 0,384 | -0,001 |

Valori superioare în ceea ce privește intensitatea și semnificația corelațiilor s-au obținut pentru valorile indicelui IDIV_min. Acesta stabilește legături de intensitate cel puțin slabă cu toți indicii, remarcându-se relațiile de intensitate medie, cu asigurare statistică, cu indicii Simpson, McIntosh și echitatea Shannon. Relația cu indicele Berger-Parker indică în cazul pasului de 50 cm o legătură puternică, foarte semnificativă. Legăturile mai strânse dintre IDIV_min și ceilalți indicatori ai diversității structurale vin în sprijinul observațiilor anterioare privitoare la proprietatea acestuia de a surprinde preponderent diversitatea structurală a parametrului analizat, neglijând o parte a informațiilor legate de distribuția în spațiu.

Indicii IDIV și IDIV_max sunt influențați într-o măsură mai mare de organizarea spațială a variabilei studiate, valorile lor reflectând acest aspect. Această situație poate explica corelațiile slabe cu valorile indicatorilor clasici ai diversității structurale.

S-au analizat și relațiile dintre indicele IDIV și indicii de diferențiere și dominanță dimensională, aceștia din urmă folosind, chiar dacă nu direct în relația lor de calcul, informații referitoare la poziționarea în spațiu a puieților. Valorile coeficienților de corelație dintre IDIV, IDIV_min, IDIV_max și indicii T și U calculați pentru 1-5 vecini sunt prezentate în tabelul 6.9.

Între IDIV_min și indicele de diferențiere T se constată o relație pozitivă de intensitate medie, caracterizată de un coeficient de corelație semnificativ. Numărul de vecini nu afectează caracterul relației, dar se observă că odată cu creșterea acestuia se înregistrează o ușoară scădere a intensității legăturii. Nu este surprinzătoare corelația mai puternică dintre indicele T și valoarea minimă a IDIV, acesta din urmă, după cum remarcam și anterior, fiind mai expresiv în caracterizarea diversității dimensionale care nu este afectată de distribuția în spațiu.

În ceea ce privește relația IDIV cu cele două variante ale indicelui de dominanță U, se constată legături de intensitate medie, dar mai puternice decât în cazul relației cu indicele T.

Tabelul 6.9

**Coeficienții de corelație dintre IDIV și indicii de diferențiere și dominanță dimensională**

relația IDIV - T

|  | IDIV 25 | IDIV min 25 | IDIV max 25 | IDIV 50 | IDIV_min 50 | IDIV_max 50 |
|---|---|---|---|---|---|---|
| T1 | 0,457 | **0,571\*** | 0,407 | 0,434 | **0,570\*** | 0,385 |
| T2 | 0,490 | **0,543\*** | 0,435 | 0,468 | **0,536\*** | 0,418 |
| T3 | 0,459 | **0,531\*** | 0,411 | 0,441 | **0,528\*** | 0,395 |
| T4 | 0,469 | **0,547\*** | 0,426 | 0,452 | **0,544\*** | 0,410 |
| T5 | 0,438 | **0,524\*** | 0,397 | 0,423 | **0,526\*** | 0,382 |
| relația IDIV - U varianta 1 Hui et al (1998) | | | | | | |
|  | IDIV 25 | IDIV min 25 | IDIV max 25 | IDIV 50 | IDIV_min 50 | IDIV_max 50 |
| U1 | -0,282 | -0,087 | -0,195 | -0,262 | -0,092 | -0,194 |
| U2 | -0,508 | -0,517 | **-0,556\*** | -0,498 | -0,480 | **-0,526\*** |
| U3 | **-0,540\*** | -0,459 | **-0,542\*** | **-0,542\*** | -0,455 | -0,516 |
| U4 | **-0,620\*\*** | -0,397 | **-0,656\*\*** | **-0,639\*\*** | -0,396 | **-0,642\*\*** |
| U5 | **-0,524\*** | -0,355 | **-0,551\*** | **-0,529\*** | -0,346 | **-0,532\*** |
| relația IDIV - U varianta 2 Gadow și Hui (1999) și Aguirre et al (2003) | | | | | | |
|  | IDIV 25 | IDIV min 25 | IDIV max 25 | IDIV 50 | IDIV_min 50 | IDIV_max 50 |
| U1 | 0,473 | 0,212 | 0,429 | 0,456 | 0,205 | 0,426 |
| U2 | **0,612\*\*** | 0,468 | **0,679\*\*** | **0,612\*\*** | 0,437 | **0,659\*\*** |
| U3 | **0,655\*\*** | 0,388 | **0,686\*\*** | **0,669\*\*** | 0,387 | **0,673\*\*** |
| U4 | **0,677\*\*** | 0,267 | **0,740\*\*\*** | **0,702\*\*** | 0,257 | **0,742\*\*\*** |
| U5 | **0,563\*** | 0,114 | **0,608\*\*** | **0,578\*** | 0,094 | **0,613\*\*** |

Legătura este mult mai strânsă cu cea de a doua variantă a indicelui U (U var 2), coeficienții de corelație fiind distinct semnificativi sau chiar foarte semnificativi. Relația cu dominanța se realizează de această dată prin intermediul IDIV și IDIV_max, valorile coeficienților de corelație cu IDIV_min fiind inferioare. Dominanța nu pune în valoare diferențele dimensionale dintre arbori dar surprinde mai bine informația spațială datorită relației simple ce caracterizează superioritatea sau inferioritatea dimensiunilor față de o referință într-un spațiu limitat bine definit.

Fundamentarea teoretică și informația surprinsă de IDIV este complexă și diferită de a celorlalți indicatori, drept urmare rezultatele comparațiilor confirmă așteptările. Singurul indice apropiat ca mod de fundamentare de IDIV prin capacitatea de a surprinde particularităţile diversităţii spaţiale ale unui parametru este cel conceput de Zenner şi Hibbs (2000), modalitatea greoaie de calcul a acestuia împiedicând deocamdată comparaţia directă a valorilor caracteristice fiecărei suprafeţe. În viitor se va studia posibilitatea algoritmizării şi implementării într-un program informatic a acestui indice spaţial, pentru a studia diferenţele faţă de rezultatele oferite de IDIV.

A fost acordată o atenţie sporită diversităţii structurale a înălţimilor datorită importanţei deosebite a acestui parametru în dinamica seminţişului precum şi în dezvoltarea ulterioară a arboretului. Unele cercetări (Wichmann, 2002) au arătat că ierarhia dimensională în cazul înălţimii se formează în populaţiile de arbori din stadiile incipiente de creştere şi dezvoltare, neapărând modificări majore ale acestei ierarhii decât în cazul derulării unor evenimente perturbatoare majore. Corelarea informaţiilor privitoare la diversitatea structurii înălţimii cu cele referitoare la distribuţia în spaţiu a puieţilor poate fi determinantă în realizarea unor modele individuale, dependente de distanţă, de evaluare şi apreciere a dinamicii regenerării arboretelor.

# Capitolul 7
# Evaluarea modelelor de organizare spațială a puieților instalați pe cale naturală

## 7.1. Introducere

Tobler (1970) a formulat prima lege a geografiei prin celebra sintagmă "*Everything is related to everything else, but near things are more related than distant things.*", în traducere „*Lucrurile depind de toate celelalte lucruri, dar mai mult de cele situate în apropiere*". A fost un concept rapid adoptat și recunoscut de numeroase alte științe, ecologia fiind una dintre acestea. Acest principiu acordă o mare importanță informației privitoare la organizarea în spațiu a unor evenimente, fiind dezvoltate numeroase metode de testare a autocorelației, multe din aceste metode fiind adaptate analizei modului de organizare în spațiu - indicele Moran, indicele Geary sau testul Mantel.

Distribuția în spațiu a indivizilor influențează o gamă largă de procese specifice unui ecosistem, de aceea s-a acordat o atenție sporită studierii acesteia. Cele mai utilizate metode de cercetare sunt cele ale analizei proceselor punctiforme, care au drept model de reprezentare așa numitul „*spatial point pattern*", termen utilizat la noi în diverse forme: tipar spațial, model de aranjare în spațiu sau model de organizare spațială.

Diggle (1983) definește tiparul spațial ca fiind o mulțime formată din pozițiile evenimentelor situate într-o zonă de interes. Accentuându-i importanța, Goreaud și Pelissier (2003) afirmă că tiparul spațial se constituie într-o veritabilă

„amprentă ecologică" specifică fiecărui ecosistem. În ceea ce privește importanța acestor informații în cercetarea forestieră, Nigh (1997) consideră coordonatele poziției arborilor instrumente utile ce permit specialistului silvic să rezolve aspecte privitoare la dinamica arboretelor – evaluarea competiției, creșterii și dezvoltării.

Probabil una din primele cercetări care a folosit informațiile privitoare la pozițiile în spațiu ale indivizilor a fost estimarea densității în baza distanțelor calculate dintre arbori, prezentată de către Konig (1835) în lucrarea „*Die Forst-Mathematik*". Ulterior Svedberg în 1922 (citat de Stoyan, Penttinen, 2000) a folosit procesele Poisson pentru a explica modul de organizare în spațiu a populațiilor de plante. Introducerea unor metode moderne de analiză a proceselor punctiforme în populațiile de arbori a fost realizată de către Matern (1960) și Warren (1972) (citați de Stoyan, Penttinen, 2000). Trebuie amintite și lucrările devenite clasice ale lui Ripley (1981), Diggle (1983), Cressie (1991, citat de Haining, 2004) sau lucrări importante apărute mai recent: Dale (2004), Fortin și Dale (2005), Illian et al (2008).

Principalele modele de organizare în spațiu (modelul aleatoriu, agregat și uniform) au fost prezentate în subcapitolul 3.4.2, fiind amintite și metodele de investigare ce sunt folosite în stabilirea abaterii unui tipar de la ipoteza CSR (*Complete spatial randomness*), ipoteza centrală a teoriei analizei proceselor punctiforme bazate pe distribuția omogenă Poisson.

Numărul indivizilor și modul de poziționare al acestora în spațiu nu sunt suficiente pentru a caracteriza cât mai complet o populație de arbori. Specia și caracteristicile biometrice măsurate sau calculate sunt importante în definirea structurii, fapt ce a condus la folosirea așa numitelor procese etichetate sau marcate (*marked point processes/fields*) (Penttinen et al., 1992; Stoyan, Stoyan, 1994). În acest caz fiecărui punct îi este atașată o etichetă, ce poate fi folosită în analizele ulterioare, trecându-se în acest mod de la analizele univariate (ce testează abaterea de la CSR) la analize bivariate (ce testează abaterea de la independența sau de la randomizarea etichetării) (Goreaud și Pelissier, 2003). Procesele etichetate pot să facă trecerea spre indicii de competiție dependenți de distanțe, care folosesc în aceeași măsură elementele biometrice și informația spațială dar într-o altă formă.

Metodele specifice analizei spațiale sunt extrem de adecvate aplicării în cercetările privitoare la regenerarea arboretelor datorită numărului mare de indivizi pe suprafață și numeroaselor tipare de organizare întâlnite. Astfel se explică și faptul că una dintre primele aplicații ale proceselor punctiforme a vizat chiar modul de organizare în spațiu a puieților (Ripley, 1976), observându-se că puieții de sequoia nu sunt răspândiți conform unui proces Poisson. În ultimii ani au fost efectuate numeroase cercetări care au urmărit acest tip de aspecte în regenerări (Awada, et al., 2004; Camarero et al., 2005; Fajardo et al., 2006;  Gratzer, Rai, 2004; Hofmeister et al., 2008; Maltez-Mouro et al., 2007; McDonald et al., 2003; Montes et al., 2007; Nigh, 1997; Paluch, 2005).

În cazul de față obiectivele propuse în acest capitol se referă la:

- stabilirea tiparului de organizare spațială a  puieților în general sau a unor clase de puieți prin folosirea unor metode diferite (inclusiv stabilirea unor parametri ce caracterizează agregarea);

- testarea unor ipoteze privitoare la asocierea dintre puieții unor specii diferite;

- evaluarea relațiilor competiționale dintre puieți prin intermediul unor metode care să țină cont de informația spațială.

## 7.2. Determinarea modelelor de organizare spațială prin metode independente de distanțe

Metodele independente de distanță care urmăresc determinarea tipului de distribuție în spațiu folosesc informații privitoare la variația densității evenimentelor pentru a evalua situația la nivelul unei suprafețe. Datele legate de modul de organizare în spațiu a indivizilor sunt folosite în mod indirect în evaluări, de aceea metodele sunt denumite generic independente de distanță. Cea mai cunoscută abordare a acestui concept este metoda quadratelor (*quadrat sampling*), dezvoltată în urma cercetărilor de la începutul secolului trecut ale lui Gleason (1920). Metoda se bazează pe împărțirea ariei de studiu în unități statistice elementare (*quadrate*) de dimensiune egală în care se înregistrează frecvența evenimentelor și apoi se compară raportul dintre varianța frecvențelor experimentale și medie cu raportul unei distribuții randomizate, care respectă ipoteza CSR (*Complete spatial randomness*). Analiza datelor urmărește detectarea tipului modelului de organizare spațială - randomizat, agregat sau uniform prin intermediul valorilor obținute pentru o serie de indici sintetici.

La noi în țară, deși o parte din indicii de distribuție spațială sunt menționați de câteva decenii în literatura de specialitate (Botnariuc și Vădineanu, 1982) aplicarea lor a fost posibilă mult mai târziu (Cenușă, 1996 a; Popa, 2001; Avăcăriței, 2005). Acest lucru a fost datorat nu caracterului modern de analiză ci faptului că prelucrările numerice ce trebuie efectuate în vederea stabilirii valorii indicilor respectivi sunt relativ complexe, calculatorul fiind în acest caz un instrument extrem de folositor. Dezvoltarea unor aplicații de calcul tabelar a ușurat într-o oarecare măsură calculul acestor indici. Totuși, chiar cu ajutorul unei aplicații de calcul tabelar, multe din aceste prelucrări ale datelor ar fi imposibil de efectuat fără cunoștințe de programare în VBA (*Visual Basic for Applications*).

Pentru a putea prelucra datele conform acestei tehnici de analiză, precum și pentru a veni în întâmpinarea cercetătorilor care doresc să efectueze studii asupra proceselor spațiale, dar nu posedă toate cunoștințele necesare prelucrării datelor am realizat o aplicație software proprie de analiză și interpretare a distribuțiilor spațiale. Prima versiune a aplicației SPATIAL (Palaghianu, Horodnic, 2007) avea o aplicabilitate restrânsă, dar ulterior au fost adăugate module noi de analiză și interpretare a datelor. Principalele îmbunătățiri se referă la integrarea unor noi module de prelucrare a datelor prin metode dependente de distanțe, înglobarea unor subrutine de eliminare a efectului de margine, automatizarea interpretărilor statistice ale indicilor calculați și introducerea posibilității de creare a unor fișiere de configurare care rețin preferințele de analiză ale utilizatorilor.

Aplicația a fost realizată în mediul de dezvoltare *Microsoft Visual Basic*, fiind concepută sub forma unui fișier executabil independent care preia datele dintr-un registru de lucru de tip *Microsoft Excel*.

SPATIAL posedă patru module, primul dintre ele efectuând o prelucrare prin metoda quadratelor iar celelalte trei furnizează rezultate obținute în urma unor prelucrări prin metode dependente de distanță. Modulul de analiză prin metoda quadratelor are ca date de intrare coordonatele arborilor, dimensiunile suprafeței studiate și dimensiunile unui quadrat. În baza acestor date, programul calculează o serie de indici și stabilește tipul de structură spațială prin interpretarea statistică a valorii acestora la un prag de semnificație de 5%.

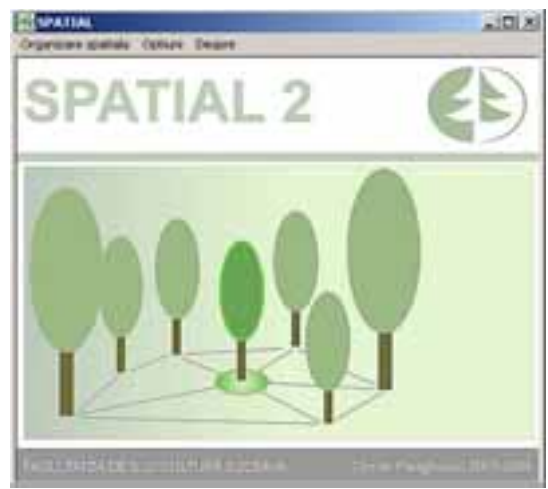

**Figura 7.1 Interfața aplicației SPATIAL**

Sunt determinați următorii indici: indicele de dispersie (ID), indicele mărimii agregatului (ICS), indicele Morisita ($I_\delta$) și indicele Green (C), detaliile privitoare la relațiile de calcul și modul de interpretare a acestora fiind prezentate în continuare.

**Indicele de dispersie (ID)** - indicele cu cea mai simplă relație de calcul se bazează pe raportul dintre varianță și medie, indicând o structură aleatoare la valori egale ale celor două caracteristici (ID=1), o structură agregată la valori supraunitare (ID>1), respectiv uniformă pentru un raport subunitar (ID<1), (Botnariuc și Vădineanu, 1982).

$$ID = \frac{s^2}{\overline{x}} \qquad (7.1)$$

Testarea semnificației pentru pragul de 5% se face cu ajutorul testului $\chi^2$ la *(n-1)* grade de libertate, ipoteza nulă CSR fiind respinsă în cazul în care:

$$\frac{s^2}{\overline{x}} \cdot (n-1) \not\subset \left[ \chi^2_{0.975} \ldots \chi^2_{0.025} \right] \qquad (7.2)$$

*n* - numărul de quadrate;

*$s^2$* - varianța numărului de evenimente pe quadrate;

$\overline{x}$ – media numărului de evenimente pe quadrat.

**Indicele mărimii agregatului (ICS)** – este un indice prezentat în 1954 de către David și Moore (citați de Green, 1966) similar ca și concepție indicelui de dispersie. Interpretarea este asemănătoare, comparația efectuându-se în acest caz față de 0, valoarea la care structura este aleatoare. O valoare pozitivă indică o structură agregată, iar valorile negative identifică o structură uniformă.

$$ICS = \frac{s^2}{\overline{x}} - 1 \qquad (7.3)$$

Interpretarea statistică se face prin intermediul testului u bilateral. Se calculează:

$$z = (\frac{s^2}{\overline{x}} - 1) \cdot \sqrt{\frac{(n-1)}{2}} \qquad (7.4)$$

și se compară cu valoarea critică u pentru un prag de semnificație (u=±1,96 pentru pragul de 5%). În cazul în care $z \not\subset [-1,96\ldots+1,96]$ se respinge ipoteza nulă CSR și se realizează interpretarea structurii prin compararea valorii indicelui cu 0.

**Indicele Morisita** – este unul dintre cei mai cunoscuți indici de caracterizare a tiparului distribuției spațiale, conceput de către cercetătorul japonez Morisita (1962). Hurlbert (1990) consideră că este cel mai stabil indice, fiind extrem de versatil și foarte puțin afectat de dimensiunea quadratului.

$$I_\delta = n \cdot \frac{\sum \left(x_i^{\,2}\right) - \sum x_i}{\left(\sum x_i\right)^2 - \sum x_i} \tag{7.5}$$

$n$ – numărul de quadrate,

$x_i$ – numărul de evenimente din quadratul *i*.

Semnificația abaterii de la distribuția Poisson se apreciază prin intermediul testului $\chi^2$, după relația:

$$\chi^2_{\exp} = I_\delta \left(\sum x_i - 1\right) + n - \sum x_i \tag{7.6}$$

Ipoteza nulă (CSR) va fi admisă dacă valoarea experimentală a lui $\chi^2$ corespunzătoare unui număr de *n-1* grade de libertate se încadrează în intervalul: $\left[\chi^2_{0.975}\ldots\chi^2_{0.025}\right]$. În cazul abaterii de la tipul de distribuție aleatoare, valoarea indicelui $I_\delta$ mai mare de 1 indică o structură agregată, iar o valoare mai mică de 1 o structură uniformă.

Indicele Morisita este singurul indice care este dublat de o justificare a valorii sale, unii autori (Hurlbert, 1990) reproșând celorlalți indici lipsa explicației valorii lor. $I_\delta$ măsoară de câte ori este mai mare probabilitatea ca doi indivizi selectați la întâmplare să aparțină aceluiași quadrat față de probabilitatea ca x indivizi din populație să fie distribuiți aleatoriu în spațiu. În aceste condiții o valoare a indicelui de 1,5 semnifică faptul că probabilitatea ca doi indivizi aleși întâmplător să aparțină aceluiași quadrat este cu 50% mai mare față de cazul în care în care distribuția indivizilor ar fi aleatoare.

**Indicele Green** - concept de Green (1966), a fost considerat de către autorul său un indice puțin afectat de numărul de quadrate sau de mărimea acestora. Green l-a realizat pentru a surprinde situații pe care ceilalți indici, în opinia sa, nu reușeau să le caracterizeze în mod adecvat. Relația de calcul este:

$$C = (\frac{s^2}{x} - 1) \cdot \frac{1}{\sum x_i - 1} \qquad (7.7)$$

$s^2$ - varianța numărului de evenimente pe quadrate;

$x_i$ – numărul de evenimente din quadratul *i*.

Interpretarea se face prin intermediul variabilei normale u, după relația:

$$u = \sqrt{2z^2} - \sqrt{2f - 1} \qquad (7.8)$$

iar valoarea experimentală a testului statistic este:  $z = \dfrac{s^2(n-1)}{x} \qquad (7.9)$

*f* = *n-1* - reprezintă numărul gradelor de libertate;

*n* - reprezintă numărul de unități statistice elementare.

În cazul în care $|u| > 1,96$ se respinge ipoteza CSR a structurii aleatoare și se poate stabili modelul organizării spațiale în funcție de valoarea indicelui C. Green a stabilit că valorile pozitive (C poate atinge maxim valoarea 1) indică o structură agregată (numită de către autor *contagiozitate pozitivă* conform terminologiei introduse de Cole, 1946), iar valorile negative (valoarea minimă: -*1/(Σx - 1))* indică o structură uniformă (*contagiozitate negativă*).

În lucrarea de prezentare a indicelui C, „*Measurement of non-randomness in spatial distributions*", Green afirmă că nu există nici un indice perfect, capabil să identifice toate situațiile particulare ale structurii spațiale, considerând propriul indice mai potrivit identificării structurilor agregate decât celor uniforme.

Un avantaj al folosirii programului SPATIAL pentru analiza datelor constă și în faptul că se pot preciza dimensiuni diferite pentru quadrate. Acest lucru este important deoarece prin metoda quadratelor se pot obține rezultate diferite în ceea ce privește organizarea spațială la o variație a dimensiunilor quadratului. La creșterea mărimii quadratului se pot obține valori mai mari ale indicelui de dispersie, ceea ce poate conduce (pentru unii indici) spre o structură de tip agregat.

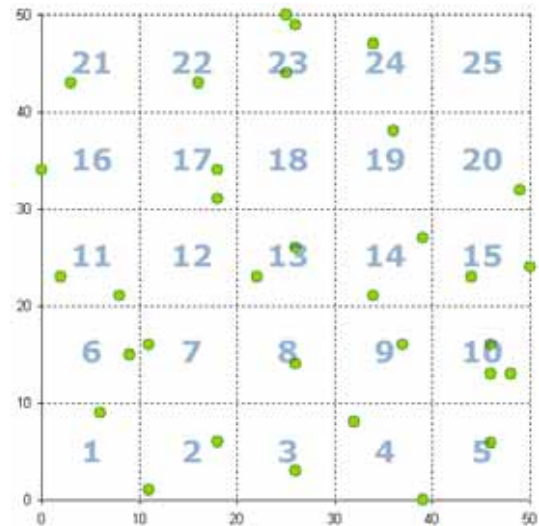

**Figura 7.2 Exemplu de grafic de repartiție a evenimentelor pe quadrate și modul de numerotare a suprafețelor statistice elementare (SPATIAL)**

Alegerea mărimii quadratului este, așadar, de maximă importanță în interpretarea rezultatelor obținute. Aplicația SPATIAL permite efectuarea rapidă a calculelor pentru diferite dimensiuni ale quadratului și astfel dă posibilitatea cercetătorului să stabilească care este dimensiunea optimă a acestuia din perspectiva obiectivelor propuse. Programul realizează încadrarea automată a fiecărui punct în quadratul corespunzător și întocmește un grafic pentru verificarea corectitudinii încadrării, calculează indicii spațiali menționați anterior și oferă interpretarea statistică a valorii acestora, calculează frecvența evenimentelor pe fiecare quadrat și de asemenea efectuează o distribuție a quadratelor pe clase ale numărului de evenimente pe quadrat.

Pentru o înțelegere facilă a analizelor ulterioare și pentru a clarifica modalitatea de interpretare a informațiilor spațiale se prezintă un exemplu folosind cele trei situații corespunzătoare principalelor tipare de distribuție în spațiu – uniform, aleatoriu și agregat. Se prezintă grafic distribuția în spațiu, valorile indicilor prezentați anterior precum și caracteristicile distribuției pe clase ale numărului de evenimente pe quadrat.

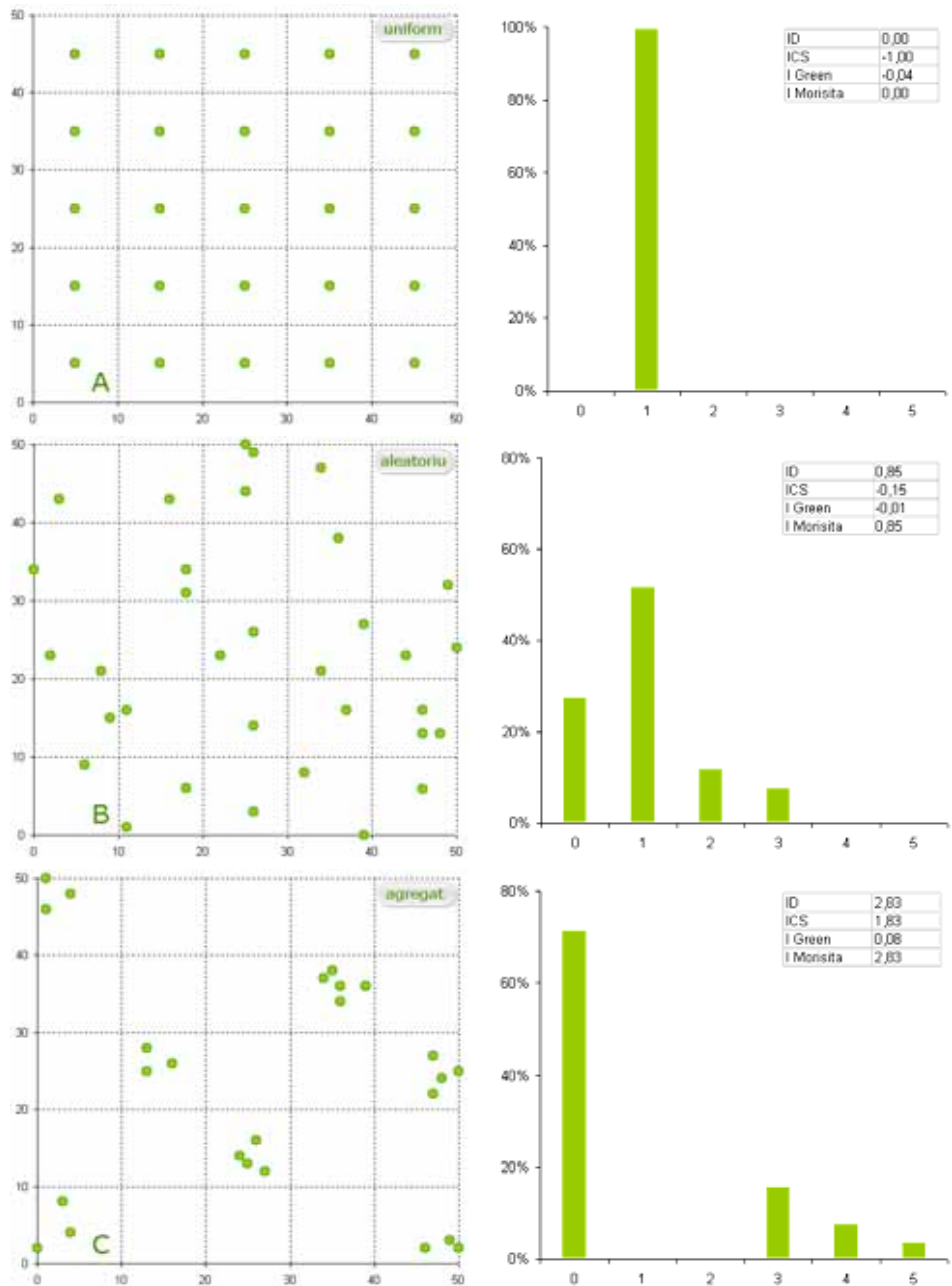

**Figura 7.3 Exemplu de structură spațială uniformă (A), aleatoare (B), și agregată (C) - valorile indicilor de caracterizare și distribuția frecvențelor relative ale quadratelor pe clase ale numărului de evenimente pe quadrat**

Legat de numărul adecvat de quadrate care trebuie utilizat în analiză, Green (1966) sugerează un minim situat în jurul valorii de 50 de quadrate pentru calculul indicilor specifici acestei metode (pentru situația dată latura maximă a quadratului ar trebui să fie de 100 cm, pentru care s-ar obține 49 de quadrate). Acest fapt trebuie corelat și cu afirmațiile lui Gratzer și Rai (2004) care consideră că metoda quadratelor trebuie efectuată pentru un număr de minim 15 evenimente pe quadrat (în situația dată latura minimă a quadratului ar trebui să fie de 100 cm, pentru care s-ar obține o densitate a evenimentelor pe suprafața statistică elementară de 14,8). Unii au sugerat că aria unui quadrat ar trebui să fie apropiată de dublul suprafeței medii aferente unui eveniment (Lembo, 2007) – în cazul de față, ținând cont de densitatea puieților pe suprafață s-ar obține latura quadratului de circa 40 cm în ipoteza unor suprafețe statistice elementare de formă pătrată. Ținând cont de aceste aspecte, de dimensiunile suprafețelor de probă și de densitatea evenimentelor s-a decis folosirea în analiză a trei variante dimensionale – 50x50 cm, 70x70 cm și 100x100 cm, care să acopere intervalul calculat anterior.

Încadrarea în cele trei tipuri de structură s-a realizat prin folosirea testelor statistice specifice fiecărui indicator pentru a se testa dacă abaterile de la CSR sunt semnificative (s-a folosit pragul de semnificație de 5%) iar apoi s-au interpretat valorile individuale ale indicilor calculați. Valorile indicilor și testelor statistice folosite, precum și graficele pentru toate suprafețele în cele trei variante dimensionale de analiză sunt prezentate în anexa 7 și 8.

În toate cele 10 suprafețe studiate s-au obținut diferențe statistic semnificative față de ipoteza distribuției aleatoare în spațiu a puieților, în cazul tuturor variantelor dimensionale ale quadratelor. Mai mult, valorile indicilor au evidențiat același tipar de distribuție în spațiu a puieților – tiparul agregat. Aceste rezultate sunt în concordanță cu cele ale altor cercetări efectuate în suprafețele cu regenerare, menționate în literatura de specialitate (McDonald et al., 2003; Awada, et al., 2004; Gratzer, Rai, 2004; Camarero et al., 2005; Paluch, 2005; Fajardo et al., 2006; Maltez-Mouro et al., 2007; Montes et al., 2007).

Rezultatele sunt prezentate grafic în figura 7.4 în relație cu desimea puieților, la dimensiunea quadratului de 100x100 cm, pentru valorile indicelui Morisita. A fost aleasă această dimensiune datorită faptului că diferențele dintre cele trei alternative sunt reduse, iar varianta de 100x100 cm este cea care răspunde cerințelor la două dintre cele trei criterii utilizate în stabilirea suprafeței adecvate a unui quadrat. În privința indicatorului ales, s-a optat pentru indicele Morisita care are un comportament mai stabil decât ceilalți indici la variația dimensiunii quadratului (Hurlbert, 1990). În mod normal se consideră că o creștere a dimensiunii quadratului conduce în mod artificial la creșterea tendinței de agregare în cazul evaluărilor bazate pe analiza raportului dintre varianță și medie. Indicele Morisita nu este afectat de acest fenomen, în suprafețele studiate înregistrând chiar o ușoară scădere la creșterea suprafeței quadratului. Un motiv în plus pentru alegerea indicelui este faptul că este singurul care oferă o explicație a valorii sale.

Legătura dintre desimea puieților și indicele Morisita este caracterizată prin coeficienți de corelație negativi ce indică o intensitate medie și chiar puternică: -0,77 ** (în varianta 50x50cm), -0,70* (70x70cm) și -0,69 * (100x100cm). Scăderea intensității de agregare odată cu creșterea desimii confirmă rezultatele înregistrate anterior la punctul 5.1 (redate sintetic în figura 5.19).

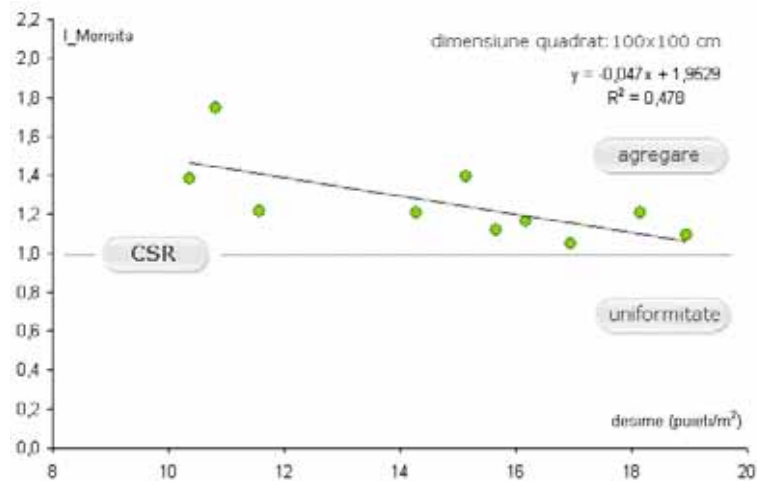

**Figura 7.4 Variația tendinței de agregare exprimate prin indicele Morisita, în relație cu desimea puieților în suprafețele studiate**

Creșterea concurenței determină o valorificare superioară a spațiului printr-o răspândire a indivizilor ce tinde să urmeze un model uniform de organizare spațială (Cole, 1946). Rezultate similare a înregistrat Pretzsch (1997) în cazul unor arborete de fag și larice.

Metoda quadratelor este una dintre cele mai simple și mai rapide metode utilizate în determinarea organizării în spațiu a evenimentelor dar prezintă anumite puncte slabe:

- mărimea quadratului poate influența rezultatele obținute;
- nu s-a definit o mărime optimă a quadratului, fiind nevoie în unele cazuri de analiza unei multitudini de variante dimensionale;
- se măsoară dispersia bazată pe densitatea evenimentelor și nu se identifică relațiile dintre elementele individuale;
- încadrarea într-un anumit tipar se face pentru toată suprafața, neputând fi identificate variații în interiorul acesteia; probleme de fundamentare și interpretare a unor indici - a se vedea considerațiile lui Hurlbert (1990) asupra distribuțiilor unicorniene (distribuții cu raportul dintre varianță și medie unitar dar care diferă semnificativ de distribuțiile Poisson) și asupra ambiguității conceptului de agregare.

Cu toate acestea reprezintă o metodă accesibilă și adecvată de analiză, frecvent folosită de numeroși cercetători, care poate să ofere informații utile asupra naturii organizării în spațiu a elementelor unui sistem.

## 7.3. Determinarea modelelor de organizare spațială și asociere prin metode bazate pe distanțele dintre puieți

Pentru a compensa o parte din dezavantajele metodelor independente de distanță, au fost dezvoltate metode de analiză care includ explicit informația spațială în prelucrarea datelor. Acestea sunt denumite generic metode dependente de distanțe datorită faptului că se bazează pe media distanțelor dintre cei mai apropiați vecini și prezintă avantajul că interpretarea se poate face nu doar la nivelul întregii suprafețe, ci și în interiorul ariilor studiate.

Cea mai folosită categorie de metode dependente de distanță este metoda celui mai apropiat vecin (*nearest neighbour*) (Clark și Evans, 1954) ce compară distribuția distanțelor dintre cele mai apropiate evenimente cu distribuția teoretică Poisson, corespunzătoare unui proces spațial aleator. Interpretarea abaterii de la ipoteza CSR (*Complete spatial randomness*) se face prin metode variate – prin intermediul valorilor unor indici sintetici sau prin analiza grafică a distribuției cumulative a distanțelor.

Și în cazul aplicării acestei metode s-a folosit aplicația proprie SPATIAL, căreia i-au fost aduse îmbunătățiri substanțiale comparativ cu varianta inițial concepută. Aplicația SPATIAL include în prezent trei module de prelucrare a datelor prin metoda  distanțelor ce folosesc drept date de intrare coordonatele arborilor, dimensiunile suprafeței studiate și lățimea benzii de tampon în cazul în care se dorește eliminarea erorilor provocate de efectul de margine. Datele sunt preluate dintr-o foaie de calcul de tip Microsoft Excel iar rezultatele sunt salvate în registrul de lucru inițial. Pentru automatizarea completă a prelucrărilor, toate setările inițiale referitoare la prelucrarea datelor pot fi salvate într-un fișier de configurare extern.

Metodele dependente de distanță prezintă și anumite dezavantaje, care țin de stabilirea limitelor ferestrei de observație și de efectul de margine care poate afecta rezultatele prelucrărilor.

Efectul de margine în analizele spațiale se referă la erorile apărute în rezultate ca urmare a ignorării elementelor situate în afara marginilor așa numitei *ferestre de observație* (Illian et al., 2008), elemente care pot stabili relații cu cele din interiorul ariei studiate. Au fost propuse mai multe soluții de corectare a acestui efect (Ripley, 1981; Besag, 1977 ; Getis, Franklin, 1987; Stoyan, Stoyan, 1994; Haase, 1995; Goreaud, Pellisier, 1999; Pommerening, Stoyan, 2006), în general fiind folosite patru categorii de metode de corecție (Tokola, 2004; Illian et al., 2008):

- inventariere suplimentară (*plus sampling*) – pentru a avea informații complete despre vecini;

- folosirea unei zone tampon (*minus sampling*) – deși se inventariază toată suprafața, se lasă o zonă de tampon, calculele efectuându-se doar pentru zona centrală (*core area*);

- inventarierea în oglindă sau folosind înfășurarea pe un toroid (*toleroid edge correction sau mirror sampling*) – calculele se efectuează în baza datelor obținute prin copierea unei zone de margine în zona tampon opusă;

- acordarea unor ponderi observațiilor (*weighting of observations*) – datele privitoare la vecinătăți sunt ponderate în funcție de probabilitatea existenței informațiilor în respectiva zonă.

Metodele de corectare a efectului de margine vor fi alese în funcție de tipul datelor, modalitatea de eșantionare și tehnicile de analiză spațială utilizate (Dale, 2004; Fortin, Dale, 2005). În cazul unor dimensiuni mari ale ferestrei de observație, cu un număr mare de evenimente, efectul de margine poate fi ignorat (Pommerening, Stoyan, 2006; Illian, 2008).

În aplicația SPATIAL s-a ținut cont de efectul de margine, utilizatorul având posibilitatea de a corecta consecințele acestuia prin introducerea lățimii benzii de tampon care va fi folosită în calcule. Această valoare trebuie stabilită în funcție de numărul considerat al celor mai apropiați vecini și de distanța medie față de aceștia.

### 7.3.1. Analiza prin intermediul indicilor dependenți de distanță

Una din abordările cele mai rapide și mai utilizate ale metodelor dependente de distanță este reprezentată de interpretarea tiparului spațial prin intermediul valorii unor indici sintetici. Indicii sunt calculați în funcție de media distanțelor față de cel mai apropiat vecin al fiecărui eveniment, iar apoi se efectuează o comparație cu media distanțelor față de cel mai apropiat vecin în ipoteza unei distribuiri aleatoare în spațiu.

Aplicația SPATIAL calculează o serie de indicatori menționați în literatura de specialitate, care sunt folosiți în scopul stabilirii tipului de organizare. Se efectuează și o interpretare statistică a valorilor obținute, pentru a vedea dacă se poate respinge $H_0$ - ipoteza de nul CSR (conform căreia evenimentele sunt distribuite aleator în suprafață). Conform ipotezei alternative $H_1$ evenimentele sunt dispuse nerandomizat (dispunere agregată sau uniformă). Au fost folosite în funcție de recomandările din literatura de specialitate referitoare la fiecare tip de indice, testele $\chi^2$ și testul u bilateral, la un prag de semnificație de 5%.

Se prezintă relațiile de calcul ale indicilor determinați de aplicația SPATIAL, precum și interpretarea valorilor acestora și a modului de testare a semnificației lor statistice.

**Media distanțelor față de cel mai apropiat vecin**

$$\overline{d} = \frac{\sum_{i=1}^{n} d_i}{n} \tag{7.10}$$

$d_i$ – distanța de la evenimentul $i$ la cel mai apropiat vecin;

$n$ – numărul de evenimente din suprafață.

**Media distanțelor față de cel mai apropiat vecin în ipoteza unei distribuiri aleatoare în spațiu ($d_{CSR}$)**

$$\overline{d}_{CSR} = \frac{1}{2\sqrt{\lambda}} = \frac{1}{2\sqrt{\dfrac{n}{A}}} \tag{7.11}$$

$n$ – numărul de evenimente din suprafață;

*A* – aria suprafeței analizate;

$\lambda$ - densitatea de evenimente pe suprafață.

**Dispersia distanței față de cel mai apropiat vecin**

$$s^2 = \frac{\sum_{i=1}^{n}\left(d_i - \overline{d}\right)^2}{n-1} \tag{7.12}$$

$d_i$ – distanța de la evenimentul *i* la cel mai apropiat vecin;

*n* – numărul de evenimente din suprafață.

**Indicele Fisher** – este mai puțin folosit, fiind derivat din testul introdus de Fisher (et al. 1922) pentru testarea asocierii neîntâmplătoare a evenimentelor. Valoarea sa este bazată pe controversatul raport dintre varianță și medie.

$$I_f = \frac{s^2}{\overline{d}} \tag{7.13}$$

O valoare apropiată de 1 indică o structură aleatoare ($I_f \approx 1$), valorile subunitare o structură uniformă ($I_f < 1$), iar cele supraunitare o structură agregată ($I_f > 1$).

Pentru testarea semnificației se folosește produsul $I_f \cdot$ *(n-1)* care se compară cu valoarea lui $\chi^2$ corespunzătoare la *(n-1)* grade de libertate. Ipoteza nulă se respinge pentru pragul de semnificație de 5% (test bilateral) dacă $I_f$ este mai mic decât valoarea critică $\chi^2$ pentru p=0,975 sau este mai mare decât valoarea critică $\chi^2$ pentru p=0,025.

**Indicele de agregare Clark–Evans** – este unul din cei mai cunoscuți și utilizați indici de testare a abaterii de la distribuția aleatoare, prezentat de Clark și Evans (1954) în lucrarea devenită clasică „*Distance to nearest neighbour as a measure of spatial relationships in populations*". Cei doi autori menționează o aplicație a indicelui în mediul forestier, identificând un tipar uniform în stejăretele din Michigan. Valoarea indicelui reprezintă o standardizare a mediei distanței față de cel mai apropiat vecin, reprezentând proporția acesteia din media distanțelor față de cel mai apropiat vecin în ipoteza unei distribuiri aleatoare în spațiu a evenimentelor.

$$R = \frac{\overline{d}}{\overline{d}_{CSR}} = \frac{\overline{d}}{\left(\dfrac{1}{2\sqrt{\lambda}}\right)} = 2 \cdot \overline{d} \cdot \sqrt{\lambda} \tag{7.14}$$

unde: $\lambda = \dfrac{n}{A} = \dfrac{n}{(x_{max} - x_{min})(y_{max} - y_{min})}$ (7.15)

x, y - valorile extreme ale pozițiilor carteziene ale arborilor;

$\lambda$ - densitatea de evenimente pe suprafață;

n – numărul de evenimente din suprafață;

A – aria suprafeței analizate.

Pentru a testa semnificația valorii indicelui și a stabili dacă se respinge ipoteza nulă s-a folosit testul propus de Clark și Evans. S-a calculat variabila repartizată normal:

$$c = \frac{\overline{d} - \overline{d}_{CSR}}{\sigma_{\overline{d}_{CSR}}} \tag{7.16}$$

unde $\sigma_{\overline{d}_{CSR}}$ reprezintă abaterea standard a mediei aritmetice a distanței dintre cei mai apropiați vecini în ipoteza unei distribuții aleatoare de densitate $\lambda$.

$$\sigma_{\overline{d}_{CSR}} = \sqrt{\frac{4 - \pi}{4 \cdot \pi \cdot n \cdot \lambda}} = \frac{0.26136}{\sqrt{n \cdot \lambda}} \tag{7.17}$$

Dacă valoarea absolută|c| > 1,96 (la pragul de semnificație de 5%) sau 2,58 (la pragul de semnificație de 1%) se respinge ipoteza nulă și se va stabili tiparul spațial în funcție de valoarea indicelui R. Valorile mai mici de 1 indică o structură agregată, iar cele mai mari de 1 o structură uniformă (maximul R este de 2,1491 pentru distribuția punctelor într-o rețea hexagonală perfectă). Valoarea R=1 este caracteristică unei structuri perfect aleatoare.

**Indicele Donnelly** - folosind o serie de simulări, Donnelly în 1978 (citat de Ripley, 1981) a efectuat modificări formulei inițial concepute de Clark și Evans, în vederea corectării erorilor determinate de efectul de margine.

$$IC = \frac{\overline{d}}{\overline{d}_{CSR\_corectata}} \qquad (7.18)$$

$$\overline{d}_{CSR\_corectata} = 0.5 \cdot \sqrt{\frac{A}{n}} + 0.0514 \cdot \frac{P}{n} + \frac{0.041 \cdot P}{\sqrt{n^3}} \qquad (7.19)$$

$P$ - perimetrul suprafeței experimentale;

$n$ – numărul de evenimente din suprafață;

$A$ – aria suprafeței analizate.

Testarea statistică a semnificației se face folosind aceeași relație introdusă de Clark și Evans, folosind însă distanța corectată conform formulei (7.19). Abaterea standard a mediei aritmetice a distanței dintre cei mai apropiați vecini în ipoteza unei distribuții aleatoare de densitate $\lambda$ este:

$$\sigma_{\overline{d}_{CSR}} = \frac{1}{n} \cdot \sqrt{0.07\,A + 0.037\,P\sqrt{\frac{A}{n}}} \qquad (7.20)$$

Ipoteza nulă se respinge la valori ale |c| mai mari de 1,96 (la pragul de semnificație de 5%) sau 2,58 (la pragul de semnificație de 1%). Valorile mai mici de 1 indică o structură agregată, iar cele mai mari de 1 o structură uniformă.

**Indicele Skellam** – a fost inițial fundamentat de către Skellam (1952), fiind modificat apoi de către Hopkins și Skellam (1954) (citați de Lesseps, 1975).

$$IS \; = \; 2\,\pi\lambda \; \cdot \sum_{i=1}^{n} d_i^2 \qquad (7.21)$$

$d_i$ – distanța de la evenimentul $i$ la cel mai apropiat vecin;

$n$ – numărul de evenimente din suprafață;

$\lambda$ - densitatea de evenimente pe suprafață.

Valoarea pentru acest indice în cazul unei distribuții aleatoare este de *2n*, de aceea pentru interpretare se folosește raportul *IS/2n*. Dacă acest raport este mai mare de 1, structura este uniformă, iar pentru valori subunitare structura este agregată. Pentru testarea semnificației se folosește valoarea indicelui *IS* care se compară cu valoarea testului $\chi^2$ pentru *2n* grade de libertate. Ipoteza nulă se respinge pentru pragul de semnificație de 5% (test bilateral) dacă *IS* este mai mic decât valoarea critică $\chi^2$ pentru p=0,975 sau este mai mare decât valoarea critică $\chi^2$ pentru p=0,025.

**Indicele Pielou** - conceput de Pielou (1959), este asemănător ca fundamentare cu indicele Skellam (valoarea indicelui Pielou este echivalentă cu valoarea testului *IS/2n*). Pielou recomanda în formularea inițială utilizarea distanțelor dintre puncte aleatoriu alese și evenimente pentru a ușura calculele. În multe cazuri este însă folosită formula ce utilizează direct distanțele dintre evenimente:

$$IP = \pi\lambda\varpi \qquad \text{unde:} \quad \varpi = \sum_{i=1}^{n} d_i^2 \cdot \frac{1}{n} \tag{7.22}$$

$d_i$ – distanța de la evenimentul $i$ la cel mai apropiat vecin;

$n$ – numărul de evenimente din suprafață;

$\lambda$ - densitatea de evenimente pe suprafață;

$\varpi$ - distanța pătratică medie dintre evenimente.

Valoarea unitară (IP =1) corespunde unui proces aleatoriu, o valoare mai mare de 1 corespunde unei structuri uniforme, iar mai mică de 1 unei structuri agregate (în multe lucrări interpretarea este dată exact invers datorită faptului că nu se ține cont de formularea inițială a lui Pielou care se referea la distanța dintre puncte aleatoriu alese și evenimente). Semnificația se testează prin compararea

valorii: $D = 2\pi\lambda \cdot \sum_{i=1}^{n} d_i^2$ \hfill (7.23)

cu valoarea testului $\chi^2$ pentru *2n* grade de libertate. Ipoteza nulă se respinge dacă valoarea D nu aparține intervalului $\left[\chi_{0.975}^2 \ldots \chi_{0.025}^2\right]$.

Datorită relației dintre indicele Pielou și Skellam aplicația SPATIAL precizează o valoare comună pentru indicele Pielou și testul Skellam. Pentru calculul valorii testului $\chi^2$ a fost adaptat și implementat algoritmul propus de Ibbetson (1963) în lucrarea „*Algorithm 209: Gauss*" deoarece funcția CHIINV a *Microsoft Excel* nu este definită pe anumite intervale în cazul unor valori mari ale numărului de grade de libertate.

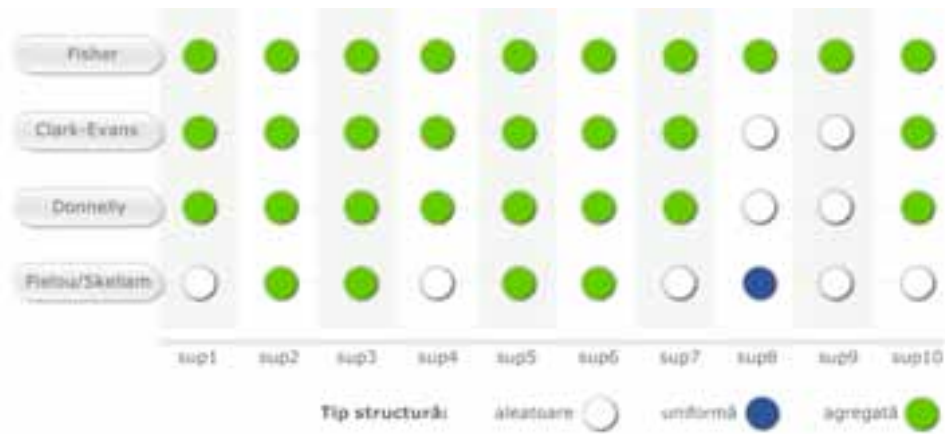

**Figura 7.5 Identificarea tipului de structură spațială în suprafețele studiate pe baza interpretării valorii și semnificației statistice a indicilor dependenți de distanță**

Valorile indicilor au fost calculate fără a corecta efectul de margine datorită faptului că literatura de specialitate recomandă folosirea corecțiilor pentru acest tip de prelucrări doar în cazul unui volum mic de evenimente (Illian, 2008).

În figura 7.5 este prezentată sintetic situația pe suprafețe a tipului de structură spațială determinată prin intermediul indicilor dependenți de distanță. Valorile indicilor și a testelor statistice utilizate sunt prezentate în anexa 9. Comparativ cu situația identificată prin metoda quadratelor se observă unele mici diferențe, cu precădere în suprafețele 8 și 9, unde doar indicele Fisher a indicat o abatere semnificativă de la CSR în direcția unei contagiozități pozitive. În suprafețele 8 și 9 au fost înregistrate tiparele cele mai apropiate de distribuția Poisson, în restul suprafețelor fiind înregistrate majoritar tipare agregate.

În ceea ce privește capacitatea indicilor de a surprinde informația spațială se constată diferențe mai ales în privința indicilor Fisher și Pielou/Skellam, indici cu relații de calcul simpliste. În literatura de specialitate indicele Clark-Evans (sau varianta Donnelly a acestui indice care include o corecție a efectului de margine) este considerat cel mai expresiv, datorită modului de fundamentare - faptul că include direct în relația de calcul comparația cu distanța medie în ipoteza CSR.

Pentru a explica diferențele dintre suprafețe privitoare la tendința procesului contagios de distribuție în spațiu au fost studiate relațiile dintre indicii de determinare a tiparului spațial și o serie de caracteristici structurale ale suprafețelor analizate. În anexa 10 se prezintă situația completă a valorilor coeficienților de corelație.

În continuare se face referire doar la relația indicelui de agregare Clark-Evans cu atributele structurale selectate (figura 7.6). S-a ales acest indice datorită versatilității și intensei utilizări în lucrările de cercetare. În plus comportamentul celorlalți indici în relația cu atributele structurale este similar.

Desimea influențează în mod cert agregarea puieților, acest aspect fiind evidențiat în unele cercetări (Pretzsch, 1997) precum și în analiza precedentă prin metoda quadratelor. Valorile coeficienților de corelație indică legături puternice, distinct semnificative și foarte semnificative. Pentru indicele Clark-Evans s-a înregistrat un coeficient de corelație de -0,805**, cele mai puternice legături fiind cu indicele Fisher (-0,918***) și Skellam (0,967***), în relația de calcul a celui din urmă intervenind direct desimea. Se confirmă încă odată că la creșterea desimii, datorită competiției intense se accentuează procesele repulsive (contagiozitatea negativă).

Parametrii biometrici influențează în mică măsură distribuția spațială surprinsă de indici, singurele relații evidente fiind surprinse în cazul înălțimii și volumului coroanei. Acestea sunt în general de intensitate slabă, fără acoperire statistică. Singura relație semnificativă s-a înregistrat între volumul coroanei și indicele Fisher (0,710*). Creșterea înălțimii medii poate conduce la o intensificare a proceselor repulsive dintre puieți, aspect natural și firesc. Este însă interesant în cazul volumului coroanei (nu și a suprafeței exterioare a coroanei) că la o creștere a mediei acestui parametru se constată o amplificare a tendințelor de agregare. În această situație consider că trebuie analizată cu prudență relația dintre cele două elemente, media volumului coroanei fiind un parametru cu o reprezentativitate extrem de scăzută ținând cont de neomogenitatea sa (coeficienți de variație în general de peste 150-200%, chiar de peste 400% în suprafețele 3 și 6). În sprijinul acestei păreri vine și faptul că nu s-a obținut o corelație asemănătoare cu suprafața exterioară a coroanei, un parametru mult mai omogen şi care exprimă într-o măsură similară gradul de expansiune al coroanei puieţilor pe verticală şi orizontală.

S-a studiat şi legătura dintre intensitatea agregării şi indicatorii diversităţii structurale (calculaţi pentru înălţimi). S-au constatat legături slabe cu indicii clasici de apreciere a neomogenităţii structurale (Shannon, Simpson), singurul indice pentru care s-au obţinut intensităţi slabe spre medii ale corelaţiei (dar nesemnificative din punct de vedere statistic) fiind indicele Gleason. Legături mai intense (de intensitate slabă şi medie) şi în anumite cazuri semnificative (0,659* pentru indicele Skellam) au fost stabilite cu indicele fundamentat în această lucrare – IDIV. Cele mai intense legături s-au realizat cu valoarea medie IDIV observată şi nu cu valorile potenţiale IDIV_min sau IDIV_max. Valori superioare ale coeficienţilor de corelaţie au fost înregistrate pentru legătura cu IDIV_50 (pasul de analiză de 50cm) comparativ cu IDIV_25.

Organizarea spaţială stabileşte legături slabe sau chiar neconcludente cu indicii de diferenţiere dimensională T, indiferent de numărul de vecini consideraţi în relaţia de calcul, fiind mai intense relaţiile cu indicii de dominanţă, în special U var 2 – varianta a doua de calcul, propusă de Gadow şi Hui (1999) şi Aguirre (et al., 2003).

Rezultatele indică faptul că o creştere a diversităţii structurale a înălţimilor conduce la creşterea uniformităţii distribuţiei în spaţiu a puieţilor. Observaţia este foarte interesantă şi se poate explica prin tendinţa unei utilizări optime a spaţiului – la o utilizare optimă a spaţiului pe verticală (indicată de valorile superioare ale indicilor de diversitate structurală) se produce şi o utilizare optimă a spaţiului disponibil pe orizontală, datorită presiunii competiţionale exercitate de puieţii din stratul superior asupra celor din straturile inferioare. O valorificare inadecvată a spaţiului disponibil pe verticală (omogenitatea înălţimilor) implică o valorificare defectuoasă şi a spaţiului în plan orizontal. În aceste condiţii este de înţeles tendinţa arboretelor în general de a tinde în timp spre tipare uniforme de distribuţie în suprafaţă (Gratzer, Rai, 2004; Houle, Duchesne, 1999; McDonald et al., 2003; Wolf, 2005), structuri mai stabile şi din punctul organizării spaţiale tridimensionale.

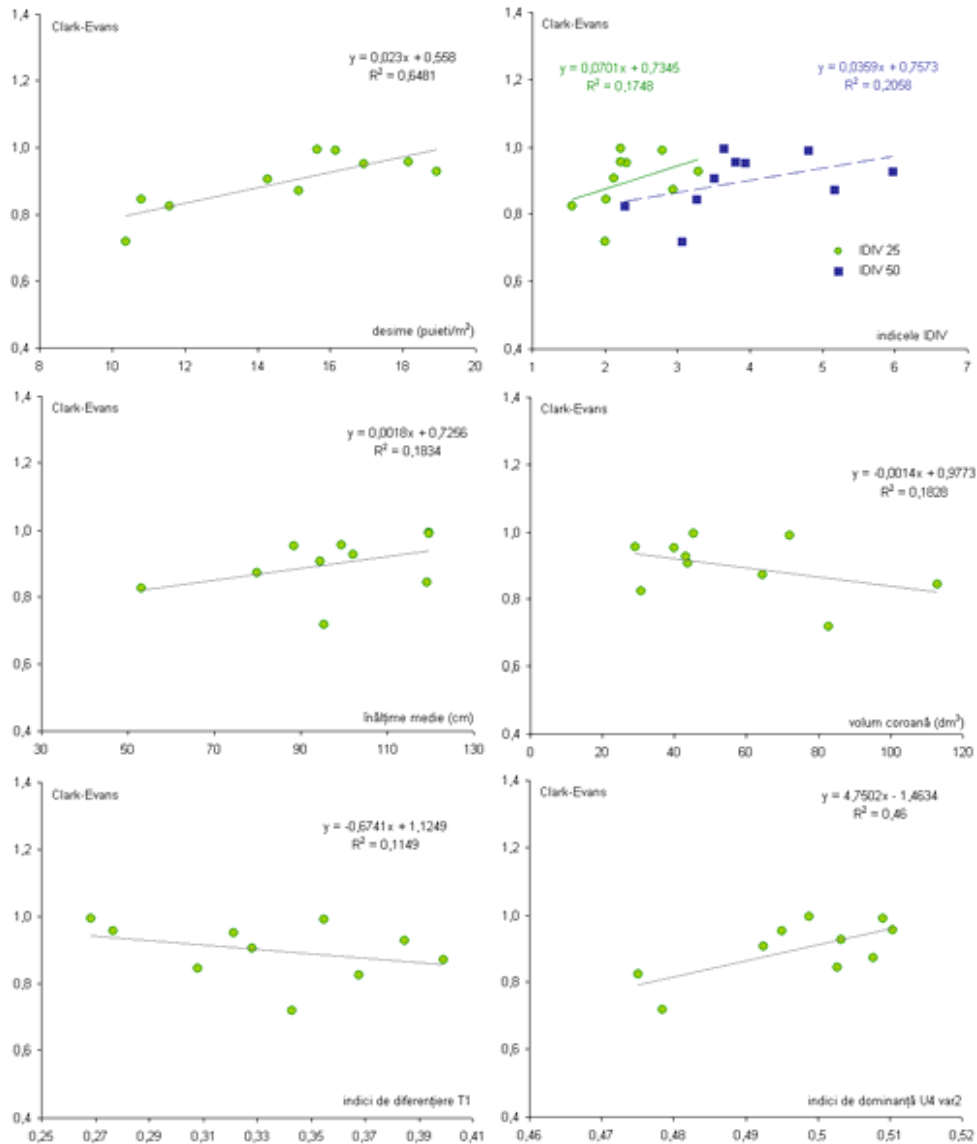

**Figura 7.6 Legătura dintre indicele Clark-Evans de determinare a structurii spațiale și diverși parametri structurali ai suprafețelor studiate**

## 7.3.2. Analiza prin intermediul funcţiilor bazate pe distribuţiile cumulative ale distanţelor

O altă metodă din gama celor dependente de distanţe este aşa numita analiză rafinată a distanţelor faţă de cel mai apropiat vecin (*Refined nearest neighbour analysis*) - o procedură ce compară un set de valori obţinute pe baza datelor experimentale cu un alt set obţinut printr-un număr de simulări Monte Carlo ale distribuţiei aleatoare a evenimentelor, folosind aceeaşi mărime şi formă a suprafeţei şi acelaşi număr de evenimente. Diggle (1979) propune analiza cu ajutorul unor funcţii dependente de distanţa faţă de cel mai apropiat vecin. Pentru fiecare distanţă dată w se calculează proporţia evenimentelor care se află la o distanţă mai mare de w faţă de orice limită exterioară pentru care distanţa până la cel mai apropiat vecin este mai mică decât w. Această funcţie/proporţie se notează cu G(w) şi se estimează calculând frecvenţele relative cumulate pentru distribuţia distanţelor faţă de primul vecin.

G(w) este o funcţie capabilă să identifice procesele contagioase pozitive sau negative prin comparaţia cu valoarea G(w) obţinută în ipoteza unei distribuţii aleatoare în suprafaţă. În ipoteza CSR, notăm această valoare G(w)$_{CSR}$ şi o calculăm folosind relaţia:

$$G(w)\_{CSR} = 1 - e^{-\lambda \cdot \pi \cdot w^2} \qquad (7.24)$$

$\lambda$ - densitatea de evenimente pe suprafaţă;

$w$ – clasa de distanţe până la cel mai apropiat vecin.

Analiza grafică a valorilor G(w) permite identificarea tiparului spaţial folosind două moduri de interpretare (Bailey, Gatrell, 2008):

- o creştere rapidă a G(w) la valori mici ale w indică agregarea iar o creştere bruscă spre final indică uniformitatea;
- valorile G(w) superioare G(w)$_{CSR}$ indică o structură agregată iar valorile inferioare G(w)$_{CSR}$ indică o structură uniformă.

Testarea semnificaţiei abaterii de la CSR se realizează prin generarea de *m* simulări Monte Carlo ale poziţiilor unui număr identic de evenimente într-o suprafață similară ca formă și dimensiuni. În baza analizelor datelor furnizate de simulări se poate stabili o *înfășurătoare de încredere* (*confidence envelope*) a G(w)$_{CSR}$ mărginită de valorile minime și maxime ale funcției G(w) calculate pentru cele *m* situații de simulare a distribuției aleatoare. Ipoteza CSR va fi respinsă doar în cazul în care valorile G(w) depășesc valorile funcției înfășurătoare pentru un anumit prag de semnificație. În literatura de specialitate termenul de înfășurătoare (*confidence envelope*) se folosește în locul intervalului sau limitei de încredere pentru a evidenția faptul că valorile acesteia au fost obținute prin simulare.

Pragul de semnificație depinde de numărul de simulări efectuate. Deși mulți cercetători obișnuiesc să folosească „acoperitor" un număr mare de simulări (de ordinul sutelor), din punct de vedere statistic este suficient un număr de 19 simulări pentru un prag de semnificație de 5%.

Pentru un număr de *m* simulări probabilitatea de transgresiune se poate calcula conform relației (Leemans, 1991; Bailey, Gatrell, 2008) :

$$q = 1/(m+1) \qquad (7.25)$$

În figura 7.7 se prezintă un exemplu de folosire a funcției G(w) pentru identificarea celor trei tipuri de structuri. Pentru generarea celor 19 simulări pe baza cărora a fost construită înfășurătoarea s-a folosit instrumentul software *SpPack* (Perry, 2004). Coordonatele puieților distribuiți aleatoriu în spațiu de către SpPack au fost încărcate în programul propriu SPATIAL și au fost prelucrate prin modulul de calcul al distanțelor.

Datorită faptului că uneori reprezentarea grafică a variației funcției G(w) este dificil de interpretat, s-a conceput o modalitate alternativă ce constă în calcularea unei valori $d_w$ obținută prin aflarea maximului diferenței dintre G(w) și G(w)$_{CSR}$:

$$d_w = \max \left| G(w) - G(w)_{CSR} \right| \qquad (7.26)$$

Rezultatul se compară cu valorile obținute prin mai multe simulări ale aceluiași număr de evenimente într-o suprafață de aceeași dimensiune cu cea studiată.

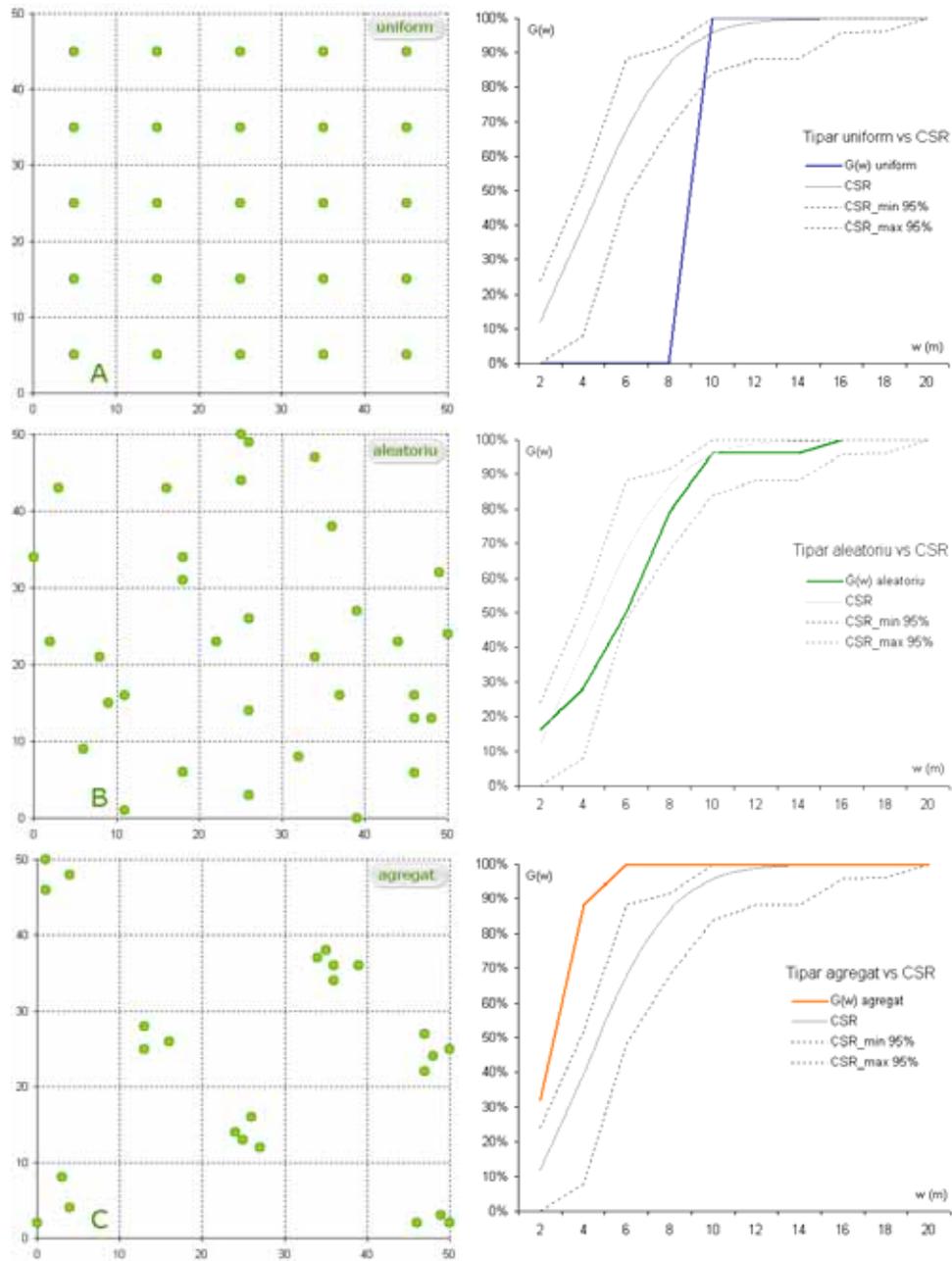

**Figura 7.7** Reprezentarea funcţiei G(w) şi a înfăşurătoarelor pentru probabilitatea de acoperire de 95% pentru un caz de structură spaţială uniformă (A), aleatoare (B), şi agregată (C)

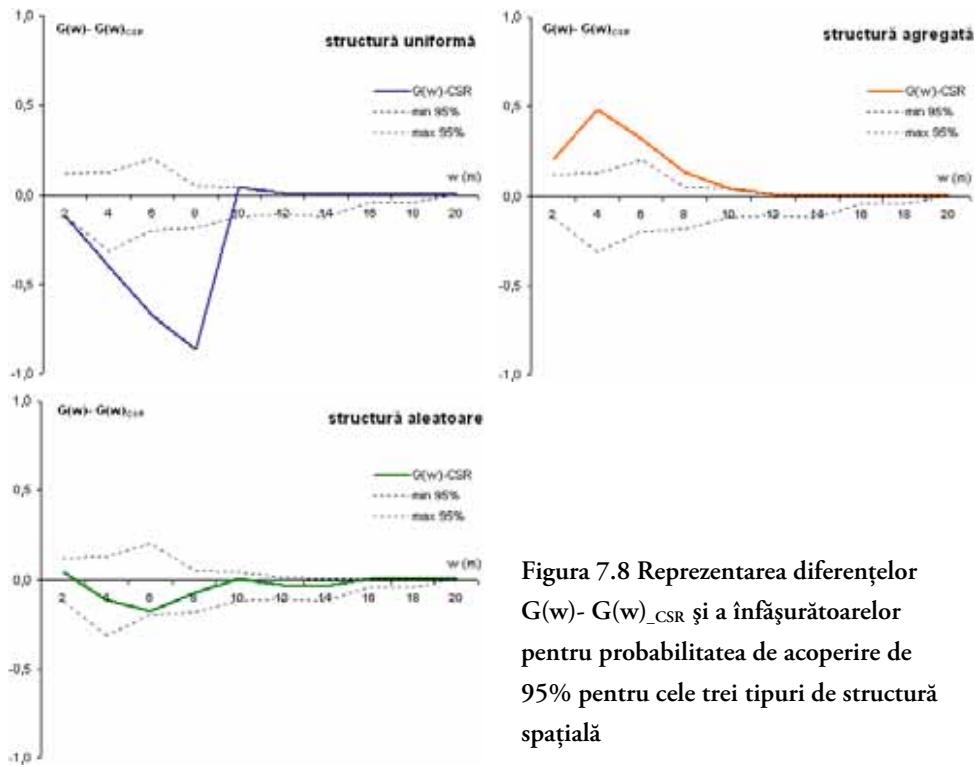

**Figura 7.8 Reprezentarea diferențelor G(w)- G(w)_CSR și a înfășurătoarelor pentru probabilitatea de acoperire de 95% pentru cele trei tipuri de structură spațială**

Analiza poate fi îmbunătățită prin reprezentarea grafică a diferențelor dintre G(w) și G(w)_CSR. Semnificația statistică se apreciază tot în baza simulărilor Monte Carlo efectuate conform ipotezei de distribuire aleatoare în spațiu (CSR) prin stabilirea valorilor ce definesc înfășurătoarea la pragul de semnificație dorit. Valorile pentru noua funcție de înfășurare se obțin prin calcularea diferențelor dintre valorile înfășurătoarei anterior determinate și G(w)_CSR și se reprezintă grafic.

În cazul în care diferențele G(w)- G(w)_CSR depășesc valorile înfășurătoarei, din punct de vedere statistic abaterile de la CSR sunt semnificative și se poate aprecia sensul acestora. Diferențele pozitive indică o structură agregată iar cele negative o structură uniformă. Pentru situația prezentată anterior în figura 7.7 s-a realizat și graficul diferențelor G(w)- G(w)_CSR pentru a putea observa îmbunătățirea percepției asupra abaterilor de la structura aleatoare.

Pentru a analiza caracteristicile tiparului spațial în suprafețele studiate s-au determinat funcțiile G(w) - frecvențele relative cumulate pentru distribuțiile distanțelor față de cel mai apropiat vecin în baza prelucrărilor efectuate de programul SPATIAL. Funcțiile G(w)_CSR și funcțiile înfășurătoare pentru pragul de semnificație de 5% s-au determinat prin prelucrările efectuate asupra a 190 de simulări (19 simulări pentru fiecare din cele 10 suprafațe de probă). Pentru simularea răspândirii aleatoare a puieților în spațiu s-a utilizat aplicația *SpPack* (Perry, 2004). Chiar dacă toate suprafețele au aceeași formă și dimensiune nu s-a putut folosi aceeași funcție G(w)_CSR pentru toate ariile studiate deoarece desimea puieților este diferită, iar distribuția Poisson are drept parametru densitatea evenimentelor.

Rezultatele prelucrărilor sunt prezentate grafic în calupul de figuri 7.9, sub forma funcției G(w) și a diferențelor G(w)-G(w)_CSR. Interpretările diagramelor susțin determinările efectuate anterior privitoare la distribuția în spațiu a puieților, dar evidențiază aspecte privitoare la contagiozitate la o altă rezoluție. Analizele confirmă abateri semnificative din punct de vedere statistic de la ipoteza CSR în toate suprafețele (mai puțin evidente în suprafețele 8, 9 și 10) în direcția unei contagiozități pozitive. Metoda analizei rafinate a distanțelor față de cel mai apropiat vecin oferă în plus față de metodele anterioare informații cu privire la variabilitatea în suprafață a tiparului spațial. Mai precis se permite identificarea abaterilor de la CSR pe o distanță teoretic echivalentă cu maximul distanței față de primul vecin – în realitate distanța relevantă pentru care se pot face analize pertinente este mai mică deoarece la apropierea de distanța maximă valorile G(w) se apropie foarte mult de cele ale G(w)_CSR.

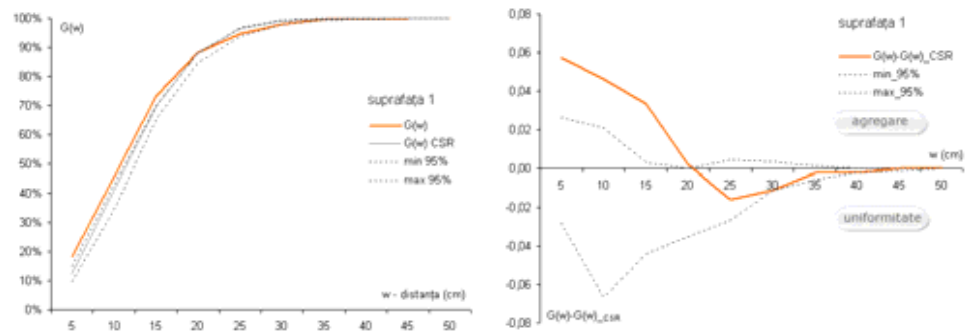

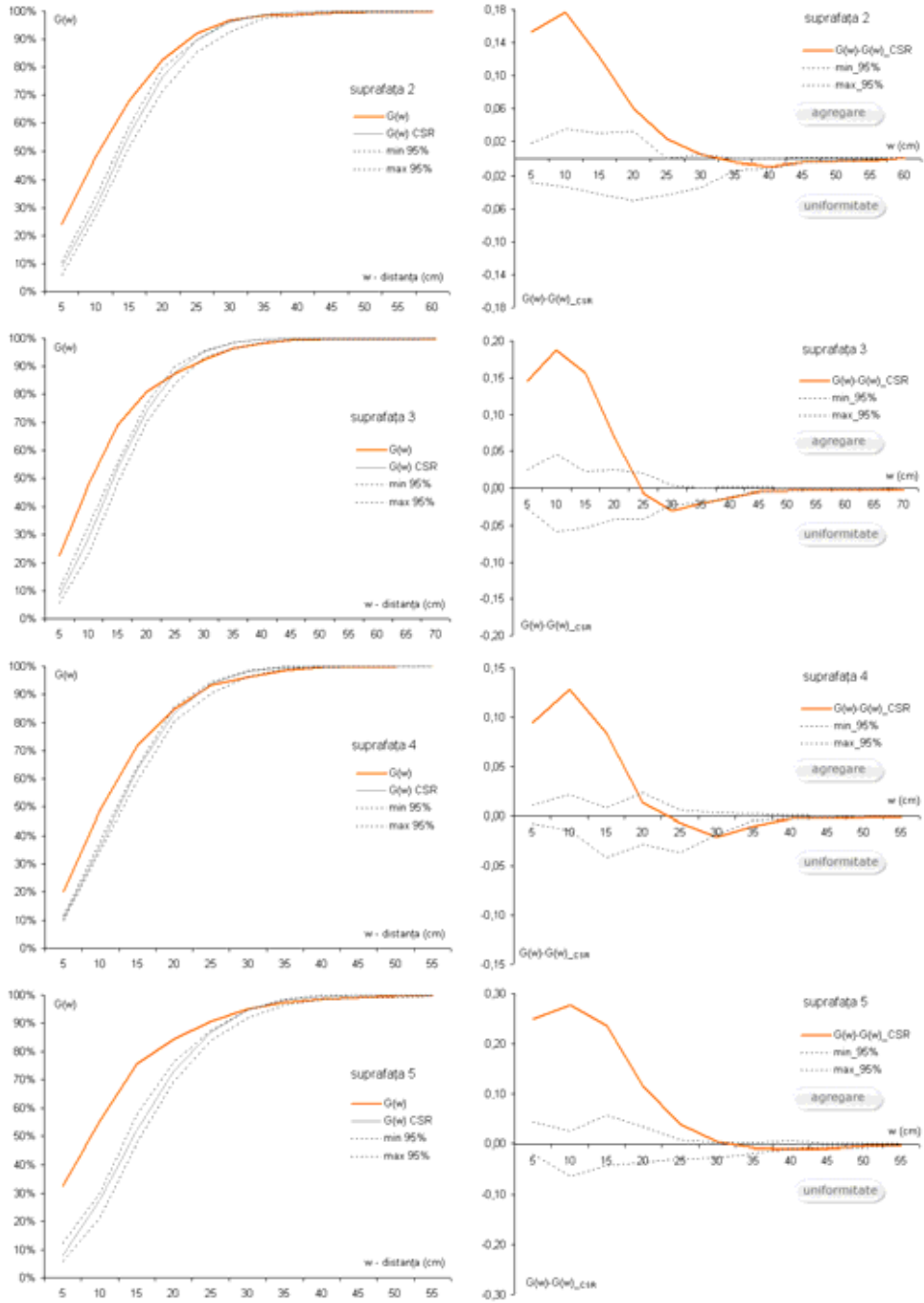

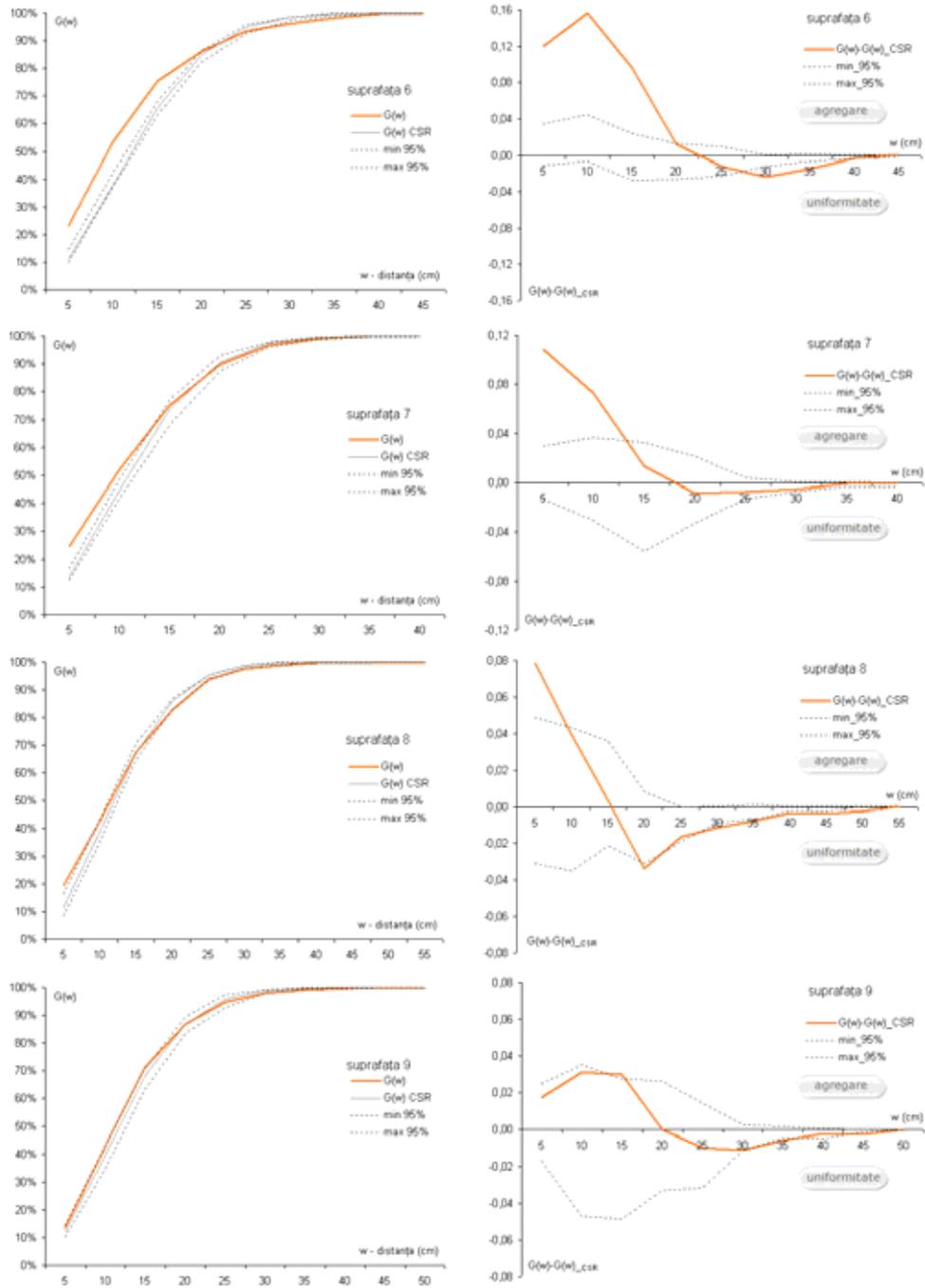

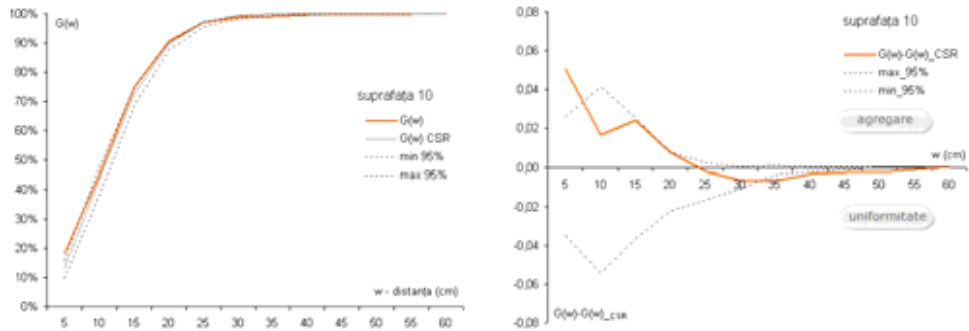

**Figura 7.9 Reprezentarea funcției G(w), a diferențelor G(w)- G(w)_CSR și a înfășurătoarelor pentru probabilitatea de acoperire de 95% pentru suprafețele studiate**

Se constată că procentul arborilor care au distanța până la cel mai apropiat vecin mai mică de 10-30 cm este mai mare decât în cazul unei distribuiri aleatoare în spațiu (la un prag de acoperire statistică de 5%), fapt care poate să conducă la ideea agregării puieților. Precauția manifestată în afirmația precedentă este dictată de faptul că analizele se referă doar la cel mai apropiat vecin și nu se ține cont de poziția celorlalți puieți învecinați. Distanța de agregare surprinsă de funcția G(w) variază în suprafață de la 5-10 cm (suprafața 10) până la un maxim de 30 cm (suprafețele 2 și 5), fiind corelată negativ cu desimea (r = -0.84**).

Pentru a îmbunătăți rezultatele acestui tip de analiză se poate folosi distanța medie dintre evenimente și primii *k* vecini. Fortin și Dale (2005) susțin că în general distanțele calculate pentru primul sau primii 2 vecini sunt foarte importante, fiind foarte sensibile la abaterile semnificative ale structurii de la CSR.

Un avantaj al acestei metode este faptul că testul se poate adapta oricărui alt tip de distribuție pentru a formula ipoteza nulă și nu necesită limitarea doar la ipoteza CSR, corespunzătoare distribuției omogene Poisson.

Dar există și dezavantaje ale acestei metode: i) prelucrările sunt influențate direct de numărul vecinilor care prin poziția lor determină distanțele; ii)nivelul analizei este în general prea detaliat și pe distanțe prea mici pentru o interpretare a grupurilor de indivizi; iii) analiza are relevanță doar pentru valori ale distanței w mai mici decât distanța maximă față de evenimentul vecin înregistrată în suprafață – peste această distanță G(w) are o valoare constantă (egală cu 1).

### 7.3.3 Analiza prin metoda celor mai apropiați k vecini (KNN)

Metodele din gama celui mai apropiat vecin (*NN – nearest neighbour*) derivate din cercetările lui Clark și Evans (1954) se referă doar la distribuția distanțelor față de primul vecin. Este interesant și util să se determine distanța față de al doilea, al treilea sau al $k$-lea vecin prin utilizarea așa numitelor metode ale celor mai apropiați k vecini (*KNN – $k^{th}$ nearest neighbours*).

Aplicația SPATIAL efectuează prelucrările asupra coordonatelor puieților, fiind calculată implicit distanța față de cel mai apropiat puiet, dar este oferită prin cel de al doilea modul de analiză și posibilitatea calculului distanței medii față de un număr maxim de vecini introdus de la consolă. S-a considerat că distanța față de primul vecin poate să ofere informații asupra spațiului propriu de dezvoltare al fiecărui individ iar distanța față de ceilalți vecini poate să conducă la informații relevante privitoare la relațiile de grup.

Thompson (1956) a prezentat în lucrarea „*Distribution of Distance to Nth Neighbour in a Population of Randomly Distributed Individuals*" o metodă de analiză a distribuției în spațiu a evenimentelor pe baza distanțelor față de un număr $k>1$ al celor mai apropiați vecini, autorul considerând că în acest mod se pot determina caracteristici spațiale la o scară mai largă decât în cazul folosirii celui mai apropiat vecin.

În această situație distanța față de primii $k$ vecini în ipoteza CSR, a răspândirii randomizate în spațiu se poate calcula folosind formula:

$$d_{K\_CSR} = \frac{k \cdot (2 \cdot k)!}{(2^k \cdot k!)^2 \cdot \sqrt{\lambda}} \qquad (7.27)$$

cu $\lambda$ - densitatea de evenimente pe suprafață $\lambda = \dfrac{n}{A}$ \qquad (7.28)

$k$ – numărul de vecini considerat;

$n$ – numărul de evenimente din suprafață;

$A$ – aria suprafeței.

Pentru densitatea $\lambda$ a evenimentelor și distanțele $d_i$ către cel mai apropiat al k-lea vecin, se poate nota: $x_n = 2 \cdot \pi \cdot \lambda \cdot d_i^2$ (7.29)

Distribuția valorilor $x_n$ este o distribuție de tip $\chi^2$ cu *2·n·k* grade de libertate pentru care se poate defini un test de abatere de la distribuția aleatoare. Thompson recomandă calcularea valorilor $x_n$ și a mediei acestora,, urmând a fi comparate cu limitele intervalului de încredere:

$$\left[ \frac{(\sqrt{4 \cdot n \cdot k - 1} - 1.96)^2}{2n} \cdots \frac{(\sqrt{4 \cdot n \cdot k - 1} + 1.96)^2}{2n} \right]$$ (7.30)

Această modalitate de interpretare statistică este una discutabilă datorită unor greşeli strecurate în lucrarea originală a lui Thompson, care au fost ulterior preluate de numeroşi cercetători (Mumme et al., 1983).

Thompson propune şi o modalitate alternativă de testare a semnificației, asemănătoare cu cea propusă de Clark şi Evans (1954), prin care se compară media distanțelor observate $\overline{d}_k$ cu cea teoretică $\overline{d}_{K\_CSR}$ (media distanțelor în ipoteza CSR). Raportul $R_k = \dfrac{\overline{d}_k}{\overline{d}_{K\_CSR}}$ va fi egal sau apropiat de 1 în cazul unei distribuții aleatoare, va fi subunitar în cazul unei distribuții agregate şi supraunitar în cazul uneia uniforme. Limitele de încredere ale $\overline{d}_{K\_CSR}$, se află în interval:

$$\left[ \overline{d}_{K\_CSR} - \frac{1.96 \cdot \sigma_{\overline{d}_{K\_CSR}}}{\sqrt{n}} \cdots \overline{d}_{K\_CSR} + \frac{1.96 \cdot \sigma_{\overline{d}_{K\_CSR}}}{\sqrt{n}} \right]$$ (7.31)

$\sigma_{\overline{d}_{K\_CSR}}$ fiind abaterea standard a mediei $\overline{d}_{K\_CSR}$

Distanțele față de primii 2, 3 şi 4 vecini în ipoteza răspândirii aleatoare sunt date de formulele:

$$\overline{d}_{2\_CSR} = \frac{0.75}{\sqrt{\lambda}} \qquad \overline{d}_{3\_CSR} = \frac{0.9375}{\sqrt{\lambda}} \qquad \overline{d}_{4\_CSR} = \frac{1.09375}{\sqrt{\lambda}}$$ (7.32)

iar $\sigma_{\overline{d}_k}$ - abaterile standard ale mediilor aritmetice sunt egale cu:

$$\sigma_{\overline{d}_{2\_CSR}} = \frac{0.2723}{\sqrt{\lambda}} \qquad \sigma_{\overline{d}_{3\_CSR}} = \frac{0.2757}{\sqrt{\lambda}} \qquad \sigma_{\overline{d}_{4\_CSR}} = \frac{0.2777}{\sqrt{\lambda}}$$ (7.33)

În cazul unor valori mai mari ale numărului de vecini (ce creşte implicit dificultatea calculării factorialului) formula 7.27 poate fi aproximată prin relaţia:

$$d_{K\_CSR} \approx \frac{\sqrt{k}}{\sqrt{\pi \cdot \lambda}} \qquad (7.34)$$

iar abaterea standard a mediei devine: $\sigma_{\bar{d}_4} = \frac{0.2821}{\sqrt{\lambda}}$ (7.35)

O altă modalitate de testare a semnificaţiei constă în generarea unei înfăşurătoare pentru raportul R prin simulări Monte Carlo ale distribuţiei aleatoare a evenimentelor.

Programul SPATIAL calculează prin modulul KNN raportul $R_k$ şi efectuează interpretarea statistică în baza intervalului de încredere al $\bar{d}_{K\_CSR}$, conform formulei 7.31. Dacă valoarea $\bar{d}_k$ observată este mai mică decât limita intervalului de încredere se respinge ipoteza CSR în favoarea unei distribuţii agregate, iar dacă este mai mare decât limita superioară a intervalului se poate trage concluzia că tiparul spaţial este unul uniform.

Rezultatele prelucrărilor sunt prezentate în continuare într-o formă grafică (figura 7.10), fiind reprezentat în diagrame raportul R şi limitele inferioare şi superioare ale acestuia corespunzătoare distribuţiei aleatoare. Valorile distanţelor medii faţă de cei mai apropiaţi 1-15 vecini şi distanţele corespunzătoare în ipoteza CSR, precum şi limitele inferioare şi superioare ale intervalului de încredere calculat se regăsesc în anexa 11.

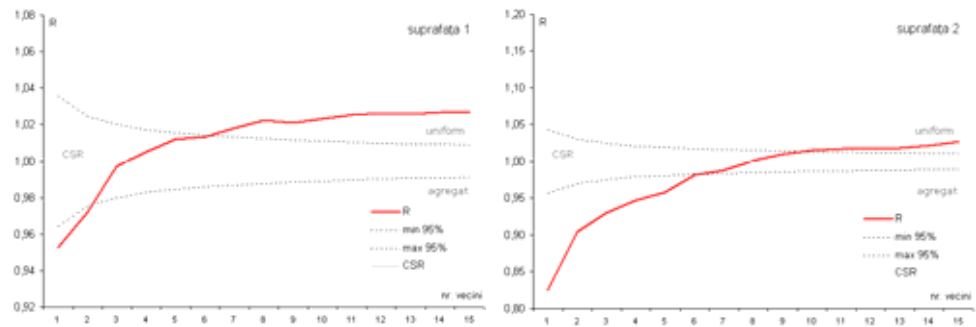

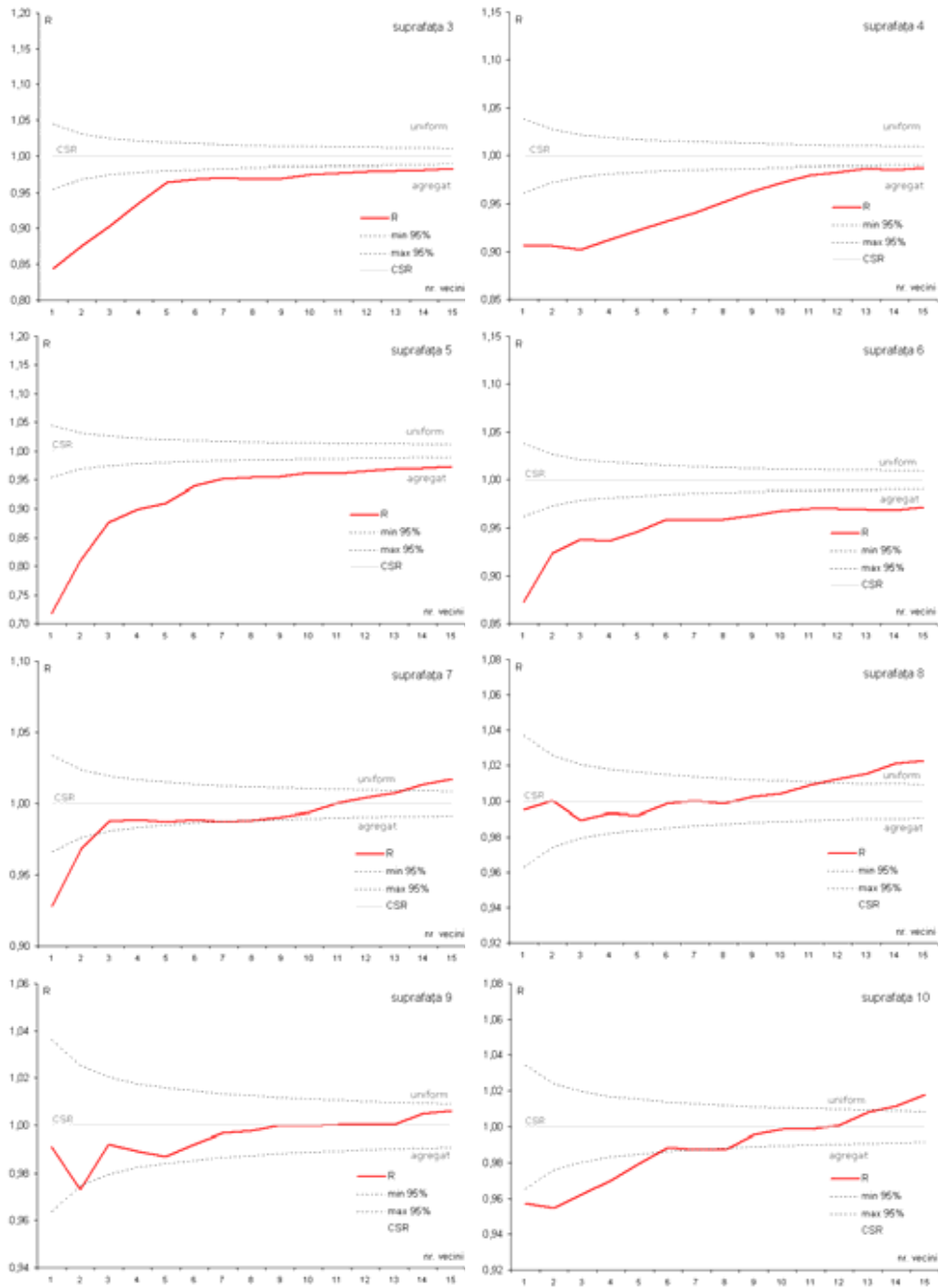

**Figura 7.10 Reprezentarea raportului $R_k$ și a limitelor acestuia inferioare și superioare la probabilitatea de acoperire de 95% pentru suprafețele studiate**

Analiza celor mai apropiați k vecini (KNN) are avantajul că permite identificarea agregării nu la diferite variante ale valorilor distanței dintre puieți precum metodele anterioare ci la diferite valori ale numărului de vecini. În acest mod se poate stabili numărul de indivizi din aglomerările de puieți. O valoare a raportului $R_k$ subunitară indică faptul că distanța $\overline{d}_k$ față de al k-lea puiet este mai mică față de distanța teoretică $\overline{d}_{K\_CSR}$ care s-ar înregistra în ipoteza unei distribuții aleatoare, drept urmare în suprafața analizată primii $k$ vecini ai puieților sunt mai aglomerați decât în mod normal. Numărul de puieți din nucleul identificat ca având o structură agregată va fi de $k+1$.

În general în arboretele mature analizele KNN implică valori ale lui k de maxim 5, dar pentru analiza suprafețelor regenerate această valoare este mică dată fiind densitatea în suprafață a puieților. Drept urmare, pentru suprafețele studiate s-a ales să se efectueze această analiză pentru un număr de vecini de până la 15.

În cele zece suprafețe amplasate agregarea puieților implică un număr diferit de indivizi, începând de la grupări de 2-3 indivizi (în suprafețele 1, 7, 9), grupări de până la 6-7 indivizi (suprafețele 2 și 10) sau chiar aglomerări de peste 16 puieți (cazul suprafețelor 3, 4, 5, 6). Chiar și în cazul situațiilor din urmă se poate constata un maxim al tendinței procesului contagios pozitiv la anumite valori ale dimensiunii grupului, după care are loc o stabilizare a acestei tendințe - maxim la 2-6 puieți (suprafața 3), 2-4 puieți (suprafața 4), 2-7 puieți (suprafața 5 și 6).

Singura suprafață în care nu s-au înregistrat abateri semnificative statistic în direcția agregării, indiferent de numărul de vecini considerat, este suprafața 8. În această suprafață abateri semnificative de la CSR s-au constatat doar în direcția uniformității, la valori ale lui $k$ mai mari de 12 – situația diferită a suprafeței 8 a fost evidențiată și prin analizele anterioare dependente de distanța față de primul vecin (atât prin indicii sintetici cât și prin funcția G(w)). Odată cu raportarea la un vecin mai îndepărtat (creșterea valorii $k$), se reduce diferența dintre $\overline{d}_k$ și $\overline{d}_{K\_CSR}$, ba chiar apare tendința de uniformitate, fenomen manifestat în suprafețele 1 (pentru $k \geq 7$), 2 ($k \geq 10$), 7 ($k \geq 14$), 8 ($k \geq 12$) și 10 ($k \geq 14$).

Datorită capacității metodei de analiză de a identifica nu doar tiparul spațial ci și numărul de puieți din cadrul unei grupări s-a considerat oportună analiza a trei procese punctiforme diferențiate pe straturi dimensionale ale puieților. S-a folosit înălțimea drept parametru dimensional, fiind utilizate straturile definite anterior în capitolul 5.2.2, cu o modificare – straturile mediane 2 și 3 au fost grupate astfel că s-au obținut în acest caz doar 3 straturi:

- stratul 1 – al plantulelor (înălțimea 0 .. 25 cm);
- stratul 2 – al puieților din clasa medie (înălțimea 26 .. 150 cm);
- stratul 3 – al puieților de mari dimensiuni (înălțimea > 150 cm).

În figura 7.11 este prezentată sinteza acestei analize.

Harta completă a proceselor punctiforme și graficele variației raportului $R_k$ (pe baza cărora s-a generat situația sintetică) în cele zece suprafețe se pot consulta în anexa 12. În suprafețele 8 și 9 s-a înregistrat un număr de plantule prea mic pentru efectuarea prelucrărilor, iar în suprafața 2 un număr prea mic al puieților de mari dimensiuni.

Analiza situației sintetice relevă informații interesante asupra organizării spațiale a celor trei straturi în funcție de numărul $k$ al celor mai apropiați vecini la care se realizează prelucrările. În stratul plantulelor agregarea se produce doar la nivelul grupurilor de două plantule (mai rar în grupuri de 5-7 plantule – în suprafețele 7 și 10). Preponderentă este distribuția uniformă a plantulelor, în majoritatea suprafețelor acest tipar spațial devenind vizibil începând cu analiza grupurilor de 5-6 plantule.

Analiza grafică a hărților procesului spațial corespunzător stratului 1 (anexa 12) relevă o distribuție relativ uniformă în zona acoperită de plantule dar foarte neomogenă la nivelul întregii suprafețe – aspect foarte evident în suprafețele 3, 4, 5, 6, 10. Chiar dacă răspândirea este neomogenă, se poate afirma că între plantule nu apar procese de agregare decât în grupuri foarte mici, pentru grupurile de peste 5-6 indivizi fiind semnificative abaterile de la CSR în direcția evidențierii proceselor contagioase repulsive.

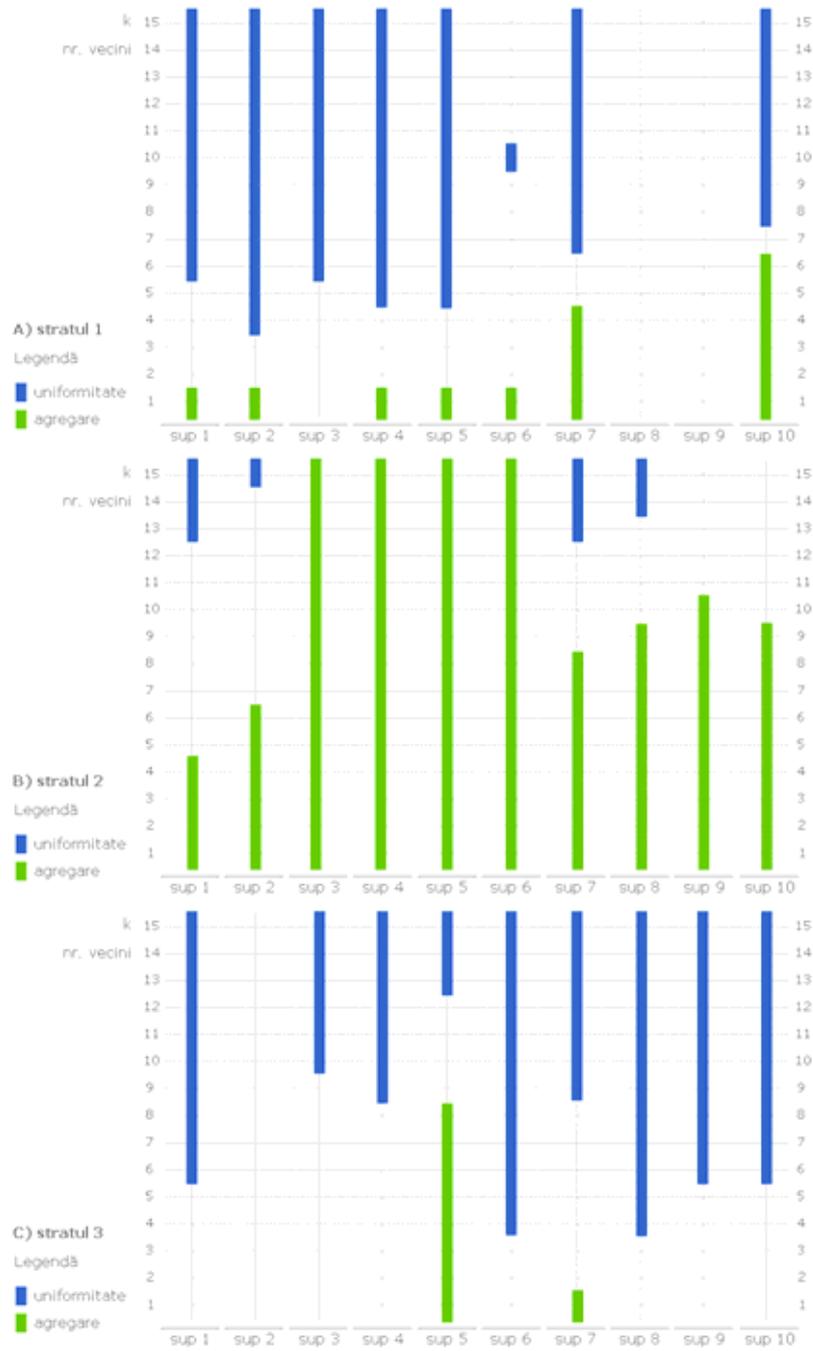

**Figura 7.11** Situația sintetică a identificării tiparului spațial pe baza raportului $R_k$ pentru procesele punctiforme corespunzătoare celor 3 straturi dimensionale definite

În cel de al doilea strat agregarea este prezentă în toate suprafețele analizate, începând de la grupuri de 5-11 indivizi și chiar până la grupuri de peste 16 indivizi (suprafețele 3, 4, 5, 6).

Procesele contagioase pozitive evidențiate arată că majoritatea puieților (stratul al doilea având o pondere de peste 77% din totalul puieților) sunt organizați conform unui tipar spațial agregat, fapt ce corespunde observațiilor făcute în numeroase lucrări de cercetare (McDonald et al., 2003; Awada, et al., 2004; Gratzer, Rai, 2004; Camarero et al., 2005; Paluch, 2005; Fajardo et al., 2006; Maltez-Mouro et al., 2007; Montes et al., 2007).

Tipul de organizare spațială arată că în rândul acestui strat nu se înregistrează fenomene repulsive intense, agregarea putând fi explicată și prin particularitățile de reproducere a speciilor (prezența în suprafață a unor specii cu o mare capacitate de regenerare vegetativă), mecanismele dispersiei sau neomogenitatea condițiilor pedologice.

În cel de al treilea strat predomină caracterul uniform, procesele repulsive fiind semnificative în toate suprafețele de la valori ale dimensiunii grupului de peste 5 indivizi. Au fost înregistrate structuri agregate în doar două suprafețe (5 și 7), la valori reduse ale dimensiunii grupurilor. Situația este normală, puieții de dimensiuni mari inhibând dezvoltarea altora de dimensiuni similare în apropiere, rezultatul fiind o distribuire de tip uniform în suprafață.

Informațiile rezultate în urma analizelor de tip *KNN* pot conduce la elucidarea unor aspecte interesante, cu implicații asupra practicii forestiere, în special în stabilirea dimensiunii optime a biogrupelor (a numărului de puieți și a spațierii acestora).

## 7.3.4. Analiza modului de organizare spațială prin metode ale statisticii spațiale de ordinul al II–lea

Metodele quadratelor și chiar cele bazate pe distanțele dintre evenimentele vecine prezintă anumite dezavantaje, menționate anterior în lucrare. Pentru îmbunătățirea determinărilor tiparelor spațiale unii autori recomandă utilizarea concomitentă a mai multor metode complementare de analiză pentru a identifica abaterile semnificative de la ipoteza CSR (Fortin și Dale 2005). Au fost concepute și soluții alternative de analiză spațială de ordin I care combină tehnica quadratelor cu metodele distanțelor, un bun exemplu în acest sens fiind tehnicile de tip SADIE (*Spatial Analysis by Distance IndicEs*) (Perry, 1996; Perry et al 1996, 1999), care, deși sunt apărute recent au fost incluse de Fortin et al (2002) între tehnicile adecvate cercetărilor ecologice. Analiza SADIE (sau tehnica fragmentelor roșii-albastre – *red-blue plots*) determină intensitatea agregării și este capabilă să identifice forma și dimensiunea pâlcurilor, respectiv a ochiurilor din interiorul suprafeței studiate.

În prezent se consideră că cele mai potrivite metode de analiză a proceselor punctiforme sunt cele care aparțin statisticii spațiale de ordinul al II-lea (Dale, 2004). Metodele statisticii spațiale de ordinul al II-lea evaluează tiparul de organizare la un nivel secundar – nivelul local, fiind bazate pe varianța datelor, pe când metodele de ordinul I sunt considerate acele metode care oferă informații referitoare la atributele spațiale ale suprafeței luate ca întreg, pe baza analizei mediei datelor. Analiza spațială de ordinul al II-lea are o bază teoretică solidă și este capabilă să ofere informații mult mai detaliate privitoare la organizarea în suprafață, la diferite rezoluții.

Funcțiile de tip K sau K-Ripley au fost fundamentate în urmă cu trei decenii de către Ripley (1976, 1977) în lucrările devenite clasice „*The Second-Order Analysis of Stationary Point Processes*", respectiv „*Modelling spatial patterns*" și au cunoscut transformări și adaptări care le-au sporit gradul de adaptabilitate la situațiile diferite în care sunt folosite.

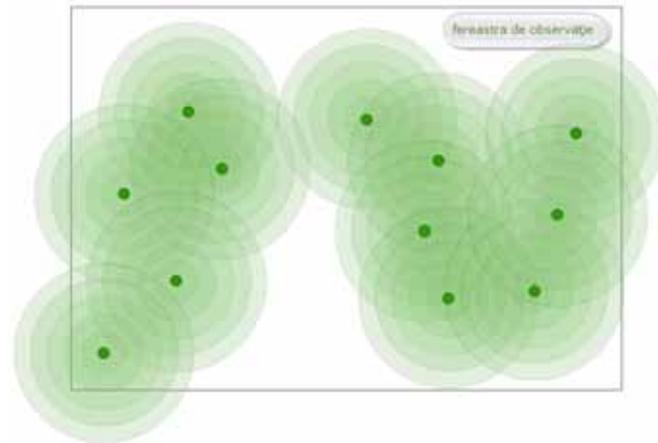

**Figura 7.12 Principiul de determinare a funcției *K(t)***

Fundamentarea funcției K-Ripley este simplă – valoarea sa reprezintă în principiu numărul de evenimente situate la o distanță *t* față de un eveniment ales aleatoriu drept referință. Pentru calculul valorii *K(t)* se numără toate evenimentele situate într-un cerc cu raza egală cu *t* și centrul în poziția evenimentului referință iar apoi se mărește iterativ valoarea razei *t* cu un pas dat și se contorizează numărul evenimentelor situate la noua distanță *t* față de centrul cercului (figura 7.12). În final media evenimentelor pentru fiecare clasă de distanțe se raportează la densitatea *λ* a evenimentelor din suprafață pentru a se obține valoarea funcției.

Spre deosebire de indicatorii metodelor bazate pe distanța față de cel mai apropiat vecin, funcția *K(t)* înregistrează valori mai mari în cazul tiparelor spațiale agregate și mai mici în cazul celor uniforme. Dacă se ignoră efectul de margine apărut atunci când o parte din cercuri se situează parțial în afara ferestrei de observație se poate considera că media valorilor în cazul unui proces Poisson omogen staționar este de $\lambda \cdot \pi \cdot h^2$.

În situația unei distribuiri a unui număr de *n* evenimente în concordanță cu ipoteza CSR numărul evenimentelor situate într-un cerc cu raza egală cu *t* este:

$$NE = \lambda \cdot K(t) = \frac{n}{A} \cdot K(t) \tag{7.36}$$

unde *λ* este densitatea evenimentelor în suprafața de arie *A*, iar *n* reprezintă numărul de evenimente din suprafață.

Definiția formală se bazează pe o funcție indicatoare $I$, descrisă prin:

$$I_{ij}(t) = \begin{cases} 1, & dacă \ d_{ij} \leq t \\ 0, & dacă \ d_{ij} > t \end{cases} \tag{7.37}$$

unde $d_{ij}$ reprezintă distanța de la evenimentul $i$ la evenimentul $j$.

Din punct de vedere matematic funcția $K(t)$ se estimează prin formula:

$$\hat{K}(t) = \frac{1}{\lambda \cdot n} \cdot \sum_i \sum_j \frac{I_t}{w_{ij}} \tag{7.38}$$

unde $w_{ij}$ reprezintă o pondere de corectare a efectului de margine, fiind proporția din aria cercului de rază $t$ care se află în afara ferestrei de observație.

Pentru corecția efectului de margine au fost propuse mai multe variante (Ripley, 1981; Besag, 1977; Getis, Franklin, 1987; Stoyan, Stoyan, 1994; Haase, 1995; Goreaud și Pellisier, 1999), fiecare dintre acestea având o modalitate diferită de determinare a valorii proporției $w_{ij}$:

$$w_{ij} = 1 - \cos^{-1}(e/d_{ij})/\pi \qquad \text{(Getis, Franklin, 1987)} \quad (7.39)$$

$$w_{ij} = 1 - \left[ \cos^{-1}(e_1/d_{ij}) + \cos^{-1}(e_2/d_{ij}) + \pi/2 \right]/2\pi \quad \text{(Getis, Franklin, 1987)} \quad (7.40)$$

$$w_{ij} = 1 - \left[ 2 \cdot \cos^{-1}(e_1/d_{ij}) + 2 \cdot \cos^{-1}(e_2/d_{ij}) \right]/2\pi \qquad \text{(Haase, 1995)} \quad (7.41)$$

S-a notat cu $e$ distanța de la evenimentul $i$ la cea mai apropiată limită, iar cu $e_1$ și $e_2$ distanțele de la evenimentul $i$ la cele mai apropiate două margini. Formula 7.39 este folosită în cazul în care $d_{ij} > e$, varianta 7.40 în situația în care distanța de la evenimentul $i$ la cel mai apropiat colț (pentru suprafețele rectangulare) este mai mică decât $d_{ij}$, iar ultima variantă în cazul în care distanța de la evenimentul $i$ la cel mai apropiat colț (pentru suprafețele rectangulare) este mai mare decât $d_{ij}$.

Interpretarea funcției se bazează pe comparația acesteia cu valoarea $K(t)$ în ipoteza CSR. Dacă valoarea estimată este mai mare decât $\pi \cdot t^2$ se indică un tipar agregat, iar dacă este mai mică un tipar uniform. Funcția $K(t)$ are o creștere exponențială în funcție de distanță, fiind uneori dificil de interpretat reprezentarea sa grafică. În plus, pentru comparația la diferite valori ale distanței $t$ s-a dorit obținerea unei funcții care să fie constantă în ipoteza CSR.

Besag (1977) a propus standardizarea funcției prin liniarizarea sa, această variantă a funcției *K(t)* fiind cunoscută drept funcția *L(t)*, cu relația:

$$L(t) = \sqrt{\frac{K(t)}{\pi}} \qquad\qquad (7.42)$$

Frecvent este folosită și o altă formă de standardizare:

$$L(t) = \sqrt{\frac{K(t)}{\pi}} - t \qquad\qquad (7.43)$$

Interpretarea grafică a funcției *L(t)* (în varianta dată de formula 7.43) se realizează mult mai ușor decât în cazul funcției *K(t)* (figura 7.13). Valorile pozitive ale *L(t)* indică un model spațial agregat, iar cele negative un model regulat (uniform). Valorile apropiate de zero indică o distribuție aleatoare în spațiu a evenimentelor.

Testarea semnificației se poate face prin aproximarea unui interval de încredere echivalent cu $\pm 1.42 \cdot \sqrt{A}/(n-1)$ pentru pragul de semnificație de 5% și $\pm 1.68 \cdot \sqrt{A}/(n-1)$ pentru cel de 1%.

O metodă mai des folosită presupune generarea înfășurătoarelor de încredere prin simulări Monte Carlo. Aceasta este o tehnică frecvent utilizată în testele neparametrice, fiind o modalitate acceptată de testare a semnificației abaterilor de la ipoteza nulă în cazul funcțiilor de tip K-Ripley (Besag și Diggle, 1977).

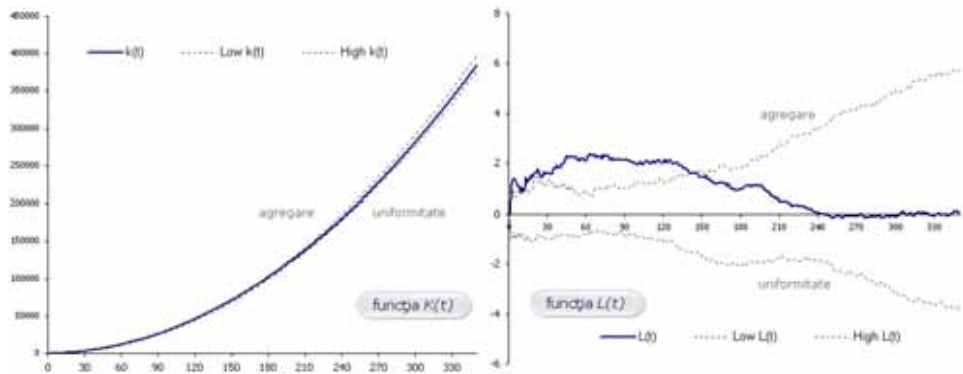

**Figura 7.13 Diferența dintre funcția K(t) și L(t) în reprezentarea aceluiași proces punctiform (toți puieții - suprafața 1)**

Numărul de simulări ce vor fi efectuate depinde de pragul de semnificaţie impus. Teoretic, din perspectivă statistică este suficient un număr de 19 simulări pentru asigurarea unui prag de semnificaţie de 5%. Pentru un număr de $m$ simulări probabilitatea de transgresiune se calculează conform relaţiei (Leemans, 1991):

$$m = (1/q)-1 \qquad\qquad (7.44)$$

În unele cazuri valorile limită necesare calculului înfăşurătorii de încredere se determină nu pe baza valorilor extreme ci folosind cele mai apropiate $r$ valori (de obicei $r = 5$) de limită (Perry, 2004). În această situaţie numărul de simulări necesare creşte:

$$m= (r/q)-1 \qquad\qquad (7.45)$$

Personal consider că folosind a doua variantă (neincluzând direct extremele în înfăşurătoare) este teoretic posibil să se respingă ipoteza nulă chiar şi pentru datele produse de o simulare, date care au fost generate folosind legea de distribuţie Poisson. Conform formulei 7.45 pentru un prag de semnificaţie de 5% sunt necesare 99 de simulări. Desigur, un număr mai mare de simulări nu dăunează, dar ar trebui specificat pentru fiecare analiză numărul de simulări sau pragul de semnificaţie.

În cercetarea silvică din străinătate, identificarea tiparului spaţial prin metode ale statisticii spaţiale de ordinul al doilea a devenit una din analizele cele mai frecvente în ultimii ani, numeroase cercetări fiind desfăşurate chiar în seminţişuri (Leemans, 1991; Awada, et al., 2004; Camarero et al., 2005; Fajardo et al., 2006; Gratzer, Rai, 2004; Hofmeister et al., 2008; Maltez-Mouro et al., 2007; McDonald et al., 2003; Montes et al., 2007; Nigh, 1997; Paluch, 2005). În România au fost realizate puţine cercetări în acest domeniu, metodele analizei spaţiale de ordinul al II-lea fiind folosite abia în ultimul deceniu (Popa, 2006; Roibu, Popa, 2007), dar nefiind utilizate în regenerări.

În continuare sunt prezentate rezultatele analizelor efectuate cu ajutorul funcţiei de tip Ripley $L(t)$. Pentru prelucrarea datelor s-au folosit aplicaţiile software *SpPack* (Perry, 2004) şi *SPPA* (Haase, 2002), ambele cu acordul autorilor. Rezultatele furnizate de cele două programe informatice au fost similare, existând mici diferenţe între ele doar în modul de concepere a testelor de semnificaţie şi în ergonomia de utilizare. În final a fost preferat *SpPack* datorită integrării în sistemul de calcul tabelar *Microsoft Excel*. Prelucrările au fost efectuate pe distanţa de maxim 350 cm - jumătate din latura minimă a suprafeţei (Haase, 1995), cu un pas de analiză de 5 cm, aplicând corecţii ponderate pentru înlăturarea efectului de margine, pentru determinarea înfăşurătoarelor de încredere fiind generate 99 de simulări.

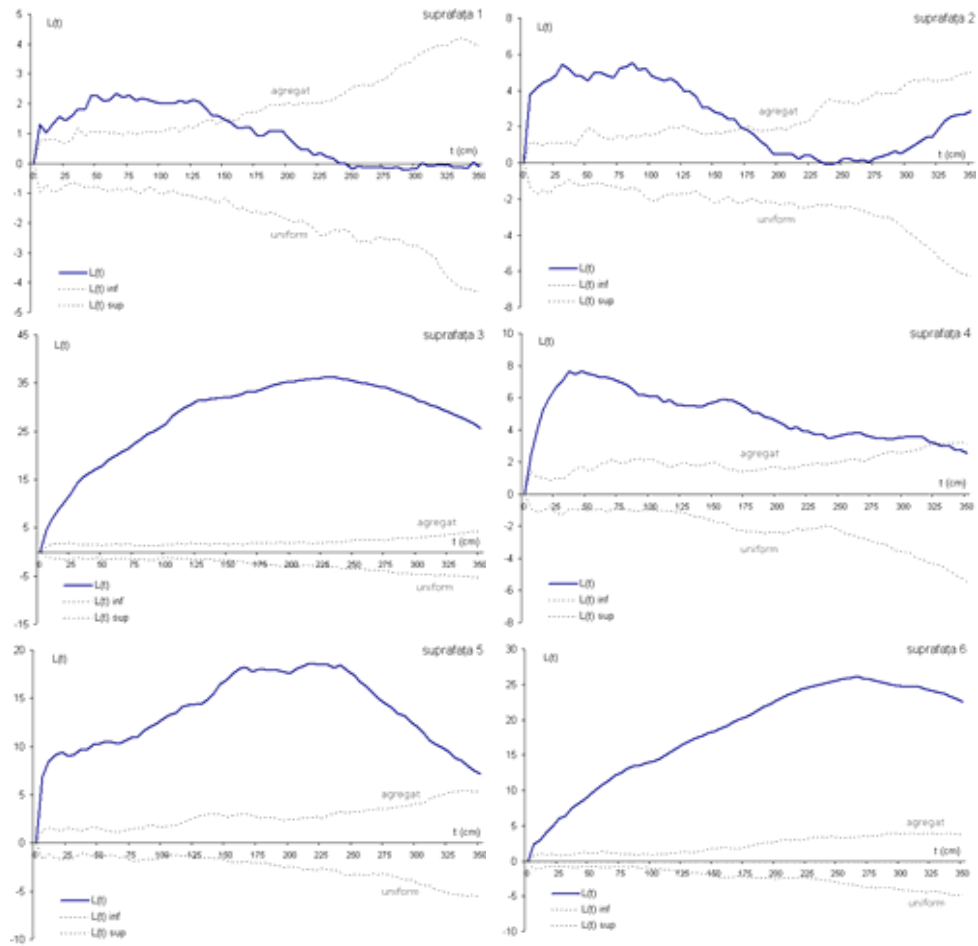

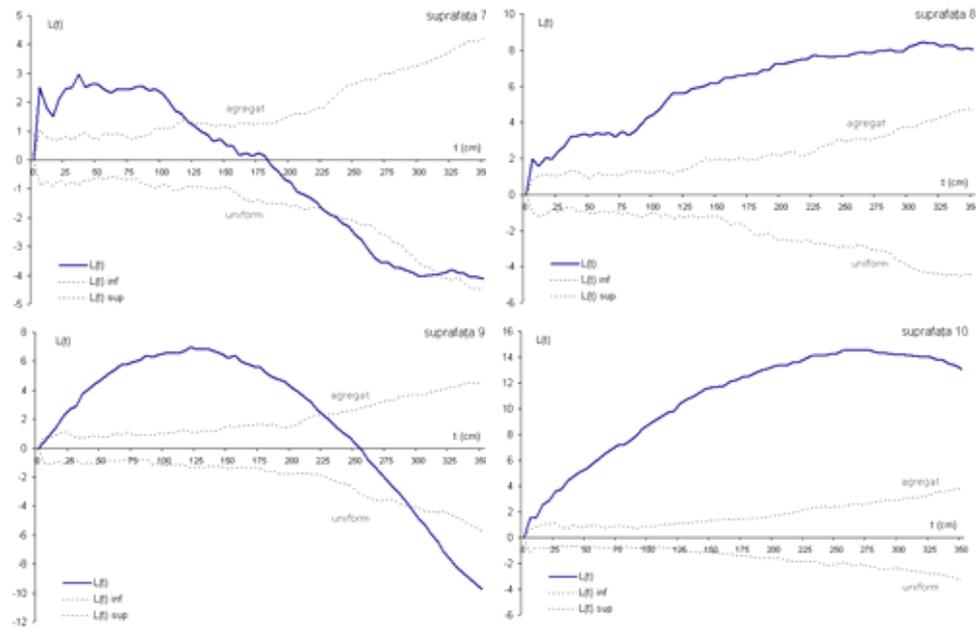

**Figura 7.14 Determinarea tiparului spaţial al distribuţiei puieţilor în suprafeţele studiate prin interpretarea grafică a funcţiei L(t)**

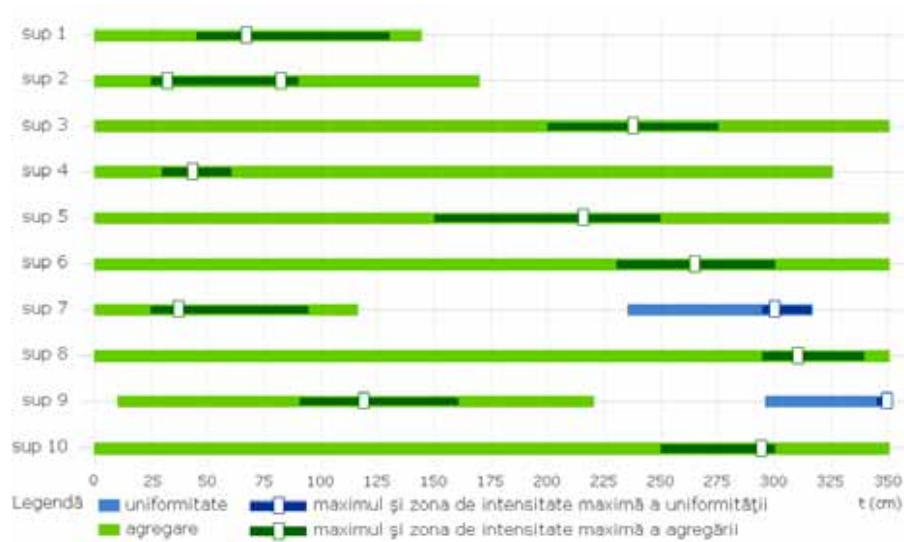

**Figura 7.15 Centralizatorul determinărilor tiparului spaţial prin intermediul L(t) pe suprafeţe**

Rezultatele arată unele similitudini cu situația determinată prin metoda *KNN*, dar se observă rezoluția spațială superioară a funcției *L(t)* precum și sensibilitatea sporită în cazul distribuțiilor apropiate de cea aleatoare. Multe din situațiile pentru care celelalte metode nu au putut determina faptul că abaterile de la CSR erau semnificative au fost soluționate de către analiza spațială de ordinul al II-lea (aspect evident în suprafețele 8, 9 sau 10). Tiparul spațial agregat a fost identificat în absolut toate suprafețele până la valori ale distanței de 120-220 cm sau în unele cazuri chiar pe toată lungimea analizată (suprafețele 3, 5, 6, 8, 10). Abateri spre structura uniformă au fost observate doar în suprafața 7 și 9, la distanțe de peste 230 cm.

În centralizatorul din figura 7.15 au fost marcate și intervalele cu intensitatea maximă de agregare și maximele absolute înregistrate în fiecare suprafață. Sunt evidente în diagramă două intervale în care agregarea este maximă – de 30-90 cm (în suprafețele 1, 2, 4, 7), respectiv de 220-310 cm (în suprafețele 3, 5, 6, 8, 10). Acest fapt conduce la ideea că puieții au tendința de a forma fie nuclee mai mici cu diametrul de 30 până la 90 cm fie aglomerări de dimensiuni mai mari – de 2-3 m (pentru vizualizarea hărților proceselor punctiforme în suprafețele elementare se poate consulta anexa 8).

S-a considerat oportună și efectuarea analizei la nivelul celor trei procese punctiforme diferențiate pe straturi dimensionale ale puieților în funcție de înălțime. Au fost utilizate cele trei straturi definite în capitolul 7.3.3: stratul 1 – al plantulelor (înălțimea 0 .. 25 cm); stratul 2 – al puieților din clasa medie (înălțimea 26 .. 150 cm) și stratul 3 – al puieților de mari dimensiuni (înălțimea > 150 cm). S-a efectuat atât o analiză individuală univariată a acestora cât și analiza bivariată pentru a observa eventualele interacțiuni dintre cele trei straturi.

Graficele variației funcției L(t) folosite pentru interpretarea tiparului spațial în cazul proceselor punctiforme corespunzătoare celor 3 straturi de puieți definite în funcție de înălțime pot fi consultate în anexa 13. Pe baza acestora au fost realizate diagramele centralizatoare prezentate în figura 7.16.

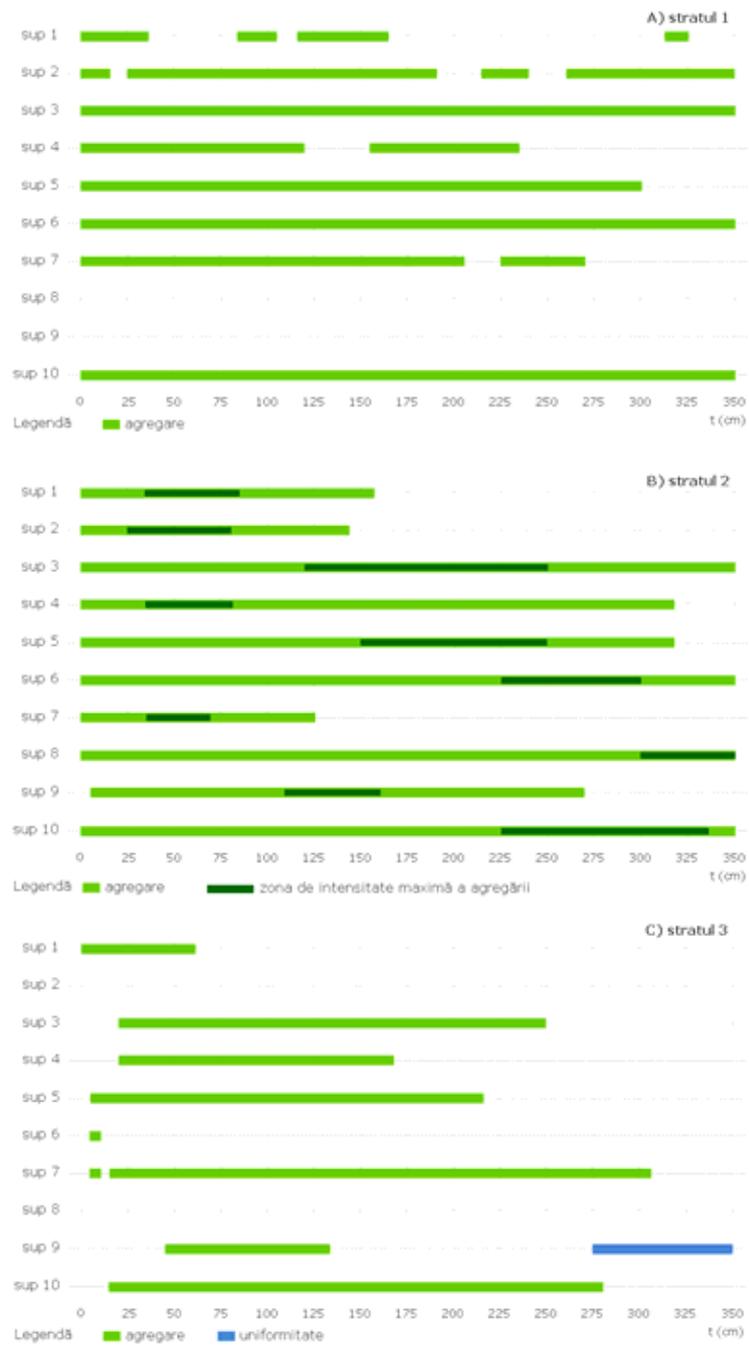

**Figura 7.16 Centralizatorul de determinare a modelului de distribuție spațială a puieților din cele trei straturi dimensionale cu ajutorul funcției L(t) – analiza univariată**

În toate straturile şi toate suprafeţele este preponderent fenomenul de agregare, tiparul uniform fiind întâlnit într-o singură suprafaţă (nr. 9) în stratul 3 (al puieţilor de mari dimensiuni), pentru valori ale distanţei de peste 275 cm. În stratul puieţilor de mari dimensiuni se remarcă o agregare care apare pe intervale mai reduse decât în celelalte straturi şi pentru valori ale distanţei în general mai mari de 15-25 cm (stratul 3 fiind singurul strat în care agregarea nu se produce începând de la cele mai mici distanţe). Puieţii de mari dimensiuni manifestă uneori chiar o tendinţă de repulsie (valori $L(t)$ negative dar situate în intervalul înfăşurătoarei) pe distanţa de până la 15-25 cm (evidenţiată în graficele pentru suprafeţele 4, 8 şi 10).

Agregarea se manifestă pregnant în stratul 2, al puieţilor de dimensiuni medii, intensitatea fiind maximă în grupurile cu dimensiunea de 30-80 cm (suprafaţa 1, 2, 4, 7), 120-250 cm (suprafaţa 3, 5, 9) sau peste 250 cm (6, 8, 10).

În stratul 1, al plantulelor, agregarea are un caracter fragmentat, fiind întâlnită în grupuri foarte variate ca dimensiune, un exemplu elocvent fiind cel din suprafaţa 1, în care procesele contagioase pozitive se întâlnesc în grupuri cu dimensiunea mai mică de 35 cm, precum şi în grupuri de 85-105 cm, 115-165 cm, 315-325 cm. În această situaţie grupurile mici sunt dispuse aleatoriu în grupuri mai mari care la rândul lor sunt dispuse aleatoriu în grupe şi mai mari. Rezultatele conduc la ideea că modelarea distribuţiei în suprafaţă a plantulelor ar putea fi analizată folosind şi alte tipuri de procese punctiforme – procese agregate sau compuse (e.g. procesele Neyman-Scott).

Pentru efectuarea analizei bivariate determinările au fost efectuate folosind conceptul proceselor etichetate sau marcate (Penttinen et al., 1992; Stoyan, Stoyan, 1994), straturile diferite constituind în acest context etichetele. Funcţiile de tip K-Ripley permit efectuarea analizelor bivariate (între două procese etichetate diferit) pe baza metodologiei şi relaţiilor fundamentate de Lotwick şi Silverman (1982).

Abordarea bivariată presupune folosirea a cel puţin două ipoteze nule, cercetătorul fiind nevoit să adopte metode diferite de testare a semnificaţiei pentru a nu confunda interpretările (Diggle, Cox, 1983; Goreaud, Pelissier, 2003).

Prima dintre ipoteze presupune distribuția independentă în spațiu a celor două procese, fiind folosită în cazul aprecierii interacțiunii dintre două specii diferite sau două categorii sociale/dimensionale diferite. Cea de a doua ipoteză nulă presupune etichetarea aleatoare a evenimentelor, aplicațiile ei făcând referire de exemplu la mortalitatea dintr-o populație de arbori sau la acțiunea unui fenomen perturbator accidental. În cazul de față a fost folosită prima ipoteză nulă, a independenței proceselor studiate.

Pentru analiza unui proces bivariat cu $n_1$ evenimente de tipul 1 și $n_2$ evenimente de tip 2 într-o suprafață de arie $A$, Lotwick și Silverman (1982) au definit funcțiile inter-tipuri:

$$\widetilde{K}_{12}(t) = \frac{1}{\lambda_2 \cdot n_1} \cdot \sum_{i=1}^{n_1} \sum_{j=1}^{n_2} \frac{I_{t_{12}}}{w'_{ij}} \tag{7.46}$$

$$\widetilde{K}_{21}(t) = \frac{1}{\lambda_1 \cdot n_2} \cdot \sum_{i=1}^{n_2} \sum_{j=1}^{n_1} \frac{I_{t_{21}}}{w''_{ij}} \tag{7.47}$$

unde $\lambda_1 = \frac{n_1}{A}$ și $\lambda_2 = \frac{n_2}{A}$ sunt densitățile evenimentelor iar $w'_{ij}$ și $w''_{ij}$ sunt ponderile asociate funcțiilor indicatoare în vederea corecției efectului de margine. Funcția indicatoare $I_{t12}$ are valoarea 1 dacă distanța dintre evenimentul $i$ de tip 1 și cel $j$ de tip 2 este mai mică sau egală cu $t$ și zero altfel. Similar se definește funcția $I_{t21}$ ținând cont de distanța dintre evenimentul $i$ de tip 2 și cel $j$ de tip 1. Cele două estimări ale funcției $K(t)$ se bazează pe contorizarea evenimentelor de un anumit tip care se află la distanța $t$ de evenimentele de celălalt tip.

$\widetilde{K}_{12}(t)$ și $\widetilde{K}_{21}(t)$ sunt estimări ale aceleiași funcții, astfel că Upton și Fingleton (1985, citați de Dale, 2004) au definit un estimator comun:

$$\hat{K}_{12}(t) = \frac{n_2 \cdot \widetilde{K}_{12}(t) + n_1 \cdot \widetilde{K}_{21}(t)}{n_1 + n_2} \tag{7.48}$$

Frecvent în analiza proceselor bivariate este folosită o standardizare a acestui estimator, similară celei definite de Besag (1977) pentru funcția $K(t)$:

$$\hat{L}_{12}(t) = \sqrt{\frac{\widetilde{K}_{12}(t)}{\pi}} - t \qquad (7.49)$$

Dacă valoarea estimată a funcției $L_{12}(t)$ este nulă distribuția în spațiu a evenimentelor aparținând unui tip este complet independentă de distribuția celuilalt tip. Valorile estimate mai mari decât zero indică o asociere pozitivă a celor două tipuri de evenimente, fenomen interpretat din punct de vedere ecologic drept efect de atracție. Valorile negative indică asocierea negativă sau segregarea, interpretată ecologic ca rezultat al efectului de repulsie (Goreaud, Pelissier, 2003).

Pentru interpretarea semnificației statistice a abaterilor s-au folosit înfășurătoarele de încredere ale valorii corespunzătoare ipotezei nule de independență, generate prin simulări Monte Carlo. Prelucrările au fost efectuate cu ajutorul aplicației software *SpPack* (Perry, 2004), pentru distanța de 350 cm, un pas de analiză de 5 cm, corecții ponderate pentru înlăturarea efectului de margine, fiind generate 99 de simulări pentru determinarea înfășurătoarelor de încredere.

Cele 24 de grafice folosite pentru determinarea interacțiunii dintre cele trei straturi dimensionale în cele 10 suprafețe se regăsesc în anexa 14. În baza acestora au fost concepute graficele centralizatoare prezentate în figura 7.17.

*Interacțiunea în spațiu a stratului plantulelor (stratul 1) cu celelalte straturi*

Plantulele realizează în toate cazurile analizate asocierea cu puieții de dimensiuni medii. Rezultatele înregistrate au arătat abateri semnificative de la ipoteza nulă a independenței distribuției în spațiu. Privitor la distanțele de asociere, acestea se pot situa în intervale fragmentate, mai reduse (suprafața 1, 2, 4) sau asocierea poate avea loc pe tot intervalul analizat (0-350 cm). Valorile cele mai intense ale asocierii nu se produc la distanțe mici, ci mai frecvent în intervalul 50-200 cm. A fost constatat și un caz ușor atipic în suprafața 7 unde asocierea de intensitate maximă s-a produs la distanțe mai mici (20-90 cm), dar la o distanță de peste 185 cm apare fenomenul de repulsie.

Rezultatele conduc la concluzia că în general puieții de dimensiuni medii manifestă toleranță și chiar suport pentru instalarea și dezvoltarea plantulelor, chiar dacă nu în imediata proximitate.

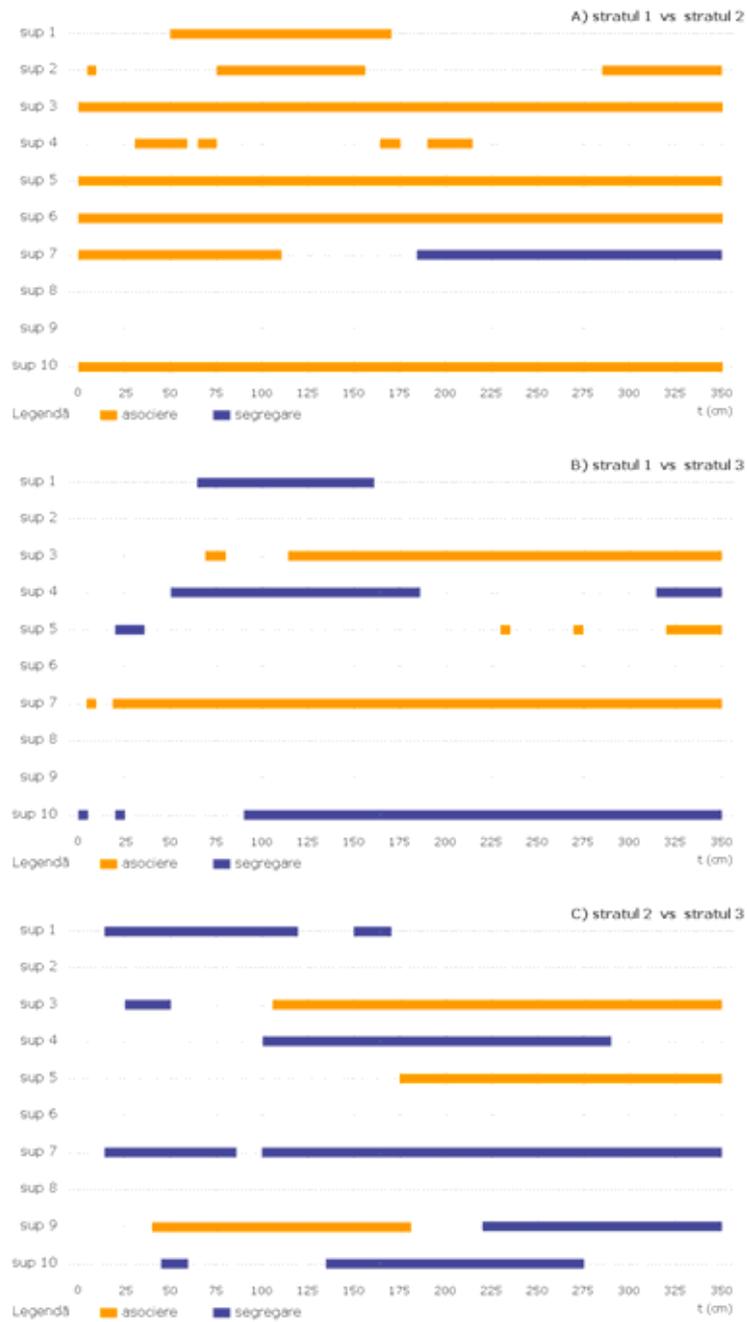

**Figura 7.17 Centralizatorul de determinare a tipului de interacțiune în spațiu dintre cele trei straturi de puieți definite în funcție de înălțime prin intermediul funcției $L_{12}(t)$**

Nu același lucru se poate afirma despre puieții de dimensiuni mari, rezultatele indicând mai degrabă fenomenul de segregare în relația acestora cu plantulele. Chiar dacă situația din centralizator arată doar un ușor avantaj al fenomenului de segregare analiza graficelor individuale (anexa 14) indică în toate suprafețele (mai puțin în suprafața 7) tendința de segregare dar la un nivel situat de multe ori în interiorul înfășurării de încredere. Puieții de mari dimensiuni acționează în general repulsiv față de plantule, deși există și situații (suprafața 3 și 7) în care s-a constatat o asociere pozitivă.

*Interacțiunea în spațiu a stratului puieților de dimensiuni medii (stratul 2) cu stratul puieților de dimensiuni mari (stratul 3)*

În această situație caracterul repulsiv este mai evident, în șase suprafețe abaterile semnificative de la ipoteza nulă de independență a poziționării indicând segregarea puieților din cele două straturi. Acțiunea repulsivă se declanșează de la distanțe de 20 cm și poate să se desfășoare chiar și până la distanțe de peste 350 cm. Poziționarea puieților de talie mare pare să influențeze mai pregnant distribuția în spațiu a puieților de dimensiuni medii față de cea a plantulelor. Explicația constă în faptul că puieții de talie medie au cerințe superioare plantulelor în ceea ce privește necesarul de spațiu individual de dezvoltare, fiind mai afectați de presiunea dominanței dimensionale a puieților din stratul 3. Există și în cazul interacțiunii dintre straturile 2 și 3 situații izolate în care s-a constatat asocierea dintre puieții celor două categorii. Coerența determinărilor este posibil să fie influențată de încadrarea în cele trei straturi, care s-a efectuat luând în considerație valorile înălțimii totale a puieților din toate suprafețele. S-a sacrificat în acest mod surprinderea unor situații particulare în suprafețele în care înălțimea puieților se abate de la media generală, dar s-a asigurat posibilitatea comparării rezultatelor obținute pentru suprafețe diferite. De asemenea o stratificare cu un nivel mai mare de detaliu ar putea oferi informații suplimentare, dar un număr mai mare de straturi conduce la creșterea exponențială a variantelor ce trebuie prelucrate și comparate. În condițiile a 4 straturi și 10 suprafețe este nevoie de 60 de analize individuale, iar pentru 5 straturi de 100 de analize individuale. Volumul mare de date pentru fiecare suprafaţă şi prelucrările foarte dificile chiar şi în condiţiile definirii a trei straturi au condus la alegerea variantei prezentate, considerate optime în condiţiile date.

În urma efectuării în suprafeţele studiate a unei game complete de analize, prin metode diferite, putem trage concluzia că abaterile de la ipoteza nulă a distribuţiei aleatoare în spaţiu se întâlnesc în toate pieţele de probă, preponderent în favoarea unei dispuneri agregate. Acest fapt este în concordanţă cu rezultatele a numeroase cercetări de analiză spaţială întreprinse în seminţişuri menţionate anterior în această lucrare.

Aşa-zisele „abateri" de la caracterul întâmplător par să fie mai degrabă o regulă în regenerările arboretelor. Testele abaterii de la CSR se efectuează în ipoteza proceselor Poisson omogene şi staţionare. În cazul în care se consideră că aceasta nu este cea mai potrivită ipoteză sau se doreşte analiza amănunţită a detaliilor de agregare există şi alternative – procese Poisson neomogene, agregate sau compuse, procese de tip Cox, Markov, Neyman-Scott, Gibbs.

Studiul proceselor punctiforme etichetate prin intermediul analizelor bivariate efectuate cu ajutorul funcţiilor de tip K-Ripley a reuşit să identifice şi fenomene de asociere sau repulsie între anumite categorii dimensionale de puieţi, indicând abateri semnificative de la ipoteza distribuţiei independente în spaţiu a acestor categorii, în conformitate cu cercetări similare menţionate în literatura de specialitate (Penttinen, et al 1992; Szwagrzyk, Czerwczak, 1993; Boyden et al., 2005; Camarero et al., 2005; Fajardo et al., 2006).

Tiparele aparent întâmplătoare sau independente ale distribuţiei puieţilor în spaţiu sunt structuri organizate care pot fi definite şi reproduse matematic prin intermediul proceselor punctiforme. Această constatare conduce la ideea că modelele individuale ale arborilor pot beneficia de informaţiile obţinute prin analiza spaţială, fiind posibilă generarea nu doar a distribuţiilor dimensionale ci şi a distribuţiilor distanţelor şi a modului de organizare în suprafaţă a indivizilor. Se poate considera că analiza spaţială este o etapă importantă a modelării care poate aduce numeroase îmbunătăţiri capacităţii predictive a modelelor forestiere.

## 7.4 Determinarea particularităților de asociere prin metode dependente de distanțe

Identificarea unor tendințe de asociere în ceea ce privește distribuția în spațiu a speciilor are nu doar o importanță științifică ci și o relevanță deosebită pentru practica silvică, informațiile putând fi folosite în alcătuirea compozițiilor de regenerare, precum și în stabilirea formulelor de împădurire, a modului de asociere și a dispozitivului de amplasare a speciilor în cazul intervențiilor de instalare artificială a vegetației forestiere.

În subcapitolul 5.1, pe baza corelațiilor realizate între desimile speciilor pe quadrate de 1x1 m, au fost formulate câteva ipoteze privitoare la identificarea unor tendințe pozitive și negative ale modului de asociere a speciilor. Obiectivul acestui subcapitol îl reprezintă testarea ipotezelor formulate anterior prin metode adecvate oferite de statistica spațială.

În ecologie, interesul pentru problema modului de asociere în spațiu a speciilor este manifestat de foarte mult timp. Primele analize în acest sens s-au desfășurat folosind împărțirea în quadrate a suprafeței și contorizarea apariției concomitente a speciilor de interes. Această metodă „rudimentară" este influențată foarte mult de alegerea dimensiunii quadratului. Pielou (1961) a dezvoltat prima metodă bazată pe distanțele dintre indivizi, folosind tabele de contingență create pe baza informațiilor referitoare la specia individului referință și a primului său vecin. Autoarea considera că speciile pot fi considerate segregate când au tendința de a fi grupate în pâlcuri, astfel încât orice individ este mai probabil să fie găsit lângă membrii propriei sale specii decât lângă indivizii celeilalte specii.

Pielou a explicat segregarea indivizilor, menționând două cauze:

- agregarea indivizilor de aceeași specie determinată de reproducerea vegetativă, dispersia grupată a semințelor și competiția inter-specifică;
- heterogenitatea habitatului - percepută ca variație spațială a unor factori de mediu ce poate interacționa cu cerințele ecologice diferite ale speciilor și toleranța lor distinctă față de anumiți factori.

Indicele de segregare S definit de Pielou se calculează folosind un raport între numărul perechilor mixte (constituite din specii diferite) observate și cel așteptat la o distribuție independentă. Pentru o suprafață cu $N$ indivizi, dintre care $N_A$ indivizi ai speciei A și $N_B$ indivizi ai speciei B ($N=N_A+N_B$) se poate defini tabelul de contingență:

| Specia individului de referință | Specia celui mai apropiat vecin | | Total |
|---|---|---|---|
| | A | B | |
| A | $f_{AA}$ | $f_{AB}$ | $N_A$ |
| B | $f_{BA}$ | $f_{BB}$ | $N_B$ |
| Total | $n_A$ | $n_B$ | $N$ |

$$S = 1 - \frac{f_{AB} + f_{BA}}{N_A \cdot n_B + N_B \cdot n_A} \qquad (7.50)$$

Valorile $S$ apropiate de zero indică o populație nesegregată, independentă din punctul de vedere al distribuției în spațiu. Valorile pozitive indică tendința de segregare, maximul atingând unitatea în condițiile de repulsie perfectă între cele două specii (nici un individ nu are cel mai apropiat vecin un individ aparținând celeilalte specii). Valorile negative sunt caracteristice asocierii pozitive dintre specii, minimul fiind atins la -1, în cazul în care numărul de indivizi ai speciei A este egal cu cel al speciei B iar distribuția în spațiu este formată din perechi perfect izolate de tip AB (e.g. o specie parazitată de o alta). Pentru testarea semnificației statistice Pielou recomandă folosirea unui test $\chi^2$.

Deși indicele de segregare Pielou a fost și este încă folosit frecvent pentru interpretarea raporturilor de interacțiune în spațiu, Dixon (1994) a subliniat câteva probleme de interpretare și concepere a acestui indice, indicând situații în care indicele S nu este capabil să distingă segregarea. Dixon a efectuat de asemenea câteva modificări și recomandări cu privire la metodologia propusă de Pielou de folosire a tabelelor de contingență în aprecierea segregării.

Cercetări privitoare la aspectele de asociere au mai fost efectuate de Fuldner (1995) (citat de Aguirre et al., 2003) și von Gadow (1999). Ei au definit variante diferite ale unui indice de amestec al speciilor (*species mingling index*) care reflectă capacitatea unei specii de a forma grupuri pure. Fundamentarea indicilor este bazată pe metode dependente de distanță, fiind folosită identitatea specifică a celor mai apropiați vecini.

Datorită inconvenientelor pe care le implică metodele celor mai apropiați vecini se consideră că cele mai adecvate modalități de studiere a asocierii sunt metodele bivariate de analiză spațială de ordinul al II-lea (Dale, 2004). Funcțiile de tip Ripley pentru analiză bivariată $K_{12}(t)$ și $L_{12}(t)$ sunt capabile să ofere o imagine detaliată a proceselor de interacțiune dintre două specii raportate la distribuția în spațiu, fiind menționate în literatura forestieră de specialitate numeroase cercetări în acest domeniu (Szwagrzyk, Czerwczak, 1993; Beland et al., 2003; Dimov, 2004).

În lucrarea de față s-a optat pentru studierea asocierii/segregării speciilor prin analiza bivariată cu ajutorul funcției $L_{12}(t)$ (Lotwick și Silverman, 1982). Principiul acesteia și modalitatea de interpretare au fost prezentate în subcapitolul anterior. Pentru prelucrarea datelor a fost folosită componenta *SpPack* (Perry, 2004). Interpretarea semnificației statistice a abaterilor s-a efectuat prin compararea valorilor obținute cu cele ale înfășurătoarelor de încredere corespunzătoare ipotezei nule de independență, generate prin 99 de simulări Monte Carlo. Analizele au fost efectuate pentru distanța de 350 cm cu un pas de analiză de 5 cm și corecții ponderate pentru înlăturarea efectului de margine.

Graficele utilizate pentru determinarea interacțiunii spațiale dintre specii se regăsesc în anexa 15. Cu ajutorul acestora au fost concepute diagramele centralizatoare prezentate în figura 7.18. Suprafețele care nu se regăsesc în situația prezentată sintetic au înregistrat un număr prea mic de puieți dintr-o specie implicată în analiză, în consecință prelucrările nu au mai fost efectuate pentru acestea. Perechile de specii pentru care s-a efectuat analiza au fost alese conform ipotezelor de asociere a speciilor formulate în subcapitolul 5.1.

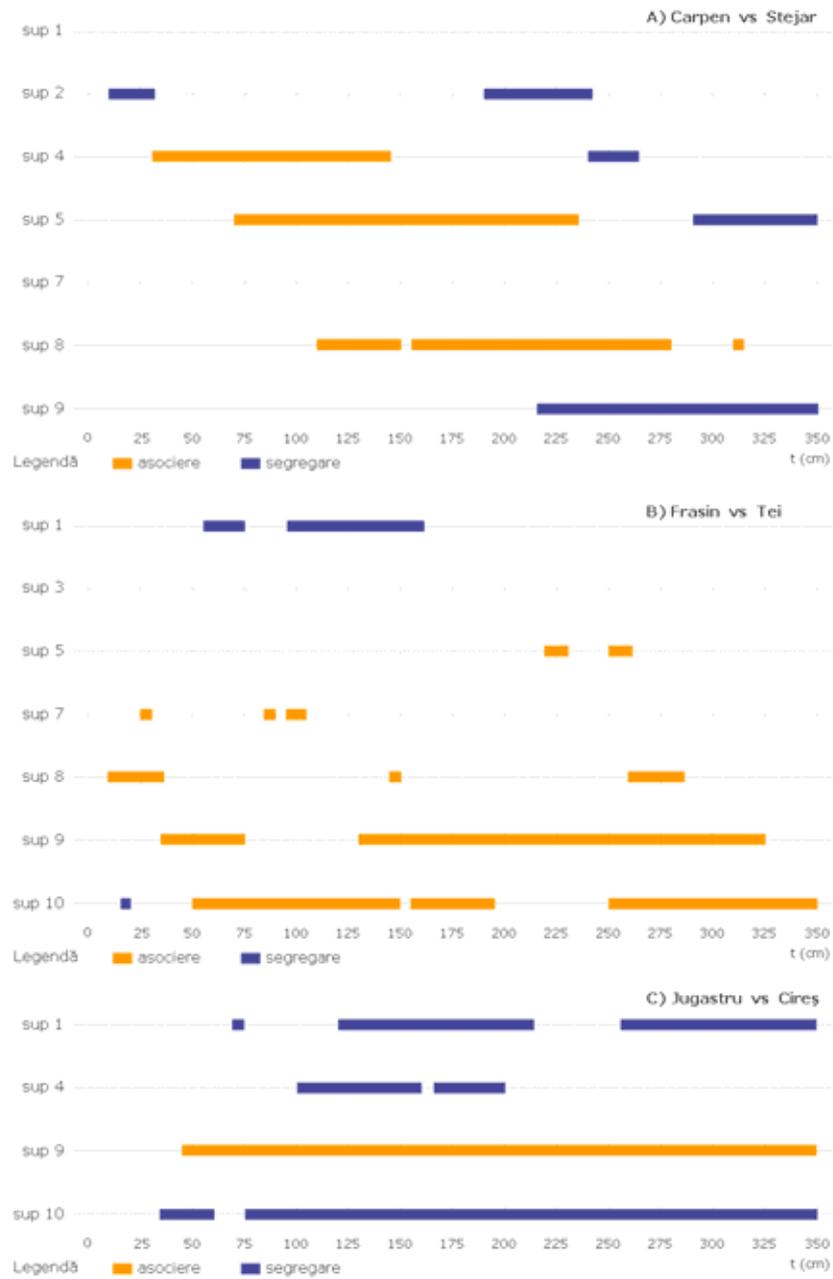

**Figura 7.18 Centralizatorul de determinare a tipului de interacțiune în spațiu între specii prezente în suprafețele studiate, prin intermediul funcției $L_{12}(t)$**

*Relația dintre distribuția în spațiu a carpenului și stejarului*

În cazul acestei relații, rezultatele inițiale obținute din corelația desimilor pe quadrate indicau o ușoară tendință de segregare la dimensiunea de 50x50 cm și mai intensă dar nesemnificativă la valori ale quadratului de 350x350 cm. Analiza bivariată permite nuanțarea unor aspecte legate de distribuția în spațiu a celor două specii. Se înregistrează și prin această metodă o slabă tendință de repulsie la distanțe de 25-50 cm (vizibilă în graficele din anexa 15 – suprafețele 1, 2, 4, 5, 7) dar nesemnificativă. Abaterile semnificative de la ipoteza nulă a independenței poziționării în spațiu arată în general o asociere în zona mediană a intervalului distanțelor iar spre sfârșitul acestuia, la distanțe de peste 200-250 cm apar frecvent procese repulsive. Cele două specii par să manifeste o asociere, dar nu în imediata apropiere, și nu pe distanțe foarte mari, odată cu creșterea intervalului de analiză fiind înregistrată chiar segregarea lor.

*Relația dintre distribuția în spațiu a frasinului și teiului*

Pentru cele două specii s-au înregistrat în analizele inițiale legăturile corelative pozitive cele mai intense, foarte semnificative la toate dimensiunile quadratelor. Rezultatele analizei bivariate susțin ipoteza formulată anterior, asocierea pozitivă semnificativă fiind înregistrată în aproape toate suprafețele (mai puțin în suprafața 1). Nu s-a înregistrat o tendință de asociere la o anumită distanță, fenomenul fiind semnificativ pe distanțe scurte, în intervale fragmentate sau chiar pe intervale întinse.

*Relația dintre distribuția în spațiu a jugastrului și cireșului*

Chiar dacă ipoteza inițială susținea asocierea celor două specii, rezultatele analizei bivariate indică mai degrabă contrariul, în 3 din cele 4 suprafețe analizate fiind semnalată segregarea semnificativă pe intervale întinse. Tendința de variație a funcției $L_{12}(t)$ chiar și în interiorul spațiului definit de înfășurătoarele de încredere este tot negativă indicând repulsia indivizilor din cele două specii începând de la circa 30 cm și continuând chiar și până la 350 cm.

Faptul că rezultatele nu confirmă în totalitate ipotezele formulate anterior prin analiza pe quadrate confirmă încă odată afirmațiile lui Hurlbert (1990), care sublinia că scara de analiză este foarte importantă și poate influența rezultatele obținute. Funcțiile de tip Ripley sunt mai puțin influențate de scara de analiză, oferind și informații mult mai detaliate. În concluzie se poate afirma că fenomenele de asociere și agregare se produc, dar nu manifestă constanță la nivelul tuturor suprafețelor analizate, puieții fiind mai sensibili la dinamica mai intensă a proceselor și fenomenelor specifică acestui stadiu de dezvoltare.

O explicație interesantă a interacțiunilor dintre specii este dată de Szwagrzyk și Czerwczak (1993), care sugerează că valorile semnificative ale asocierii – în direcția pozitivă sau negativă pot fi determinate de influențe directe între indivizii speciilor, prin fenomenul de alelopatie, considerat de autori unul din factorii esențiali care modelează distribuția în spațiu a speciilor.

## 7.5. Folosirea informaţiei spaţiale în evaluarea relaţiilor competiţionale

Competiţia este procesul care face referire la presiunea exercitată asupra resurselor unui spaţiu limitat de către indivizi care se bazează pe acestea pentru supravieţuire, creştere şi dezvoltare. Prin intensitatea competiţiei se exprimă intensitatea presiunii exercitate asupra resurselor comune.

Datorită legăturii dintre accesul individual la resursele disponibile şi creştere, competiţia dintre indivizi este frecvent folosită în fundamentarea modelelor dinamicii ecosistemelor forestiere (Botkin et al., 1972; Stage, 1973; Ek, Monserud, 1974; Ek, Monserud, 1974; Pretzsch, 1992; Hasenauer, 1994; Monserud, Sterba, 1996).

Pentru evaluarea intensităţii competiţiei pot fi folosiţi indicatori absoluţi caracteristici unei suprafeţe (densitatea în suprafaţă, suprafaţa de bază) sau pot fi folosiţi indicatori care determină intensitatea procesului la nivel individual. Indicii individuali oferă un grad de detaliu mai mare, o rezoluţie de apreciere superioară celor absoluţi, exprimând intensitatea relativă a competiţiei.

Acest concept de intensitate relativă a fost utilizat iniţial de către Newnham (1964, citat de Vanclay, 1992), care afirma că în mediul forestier un arbore fără competitori are aceleaşi rate de creştere cu un arbore de acelaşi diametru care creşte în câmp deschis. În aceeaşi idee considera că un arbore afectat de competiţie va avea creşterea în diametru redusă proporţional cu intensitatea competiţiei.

Indicatorii individuali folosiţi în aprecierea competiţiei au avantajul aprecierii cantitative a acestui proces, dar valoarea lor nu poate fi explicată decât în termeni relativi, prin comparaţie cu valori calculate similar pentru alţi arbori.

A fost conceput un număr foarte mare de indici de competiţie, mai mult sau mai puţin diferiţi în modul de fundamentare şi în relaţiile de calcul. În general este acceptată clasificarea dată de Munro (1974) în indici independenţi de distanţă şi indici dependenţi de distanţă.

Prezentarea acestor două categorii a fost efectuată în capitolul 4.1, fiind detaliate tipurile de indici ce fac parte din cea de a doua categorie. Acești indici sunt capabili să surprindă mai fidel relațiile de competiție dintre indivizi deoarece includ în relația de calcul nu doar de caracteristicile biometrice, ci și pozițiile arborilor raportate la vecinii lor.

Mulți autori au efectuat comparații între categoria indicilor independenți de distanță și a celor dependenți de distanță, părerile privind superioritatea celor din urmă nefiind unanime (Biging, Dobbertin, 1995). Reticența unor cercetători privitoare la folosirea indicilor dependenți de distanță se referă la dificultatea culegerii poziției tuturor arborilor ce nu compensează întotdeauna sporul de precizie obținut. În general se consideră că modelele individuale de creștere pot să beneficieze de pe urma informațiilor privitoare la variația în spațiu a procesului competițional, fiind efectuate numeroase cercetări în sprijinul acestei idei (Pukkala, Kolstrom, 1991; Pretzsch, 1997; Nigh, 1997, Aitkenhead et al., 2004; Little, 2002; Fox et al., 2007, a, b).

Cu toate că interacțiunile competiționale dintre arbori au loc atât la nivel supra cât și subteran, majoritatea cercetărilor au urmărit procesul competițional desfășurat în sfera vizibilă. Majoritatea indicilor de competiție includ în relația de calcul diametrul de bază sau suprafața de bază a arborilor, datorită simplității de prelevare a datelor și importanței acestor parametri în prognoza creșterilor dar adesea sunt utilizați și alți parametri: înălțimea sau dimensiunile coroanei (diametru, volum, suprafață exterioară).

Fundamentarea indicilor dependenți de distanță se bazează pe ipoteza conform căreia concurența resimțită de un individ este invers proporțională cu distanța dintre arborele de referință și vecinii săi și direct proporțională cu raportul dintre dimensiunile arborilor vecini și dimensiunile arborelui de referință.

Foarte multe discuții contradictorii sunt legate de modul de alegere al competitorilor direcți. Informațiile asupra vecinătăților se pot raporta la diferite criterii de definire a vecinilor (Haining, 2004):

- distanța fixă în linie dreaptă dintre puncte – sunt considerați vecini toți indivizii care se încadrează într-o distanță limită precizată;
- cel mai apropiat număr de vecini – se consideră că fiecare individ are un număr egal cu $K$ al celor mai apropiați vecini;
- grafuri Gabriel – două elemente situate în poziția A și B sunt vecine dacă și numai dacă toate celelalte elemente sunt poziționate în afara cercului de circumferință AB (figura 7.19 a);
- triangulația Delauney – toți indivizii care au o muchie comună în partiționarea Dirichlet a spațiului respectiv sunt considerați vecini. O partiție Dirichlet a unui spațiu consideră că aria specifică a unui punct A constă din mulțimea tuturor punctelor ce sunt mai apropiate de punctul A față de orice alt punct din spațiu – aceasta este metoda cea mai naturală de definire a vecinilor dar din păcate și cea mai complicată (figura 7.19 b).

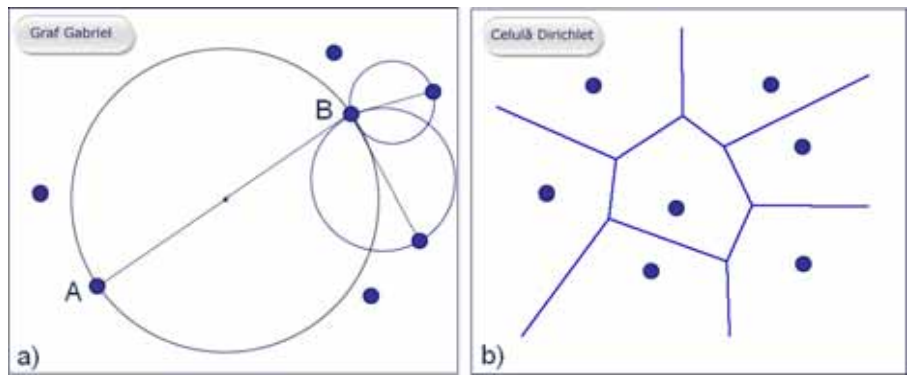

**Figura 7.19  Definirea vecinilor prin grafuri Gabriel (a) sau folosind partiționarea Dirichlet (b)**

### 7.5.1. Evaluarea competiției prin indici sintetici dependenți de distanță

Au fost concepuți foarte mulți indici de apreciere a competiției dependenți de distanță, aceștia fiind preferați datorită capacității lor de a surprinde variația în spațiu a intensității acestui proces. Deși au fost efectuate numeroase cercetări în care s-a comparat precizia de evaluare a diferiților indici (Vanclay, 1992; Biging, Dobbertin, 1995; Stadt et al., 2002; Popa, 2004) nici un indice nu și-a dovedit superioritatea universală, fiind dificil de ales din multitudinea de indicatori menționați de literatura de specialitate.

În cazul de față s-a optat pentru evaluarea competiției prin indicii Hegyi (1974, citat Schaer, 1981) și Schutz (1989, citat de Ung et al., 1997), doi indici foarte apreciați pentru versatilitatea lor și acuratețea cu care surprind relațiile dintre indivizi (Popa, 2004).

Indicele Hegyi este caracterizat prin relația de calcul:

$$CI = \sum_{j=1}^{n} \frac{d_j}{d_i} \cdot \frac{1}{l_{ij}} \qquad (7.51)$$

unde $d_i$ – diametrul arborelui referință; $d_j$ – diametrul arborelui competitor; $l_{ij}$ – distanța dintre arborele de referință și vecinul său competitor; $n$ – numărul de competitori considerați. Acest indice ponderează raportul dintre diametrele arborilor aflați în competiție cu distanța dintre ei, însumând componentele elementare obținute pentru un număr de vecini dat sau obținut în funcție de o distanță limită impusă. Inițial Hegyi a definit numărul de vecini competitori în funcție de o distanță limită de 3 m, dar există numeroase alte variante de identificare a vecinilor.

Indicele Schutz folosește o relație de calcul complexă și include o modalitate proprie de selectare a arborilor vecini competitori:

$$S = 0.65 \cdot S_V + S_O = 0.65 \cdot \frac{H_V - H_C}{D_{VC}} + (0.5 - \frac{E_{VC}}{D_{VC}}) \qquad (7.52)$$

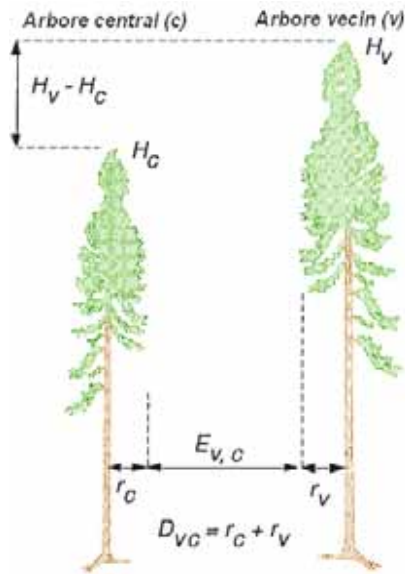

un arbore fiind considerat competitor dacă:

$$E_{VC} \leq 0.5 \cdot D_{VC} + 0.65 \cdot (H_V - H_C)$$

$$(7.53)$$

semnificația notațiilor:

$S_V$, $S_O$ – componenta verticală, respectiv orizontală a indicelui; $H_V$, $H_C$ – înălțimea arborelui vecin, respectiv a arborelui referință; $D_{VC}$ – media diametrelor coroanelor celor doi arbori; $E_{VC}$ – distanța dintre limitele coroanelor celor doi arbori.

**Figura 7.20  Indicele de competiție Schutz** (după Ung et al., 1997)

Pentru calculul indicilor Hegyi am propus o modificare referitoare la alegerea vecinilor cu care un arbore intră în competiție. Indicele Hegyi nu include o modalitate de determinare a competitorilor, drept urmare se folosesc metode care sunt limitative și nenaturale din punctul de vedere al relațiilor dintre arbori. Cel mai frecvent se impune un număr fix de vecini cu care fiecare arbore dintr-o suprafață poate interacționa (se folosesc 3, 4, 5, 8 vecini) sau o distanță limită în care cercetătorul crede că se restrâng procesele concurențiale. Ambele variante sunt la fel de absurde din perspectiva realităților biologice, problema stabilirii vecinilor fiind cu adevărat o problemă foarte complicată. Se poate recurge la grafuri Gabriel sau la partiționarea Dirichlet a spațiului, dar algoritmii specifici unor astfel de soluții sunt dificil de implementat mai ales pentru un număr mare de indivizi.

Rezolvarea pe care o propun ține cont de gradul de umbrire al coroanei unui arbore de către ceilalți vecini. Un arbore vecin va fi considerat competitor doar dacă va umbri coroana arborelui de referință. Am plecat de la ipoteza că umbrirea maximă care se realizează în perioada diurnă corespunde unui unghi de incidență ($\gamma$) al razelor solare cu suprafața terestră de maxim 45°. Aceasta este o valoare care se poate modifica în funcție de poziționarea regiunii în care se desfășoară analiza. În general peste această valoare a unghiului de incidență, intensitatea radiației solare scade fiind în general considerată inutilizabilă.

Folosind ipoteza anterior menționată am stabilit că un arbore vecin va fi considerat competitor doar dacă la un unghi maxim (de 45°) realizează (chiar și parțial) umbrirea coroanei arborelui de referință (fig. 7.21, situația 2 și 3).

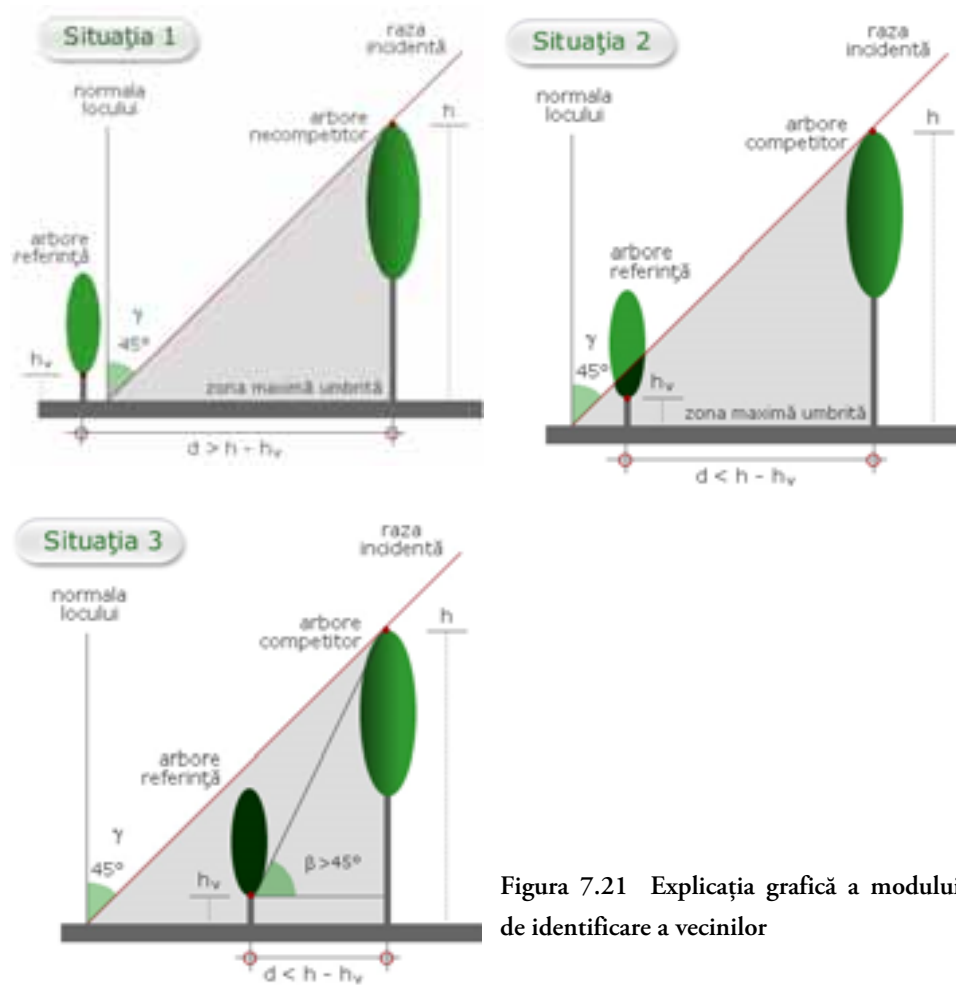

**Figura 7.21   Explicația grafică a modului de identificare a vecinilor**

Această condiție se traduce prin inegalitatea:

$$\arctan\ (\frac{h_j - h_{V\_i}}{d_{ij}}) \cdot \frac{180°}{\pi} \geq 90° - \gamma \tag{7.54}$$

dacă rezultatul funcției arctan este dat în radiani, iar $\gamma$ unghiul de incidență al razelor solare.

În ipoteza unghiului de 45° inegalitatea se reduce la:

$$d_{ij} \leq h_j - h_{V\_i} \tag{7.55}$$

unde $d_{ij}$ – distanța dintre arbori; $h_j$ – înălțimea arborelui vecin; $h_{V\_i}$ – înălțimea până la punctul de inserție a coroanei pentru arborele de referință.

Indicele Hegyi s-a calculat folosind atât relația propusă inițial de autor – prin raportarea diametrelor arborilor competitori, cât și printr-o relație care evaluează intensitatea procesului competițional la nivelul înălțimilor. Varianta din urmă nu este folosită în cazul arboretelor, dar personal consider că este o variantă mai potrivită pentru evaluarea relațiilor dintre puieți, înălțimea, după cum am evidențiat în capitolul 5, putând fi considerată parametrul central în evaluarea structurii semințișului.

Pentru justificarea fiziologică a legăturii dintre competiție și creștere am propus o variantă care include în relația de calcul aria suprafeței exterioare a coroanei, aceasta fiind suprafața receptoare a luminii folosită în procesul de fotosinteză.

Suplimentar a fost calculată o variantă a indicelui care utilizează volumul coroanei, pentru a vedea care din cele două atribute ale coroanei influențează mai mult procesul de creștere. Assman (1970) acordă o importanță deosebită ambilor parametri ai coroanei menționați anterior, considerând suprafața exterioară a coroanei mai adecvată în studierea proceselor de creștere și dezvoltare. Atât volumul coroanei, cât și suprafața exterioară a acesteia au fost calculate prin asimilarea cu un elipsoid, folosind formulele (5.3) și (5.4) prezentate în capitolul 5.

Calculul valorilor indicilor s-a realizat cu aplicații software proprii, realizate în mediul de dezvoltare *Microsoft Visual Basic*.

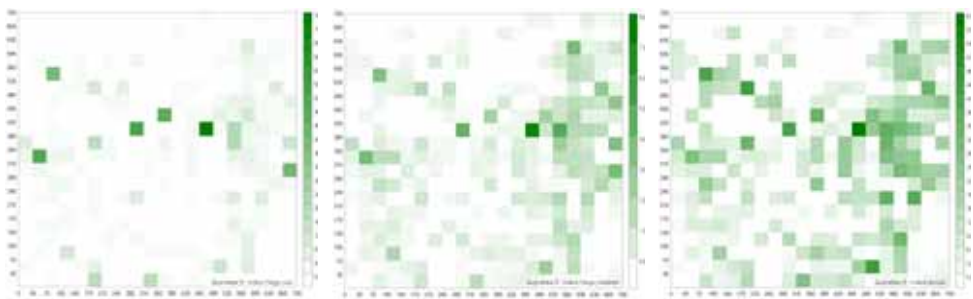

a) indicele Hegyi_sup          b) indicele Hegyi_înălţime          c) indicele Schutz

**Figura 7.22  Identificarea nucleelor de intensitate maximă a competiţiei prin analiza grafică a valorilor obţinute pentru indicii Hegyi şi Schutz (ex. suprafaţa 5)**

Primul modul determină valorile indicelui Hegyi folosind diferenţele dintre diametre, înălţimi, volumele coroanei şi suprafeţele exterioare ale acestora, selectând competitorii pe baza principiului fundamentat anterior.

Cel de al doilea modul determină valorile indicelui Schutz, precum şi a componentelor verticale şi orizontale ale acestuia.  Utilizatorul are posibilitatea de a introduce constanta k ce reprezintă ponderea acordată componentei verticale a indicelui.

Prelucrarea datelor s-a efectuat pentru toate suprafeţele şi au fost întocmite cu ajutorul aplicaţiei personale CARTOGRAMA diagramele ce prezintă variaţia în spaţiu a intensităţii proceselor competiţionale (anexa 16). Analiza cartogramelor ce prezintă variabilitatea în spaţiu a proceselor competiţionale permite identificarea nucleelor cu intensitatea maximă a competiţiei.

Tabelul 7.1

Matricea corelaţiilor pentru indicii de evaluare a competiţiei

| | *Hegyi_d* | *Hegyi_vol* | *Hegyi_sup* | *Hegyi_h* | *Schutz* |
|---|---|---|---|---|---|
| *Hegyi_d* | 1 | | | | |
| *Hegyi_vol* | 0,427 *** | 1 | | | |
| *Hegyi_sup* | 0,593 *** | 0,929 *** | 1 | | |
| *Hegyi_h* | 0,836 *** | 0,585 *** | 0,771 *** | 1 | |
| *Schutz* | 0,776 *** | 0,486 *** | 0,685 *** | 0,874 *** | 1 |

Studiul acestor grafice (prezentate pentru fiecare suprafață în anexa 16) relevă faptul că indicii sintetici folosiți evidențiază aceleași centre ale competiției (fig. 7.22), diferențe înregistrându-se în zonele cu o presiune mai scăzută. Indicele Hegyi_sup, calculat folosind suprafața exterioară a coroanei, pare să accentueze foarte mult zonele de intensitate maximă, estompând restul valorilor deoarece are cel mai mare coeficient de variație al valorilor (anexa 17). Varianta calculată folosind înălțimea evidențiază maximele, dar pare mai sensibil grafic în aprecierea valorilor intermediare, iar indicele Schutz, deși diferit ca și concepție de cei anteriori, identifică nuclee de intensitate maximă similare dar pare să surprindă grafic foarte bine amplitudinea intensităților.

Relațiile dintre indicii folosiți, observate prin comparația cartogramelor, sunt confirmate de valorile ridicate ale coeficienților de corelație (tabelul 7. 1).

În general obiectivul evaluării competiției îl constituie folosirea informațiilor în corectarea modelelor funcțiilor de creștere a arborilor sau chiar determinarea unor funcții de estimare a creșterii. Pentru a aprecia relevanța informațiilor furnizate de indici s-a studiat legătura dintre valorile acestora și creșterile puieților în înălțime. În teren s-a măsurat prin sondaj ultima creștere în înălțime pentru puieții aparținând speciilor cu ponderea cea mai mare de participare (carpen, stejar, tei și frasin). Primul pas a constat în determinarea coeficienților de corelație dintre indici și creșterea în înălțime.

Tabelul 7.2

**Coeficienții de corelație ai legăturii dintre creșterea în înălțime și indicii de competiție**

|                 | Toți puieții    | Ca              | St              | Te              | Fr              |
|-----------------|-----------------|-----------------|-----------------|-----------------|-----------------|
| Hegyi_d         | -0,298 ***      | -0,244 ***      | -0,339 ***      | -0,559 ***      | -0,445 ***      |
| Hegyi_vol       | -0,194 ***      | -0,170 ***      | -0,294 ***      | -0,386 ***      | -0,156          |
| Hegyi_sup       | -0,309 ***      | **-0,401 *****  | -0,362 ***      | -0,475 ***      | -0,391 ***      |
| Hegyi_h         | -0,331 ***      | -0,312 ***      | -0,370 ***      | **-0,599 *****  | -0,566 ***      |
| Schutz          | **-0,386 *****  | -0,358 ***      | **-0,426 *****  | -0,577 ***      | **-0,637 *****  |
| Schutz vertical | -0,348 ***      | -0,339 ***      | -0,409 ***      | -0,518 ***      | -0,589 ***      |
| Schutz orizontal| 0,242 ***       | 0,204 ***       | 0,300 ***       | 0,376 ***       | 0,510 ***       |

Din analiza valorilor prezentate în tabelul 7.2 se remarcă faptul că toți indicii calculați stabilesc legături foarte semnificative de intensitate slabă și medie cu creșterea în înălțime. Există însă diferențe în ceea ce privește capacitatea indicilor de a explica variația creșterilor.

Valorile coeficienților de corelație au fost prezentate atât pentru toți arborii cât și pentru principalele patru specii, considerând că fiecare specie are propriile mecanisme auxologice și caracteristici distincte de reacție la competiție. Corelațiile obținute pe specii diferă pentru același indice și au fost obținute în general intensități mai mari decât cele obținute pentru toți puieții, fapt ce confirmă ideea că speciile reacționează diferit la competiție. Drept urmare, includerea informațiilor legate de procesul competițional în modelele de creștere trebuie să se facă diferențiat pe specii pentru a crește acuratețea predicțiilor.

În ceea ce privește capacitatea diferențiată a indicilor de a surprinde caracteristicile auxologice ale speciilor, se remarcă anumiți indici cu un comportament predictiv superior: indicele Schutz (intensitatea legăturii a fost maximă pentru St, Fr și per total), indicele bazat pe raportarea înălțimilor Hegyi_h (intensitate maximă pentru Te) și indicele bazat pe raportarea suprafețelor exterioare ale coroanei Hegyi_sup (intensitate maximă pentru Ca).

Componenta verticală a indicelui Schutz explică mult mai bine variația creșterii prin comparație cu componenta orizontală, fapt determinat pe de o parte de comparația realizată cu creșterea în înălțime și pe de altă parte de importanța acestui parametru biometric în caracterizarea structurii semințișului.

Surprinde faptul că deși volumul coroanei și suprafața coroanei sunt parametri bazați pe aceleași elemente, sunt diferențe mari între Hegyi_sup și Hegyi_vol în ceea ce privește relația cu creșterea, cu atât mai mult cu cât coeficientul de corelație dintre cei doi indici este de 0,929 ***. Explicația se regăsește în fundamentarea fiziologică a Hegyi_sup, fapt confirmat de opiniile lui Assman (1970) privitoare la superioritatea suprafeței exterioare a coroanei în explicarea proceselor de creștere.

Pentru a identifica eventuala superioritate a capacității predictive a unuia dintre indicii folosiți, legăturile corelative au fost analizate și grafic deoarece dependența dintre elemente nu pare să fie liniară, caz în care coeficienții de corelație determinați nu sunt pe deplin sugestivi.

În vederea determinării formei legăturii corelative au fost folosite ecuații de regresie logaritmice, exponențiale sau de tip putere, fiind selectate și prezentate grafic acele ecuații pentru care s-au obținut coeficienții de determinare maximi, în acest fel explicându-se cea mai mare parte a variației datelor.

Este evident în toate diagramele din figura 7.23 faptul că cele mai mari creșteri se înregistrează în condițiile unor valori reduse ale competiției. Se remarcă descreșterea exponențială a valorilor indicilor de competiție odată cu majorarea creșterilor.

Ecuațiile de regresie care modelează cel mai bine legătura dintre competiție și creșterea în înălțime sunt ecuațiile exponențiale și cele de tip putere. Valorile maxime ale coeficienților de determinare au fost obținute și în acest caz la analiza pe specii, în cazul teiului și frasinului modelele matematice reușind să explice peste 60% din varianța datelor.

Indicele sintetic pentru care s-a obținut cea mai puternică legătură între creștere și intensitatea competiției este indicele Hegyi_sup, varianta pentru care s-a folosit în definirea raportului dimensional aria suprafeței exterioare a coroanei. S-a dovedit astfel că acest parametru al coroanei nu este superior doar teoretic (prin justificarea sa fiziologică) ci și practic, oferind informații utile care ar putea fi folosite în activitatea de modelare.

Un comportament similar, dar inferior din punctul de vedere al varianței explicate l-a avut indicele Hegyi_h. În cazul indicelui Schutz ecuațiile de regresie descriu mai greu relația acestuia cu creșterile în special datorită numărului mare al valorilor nule pe care acest indice le determină. Numărul puieților pentru care nu se identifică vecini competitori (și implicit valoarea intensității competiției este nulă) reprezintă o diferență majoră între cele două tipuri de indici folosiți.

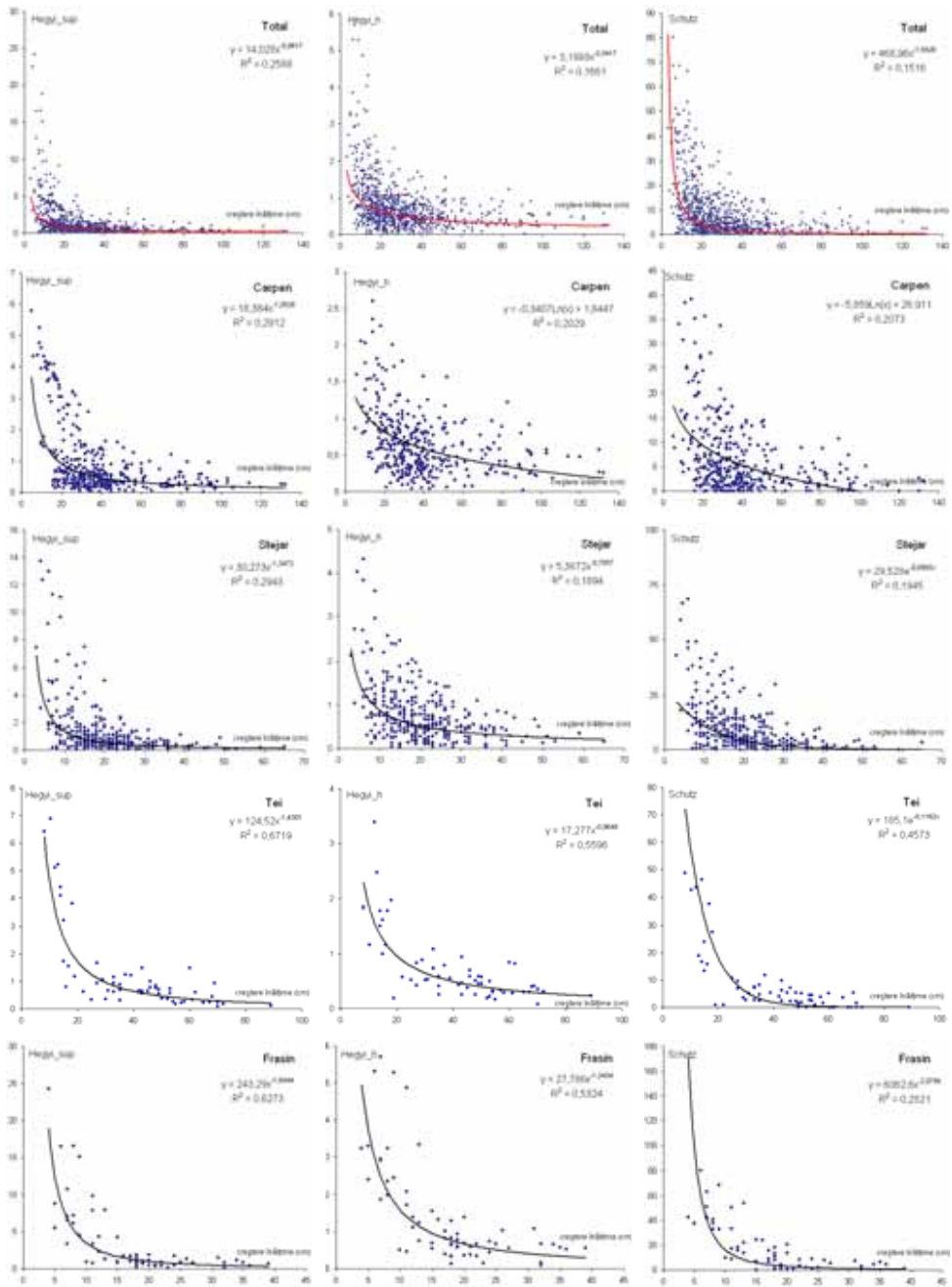

**Figura 7.23  Analiza grafică a legăturilor corelative dintre indicii de competiţie şi creşterea în înălţime – cazul tuturor puieţilor şi separat pe specii**

În cazul folosirii indicelui Schutz au fost identificați 3% puieți fără competitori direcți, pe când în cazul utilizării pentru indicele Hegyi a criteriului de selecție a vecinilor competitori fundamentat anterior s-au identificat doar 0,03% puieți din această categorie. Diferența este determinată de criteriul de selecție al vecinilor, pentru indicele Schutz acest criteriu fiind foarte limitativ, fiind acceptate situații evidente de umbrire a coroanelor drept cazuri în care nu intervine competiția. Criteriul de selecție al vecinilor fundamentat în această lucrare a permis identificarea de peste 100 de ori mai multe cazuri în care competiția nu poate fi exclusă comparativ cu varianta de selecție a competitorilor propusă de Schutz. Indicele Schutz exclude în acest fel date privitoare la relațiile dintre puieți, pe când utilizarea criteriului nou propus include date suplimentare, crescând relevanța indicilor de competiție care îl folosesc.

Trebuie amintit faptul că toți indicii folosiți au avut rezultate comune în privința aprecierii speciilor cel mai puțin afectate de competiție – valorile medii calculate pe specii (anexa 17) indică aceeași ordine – paltinul, cireșul și teiul sunt speciile care au puieții cel mai puțin afectați de competiție. Informația este interesantă dar trebuie privită cu rezervă deoarece valorile mari ale coeficienților de variație indică o structură foarte eterogenă din punctul de vedere al indicilor calculați, media fiind în acest caz prea puțin semnificativă.

Relațiile de competiție dintre puieți permit obținerea unor informații utile din punctul de vedere al modelării creșterilor în înălțime și ulterior de determinare a dinamicii semințișului, dar metodele de obținere a acestor informații trebuie selectate cu multă atenție deoarece insuficienta fundamentare a unora dintre ele poate să conducă la rezultate care să nu reflecte foarte obiectiv situația reală din teren.

### 7.5.2. Determinarea suprafeței potențiale de dezvoltare a puieților

Încercările cercetătorilor de a explica mecanismele competiției dintre indivizi au pătruns de multe ori în domenii ale matematicii pentru a găsi metodele cele mai potrivite de evaluare a intensității acestei relații.

O abordare mai puțin întâlnită dar foarte interesantă se referă la suprafața potențială de dezvoltare sau aria potențial disponibilă a unui individ, conceptul de APA (*area potentially available*) fiind introdus în ecologie de către Brown (1965, citat Kenkel et al., 1989). Același concept se pare că a fost dezvoltat independent și de către Mead (1966, citat Kenkel et al., 1989), dar cercetări în domeniul spațiului de dezvoltare sunt menționate încă din 1835 în lucrarea *„Die Forst-Mathematik"* a lui Konig.

Prin APA se definește suprafața pe care un individ o are la dispoziție pentru accesul la resursele necesare creșterii și dezvoltării. Includerea APA în evaluarea competiției încearcă rezolvarea unei probleme de referință circulară care frământă cercetătorii interesați de modelarea creșterilor (Wichmann, 2002; Garcia 2008) - arborii au dimensiuni mai mari pentru că au acces la mai multe resurse sau au acces la resurse mai multe pentru că sunt mai mari? Există relativ puține cercetări în acest domeniu (Mark, Esler, 1970; Moore et al., 1973; Kenkel et al., 1989; Mercier, Baujard, 1997), Smith (1987) considerând că această abordare a fost în general ignorată sau chiar evitată datorită neînțelegerii fundamentării geometrice a APA și datorită dificultăților de calcul și algoritmizare. Ultimii ani au condus la o dezvoltare fără precedent a computerelor și s-au realizat numeroase progrese în cazul algoritmilor specifici geometriei computaționale, astfel că problema ariei potențiale de dezvoltare revine în atenția cercetătorilor.

Soluționarea problemelor prin utilizarea acestui concept nu se rezumă doar la aprecierea competiției ci implică și alte aspecte – cercetări de apreciere a mortalității plantulelor efectuate de Owens și Norton (1989) sau determinarea tiparului spațial al populațiilor de arbori (Mercier, Baujard, 1997; Palaghianu, 2012). Asupra acestui ultim aspect Garcia (2008) sublinia faptul că se poate vorbi de o interacțiune între spațiul de dezvoltare al arborilor apropiați ca efect al autocorelației. Vecinii foarte apropiați (între care există o distanță mai mică decât media distanțelor din suprafață) vor avea amândoi APA mai mică decât media spațiilor de dezvoltare din suprafața analizată și viceversa, APA puieților mai îndepărtați va fi mai mare decât media.

Winsauer și Mattson (1992) menționau câteva din avantajele utilizării ariei potențiale de dezvoltare în cercetările forestiere - suprafețele potențial disponibile ale arborilor se exclud reciproc, sunt sensibile la dinamica populației și se înregistrează corelații bune ale ariei acestora cu ratele de creștere. Această ultimă observație stă la baza utilizării APA drept metodă de evaluare a intensității competiției - cu cât suprafața corespunzătoare unui individ este mai mare, cu atât mai mică va fi presiunea concurențială ce acționează asupra lui.

Spațiul potențial de dezvoltare al arborilor a cunoscut diverse forme de exprimare similare conceptului formulat de Brown – e.g. zona de influență folosită de Staebler (1951) și Bella (1971) (citați de Schaer, 1981) sau Moore (et al., 1973). Cea mai corectă formulare rămâne totuși cea inițială, bazată pe un concept matematic al partiționării unui spațiu – partiționarea în poligoane Voronoi sau în celule Dirichlet. Se admite că aria potențial disponibilă este echivalentă cu suprafața poligonului Voronoi asociat unui individ.

În matematică diagramele Voronoi sau Dirichlet sunt un tip particular de partiționare a spațiului metric prin distanțele către o mulțime finită de obiecte din spațiu. Primele preocupări și referiri la acest tip de partiționare a spațiului aparțin lui Descartes (1644, citat de Aurenhammer și Klein, 1999), dar acesta nu a definit explicit metoda. Numele ei provine de la matematicianul ucrainean Georgy Fedoseevich Voronoi, respectiv de la cel german Peter Gustav Dirichlet care au definit geometric acest proces particular de „mozaicare" (Aurenhammer și Klein, 1999).

În spațiul bidimensional un poligon Voronoi sau o celulă Dirichlet a unui element conține toate punctele din plan mai apropiate de acel element decât de orice alt element din plan (figura 7.24 a, b).

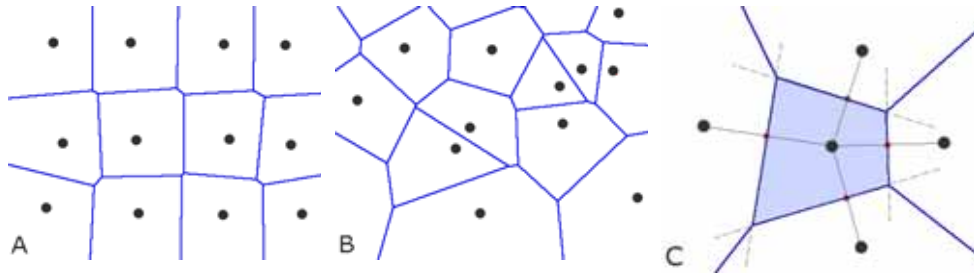

**Figura 7.24 Partiţionarea spaţiului în poligoane Voronoi (a, b) şi modul geometric de construcţie a unui astfel de poligon (c)**

Muchiile poligonului sunt formate din punctele situate la distanţe egale de două elemente considerate învecinate. Vârfurile poligonului sunt formate din puncte situate la distanţe egale de minim 3 elemente. În această situaţie două elemente sunt considerate vecine dacă au o latură comună a poligonului Voronoi asociat.

Din punct de vedere geometric posibilitatea construirii unui astfel de poligon se reduce la intersectarea mediatoarelor segmentelor care unesc elementele vecine (figura 7.24 c). Odată obţinută diagrama Voronoi pentru o suprafaţă, dacă se unesc prin segmente elementele considerate vecine din interiorul înfăşurătoarei convexe, se obţine triangulaţia Delauney. Datorită dificultăţilor de construire a partiţionării Dirichlet pentru un număr mare de elemente, s-a încercat automatizarea procesului, în ultimii ani fiind concepuţi numeroşi algoritmi, de referinţă în acest domeniu fiind lucrarea *„Spatial Tessellations: Concepts and Applications of Voronoi Diagrams"* (Okabe et al., 2000).

În vederea stabilirii ariei potenţiale a puieţilor din suprafeţele studiate a fost realizat un program în mediul de dezvoltare Microsoft Visual Basic, folosind un algoritm prezentat de Ohyama (2008). Complexitatea algoritmului nu este cea mai potrivită (O ($n^2$)) existând şi metode mai rapide - cunoscutul algoritm al lui Fortune (1987) are o complexitate de O(n log n), dar este dificil de implementat.

Programul VORONOI preia coordonatele puieților dintr-o foaie de calcul tabelar de tip Excel și construiește diagrama Voronoi pentru suprafața dată.

Analiza vizuală a diagramelor permite obținerea de informații privitoare la vecinii fiecărui puiet – ceea ce poate constitui o metodă naturală de selectare a vecinilor în cazul determinării diverșilor indici de competiție. De asemenea se pot face aprecieri asupra spațierii indivizilor și asupra tiparului de organizare spațială, identificându-se mai ușor zonele de agregare sau cele care tind spre un tipar uniform.

Se pot efectua numeroase analize folosind aria poligoanelor Voronoi – coeficientul de variație poate constitui un test care să aprecieze îndepărtarea de la ipoteza distribuției uniforme în spațiu, valorile mari ale acestuia indicând abaterea de la ipoteza CSR în direcția agregării (Palaghianu, 2012). A fost constituită o aplicație de calcul a valorii ariei potențiale de dezvoltare și s-a determinat pe fiecare suprafață și coeficientul de variație al APA. Legăturile corelative ale acestui parametru cu principalii indici de apreciere ai organizării spațiale (r = 0.88 ** în cazul indicelui Morisita și r = -0.58 în cazul indicelui Clark-Evans) arată potențialul folosirii coeficientului de variație al APA drept indicator de evaluare a tiparului spațial al puieților (figura 7.25). Ideea este susținută și de rezultatele obținute în cazul folosirii simulărilor Monte-Carlo pentru a genera o înfășurătoare de încredere a CSR pentru parametrul constituit de coeficientul de variație al APA.

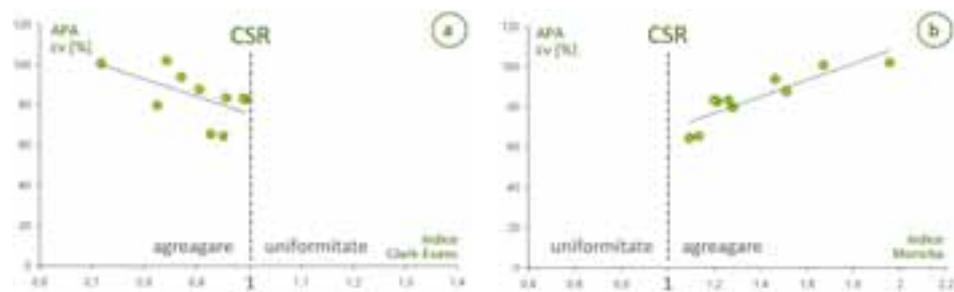

**Figura 7.25 Corelația dintre coeficientul de variație ala APA și (a) indicele Clark-Evans, respectiv (b) indicele Morisita**

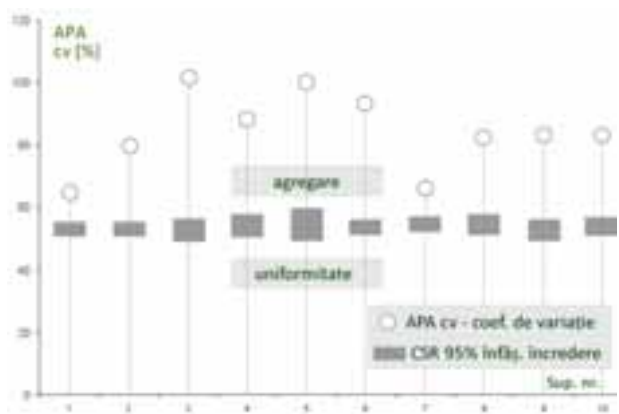

**Figura 7.26 Valorile coeficientului de variație al APA indică agregarea în toate suprafețele studiate - depășește valorile înfășurătoarei de încredere**

Rezultatele obținute arată abateri semnificative de la ipoteza CSR pentru toate suprafețele considerând pragul de semnificație de 5% (figura 7.26).

Aria poligoanelor poate fi folosită și drept indice de competiție, valorile mici fiind asociate cu o intensitate mare a proceselor concurențiale. Structurile complexe de reprezentare a datelor pentru poligoanele Voronoi fac totuși dificilă folosirea ariei în prelucrări numerice uzuale.

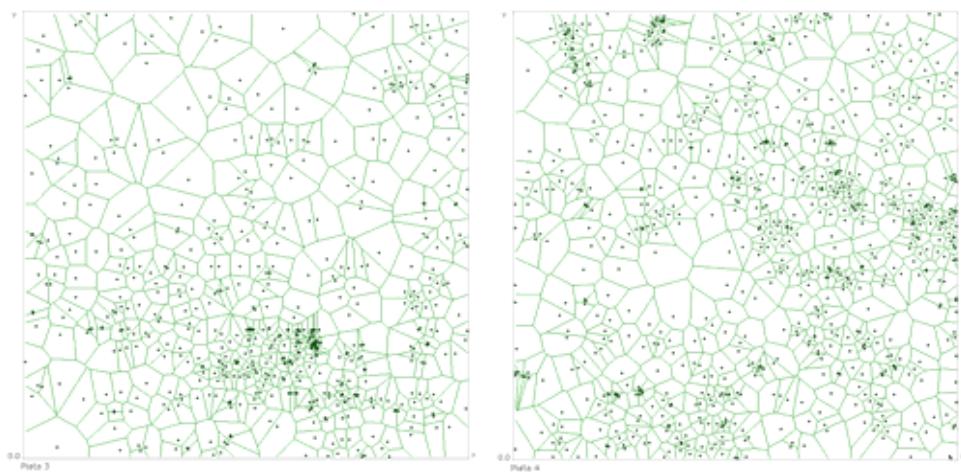

**Figura 7.27 Diagramele Voronoi generate pentru suprafețele 3 și 4**

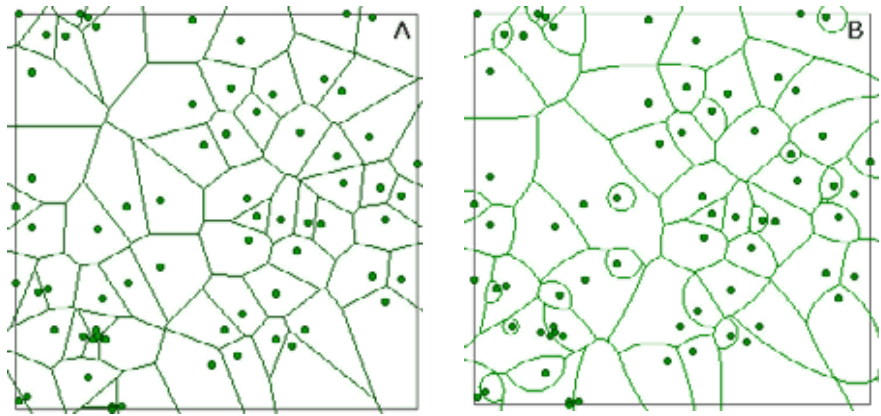

**Figura 7.28 Diferența dintre diagramele Voronoi neponderate (a) și ponderate (b), detaliu din suprafața 4**

În figura 7.27 sunt prezentate diagramele Voronoi pentru două din suprafețele studiate, diagramele pentru restul suprafețelor regăsindu-se în anexa 18. Prin analiza diagramelor Voronoi (figura 7.27) devin evidente zonele de agregare din suprafețe, iar prin prin compararea acestora cu cartogramele ce prezintă intensitatea competiției se observă că multe din nucleele de intensitate maximă a competiției corespund zonelor de agregare identificate în diagramele Voronoi. Spațiul de dezvoltare influențează în mod evident relațiile dintre tinerele exemplare cu efect direct asupra creșterilor acestora. În cazul arborilor literatura de specialitate menționează corelații puternice semnificative între APA și creșterea în suprafața de bază (Moore et al., 1973; Winsauer, Mattson, 1992).

În unele situații se consideră că suprafața potențială de dezvoltare nu surprinde suficient de realist spațiul aferent unui individ deoarece nu se iau în calcul datele biometrice caracteristice (înălțimea, dimensiunea coroanei) care modelează diferit configurația spațiului care asigură accesul la resurse. Au apărut soluții matematice de corectare a acestei probleme - ponderarea ariilor Voronoi cu valori atribuite fiecărui punct. În acest fel ariile potențiale determinate inițial pot fi corectate prin intermediul unor ponderi individuale. Lucrările de cercetare menționează că ponderarea se poate face cu ajutorul diametrului, suprafeței de bază sau razei coroanei (Moore et al., 1973; Smith, 1987; Winsauer, Mattson, 1992).

În cazul semințișului consider că ar fi mai indicată folosirea înălțimii (fiind evidențiată în subcapitolul 7.5.1 importanța acesteia în competiție) sau a parametrilor coroanei (raza medie sau volumul coroanei).

Aplicația VORONOI prezintă un modul separat pentru generarea diagramelor ponderate (figura 7.28 b). Utilizatorul poate alege caracteristica biometrică pe baza căreia să se realizeze corecția spațiului individual de dezvoltare. Suprafețele obținute sunt neregulate și este dificil de calculat aria mai ales în cazul suprafețelor cu o desime mare.

Potențialul de utilizare al diagramelor Voronoi este foarte mare dar insuficient exploatat datorită prelucrărilor dificile, de aceea consider aplicația concepută un instrument foarte util cercetătorilor care intenționează să folosească această tehnică de investigare.

# Capitolul 8
# Concluzii. Contribuții originale

## 8.1. Sinteza cercetărilor efectuate

Concluziile sunt redactate sub forma unei sinteze a rezultatelor obținute, prezentarea fiind făcută în conformitate cu obiectivele principale propuse inițial. Astfel s-au obținut rezultate ale cercetărilor cu privire la:

### 1. Evaluarea structurii semințișului

Pentru determinarea particularităților structurii semințișului au fost efectuate analize statistice ale principalelor elemente structurale. Datele biometrice ale celor 7253 de puieți inventariați au constituit suportul determinării distribuțiilor, indicatorilor statistici și relațiilor dintre parametrii analizați. S-a determinat grafic, cu ajutorul aplicației proprii CARTOGRAMA, variația spațială a desimii, precum și a valorilor atributelor biometrice analizate, ceea ce a permis stabilirea gradului de neomogenitate din suprafețele studiate. Pentru modelarea matematică a distribuțiilor experimentale cele mai bune rezultate au fost obținute prin folosirea funcțiilor de densitate Beta, Gamma și Weibull.

**a)** Referitor la situația numărului de puieți pe specii, au fost formulate primele ipoteze privitoare la asocierea/repulsia unor specii, ca urmare a analizei corelațiilor dintre desimile speciilor pe quadrate de diferite mărimi, ipoteze care au fost testate ulterior prin metode ale analizei spațiale. Distribuția suprafețelor statistice elementare de 1x1m pe categorii ale numărului de puieți la metrul pătrat este unimodală, cu asimetrie de stânga, și a fost bine ajustată de distribuția teoretică Weibull și Beta. S-a înregistrat un număr mediu de 14 puieți la metrul pătrat și un coeficient de variație de 57%.

**b)** Au fost de asemenea calculați indicatorii statistici (pentru toți puieții și pe specii) și analizate distribuțiile experimentale pentru: diametrul la colet (media de 8,3 mm /coeficientul de variație 63,8%), înălțime (96,7 cm/54,9%), înălțimea până la prima ramură verde (26,2 cm/55,3%) și diametrul mediu al coroanei (23,2 cm/65,2%). Distribuțiile sunt în general bine ajustate de funcțiile de densitate folosite, cu toate că s-au remarcat deficiențe ale testului $\chi^2$ la valori mari ale frecvențelor pe clase. Deși aprecierea vizuală și testul Kolmogorov–Smirnov indică o bună ajustare, la folosirea testului $\chi^2$ se respinge ipoteza nulă.

S-a efectuat și o stratificare a puieților în patru straturi diferențiate după înălțime, pe specii și s-a arătat că instalarea noilor generații de puieți se realizează greu în condițiile în care generațiile anterioare au atins dimensiuni apreciabile. Distribuția pe categorii de înălțimi, cât și cea pe straturile definite relevă o foarte bună valorificare a spațiului pe verticală datorită competiției pentru lumină. Capacitatea speciilor de a se situa preponderent în straturile superioare sau inferioare este dată de mecanismul competitiv al fiecărei specii - frasinul are peste 70% din puieți în cele mai de jos două straturi, iar teiul, cireșul, jugastrul au peste 70% puieți în straturile superioare (3 și 4).

Din analiza indicatorilor statistici ai parametrilor coroanei se desprinde tendința speciilor de a-și forma încă din stadiile foarte tinere un tipar al coroanei. Atât proporția înălțimii până la punctul de inserție al coroanei din înălțimea totală (hv/h) cât și diametrul mediu al coroanei arată diferențe foarte mari înregistrate între specii. Media raportului hv/h este de 33% pentru toți puieții, dar frasinul înregistrează valori cu peste 27% peste medie pe când jugastrul cu mai mult de 21% sub medie. Similar se înregistrează diferențe majore și în cazul diametrului coroanei - cireșul are valori cu 30% peste media generală, pe când frasinul cu 34% sub medie. S-a folosit termenul de coroană a puieților, chiar dacă puieții de 1-2 ani nu au această structură bine diferențiată. Geometric s-a determinat un corp al „coroanei" ținând cont de aparatul foliar, deschiderea dintre lujeri și punctul de inserție pe tulpină al primului lujer.

**c)** Determinarea precisă în teren a poziției fiecărui puiet precum și a caracteristicilor biometrice care definesc arhitectura sa a permis reconstituirea modelului tridimensional al fiecărei suprafețe. Metoda completează informațiile referitoare la structură și reprezintă o formă intuitivă de prezentare și interpretare a informațiilor. Reprezentările tridimensionale au fost realizate cu ajutorul programului SVS (McGaughey, 1997). Pentru a ușura transferul datelor din foile de calcul tabelar în formatul complex impus de SVS s-a conceput un program (SVS Export) care realizează automat această operație.

**d)** S-au studiat și relațiile dintre elementele biometrice ale puieților. S-a demonstrat că spre deosebire de situația arboretelor, unde parametrul central al structurii este reprezentat de diametrul de bază, în cazul semințișului diametrul la colet este un parametru important, dar greu măsurabil, instabil și cu o reprezentativitate scăzută. Înălțimea se măsoară ușor, precizia determinărilor este mare, este mai stabilă din punct de vedere statistic (are cel mai mic coeficient de variație) și influențează esențial dinamica semințișului datorită competiției intense pentru lumină. În urma acestor aprecieri se consideră înălțimea totală a puieților atributul central al structurii semințișului.

Valorile coeficienților de corelație dintre elementele biometrice indică legături de intensitate medie și puternică, foarte semnificative, între toți parametrii. Se poate menționa legătura dintre *d (diametru)* și *h (înălțime)* - 0,761***, dintre *h* și *dc (diametrul mediu al coroanei)* - 0,644***, respectiv *d* și *dc* - 0,654***. Dacă în cazul arboretelor se practică modelarea elementelor biometrice în funcție de diametrul de bază, acesta putând fi ușor măsurat, în cazul semințișurilor se recomandă modelarea celorlalte elemente, mai greu măsurabile, în funcție de înălțimea puieților. Au fost determinate ecuațiile de regresie ce descriu variația parametrilor biometrici în funcție de înălțime și s-a constatat că variabila independentă (*h*) explică 61,3% din variația diametrelor (*d*), 44,4% din cea a diametrului mediu al coroanei (*dc*) și 29,9% din cea a înălțimii punctului de inserție a coroanei (*hv*). Stratificarea pe specii a permis îmbunătățirea substanțială a capacităților predictive ale modelelor matematice. Acuratețea de estimare a modelelor propuse s-a testat prin analize ale abaterilor reziduale. Parametrul pentru care modelele matematice dependente de înălţime explică cea mai mare parte a varianţei este diametrul la colet (61% per total, 72% în cazul stejarului).

e) În scopul îmbunătăţirii capacităţii predictive a modelelor s-au introdus variabile independente suplimentare înălţimii şi s-a efectuat o analiză a regresiei multiple. Au fost înregistrate progrese minore, care nu justifică efortul prelevării datelor suplimentare.

f) S-a determinat volumul şi suprafaţa exterioară a coroanei, doi parametri consideraţi importanţi în definirea relaţiilor dintre arbori şi în fundamentarea eficienţei creşterilor (Assman, 1970). Pentru estimarea suprafeţei exterioare a elipsoidului asociat coroanei s-a folosit o formulă de aproximare recentă (Cantrell, 2004). Au fost determinate şi modele matematice de estimare a celor doi parametri ai coroanei în funcţie de înălţime şi diametrul la colet. Înălţimea a reprezentat variabila independentă ce a oferit predicţii superioare celor oferite de modelele bazate pe diametrul la colet (prin comparaţia abaterilor reziduale), explicând peste 80% din varianţa suprafeţei exterioare a coroanei şi peste 75% din cea a volumului coroanei. Suprafaţa exterioară a coroanei este un element mai expresiv comparativ cu volumul acesteia, atât în relaţia cu diametrul cât şi cu înălţimea, fiind obţinuţi coeficienţi de determinare superiori. Pentru creşterea preciziei de estimare au fost determinate ecuaţiile de regresie (în funcţie de înălţime) pe specii, dar s-au obţinut îmbunătăţiri doar în cazul puieţilor de stejar.

g) S-a analizat şi relaţia parametrilor biometrici cu desimea, la diferite scări de analiză, pentru prelucrarea datelor fiind folosită aplicaţia proprie CARTOGRAMA. Toate variantele dimensionale de analiză (quadrate de 1x1 m, 1,75x1,75 m, 3,5x3,5 m) au evidenţiat legături corelative negative, de intensitate medie, foarte semnificative, parametrii cei mai afectaţi de desime fiind în ordine diametrul la colet, diametrul mediu al coroanei, volumul coroanei şi suprafaţa exterioară a acesteia. Corelaţia dintre desime şi înălţime este de intensitate slabă astfel că s-a propus folosirea concomitentă a acestor două variabile în modelare, riscul ca rezultatele să fie afectate de multicoliniaritate fiind redus. Analiza regresiei multiple a arătat că folosirea înălţimii şi desimii drept variabile independente în modele matematice de determinare a parametrilor biometrici conduce în unele cazuri la îmbunătăţirea consistentă a preciziei modelelor.

## 2. Evaluarea diversităţii structurale dimensionale

**a)** Heterogenitatea structurii dimensionale a fost apreciată pentru distribuţia diametrelor, înălţimilor, volumelor şi suprafeţelor exterioare ale coroanei prin indici sintetici consacraţi: Shannon, Simpson, Brillouin, Berger-Parker, McIntosh, Gleason, Margalef şi Menhinick, calculaţi cu ajutorul aplicaţiei software proprii BIODIV. Analiza valorilor obţinute şi a corelaţiilor a permis identificarea a două grupe, ce conţin indici similari ca expresivitate sau chiar redundanţi: prima grupă formată din indicele Shannon, Simpson, Brillouin, Berger-Parker, McIntosh, iar cea de a doua reprezentată de indicii Gleason, Margalef şi Menhinick.

Rezultatele confirmă stabilitatea structurii înălţimilor, pentru care s-au înregistrat valorile maxime ale diversităţii structurale (cea mai echilibrată răspândire în clase). În cazul speciilor valorile maxime au fost obţinute pentru carpen şi tei, subliniindu-se capacitatea deosebită a acestora de a se impune în comunităţi.

**b)** Diversitatea structurală a fost apreciată pentru înălţime şi prin intermediul indicilor de diferenţiere dimensională T (Gadow, 1997) şi dominanţă U (Hui et al., 1998; Gadow şi Hui, 1999). Determinarea acestora a fost efectuată pentru un număr de 1-5 vecini, cu ajutorul programului propriu DIFDOM.

Rezultatele indică o diferenţiere dimensională slabă spre medie în grupurile de 2-6 indivizi crescând odată cu mărirea grupului. Şi indicele de dominanţă U indică o diversitate structurală redusă după înălţime. Luând în calcul valorile medii pe suprafaţă s-a arătat că indicele de diferenţiere dimensională T este puţin influenţat de numărul de vecini luat în calcul, indicele de dominanţă U fiind însă mai sensibil. S-a stabilit că indicele T nu este influenţat de desimea puieţilor dar este corelat negativ cu înălţimea medie.

**c)** Datorită faptului că indicii sintetici de apreciere a diversităţii structurale nu includ date explicite privitoare la distribuţia în spaţiu a atributelor dimensionale a fost fundamentat un indice propriu care corectează această deficienţă a indicilor.

Indicele IDIV se bazează pe un principiu simplu de comparare a sumei distanţei dintre vârfurile puieţilor apropiaţi cu suma distanţelor reduse la orizont dintre puieţi. Cu cât raportul se abate mai mult de la valoarea unitară, cu atât diversitatea este mai mare. Fundamentarea include şi definirea unui maxim şi minim teoretic, precum şi a unui maxim şi minim posibil de realizat pentru aceleaşi valori ale înălţimii puieţilor. Pentru acoperirea întregii suprafeţe aceasta se împarte în benzi paralele de lăţime egală şi se definesc două componente axiale ale indicelui IDIV – IDIV_Ox şi IDIV_Oy. Indicele IDIV este foarte flexibil şi adaptabil şi permite identificarea unor fenomene direcţionale prin analiza diferenţelor dintre valorile componentelor axiale. Pentru calculul componentelor indicelui a fost concepută o aplicaţie software proprie, numită IDIV.

Rezultatele arată influenţa scării de analiză în cercetările ce implică distribuţia spaţială a evenimentelor. A fost observată o tendinţă de maximizare a diversităţii structurale odată cu creşterea pasului de analiză explicabilă prin tendinţa de agregare a puieţilor de dimensiuni apropiate pe spaţii reduse. Creşterea heterogenităţii se datorează diferenţelor dintre aglomerările de puieţi odată cu mărirea  perspectivei de analiză. În urma studierii diferenţelor dintre componentele axiale IDIV_Ox şi IDIV_Oy nu s-au constatat decalaje care să conducă la ipoteza unor fenomene direcţionale.

Corelaţiile cu desimea au arătat că indicii IDIV şi IDIV_max sunt puternic influenţaţi de desime, dar au arătat şi o componentă independentă de desime: IDIV_min. Corelaţiile semnificative, de intensitate medie cu indicii clasici ai diversităţii structurale precum şi cu cei de dominanţă şi diferenţiere dimensională dovedesc capacitatea indicelui IDIV de a surprinde heterogenitatea dar nu se poate stabili superioritatea sa fără analize suplimentare.

### 3. Evaluarea particularităţilor de organizare în spaţiu a puieţilor

a) Au fost folosite analize diferite care au urmărit identificarea unor tipare care nu sunt evidente la o simplă analiză vizuală şi confirmarea faptului că un tipar spaţial diferă statistic semnificativ de o distribuţie aleatoare în spaţiu. Prelucrările datelor au fost realizate cu ajutorul unei versiuni îmbunătăţite a aplicaţiei proprii

SPATIAL (Palaghianu, Horodnic, 2007) și a modulului *SpPack* (Perry, 2004).

Primele analize au fost realizate printr-o metodă independentă de distanțe - metoda quadratelor. Au fost determinați: indicele de dispersie, indicele mărimii agregatului, indicele Morisita și indicele Green în situația a trei variante dimensionale ale quadratului (50 cm, 70 cm și 100 cm). În toate suprafețele și variantele dimensionale s-au obținut abateri statistic semnificative față de ipoteza CSR în direcția tiparului agregat. S-a remarcat scăderea intensității de agregare la creșterea desimii. Legat de expresivitatea indicilor, recomand folosirea indicelui Morisita datorită faptului că este singurul indice care oferă o explicație a valorii sale și este mai puțin afectat de variația dimensiunii quadratului (Hurlbert, 1990).

**b)** S-au determinat o serie de indici sintetici dependenți de distanță: indicele Fisher, Clark-Evans, Donnelly, Skellam, Pielou. Rezultatele indică în 75% din cazuri procese contagioase pozitive. Influența desimii asupra modului de organizare spațială s-a confirmat și prin această metodă. Creșterea desimii și implicit a competiției accentuează procesele repulsive, determinând apariția unui tipar spațial uniform. Corelația dintre indicii de apreciere a agregării și cei ai diversității arată că la creșterea diversității structurale a înălțimilor se accentuează uniformitatea în spațiu. Se dovedește în acest fel tendința puieților de a utiliza optim spațiul atât pe verticală (prin distribuirea echilibrată în clase de înălțimi) cât și pe orizontală (prin tendința de a se distribui uniform).

**c)** S-a inclus în seria analizelor efectuate și metoda rafinată a distanțelor față de cel mai apropiat vecin, fiind folosită funcția G(w) (Diggle, 1979) pentru identificarea proceselor contagioase. Au fost realizate diagrame pentru toate suprafețele pentru a identifica diferențele dintre valoarea experimentală a G(w) și valoarea teoretică, obținută în condițiile unei distribuții aleatoare în spațiu. Testarea semnificației statistice s-a realizat pentru pragul de semnificație de 5% prin intermediul generării înfășurătoarelor de încredere prin simulări. Și în acest caz s-a constatat preponderent agregarea, dar metoda se bazează pe analiza distanțelor față de primul vecin și nu oferă informații privitoare la spațiere pentru distanțe mai mari decât distanța maximă față de primul vecin.

**d)** S-au efectuat și analize prin metoda celor mai apropiați k vecini - (*KNN – k^{th} nearest neighbours*). Indicatorii specifici acestei metode au fost calculați pentru cei mai apropiați 1-15 vecini. Pentru testarea semnificației statistice limitele intervalului de încredere au fost determinate analitic, deși există posibilitatea folosirii unor înfășurătoare de încredere generate prin simulări Monte Carlo. În nouă din cele zece suprafețe au fost identificate abateri semnificative de la ipoteza distribuției aleatoare în direcția agregării. S-a constatat agregarea puieților pentru grupări de 2-3 indivizi (în suprafețele 1, 7, 9), grupări de până la 6-7 indivizi (suprafețele 2 și 10) sau chiar aglomerări de peste 16 puieți (suprafețele 3, 4, 5, 6).

**e)** Ultima dintre analize a constat în folosirea metodelor analizei spațiale de ordinul al II-lea. S-a folosit funcția *L(t)* (Besag, 1977), o variantă standardizată a funcției *K(t)* (Ripley, 1977). Testarea semnificației abaterilor de la ipoteza CSR s-a făcut cu ajutorul generării înfășurătoarelor de încredere prin efectuarea a 99 de simulări. Tendința de agregare este majoritară, fiind prezentă în toate suprafețele studiate pe diferite distanțe. Au fost identificate două intervale în care agregarea este maximă: 30-90 cm (suprafețele 1, 2, 4, 7) și 220-310 cm (suprafețele 3, 5, 6, 8, 10). Puieții au tendința de a forma nuclee mai mici cu diametrul de 30 până la 90 cm incluse în aglomerări de dimensiuni mai mari (de 2-3 m).

S-a efectuat și o analiză a straturilor dimensionale ale puieților în funcție de înălțime. Agregarea este preponderentă în toate straturile. Puieții de mari dimensiuni se agregă pe intervale mai reduse comparativ cu celelalte straturi iar agregarea nu se produce de la cele mai mici distanțe. Puieții de dimensiuni medii manifestă cele mai intense procese contagioase pozitive, cu agregarea maximă în grupurile cu dimensiunea de 30-80 cm, 120-250 cm, sau chiar peste 250 cm. Agregarea plantulelor are un caracter fragmentat, formându-se grupuri foarte variate ca dimensiune (grupuri mici dispuse aleatoriu în grupuri mai mari).

**f)** Analiza bivariată prin funcția *L₁₂(t)* a evidențiat faptul că plantulele se asociază în general pozitiv cu puieții de dimensiuni medii iar puieții de talie mare acționează repulsiv atât față de plantule cât mai ales față de puieții de dimensiuni medii. Cu ajutorul analizei bivariate s-au testat și ipotezele privitoare la asocierea speciilor. S-a infirmat ipoteza relaţiei repulsive dintre carpen şi stejar, dar a fost confirmată ipoteza asocierii pozitive dintre frasin şi tei, identificându-se şi că relaţia dintre cireş şi jugastru are tendinţe preponderent repulsive.

### 4. Evaluarea relaţiilor de competiţie în seminţişuri

**a)** Competiţia este unul din fenomenele care se produc cu cea mai mare intensitate în seminţişuri, având un impact deosebit asupra creşterii puieţilor. Din gama largă de indici de competiţie menţionaţi de literatura de specialitate au fost selectaţi pentru evaluarea intensităţii fenomenului în suprafeţele studiate indicele Hegyi şi Schutz, foarte apreciaţi pentru acurateţea determinărilor. Datorită faptului că indicele Hegyi nu are un sistem de alegere a vecinilor competitori s-a fundamentat matematic un criteriu propriu de selectare a competitorilor unui puiet, pentru a nu se impune un număr fix de vecini competitori sau o distanţă limită pe care puieţii să relaţioneze. Un puiet vecin este considerat competitor doar dacă umbreşte (chiar şi parţial) coroana puietului de referinţă. Pentru adaptarea indicelui Hegyi particularităţilor de structură ale seminţişului s-au propus trei variante noi de indici, care să includă înălţimea puieţilor, volumul coroanei şi suprafaţa exterioară a acesteia. Pentru calcularea valorilor indicilor au fost create două programe informatice: HEGYI şi SCHUTZ. Rezultatele prelucrărilor pentru fiecare suprafaţă sunt prezentate grafic prin intermediul unor diagrame generate cu ajutorul aplicaţiei proprii CARTOGRAMA. Analiza vizuală a acestora relevă faptul că toţi indicii de competiţie folosiţi evidenţiază aceleaşi nuclee de intensitate maximă a competiţiei dar surprind diferit amplitudinea de variaţie a fenomenului.

**b)** Testarea relevanţei rezultatelor obţinute s-a studiat prin analiza legăturii dintre creşterea în înălţime a puieţilor şi valorile indicilor de competiţie. S-a constatat că toţi indicii selectaţi stabilesc legături foarte semnificative cu creşterea în înălţime dar au o capacitate diferită de a explica variaţia creşterilor. Creşterile în înălţime descresc exponenţial odată cu majorarea intensităţii competiţiei. Au fost determinate ecuaţiile de regresie care exprimă legătura dintre intensitatea competiţiei şi creşterea în înălţime şi s-a identificat în indicele Hegyi_sup (pentru care s-a folosit în definirea raportului dimensional aria suprafeţei exterioare a coroanei) varianta predictivă optimă. Varianta Hegyi_h (definită prin raportarea înălţimilor) a avut un comportament predictiv bun, superior indicelui Schutz. Criteriul de identificare a vecinilor competitori fundamentat în lucrare este mult mai sensibil în aprecierea intensităţii competiţiei comparativ cu cel al indicelui Schutz (care se dovedeşte limitativ în seminţişuri), reuşind să identifice de peste 100 de ori mai multe cazuri în care competiţia nu poate fi exclusă.

Valorile calculate pe specii indică faptul că paltinul, cireşul şi teiul sunt specii cu mecanisme concurenţiale bine dezvoltate, având puieţii cel mai puţin afectaţi de competiţie.

c) În aprecierea competiţiei a fost folosit şi conceptul de suprafaţă potenţială de dezvoltare (APA - *area potentially available*), fiind determinat pentru fiecare puiet acest spaţiu prin intermediul programului propriu VORONOI, capabil să genereze inclusiv diagrame Voronoi ponderate cu o anumită variabilă. A fost de asemenea susţinut şi potenţialul folosirii coeficientului de variaţie al APA drept indicator al structurii spaţiale a indivizilor.

Faţă de cele evidenţiate mai sus, considerăm că **obiectivul secundar** al cercetărilor a fost îndeplinit, pentru fiecare metodă de analiză fiind specificate particularităţile care apar în cazul seminţişurilor.

Totodată au fost îndeplinite **obiectivele adiţionale** privitoare la mijloacele informatice. Au fost realizate aplicaţii software proprii care constituie instrumente utile în prelucrarea, analiza şi prezentarea datelor. Au fost concepute nouă aplicaţii informatice independente: CARTOGRAMA, BIODIV, DIFDOM, IDIV, SPATIAL, SVS_Export, VORONOI, HEGYI, SCHUTZ şi alte module pentru aplicaţia de calcul tabelar Microsoft Excel – testul Grubbs pentru un volum mare de date, calculul valorilor teoretice $\chi^2$ pentru un număr foarte mare de grade de libertate, determinarea volumului şi suprafeţei exterioare a coroanei. Toate aceste aplicaţii informatice se adaugă celor concepute în afara temei de doctorat (Palaghianu, 2004; Palaghianu, 2014). Toate programele informatice concepute cât şi codul sursă sunt puse la dispoziţia cercetătorilor interesaţi în folosirea lor.

## 8.2. Contribuții originale

În urma cercetărilor efectuate se pot remarca o serie de contribuții originale:

- s-a demonstrat că înălțimea poate fi considerată parametrul central de structură a semințișurilor;

- s-a arătat că suprafața exterioară a elipsoidului asociat coroanei este un atribut biometric important în semințișuri;

- s-a fundamentat un indice de apreciere a diversității structurale dimensionale care include poziția în spațiu a puieților (indicele IDIV);

- a fost fundamentat matematic un criteriu de identificare a vecinilor competitori ai unui puiet;

- s-au realizat nouă aplicații informatice independente de analiză a datelor: CARTOGRAMA, BIODIV, DIFDOM, IDIV, SPATIAL, SVS_Export, VORONOI, HEGYI, SCHUTZ, plus alte componente de analiză;

- s-a efectuat un amplu studiu bibliografic privitor la relația ciberneticii cu pădurea, în special cu regenerarea arboretelor;

- au fost identificate tendințe (statistic semnificative) asociative sau repulsive ale speciilor ce vegetează în suprafețele studiate;

- s-a arătat că speciile tind să formeze din fazele foarte tinere tipare particulare ale coroanei;

- au fost semnalate deficiențe ale testului statistic $\chi^2$ la valori foarte mari ale frecvențelor experimentale;

- a fost evidențiată suprafața potențial disponibilă (APA) a puieților prin diagrame de tip Voronoi și s-a evidențiat capacitatea coeficientului de variație al valorilor APA de a decela tiparul spațial al evenimentelor;

- s-a determinat valoarea intensității de competiție prin indici consacrați și adaptări proprii ale acestora, adecvate folosirii în semințișuri;

- au fost determinate ecuațiile de regresie care descriu variația parametrilor biometrici în funcție de  înălțime și a fost testată acuratețea de estimare a acestora prin analiza abaterilor reziduale;

- s-a arătat că înălțimea este variabila independentă cu cea mai bună capacitate predictivă a volumului coroanei și a suprafeței exterioare a acesteia;

- au fost identificate legături negative semnificative, de intensitate medie, între desime și diverși parametrii biometrici;

- s-a arătat că utilizarea concomitentă a desimii și înălțimii ca variabile predictive ale caracteristicilor biometrice conduce la îmbunătățiri substanțiale ale determinărilor comparativ cu modelele bazate pe o singură variabilă;

- s-a efectuat o analiză a indicilor de apreciere a diversității structurale;

- s-a identificat înălțimea drept parametrul structural cel mai stabil, pentru care s-au obținut valorile maxime ale diversității dimensionale;

- s-a constatat creșterea diversității structurale odată cu lărgirea perspectivei de analiză și odată cu creșterea desimii;

- au fost identificate abateri statistic semnificative de la ipoteza distribuirii aleatoare în spațiu a puieților în direcția agregării prin metode diferite ale analizei spațiale;

- s-a constatat o creștere a uniformității distribuției în spațiu odată cu creșterea desimii;

- a fost identificat numărul mediu de puieți din nucleele de agregare prin metoda celor mai apropiați k vecini;

- au fost folosite metode de analiză spațială bivariată de ordinul al II-lea pentru determinarea tendințelor de asociere sau segregare;

- s-a determinat faptul că plantulele se asociază cu puieții de dimensiuni medii iar puieții de talie mare au un comportament repulsiv atât față de plantule cât și față de ceilalți puieți.

# Bibliografie

# Anexe

## Anexa 1. Valoarea indicilor de diferențiere și dominanță dimensională obținuți în suprafețele studiate

Valori calculate pentru primii 1-5 vecini, în condițiile utilizării unei zone tampon cu lățimea de 30 cm.

| Nr. vecini: 1 | T1 | U1 var1 | U1 var2 | distanța medie 1v |
|---|---|---|---|---|
| suprafața 1 | 0,32 | 0,48 | 0,50 | 11,25 |
| suprafața 2 | 0,37 | 0,50 | 0,47 | 12,51 |
| suprafața 3 | 0,31 | 0,48 | 0,50 | 12,03 |
| suprafața 4 | 0,33 | 0,47 | 0,49 | 12,00 |
| suprafața 5 | 0,34 | 0,49 | 0,49 | 10,82 |
| suprafața 6 | 0,40 | 0,48 | 0,51 | 11,29 |
| suprafața 7 | 0,38 | 0,48 | 0,50 | 10,63 |
| suprafața 8 | 0,27 | 0,44 | 0,54 | 12,46 |
| suprafața 9 | 0,35 | 0,48 | 0,51 | 12,34 |
| suprafața 10 | 0,28 | 0,50 | 0,49 | 11,64 |
| **medie** | 0,34 | 0,48 | 0,50 | 11,70 |
| Nr. vecini: 2 | T2 | U2 var1 | U2 var2 | distanța medie 2v |
| suprafața 1 | 0,35 | 0,49 | 0,49 | 14,29 |
| suprafața 2 | 0,38 | 0,52 | 0,46 | 16,50 |
| suprafața 3 | 0,32 | 0,47 | 0,50 | 15,39 |
| suprafața 4 | 0,33 | 0,48 | 0,49 | 14,81 |
| suprafața 5 | 0,36 | 0,48 | 0,50 | 14,61 |
| suprafața 6 | 0,41 | 0,48 | 0,51 | 14,57 |
| suprafața 7 | 0,40 | 0,48 | 0,51 | 13,56 |
| suprafața 8 | 0,28 | 0,48 | 0,51 | 15,40 |
| suprafața 9 | 0,37 | 0,49 | 0,49 | 15,17 |
| suprafața 10 | 0,29 | 0,49 | 0,51 | 14,43 |
| **medie** | 0,35 | 0,49 | 0,50 | 14,87 |
| Nr. vecini: 3 | T3 | U3 var1 | U3 var2 | distanța medie 3v |
| suprafața 1 | 0,36 | 0,49 | 0,49 | 16,91 |
| suprafața 2 | 0,39 | 0,52 | 0,47 | 19,65 |
| suprafața 3 | 0,32 | 0,49 | 0,50 | 18,17 |

| suprafaţa 4 | 0,34 | 0,47 | 0,50 | 17,19 |
|---|---|---|---|---|
| suprafaţa 5 | 0,36 | 0,50 | 0,49 | 17,97 |
| suprafaţa 6 | 0,42 | 0,47 | 0,51 | 17,15 |
| suprafaţa 7 | 0,41 | 0,49 | 0,50 | 16,06 |
| suprafaţa 8 | 0,29 | 0,48 | 0,50 | 17,78 |
| suprafaţa 9 | 0,37 | 0,48 | 0,50 | 17,72 |
| suprafaţa 10 | 0,29 | 0,49 | 0,51 | 16,80 |
| **medie** | 0,36 | 0,49 | 0,50 | 17,54 |
| **Nr. vecini: 4** | **T4** | **U4 var1** | **U4 var2** | **distanţa medie 4v** |
| suprafaţa 1 | 0,37 | 0,49 | 0,50 | 19,17 |
| suprafaţa 2 | 0,40 | 0,51 | 0,48 | 22,36 |
| suprafaţa 3 | 0,33 | 0,48 | 0,50 | 20,72 |
| suprafaţa 4 | 0,34 | 0,48 | 0,49 | 19,39 |
| suprafaţa 5 | 0,37 | 0,51 | 0,48 | 20,83 |
| suprafaţa 6 | 0,43 | 0,47 | 0,51 | 19,34 |
| suprafaţa 7 | 0,42 | 0,49 | 0,50 | 18,17 |
| suprafaţa 8 | 0,29 | 0,49 | 0,50 | 19,95 |
| suprafaţa 9 | 0,37 | 0,48 | 0,51 | 19,93 |
| suprafaţa 10 | 0,30 | 0,48 | 0,51 | 18,87 |
| **medie** | 0,36 | 0,49 | 0,50 | 19,87 |
| **Nr. vecini: 5** | **T5** | **U5 var1** | **U5 var2** | **distanţa medie 5v** |
| suprafaţa 1 | 0,37 | 0,49 | 0,49 | 21,20 |
| suprafaţa 2 | 0,41 | 0,50 | 0,48 | 24,78 |
| suprafaţa 3 | 0,33 | 0,48 | 0,50 | 23,13 |
| suprafaţa 4 | 0,34 | 0,48 | 0,49 | 21,40 |
| suprafaţa 5 | 0,37 | 0,50 | 0,49 | 23,35 |
| suprafaţa 6 | 0,43 | 0,48 | 0,50 | 21,35 |
| suprafaţa 7 | 0,43 | 0,49 | 0,50 | 20,04 |
| suprafaţa 8 | 0,29 | 0,49 | 0,50 | 21,91 |
| suprafaţa 9 | 0,37 | 0,47 | 0,51 | 21,88 |
| suprafaţa 10 | 0,30 | 0,48 | 0,51 | 20,78 |
| **medie** | 0,36 | 0,49 | 0,50 | 21,98 |

**Anexa 2. Coeficienții de corelație dintre indicii de diferențiere (T) și dominanță (U) și înălțimea medie, respectiv desimea puieților**

| | T1 | T2 | T3 | T4 | T5 |
|---|---|---|---|---|---|
| Corelația cu înălțimea medie | -0,492 | -0,474 | -0,532 | -0,529 | -0,564* |
| | **U1 var1** | **U2 var1** | **U3 var1** | **U4 var1** | **U5 var1** |
| | -0,556* | -0,702** | -0,510 | -0,500 | -0,605* |
| | **U1 var2** | **U2 var2** | **U3 var2** | **U4 var2** | **U5 var2** |
| | 0,690** | 0,674** | 0,583* | 0,587* | 0,694** |
| Corelația cu desimea puieților | **T1** | **T2** | **T3** | **T4** | **T5** |
| | -0,048 | 0,005 | 0,005 | 0,014 | 0,001 |
| | **U1 var1** | **U2 var1** | **U3 var1** | **U4 var1** | **U5 var1** |
| | -0,170 | -0,192 | -0,363 | -0,486 | -0,357 |
| | **U1 var2** | **U2 var2** | **U3 var2** | **U4 var2** | **U5 var2** |
| | 0,356 | 0,430 | 0,614* | 0,660** | 0,602* |

În anexa 1 și 2 au fost folosite următoarele notații:

**T**k – reprezintă indicele de diferențere dimensională calculat în condițiile comparației făcute cu primii k vecini ai puietului de referință (Gadow, 1997);

**U**k var 1 – reprezintă indicele de dominanță dimensională calculat în condițiile comparației făcute cu primii k vecini ai puietului de referință - varianta de calcul propusă de Hui et al (1998);

**U**k var 2 – reprezintă indicele de dominanță dimensională calculat în condițiile comparației făcute cu primii k vecini ai puietului de referință - varianta de calcul propusă de Gadow și Hui (1999) și Aguirre et al (2003);

**distanța medie** Kv – reprezintă media distanțelor medii de la puietul de referință la primii k vecini.

**Anexa 3. Coeficienții de corelație dintre indicii de diferențiere (T) și dominanță (U) și cei mai utilizați indici de diversitate structurală**

Indicele de diferențiere dimensională T

| Nr. vecini | Shannon | Simpson (D) | Berger Parker | Margalef | Menhinick |
|---|---|---|---|---|---|
| 1 vecin | -0,192 | 0,348 | 0,520 | 0,035 | 0,059 |
| 2 vecini | -0,194 | 0,339 | 0,480 | -0,031 | -0,005 |
| 3 vecini | -0,228 | 0,371 | 0,508 | -0,071 | -0,025 |
| 4 vecini | -0,225 | 0,376 | 0,518 | -0,067 | -0,026 |
| 5 vecini | -0,228 | 0,377 | 0,526 | -0,074 | -0,022 |

Indicele de dominanță dimensională U – varianta 1, Hui et al (1998);

| Nr. vecini | Shannon | Simpson (D) | Berger Parker | Margalef | Menhinick |
|---|---|---|---|---|---|
| 1 vecin | -0,003 | -0,044 | -0,007 | 0,006 | 0,142 |
| 2 vecini | -0,029 | -0,060 | -0,039 | -0,253 | -0,033 |
| 3 vecini | 0,166 | -0,245 | -0,160 | -0,030 | 0,227 |
| 4 vecini | 0,089 | -0,155 | -0,061 | 0,033 | 0,336 |
| 5 vecini | -0,074 | 0,003 | 0,038 | -0,191 | 0,144 |

Indicele de dominanță dimensională U – varianta 2, Gadow și Hui (1999) și Aguirre et al (2003);

| Nr. vecini | Shannon | Simpson (D) | Berger Parker | Margalef | Menhinick |
|---|---|---|---|---|---|
| 1 vecin | 0,264 | -0,119 | -0,046 | 0,190 | -0,132 |
| 2 vecini | 0,286 | -0,113 | -0,026 | 0,307 | -0,073 |
| 3 vecini | 0,112 | 0,035 | 0,040 | 0,080 | -0,343 |
| 4 vecini | 0,259 | -0,146 | -0,145 | 0,063 | -0,376 |
| 5 vecini | 0,438 | -0,352 | -0,330 | 0,195 | -0,275 |

**Anexa 4. Variația grafică a indicelui IDIV în suprafețele studiate**

**a) pasul de analiză - 25 cm**

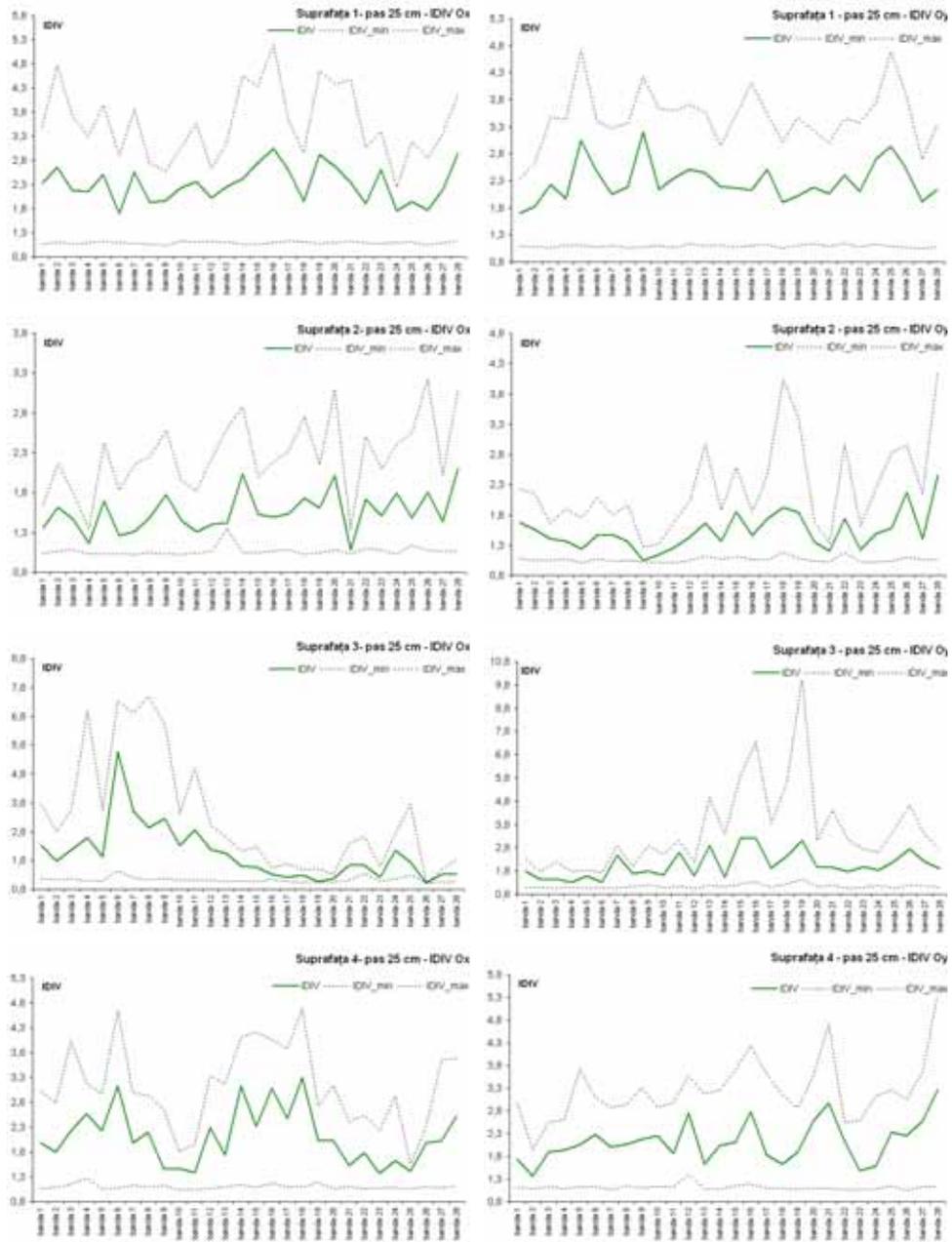

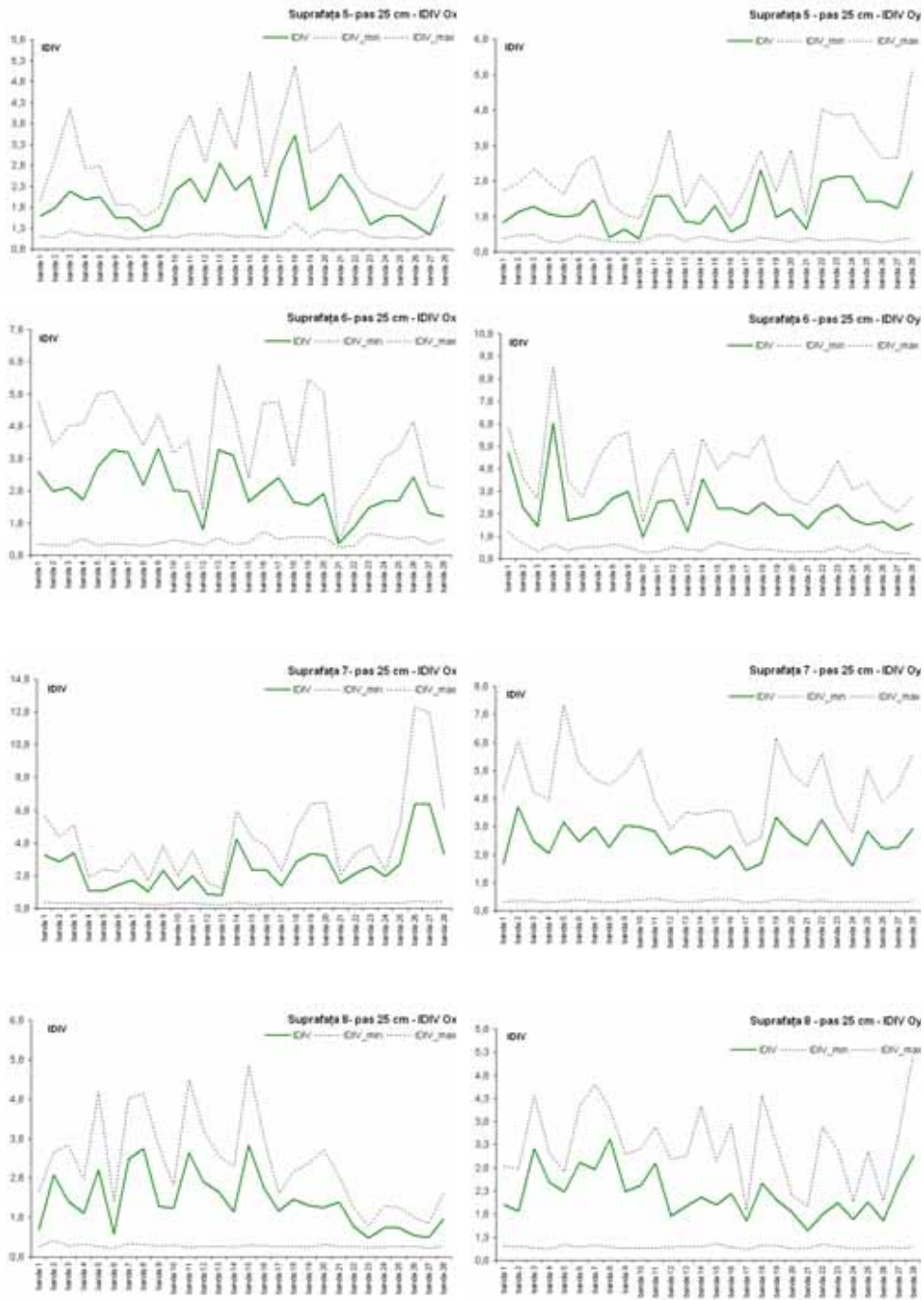

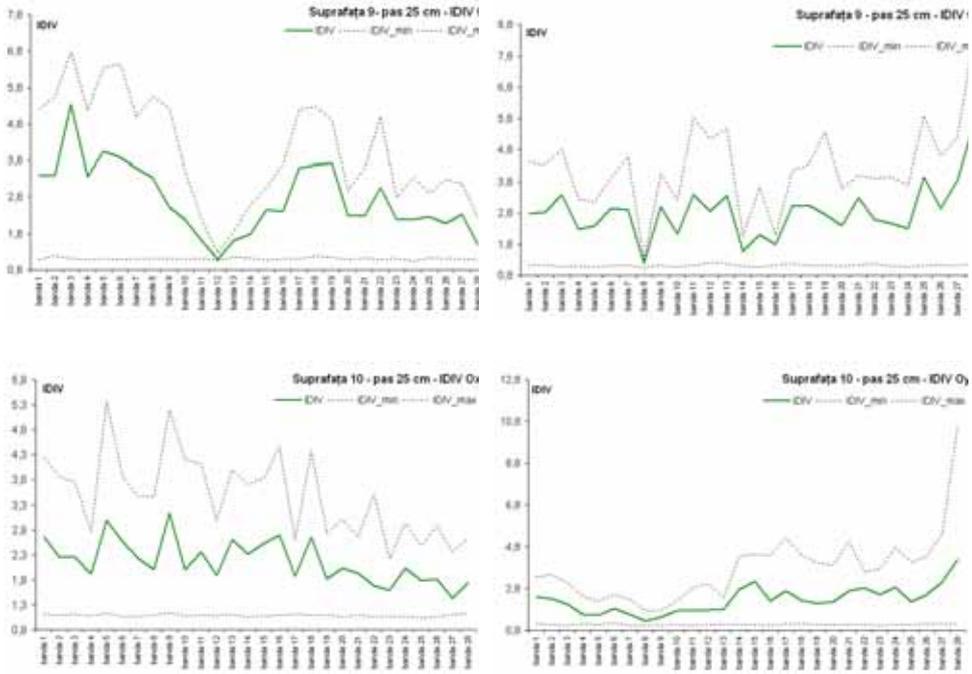

**b) pasul de analiză - 50 cm**

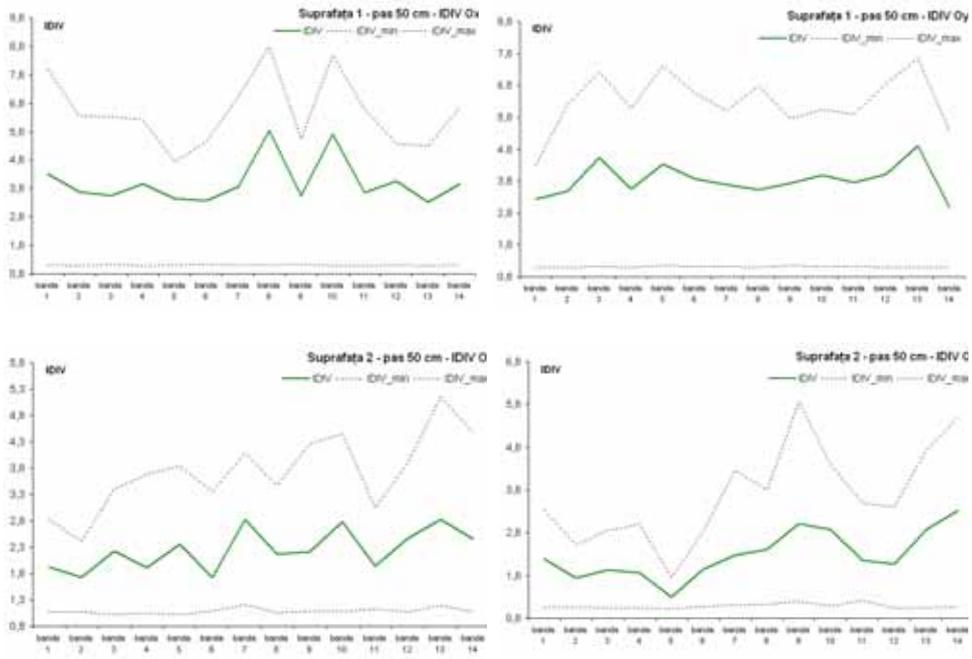

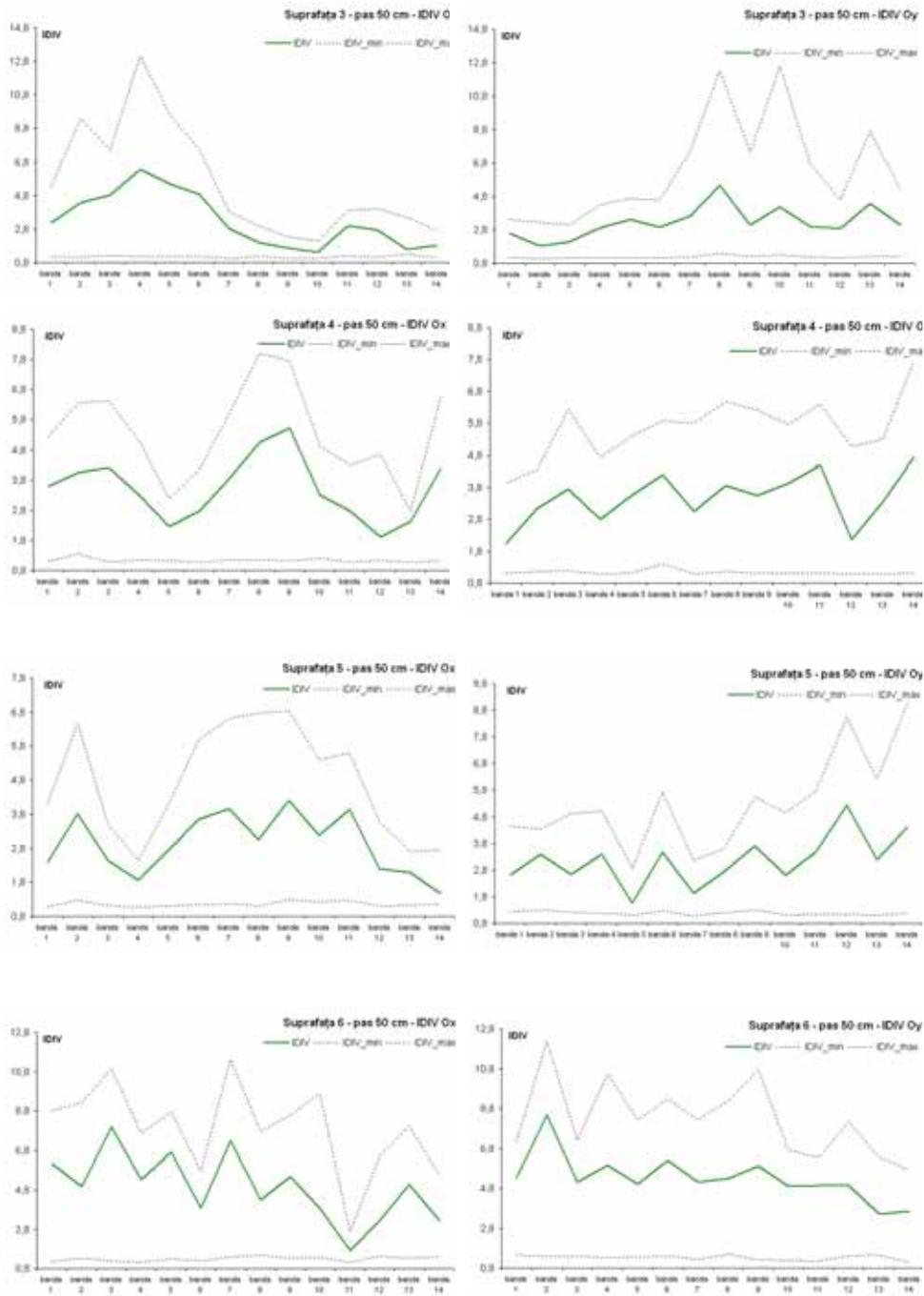

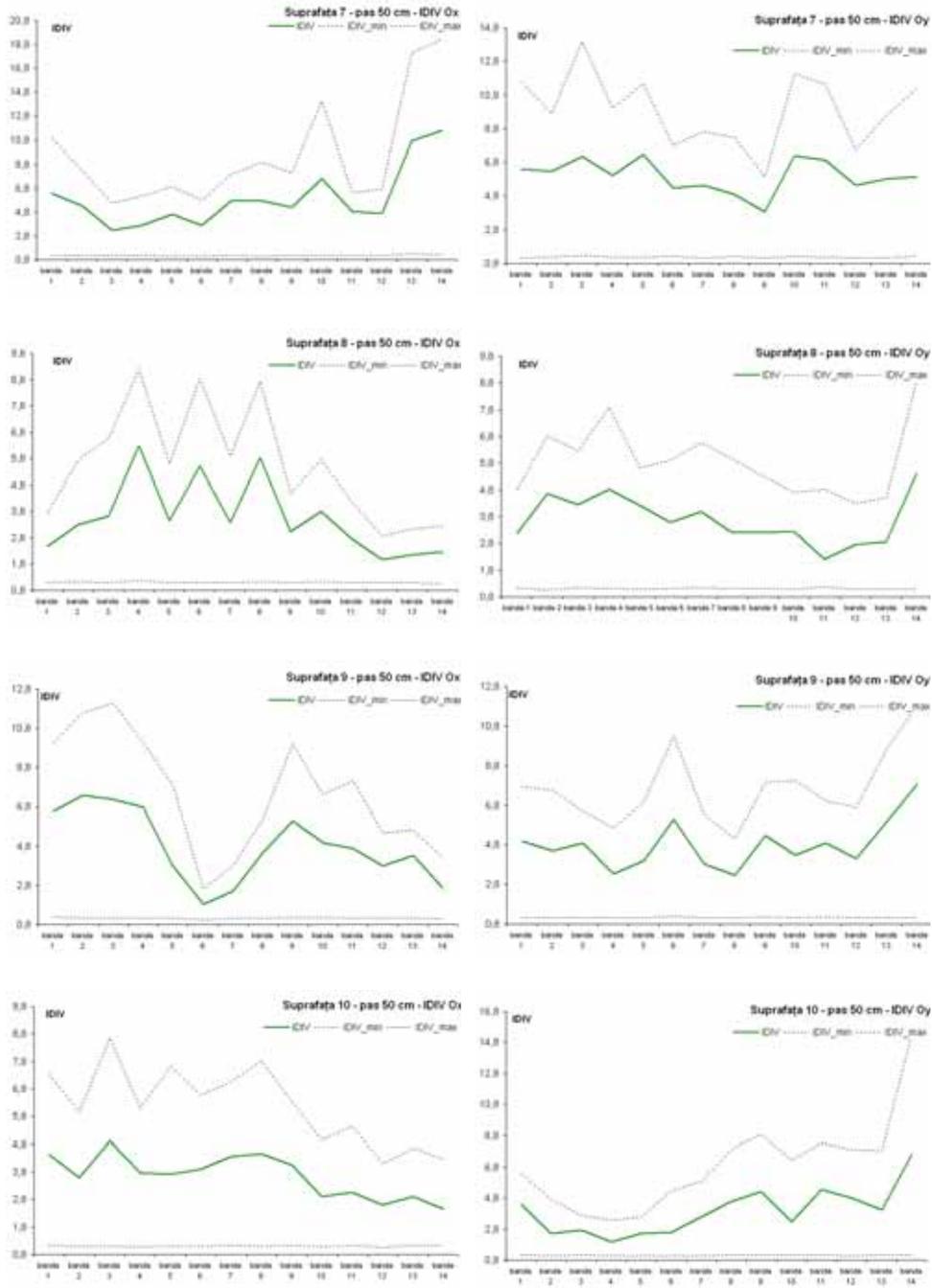

## c) pasul de analiză – 100 cm

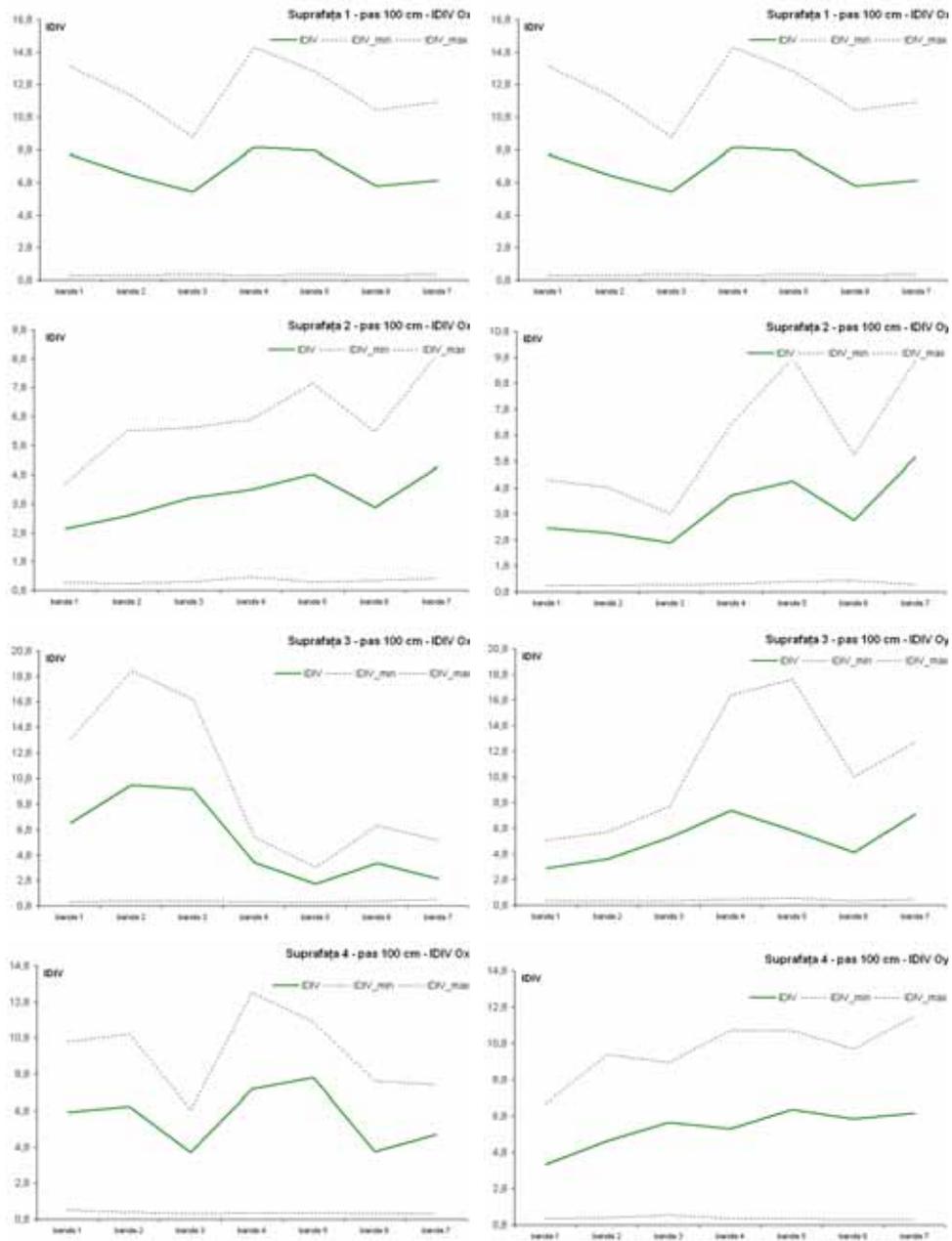

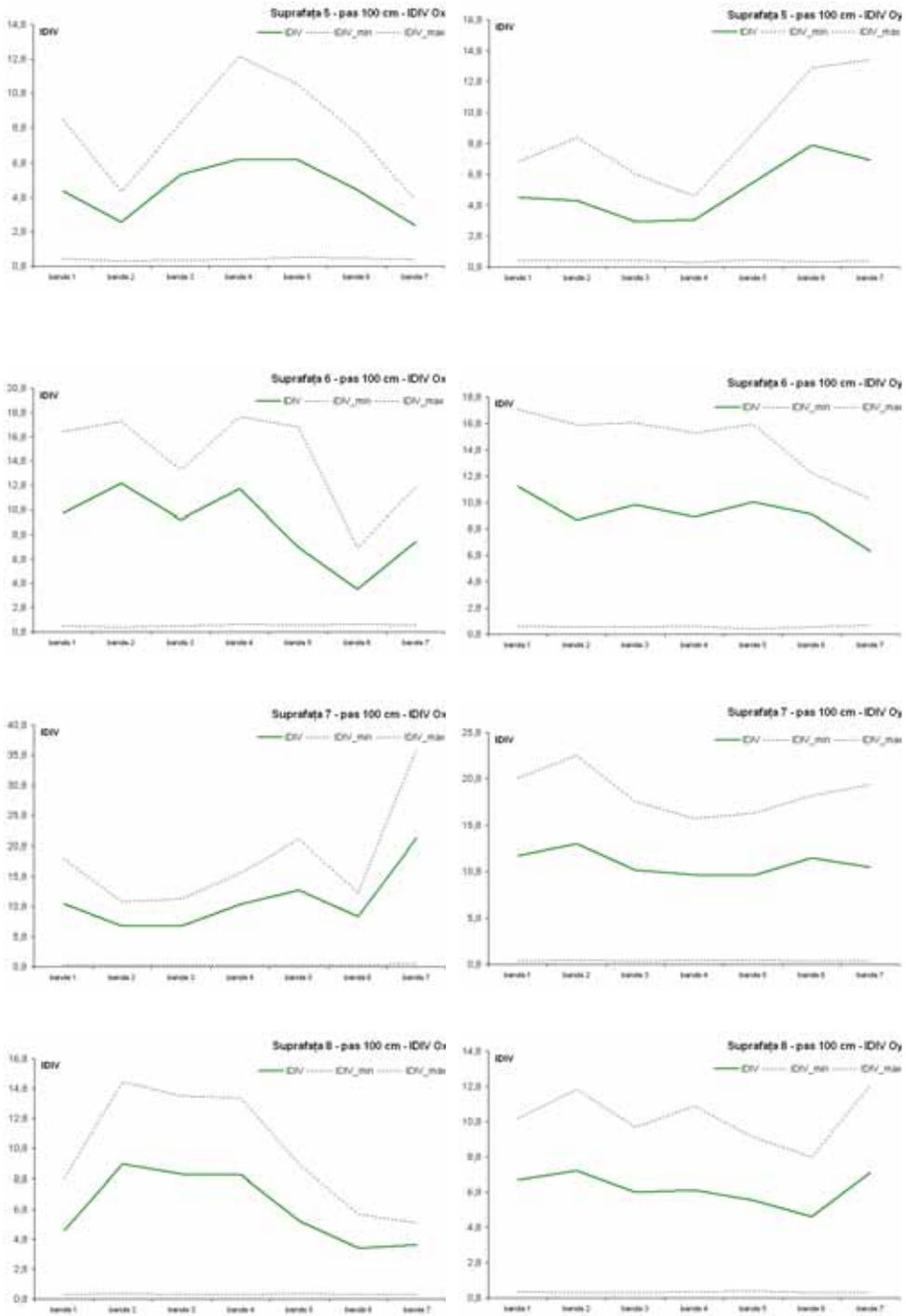

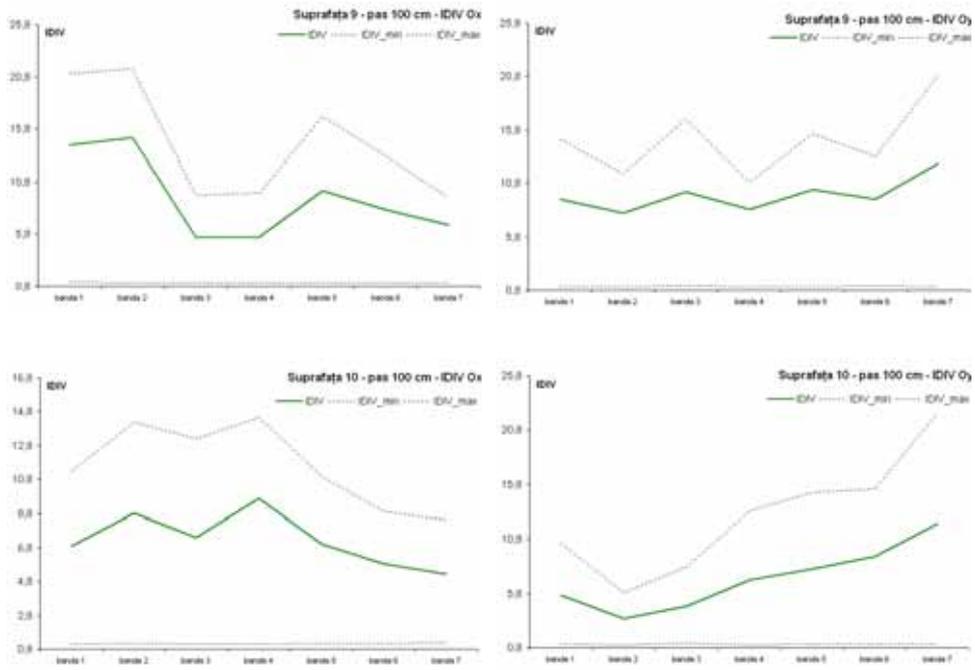

## Anexa 5. Indicatorii obținuți prin prelucrarea datelor cu ajutorul aplicației IDIV

| Nr. bandă | Suprafața 1 componenta Ox | | | | | | Suprafața 1 componenta Oy | | | | | |
|---|---|---|---|---|---|---|---|---|---|---|---|---|
| | IDIV | IDIV min | IDIV max | min teor. | max teor. | nr. elem | IDIV | IDIV min | IDIV max | min teor. | max teor. | nr. elem. |
| **pas 25 cm** IDIV Ox 2,311 (coef. var. 16,42 %); IDIV Oy 2,286 (coef. var. 15,82 %); media distanței medii față de primii 3/5/7 vecini: 17,328/21,788/25,504 cm | | | | | | | | | | | | |
| 1 | **2,330** | 1,080 | 3,496 | 1,029 | 8,302 | 32 | **1,701** | 1,098 | 2,335 | 1,027 | 5,455 | 21 |
| 2 | **2,647** | 1,089 | 4,769 | 1,039 | 12,811 | 44 | **1,823** | 1,074 | 2,638 | 1,019 | 5,680 | 26 |
| 3 | **2,181** | 1,059 | 3,719 | 1,023 | 7,547 | 32 | **2,233** | 1,060 | 3,487 | 1,024 | 6,730 | 28 |
| 4 | **2,154** | 1,098 | 3,306 | 1,026 | 7,456 | 30 | **1,962** | 1,107 | 3,432 | 1,029 | 8,296 | 32 |
| 5 | **2,505** | 1,118 | 3,951 | 1,039 | 9,521 | 32 | **3,048** | 1,105 | 4,736 | 1,028 | 10,642 | 42 |
| 6 | **1,699** | 1,101 | 2,904 | 1,027 | 5,438 | 21 | **2,474** | 1,082 | 3,402 | 1,029 | 8,586 | 33 |
| 7 | **2,558** | 1,082 | 3,843 | 1,030 | 7,168 | 27 | **2,048** | 1,107 | 3,289 | 1,025 | 7,609 | 31 |
| 8 | **1,928** | 1,062 | 2,759 | 1,017 | 5,895 | 28 | **2,195** | 1,054 | 3,356 | 1,018 | 6,703 | 32 |
| 9 | **1,964** | 1,035 | 2,585 | 1,013 | 4,463 | 23 | **3,207** | 1,080 | 4,225 | 1,019 | 8,736 | 41 |
| 10 | **2,226** | 1,124 | 3,054 | 1,044 | 7,608 | 24 | **2,141** | 1,109 | 3,636 | 1,035 | 8,238 | 29 |
| 11 | **2,364** | 1,111 | 3,561 | 1,037 | 8,479 | 29 | **2,358** | 1,071 | 3,596 | 1,027 | 7,988 | 32 |
| 12 | **2,013** | 1,117 | 2,626 | 1,036 | 5,917 | 20 | **2,513** | 1,130 | 3,707 | 1,034 | 8,656 | 31 |
| 13 | **2,252** | 1,114 | 3,169 | 1,029 | 6,893 | 26 | **2,450** | 1,104 | 3,573 | 1,036 | 8,895 | 31 |
| 14 | **2,422** | 1,064 | 4,543 | 1,022 | 7,383 | 32 | **2,199** | 1,100 | 2,956 | 1,028 | 5,593 | 21 |
| 15 | **2,744** | 1,059 | 4,326 | 1,021 | 9,561 | 43 | **2,167** | 1,074 | 3,525 | 1,029 | 7,605 | 29 |
| 16 | **3,053** | 1,096 | 5,211 | 1,038 | 11,86 | 41 | **2,133** | 1,100 | 4,116 | 1,032 | 8,456 | 31 |
| 17 | **2,595** | 1,142 | 3,625 | 1,049 | 8,660 | 26 | **2,506** | 1,124 | 3,532 | 1,041 | 8,855 | 29 |
| 18 | **1,949** | 1,104 | 2,961 | 1,028 | 6,518 | 25 | **1,900** | 1,047 | 3,020 | 1,018 | 6,739 | 32 |
| 19 | **2,914** | 1,076 | 4,655 | 1,032 | 11,42 | 43 | **2,025** | 1,104 | 3,466 | 1,029 | 7,633 | 29 |
| 20 | **2,686** | 1,101 | 4,390 | 1,028 | 8,677 | 34 | **2,184** | 1,129 | 3,231 | 1,038 | 7,433 | 25 |
| 21 | **2,351** | 1,119 | 4,467 | 1,042 | 11,83 | 39 | **2,061** | 1,088 | 3,010 | 1,046 | 6,886 | 21 |
| 22 | **1,901** | 1,094 | 3,068 | 1,033 | 6,733 | 24 | **2,405** | 1,144 | 3,453 | 1,053 | 8,637 | 25 |
| 23 | **2,613** | 1,085 | 3,409 | 1,035 | 7,187 | 25 | **2,103** | 1,078 | 3,375 | 1,029 | 7,557 | 29 |
| 24 | **1,750** | 1,096 | 2,248 | 1,024 | 5,695 | 23 | **2,694** | 1,127 | 3,727 | 1,048 | 9,173 | 28 |
| 25 | **1,950** | 1,112 | 3,195 | 1,029 | 7,535 | 29 | **2,939** | 1,086 | 4,685 | 1,028 | 9,144 | 36 |
| 26 | **1,765** | 1,043 | 2,839 | 1,016 | 5,510 | 27 | **2,528** | 1,067 | 3,895 | 1,027 | 8,906 | 36 |
| 27 | **2,203** | 1,098 | 3,379 | 1,037 | 7,102 | 24 | **1,910** | 1,053 | 2,693 | 1,022 | 5,470 | 23 |
| 28 | **2,949** | 1,127 | 4,171 | 1,055 | 9,485 | 27 | **2,137** | 1,081 | 3,329 | 1,039 | 8,117 | 27 |
| **pas 50 cm** IDIV Ox 4,023 (coef. var. 19,77%); IDIV Oy 3,830 (coef. var. 13,29%); | | | | | | | | | | | | |
| 1 | **4,333** | 1,110 | 8,007 | 1,038 | 21,75 | 76 | **3,231** | 1,108 | 4,295 | 1,027 | 11,514 | 47 |
| 2 | **3,675** | 1,090 | 6,364 | 1,027 | 14,93 | 62 | **3,485** | 1,103 | 6,202 | 1,029 | 15,079 | 60 |
| 3 | **3,558** | 1,124 | 6,320 | 1,039 | 15,39 | 53 | **4,540** | 1,114 | 7,216 | 1,033 | 19,831 | 75 |
| 4 | **3,974** | 1,085 | 6,230 | 1,030 | 14,17 | 55 | **3,552** | 1,087 | 6,061 | 1,024 | 14,550 | 63 |
| 5 | **3,440** | 1,108 | 4,738 | 1,038 | 13,52 | 47 | **4,327** | 1,132 | 7,426 | 1,034 | 18,868 | 70 |
| 6 | **3,374** | 1,126 | 5,464 | 1,037 | 13,99 | 49 | **3,864** | 1,113 | 6,559 | 1,032 | 16,662 | 63 |
| 7 | **3,862** | 1,104 | 7,020 | 1,028 | 14,38 | 58 | **3,698** | 1,123 | 6,005 | 1,036 | 14,567 | 52 |

| Nr. bandă | Suprafața 1 componenta Ox | | | | | | Suprafața 1 componenta Oy | | | | | |
|---|---|---|---|---|---|---|---|---|---|---|---|---|
| | IDIV | IDIV min | IDIV max | min teor. | max teor. | nr. elem. | IDIV | IDIV min | IDIV max | min teor. | max teor. | nr. elem. |
| 8 | 5,837 | 1,105 | 8,796 | 1,033 | 22,25 | 84 | 3,518 | 1,096 | 6,783 | 1,028 | 14,959 | 60 |
| 9 | 3,557 | 1,143 | 5,556 | 1,042 | 15,47 | 51 | 3,725 | 1,135 | 5,750 | 1,041 | 18,115 | 61 |
| 10 | 5,716 | 1,093 | 8,487 | 1,031 | 19,73 | 77 | 3,997 | 1,129 | 6,056 | 1,039 | 15,652 | 54 |
| 11 | 3,650 | 1,095 | 6,582 | 1,038 | 18,04 | 63 | 3,753 | 1,128 | 5,881 | 1,052 | 15,457 | 46 |
| 12 | 4,066 | 1,099 | 5,383 | 1,032 | 12,73 | 48 | 4,016 | 1,102 | 6,850 | 1,037 | 16,224 | 57 |
| 13 | 3,331 | 1,086 | 5,307 | 1,026 | 13,8 | 56 | 4,901 | 1,091 | 7,660 | 1,027 | 17,551 | 72 |
| 14 | 3,954 | 1,116 | 6,650 | 1,044 | 15,77 | 51 | 2,982 | 1,101 | 5,360 | 1,038 | 14,323 | 50 |
| **pas 100 cm** IDIV Ox 7,586 (coef. var. 14,65%); IDIV Oy 7,517 (coef. var. 6,82%); | | | | | | | | | | | | |
| 1 | 8,505 | 1,116 | 13,89 | 1,038 | 38,78 | 138 | 6,602 | 1,124 | 10,22 | 1,031 | 27,20 | 107 |
| 2 | 7,225 | 1,128 | 12,09 | 1,041 | 31,66 | 108 | 8,147 | 1,136 | 12,75 | 1,036 | 37,61 | 138 |
| 3 | 6,246 | 1,137 | 9,55 | 1,037 | 26,80 | 96 | 7,492 | 1,115 | 13,25 | 1,032 | 34,60 | 133 |
| 4 | 8,953 | 1,117 | 15,09 | 1,033 | 37,27 | 142 | 7,333 | 1,117 | 12,24 | 1,035 | 30,46 | 112 |
| 5 | 8,745 | 1,134 | 13,61 | 1,041 | 37,59 | 128 | 7,433 | 1,123 | 11,08 | 1,039 | 32,76 | 115 |
| 6 | 6,579 | 1,094 | 11,24 | 1,034 | 29,85 | 111 | 7,521 | 1,138 | 12,35 | 1,050 | 33,30 | 103 |
| 7 | 6,926 | 1,129 | 11,71 | 1,049 | 34,30 | 107 | 8,065 | 1,102 | 12,45 | 1,035 | 32,97 | 122 |

| Nr. bandă | Suprafața 2 componenta Ox | | | | | | Suprafața 2 componenta Oy | | | | | |
|---|---|---|---|---|---|---|---|---|---|---|---|---|
| | IDIV | IDIV min | IDIV max | min teor. | max teor. | nr. elem. | IDIV | IDIV min | IDIV max | min teor. | max teor. | nr. elem. |
| **pas 25 cm** IDIV Ox 1,563(coef. var. 16,09 %); IDIV Oy 1,537 (coef. var. 20,51 %); media distanței medii față de primii 3/5/7 vecini: 19,208/ 24,536/ 29,180cm | | | | | | | | | | | | |
| 1 | 1,355 | 1,044 | 1,634 | 1,010 | 2,293 | 11 | 1,680 | 1,078 | 2,232 | 1,014 | 3,842 | 19 |
| 2 | 1,619 | 1,067 | 2,165 | 1,019 | 3,973 | 17 | 1,553 | 1,052 | 2,150 | 1,013 | 3,554 | 18 |
| 3 | 1,466 | 1,080 | 1,799 | 1,012 | 3,340 | 17 | 1,404 | 1,047 | 1,669 | 1,012 | 2,903 | 14 |
| 4 | 1,165 | 1,031 | 1,348 | 1,008 | 2,515 | 14 | 1,368 | 1,073 | 1,899 | 1,020 | 3,820 | 16 |
| 5 | 1,696 | 1,029 | 2,420 | 1,006 | 4,387 | 34 | 1,237 | 1,017 | 1,749 | 1,006 | 3,153 | 21 |
| 6 | 1,264 | 1,031 | 1,826 | 1,009 | 2,983 | 17 | 1,460 | 1,076 | 2,082 | 1,020 | 4,679 | 20 |
| 7 | 1,318 | 1,024 | 2,158 | 1,006 | 3,284 | 22 | 1,471 | 1,035 | 1,810 | 1,014 | 3,001 | 14 |
| 8 | 1,491 | 1,050 | 2,258 | 1,010 | 4,315 | 25 | 1,353 | 1,045 | 1,955 | 1,010 | 3,327 | 18 |
| 9 | 1,778 | 1,030 | 2,586 | 1,012 | 4,722 | 26 | 1,052 | 1,020 | 1,271 | 1,010 | 1,707 | 7 |
| 10 | 1,459 | 1,026 | 1,956 | 1,008 | 3,347 | 20 | 1,149 | 1,019 | 1,318 | 1,005 | 1,833 | 10 |
| 11 | 1,308 | 1,036 | 1,823 | 1,020 | 3,047 | 12 | 1,256 | 1,019 | 1,687 | 1,006 | 2,989 | 20 |
| 12 | 1,402 | 1,070 | 2,243 | 1,025 | 4,152 | 16 | 1,443 | 1,052 | 2,044 | 1,021 | 3,692 | 15 |
| 13 | 1,422 | 1,353 | 2,613 | 1,196 | 5,127 | 8 | 1,662 | 1,117 | 2,975 | 1,028 | 5,316 | 20 |
| 14 | 2,031 | 1,052 | 2,875 | 1,010 | 5,785 | 35 | 1,366 | 1,071 | 1,882 | 1,018 | 3,306 | 14 |
| 15 | 1,527 | 1,044 | 1,994 | 1,010 | 4,344 | 26 | 1,856 | 1,103 | 2,583 | 1,036 | 6,477 | 22 |
| 16 | 1,492 | 1,064 | 2,195 | 1,015 | 3,954 | 19 | 1,469 | 1,066 | 1,836 | 1,016 | 3,893 | 18 |
| 17 | 1,534 | 1,082 | 2,318 | 1,032 | 6,377 | 23 | 1,732 | 1,055 | 2,463 | 1,020 | 4,636 | 20 |
| 18 | 1,731 | 1,024 | 2,754 | 1,009 | 4,452 | 27 | 1,928 | 1,184 | 4,033 | 1,063 | 10,09 | 27 |
| 19 | 1,607 | 1,046 | 2,150 | 1,008 | 3,978 | 25 | 1,847 | 1,079 | 3,418 | 1,022 | 7,580 | 33 |
| 20 | 2,017 | 1,086 | 3,097 | 1,021 | 5,979 | 26 | 1,351 | 1,041 | 1,694 | 1,008 | 2,649 | 15 |
| 21 | 1,083 | 1,033 | 1,330 | 1,009 | 1,831 | 8 | 1,204 | 1,026 | 1,332 | 1,003 | 1,857 | 14 |

| | Suprafața 2 componenta Ox | | | | | | Suprafața 2 componenta Oy | | | | | |
|---|---|---|---|---|---|---|---|---|---|---|---|---|
| Nr. bandă | IDIV | IDIV min | IDIV max | min teor. | max teor. | nr. elem. | IDIV | IDIV min | IDIV max | min teor. | max teor. | nr. elem. |
| 22 | **1,713** | 1,100 | 2,508 | 1,030 | 4,949 | 18 | **1,734** | 1,180 | 2,977 | 1,076 | 7,433 | 18 |
| 23 | **1,513** | 1,078 | 2,098 | 1,017 | 4,205 | 19 | **1,227** | 1,030 | 1,617 | 1,011 | 3,225 | 17 |
| 24 | **1,792** | 1,030 | 2,411 | 1,013 | 3,923 | 20 | **1,492** | 1,030 | 2,234 | 1,010 | 3,797 | 22 |
| 25 | **1,487** | 1,135 | 2,551 | 1,071 | 4,905 | 12 | **1,577** | 1,035 | 2,832 | 1,015 | 5,996 | 31 |
| 26 | **1,801** | 1,071 | 3,220 | 1,027 | 5,980 | 23 | **2,171** | 1,091 | 2,952 | 1,024 | 6,808 | 28 |
| 27 | **1,438** | 1,066 | 2,019 | 1,014 | 3,337 | 16 | **1,409** | 1,063 | 2,151 | 1,012 | 3,636 | 19 |
| 28 | **2,104** | 1,062 | 3,099 | 1,019 | 6,766 | 31 | **2,461** | 1,054 | 4,187 | 1,016 | 10,909 | 57 |

**pas 50 cm** IDIV Ox 2,260 (coef. var. 17,08 %); IDIV Oy 2,286 (coef. var. 24,53%);

| Nr. bandă | IDIV | IDIV min | IDIV max | min teor. | max teor. | nr. elem. | IDIV | IDIV min | IDIV max | min teor. | max teor. | nr. elem. |
|---|---|---|---|---|---|---|---|---|---|---|---|---|
| 1 | **1,934** | 1,065 | 2,840 | 1,020 | 6,179 | 28 | **2,184** | 1,064 | 3,336 | 1,013 | 6,625 | 37 |
| 2 | **1,737** | 1,069 | 2,412 | 1,012 | 5,513 | 31 | **1,735** | 1,052 | 2,518 | 1,014 | 5,695 | 30 |
| 3 | **2,231** | 1,034 | 3,388 | 1,007 | 6,621 | 51 | **1,938** | 1,046 | 2,863 | 1,013 | 7,355 | 41 |
| 4 | **1,917** | 1,047 | 3,685 | 1,011 | 7,673 | 47 | **1,871** | 1,046 | 2,998 | 1,013 | 5,941 | 32 |
| 5 | **2,365** | 1,034 | 3,837 | 1,010 | 7,372 | 46 | **1,290** | 1,031 | 1,763 | 1,010 | 3,169 | 17 |
| 6 | **1,729** | 1,086 | 3,353 | 1,025 | 6,893 | 28 | **1,955** | 1,078 | 2,837 | 1,017 | 7,202 | 35 |
| 7 | **2,836** | 1,222 | 4,088 | 1,074 | 17,17 | 43 | **2,273** | 1,104 | 4,265 | 1,027 | 8,505 | 34 |
| 8 | **2,169** | 1,053 | 3,458 | 1,015 | 8,513 | 45 | **2,410** | 1,118 | 3,794 | 1,030 | 10,44 | 40 |
| 9 | **2,214** | 1,086 | 4,254 | 1,024 | 11,62 | 50 | **3,006** | 1,199 | 5,851 | 1,063 | 17,23 | 47 |
| 10 | **2,791** | 1,081 | 4,448 | 1,019 | 10,72 | 51 | **2,868** | 1,085 | 4,396 | 1,023 | 10,96 | 48 |
| 11 | **1,940** | 1,134 | 3,031 | 1,030 | 6,909 | 26 | **2,149** | 1,215 | 3,486 | 1,063 | 11,86 | 32 |
| 12 | **2,477** | 1,070 | 3,894 | 1,017 | 7,845 | 39 | **2,074** | 1,041 | 3,401 | 1,011 | 6,473 | 39 |
| 13 | **2,835** | 1,200 | 5,159 | 1,068 | 13,45 | 35 | **2,879** | 1,046 | 4,719 | 1,016 | 11,25 | 59 |
| 14 | **2,467** | 1,077 | 4,472 | 1,018 | 9,644 | 47 | **3,325** | 1,077 | 5,491 | 1,016 | 14,27 | 76 |

**pas 100 cm** IDIV Ox 4,018 (coef. var. 19,03%); IDIV Oy 4,014 (coef. var. 29,82%);

| Nr. bandă | IDIV | IDIV min | IDIV max | min teor. | max teor. | nr. elem. | IDIV | IDIV min | IDIV max | min teor. | max teor. | nr. elem. |
|---|---|---|---|---|---|---|---|---|---|---|---|---|
| 1 | **2,930** | 1,080 | 4,484 | 1,019 | 12,23 | 59 | **3,265** | 1,051 | 5,113 | 1,014 | 11,93 | 67 |
| 2 | **3,375** | 1,050 | 6,316 | 1,011 | 15,22 | 98 | **3,065** | 1,044 | 4,808 | 1,013 | 12,37 | 73 |
| 3 | **3,985** | 1,096 | 6,404 | 1,021 | 15,88 | 74 | **2,686** | 1,079 | 3,795 | 1,017 | 10,39 | 52 |
| 4 | **4,296** | 1,261 | 6,706 | 1,066 | 32,85 | 88 | **4,509** | 1,114 | 7,243 | 1,030 | 18,85 | 74 |
| 5 | **4,831** | 1,105 | 7,950 | 1,024 | 22,84 | 101 | **5,065** | 1,222 | 9,754 | 1,062 | 34,38 | 95 |
| 6 | **3,687** | 1,146 | 6,270 | 1,029 | 16,35 | 65 | **3,554** | 1,234 | 6,058 | 1,064 | 26,09 | 71 |
| 7 | **5,052** | 1,228 | 8,943 | 1,064 | 30,19 | 82 | **5,969** | 1,067 | 9,668 | 1,015 | 24,40 | 135 |

| | Suprafața 3 componenta Ox | | | | | | Suprafața 3 componenta Oy | | | | | |
|---|---|---|---|---|---|---|---|---|---|---|---|---|
| Nr. bandă | IDIV | IDIV min | IDIV max | min teor. | max teor. | nr. elem. | IDIV | IDIV min | IDIV max | min teor. | max teor. | nr. elem. |

**pas 25 cm** IDIV Ox 1,982 (coef. var. 47,57%); IDIV Oy 2,040 (coef. var. 28,06%); media distanței medii față de primii 3/5/7 vecini: 19,508/ 25,149/ 29,844 cm

| Nr. bandă | IDIV | IDIV min | IDIV max | min teor. | max teor. | nr. elem. | IDIV | IDIV min | IDIV max | min teor. | max teor. | nr. elem. |
|---|---|---|---|---|---|---|---|---|---|---|---|---|
| 1 | **2,311** | 1,173 | 3,751 | 1,063 | 8,649 | 23 | **1,791** | 1,107 | 2,340 | 1,050 | 4,555 | 13 |
| 2 | **1,775** | 1,133 | 2,796 | 1,059 | 4,899 | 13 | **1,428** | 1,107 | 1,778 | 1,054 | 3,392 | 9 |
| 3 | **2,171** | 1,158 | 3,563 | 1,054 | 6,044 | 17 | **1,466** | 1,082 | 2,209 | 1,032 | 4,078 | 14 |
| 4 | **2,581** | 1,097 | 6,995 | 1,050 | 15,461 | 47 | **1,320** | 1,092 | 1,761 | 1,049 | 2,952 | 8 |
| 5 | **1,931** | 1,105 | 3,549 | 1,040 | 8,219 | 27 | **1,582** | 1,071 | 1,851 | 1,041 | 3,348 | 10 |

| Nr. bandă | Suprafața 3 componenta Ox | | | | | | Suprafața 3 componenta Oy | | | | | |
| | IDIV | IDIV min | IDIV max | min teor. | max teor. | nr. elem. | IDIV | IDIV min | IDIV max | min teor. | max teor. | nr. elem. |
|---|---|---|---|---|---|---|---|---|---|---|---|---|
| 6 | **5,549** | 1,450 | 7,323 | 1,260 | 25,039 | 33 | **1,318** | 1,060 | 1,759 | 1,012 | 2,852 | 14 |
| 7 | **3,494** | 1,192 | 6,900 | 1,077 | 20,677 | 51 | **2,462** | 1,087 | 2,915 | 1,038 | 6,028 | 20 |
| 8 | **2,943** | 1,154 | 7,505 | 1,067 | 17,767 | 47 | **1,676** | 1,122 | 2,000 | 1,039 | 4,153 | 13 |
| 9 | **3,253** | 1,178 | 6,515 | 1,063 | 13,362 | 36 | **1,815** | 1,204 | 2,903 | 1,075 | 5,027 | 12 |
| 10 | **2,329** | 1,110 | 3,436 | 1,040 | 6,736 | 22 | **1,610** | 1,102 | 2,495 | 1,040 | 5,605 | 18 |
| 11 | **2,854** | 1,121 | 5,022 | 1,069 | 10,125 | 26 | **2,584** | 1,131 | 3,124 | 1,091 | 6,770 | 15 |
| 12 | **2,178** | 1,108 | 3,013 | 1,048 | 6,360 | 19 | **1,581** | 1,072 | 2,206 | 1,032 | 3,831 | 13 |
| 13 | **2,059** | 1,067 | 2,609 | 1,037 | 4,906 | 16 | **2,917** | 1,170 | 4,985 | 1,059 | 12,170 | 34 |
| 14 | **1,619** | 1,101 | 2,131 | 1,029 | 4,188 | 15 | **1,525** | 1,123 | 3,358 | 1,040 | 5,871 | 19 |
| 15 | **1,566** | 1,065 | 2,281 | 1,029 | 4,680 | 17 | **3,223** | 1,219 | 5,912 | 1,086 | 12,117 | 28 |
| 16 | **1,314** | 1,134 | 1,539 | 1,048 | 2,623 | 7 | **3,248** | 1,344 | 7,341 | 1,176 | 17,883 | 29 |
| 17 | **1,215** | 1,072 | 1,690 | 1,015 | 2,224 | 9 | **1,916** | 1,107 | 3,837 | 1,049 | 7,086 | 21 |
| 18 | **1,299** | 1,039 | 1,483 | 1,016 | 2,471 | 10 | **2,451** | 1,250 | 5,573 | 1,092 | 12,061 | 27 |
| 19 | **1,061** | 1,054 | 1,524 | 1,020 | 2,023 | 7 | **3,120** | 1,425 | 10,094 | 1,192 | 27,158 | 42 |
| 20 | **1,177** | 1,037 | 1,314 | 1,020 | 2,415 | 9 | **1,960** | 1,131 | 3,111 | 1,057 | 5,162 | 14 |
| 21 | **1,633** | 1,126 | 2,392 | 1,065 | 4,716 | 12 | **1,953** | 1,166 | 4,442 | 1,048 | 8,235 | 25 |
| 22 | **1,627** | 1,338 | 2,617 | 1,210 | 4,619 | 7 | **1,787** | 1,088 | 3,160 | 1,039 | 6,145 | 20 |
| 23 | **1,220** | 1,097 | 1,596 | 1,028 | 2,225 | 7 | **1,994** | 1,091 | 2,765 | 1,042 | 4,251 | 13 |
| 24 | **2,147** | 1,140 | 2,743 | 1,035 | 5,582 | 19 | **1,826** | 1,170 | 2,602 | 1,040 | 3,605 | 11 |
| 25 | **1,729** | 1,302 | 3,738 | 1,136 | 4,905 | 9 | **2,231** | 1,087 | 3,618 | 1,032 | 7,162 | 26 |
| 26 | **1,016** | 1,016 | 1,016 | 1,016 | 1,016 | 2 | **2,741** | 1,200 | 4,634 | 1,075 | 10,527 | 26 |
| 27 | **1,303** | 1,034 | 1,488 | 1,010 | 1,995 | 9 | **2,195** | 1,154 | 3,358 | 1,084 | 7,359 | 17 |
| 28 | **1,322** | 1,080 | 1,820 | 1,022 | 3,539 | 14 | **1,892** | 1,115 | 2,718 | 1,054 | 6,702 | 19 |
| **pas 50 cm** IDIV Ox 3,303 (coef. var. 48,66 %); IDIV Oy 3,221 (coef. var. 28,85%); | | | | | | | | | | | | |
| 1 | **3,166** | 1,116 | 5,306 | 1,046 | 11,448 | 36 | **2,598** | 1,142 | 3,423 | 1,053 | 7,661 | 22 |
| 2 | **4,362** | 1,150 | 9,406 | 1,050 | 20,904 | 64 | **1,854** | 1,091 | 3,259 | 1,047 | 7,267 | 22 |
| 3 | **4,813** | 1,204 | 7,549 | 1,097 | 27,254 | 60 | **2,083** | 1,147 | 3,083 | 1,040 | 7,331 | 24 |
| 4 | **6,371** | 1,188 | 13,124 | 1,075 | 38,981 | 98 | **2,907** | 1,128 | 4,319 | 1,044 | 10,300 | 33 |
| 5 | **5,498** | 1,154 | 9,590 | 1,062 | 21,139 | 58 | **3,422** | 1,154 | 4,668 | 1,056 | 10,526 | 30 |
| 6 | **4,853** | 1,169 | 7,536 | 1,066 | 16,992 | 45 | **2,965** | 1,149 | 4,576 | 1,060 | 10,207 | 28 |
| 7 | **2,826** | 1,066 | 3,817 | 1,029 | 8,076 | 31 | **3,647** | 1,167 | 7,511 | 1,059 | 18,768 | 53 |
| 8 | **2,012** | 1,162 | 3,022 | 1,048 | 7,927 | 24 | **5,478** | 1,369 | 12,356 | 1,162 | 33,700 | 57 |
| 9 | **1,683** | 1,070 | 2,341 | 1,022 | 4,600 | 19 | **3,101** | 1,197 | 7,425 | 1,063 | 17,648 | 48 |
| 10 | **1,422** | 1,037 | 2,103 | 1,020 | 3,811 | 16 | **4,183** | 1,309 | 12,594 | 1,176 | 34,588 | 56 |
| 11 | **2,989** | 1,235 | 3,932 | 1,112 | 9,367 | 19 | **3,024** | 1,157 | 6,766 | 1,045 | 14,148 | 45 |
| 12 | **2,765** | 1,133 | 4,014 | 1,045 | 8,359 | 26 | **2,864** | 1,117 | 4,573 | 1,040 | 7,335 | 24 |
| 13 | **1,603** | 1,282 | 3,488 | 1,101 | 5,250 | 11 | **4,374** | 1,216 | 8,706 | 1,080 | 21,436 | 52 |
| 14 | **1,843** | 1,086 | 2,715 | 1,029 | 6,167 | 23 | **3,081** | 1,217 | 5,234 | 1,084 | 15,339 | 36 |
| **pas 100 cm** IDIV Ox 5,916 (coef. var. 55,23%); IDIV Oy 5,973 (coef. var. 28,85%); | | | | | | | | | | | | |
| 1 | **7,281** | 1,148 | 13,85 | 1,050 | 32,433 | 100 | **3,700** | 1,144 | 5,894 | 1,049 | 14,362 | 44 |

**Suprafața 3** componenta Ox                 **Suprafața 3** componenta Oy

| Nr. bandă | IDIV | IDIV min | IDIV max | min teor. | max teor. | nr. elem. | IDIV | IDIV min | IDIV max | min teor. | max teor. | nr. elem. |
|---|---|---|---|---|---|---|---|---|---|---|---|---|
| 2 | **10,328** | 1,205 | 19,25 | 1,076 | 63,12 | 158 | **4,419** | 1,131 | 6,520 | 1,045 | 17,704 | 57 |
| 3 | **9,965** | 1,174 | 17,01 | 1,062 | 37,27 | 103 | **6,092** | 1,151 | 8,567 | 1,059 | 20,565 | 58 |
| 4 | **4,220** | 1,146 | 6,252 | 1,046 | 17,35 | 55 | **8,210** | 1,257 | 17,180 | 1,103 | 51,482 | 110 |
| 5 | **2,510** | 1,068 | 3,814 | 1,022 | 7,97 | 35 | **6,669** | 1,374 | 18,386 | 1,150 | 59,025 | 104 |
| 6 | **4,173** | 1,212 | 7,138 | 1,083 | 18,92 | 45 | **4,936** | 1,148 | 10,762 | 1,048 | 22,136 | 69 |
| 7 | **2,943** | 1,291 | 5,980 | 1,107 | 16,33 | 34 | **7,916** | 1,230 | 13,487 | 1,083 | 36,796 | 88 |

**Suprafața 4** componenta Ox                 **Suprafața 4** componenta Oy

| Nr. bandă | IDIV | IDIV min | IDIV max | min teor. | max teor. | nr. elem. | IDIV | IDIV min | IDIV max | min teor. | max teor. | nr. elem. |
|---|---|---|---|---|---|---|---|---|---|---|---|---|
| **pas 25 cm** IDIV Ox 2,121 (coef. var. 26,48 %); IDIV Oy 2,123 (coef. var. 21,54 %); media distanței medii față de primii 3/5/7 vecini: 17,429/21,736/25,505 cm ||||||||||||
| 1 | **1,986** | 1,066 | 3,029 | 1,020 | 7,466 | 34 | **1,738** | 1,111 | 2,973 | 1,029 | 4,922 | 18 |
| 2 | **1,804** | 1,088 | 2,784 | 1,039 | 6,998 | 23 | **1,368** | 1,100 | 1,937 | 1,057 | 2,425 | 6 |
| 3 | **2,228** | 1,158 | 4,038 | 1,043 | 9,615 | 31 | **1,893** | 1,126 | 2,550 | 1,037 | 5,726 | 19 |
| 4 | **2,573** | 1,268 | 3,165 | 1,133 | 12,29 | 23 | **1,933** | 1,090 | 2,606 | 1,021 | 5,751 | 25 |
| 5 | **2,222** | 1,055 | 2,969 | 1,021 | 6,388 | 28 | **2,045** | 1,134 | 3,721 | 1,041 | 8,601 | 28 |
| 6 | **3,135** | 1,074 | 4,643 | 1,022 | 8,274 | 36 | **2,282** | 1,122 | 3,091 | 1,056 | 7,186 | 20 |
| 7 | **1,979** | 1,134 | 2,979 | 1,055 | 7,124 | 20 | **2,010** | 1,070 | 2,879 | 1,025 | 5,324 | 21 |
| 8 | **2,202** | 1,104 | 2,938 | 1,033 | 9,098 | 33 | **2,062** | 1,148 | 2,947 | 1,045 | 4,687 | 14 |
| 9 | **1,463** | 1,122 | 2,653 | 1,059 | 5,248 | 14 | **2,173** | 1,108 | 3,316 | 1,027 | 8,571 | 34 |
| 10 | **1,473** | 1,044 | 1,797 | 1,010 | 2,973 | 16 | **2,254** | 1,122 | 2,876 | 1,022 | 6,462 | 28 |
| 11 | **1,384** | 1,050 | 1,970 | 1,016 | 3,341 | 15 | **1,869** | 1,134 | 2,985 | 1,044 | 5,532 | 17 |
| 12 | **2,281** | 1,073 | 3,326 | 1,023 | 6,614 | 28 | **2,755** | 1,414 | 3,559 | 1,157 | 10,460 | 18 |
| 13 | **1,747** | 1,093 | 3,173 | 1,035 | 4,792 | 16 | **1,628** | 1,087 | 3,182 | 1,028 | 5,820 | 22 |
| 14 | **3,127** | 1,137 | 4,110 | 1,048 | 9,220 | 28 | **2,039** | 1,081 | 3,256 | 1,028 | 6,939 | 27 |
| 15 | **2,317** | 1,091 | 4,214 | 1,040 | 11,34 | 38 | **2,111** | 1,154 | 3,706 | 1,045 | 9,234 | 29 |
| 16 | **3,084** | 1,175 | 4,063 | 1,054 | 13,35 | 39 | **2,791** | 1,195 | 4,238 | 1,067 | 11,49 | 30 |
| 17 | **2,476** | 1,106 | 3,867 | 1,037 | 9,260 | 32 | **1,826** | 1,094 | 3,612 | 1,034 | 7,560 | 27 |
| 18 | **3,302** | 1,104 | 4,687 | 1,035 | 12,24 | 44 | **1,626** | 1,095 | 3,139 | 1,031 | 5,809 | 21 |
| 19 | **2,037** | 1,185 | 2,732 | 1,080 | 5,560 | 13 | **1,894** | 1,082 | 2,863 | 1,033 | 7,734 | 28 |
| 20 | **2,023** | 1,074 | 3,139 | 1,030 | 7,207 | 27 | **2,581** | 1,092 | 3,575 | 1,040 | 8,451 | 28 |
| 21 | **1,526** | 1,102 | 2,414 | 1,039 | 4,707 | 15 | **2,969** | 1,098 | 4,714 | 1,038 | 11,39 | 39 |
| 22 | **1,792** | 1,073 | 2,543 | 1,028 | 4,606 | 17 | **2,130** | 1,066 | 2,551 | 1,031 | 6,804 | 25 |
| 23 | **1,367** | 1,068 | 2,222 | 1,033 | 4,648 | 16 | **1,483** | 1,068 | 2,587 | 1,024 | 6,139 | 25 |
| 24 | **1,625** | 1,089 | 2,934 | 1,042 | 5,711 | 18 | **1,579** | 1,076 | 3,133 | 1,026 | 7,247 | 29 |
| 25 | **1,417** | 1,058 | 1,564 | 1,011 | 2,631 | 13 | **2,338** | 1,145 | 3,255 | 1,047 | 6,930 | 21 |
| 26 | **1,990** | 1,099 | 2,331 | 1,027 | 5,206 | 20 | **2,257** | 1,057 | 3,055 | 1,026 | 7,190 | 29 |
| 27 | **2,027** | 1,086 | 3,657 | 1,022 | 7,156 | 31 | **2,587** | 1,122 | 3,653 | 1,034 | 7,870 | 28 |

**Suprafața 4** componenta Ox  **Suprafața 4** componenta Oy

| Nr. bandă | IDIV | IDIV min | IDIV max | min teor. | max teor. | nr. elem. | IDIV | IDIV min | IDIV max | min teor. | max teor. | nr. elem. |
|---|---|---|---|---|---|---|---|---|---|---|---|---|
| 28 | **2,529** | 1,111 | 3,690 | 1,029 | 8,334 | 32 | **3,270** | 1,134 | 5,410 | 1,048 | 14,168 | 44 |
| **pas 50 cm** IDIV Ox 3,534 (coef. var. 29,83 %); IDIV Oy 3,478 (coef. var. 22,77 %); | | | | | | | | | | | | |
| 1 | **3,587** | 1,111 | 5,218 | 1,035 | 15,696 | 57 | **2,033** | 1,107 | 3,938 | 1,029 | 6,382 | 24 |
| 2 | **4,055** | 1,364 | 6,361 | 1,133 | 28,845 | 54 | **3,138** | 1,150 | 4,365 | 1,037 | 12,55 | 44 |
| 3 | **4,217** | 1,068 | 6,440 | 1,019 | 13,010 | 64 | **3,748** | 1,185 | 6,226 | 1,064 | 17,76 | 48 |
| 4 | **3,308** | 1,140 | 5,111 | 1,058 | 18,619 | 53 | **2,814** | 1,085 | 4,712 | 1,026 | 8,545 | 35 |
| 5 | **2,257** | 1,117 | 3,174 | 1,031 | 8,041 | 30 | **3,556** | 1,117 | 5,401 | 1,025 | 14,62 | 62 |
| 6 | **2,783** | 1,076 | 4,164 | 1,021 | 9,526 | 43 | **4,198** | 1,419 | 5,877 | 1,155 | 20,24 | 35 |
| 7 | **3,861** | 1,156 | 6,030 | 1,048 | 14,208 | 44 | **3,050** | 1,098 | 5,807 | 1,031 | 12,85 | 49 |
| 8 | **5,043** | 1,159 | 8,003 | 1,040 | 22,499 | 77 | **3,871** | 1,150 | 6,471 | 1,049 | 19,15 | 59 |
| 9 | **5,532** | 1,125 | 7,762 | 1,033 | 20,301 | 76 | **3,546** | 1,125 | 6,211 | 1,035 | 13,28 | 48 |
| 10 | **3,299** | 1,193 | 4,924 | 1,066 | 15,079 | 40 | **3,922** | 1,113 | 5,763 | 1,037 | 15,91 | 56 |
| 11 | **2,762** | 1,082 | 4,301 | 1,038 | 9,353 | 32 | **4,498** | 1,113 | 6,411 | 1,038 | 18,37 | 64 |
| 12 | **1,912** | 1,131 | 4,659 | 1,041 | 10,257 | 54 | **2,176** | 1,089 | 5,058 | 1,025 | 12,77 | 54 |
| 13 | **2,420** | 1,068 | 2,772 | 1,021 | 7,413 | 33 | **3,341** | 1,098 | 5,302 | 1,033 | 13,45 | 50 |
| 14 | **4,181** | 1,135 | 6,584 | 1,037 | 17,872 | 63 | **4,743** | 1,110 | 7,771 | 1,036 | 19,96 | 72 |
| **pas 100 cm** IDIV Ox 6,447 (coef. var. 25,19%); IDIV Oy 6,131 (coef. var. 17,13%); | | | | | | | | | | | | |
| 1 | **6,697** | 1,321 | 10,57 | 1,112 | 54,131 | 111 | **4,139** | 1,151 | 7,49 | 1,037 | 19,13 | 68 |
| 2 | **7,041** | 1,210 | 11,02 | 1,057 | 40,542 | 117 | **5,424** | 1,185 | 10,16 | 1,064 | 30,47 | 83 |
| 3 | **4,508** | 1,107 | 6,83 | 1,033 | 19,316 | 73 | **6,468** | 1,362 | 9,71 | 1,106 | 45,93 | 97 |
| 4 | **7,971** | 1,146 | 13,29 | 1,044 | 36,625 | 121 | **6,104** | 1,152 | 11,51 | 1,049 | 34,71 | 108 |
| 5 | **8,611** | 1,174 | 11,69 | 1,049 | 37,255 | 116 | **7,159** | 1,143 | 11,48 | 1,042 | 30,99 | 104 |
| 6 | **4,553** | 1,128 | 8,399 | 1,041 | 19,474 | 66 | **6,659** | 1,112 | 10,46 | 1,032 | 30,73 | 118 |
| 7 | **5,501** | 1,128 | 8,198 | 1,031 | 24,565 | 96 | **6,965** | 1,126 | 12,27 | 1,036 | 33,48 | 122 |

**Suprafața 5** componenta Ox  **Suprafața 5** componenta Oy

| Nr. bandă | IDIV | IDIV min | IDIV max | min teor. | max teor. | nr. elem. | IDIV | IDIV min | IDIV max | min teor. | max teor. | nr. elem. |
|---|---|---|---|---|---|---|---|---|---|---|---|---|
| **pas 25 cm** IDIV Ox 1,938 (coef. var. 28,76 %); IDIV Oy 2,080 (coef. var. 27,13 %); media distanței medii față de primii 3/5/7 vecini: 18,404/ 24,031/ 28,933 cm | | | | | | | | | | | | |
| 1 | **1,587** | 1,096 | 1,961 | 1,039 | 3,587 | 11 | **1,641** | 1,190 | 2,531 | 1,042 | 4,548 | 14 |
| 2 | **1,786** | 1,076 | 2,974 | 1,026 | 5,180 | 20 | **1,922** | 1,265 | 2,742 | 1,133 | 5,399 | 10 |
| 3 | **2,181** | 1,241 | 4,153 | 1,104 | 8,109 | 17 | **2,090** | 1,276 | 3,124 | 1,180 | 8,688 | 14 |
| 4 | **1,975** | 1,129 | 2,724 | 1,041 | 5,970 | 19 | **1,861** | 1,097 | 2,758 | 1,033 | 6,723 | 24 |
| 5 | **2,041** | 1,152 | 2,787 | 1,041 | 6,546 | 21 | **1,800** | 1,087 | 2,418 | 1,031 | 5,301 | 19 |
| 6 | **1,552** | 1,095 | 1,860 | 1,032 | 3,839 | 13 | **1,851** | 1,259 | 3,255 | 1,113 | 9,908 | 20 |
| 7 | **1,538** | 1,054 | 1,874 | 1,012 | 3,166 | 16 | **2,275** | 1,182 | 3,499 | 1,068 | 7,429 | 19 |
| 8 | **1,225** | 1,084 | 1,574 | 1,016 | 2,635 | 11 | **1,198** | 1,108 | 2,174 | 1,026 | 3,097 | 11 |
| 9 | **1,397** | 1,104 | 1,832 | 1,032 | 4,357 | 15 | **1,440** | 1,067 | 1,845 | 1,040 | 3,029 | 9 |
| 10 | **2,193** | 1,079 | 3,294 | 1,030 | 6,442 | 24 | **1,172** | 1,089 | 1,743 | 1,033 | 2,321 | 7 |
| 11 | **2,483** | 1,163 | 4,004 | 1,036 | 10,01 | 35 | **2,373** | 1,254 | 2,731 | 1,105 | 5,328 | 11 |

| | Suprafața 5 componenta Ox | | | | | | Suprafața 5 componenta Oy | | | | | |
|---|---|---|---|---|---|---|---|---|---|---|---|---|
| Nr. bandă | IDIV | IDIV min | IDIV max | min teor. | max teor. | nr. elem. | IDIV | IDIV min | IDIV max | min teor. | max teor. | nr. elem. |
| 12 | **1,914** | 1,146 | 2,850 | 1,049 | 6,103 | 18 | **2,375** | 1,283 | 4,231 | 1,139 | 7,693 | 14 |
| 13 | **2,842** | 1,163 | 4,173 | 1,066 | 11,03 | 29 | **1,665** | 1,108 | 2,068 | 1,044 | 3,151 | 9 |
| 14 | **2,203** | 1,089 | 3,205 | 1,039 | 6,093 | 20 | **1,597** | 1,239 | 2,963 | 1,127 | 4,766 | 9 |
| 15 | **2,537** | 1,111 | 5,033 | 1,037 | 9,319 | 32 | **2,100** | 1,159 | 2,491 | 1,063 | 3,958 | 10 |
| 16 | **1,271** | 1,078 | 2,504 | 1,036 | 3,753 | 12 | **1,368** | 1,093 | 1,776 | 1,035 | 2,911 | 9 |
| 17 | **2,728** | 1,118 | 3,891 | 1,039 | 8,370 | 28 | **1,621** | 1,114 | 2,637 | 1,040 | 4,174 | 13 |
| 18 | **3,516** | 1,443 | 5,172 | 1,247 | 19,15 | 26 | **3,104** | 1,207 | 3,663 | 1,100 | 9,353 | 20 |
| 19 | **1,735** | 1,097 | 3,098 | 1,031 | 6,594 | 24 | **1,761** | 1,152 | 2,488 | 1,086 | 4,484 | 10 |
| 20 | **1,995** | 1,297 | 3,347 | 1,088 | 7,073 | 16 | **2,024** | 1,096 | 3,666 | 1,032 | 6,615 | 24 |
| 21 | **2,584** | 1,232 | 3,802 | 1,099 | 8,389 | 18 | **1,434** | 1,198 | 1,842 | 1,047 | 3,846 | 11 |
| 22 | **2,111** | 1,251 | 2,640 | 1,083 | 6,472 | 15 | **2,792** | 1,112 | 4,813 | 1,046 | 10,513 | 33 |
| 23 | **1,388** | 1,094 | 2,166 | 1,020 | 3,420 | 14 | **2,922** | 1,154 | 4,645 | 1,058 | 10,381 | 29 |
| 24 | **1,585** | 1,067 | 2,019 | 1,023 | 4,468 | 18 | **2,941** | 1,162 | 4,707 | 1,045 | 10,431 | 33 |
| 25 | **1,598** | 1,101 | 1,850 | 1,038 | 4,102 | 13 | **2,209** | 1,123 | 3,997 | 1,044 | 8,563 | 27 |
| 26 | **1,377** | 1,046 | 1,729 | 1,012 | 2,424 | 11 | **2,218** | 1,066 | 3,439 | 1,019 | 7,656 | 36 |
| 27 | **1,150** | 1,188 | 2,094 | 1,047 | 2,912 | 8 | **2,016** | 1,147 | 3,445 | 1,043 | 6,951 | 22 |
| 28 | **2,085** | 1,507 | 2,664 | 1,357 | 3,193 | 4 | **3,050** | 1,182 | 5,955 | 1,059 | 14,624 | 41 |
| **pas 50 cm** IDIV Ox  2,929 (coef. var. 30,19 %);  IDIV Oy  3,197 (coef. var.  29,83 %); | | | | | | | | | | | | |
| 1 | **2,375** | 1,093 | 4,114 | 1,034 | 8,670 | 31 | **2,619** | 1,248 | 4,456 | 1,103 | 11,37 | 24 |
| 2 | **3,824** | 1,262 | 6,460 | 1,104 | 16,98 | 36 | **3,390** | 1,299 | 4,349 | 1,110 | 18,49 | 38 |
| 3 | **2,447** | 1,127 | 3,516 | 1,032 | 9,218 | 34 | **2,648** | 1,232 | 4,918 | 1,105 | 18,53 | 39 |
| 4 | **1,871** | 1,066 | 2,432 | 1,018 | 5,764 | 27 | **3,387** | 1,193 | 5,035 | 1,079 | 12,43 | 30 |
| 5 | **2,754** | 1,110 | 4,089 | 1,030 | 10,14 | 39 | **1,559** | 1,126 | 2,882 | 1,040 | 5,036 | 16 |
| 6 | **3,656** | 1,150 | 5,990 | 1,042 | 16,05 | 53 | **3,487** | 1,280 | 5,739 | 1,139 | 13,70 | 25 |
| 7 | **3,967** | 1,161 | 6,601 | 1,066 | 18,41 | 49 | **1,938** | 1,100 | 3,166 | 1,041 | 5,67 | 18 |
| 8 | **3,049** | 1,106 | 6,783 | 1,037 | 12,63 | 44 | **2,745** | 1,182 | 3,597 | 1,069 | 7,48 | 19 |
| 9 | **4,198** | 1,300 | 6,838 | 1,110 | 26,19 | 54 | **3,703** | 1,302 | 5,556 | 1,102 | 15,48 | 33 |
| 10 | **3,185** | 1,233 | 5,407 | 1,076 | 16,14 | 40 | **2,605** | 1,117 | 4,939 | 1,033 | 9,310 | 34 |
| 11 | **3,949** | 1,278 | 5,604 | 1,100 | 15,33 | 33 | **3,465** | 1,152 | 5,771 | 1,051 | 14,69 | 44 |
| 12 | **2,196** | 1,106 | 3,570 | 1,033 | 8,751 | 32 | **5,246** | 1,141 | 8,550 | 1,045 | 19,21 | 62 |
| 13 | **2,095** | 1,128 | 2,703 | 1,037 | 7,111 | 24 | **3,182** | 1,123 | 6,212 | 1,037 | 17,74 | 63 |
| 14 | **1,473** | 1,172 | 2,747 | 1,047 | 4,155 | 12 | **4,420** | 1,187 | 9,057 | 1,059 | 22,28 | 63 |
| **pas 100 cm** IDIV Ox  5,291 (coef. var.  29,60%);  IDIV Oy  5,853 (coef. var.  32,13 %); | | | | | | | | | | | | |
| 1 | **5,163** | 1,228 | 9,273 | 1,076 | 26,89 | 67 | **5,306** | 1,238 | 7,659 | 1,097 | 28,20 | 62 |
| 2 | **3,375** | 1,121 | 5,124 | 1,028 | 15,14 | 61 | **5,085** | 1,253 | 9,171 | 1,105 | 32,66 | 69 |
| 3 | **6,122** | 1,139 | 9,074 | 1,040 | 26,79 | 92 | **3,746** | 1,224 | 6,785 | 1,098 | 18,81 | 41 |
| 4 | **7,014** | 1,200 | 12,96 | 1,066 | 34,55 | 93 | **3,884** | 1,115 | 5,397 | 1,040 | 11,04 | 37 |
| 5 | **7,006** | 1,302 | 11,35 | 1,110 | 45,48 | 94 | **6,295** | 1,286 | 9,432 | 1,105 | 31,70 | 67 |
| 6 | **5,238** | 1,264 | 8,462 | 1,092 | 28,75 | 65 | **8,701** | 1,145 | 13,70 | 1,045 | 32,71 | 106 |
| 7 | **3,176** | 1,181 | 4,676 | 1,055 | 12,49 | 36 | **7,717** | 1,176 | 14,22 | 1,055 | 42,57 | 126 |

| | **Suprafața 6** componenta Ox | | | | | | **Suprafața 6** componenta Oy | | | | | |
|---|---|---|---|---|---|---|---|---|---|---|---|---|
| Nr. bandă | IDIV | IDIV min | IDIV max | min teor. | max teor. | nr. elem | IDIV | IDIV min | IDIV max | min teor. | max teor. | nr. elem. |
| **pas 25 cm** IDIV Ox 2,890 (coef. var. 27,74 %); IDIV Oy 2,992 (coef. var. 34,90 %); media distanței medii față de primii 3/5/7 vecini: 17,196/ 21,563/ 25,322 cm | | | | | | | | | | | | |
| 1 | **3,406** | 1,154 | 5,536 | 1,051 | 13,37 | 40 | **5,519** | 2,009 | 6,611 | 1,660 | 25,51 | 20 |
| 2 | **2,773** | 1,127 | 4,209 | 1,038 | 10,48 | 36 | **3,092** | 1,474 | 4,408 | 1,266 | 12,91 | 17 |
| 3 | **2,906** | 1,127 | 4,798 | 1,050 | 11,59 | 35 | **2,257** | 1,124 | 3,486 | 1,039 | 6,428 | 21 |
| 4 | **2,527** | 1,321 | 4,864 | 1,124 | 16,49 | 32 | **6,815** | 1,434 | 9,379 | 1,194 | 21,41 | 33 |
| 5 | **3,560** | 1,091 | 5,828 | 1,030 | 13,59 | 53 | **2,486** | 1,169 | 4,299 | 1,068 | 8,963 | 23 |
| 6 | **4,077** | 1,176 | 5,876 | 1,053 | 18,58 | 55 | **2,648** | 1,330 | 3,553 | 1,170 | 7,243 | 12 |
| 7 | **3,998** | 1,147 | 5,060 | 1,047 | 13,47 | 42 | **2,797** | 1,332 | 5,259 | 1,181 | 14,99 | 24 |
| 8 | **2,981** | 1,100 | 4,199 | 1,031 | 11,81 | 45 | **3,490** | 1,445 | 6,175 | 1,202 | 17,85 | 27 |
| 9 | **4,126** | 1,165 | 5,187 | 1,046 | 15,50 | 49 | **3,779** | 1,333 | 6,453 | 1,185 | 22,14 | 35 |
| 10 | **2,825** | 1,264 | 3,963 | 1,091 | 13,73 | 31 | **1,742** | 1,076 | 2,421 | 1,035 | 5,325 | 18 |
| 11 | **2,776** | 1,192 | 4,362 | 1,079 | 12,02 | 29 | **3,325** | 1,094 | 4,568 | 1,041 | 11,22 | 37 |
| 12 | **1,591** | 1,129 | 2,215 | 1,038 | 4,648 | 15 | **3,421** | 1,319 | 5,676 | 1,186 | 15,19 | 24 |
| 13 | **4,080** | 1,338 | 6,695 | 1,156 | 23,17 | 40 | **2,017** | 1,191 | 3,172 | 1,058 | 9,367 | 26 |
| 14 | **3,896** | 1,142 | 5,288 | 1,053 | 13,53 | 40 | **4,336** | 1,164 | 6,126 | 1,106 | 16,19 | 34 |
| 15 | **2,464** | 1,189 | 3,192 | 1,054 | 8,68 | 25 | **3,050** | 1,542 | 4,736 | 1,204 | 18,62 | 28 |
| 16 | **2,851** | 1,537 | 5,516 | 1,191 | 15,41 | 24 | **3,022** | 1,390 | 5,517 | 1,249 | 18,45 | 25 |
| 17 | **3,219** | 1,291 | 5,564 | 1,150 | 11,35 | 20 | **2,772** | 1,187 | 5,319 | 1,080 | 10,89 | 26 |
| 18 | **2,469** | 1,378 | 3,547 | 1,144 | 11,14 | 20 | **3,307** | 1,227 | 6,250 | 1,102 | 15,01 | 32 |
| 19 | **2,347** | 1,362 | 6,280 | 1,201 | 8,507 | 13 | **2,772** | 1,161 | 4,192 | 1,065 | 11,33 | 30 |
| 20 | **2,712** | 1,353 | 5,845 | 1,191 | 8,949 | 14 | **2,731** | 1,110 | 3,461 | 1,057 | 7,92 | 22 |
| 21 | **1,157** | 1,052 | 1,260 | 1,039 | 1,602 | 4 | **2,126** | 1,149 | 3,192 | 1,053 | 6,65 | 19 |
| 22 | **1,661** | 1,087 | 2,338 | 1,042 | 4,870 | 15 | **2,857** | 1,113 | 3,936 | 1,053 | 9,92 | 29 |
| 23 | **2,267** | 1,485 | 2,945 | 1,276 | 6,826 | 9 | **3,206** | 1,321 | 5,177 | 1,190 | 19,23 | 30 |
| 24 | **2,482** | 1,401 | 3,828 | 1,239 | 7,829 | 11 | **2,559** | 1,123 | 3,858 | 1,051 | 9,451 | 28 |
| 25 | **2,503** | 1,314 | 4,107 | 1,162 | 9,459 | 16 | **2,318** | 1,430 | 4,192 | 1,195 | 14,26 | 22 |
| 26 | **3,233** | 1,391 | 4,948 | 1,220 | 8,209 | 12 | **2,464** | 1,115 | 3,288 | 1,052 | 8,85 | 26 |
| 27 | **2,108** | 1,150 | 2,980 | 1,098 | 4,725 | 10 | **2,079** | 1,058 | 2,864 | 1,019 | 6,87 | 32 |
| 28 | **2,013** | 1,298 | 2,883 | 1,233 | 4,843 | 7 | **2,377** | 1,060 | 3,537 | 1,018 | 8,57 | 42 |
| **pas 50 cm** IDIV Ox 5,031 (coef. var. 34,79 %); IDIV Oy 5,302 (coef. var. 22,30 %); | | | | | | | | | | | | |
| 1 | **6,160** | 1,145 | 8,811 | 1,051 | 25,04 | 76 | **5,307** | 1,468 | 7,173 | 1,245 | 27,21 | 37 |
| 2 | **4,978** | 1,329 | 9,208 | 1,118 | 33,56 | 67 | **8,502** | 1,408 | 12,16 | 1,190 | 34,72 | 54 |
| 3 | **7,977** | 1,178 | 10,97 | 1,048 | 34,31 | 108 | **5,100** | 1,382 | 7,227 | 1,141 | 19,25 | 35 |
| 4 | **5,348** | 1,121 | 7,666 | 1,033 | 22,98 | 87 | **5,940** | 1,337 | 10,53 | 1,184 | 32,20 | 51 |
| 5 | **6,717** | 1,294 | 8,722 | 1,087 | 34,22 | 80 | **5,020** | 1,364 | 8,238 | 1,185 | 33,56 | 53 |
| 6 | **3,881** | 1,198 | 5,761 | 1,079 | 18,1 | 44 | **6,176** | 1,385 | 9,293 | 1,169 | 36,86 | 61 |
| 7 | **7,277** | 1,417 | 11,42 | 1,156 | 46,35 | 80 | **5,113** | 1,251 | 8,227 | 1,090 | 26,31 | 60 |
| 8 | **4,296** | 1,474 | 7,741 | 1,172 | 29,94 | 49 | **5,280** | 1,493 | 9,175 | 1,210 | 35,92 | 53 |
| 9 | **5,489** | 1,343 | 8,590 | 1,150 | 22,69 | 40 | **5,930** | 1,229 | 10,73 | 1,097 | 26,29 | 58 |
| 10 | **3,878** | 1,360 | 9,680 | 1,184 | 17,01 | 27 | **4,917** | 1,182 | 6,744 | 1,065 | 19,39 | 52 |
| 11 | **1,726** | 1,106 | 2,660 | 1,045 | 6,186 | 19 | **4,923** | 1,134 | 6,343 | 1,053 | 16,16 | 48 |

| | **Suprafața 6** componenta Ox | | | | | | **Suprafața 6** componenta Oy | | | | | |
|---|---|---|---|---|---|---|---|---|---|---|---|---|
| Nr. bandă | IDIV | IDIV min | IDIV max | min teor. | max teor. | nr. elem. | IDIV | IDIV min | IDIV max | min teor. | max teor. | nr. elem. |
| 12 | **3,254** | 1,447 | 6,495 | 1,265 | 15,20 | 20 | **4,942** | 1,424 | 8,122 | 1,178 | 36,08 | 58 |
| 13 | **5,084** | 1,339 | 8,004 | 1,182 | 17,57 | 28 | **3,529** | 1,459 | 6,373 | 1,189 | 30,82 | 48 |
| 14 | **3,236** | 1,419 | 5,534 | 1,229 | 11,95 | 17 | **3,657** | 1,070 | 5,716 | 1,019 | 15,12 | 74 |
| **pas 100 cm** IDIV Ox 9,561 (coef. var. 31,57 %); IDIV Oy 9,942 (coef. var. 15,38 %); | | | | | | | | | | | | |
| 1 | **10,52** | 1,295 | 17,24 | 1,107 | 68,03 | 143 | **12,03** | 1,391 | 17,89 | 1,192 | 58,86 | 91 |
| 2 | **12,96** | 1,179 | 18,04 | 1,048 | 61,62 | 195 | **9,48** | 1,374 | 16,68 | 1,163 | 51,06 | 86 |
| 3 | **10,06** | 1,292 | 14,03 | 1,087 | 52,92 | 124 | **10,63** | 1,381 | 16,84 | 1,166 | 68,43 | 114 |
| 4 | **12,53** | 1,402 | 18,44 | 1,160 | 75,69 | 129 | **9,70** | 1,392 | 16,10 | 1,157 | 65,79 | 113 |
| 5 | **7,769** | 1,380 | 17,62 | 1,165 | 39,95 | 67 | **10,85** | 1,203 | 16,74 | 1,091 | 48,12 | 110 |
| 6 | **4,332** | 1,426 | 7,746 | 1,189 | 24,98 | 39 | **9,92** | 1,352 | 13,04 | 1,141 | 58,18 | 106 |
| 7 | **8,240** | 1,383 | 12,69 | 1,189 | 28,83 | 45 | **7,09** | 1,444 | 11,07 | 1,169 | 73,81 | 122 |

| | **Suprafața 7** componenta Ox | | | | | | **Suprafața 7** componenta Oy | | | | | |
|---|---|---|---|---|---|---|---|---|---|---|---|---|
| Nr. bandă | IDIV | IDIV min | IDIV max | min teor. | max teor. | nr. elem. | IDIV | IDIV min | IDIV max | min teor. | max teor. | nr. elem. |
| **pas 25 cm** IDIV Ox 3,305 (coef. var. 43,17 %); IDIV Oy 3,270 (coef. var. 17,47 %); media distanței medii față de primii 3/5/7 vecini: 16,202/ 20,272/ 23,624 cm | | | | | | | | | | | | |
| 1 | **4,056** | 1,165 | 6,462 | 1,069 | 12,38 | 32 | **2,477** | 1,118 | 5,064 | 1,054 | 11,72 | 34 |
| 2 | **3,629** | 1,152 | 5,187 | 1,058 | 12,08 | 34 | **4,495** | 1,149 | 6,882 | 1,057 | 14,07 | 40 |
| 3 | **4,199** | 1,144 | 5,908 | 1,047 | 12,20 | 38 | **3,297** | 1,148 | 5,036 | 1,050 | 10,95 | 33 |
| 4 | **1,922** | 1,076 | 2,721 | 1,025 | 5,100 | 20 | **2,848** | 1,105 | 4,769 | 1,053 | 10,60 | 31 |
| 5 | **1,857** | 1,089 | 3,205 | 1,027 | 6,921 | 27 | **3,945** | 1,146 | 8,135 | 1,046 | 14,49 | 46 |
| 6 | **2,229** | 1,135 | 3,032 | 1,062 | 7,859 | 21 | **3,266** | 1,174 | 6,059 | 1,076 | 14,22 | 35 |
| 7 | **2,526** | 1,133 | 4,151 | 1,051 | 10,08 | 30 | **3,792** | 1,151 | 5,493 | 1,063 | 12,56 | 34 |
| 8 | **1,839** | 1,068 | 2,495 | 1,024 | 5,466 | 22 | **3,062** | 1,101 | 5,261 | 1,038 | 9,422 | 32 |
| 9 | **3,123** | 1,071 | 4,652 | 1,025 | 9,164 | 38 | **3,839** | 1,151 | 5,698 | 1,053 | 9,923 | 29 |
| 10 | **1,958** | 1,128 | 2,749 | 1,035 | 7,188 | 25 | **3,780** | 1,169 | 6,542 | 1,057 | 13,68 | 39 |
| 11 | **2,811** | 1,149 | 4,333 | 1,044 | 9,477 | 30 | **3,628** | 1,224 | 4,732 | 1,079 | 13,23 | 32 |
| 12 | **1,676** | 1,052 | 2,350 | 1,018 | 4,067 | 18 | **2,822** | 1,153 | 3,699 | 1,058 | 7,614 | 21 |
| 13 | **1,614** | 1,021 | 2,046 | 1,010 | 3,551 | 20 | **3,092** | 1,095 | 4,306 | 1,042 | 9,491 | 31 |
| 14 | **5,030** | 1,164 | 6,756 | 1,055 | 18,88 | 55 | **3,012** | 1,150 | 4,260 | 1,050 | 9,410 | 28 |
| 15 | **3,146** | 1,060 | 5,134 | 1,022 | 10,17 | 45 | **2,667** | 1,208 | 4,389 | 1,093 | 11,66 | 26 |
| 16 | **3,132** | 1,114 | 4,567 | 1,033 | 10,29 | 38 | **3,102** | 1,187 | 4,365 | 1,065 | 10,60 | 28 |
| 17 | **2,171** | 1,100 | 3,084 | 1,039 | 6,082 | 20 | **2,234** | 1,106 | 3,139 | 1,040 | 6,780 | 22 |
| 18 | **3,662** | 1,136 | 5,720 | 1,058 | 12,49 | 35 | **2,479** | 1,090 | 3,410 | 1,032 | 7,863 | 29 |
| 19 | **4,141** | 1,143 | 7,155 | 1,059 | 14,69 | 41 | **4,125** | 1,188 | 6,974 | 1,083 | 17,32 | 41 |
| 20 | **4,012** | 1,156 | 7,307 | 1,061 | 15,58 | 43 | **3,496** | 1,182 | 5,687 | 1,062 | 12,11 | 33 |
| 21 | **2,322** | 1,122 | 2,871 | 1,050 | 6,195 | 18 | **3,142** | 1,123 | 5,227 | 1,050 | 12,52 | 38 |
| 22 | **2,960** | 1,090 | 4,182 | 1,041 | 8,298 | 27 | **4,043** | 1,153 | 6,421 | 1,055 | 13,85 | 40 |
| 23 | **3,368** | 1,136 | 4,701 | 1,064 | 10,84 | 29 | **3,170** | 1,094 | 4,463 | 1,049 | 9,255 | 28 |
| 24 | **2,752** | 1,149 | 3,127 | 1,059 | 4,205 | 11 | **2,385** | 1,124 | 3,594 | 1,039 | 9,504 | 32 |

**Suprafața 7** componenta Ox     **Suprafața 7** componenta Oy

| Nr. bandă | IDIV | IDIV min | IDIV max | min teor. | max teor. | nr. elem. | IDIV | IDIV min | IDIV max | min teor. | max teor. | nr. elem. |
|---|---|---|---|---|---|---|---|---|---|---|---|---|
| 25 | **3,505** | 1,119 | 5,936 | 1,054 | 12,68 | 37 | **3,626** | 1,129 | 5,842 | 1,050 | 15,48 | 47 |
| 26 | **7,147** | 1,234 | 13,09 | 1,082 | 32,17 | 77 | **3,008** | 1,110 | 4,639 | 1,050 | 9,35 | 28 |
| 27 | **7,167** | 1,183 | 12,75 | 1,062 | 23,67 | 65 | **3,059** | 1,110 | 5,233 | 1,033 | 9,63 | 35 |
| 28 | **4,125** | 1,236 | 6,853 | 1,084 | 13,61 | 32 | **3,730** | 1,139 | 6,414 | 1,065 | 13,54 | 36 |

**pas 50 cm** IDIV Ox 5,940 (coef. var. 41,93 %); IDIV Oy 5,993 (coef. var. 16,30 %);

| Nr. bandă | IDIV | IDIV min | IDIV max | min teor. | max teor. | nr. elem. | IDIV | IDIV min | IDIV max | min teor. | max teor. | nr. elem. |
|---|---|---|---|---|---|---|---|---|---|---|---|---|
| 1 | **6,361** | 1,129 | 10,93 | 1,059 | 23,30 | 66 | **6,384** | 1,139 | 11,57 | 1,057 | 25,79 | 74 |
| 2 | **5,339** | 1,128 | 8,244 | 1,045 | 17,97 | 58 | **6,269** | 1,168 | 9,709 | 1,052 | 21,29 | 64 |
| 3 | **3,274** | 1,133 | 5,565 | 1,048 | 15,52 | 48 | **7,150** | 1,241 | 13,99 | 1,077 | 32,65 | 81 |
| 4 | **3,684** | 1,135 | 6,125 | 1,047 | 16,50 | 52 | **6,005** | 1,162 | 9,991 | 1,059 | 23,38 | 66 |
| 5 | **4,632** | 1,102 | 6,928 | 1,031 | 16,23 | 63 | **7,265** | 1,174 | 11,48 | 1,056 | 23,54 | 68 |
| 6 | **3,662** | 1,121 | 5,780 | 1,036 | 13,51 | 48 | **5,266** | 1,194 | 7,797 | 1,068 | 20,26 | 53 |
| 7 | **5,791** | 1,143 | 8,005 | 1,051 | 24,77 | 75 | **5,460** | 1,131 | 8,624 | 1,050 | 19,32 | 59 |
| 8 | **5,716** | 1,114 | 8,936 | 1,033 | 21,87 | 83 | **4,905** | 1,202 | 8,300 | 1,084 | 22,85 | 54 |
| 9 | **5,196** | 1,126 | 8,066 | 1,054 | 18,66 | 55 | **3,865** | 1,117 | 5,921 | 1,039 | 14,79 | 51 |
| 10 | **7,609** | 1,165 | 14,04 | 1,063 | 30,70 | 84 | **7,190** | 1,221 | 12,08 | 1,082 | 30,88 | 74 |
| 11 | **4,843** | 1,149 | 6,448 | 1,053 | 15,23 | 45 | **6,945** | 1,154 | 11,45 | 1,055 | 26,61 | 78 |
| 12 | **4,679** | 1,143 | 6,704 | 1,064 | 14,82 | 40 | **5,421** | 1,118 | 7,570 | 1,047 | 18,94 | 60 |
| 13 | **10,73** | 1,249 | 18,08 | 1,082 | 47,50 | 114 | **5,827** | 1,145 | 9,583 | 1,052 | 25,00 | 75 |
| 14 | **11,58** | 1,215 | 19,25 | 1,084 | 40,73 | 97 | **5,929** | 1,204 | 11,18 | 1,069 | 27,06 | 71 |

**pas 100 cm** IDIV Ox 11,753 (coef. var. 43,04 %); IDIV Oy 11,671 (coef. var. 10,82%);

| Nr. bandă | IDIV | IDIV min | IDIV max | min teor. | max teor. | nr. elem. | IDIV | IDIV min | IDIV max | min teor. | max teor. | nr. elem. |
|---|---|---|---|---|---|---|---|---|---|---|---|---|
| 1 | **11,23** | 1,139 | 18,71 | 1,059 | 43,47 | 124 | **12,51** | 1,166 | 20,87 | 1,056 | 47,10 | 138 |
| 2 | **7,54** | 1,138 | 11,65 | 1,048 | 31,63 | 100 | **13,83** | 1,227 | 23,29 | 1,074 | 57,95 | 147 |
| 3 | **7,58** | 1,112 | 12,08 | 1,034 | 29,58 | 111 | **10,95** | 1,186 | 18,31 | 1,063 | 44,00 | 121 |
| 4 | **11,15** | 1,157 | 16,29 | 1,049 | 50,65 | 158 | **10,47** | 1,219 | 16,49 | 1,081 | 46,54 | 113 |
| 5 | **13,49** | 1,184 | 21,85 | 1,063 | 50,59 | 139 | **10,38** | 1,229 | 17,13 | 1,077 | 50,17 | 125 |
| 6 | **9,18** | 1,144 | 13,08 | 1,064 | 31,13 | 85 | **12,31** | 1,173 | 19,02 | 1,054 | 46,39 | 138 |
| 7 | **22,14** | 1,240 | 36,71 | 1,084 | 88,32 | 211 | **11,24** | 1,182 | 20,14 | 1,066 | 54,11 | 146 |

**Suprafața 8** componenta Ox     **Suprafața 8** componenta Oy

| Nr. bandă | IDIV | IDIV min | IDIV max | min teor. | max teor. | nr. elem. | IDIV | IDIV min | IDIV max | min teor. | max teor. | nr. elem. |
|---|---|---|---|---|---|---|---|---|---|---|---|---|
| **pas 25 cm** IDIV Ox 2,190 | | | (coef. var. 32,09 %); IDIV Oy 2,253 (coef. var. 22,75 %); | | | | | | | | | |
| media distanței medii față de primii 3/5/7 vecini: 18,277/ 22,601/ 26,293 cm | | | | | | | | | | | | |
| 1 | **1,487** | 1,056 | 2,446 | 1,021 | 4,310 | 18 | **2,000** | 1,121 | 2,830 | 1,052 | 10,22 | 30 |
| 2 | **2,884** | 1,218 | 3,462 | 1,109 | 7,333 | 15 | **1,861** | 1,106 | 2,779 | 1,037 | 6,805 | 23 |
| 3 | **2,212** | 1,090 | 3,640 | 1,029 | 8,576 | 33 | **3,214** | 1,064 | 4,369 | 1,023 | 10,15 | 44 |
| 4 | **1,916** | 1,116 | 2,775 | 1,023 | 6,660 | 28 | **2,483** | 1,054 | 3,108 | 1,024 | 7,657 | 32 |
| 5 | **3,001** | 1,066 | 4,990 | 1,028 | 11,96 | 48 | **2,292** | 1,139 | 2,708 | 1,041 | 7,701 | 25 |
| 6 | **1,379** | 1,037 | 2,199 | 1,014 | 3,548 | 17 | **2,923** | 1,090 | 4,149 | 1,030 | 8,482 | 32 |
| 7 | **3,276** | 1,127 | 4,832 | 1,044 | 12,12 | 39 | **2,773** | 1,133 | 4,619 | 1,036 | 10,19 | 36 |
| 8 | **3,545** | 1,109 | 4,928 | 1,039 | 11,17 | 38 | **3,427** | 1,083 | 4,078 | 1,040 | 9,292 | 31 |
| 9 | **2,088** | 1,077 | 3,583 | 1,031 | 8,769 | 33 | **2,282** | 1,072 | 3,098 | 1,021 | 7,467 | 33 |

| | Suprafața 8 componenta Ox | | | | | | Suprafața 8 componenta Oy | | | | | |
|---|---|---|---|---|---|---|---|---|---|---|---|---|
| Nr. bandă | IDIV | IDIV min | IDIV max | min teor. | max teor. | nr. elem. | IDIV | IDIV min | IDIV max | min teor. | max teor. | nr. elem. |
| 10 | **2,037** | 1,102 | 2,615 | 1,039 | 6,649 | 22 | **2,424** | 1,069 | 3,224 | 1,021 | 7,057 | 31 |
| 11 | **3,432** | 1,052 | 5,293 | 1,020 | 11,74 | 55 | **2,909** | 1,075 | 3,684 | 1,033 | 8,860 | 32 |
| 12 | **2,693** | 1,060 | 4,009 | 1,017 | 8,547 | 43 | **1,765** | 1,089 | 2,997 | 1,035 | 6,130 | 21 |
| 13 | **2,462** | 1,060 | 3,375 | 1,032 | 5,888 | 21 | **1,964** | 1,108 | 3,056 | 1,042 | 8,049 | 26 |
| 14 | **1,939** | 1,054 | 3,081 | 1,032 | 6,944 | 25 | **2,164** | 1,096 | 4,126 | 1,029 | 8,773 | 34 |
| 15 | **3,627** | 1,097 | 5,661 | 1,040 | 13,90 | 47 | **2,008** | 1,164 | 2,930 | 1,046 | 7,513 | 23 |
| 16 | **2,535** | 1,086 | 3,799 | 1,026 | 10,02 | 41 | **2,238** | 1,079 | 3,745 | 1,046 | 8,126 | 25 |
| 17 | **1,962** | 1,065 | 2,409 | 1,020 | 4,473 | 19 | **1,661** | 1,033 | 1,879 | 1,019 | 3,933 | 17 |
| 18 | **2,263** | 1,063 | 2,966 | 1,029 | 6,346 | 24 | **2,469** | 1,128 | 4,367 | 1,080 | 13,72 | 33 |
| 19 | **2,104** | 1,043 | 3,173 | 1,015 | 6,563 | 34 | **2,113** | 1,112 | 3,289 | 1,034 | 7,563 | 27 |
| 20 | **2,057** | 1,118 | 3,503 | 1,037 | 9,239 | 32 | **1,841** | 1,057 | 2,202 | 1,017 | 3,985 | 18 |
| 21 | **2,182** | 1,063 | 2,834 | 1,032 | 6,148 | 22 | **1,447** | 1,069 | 1,975 | 1,021 | 3,274 | 13 |
| 22 | **1,573** | 1,070 | 2,023 | 1,028 | 4,812 | 18 | **1,791** | 1,141 | 3,692 | 1,070 | 7,914 | 20 |
| 23 | **1,281** | 1,054 | 1,576 | 1,022 | 2,711 | 10 | **2,053** | 1,098 | 3,202 | 1,037 | 6,491 | 22 |
| 24 | **1,534** | 1,053 | 2,088 | 1,020 | 3,846 | 16 | **1,679** | 1,057 | 2,064 | 1,022 | 4,410 | 18 |
| 25 | **1,538** | 1,065 | 2,061 | 1,020 | 4,415 | 19 | **2,061** | 1,049 | 3,158 | 1,025 | 6,661 | 27 |
| 26 | **1,340** | 1,059 | 1,798 | 1,018 | 3,693 | 16 | **1,651** | 1,078 | 2,093 | 1,025 | 4,609 | 18 |
| 27 | **1,293** | 1,031 | 1,648 | 1,012 | 2,995 | 15 | **2,473** | 1,073 | 3,547 | 1,036 | 8,350 | 29 |
| 28 | **1,772** | 1,056 | 2,441 | 1,023 | 4,713 | 19 | **3,071** | 1,087 | 5,298 | 1,025 | 11,17 | 47 |
| **pas 50 cm**  IDIV Ox  3,578  (coef. var. 38,76 %);  IDIV Oy  3,698 (coef. var.  24,46 %); | | | | | | | | | | | | |
| 1 | **2,476** | 1,098 | 3,747 | 1,026 | 8,148 | 33 | **3,168** | 1,149 | 4,798 | 1,049 | 17,25 | 53 |
| 2 | **3,316** | 1,122 | 5,735 | 1,029 | 15,39 | 61 | **4,657** | 1,073 | 6,817 | 1,024 | 17,49 | 76 |
| 3 | **3,613** | 1,082 | 6,544 | 1,027 | 15,72 | 65 | **4,241** | 1,137 | 6,252 | 1,045 | 17,68 | 57 |
| 4 | **6,281** | 1,163 | 9,220 | 1,050 | 24,95 | 77 | **4,825** | 1,112 | 7,886 | 1,037 | 18,87 | 67 |
| 5 | **3,476** | 1,106 | 5,594 | 1,039 | 15,91 | 55 | **4,223** | 1,091 | 5,624 | 1,025 | 14,89 | 64 |
| 6 | **5,521** | 1,081 | 8,845 | 1,020 | 20,41 | 98 | **3,587** | 1,117 | 5,922 | 1,034 | 14,40 | 53 |
| 7 | **3,395** | 1,093 | 5,891 | 1,034 | 12,63 | 46 | **3,981** | 1,145 | 6,567 | 1,041 | 17,88 | 60 |
| 8 | **5,844** | 1,109 | 8,743 | 1,038 | 24,83 | 88 | **3,208** | 1,109 | 5,966 | 1,046 | 15,21 | 48 |
| 9 | **3,028** | 1,081 | 4,477 | 1,026 | 10,34 | 43 | **3,211** | 1,122 | 5,305 | 1,064 | 18,49 | 50 |
| 10 | **3,810** | 1,123 | 5,756 | 1,034 | 17,79 | 66 | **3,261** | 1,095 | 4,707 | 1,034 | 12,42 | 45 |
| 11 | **2,777** | 1,084 | 4,187 | 1,035 | 11,19 | 40 | **2,196** | 1,161 | 4,820 | 1,056 | 11,54 | 33 |
| 12 | **1,982** | 1,090 | 2,885 | 1,024 | 6,345 | 26 | **2,761** | 1,089 | 4,303 | 1,027 | 9,833 | 40 |
| 13 | **2,156** | 1,105 | 3,148 | 1,028 | 8,851 | 35 | **2,855** | 1,084 | 4,497 | 1,027 | 11,16 | 45 |
| 14 | **2,262** | 1,047 | 3,251 | 1,019 | 7,351 | 34 | **5,406** | 1,095 | 8,881 | 1,031 | 19,64 | 76 |
| **pas 100 cm**  IDIV Ox  6,855 (coef. var. 34,74%);  IDIV Oy  6,999 (coef. var. 13,13%); | | | | | | | | | | | | |
| 1 | **5,400** | 1,122 | 8,829 | 1,030 | 23,56 | 94 | **7,479** | 1,143 | 11,00 | 1,045 | 39,57 | 129 |
| 2 | **9,774** | 1,160 | 15,28 | 1,049 | 45,34 | 142 | **8,017** | 1,118 | 12,63 | 1,038 | 34,87 | 124 |
| 3 | **9,093** | 1,137 | 14,29 | 1,038 | 42,97 | 153 | **6,806** | 1,095 | 10,49 | 1,031 | 29,69 | 117 |
| 4 | **9,037** | 1,115 | 14,18 | 1,037 | 37,51 | 134 | **6,926** | 1,127 | 11,68 | 1,042 | 32,02 | 108 |
| 5 | **6,014** | 1,145 | 9,721 | 1,039 | 31,04 | 109 | **6,335** | 1,182 | 9,910 | 1,063 | 34,55 | 95 |
| 6 | **4,194** | 1,104 | 6,463 | 1,035 | 18,14 | 66 | **5,429** | 1,119 | 8,785 | 1,043 | 22,17 | 73 |
| 7 | **4,446** | 1,098 | 5,896 | 1,028 | 16,91 | 69 | **7,914** | 1,107 | 12,87 | 1,031 | 30,82 | 121 |

| | **Suprafața 9** componenta Ox | | | | | | **Suprafața 9** componenta Oy | | | | | |
|---|---|---|---|---|---|---|---|---|---|---|---|---|
| Nr. bandă | IDIV | IDIV min | IDIV max | min teor. | max teor. | nr. elem | IDIV | IDIV min | IDIV max | min teor. | max teor. | nr. elem. |
| **pas 25 cm** IDIV Ox 2,779 (coef. var. 34,83 %); IDIV Oy 2,798 (coef. var. 29,31 %); media distanței medii față de primii 3/5/7 vecini: 17,863/ 22,134/ 25,770 cm | | | | | | | | | | | | |
| 1 | **3,409** | 1,085 | 5,205 | 1,031 | 10,05 | 38 | **2,772** | 1,117 | 4,422 | 1,034 | 9,667 | 35 |
| 2 | **3,398** | 1,188 | 5,541 | 1,070 | 15,17 | 39 | **2,823** | 1,146 | 4,324 | 1,049 | 9,888 | 30 |
| 3 | **5,337** | 1,108 | 6,781 | 1,039 | 15,50 | 53 | **3,389** | 1,077 | 4,808 | 1,026 | 9,027 | 37 |
| 4 | **3,376** | 1,098 | 5,172 | 1,033 | 11,08 | 41 | **2,279** | 1,086 | 3,242 | 1,031 | 7,272 | 27 |
| 5 | **4,050** | 1,114 | 6,347 | 1,048 | 13,21 | 41 | **2,385** | 1,070 | 3,139 | 1,024 | 5,684 | 23 |
| 6 | **3,891** | 1,099 | 6,452 | 1,044 | 15,12 | 49 | **2,949** | 1,122 | 3,922 | 1,043 | 9,025 | 29 |
| 7 | **3,593** | 1,111 | 4,982 | 1,038 | 10,53 | 36 | **2,895** | 1,119 | 4,571 | 1,036 | 8,317 | 29 |
| 8 | **3,342** | 1,120 | 5,551 | 1,036 | 10,82 | 38 | **1,181** | 1,040 | 1,317 | 1,012 | 1,936 | 8 |
| 9 | **2,547** | 1,105 | 5,257 | 1,043 | 10,51 | 34 | **2,992** | 1,118 | 4,055 | 1,047 | 8,183 | 25 |
| 10 | **2,190** | 1,115 | 3,532 | 1,045 | 7,694 | 24 | **2,121** | 1,065 | 3,216 | 1,031 | 6,327 | 23 |
| 11 | **1,606** | 1,108 | 2,189 | 1,025 | 3,736 | 14 | **3,376** | 1,122 | 5,803 | 1,060 | 12,30 | 34 |
| 12 | **1,082** | 1,066 | 1,261 | 1,011 | 1,604 | 6 | **2,854** | 1,213 | 5,145 | 1,083 | 12,71 | 30 |
| 13 | **1,600** | 1,144 | 1,853 | 1,038 | 3,816 | 12 | **3,354** | 1,153 | 5,470 | 1,057 | 10,96 | 31 |
| 14 | **1,784** | 1,127 | 2,605 | 1,053 | 5,029 | 14 | **1,571** | 1,084 | 2,042 | 1,030 | 4,509 | 16 |
| 15 | **2,460** | 1,077 | 3,056 | 1,034 | 5,534 | 19 | **2,094** | 1,073 | 3,618 | 1,035 | 6,937 | 24 |
| 16 | **2,406** | 1,115 | 3,721 | 1,031 | 6,799 | 25 | **1,790** | 1,131 | 2,081 | 1,041 | 3,638 | 11 |
| 17 | **3,605** | 1,104 | 5,181 | 1,043 | 10,20 | 33 | **3,014** | 1,152 | 4,142 | 1,049 | 9,272 | 28 |
| 18 | **3,684** | 1,177 | 5,282 | 1,064 | 11,56 | 31 | **3,050** | 1,123 | 4,335 | 1,052 | 9,564 | 28 |
| 19 | **3,744** | 1,153 | 4,912 | 1,049 | 9,899 | 30 | **2,758** | 1,112 | 5,348 | 1,043 | 9,948 | 32 |
| 20 | **2,313** | 1,098 | 2,977 | 1,034 | 6,302 | 22 | **2,390** | 1,093 | 3,570 | 1,030 | 7,695 | 29 |
| 21 | **2,287** | 1,131 | 3,604 | 1,050 | 7,748 | 23 | **3,283** | 1,133 | 4,007 | 1,061 | 8,498 | 23 |
| 22 | **3,048** | 1,085 | 4,997 | 1,033 | 9,065 | 33 | **2,581** | 1,173 | 3,916 | 1,070 | 10,22 | 26 |
| 23 | **2,204** | 1,104 | 2,791 | 1,038 | 7,176 | 24 | **2,472** | 1,097 | 3,947 | 1,050 | 7,463 | 22 |
| 24 | **2,194** | 1,050 | 3,341 | 1,021 | 6,627 | 29 | **2,292** | 1,064 | 3,673 | 1,030 | 6,462 | 24 |
| 25 | **2,269** | 1,130 | 2,908 | 1,040 | 7,288 | 24 | **3,924** | 1,103 | 5,892 | 1,044 | 12,68 | 41 |
| 26 | **2,084** | 1,112 | 3,291 | 1,037 | 5,974 | 20 | **2,950** | 1,128 | 4,577 | 1,044 | 10,89 | 35 |
| 27 | **2,335** | 1,104 | 3,185 | 1,036 | 5,078 | 17 | **3,834** | 1,103 | 5,189 | 1,042 | 13,96 | 46 |
| 28 | **1,500** | 1,064 | 2,237 | 1,015 | 4,648 | 23 | **5,547** | 1,154 | 8,254 | 1,069 | 17,72 | 46 |
| **pas 50 cm** IDIV Ox 4,801 (coef. var. 37,44 %); IDIV Oy 4,806 (coef. var. 25,57 %); | | | | | | | | | | | | |
| 1 | **6,538** | 1,180 | 10,05 | 1,062 | 27,92 | 77 | **4,998** | 1,113 | 7,717 | 1,039 | 18,89 | 65 |
| 2 | **7,389** | 1,130 | 11,58 | 1,039 | 27,05 | 94 | **4,492** | 1,101 | 7,540 | 1,030 | 16,39 | 64 |
| 3 | **7,155** | 1,112 | 12,09 | 1,043 | 27,01 | 90 | **4,898** | 1,119 | 6,512 | 1,040 | 15,34 | 52 |
| 4 | **6,759** | 1,108 | 10,03 | 1,038 | 21,12 | 74 | **3,346** | 1,117 | 5,613 | 1,035 | 10,33 | 37 |
| 5 | **3,808** | 1,120 | 7,872 | 1,043 | 17,61 | 58 | **4,014** | 1,104 | 6,976 | 1,046 | 15,19 | 48 |
| 6 | **1,819** | 1,065 | 2,617 | 1,020 | 4,647 | 20 | **6,071** | 1,189 | 10,30 | 1,064 | 23,65 | 64 |
| 7 | **2,523** | 1,120 | 3,814 | 1,046 | 8,428 | 26 | **3,820** | 1,108 | 6,283 | 1,037 | 13,30 | 47 |
| 8 | **4,487** | 1,116 | 6,194 | 1,038 | 12,75 | 44 | **3,248** | 1,129 | 5,110 | 1,041 | 10,63 | 35 |
| 9 | **6,088** | 1,159 | 9,986 | 1,060 | 22,95 | 64 | **5,259** | 1,160 | 7,944 | 1,050 | 18,41 | 56 |
| 10 | **4,970** | 1,170 | 7,397 | 1,051 | 17,21 | 52 | **4,285** | 1,111 | 8,051 | 1,038 | 17,45 | 61 |

**Suprafața 9** componenta Ox   **Suprafața 9** componenta Oy

| Nr. bandă | IDIV | IDIV min | IDIV max | min teor. | max teor. | nr. elem. | IDIV | IDIV min | IDIV max | min teor. | max teor. | nr. elem. |
|---|---|---|---|---|---|---|---|---|---|---|---|---|
| | | | | | | 5 | | | | | | 6 |
| 11 | **4,701** | 1,134 | 8,112 | 1,049 | 18,10 | 56 | **4,868** | 1,169 | 7,036 | 1,070 | 18,985 | 49 |
| 12 | **3,808** | 1,127 | 5,471 | 1,041 | 15,78 | 53 | **4,114** | 1,107 | 6,701 | 1,038 | 13,316 | 46 |
| 13 | **4,336** | 1,107 | 5,630 | 1,040 | 12,97 | 44 | **5,958** | 1,116 | 9,615 | 1,042 | 22,840 | 76 |
| 14 | **2,649** | 1,083 | 4,243 | 1,024 | 9,385 | 40 | **7,878** | 1,134 | 11,900 | 1,049 | 29,566 | 92 |
| **pas 100 cm** IDIV Ox 9,318 (coef. var. 42,82%); IDIV Oy 9,690 (coef. var. 15,84%); | | | | | | | | | | | | |
| 1 | **14,29** | 1,225 | 21,09 | 1,066 | 63,29 | 171 | **9,283** | 1,131 | 14,99 | 1,039 | 36,72 | 129 |
| 2 | **14,99** | 1,136 | 21,53 | 1,041 | 47,62 | 164 | **7,999** | 1,134 | 11,71 | 1,039 | 25,54 | 89 |
| 3 | **5,469** | 1,115 | 9,532 | 1,038 | 22,30 | 78 | **9,967** | 1,198 | 16,78 | 1,062 | 40,40 | 112 |
| 4 | **5,551** | 1,142 | 9,727 | 1,046 | 21,96 | 70 | **8,388** | 1,117 | 10,87 | 1,039 | 23,52 | 82 |
| 5 | **9,975** | 1,182 | 17,02 | 1,060 | 41,19 | 116 | **10,19** | 1,168 | 15,45 | 1,056 | 39,95 | 117 |
| 6 | **8,110** | 1,150 | 13,23 | 1,047 | 34,27 | 109 | **9,334** | 1,182 | 13,36 | 1,070 | 36,46 | 95 |
| 7 | **6,641** | 1,124 | 9,245 | 1,039 | 24,15 | 84 | **12,68** | 1,150 | 20,94 | 1,049 | 53,65 | 168 |

**Suprafața 10** componenta Ox   **Suprafața 10** componenta Oy

| Nr. bandă | IDIV | IDIV min | IDIV max | min teor. | max teor. | nr. elem. | IDIV | IDIV min | IDIV max | min teor. | max teor. | nr. elem. |
|---|---|---|---|---|---|---|---|---|---|---|---|---|
| **pas 25 cm** IDIV Ox 2,161 (coef. var. 19,80%); IDIV Oy 2,282 (coef. var. 28,48 %); media distanței medii față de primii 3/5/7 vecini: 16,360/ 20,431/ 23,913 cm | | | | | | | | | | | | |
| 1 | **2,655** | 1,101 | 4,242 | 1,053 | 10,923 | 32 | **2,397** | 1,107 | 3,358 | 1,039 | 7,526 | 25 |
| 2 | **2,258** | 1,089 | 3,871 | 1,031 | 9,531 | 36 | **2,291** | 1,084 | 3,455 | 1,032 | 7,887 | 29 |
| 3 | **2,253** | 1,105 | 3,752 | 1,035 | 8,989 | 32 | **2,052** | 1,059 | 3,092 | 1,019 | 4,718 | 21 |
| 4 | **1,916** | 1,075 | 2,758 | 1,024 | 6,545 | 27 | **1,560** | 1,099 | 2,421 | 1,039 | 4,409 | 14 |
| 5 | **2,983** | 1,117 | 5,366 | 1,040 | 13,668 | 46 | **1,552** | 1,068 | 2,195 | 1,031 | 3,802 | 13 |
| 6 | **2,578** | 1,061 | 3,877 | 1,021 | 9,001 | 41 | **1,843** | 1,146 | 2,517 | 1,062 | 5,006 | 13 |
| 7 | **2,234** | 1,066 | 3,465 | 1,018 | 8,131 | 39 | **1,579** | 1,056 | 2,321 | 1,027 | 4,042 | 15 |
| 8 | **1,999** | 1,091 | 3,450 | 1,034 | 7,078 | 25 | **1,247** | 1,030 | 1,762 | 1,015 | 3,429 | 16 |
| 9 | **3,130** | 1,130 | 5,197 | 1,051 | 11,073 | 33 | **1,426** | 1,061 | 1,737 | 1,016 | 3,866 | 18 |
| 10 | **2,005** | 1,078 | 4,197 | 1,018 | 7,778 | 27 | **1,730** | 1,078 | 2,246 | 1,034 | 5,259 | 18 |
| 11 | **2,358** | 1,093 | 4,110 | 1,028 | 7,997 | 31 | **1,738** | 1,053 | 2,850 | 1,015 | 6,068 | 31 |
| 12 | **1,879** | 1,081 | 2,988 | 1,018 | 5,814 | 27 | **1,767** | 1,075 | 3,030 | 1,019 | 6,749 | 31 |
| 13 | **2,601** | 1,109 | 4,000 | 1,032 | 10,216 | 38 | **1,826** | 1,067 | 2,363 | 1,028 | 5,056 | 19 |
| 14 | **2,317** | 1,050 | 3,704 | 1,018 | 8,836 | 43 | **2,748** | 1,076 | 4,338 | 1,031 | 10,308 | 39 |
| 15 | **2,525** | 1,074 | 3,842 | 1,025 | 8,433 | 35 | **3,149** | 1,074 | 4,453 | 1,036 | 9,959 | 35 |
| 16 | **2,693** | 1,090 | 4,449 | 1,029 | 9,709 | 38 | **2,215** | 1,059 | 4,375 | 1,022 | 8,585 | 38 |
| 17 | **1,874** | 1,102 | 2,600 | 1,023 | 6,865 | 29 | **2,696** | 1,101 | 5,224 | 1,033 | 12,415 | 46 |

| Nr. bandă | Suprafața 10 componenta Ox | | | | | | Suprafața 10 componenta Oy | | | | | |
|---|---|---|---|---|---|---|---|---|---|---|---|---|
| | IDIV | IDIV min | IDIV max | min teor. | max teor. | nr. elem. | IDIV | IDIV min | IDIV max | min teor. | max teor. | nr. elem. |
| 18 | **2,654** | 1,091 | 4,375 | 1,032 | 11,24 | 42 | **2,255** | 1,124 | 4,415 | 1,044 | 10,32 | 33 |
| 19 | **1,812** | 1,087 | 2,716 | 1,020 | 5,014 | 22 | **2,094** | 1,068 | 4,036 | 1,022 | 8,632 | 38 |
| 20 | **2,028** | 1,071 | 3,018 | 1,025 | 5,743 | 23 | **2,177** | 1,081 | 3,907 | 1,022 | 8,057 | 35 |
| 21 | **1,925** | 1,089 | 2,662 | 1,032 | 6,610 | 24 | **2,687** | 1,090 | 5,081 | 1,036 | 11,81 | 42 |
| 22 | **1,679** | 1,064 | 3,495 | 1,023 | 7,059 | 30 | **2,824** | 1,089 | 3,607 | 1,032 | 10,19 | 38 |
| 23 | **1,589** | 1,059 | 2,221 | 1,014 | 5,379 | 28 | **2,511** | 1,059 | 3,763 | 1,020 | 9,056 | 42 |
| 24 | **2,030** | 1,060 | 2,933 | 1,018 | 6,597 | 31 | **2,884** | 1,072 | 4,751 | 1,024 | 10,07 | 43 |
| 25 | **1,784** | 1,031 | 2,476 | 1,009 | 5,392 | 34 | **2,184** | 1,068 | 4,076 | 1,025 | 9,730 | 41 |
| 26 | **1,807** | 1,066 | 2,879 | 1,030 | 6,455 | 24 | **2,529** | 1,121 | 4,328 | 1,047 | 11,55 | 36 |
| 27 | **1,436** | 1,107 | 2,361 | 1,032 | 6,403 | 23 | **3,121** | 1,106 | 5,426 | 1,037 | 11,67 | 41 |
| 28 | **1,740** | 1,126 | 2,632 | 1,047 | 6,320 | 19 | **4,200** | 1,132 | 10,63 | 1,043 | 23,89 | 79 |
| **pas 50 cm** IDIV Ox 3,639 (coef. var. 21,00%); IDIV Oy 3,959 (coef. var. 38,57 %); | | | | | | | | | | | | |
| 1 | **4,446** | 1,129 | 7,325 | 1,053 | 22,77 | 68 | **4,397** | 1,113 | 6,382 | 1,043 | 16,51 | 54 |
| 2 | **3,598** | 1,109 | 5,960 | 1,042 | 17,79 | 59 | **2,507** | 1,081 | 4,714 | 1,030 | 9,15 | 35 |
| 3 | **4,928** | 1,112 | 8,645 | 1,042 | 25,94 | 87 | **2,733** | 1,110 | 3,681 | 1,047 | 8,525 | 26 |
| 4 | **3,769** | 1,088 | 6,107 | 1,028 | 15,72 | 64 | **1,977** | 1,068 | 3,354 | 1,024 | 7,440 | 31 |
| 5 | **3,724** | 1,100 | 7,614 | 1,027 | 16,98 | 70 | **2,532** | 1,087 | 3,601 | 1,034 | 10,01 | 36 |
| 6 | **3,893** | 1,104 | 6,570 | 1,028 | 14,49 | 58 | **2,608** | 1,069 | 5,256 | 1,018 | 12,44 | 62 |
| 7 | **2,910** | 1,120 | 7,052 | 1,034 | 21,87 | 81 | **3,572** | 1,080 | 5,895 | 1,031 | 15,07 | 58 |
| 8 | **4,455** | 1,101 | 7,825 | 1,028 | 18,08 | 73 | **4,584** | 1,101 | 7,911 | 1,036 | 20,37 | 73 |
| 9 | **4,056** | 1,119 | 6,349 | 1,036 | 19,65 | 71 | **5,225** | 1,106 | 8,866 | 1,038 | 22,53 | 79 |
| 10 | **2,901** | 1,071 | 4,950 | 1,022 | 10,06 | 45 | **3,286** | 1,101 | 7,239 | 1,028 | 17,84 | 73 |
| 11 | **3,059** | 1,130 | 5,459 | 1,038 | 15,44 | 54 | **5,355** | 1,133 | 8,304 | 1,044 | 24,33 | 80 |
| 12 | **2,602** | 1,060 | 4,116 | 1,015 | 11,02 | 59 | **4,799** | 1,077 | 7,941 | 1,023 | 18,91 | 85 |
| 13 | **2,910** | 1,121 | 4,670 | 1,028 | 14,39 | 58 | **4,038** | 1,111 | 7,829 | 1,043 | 23,25 | 77 |
| 14 | **2,448** | 1,131 | 4,263 | 1,039 | 12,35 | 42 | **7,595** | 1,135 | 15,31 | 1,044 | 36,41 | 120 |
| **pas 100 cm** IDIV Ox 7,269 (coef. var. 21,46 %); IDIV Oy 7,198 (coef. var. 41,42 %); | | | | | | | | | | | | |
| 1 | **6,860** | 1,123 | 11,30 | 1,040 | 36,88 | 127 | **5,643** | 1,116 | 10,39 | 1,043 | 26,93 | 89 |
| 2 | **8,830** | 1,158 | 14,18 | 1,042 | 44,69 | 151 | **3,457** | 1,067 | 5,941 | 1,030 | 14,68 | 57 |
| 3 | **7,385** | 1,113 | 13,23 | 1,027 | 30,77 | 128 | **4,606** | 1,174 | 8,259 | 1,042 | 29,21 | 98 |
| 4 | **9,681** | 1,119 | 14,46 | 1,034 | 41,05 | 154 | **7,071** | 1,099 | 13,40 | 1,036 | 35,96 | 131 |
| 5 | **7,018** | 1,135 | 10,98 | 1,036 | 31,69 | 116 | **8,059** | 1,128 | 15,06 | 1,033 | 39,89 | 152 |
| 6 | **5,855** | 1,127 | 8,915 | 1,035 | 30,84 | 113 | **9,222** | 1,129 | 15,37 | 1,038 | 46,21 | 165 |
| 7 | **5,243** | 1,174 | 8,450 | 1,039 | 28,75 | 100 | **12,21** | 1,161 | 22,40 | 1,046 | 60,98 | 197 |

## Anexa 6. Indicatorii ai diversității structurale a înălțimii calculați pe fiecare suprafață

| Indici | sup 1 | sup 2 | sup 3 | sup 4 | sup 5 | sup 6 | sup 7 | sup 8 | sup 9 | sup 10 |
|---|---|---|---|---|---|---|---|---|---|---|
| indicele Simpson (D) | 0,163 | 0,144 | 0,135 | 0,158 | 0,142 | 0,179 | 0,123 | 0,132 | 0,127 | 0,129 |
| indicele Simpson (1-D) | 0,837 | 0,856 | 0,865 | 0,842 | 0,858 | 0,821 | 0,877 | 0,868 | 0,873 | 0,871 |
| indicele Simpson (1/D) | 6,148 | 6,940 | 7,404 | 6,331 | 7,061 | 5,576 | 8,109 | 7,583 | 7,897 | 7,774 |
| indicele Shannon | 1,934 | 2,065 | 2,109 | 1,971 | 2,093 | 1,976 | 2,206 | 2,155 | 2,145 | 2,155 |
| echitatea ptr. Shannon | 0,840 | 0,861 | 0,849 | 0,822 | 0,842 | 0,795 | 0,888 | 0,867 | 0,863 | 0,867 |
| indicele Brillouin | 1,905 | 2,021 | 2,061 | 1,936 | 2,042 | 1,938 | 2,173 | 2,117 | 2,110 | 2,122 |
| indicele Berger-Parker | 0,196 | 0,218 | 0,189 | 0,213 | 0,208 | 0,304 | 0,181 | 0,188 | 0,154 | 0,170 |
| indicele McIntosh | 0,618 | 0,648 | 0,661 | 0,626 | 0,653 | 0,599 | 0,671 | 0,661 | 0,668 | 0,664 |
| indicele Margalef | 1,339 | 1,579 | 1,755 | 1,527 | 1,767 | 1,673 | 1,610 | 1,656 | 1,648 | 1,620 |
| indicele Menhinick | 0,347 | 0,464 | 0,522 | 0,416 | 0,533 | 0,448 | 0,394 | 0,433 | 0,426 | 0,402 |
| indicele Gleason | 0,928 | 1,094 | 1,216 | 1,058 | 1,225 | 1,159 | 1,116 | 1,148 | 1,142 | 1,123 |
| IDIV (pas 25 cm) | 2,298 | 1,550 | 2,011 | 2,122 | 2,009 | 2,941 | 3,287 | 2,222 | 2,789 | 2,221 |
| IDIV_min (pas 25 cm) | 1,093 | 1,064 | 1,138 | 1,111 | 1,157 | 1,254 | 1,134 | 1,083 | 1,112 | 1,083 |
| IDIV_max (pas 25 cm) | 3,530 | 2,259 | 3,414 | 3,196 | 3,016 | 4,543 | 5,148 | 3,241 | 4,147 | 3,632 |
| IDIV (pas 50 cm) | 3,926 | 2,273 | 3,262 | 3,506 | 3,063 | 5,167 | 5,967 | 3,638 | 4,803 | 3,799 |
| IDIV_min (pas 50 cm) | 1,109 | 1,088 | 1,165 | 1,139 | 1,178 | 1,313 | 1,158 | 1,106 | 1,125 | 1,102 |
| IDIV_max (pas 50 cm) | 6,393 | 3,716 | 5,944 | 5,529 | 5,039 | 8,119 | 9,728 | 5,728 | 7,586 | 6,542 |

## Anexa 7. Valorile indicilor și testelor statistice folosite pentru determinarea tiparului de distribuție în spațiu a puieților prin metoda quadratelor

| Indicatori | sup 1 | sup 2 | sup 3 | sup 4 | sup 5 | sup 6 | sup 7 | sup 8 | sup 9 | sup 10 |
|---|---|---|---|---|---|---|---|---|---|---|
| Dimensiune quadrat 50x50 cm | | | | | | | | | | |
| medie evenimente | 4,235 | 2,893 | 2,704 | 3,571 | 2,592 | 3,786 | 4,735 | 3,913 | 4,041 | 4,536 |
| dispersia | 5,934 | 5,265 | 9,738 | 10,154 | 7,125 | 10,456 | 7,806 | 7,259 | 7,373 | 9,932 |
| ID | 1,401 | 1,820 | 3,601 | 2,843 | 2,749 | 2,762 | 1,649 | 1,855 | 1,825 | 2,190 |
| ID_hi² exp. | 273,26 | 354,92 | 702,21 | 554,40 | 536,04 | 538,60 | 321,50 | 361,72 | 355,78 | 427,00 |
| ICS | 0,401 | 0,820 | 2,601 | 1,843 | 1,749 | 1,762 | 0,649 | 0,855 | 0,825 | 1,190 |
| ICS_u | 3,963 | 8,098 | 25,684 | 18,199 | 17,270 | 17,399 | 6,406 | 8,442 | 8,142 | 11,748 |
| Indice Green | 0,0005 | 0,0014 | 0,0049 | 0,0026 | 0,0034 | 0,0024 | 0,0007 | 0,0011 | 0,0010 | 0,0013 |
| Green_u | 3,655 | 6,920 | 17,753 | 13,576 | 13,020 | 13,098 | 5,634 | 7,174 | 6,952 | 9,500 |
| Indice Morisita | 1,094 | 1,283 | 1,959 | 1,514 | 1,673 | 1,464 | 1,136 | 1,218 | 1,203 | 1,261 |
| Morisita _hi² exp. | 273,26 | 354,92 | 702,21 | 554,40 | 536,04 | 538,60 | 321,50 | 361,72 | 355,78 | 427,00 |
| Dimensiune quadrat 70x70 cm | | | | | | | | | | |
| medie evenimente | 8,300 | 5,670 | 5,300 | 7,000 | 5,080 | 7,420 | 9,280 | 7,670 | 7,920 | 8,890 |
| dispersia | 13,990 | 12,749 | 33,263 | 22,646 | 17,084 | 29,579 | 15,820 | 14,223 | 20,175 | 25,735 |
| ID | 1,686 | 2,248 | 6,276 | 3,235 | 3,363 | 3,986 | 1,705 | 1,854 | 2,547 | 2,895 |
| ID_hi² exp. | 166,86 | 222,59 | 621,32 | 320,28 | 332,94 | 394,65 | 168,76 | 183,58 | 252,19 | 286,59 |
| ICS | 0,686 | 1,248 | 5,276 | 2,235 | 2,363 | 2,986 | 0,705 | 0,854 | 1,547 | 1,895 |
| ICS_u | 4,823 | 8,783 | 37,120 | 15,726 | 16,626 | 21,011 | 4,958 | 6,011 | 10,887 | 13,331 |
| Indice Green | 0,0008 | 0,0022 | 0,0100 | 0,0032 | 0,0047 | 0,0040 | 0,0008 | 0,0011 | 0,0020 | 0,0021 |
| Green_u | 4,233 | 7,064 | 21,215 | 11,274 | 11,769 | 14,059 | 4,336 | 5,126 | 8,423 | 9,906 |
| Indice Morisita | 1,082 | 1,218 | 1,987 | 1,317 | 1,461 | 1,399 | 1,075 | 1,110 | 1,194 | 1,211 |
| Morisita _hi² exp. | 166,86 | 222,59 | 621,32 | 320,28 | 332,94 | 394,65 | 168,76 | 183,58 | 252,19 | 286,59 |
| Dimensiune quadrat 100x100 cm | | | | | | | | | | |
| medie evenimente | 16,939 | 11,571 | 10,816 | 14,286 | 10,367 | 15,143 | 18,939 | 15,653 | 16,163 | 18,143 |
| dispersia | 32,017 | 40,833 | 99,778 | 58,125 | 52,237 | 106,79 | 52,059 | 45,690 | 59,431 | 87,333 |
| ID | 1,890 | 3,529 | 9,225 | 4,069 | 5,039 | 7,052 | 2,749 | 2,919 | 3,677 | 4,814 |
| ID_hi² exp. | 90,728 | 169,38 | 442,79 | 195,30 | 241,85 | 338,51 | 131,94 | 140,11 | 176,49 | 231,05 |
| ICS | 0,890 | 2,529 | 8,225 | 3,069 | 4,039 | 6,052 | 1,749 | 1,919 | 2,677 | 3,814 |

| ICS_u | 4,361 | 12,389 | 40,293 | 15,034 | 19,785 | 29,650 | 8,567 | 9,401 | 13,114 | 18,683 |
|---|---|---|---|---|---|---|---|---|---|---|
| Indice Green | 0,0011 | 0,0045 | 0,0155 | 0,0044 | 0,0080 | 0,0082 | 0,0019 | 0,0025 | 0,0034 | 0,0043 |
| Green_u | 3,724 | 8,659 | 20,012 | 10,017 | 12,247 | 16,273 | 6,498 | 6,993 | 9,041 | 11,750 |
| Indice Morisita | 1,052 | 1,214 | 1,746 | 1,211 | 1,382 | 1,392 | 1,091 | 1,120 | 1,162 | 1,206 |
| Morisita _hi² exp. | 90,728 | 169,38 | 442,79 | 195,30 | 241,85 | 338,51 | 131,94 | 140,11 | 176,49 | 231,05 |

| ID | Indicele de dispersie |
|---|---|
| ID_hi² exp. | Valoarea testului $\chi^2_{exp}$ al ID |
| ICS | Indicele mărimii agregatului |
| ICS_u | Valoarea testului statistic al ICS |
| Green_u | Valoarea testului statistic a indicelui Green |
| Morisita _hi² exp. | Valoarea testului statistic $\chi^2_{exp}$ al indicelui Morisita |

**Anexa 8. Exprimarea grafică a rezultatelor obţinute în urma analizei de determinare a tiparului de distribuţie în spaţiu a puieţilor prin metoda quadratelor – varianta dimensională a quadratelor de 100x100 cm.**

Pentru toate cele 10 suprafeţe de probă se prezintă:

- încadrarea grafică a arborilor în quadratele de 100x100 cm ( aplicaţia SPATIAL)
- distribuţia experimentală a quadratelor pe categorii ale nr.de puieţi pe quadrat (conform recomandărilor din literatura de specialitate s-a evitat gruparea numărului de puieţi/quadrat în clase de peste 1 eveniment/quadrat pentru identificarea corectă a modelului de organizare spaţială)
- rezultatul interpretării statistice a indicilor de evaluare a structurii spaţiale prin metoda quadratelor

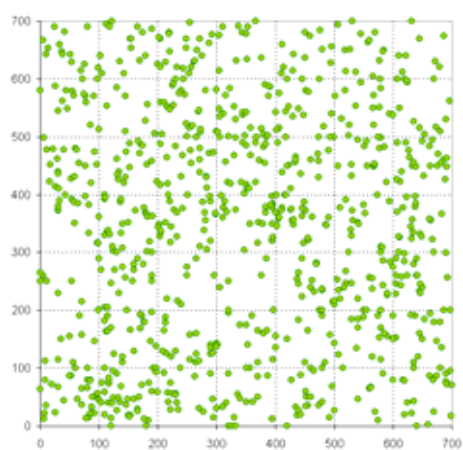

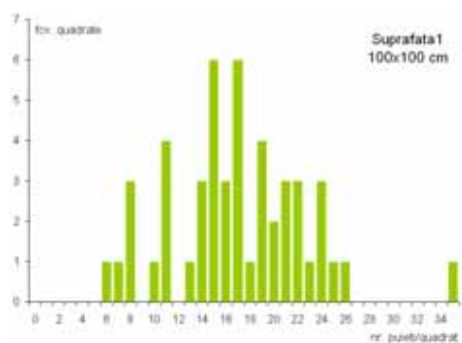

ID – structură agregată
ICS - structură agregată
Indice Morisita - structură agregată
Indice Green - structură agregată

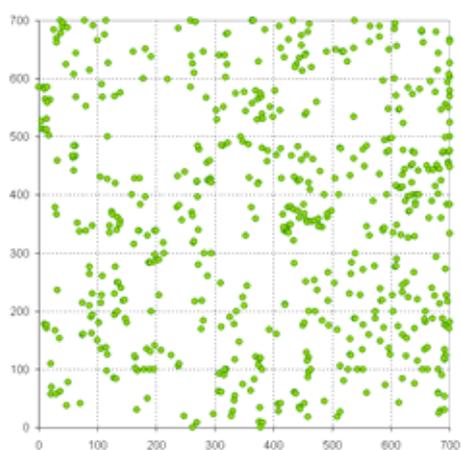

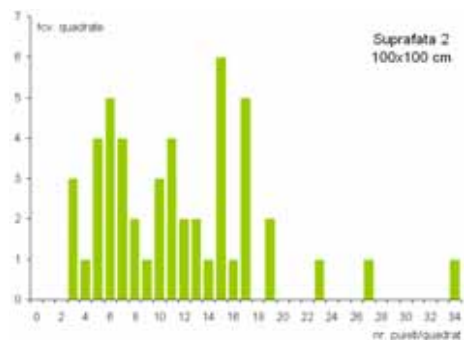

ID – structură agregată
ICS - structură agregată
Indice Morisita - structură agregată
Indice Green - structură agregată

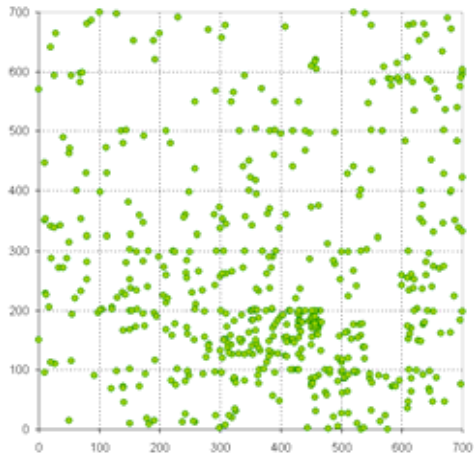
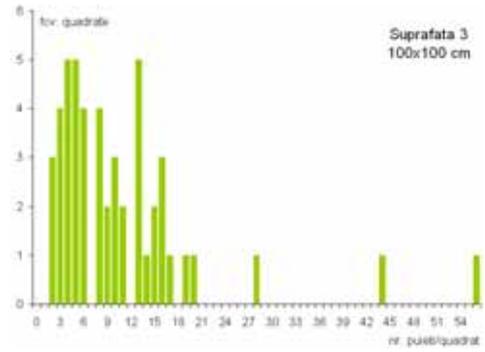

ID – structură agregată
ICS - structură agregată
Indice Morisita - structură agregată
Indice Green - structură agregată

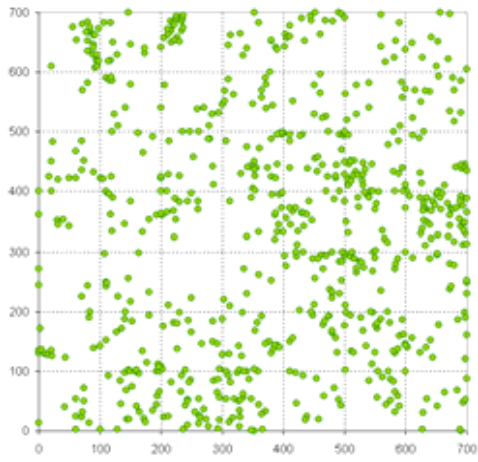
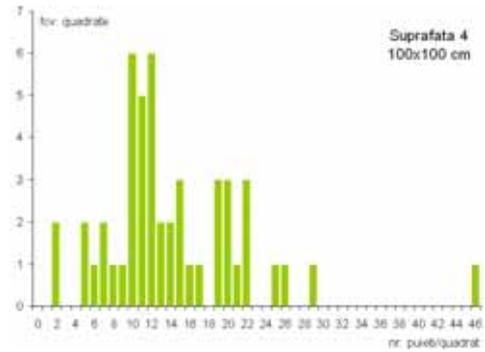

ID – structură agregată
ICS - structură agregată
Indice Morisita - structură agregată
Indice Green - structură agregată

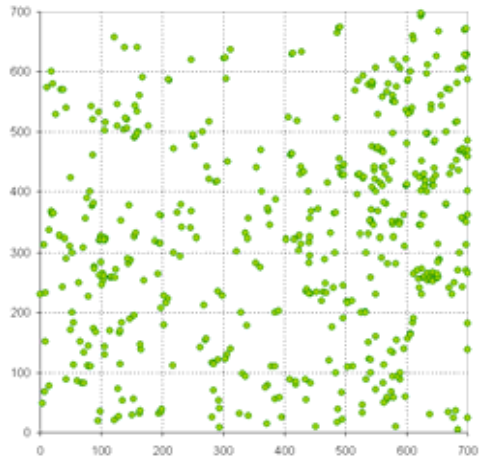
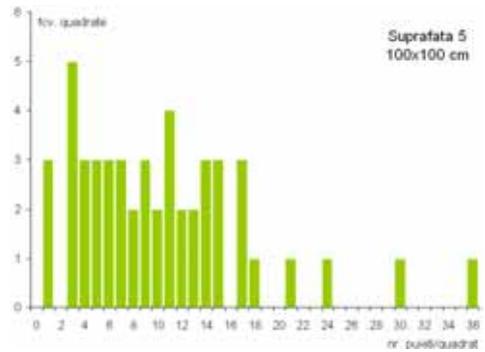

ID – structură agregată
ICS - structură agregată
Indice Morisita - structură agregată
Indice Green - structură agregată

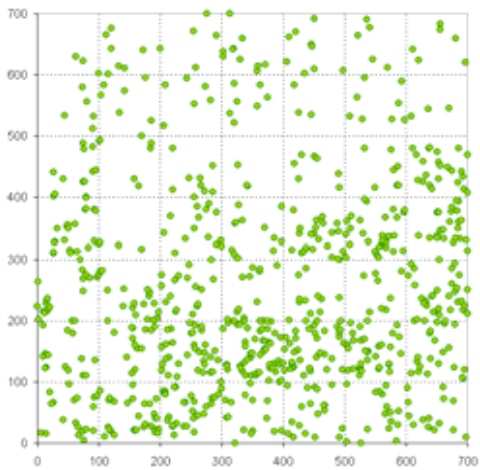

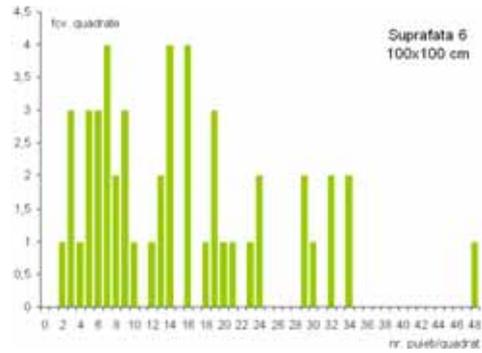

ID – structură agregată
ICS - structură agregată
Indice Morisita - structură agregată
Indice Green - structură agregată

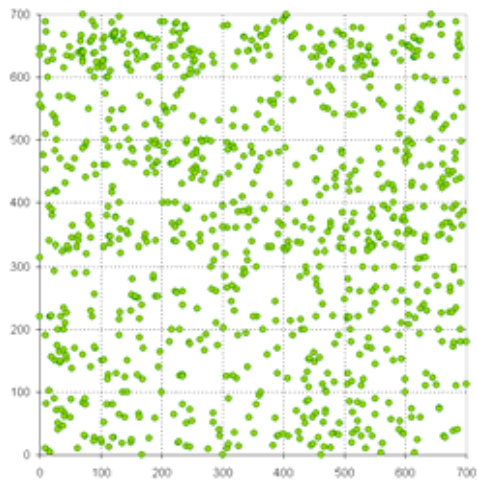

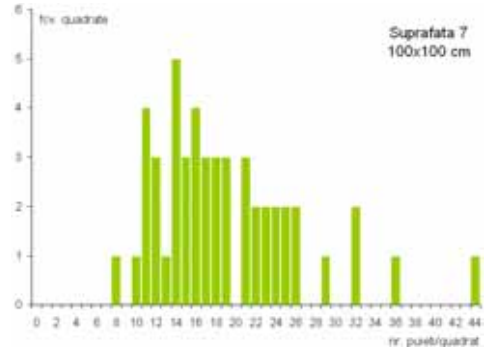

ID – structură agregată
ICS - structură agregată
Indice Morisita - structură agregată
Indice Green - structură agregată

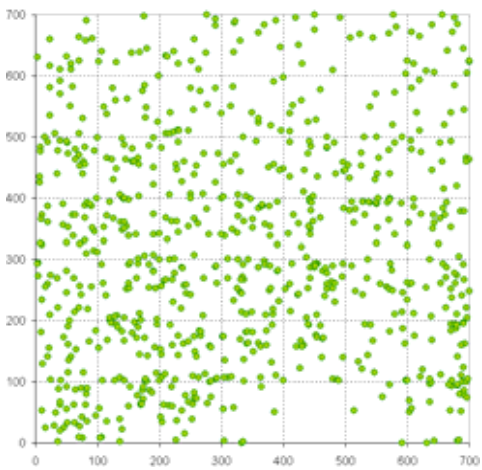

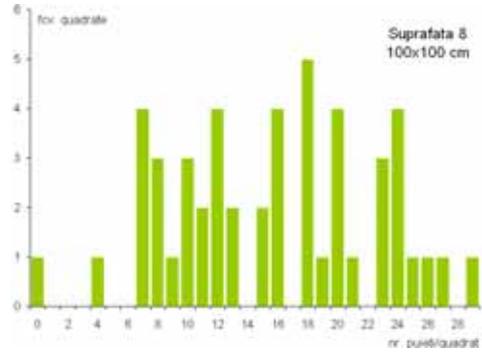

ID – structură agregată
ICS - structură agregată
Indice Morisita - structură agregată
Indice Green - structură agregată

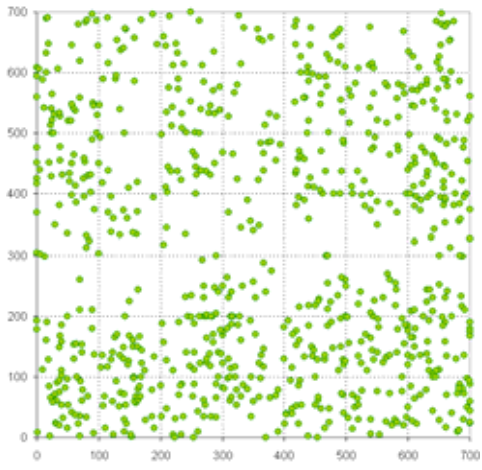 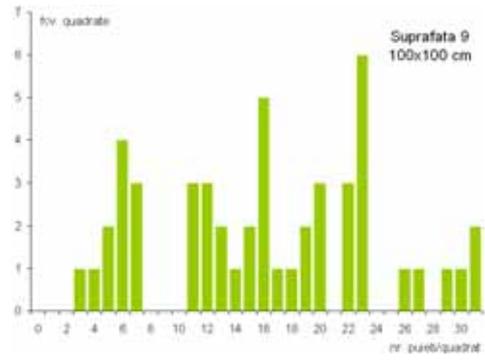

ID – structură agregată
ICS - structură agregată
Indice Morisita - structură agregată
Indice Green - structură agregată

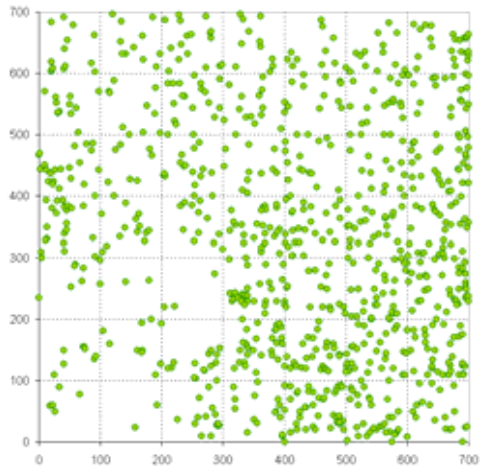 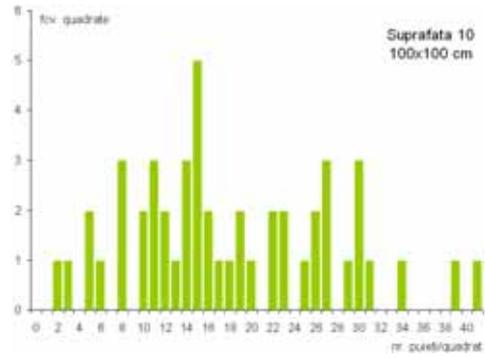

ID – structură agregată
ICS - structură agregată
Indice Morisita - structură agregată
Indice Green - structură agregată

## Anexa 9. Valorile indicilor dependenți de distanță și a testelor statistice folosite pentru aprecierea semnificației acestora – identificarea tipului de structură spațială în suprafețele studiate

| Indic. | sup 1 | sup 2 | sup 3 | sup 4 | sup 5 | sup 6 | sup 7 | sup 8 | sup 9 | sup 10 |
|---|---|---|---|---|---|---|---|---|---|---|
| d | 11,57 | 12,13 | 12,83 | 12,00 | 11,09 | 11,21 | 10,66 | 12,55 | 12,33 | 11,21 |
| d $_{CSR}$ | 12,15 | 14,70 | 15,20 | 13,23 | 15,45 | 12,85 | 11,49 | 12,61 | 12,44 | 11,71 |
| s² | 48,43 | 74,49 | 105,2 | 62,84 | 89,10 | 63,75 | 45,18 | 60,48 | 48,34 | 45,84 |
| Fisher | 4,19 | 6,14 | 8,20 | 5,24 | 8,04 | 5,69 | 4,24 | 4,82 | 3,92 | 4,09 |
| Fisher_hi² | 3470,33 | 3475,75 | 4339,51 | 3661,82 | 4074,03 | 4215,42 | 3928,56 | 3691,83 | 3101,50 | 3630,86 |
| hi² n 0.975 | 751,1 | 501,9 | 467,1 | 627,6 | 446,5 | 667,4 | 844,5 | 691,2 | 714,9 | 807,3 |
| hi² n 0.025 | 910,6 | 633,8 | 594,6 | 774,2 | 571,3 | 818,3 | 1013,3 | 844,6 | 870,8 | 972,5 |
| Clark Evans | 0,95 | 0,83 | 0,84 | 0,91 | 0,72 | 0,87 | 0,93 | 1,00 | 0,99 | 0,96 |
| CE z | -2,62 | -7,96 | -6,87 | -4,72 | -12,18 | -6,66 | -4,20 | -0,26 | -0,47 | -2,44 |
| Donnelly | 0,94 | 0,81 | 0,83 | 0,89 | 0,70 | 0,86 | 0,92 | 0,98 | 0,98 | 0,94 |
| Donnelly z | -3,27 | -8,31 | -7,27 | -5,25 | -12,27 | -7,11 | -4,79 | -1,02 | -1,22 | -3,10 |
| Pielou | 0,97 | 0,81 | 0,92 | 0,93 | 0,70 | 0,90 | 0,94 | 1,08 | 1,02 | 0,98 |
| Skellam | 1609,88 | 913,05 | 971,43 | 1298,46 | 708,13 | 1336,16 | 1753,30 | 1650,47 | 1610,82 | 1745,50 |
| hi² 2n 0.975 | 1548,97 | 1042,57 | 971,67 | 1298,20 | 929,56 | 1379,13 | 1738,49 | 1427,35 | 1475,59 | 1663,03 |
| hi² 2n 0.025 | 1774,81 | 1229,22 | 1152,12 | 1505,59 | 1106,23 | 1592,66 | 1977,30 | 1644,44 | 1696,20 | 1896,76 |
| nr. puieti | 830 | 567 | 530 | 700 | 508 | 742 | 928 | 767 | 792 | 889 |
| desime | 16,94 | 11,57 | 10,82 | 14,29 | 10,37 | 15,14 | 18,94 | 15,65 | 16,16 | 18,14 |

**Interpretarea statistică a indicilor și identificarea tiparului spațial**

| | | | | | | | | | | |
|---|---|---|---|---|---|---|---|---|---|---|
| Fisher | agreg | agreg | agreg | agreg | agreg | agreg | agreg | agreg | agreg | agreg |
| Clark Evans | agreg | agreg | agreg | agreg | agreg | agreg | agreg | aleat. | aleat. | agreg |
| Donnelly | agreg | agreg | agreg | agreg | agreg | agreg | agreg | aleat. | aleat. | agreg |
| Pielou și Skellam | aleat. | agreg | agreg | aleat. | agreg | agreg | aleat. | unif. | aleat. | aleat. |

**Notații folosite:**

| | |
|---|---|
| d | distanța observată medie față de primul vecin |
| d $_{CSR}$ | distanța teoretică medie față de primul vecin în ipoteza CSR |
| s² | dispersia distanței medii observate |
| Fisher | indicele Fisher |
| Fisher_hi² | valoarea testului statistic al indicelui Fisher |
| hi² n 0.975 | valoarea testului $\chi^2$ teoretic (n-1 grade de libertate și p=0.975) |
| hi² n 0.025 | valoarea testului $\chi^2$ teoretic (n-1 grade de libertate și p=0.025) |
| Clark Evans | indicele Clark-Evans |
| CE z | valoarea testului statistic al indicelui Clark-Evans |
| Donnelly | indicele Donnelly |
| Donnelly z | valoarea testului statistic al indicelui Donnelly |
| Pielou | Indicele Pielou (sau testul Skellam/2n) |
| Skellam | valoarea testului statistic al indicelui Pielou (echivalent cu indicele Skellam) |
| hi² 2n 0.975 | valoarea testului $\chi^2$ teoretic (2n grade de libertate și p=0.975) |
| hi² 2n 0.025 | valoarea testului $\chi^2$ teoretic (2n grade de libertate și p=0.025) |
| nr. puieti | numărul evenimentelor din suprafață |
| desime | densitatea evenimentelor în suprafață |

## Anexa 10. Valorile coeficienților de corelație ce exprimă legătura dintre indicii de determinare a structurii spațiale prin metoda distanțelor și diverși parametri structurali ai suprafețelor studiate

| | Clark-Evans | Donnelly | Pielou | Skellam | Fisher |
|---|---|---|---|---|---|
| desimea (puieti/m2) | 0,805** | 0,811** | 0,701* | 0,967*** | -0,918*** |
| diametrul mediu | 0,072 | 0,065 | 0,258 | -0,152 | 0,353 |
| înălțimea medie | 0,428 | 0,426 | 0,562 | 0,365 | -0,085 |
| diametrul mediu al coroanei | -0,178 | -0,179 | -0,234 | -0,224 | 0,107 |
| volumul coroanei | -0,428 | -0,432 | -0,252 | -0,512 | 0,710* |
| suprafața exterioară a coroanei | -0,106 | -0,110 | 0,034 | -0,207 | 0,397 |
| indicele Simpson (D) | -0,225 | -0,224 | -0,234 | -0,225 | 0,121 |
| indicele Shannon | 0,175 | 0,175 | 0,207 | 0,242 | -0,088 |
| indicele Gleason | -0,447 | -0,450 | -0,305 | -0,437 | 0,608 |
| IDIV 25 (pas 25 cm) | 0,418 | 0,424 | 0,385 | 0,622 | -0,501 |
| IDIV_min 25 | -0,349 | -0,346 | -0,275 | -0,200 | 0,316 |
| IDIV_max 25 | 0,385 | 0,391 | 0,374 | 0,604 | -0,435 |
| IDIV 50 (pas 50 cm) | 0,454 | 0,459 | 0,417 | 0,659* | -0,534 |
| IDIV_min 50 | -0,336 | -0,333 | -0,261 | -0,197 | 0,306 |
| IDIV_max 50 | 0,430 | 0,435 | 0,410 | 0,646* | -0,482 |
| Indice diferențiere $T_1$ | -0,339 | -0,335 | -0,436 | -0,211 | 0,055 |
| Indice diferențiere $T_2$ | -0,296 | -0,291 | -0,401 | -0,158 | 0,005 |
| Indice diferențiere $T_3$ | -0,316 | -0,311 | -0,423 | -0,167 | 0,012 |
| Indice diferențiere $T_4$ | -0,318 | -0,314 | -0,422 | -0,160 | 0,017 |
| $U_1$ var1 | -0,464 | -0,461 | -0,581 | -0,333 | 0,186 |
| $U_1$ var2 | 0,525 | 0,524 | 0,656* | 0,499 | -0,246 |
| $U_2$ var1 | -0,139 | -0,140 | -0,279 | -0,233 | -0,067 |
| $U_2$ var2 | 0,270 | 0,273 | 0,388 | 0,451 | -0,112 |
| $U_3$ var1 | -0,390 | -0,391 | -0,458 | -0,408 | 0,254 |
| $U_3$ var2 | 0,565 | 0,568 | 0,635* | 0,659* | -0,423 |
| $U_4$ var1 | -0,564 | -0,564 | -0,617 | -0,544 | 0,389 |
| $U_4$ var2 | 0,678* | 0,680* | 0,732* | 0,723* | -0,495 |
| $U_5$ var1 | -0,513 | -0,512 | -0,581 | -0,448 | 0,297 |
| $U_5$ var2 | 0,685* | 0,685* | 0,732* | 0,694* | -0,488 |

| | |
|---|---|
| IDIV 25, IDIV 50 | valoarea indicelui IDIV pentru un anumit pas de analiză (25/50 cm) |
| Indice de diferențiere $T_k$ | indicele T de diferențiere a înălțimilor calculat pentru $k$ vecini |
| $U_k$ var1 | indicele U de dominanță a înălțimilor calculat în varianta 1 (Hui et al., 1998) pentru k vecini |
| $U_k$ var2 | indicele U de dominanță a înălțimilor calculat în varianta 1 (Gadow și Hui, 1999; Aguirre et al., 2003) pentru k vecini |

## Anexa 11. Valorile mediilor distanțelor față de 1-15 vecini, a mediilor distanțelor în ipoteza unei distribuții aleatoare și a raportului R dintre distanța observată și distanța teoretică în ipoteza CSR.

**Notații folosite:**

| | |
|---|---|
| $\overline{d}_k$ | valoarea mediei distanțelor față de al k-lea vecin |
| $\overline{d}_{K\_CSR}$ | valoarea mediei distanțelor față de al k-lea vecin în ipoteza distribuției aleatoare (CSR) |
| inf. 95% CSR sup.95% CSR | Limitele intervalului de încredere pentru media $\overline{d}_{K\_CSR}$ |
| $R_k$ | Raportul similar Clark-Evans pentru k vecini $\quad R_k = \overline{d}_k / \overline{d}_{K\_CSR}$ |
| limita inf. 95% limita sup. 95% | Limitele intervalului de încredere ale raportului $R_k$ |
| Interpretare statistica | Determinarea sensului abaterilor de la ipoteza CSR la un prag de semnificație de 0,05 |

| k vecini | $\overline{d}_k$ | $\overline{d}_{K\_CSR}$ | inf. 95% CSR | sup.95% CSR | $R_k$ | limita sup. 95% | limita inf. 95% | Interpretare statistica |
|---|---|---|---|---|---|---|---|---|
| | | | | Suprafața 1 | | | | |
| 1 | 11,570 | 12,149 | 11,717 | 12,581 | 0,952 | 1,036 | 0,964 | agregare |
| 2 | 17,717 | 18,223 | 17,773 | 18,673 | 0,972 | 1,025 | 0,975 | agregare |
| 3 | 22,715 | 22,779 | 22,323 | 23,235 | 0,997 | 1,020 | 0,980 | aleatoare |
| 4 | 26,711 | 26,575 | 26,117 | 27,034 | 1,005 | 1,017 | 0,983 | aleatoare |
| 5 | 30,255 | 29,897 | 29,431 | 30,363 | 1,012 | 1,016 | 0,984 | aleatoare |
| 6 | 33,327 | 32,887 | 32,421 | 33,353 | 1,013 | 1,014 | 0,986 | aleatoare |
| 7 | 36,274 | 35,627 | 35,161 | 36,094 | 1,018 | 1,013 | 0,987 | uniformă |
| 8 | 39,032 | 38,172 | 37,706 | 38,639 | 1,023 | 1,012 | 0,988 | uniformă |
| 9 | 41,417 | 40,558 | 40,092 | 41,024 | 1,021 | 1,011 | 0,989 | uniformă |
| 10 | 43,796 | 42,811 | 42,345 | 43,278 | 1,023 | 1,011 | 0,989 | uniformă |
| 11 | 46,088 | 44,952 | 44,485 | 45,418 | 1,025 | 1,010 | 0,990 | uniformă |
| 12 | 48,221 | 46,995 | 46,529 | 47,461 | 1,026 | 1,010 | 0,990 | uniformă |
| 13 | 50,207 | 48,953 | 48,487 | 49,419 | 1,026 | 1,010 | 0,990 | uniformă |
| 14 | 52,190 | 50,836 | 50,370 | 51,302 | 1,027 | 1,009 | 0,991 | uniformă |
| 15 | 54,060 | 52,652 | 52,185 | 53,118 | 1,027 | 1,009 | 0,991 | uniformă |
| | | | | Suprafața 2 | | | | |
| 1 | 12,130 | 14,699 | 14,066 | 15,331 | 0,825 | 1,043 | 0,957 | agregare |
| 2 | 19,944 | 22,048 | 21,389 | 22,707 | 0,905 | 1,030 | 0,970 | agregare |
| 3 | 25,638 | 27,560 | 26,893 | 28,227 | 0,930 | 1,024 | 0,976 | agregare |

| k vecini | $\overline{d}_k$ | $\overline{d}_{K\_CSR}$ | inf. 95% CSR | sup.95% CSR | $R_k$ | limita sup. 95% | limita inf. 95% | Interpretare statistica |
|---|---|---|---|---|---|---|---|---|
| 4 | 30,446 | 32,153 | 31,482 | 32,824 | 0,947 | 1,021 | 0,979 | agregare |
| 5 | 34,674 | 36,172 | 35,490 | 36,855 | 0,959 | 1,019 | 0,981 | agregare |
| 6 | 39,052 | 39,790 | 39,107 | 40,472 | 0,981 | 1,017 | 0,983 | agregare |
| 7 | 42,588 | 43,105 | 42,423 | 43,788 | 0,988 | 1,016 | 0,984 | aleatoare |
| 8 | 46,257 | 46,184 | 45,502 | 46,867 | 1,002 | 1,015 | 0,985 | aleatoare |
| 9 | 49,548 | 49,071 | 48,388 | 49,754 | 1,010 | 1,014 | 0,986 | aleatoare |
| 10 | 52,572 | 51,797 | 51,114 | 52,480 | 1,015 | 1,013 | 0,987 | uniformă |
| 11 | 55,314 | 54,387 | 53,704 | 55,070 | 1,017 | 1,013 | 0,987 | uniformă |
| 12 | 57,834 | 56,859 | 56,176 | 57,542 | 1,017 | 1,012 | 0,988 | uniformă |
| 13 | 60,301 | 59,228 | 58,546 | 59,911 | 1,018 | 1,012 | 0,988 | uniformă |
| 14 | 62,816 | 61,506 | 60,824 | 62,189 | 1,021 | 1,011 | 0,989 | uniformă |
| 15 | 65,424 | 63,703 | 63,020 | 64,385 | 1,027 | 1,011 | 0,989 | uniformă |

**Suprafața 3**

| k vecini | $\overline{d}_k$ | $\overline{d}_{K\_CSR}$ | inf. 95% CSR | sup.95% CSR | $R_k$ | limita sup. 95% | limita inf. 95% | Interpretare statistica |
|---|---|---|---|---|---|---|---|---|
| 1 | 12,831 | 15,203 | 14,526 | 15,880 | 0,844 | 1,045 | 0,955 | agregare |
| 2 | 19,963 | 22,805 | 22,100 | 23,509 | 0,875 | 1,031 | 0,969 | agregare |
| 3 | 25,760 | 28,506 | 27,792 | 29,219 | 0,904 | 1,025 | 0,975 | agregare |
| 4 | 31,132 | 33,257 | 32,539 | 33,975 | 0,936 | 1,022 | 0,978 | agregare |
| 5 | 36,112 | 37,414 | 36,683 | 38,144 | 0,965 | 1,020 | 0,980 | agregare |
| 6 | 39,888 | 41,155 | 40,425 | 41,885 | 0,969 | 1,018 | 0,982 | agregare |
| 7 | 43,302 | 44,585 | 43,854 | 45,315 | 0,971 | 1,016 | 0,984 | agregare |
| 8 | 46,284 | 47,769 | 47,039 | 48,500 | 0,969 | 1,015 | 0,985 | agregare |
| 9 | 49,258 | 50,755 | 50,025 | 51,485 | 0,971 | 1,014 | 0,986 | agregare |
| 10 | 52,269 | 53,575 | 52,844 | 54,305 | 0,976 | 1,014 | 0,986 | agregare |
| 11 | 55,000 | 56,253 | 55,523 | 56,984 | 0,978 | 1,013 | 0,987 | agregare |
| 12 | 57,634 | 58,810 | 58,080 | 59,541 | 0,980 | 1,012 | 0,988 | agregare |
| 13 | 60,074 | 61,261 | 60,530 | 61,991 | 0,981 | 1,012 | 0,988 | agregare |
| 14 | 62,471 | 63,617 | 62,887 | 64,347 | 0,982 | 1,011 | 0,989 | agregare |
| 15 | 64,860 | 65,889 | 65,159 | 66,619 | 0,984 | 1,011 | 0,989 | agregare |

**Suprafața 4**

| k vecini | $\overline{d}_k$ | $\overline{d}_{K\_CSR}$ | inf. 95% CSR | sup.95% CSR | $R_k$ | limita sup. 95% | limita inf. 95% | Interpretare statistica |
|---|---|---|---|---|---|---|---|---|
| 1 | 11,996 | 13,229 | 12,716 | 13,741 | 0,907 | 1,039 | 0,961 | agregare |
| 2 | 17,979 | 19,843 | 19,309 | 20,377 | 0,906 | 1,027 | 0,973 | agregare |
| 3 | 22,371 | 24,804 | 24,264 | 25,344 | 0,902 | 1,022 | 0,978 | agregare |
| 4 | 26,417 | 28,938 | 28,394 | 29,482 | 0,913 | 1,019 | 0,981 | agregare |
| 5 | 30,019 | 32,555 | 32,002 | 33,108 | 0,922 | 1,017 | 0,983 | agregare |
| 6 | 33,382 | 35,811 | 35,258 | 36,364 | 0,932 | 1,015 | 0,985 | agregare |
| 7 | 36,505 | 38,795 | 38,242 | 39,348 | 0,941 | 1,014 | 0,986 | agregare |
| 8 | 39,560 | 41,566 | 41,013 | 42,119 | 0,952 | 1,013 | 0,987 | agregare |
| 9 | 42,534 | 44,164 | 43,611 | 44,717 | 0,963 | 1,013 | 0,987 | agregare |

| k vecini | $\overline{d}_k$ | $\overline{d}_{K\_CSR}$ | inf. 95% CSR | sup.95% CSR | $R_k$ | limita sup. 95% | limita inf. 95% | Interpretare statistica |
|---|---|---|---|---|---|---|---|---|
| 10 | 45,308 | 46,617 | 46,064 | 47,170 | 0,972 | 1,012 | 0,988 | agregare |
| 11 | 47,969 | 48,948 | 48,395 | 49,501 | 0,980 | 1,011 | 0,989 | agregare |
| 12 | 50,305 | 51,173 | 50,620 | 51,726 | 0,983 | 1,011 | 0,989 | agregare |
| 13 | 52,633 | 53,305 | 52,752 | 53,858 | 0,987 | 1,010 | 0,990 | agregare |
| 14 | 54,556 | 55,356 | 54,803 | 55,908 | 0,986 | 1,010 | 0,990 | agregare |
| 15 | 56,637 | 57,333 | 56,780 | 57,885 | 0,988 | 1,010 | 0,990 | agregare |

**Suprafața 5**

| k vecini | $\overline{d}_k$ | $\overline{d}_{K\_CSR}$ | inf. 95% CSR | sup.95% CSR | $R_k$ | limita sup. 95% | limita inf. 95% | Interpretare statistica |
|---|---|---|---|---|---|---|---|---|
| 1 | 11,088 | 15,451 | 14,748 | 16,153 | 0,718 | 1,045 | 0,955 | agregare |
| 2 | 18,765 | 23,176 | 22,445 | 23,908 | 0,810 | 1,032 | 0,968 | agregare |
| 3 | 25,405 | 28,970 | 28,230 | 29,711 | 0,877 | 1,026 | 0,974 | agregare |
| 4 | 30,408 | 33,799 | 33,053 | 34,544 | 0,900 | 1,022 | 0,978 | agregare |
| 5 | 34,580 | 38,024 | 37,266 | 38,782 | 0,909 | 1,020 | 0,980 | agregare |
| 6 | 39,292 | 41,826 | 41,068 | 42,584 | 0,939 | 1,018 | 0,982 | agregare |
| 7 | 43,120 | 45,312 | 44,554 | 46,070 | 0,952 | 1,017 | 0,983 | agregare |
| 8 | 46,341 | 48,548 | 47,790 | 49,306 | 0,955 | 1,016 | 0,984 | agregare |
| 9 | 49,264 | 51,582 | 50,824 | 52,340 | 0,955 | 1,015 | 0,985 | agregare |
| 10 | 52,351 | 54,448 | 53,690 | 55,206 | 0,961 | 1,014 | 0,986 | agregare |
| 11 | 54,893 | 57,170 | 56,412 | 57,929 | 0,960 | 1,013 | 0,987 | agregare |
| 12 | 57,650 | 59,769 | 59,011 | 60,527 | 0,965 | 1,013 | 0,987 | agregare |
| 13 | 60,337 | 62,260 | 61,501 | 63,018 | 0,969 | 1,012 | 0,988 | agregare |
| 14 | 62,709 | 64,654 | 63,896 | 65,412 | 0,970 | 1,012 | 0,988 | agregare |
| 15 | 65,101 | 66,963 | 66,205 | 67,721 | 0,972 | 1,011 | 0,989 | agregare |

**Suprafața 6**

| k vecini | $\overline{d}_k$ | $\overline{d}_{K\_CSR}$ | inf. 95% CSR | sup.95% CSR | $R_k$ | limita sup. 95% | limita inf. 95% | Interpretare statistica |
|---|---|---|---|---|---|---|---|---|
| 1 | 11,207 | 12,849 | 12,366 | 13,332 | 0,872 | 1,038 | 0,962 | agregare |
| 2 | 17,810 | 19,273 | 18,770 | 19,777 | 0,924 | 1,026 | 0,974 | agregare |
| 3 | 22,594 | 24,092 | 23,582 | 24,601 | 0,938 | 1,021 | 0,979 | agregare |
| 4 | 26,335 | 28,107 | 27,594 | 28,620 | 0,937 | 1,018 | 0,982 | agregare |
| 5 | 29,913 | 31,620 | 31,099 | 32,142 | 0,946 | 1,016 | 0,984 | agregare |
| 6 | 33,347 | 34,782 | 34,261 | 35,304 | 0,959 | 1,015 | 0,985 | agregare |
| 7 | 36,115 | 37,681 | 37,159 | 38,203 | 0,958 | 1,014 | 0,986 | agregare |
| 8 | 38,719 | 40,372 | 39,851 | 40,894 | 0,959 | 1,013 | 0,987 | agregare |
| 9 | 41,308 | 42,896 | 42,374 | 43,417 | 0,963 | 1,012 | 0,988 | agregare |
| 10 | 43,837 | 45,279 | 44,757 | 45,800 | 0,968 | 1,012 | 0,988 | agregare |
| 11 | 46,118 | 47,543 | 47,021 | 48,064 | 0,970 | 1,011 | 0,989 | agregare |
| 12 | 48,211 | 49,704 | 49,182 | 50,225 | 0,970 | 1,010 | 0,990 | agregare |
| 13 | 50,180 | 51,775 | 51,253 | 52,296 | 0,969 | 1,010 | 0,990 | agregare |
| 14 | 52,089 | 53,766 | 53,244 | 54,288 | 0,969 | 1,010 | 0,990 | agregare |

| k vecini | $\overline{d}_k$ | $\overline{d}_{K\_CSR}$ | inf. 95% CSR | sup.95% CSR | $R_k$ | limita sup. 95% | limita inf. 95% | Interpretare statistica |
|---|---|---|---|---|---|---|---|---|
| 15 | 54,121 | 55,686 | 55,165 | 56,208 | 0,972 | 1,009 | 0,991 | agregare |

**Suprafața 7**

| k vecini | $\overline{d}_k$ | $\overline{d}_{K\_CSR}$ | inf. 95% CSR | sup.95% CSR | $R_k$ | limita sup. 95% | limita inf. 95% | Interpretare statistica |
|---|---|---|---|---|---|---|---|---|
| 1 | 10,660 | 11,489 | 11,103 | 11,876 | 0,928 | 1,034 | 0,966 | agregare |
| 2 | 16,690 | 17,234 | 16,831 | 17,637 | 0,968 | 1,023 | 0,977 | agregare |
| 3 | 21,277 | 21,542 | 21,135 | 21,950 | 0,988 | 1,019 | 0,981 | aleatoare |
| 4 | 24,851 | 25,133 | 24,723 | 25,543 | 0,989 | 1,016 | 0,984 | aleatoare |
| 5 | 27,915 | 28,274 | 27,857 | 28,692 | 0,987 | 1,015 | 0,985 | aleatoare |
| 6 | 30,757 | 31,102 | 30,685 | 31,519 | 0,989 | 1,013 | 0,987 | aleatoare |
| 7 | 33,264 | 33,694 | 33,277 | 34,111 | 0,987 | 1,012 | 0,988 | agregare |
| 8 | 35,684 | 36,100 | 35,683 | 36,518 | 0,988 | 1,012 | 0,988 | aleatoare |
| 9 | 37,993 | 38,357 | 37,940 | 38,774 | 0,991 | 1,011 | 0,989 | aleatoare |
| 10 | 40,242 | 40,488 | 40,071 | 40,905 | 0,994 | 1,010 | 0,990 | aleatoare |
| 11 | 42,513 | 42,512 | 42,095 | 42,929 | 1,000 | 1,010 | 0,990 | aleatoare |
| 12 | 44,617 | 44,444 | 44,027 | 44,861 | 1,004 | 1,009 | 0,991 | aleatoare |
| 13 | 46,638 | 46,296 | 45,879 | 46,713 | 1,007 | 1,009 | 0,991 | aleatoare |
| 14 | 48,714 | 48,077 | 47,660 | 48,494 | 1,013 | 1,009 | 0,991 | uniformă |
| 15 | 50,628 | 49,794 | 49,377 | 50,211 | 1,017 | 1,008 | 0,992 | uniformă |

**Suprafața 8**

| k vecini | $\overline{d}_k$ | $\overline{d}_{K\_CSR}$ | inf. 95% CSR | sup.95% CSR | $R_k$ | limita sup. 95% | limita inf. 95% | Interpretare statistica |
|---|---|---|---|---|---|---|---|---|
| 1 | 12,548 | 12,611 | 12,144 | 13,077 | 0,995 | 1,037 | 0,963 | aleatoare |
| 2 | 18,919 | 18,916 | 18,430 | 19,402 | 1,000 | 1,026 | 0,974 | aleatoare |
| 3 | 23,396 | 23,645 | 23,153 | 24,137 | 0,989 | 1,021 | 0,979 | aleatoare |
| 4 | 27,400 | 27,586 | 27,091 | 28,081 | 0,993 | 1,018 | 0,982 | aleatoare |
| 5 | 30,790 | 31,034 | 30,531 | 31,538 | 0,992 | 1,016 | 0,984 | aleatoare |
| 6 | 34,088 | 34,137 | 33,634 | 34,641 | 0,999 | 1,015 | 0,985 | aleatoare |
| 7 | 36,987 | 36,982 | 36,479 | 37,486 | 1,000 | 1,014 | 0,986 | aleatoare |
| 8 | 39,571 | 39,624 | 39,120 | 40,127 | 0,999 | 1,013 | 0,987 | aleatoare |
| 9 | 42,202 | 42,100 | 41,597 | 42,604 | 1,002 | 1,012 | 0,988 | aleatoare |
| 10 | 44,616 | 44,439 | 43,936 | 44,943 | 1,004 | 1,011 | 0,989 | aleatoare |
| 11 | 47,082 | 46,661 | 46,158 | 47,165 | 1,009 | 1,011 | 0,989 | aleatoare |
| 12 | 49,394 | 48,782 | 48,279 | 49,286 | 1,013 | 1,010 | 0,990 | uniformă |
| 13 | 51,588 | 50,815 | 50,311 | 51,318 | 1,015 | 1,010 | 0,990 | uniformă |
| 14 | 53,870 | 52,769 | 52,266 | 53,273 | 1,021 | 1,010 | 0,990 | uniformă |
| 15 | 55,901 | 54,654 | 54,150 | 55,157 | 1,023 | 1,009 | 0,991 | uniformă |

**Suprafața 9**

| k vecini | $\overline{d}_k$ | $\overline{d}_{K\_CSR}$ | inf. 95% CSR | sup.95% CSR | $R_k$ | limita sup. 95% | limita inf. 95% | Interpretare statistica |
|---|---|---|---|---|---|---|---|---|
| 1 | 12,328 | 12,437 | 11,984 | 12,890 | 0,991 | 1,036 | 0,964 | aleatoare |
| 2 | 18,154 | 18,655 | 18,183 | 19,127 | 0,973 | 1,025 | 0,975 | agregare |
| 3 | 23,134 | 23,319 | 22,841 | 23,796 | 0,992 | 1,020 | 0,980 | aleatoare |
| 4 | 26,910 | 27,205 | 26,725 | 27,686 | 0,989 | 1,018 | 0,982 | aleatoare |
| 5 | 30,205 | 30,606 | 30,117 | 31,095 | 0,987 | 1,016 | 0,984 | aleatoare |
| 6 | 33,403 | 33,667 | 33,178 | 34,155 | 0,992 | 1,015 | 0,985 | aleatoare |
| 7 | 36,349 | 36,472 | 35,983 | 36,961 | 0,997 | 1,013 | 0,987 | aleatoare |
| 8 | 38,994 | 39,077 | 38,589 | 39,566 | 0,998 | 1,013 | 0,987 | aleatoare |
| 9 | 41,522 | 41,520 | 41,031 | 42,008 | 1,000 | 1,012 | 0,988 | aleatoare |
| 10 | 43,815 | 43,826 | 43,338 | 44,315 | 1,000 | 1,011 | 0,989 | aleatoare |
| 11 | 46,037 | 46,018 | 45,529 | 46,506 | 1,000 | 1,011 | 0,989 | aleatoare |
| 12 | 48,126 | 48,109 | 47,621 | 48,598 | 1,000 | 1,010 | 0,990 | aleatoare |
| 13 | 50,136 | 50,114 | 49,625 | 50,602 | 1,000 | 1,010 | 0,990 | aleatoare |
| 14 | 52,303 | 52,041 | 51,553 | 52,530 | 1,005 | 1,009 | 0,991 | aleatoare |
| 15 | 54,230 | 53,900 | 53,411 | 54,389 | 1,006 | 1,009 | 0,991 | aleatoare |

**Suprafața 10**

| k vecini | $\overline{d}_k$ | $\overline{d}_{K\_CSR}$ | inf. 95% CSR | sup.95% CSR | $R_k$ | limita sup. 95% | limita inf. 95% | Interpretare statistica |
|---|---|---|---|---|---|---|---|---|
| 1 | 11,212 | 11,713 | 11,311 | 12,116 | 0,957 | 1,034 | 0,966 | agregare |
| 2 | 16,774 | 17,570 | 17,151 | 17,990 | 0,955 | 1,024 | 0,976 | agregare |
| 3 | 21,136 | 21,963 | 21,538 | 22,387 | 0,962 | 1,019 | 0,981 | agregare |
| 4 | 24,855 | 25,623 | 25,196 | 26,050 | 0,970 | 1,017 | 0,983 | agregare |
| 5 | 28,236 | 28,826 | 28,392 | 29,260 | 0,980 | 1,015 | 0,985 | agregare |
| 6 | 31,346 | 31,709 | 31,274 | 32,143 | 0,989 | 1,014 | 0,986 | aleatoare |
| 7 | 33,919 | 34,351 | 33,917 | 34,785 | 0,987 | 1,013 | 0,987 | aleatoare |
| 8 | 36,332 | 36,805 | 36,370 | 37,239 | 0,987 | 1,012 | 0,988 | agregare |
| 9 | 38,932 | 39,105 | 38,671 | 39,539 | 0,996 | 1,011 | 0,989 | aleatoare |
| 10 | 41,210 | 41,277 | 40,843 | 41,712 | 0,998 | 1,011 | 0,989 | aleatoare |
| 11 | 43,272 | 43,341 | 42,907 | 43,776 | 0,998 | 1,010 | 0,990 | aleatoare |
| 12 | 45,345 | 45,311 | 44,877 | 45,746 | 1,001 | 1,010 | 0,990 | aleatoare |
| 13 | 47,576 | 47,199 | 46,765 | 47,634 | 1,008 | 1,009 | 0,991 | aleatoare |
| 14 | 49,567 | 49,015 | 48,580 | 49,449 | 1,011 | 1,009 | 0,991 | uniformă |
| 15 | 51,660 | 50,765 | 50,331 | 51,200 | 1,018 | 1,009 | 0,991 | uniformă |

**Anexa 12. Harta proceselor punctiforme corespunzătoare celor 3 straturi de puieți definite în funcție de înălțime și graficele variației raportului Rk (*metoda KNN*)în cele zece suprafețe**

**Stratul 1 –** stratul plantulelor (înălțimea 0 .. 25 cm)

În suprafețele 8 și 9 s-a înregistrat un număr de plantule prea mic pentru efectuarea prelucrărilor.

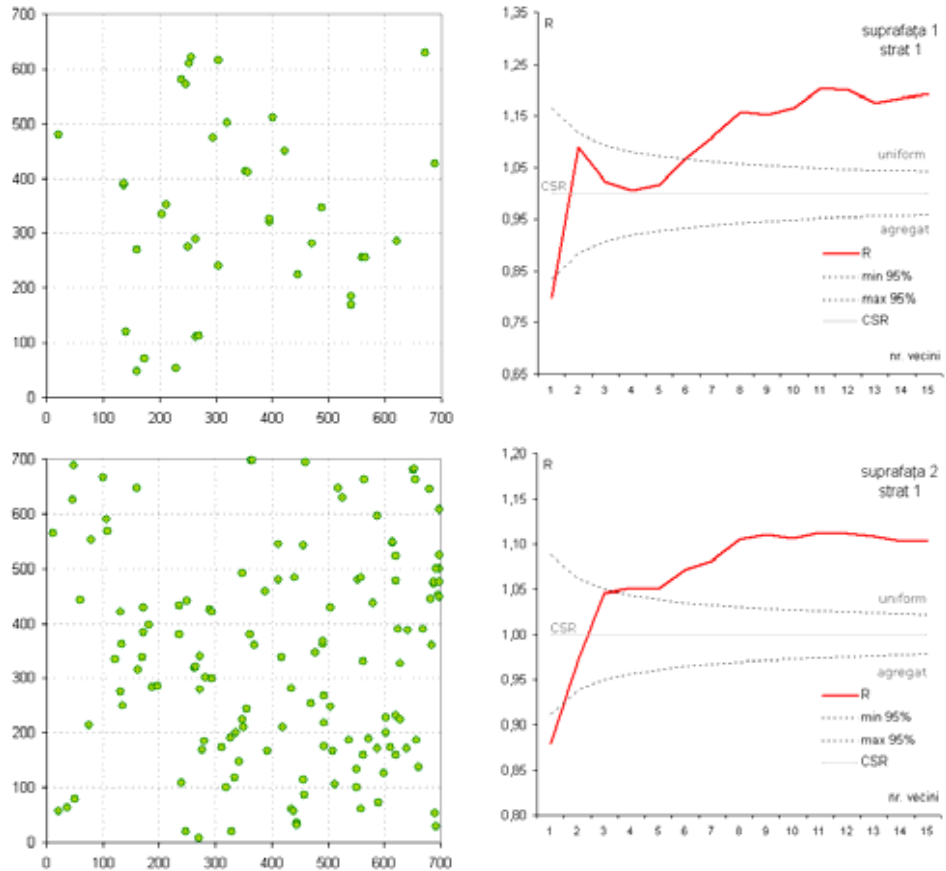

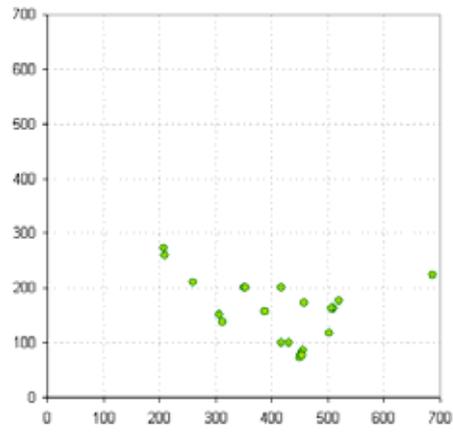
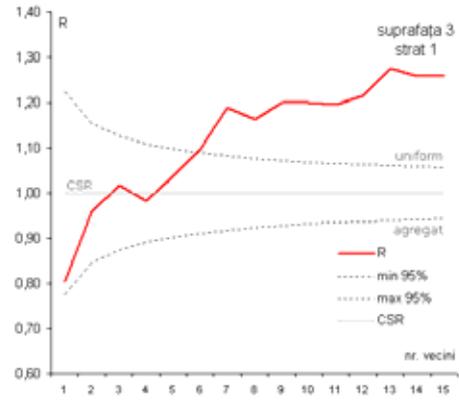
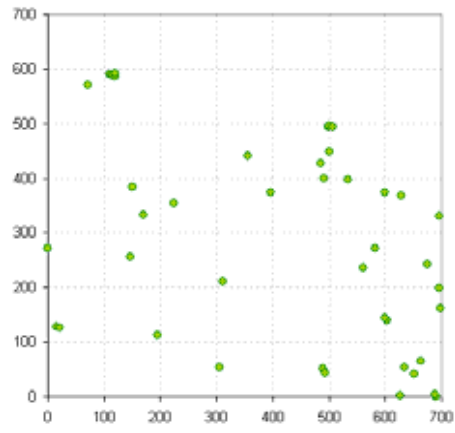
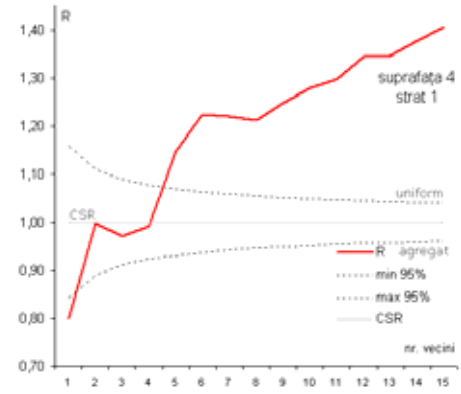
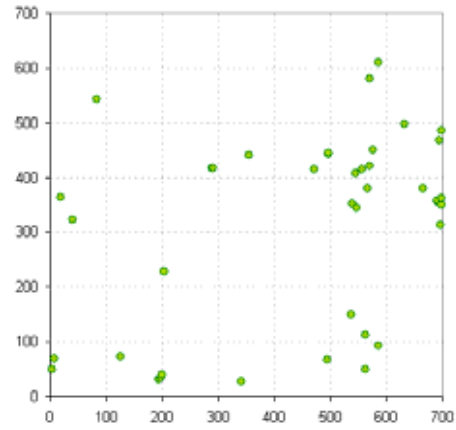
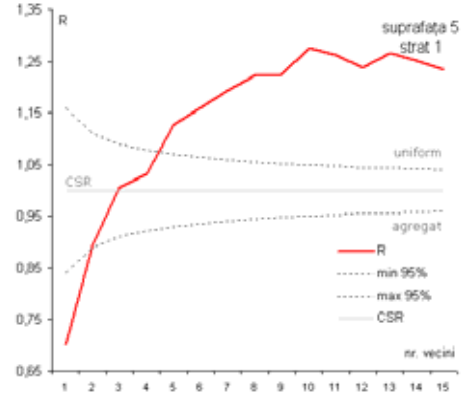

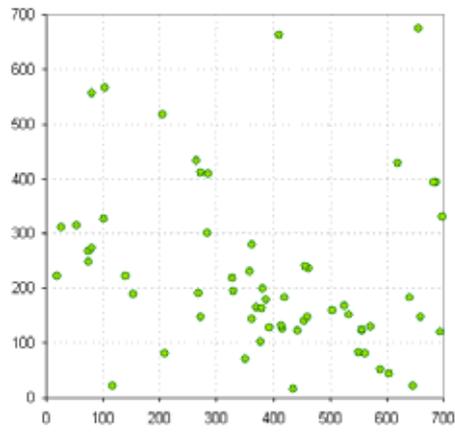
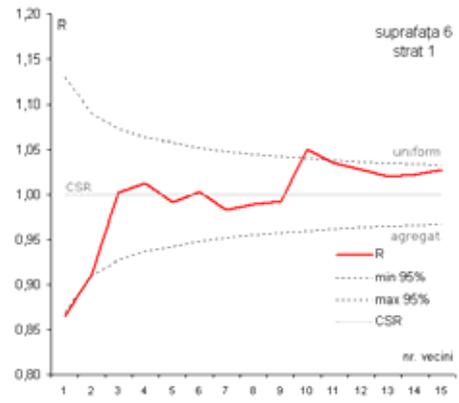

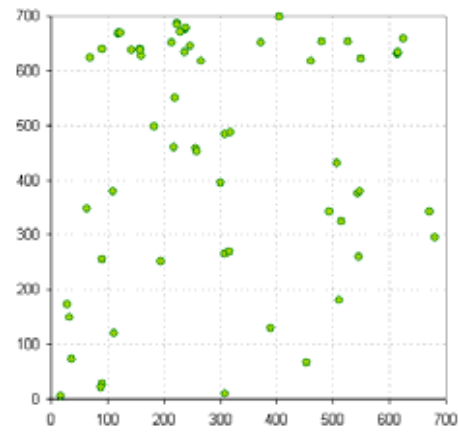
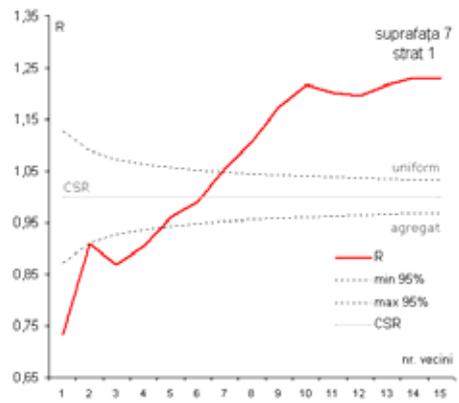

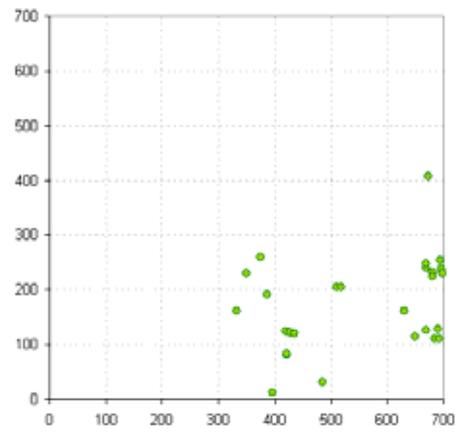
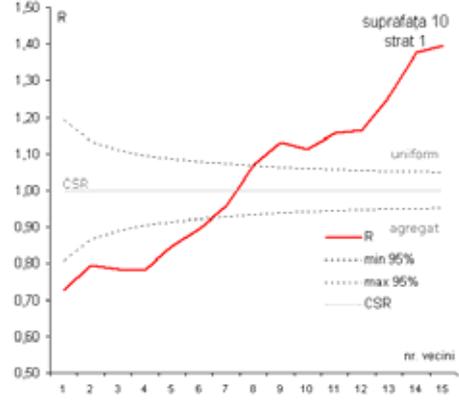

**Stratul 2** – stratul puieților din clasa medie (înălțimea 26 .. 150 cm)

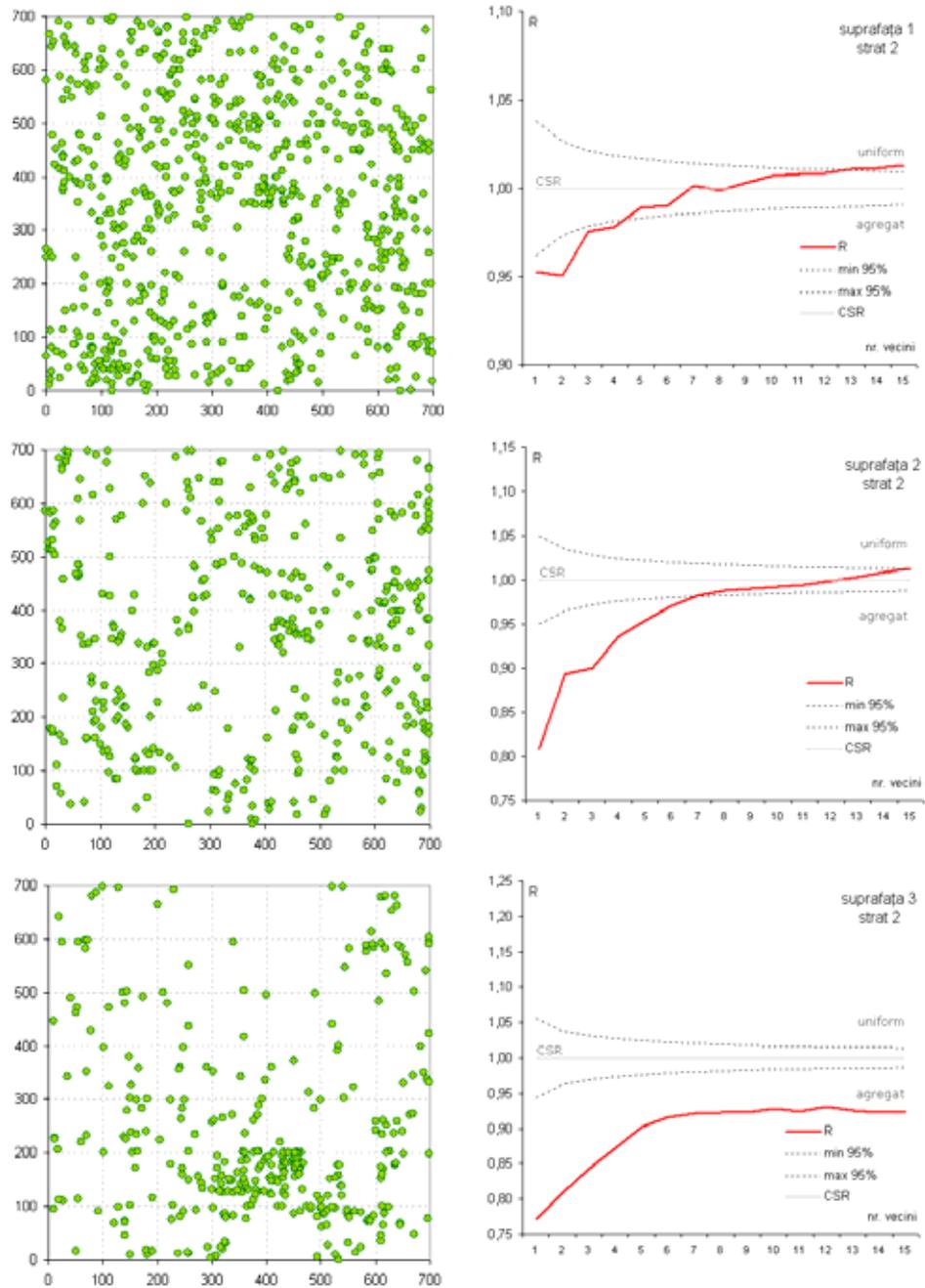

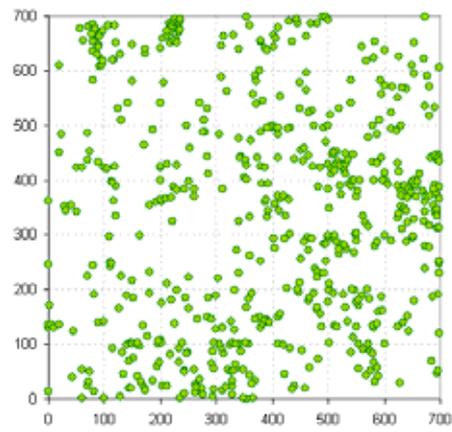
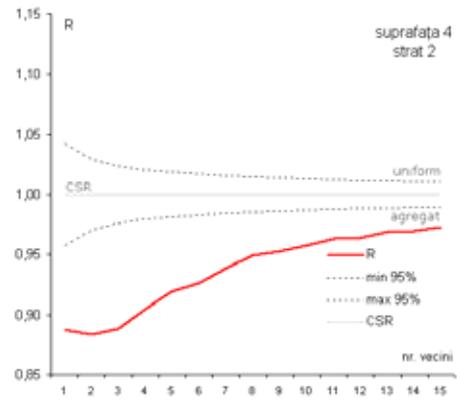
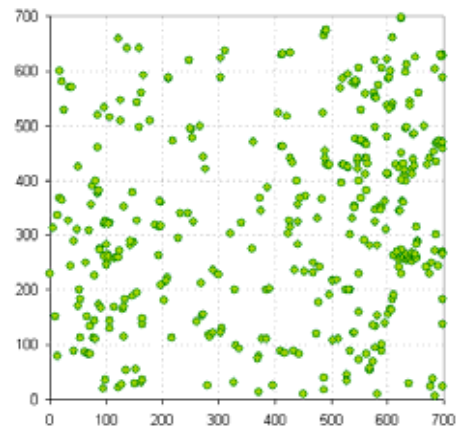
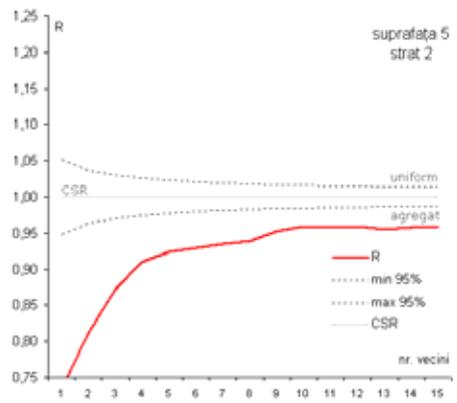
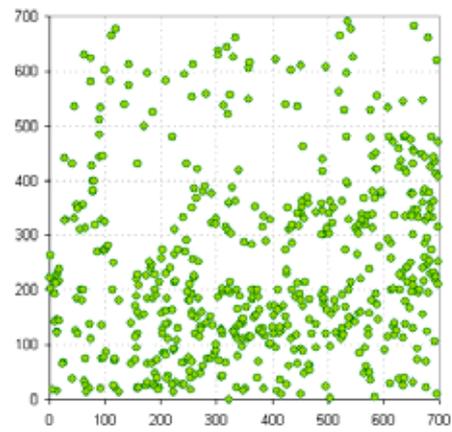
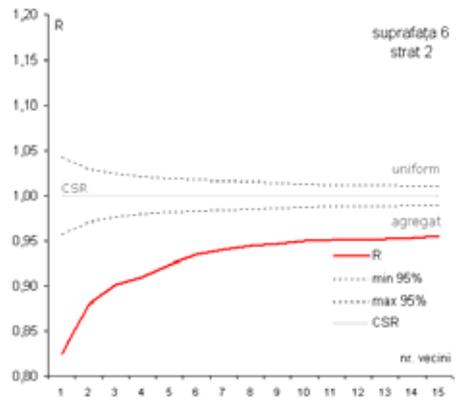

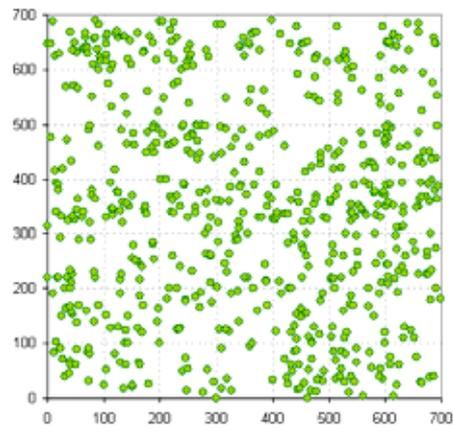
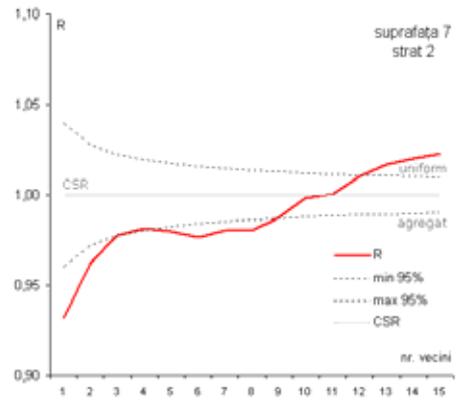

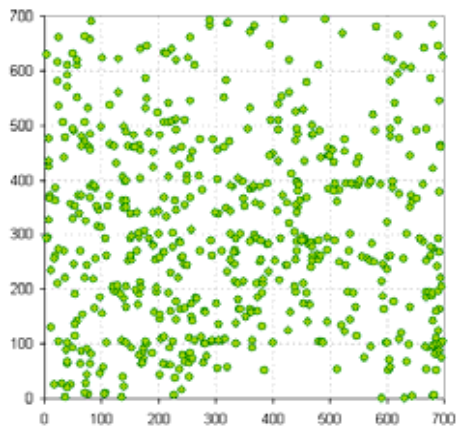
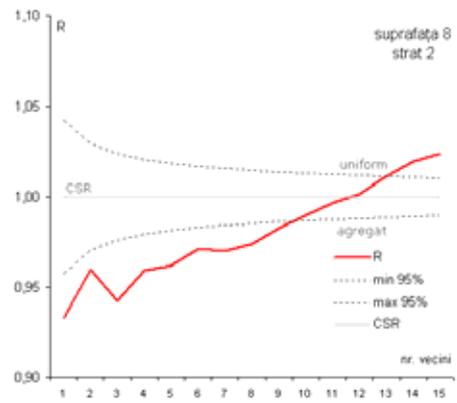

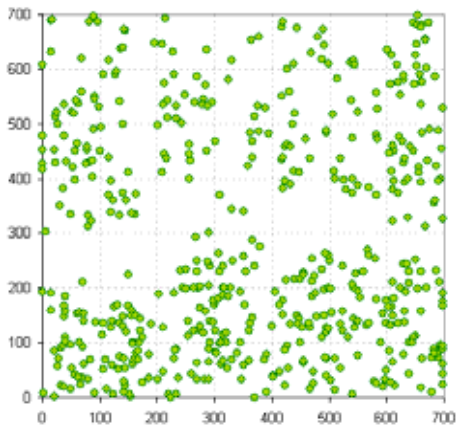
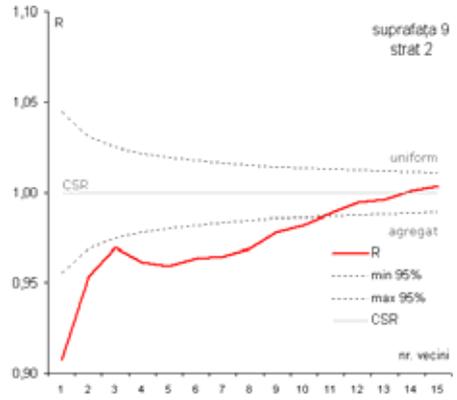

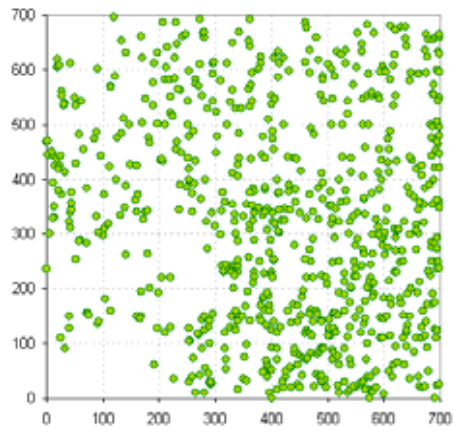
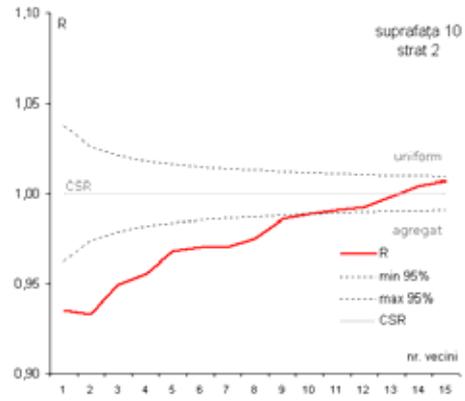

**Stratul 3 –** stratul puieţilor de mari dimensiuni (înălţimea > 150 cm)

În suprafaţa 2 s-a înregistrat un număr de puieţi prea mic pentru efectuarea prelucrărilor.

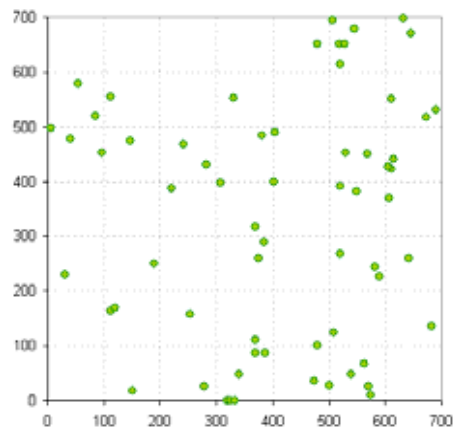
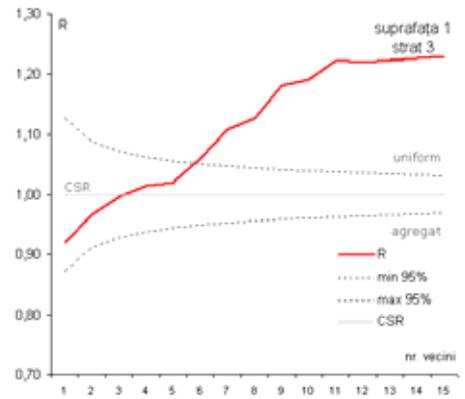

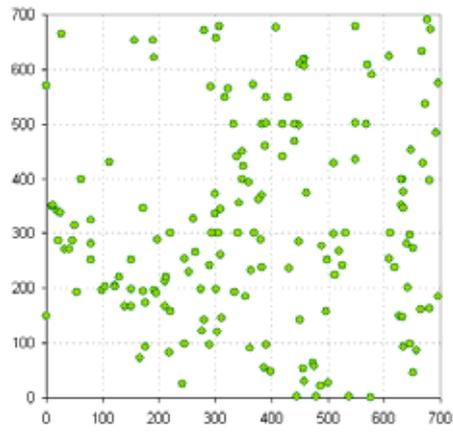
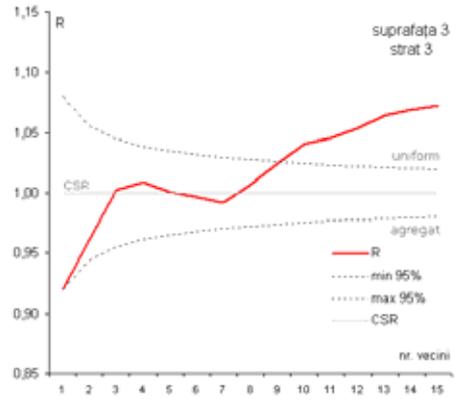

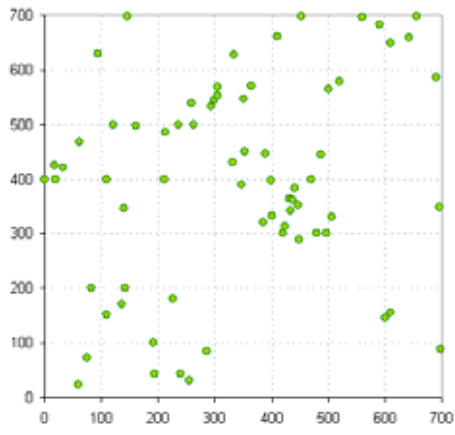
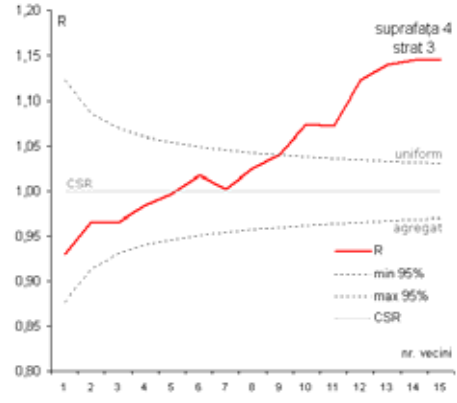

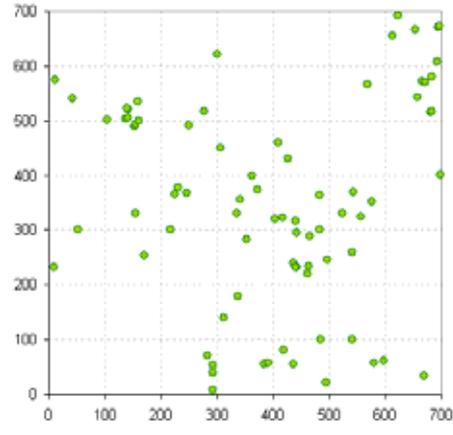
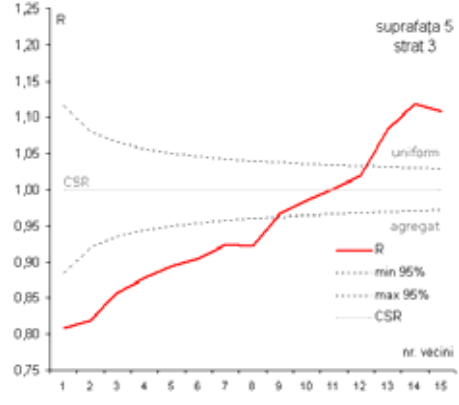

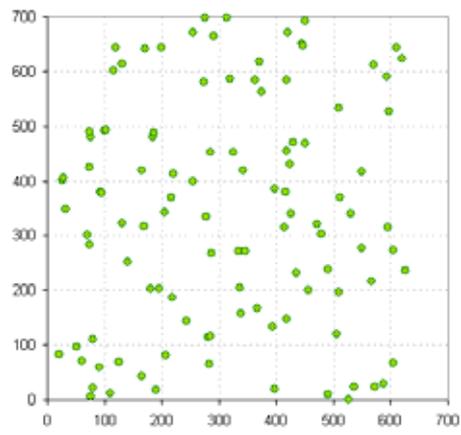
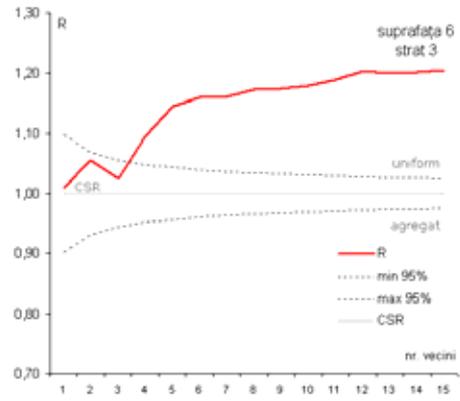

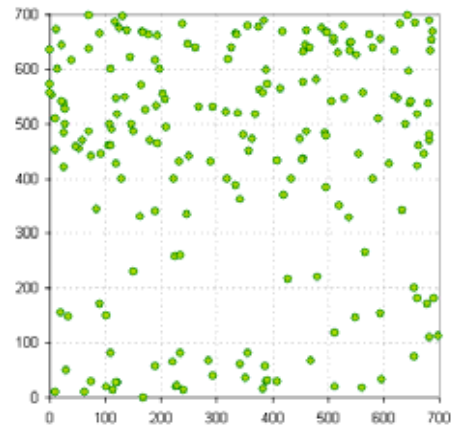
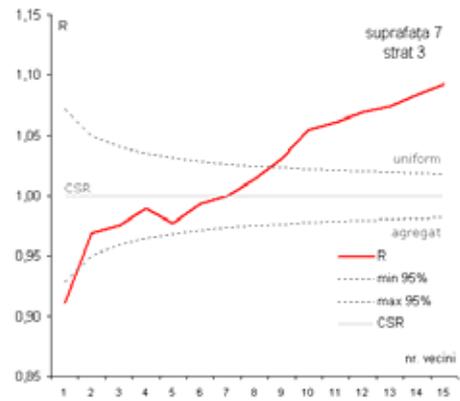

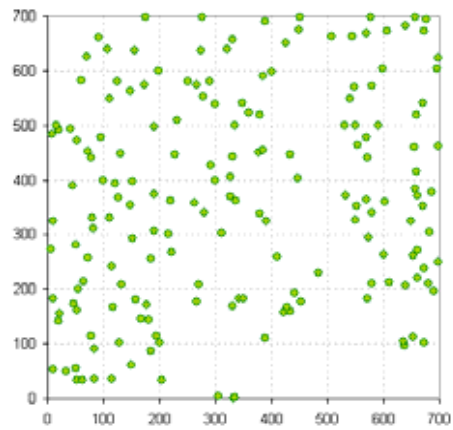
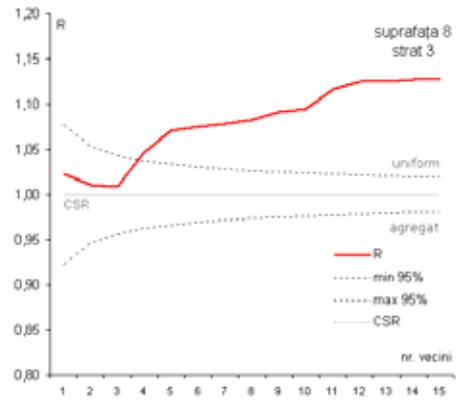

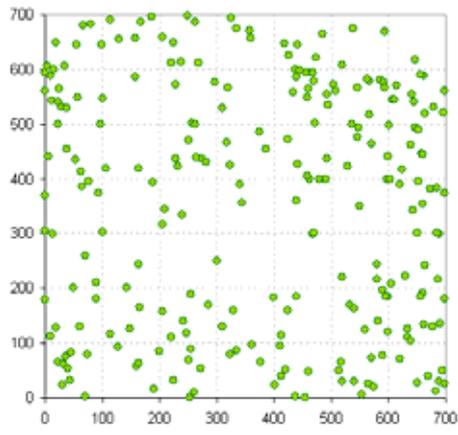
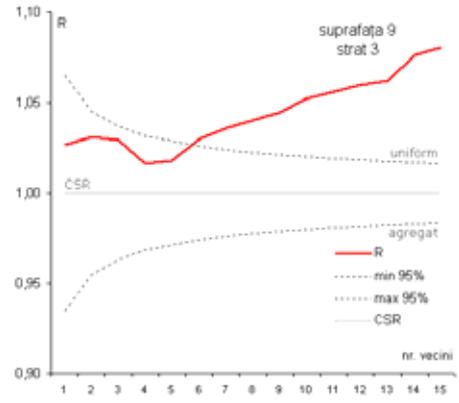

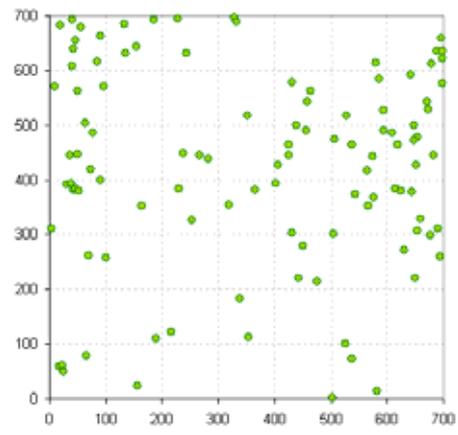
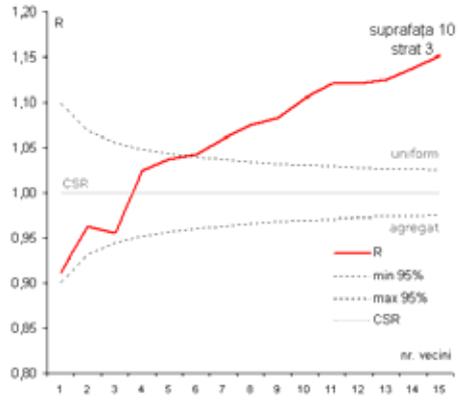

**Anexa 13. Interpretarea tiparului spațial în cazul proceselor punctiforme corespunzătoare celor 3 straturi de puieți definite în funcție de înălțime prin intermediul funcției L(t)**

**Stratul 1 –** stratul plantulelor (înălțimea 0 .. 25 cm)

În suprafețele 8 și 9 s-a înregistrat un număr de plantule prea mic pentru efectuarea prelucrărilor.

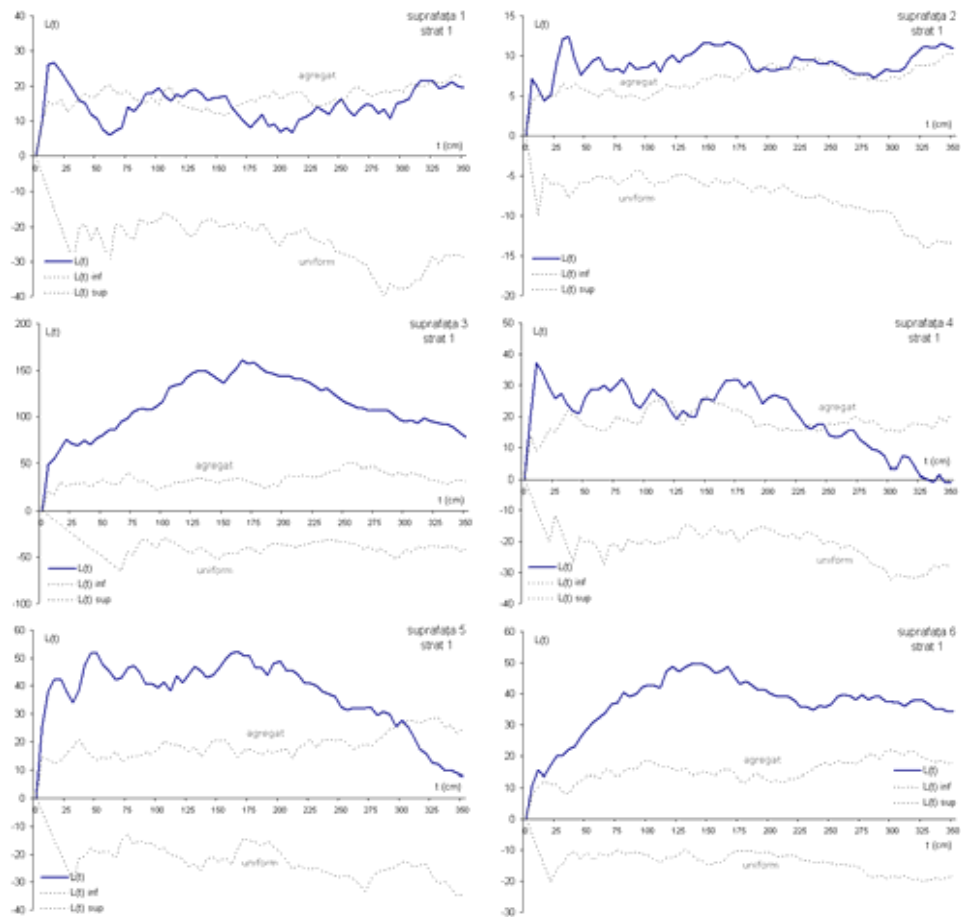

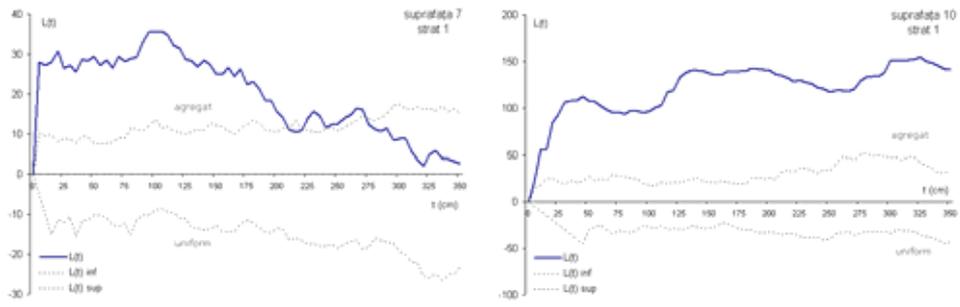

**Stratul 2** – stratul puieților din clasa medie (înălțimea 26 .. 150 cm)

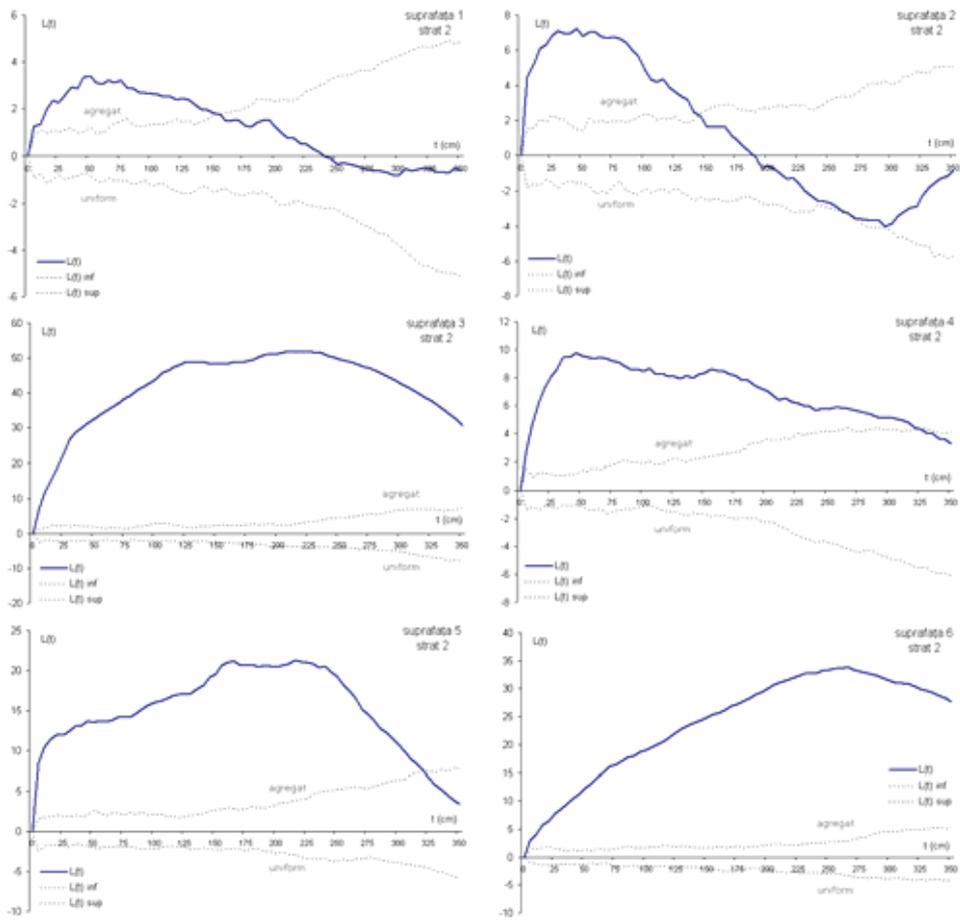

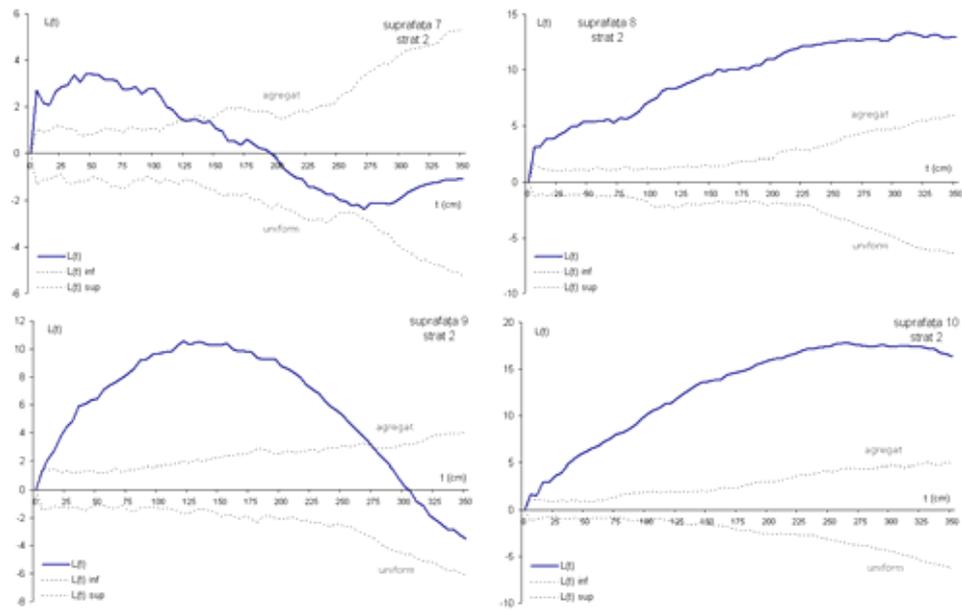

**Stratul 3** – stratul puieților de mari dimensiuni (înălțimea > 150 cm)

În suprafața 2 s-a înregistrat un număr de puieți prea mic pentru efectuarea prelucrărilor.

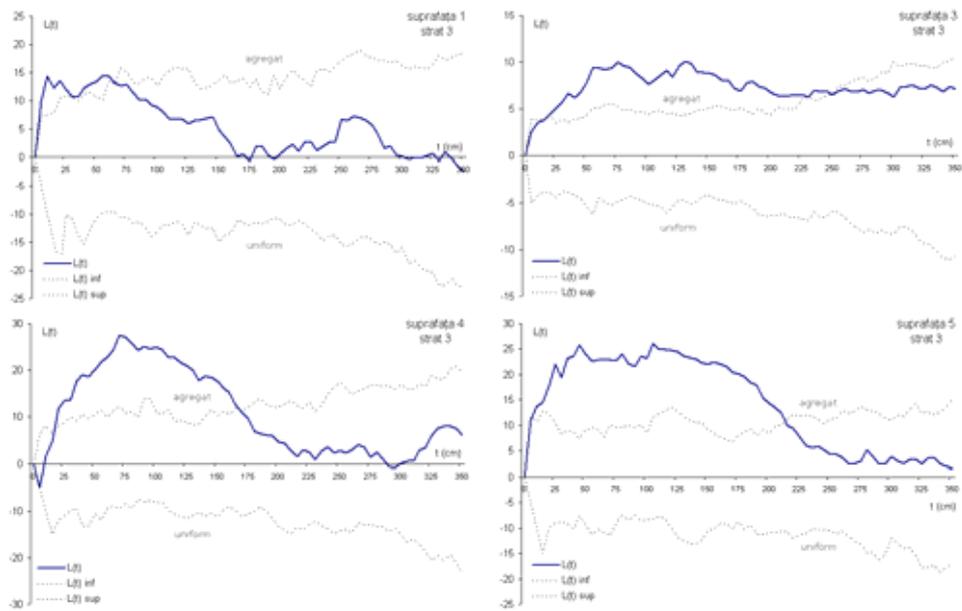

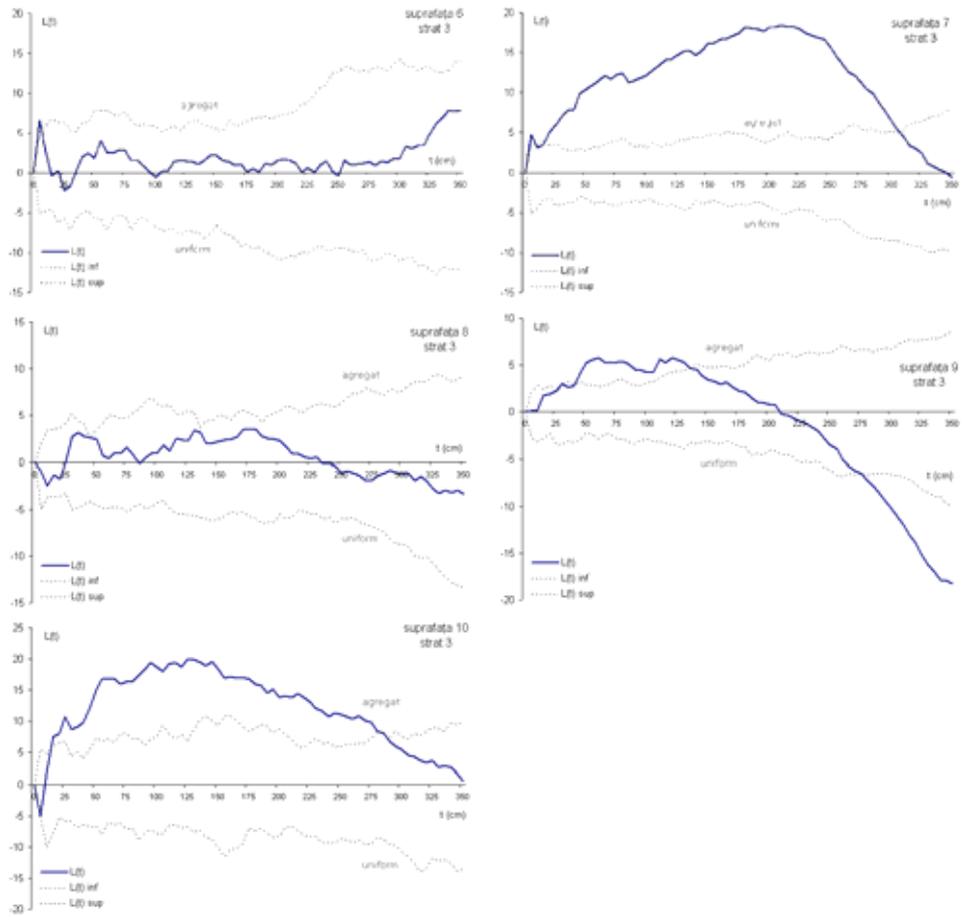

**Anexa 14. Determinarea tipului de interacțiune în spațiu dintre cele trei straturi de puieți definite în funcție de înălțime prin intermediul funcției $L_{12}(t)$**

**Interacțiunea spațială dintre stratul 1** (stratul plantulelor) **și stratul 2** (stratul puieților de dimensiuni medii)

În suprafețele 8 și 9 s-a înregistrat un număr de plantule prea mic pentru interpretarea rezultatelor.

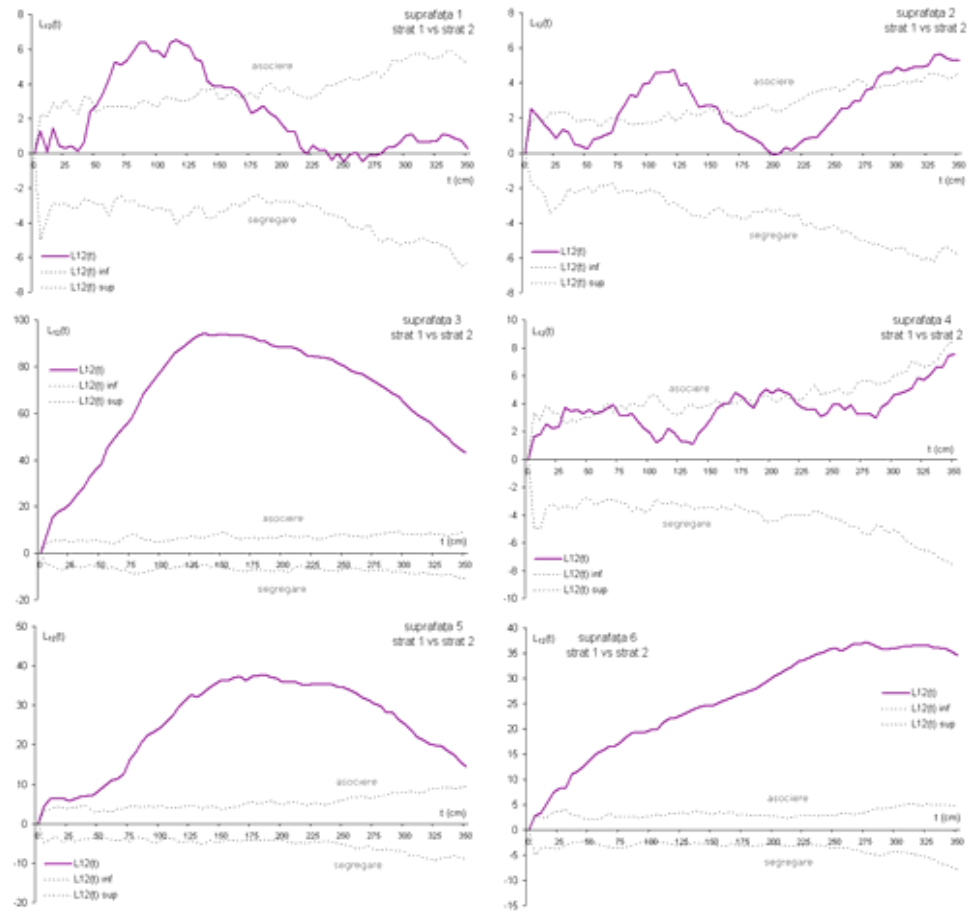

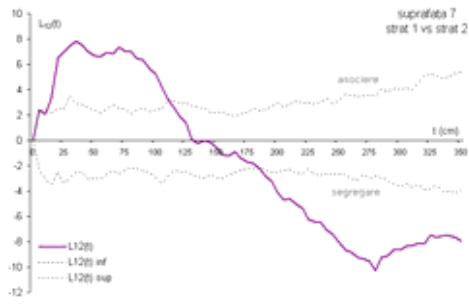 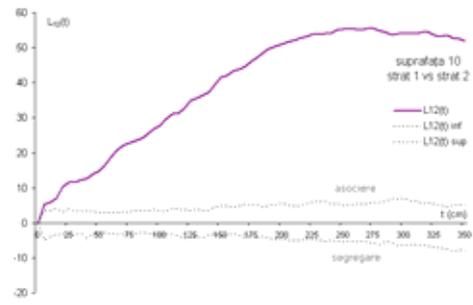

**Interacțiunea spațială dintre stratul 1** (stratul plantulelor) **și stratul 3** (stratul puieților de mari dimensiuni)

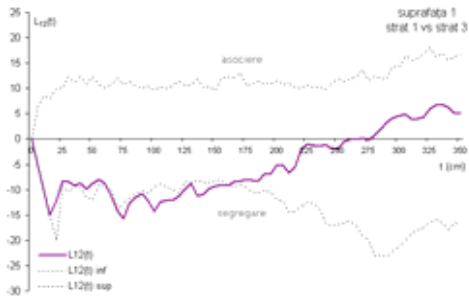 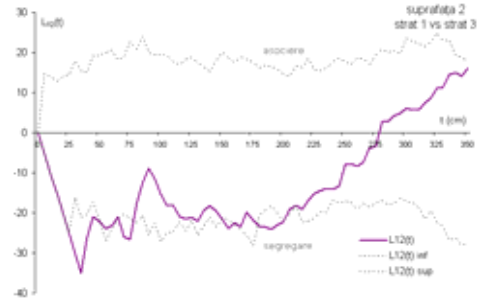

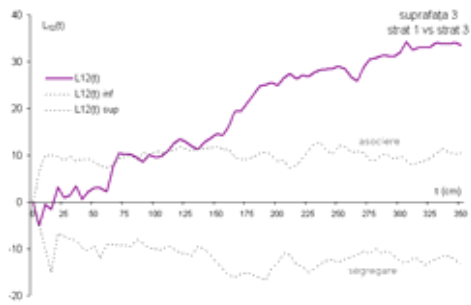 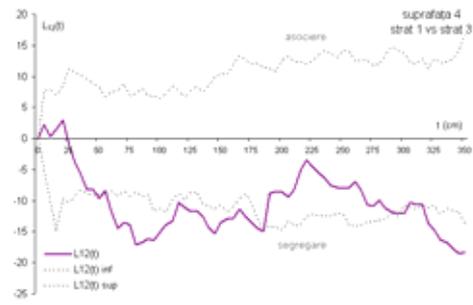

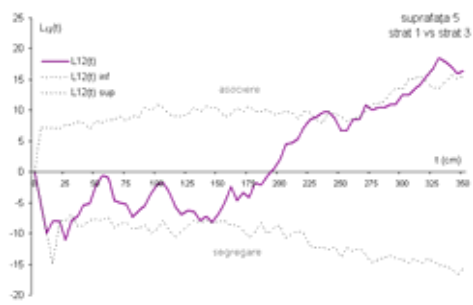 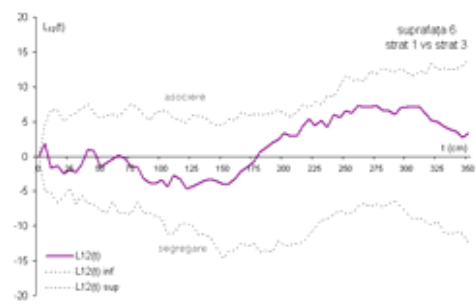

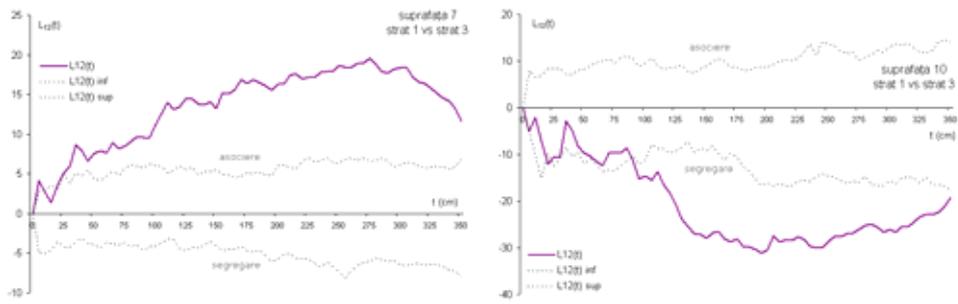

**Interacţiunea spaţială dintre stratul 2** (stratul puieţilor de dimensiuni medii) **şi stratul 3** (stratul puieţilor de mari dimensiuni)

Suprafaţa 2 are un număr de puieţi prea mic în stratul 3 pentru interpretarea rezultatelor.

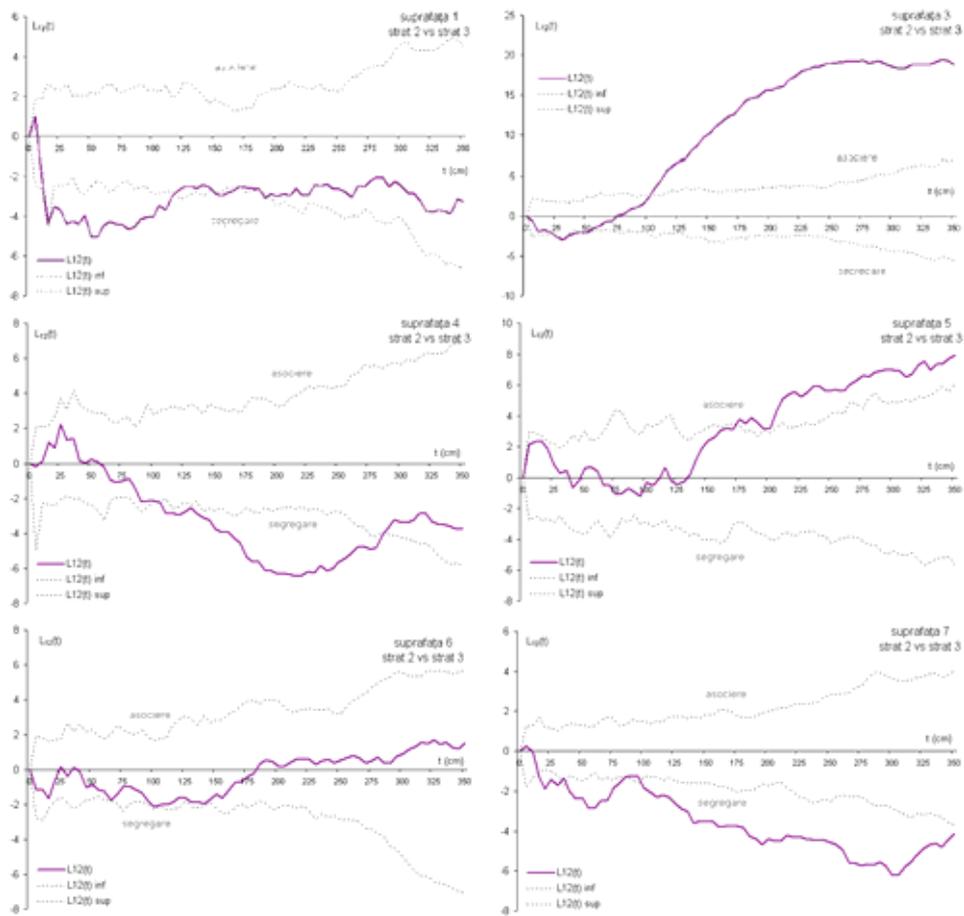

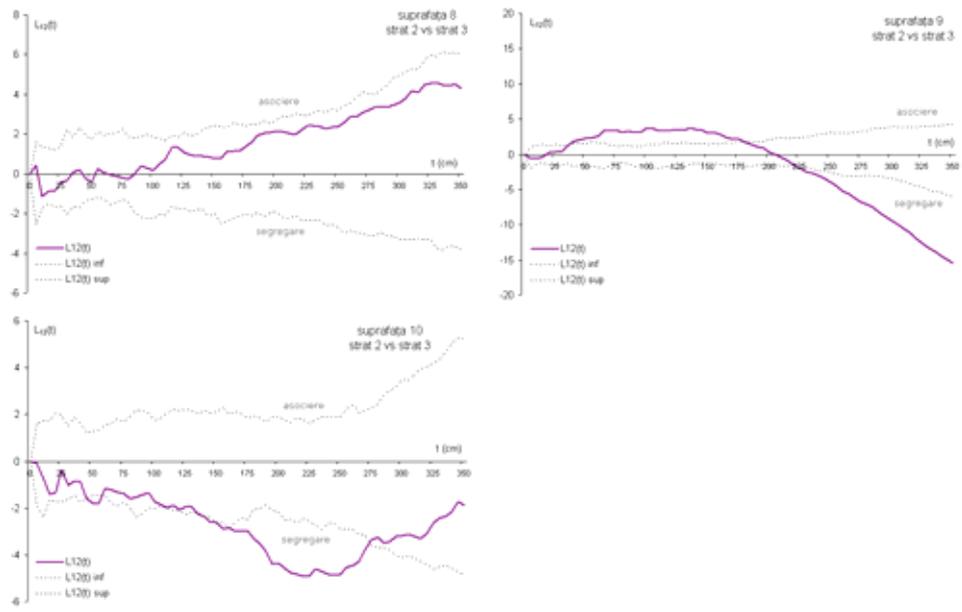

**Anexa 15. Verificarea ipotezei de asociere pozitivă/negativă a unor specii prin intermediul funcţiei $L_{12}(t)$ – analiza bivariată pentru perechile de specii: Carpen-Stejar, Frasin-Tei, Jugastru-Cireş.**

**Interacţiunea dintre carpen şi stejar**

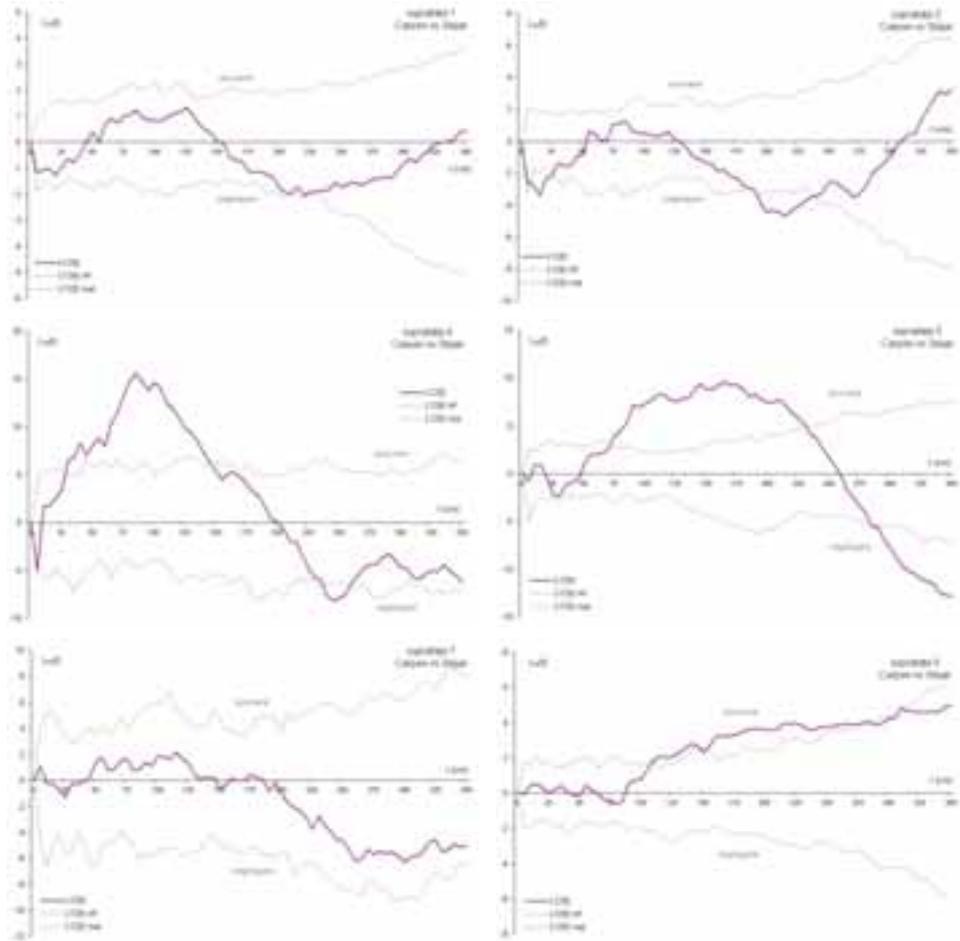

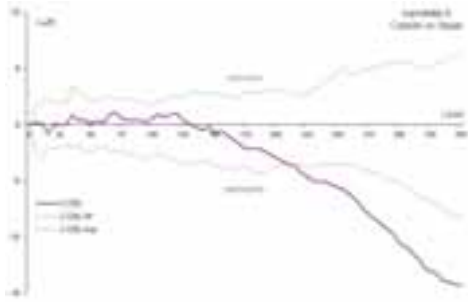

În suprafețele 3, 6, 9 nu au fost efectuate analize datorită numărului redus al puieților uneia dintre specii.

**Interacțiunea dintre frasin și tei**

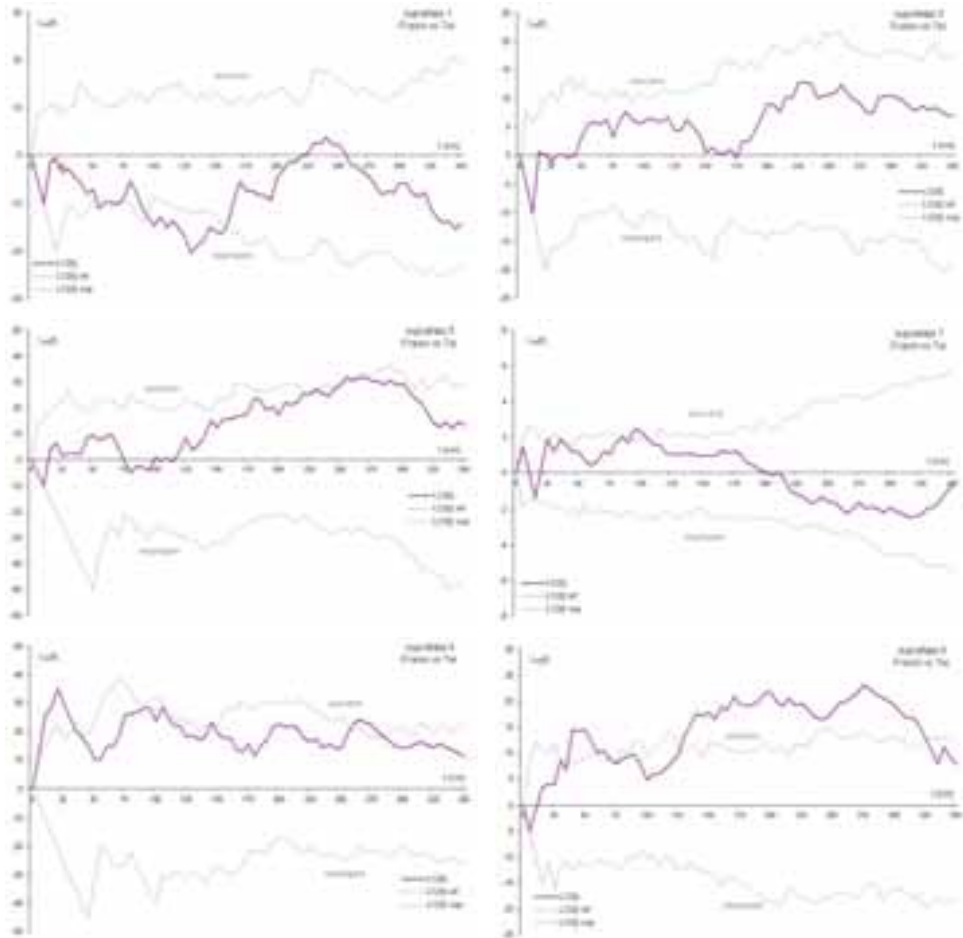

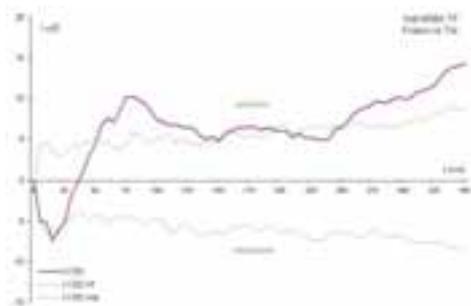

În suprafețele 2, 4, 6 nu au fost efectuate analize datorită numărului redus al puieților uneia dintre specii.

**Interacțiunea dintre jugastru și cireș**

În suprafețele 2, 3, 5, 6, 7 nu au fost efectuate analize datorită numărului redus al puieților uneia dintre specii.

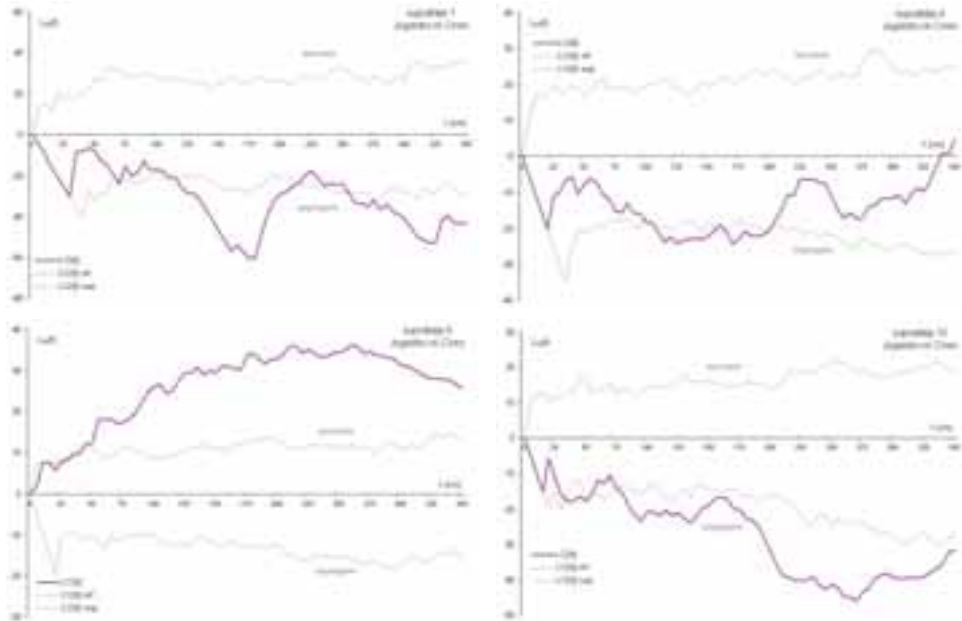

**Anexa 16 – Variația intensității concurenței în suprafețele studiate**

Intensitatea concurenței apreciată prin indicele Schutz și indicele Hegyi - varianta ce utilizează suprafața exterioară a coroanei (Hegyi_sup) și varianta ce utilizează înălțimea (Hegyi_înălțime) drept parametru de raportare. Diagramele sunt realizate cu ajutorul aplicației CARTOGRAMA.

Suprafața 1

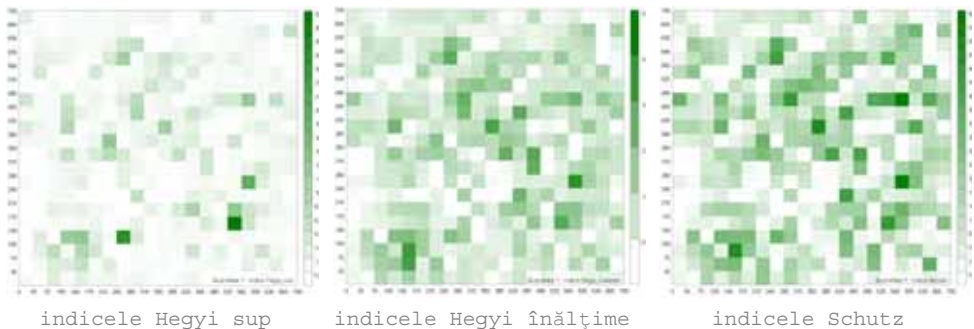

indicele Hegyi_sup          indicele Hegyi_înălțime          indicele Schutz

Suprafața 2

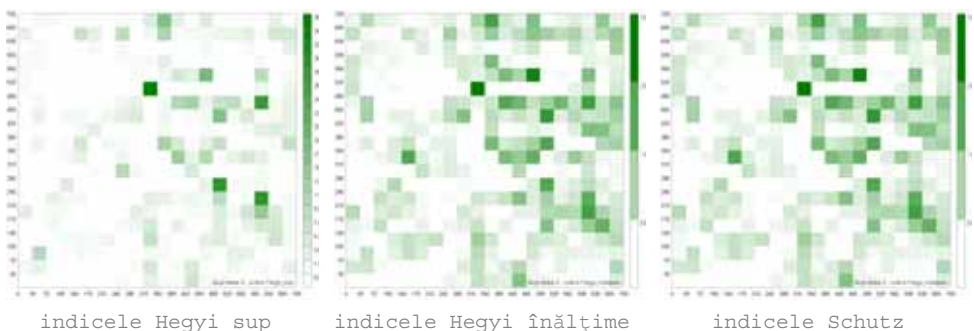

indicele Hegyi_sup          indicele Hegyi_înălțime          indicele Schutz

Suprafața 3

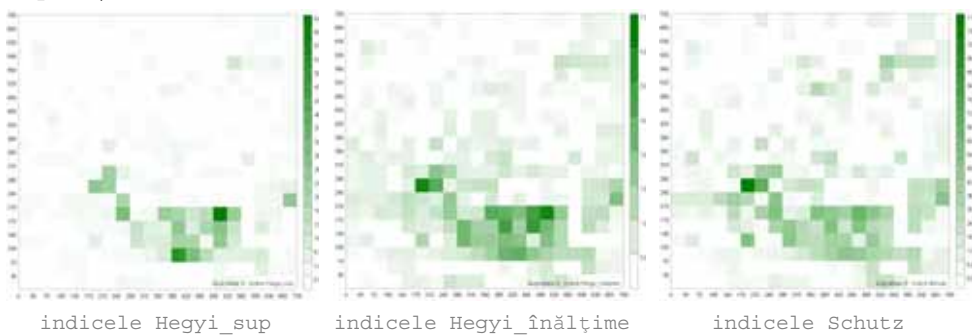

indicele Hegyi_sup          indicele Hegyi_înălțime          indicele Schutz

**Suprafaţa 4**

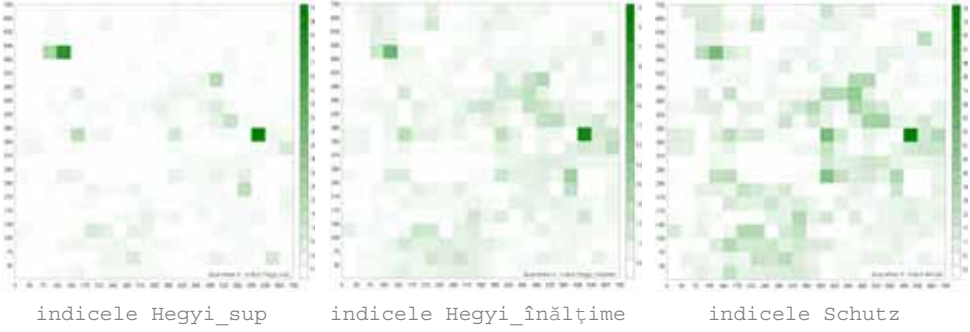

indicele Hegyi_sup          indicele Hegyi_înălţime          indicele Schutz

**Suprafaţa 5**

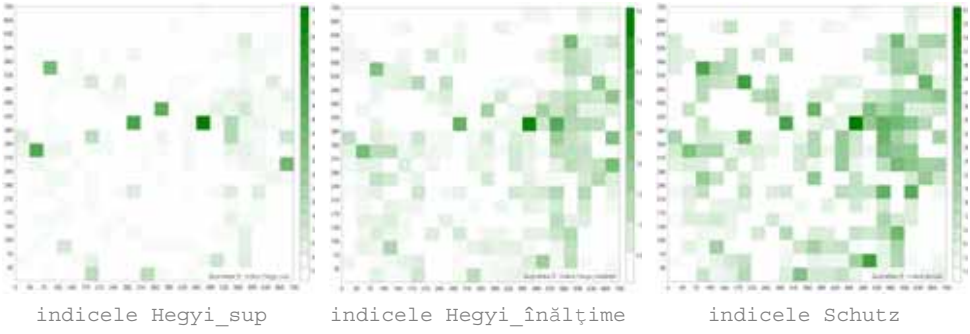

indicele Hegyi_sup          indicele Hegyi_înălţime          indicele Schutz

**Suprafaţa 6**

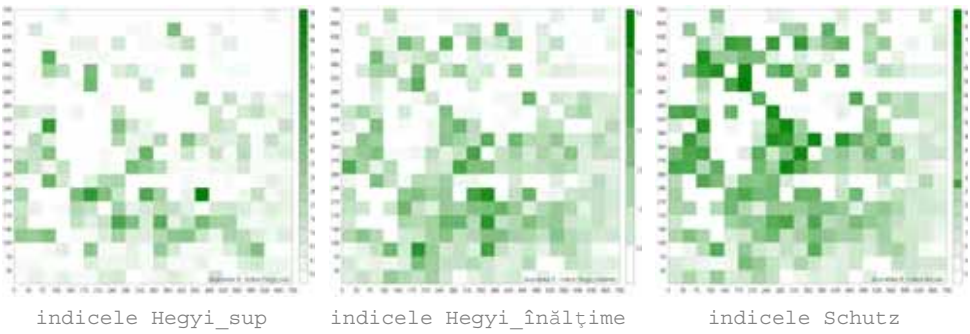

indicele Hegyi_sup          indicele Hegyi_înălţime          indicele Schutz

**Suprafaţa 7**

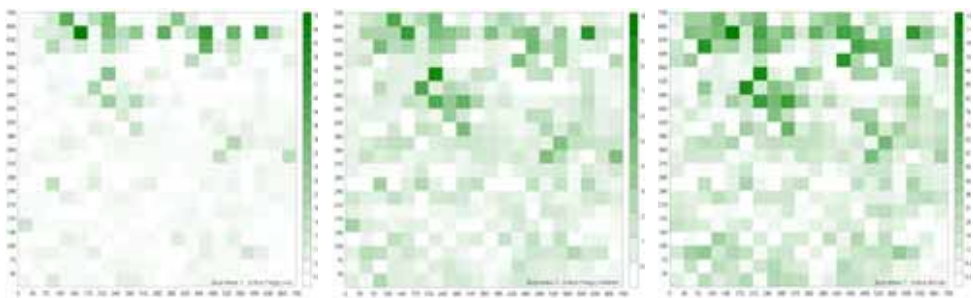

indicele Hegyi_sup      indicele Hegyi_înălţime      indicele Schutz

**Suprafaţa 8**

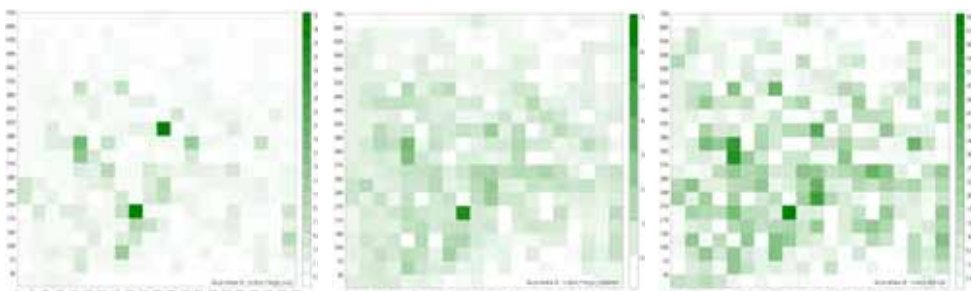

indicele Hegyi_sup      indicele Hegyi_înălţime      indicele Schutz

**Suprafaţa 9**

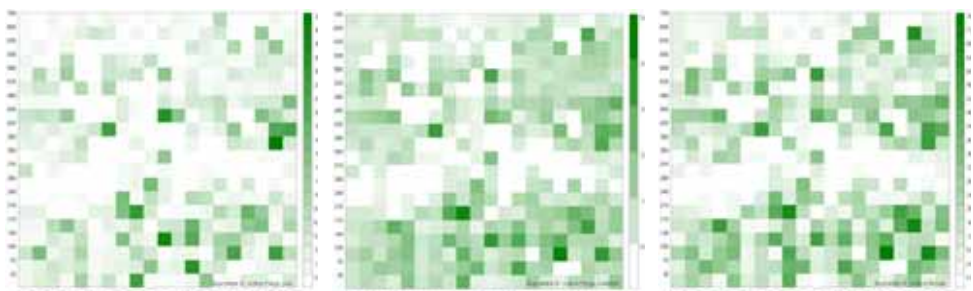

indicele Hegyi_sup      indicele Hegyi_înălţime      indicele Schutz

**Suprafaţa 10**

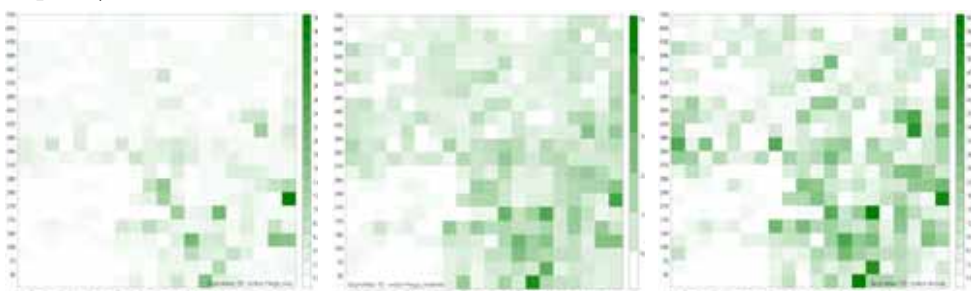

**Anexa 17 – Media şi coeficientul de variaţie al valorilor intensităţii competiţiei apreciate prin indici sintetici – situaţia pe specii şi pentru toţi puieţii**

| Indice Hegyi_sup | | | Indice Hegyi_h | | | Indice Schutz | | |
|---|---|---|---|---|---|---|---|---|
| | medie | s% | | medie | s% | | medie | s% |
| total | 5,76 | 248,37 | total | 1,42 | 110,91 | total | 18,84 | 116,99 |
| Pa | 2,05 | 187,87 | Pa | 0,66 | 86,09 | Pa | 8,09 | 108,15 |
| Ci | 2,64 | 252,77 | Ci | 1,00 | 81,72 | Ci | 12,56 | 104,86 |
| Te | 3,41 | 320,35 | Te | 1,14 | 147,55 | Te | 14,63 | 146,26 |
| St | 4,60 | 179,19 | Ca | 1,27 | 101,38 | Ca | 16,36 | 115,78 |
| Ju | 5,12 | 188,01 | Ju | 1,35 | 83,36 | Ju | 17,74 | 118,24 |
| Ca | 5,47 | 251,27 | St | 1,53 | 95,47 | St | 20,49 | 94,41 |
| Fr | 14,22 | 190,63 | Fr | 2,82 | 98,02 | Fr | 40,60 | 84,53 |

**Anexa 18 – Diagramele Voronoi generate pentru suprafețele studiate**

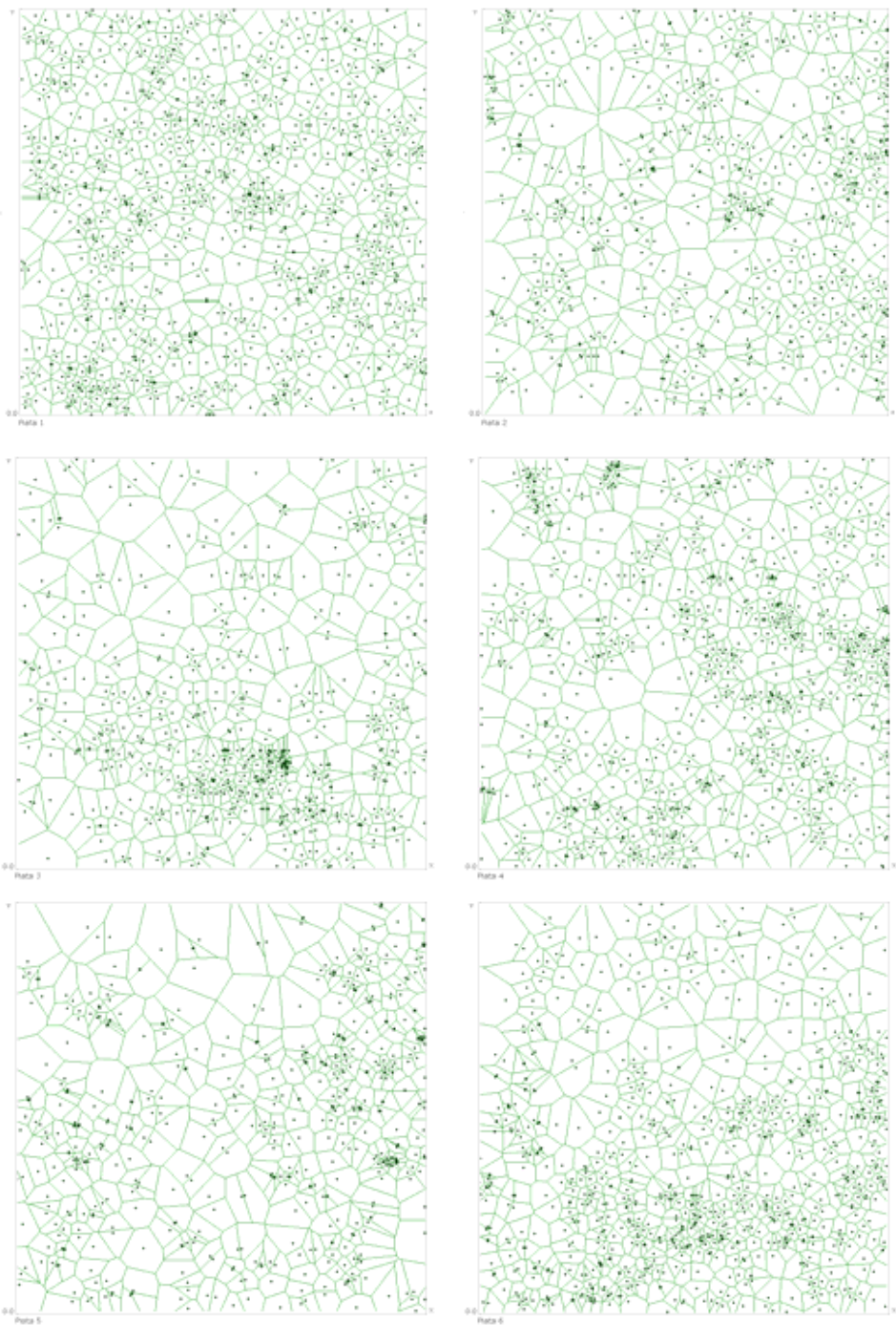

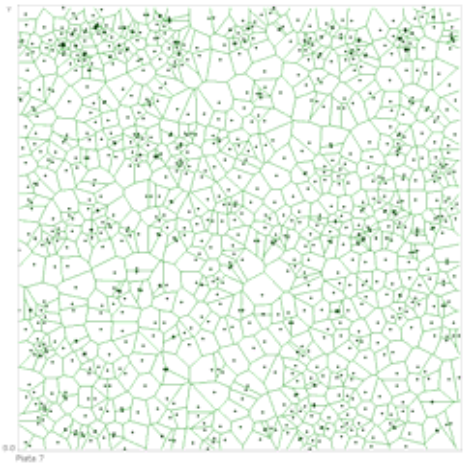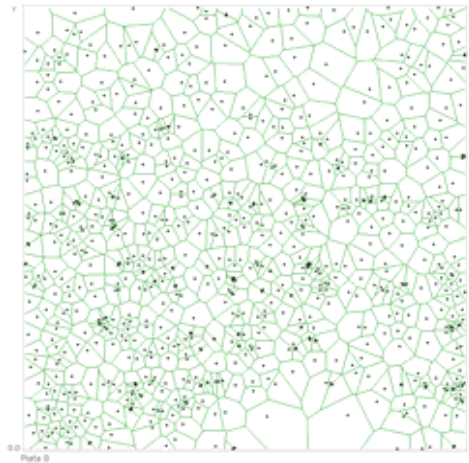

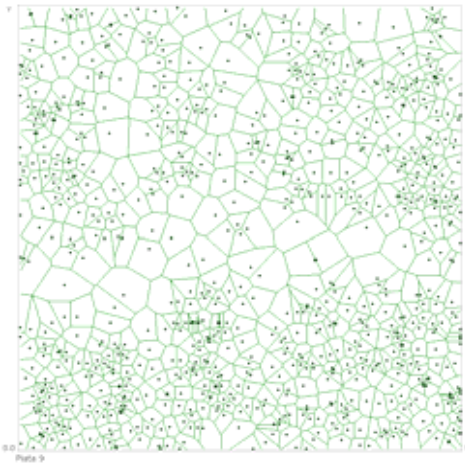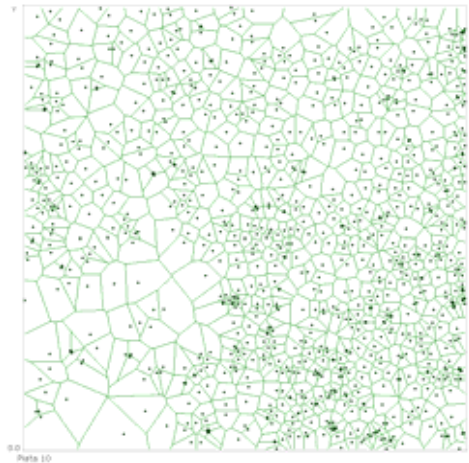

Diagramele au fost realizate cu ajutorul aplicației software proprii (VORONOI).